\documentclass[11pt]{article}
\usepackage{geometry}              
\usepackage[parfill]{parskip}    % Activate to begin paragraphs with an empty ine rather than an indent
\usepackage[dvips]{graphicx}
\usepackage{slashed}
\usepackage{amsmath}
\usepackage{amssymb}
\usepackage{verbatim}
\usepackage{url}
\usepackage{layout}
\usepackage{relsize}
\usepackage{fancyhdr}

\usepackage{slashed}
\usepackage{booktabs}
\usepackage{color}
\usepackage{mathtools}
\usepackage[
      colorlinks=true,
      linkcolor=darkblue,  
      urlcolor=blue,    
      filecolor=blue,     
      citecolor=red,
linktocpage=true,
      pdfstartview=FitV,
      bookmarksopen=true    
      ]{hyperref}

\definecolor{darkblue}{rgb}{0.2, 0, 0.8}

\newcommand{\eq}{\begin{equation}}
\newcommand{\eqe}{\end{equation}}
\newcommand{\g}{\gamma}
\newcommand{\e}{\epsilon}

\newcommand{\eqa}{\begin{eqnarray}}
\newcommand{\eqae}{\end{eqnarray}}

\newcommand{\da}{\dot{a}}
\newcommand{\db}{\dot{b}}

\newcommand{\lag}{\mathcal{L}}
\newcommand{\sgn}{\text{sgn}}

\def\a{\alpha}
\def\b{\beta}
\def\g{\gamma}
\def\c{\gamma}
\def\d{\delta}
\def\e{\epsilon}           % Also, \varepsilon
\def\g{\gamma}
\def\h{\eta}

\def\l{\lambda}
\def\m{\mu}
\def\n{\nu}

\def\th{\theta}                  %     \vartheta
\def\r{\rho}                                     %     \varrho
\def\s{\sigma}                                   %     \varsigma

\def\ga{\gamma}

\def\Tr{\mathop{\rm Tr}\nolimits}
\def\tr{\mathop{\rm tr}\nolimits}

\def\diag{\mathop{\rm diag}\nolimits}
\newcommand{\eps}{\epsilon}

\newcommand{\lra}{\leftrightarrow}

\newcommand{\be}{\begin{equation}}
\newcommand{\bea}{\begin{eqnarray}}
\newcommand{\beq}{\begin{equation}}
\newcommand{\ee}{\end{equation}}
\newcommand{\eea}{\end{eqnarray}}
\newcommand{\eeq}{\end{equation}}
\newcommand{\lsim}{\!\mathrel{\hbox{\rlap{\lower.55ex \hbox{$\sim$}} \kern-.34em \raise.4ex \hbox{$<$}}}}
\newcommand{\gsim}{\!\mathrel{\hbox{\rlap{\lower.55ex \hbox{$\sim$}} \kern-.34em \raise.4ex \hbox{$>$}}}}

\newcommand{\reef}[1]{(\ref{#1})}
\newcommand{\<}{\langle}
\renewcommand{\>}{\rangle}

\newcommand{\pa}{\partial}

\newcommand{\tQ}{\tilde{Q}}
\newcommand{\cn}{\mathcal{N}}
\newcommand{\ca}{\mathcal{A}}

\newcommand{\ds}{\displaystyle}

\def\dslash{\slash \negthinspace \negthinspace \negthinspace \negthinspace \pa}
\def\pslash{\slash \negthinspace \negthinspace \negthinspace \negthinspace  p}

\def\Dslash{\slash \negthinspace \negthinspace \negthinspace \negthinspace  D}
\def\epsslash{\slash  \negthinspace \negthinspace \negthinspace  \eps}

\newcommand{\compactsubsection}[1]{\vspace{2mm}\noindent {\bf #1}\\[1mm]}

\newcommand{\Lag}{\mathcal{L}}

\newtheorem{ex}{Exercise}[section]

\newcommand{\exercise}[2]{
\begin{itemize}
\item[$\mathlarger{\mathlarger{\blacktriangleright}}$] 
\begin{ex}\label{#1}\end{ex}\nopagebreak #2 \end{itemize}}

\newcommand{\example}[1]{\begin{itemize}
\item[$\mathlarger{\mathlarger{\mathlarger{\mathlarger{\triangleright}}}}$] {\sl Example:} #1 
$\mathlarger{\mathlarger{\mathlarger{\mathlarger{\triangleleft}}}}$\end{itemize}}

\newcommand{\refr}{r}

\begin{document}

\begin{titlepage}
\begin{flushright}
MCTP-13-21\\
\today
\end{flushright}
\vspace{1.1cm}

%%%%%%%
%%%%%%%
%%%%%%%

\begin{center}
{\bf 
{\LARGE
Scattering Amplitudes}
\\[2mm]
}
\end{center}
\vspace{4mm}
\begin{center}
{\bf Henriette Elvang and Yu-tin Huang}\\
\vspace{0.7cm}
{{\it Michigan Center for Theoretical Physics}\\
{\it Randall Laboratory of Physics, Department of Physics}\\
{\it University of Michigan, Ann Arbor, MI 48109, USA}}\\[4mm]
{\small \tt  elvang@umich.edu, yutinh@umich.edu}
\end{center}
\vspace{8mm}

\begin{abstract}
The purpose of this review is to bridge the gap between a standard course in quantum field theory and recent fascinating developments in the studies of on-shell scattering amplitudes. 
We build up the subject from basic quantum field theory, starting with Feynman rules for simple processes in Yukawa theory and QED.  
The material covered includes spinor helicity formalism, on-shell recursion relations, superamplitudes and their symmetries, twistors and momentum twistors, loops and integrands, Grassmannians, polytopes, and amplitudes in perturbative supergravity as well as 3d Chern-Simons-matter theories.
Multiple examples and exercises are included.

\vspace{20mm}
\centerline{
{\footnotesize
Part of textbook to be Published by Cambridge University Press}}

\end{abstract}

\end{titlepage}

\vspace{4mm}

\vspace{-6mm} 
\tableofcontents 
\newpage 
\setcounter{equation}{0}
\section{Introduction}
%%%%%%%%%%%%%%%%%%%%%%%%%%%%%%% 
%%%%%%%%%%%%%%%%%%%%%%%%%%%%%%% 
%%%%%%%%%%%%%%%%%%%%%%%%%%%%%%% 
 
In a traditional quantum field theory (QFT) course,  you learn to extract Feynman rules from a Lagrangian and use them to calculate a scattering amplitude ${A}$ as a sum of  Feynman diagrams organized perturbatively in the loop-expansion. 
 From the amplitude you calculate the differential cross-section, 
$\tfrac{d\s}{d\Omega}\propto |{A}|^2$, which --- if needed --- includes a suitable spin-sum average. Finally the  cross-section $\sigma$ can be found by integration of $d\s/d\Omega$ over angles, with appropriate symmetry factors included for identical final-state particles. 
The quantities $\sigma$ and $d\s/d\Omega$ are the observables of interest for particle physics experiments, but the input for computing them are the 
{\em gauge invariant on-shell scattering amplitudes} ${A}$. These on-shell amplitudes ${A}$ are the subject of this review.  

Examples of processes you have likely encountered in QFT are 
\bea
    \nonumber
    \text{Compton scattering}&& e^- + \gamma ~\to~   e^- + \gamma \,,
    \\[2mm]
    \text{M\o ller scattering}&& e^- + e^- ~\to~   e^- + e^- \,,
    \\[2mm]
    \nonumber
    \text{Bhabha scattering}&&  e^- + e^+ ~\to~   e^- + e^+\,,
\eea
and perhaps also $2 \to 2$ gluon scattering
\bea
  g + g ~\to~  g + g \,.
\eea
For instance, starting from the Quantum Electrodynamics (QED) Lagrangian you may have calculated the tree-level 
differential cross-section for Bhabha-scattering. It is typical for such a calculation that the starting point --- the Lagrangian in its most compact form --- is not too terribly complicated. And the final result can be rather compact and simple too. But the intermediate stages of the calculation often explode in an inferno of indices, contracted up-and-down and in all directions --- providing little insight of the physics and  hardly any hint of simplicity. 

Thus, while you think back at your QFT course as a class in which (hopefully!) you did a lot of long character-building calculations, you will also note that you were probably never asked to use Feynman diagrams to calculate processes that involved more than four or five particles, even at tree level: for example, 
$e^- + e^+ ~\to~   e^- + e^+ + \gamma$ \, or \, $g + g ~\to~  g + g + g$. Why not? Well, one reason is that the number of Feynman diagrams tends to grow fast with the number of particles involved: for gluon scattering at {\em tree level} we have 
\bea
  \begin{array}{lr}
  g + g ~\to~  g + g  &~~ \text{4 diagrams} \\
  g + g ~\to~  g + g + g &~~ \text{25 diagrams} \\
  g + g ~\to~  g + g  + g + g &~~ \text{220 diagrams}
  \end{array}
\eea
and for $g+ g \to 8 g$ you need more than one million diagrams \cite{Mangano:1990by}. 
Another important point is that each diagram gets significantly more complicated as the number of external particles grows.
So the reason you have not been asked to calculate the above multi-gluon processes from Feynman diagrams is that it would be awful, un-insightful, and in many cases impossible.\footnote{Using computers to do the calculation can of course be very helpful, but not in all cases. Sometimes numerical evaluation of Feynman diagrams is simply so slow that it is not realistic to do. 
Moreover, given that there are poles that can cancel between diagrams, big numerical errors can arise in this type of evaluation. Therefore compact expressions for the amplitudes are very useful in practical applications.} 

It turns out that despite the complications of the Feynman diagrams, the on-shell scattering amplitudes for multi-gluon processes can actually be written as remarkably simple expressions. This raises the questions: ``why are the on-shell amplitudes so simple?" and ``isn't there a better way to calculate amplitudes?". These are questions that have been explored in recent years and a lot of progress has been made on improving calculational techniques and gaining insight into the underlying mathematical structure.  Some of the keywords are
\begin{enumerate}
\item spinor helicity formalism
\item on-shell recursion relations (BCFW, CSW, all-line shifts,\dots)
\item on-shell superspace, superamplitudes, Ward identities
\item generalized unitarity, maximal cuts
\item dual superconformal symmetry and the Yangian 
\item twistors, zone-variables, momentum twistors
\item Leading Singularities 
\raisebox{-7mm}{\includegraphics[scale=0.3]{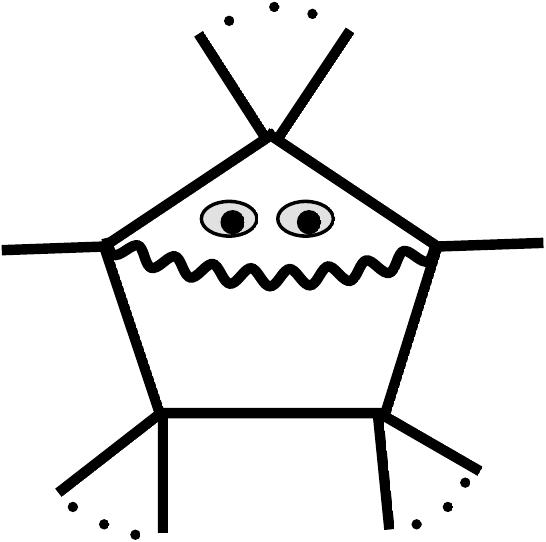}}
and on-shell blob-diagrams 
\raisebox{-7mm}{\includegraphics[scale=0.23]{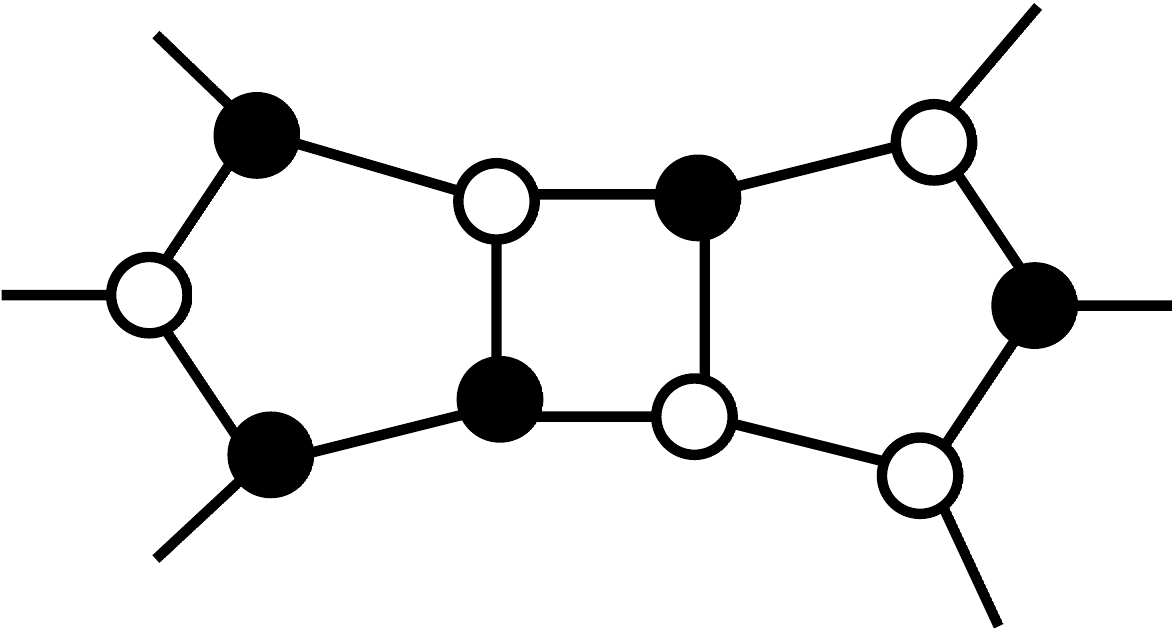}}
\item the Grassmannian, polytopes, and mathematicians
\item gravity = (gauge theory)${}^2$, KLT relations, BCJ relations
\end{enumerate}
and much more. 

The study of these methods may suggest a paradigm that can be phrased loosely as ``avoiding the (full) Lagrangian" with all its ambiguities of field redefinitions and gauge choices, and instead focus on how kinematics, symmetries, and locality impact the physical observables. Or, more strongly, we may ask if the hints from the simplicity of on-shell amplitudes allow us to find another approach to perturbative quantum field theory: one might hope for a novel formulation that captures the physics of the full perturbative S-matrix. Such a new formulation could make amplitude calculations much more efficient and one could hope that it would lead to new insights even beyond amplitudes, for example for correlation functions of gauge invariant operators and perhaps even for non-perturbative physics. 

But we are getting ahead of ourselves. The purpose of this review is to provide a practical introduction to {\em some} on-shell methods, taking as a starting point what you know after a first introductory course on quantum field theory. Indeed, much of that material in Sections 2 and 3 could be part of any modern course on quantum field theory, but as it is generally not, we hope you will find this presentation useful. We will also provide a survey, with selected details, of some of the ideas behind the keywords above; this should give you the basis for starting to pursue more advanced topics in the field and work on research projects. 

One should keep in mind that the subject of scattering amplitudes has two main motivations. 
One is practical application in particle physics: some of the on-shell methods that you learn about here are indeed already implemented in numerical codes for processes relevant in particle physics experiments. The other motivation is the fascinating internal mathematical beauty of the subject. The physical relevance and mathematical structure are both important, neither should be underestimated. They complement and benefit each other. 

The style of the presentation in this review is detailed and concrete, so that you can learn the tools. The starting point is Lagrangians and Feynman rules, and we build up the subject from there. The purpose is to be pedagogical --- but in this as well as other matters, there is no substitute for getting your own hands dirty. Therefore you'll find many exercises scattered throughout the text. Do them. It is fun.

\vspace{3mm}
\compactsubsection{Conventions}
The subject of amplitudes is often viewed as quite technical and notationally intense. We will try to avoid a long deadly-boring introduction about  $\gamma$-matrix conventions and about which indices go up and down and who is dotted and who is not. Suffice it here to say that we work in 4 dimensions (except in Sections \ref{s:Dne4}-\ref{s:BCJ}), our metric convention is mostly-plus $\eta_{\m\n} = \text{diag}(-1,+1,+1,+1)$, and 
 we  follow the spinor- and Clifford algebra conventions in Srednicki's QFT textbook \cite{MSqft}. For easy access, and to make our presentation reasonably self-contained, some conventions are collected in the short Appendix \ref{app:conv}. Appendix \ref{app:twistor} outlines the embedding formalism for twistors.

%%%%%%%%%%%%%%%%%%%%%%%%%%%%%%% 
%%%%%%%%%%%%%%%%%%%%%%%%%%%%%%% 
%%%%%%%%%%%%%%%%%%%%%%%%%%%%%%% 
\newpage
\vspace{3mm}
\compactsubsection{Acknowledgements}
It is a pleasure to thank our friends and collaborators who have worked with us and helped us learn the subject of scattering amplitudes:  Ratin Akhoury, Nima Arkani-Hamed, Zvi Bern, Freddy Cachazo, John Joseph Carrasco, Simon Caron-Huot, Tim Cohen, Scott Davies, Tristan Dennen,  Lance Dixon, Dan Freedman, David Kosower, Johannes Henn, Harald Ita, Henrik Johansson, Michael Kiermaier, Sangmin Lee, Arthur Lipstein, Thomas Lam, David McGady, Cheng Peng, Jan Plefka, Radu Roiban, Mark Srednicki, Warren Siegel, David Speyer, and Jaroslav Trnka. 

A few people have suffered early drafts of this book and we are indebted to them for their helpful comments/suggestions/corrections: Cindy Keeler, Tim Olson, Sam Roland,  David Speyer, Sri Suresh, Jonathan Walsh, John Ware, and the Harvard amplitudes reading group (communications from Marat Freytsis). Their careful readings caught multiple typos and helped us improve the presentation. We would also like to thank Michael Enciso, Karol Kampf, Joe Minahan, and Stefan Theisen for feedback on the manuscript. 

H.E.~is supported by NSF CAREER Grant PHY-0953232. 
She is also a Cottrell Scholar of the Research Corporation for Science Advancement. H.E.~is grateful for the hospitality offered by Stanford/SLAC during her visit in February/March 2013 and KITP/UCSB during January-March 2014. 

%%%%%%%%%%%%%%%%%%%%%%%%%%%%%%% 
%%%%%%%%%%%%%%%%%%%%%%%%%%%%%%% 
%%%%%%%%%%%%%%%%%%%%%%%%%%%%%%% 
\newpage
\setcounter{equation}{0}
\section{Spinor helicity formalism}
\label{s:sh}
%%%%%%%%%%%%%%%%%%%%%%%%%%%%%%% 
%%%%%%%%%%%%%%%%%%%%%%%%%%%%%%% 
%%%%%%%%%%%%%%%%%%%%%%%%%%%%%%% 
We are going to introduce the spinor helicity formalism in the context of the basic Feynman rules that you are familiar with from Yukawa interactions and QED. So we start with Dirac spinors and build up the formalism based on simple scattering problems.
\subsection{Dirac spinors}
The Lagrangian for a free massive 4-component Dirac field $\Psi$ is
\be
 \Lag = i \overline{\Psi} \gamma^\mu  \pa_\m \Psi - m \overline{\Psi} \Psi \, .
\ee
Our conventions for the Dirac conjugate $\overline{\Psi}$ and the $\gamma^\m$'s can be found in Appendix \ref{app:conv}.
The equation of motion for $\overline{\Psi}$ gives the Dirac equation
\be
  \label{DiracEq}
  (-i \,\dslash + m ) \Psi = 0 \,.
\ee
As you have seen in your QFT class, multiplying  the Dirac equation by 
$(i \,\dslash + m )$ gives the Klein-Gordon equation, $(-\pa^2 + m^2 ) \Psi = 0$.
It is solved by a plane-wave expansion 
\be
  \Psi(x) ~\sim~ u(p)\,e^{ip.x} + v(p)\,e^{-ip.x} 
\ee
provided $p^2 \equiv p^\m p_\m = - m^2$. This $\Psi(x)$ will also solve the Dirac equation \reef{DiracEq}  if
\be
  \label{DiracEqP}
  (\,\pslash + m)u(p) = 0\,
  ~~~~\text{and}~~~~~
  (-\pslash + m)v(p) = 0\,.
\ee
These are the momentum space form of the Dirac equation. 
Each of the equations in \reef{DiracEqP} has two independent solutions which we will label by a subscript $s=\pm$.  We can now write the general free field expansion of $\Psi$ as 
\be
  \Psi(x) 
  =\sum_{s=\pm} \int \widetilde{dp} 
  \,
  \Big[
     b_s(p) \, u_s(p) \,e^{ip.x}
     + 
     d^\dagger_s(p) \, v_s(p) \,e^{-ip.x}
  \Big]
  \,,
\ee
where $\widetilde{dp} = \frac{d^3p}{(2\pi)^3\, 2 E_p}$ is the 3d Lorentz-invariant momentum measure. 
For $\overline{\Psi}$ one finds a similar result involving $d_\pm(p)$ and $b_\pm^\dagger(p)$.  

When the field is quantized, $b^{(\dagger)}_\pm(p)$ and $d^{(\dagger)}_\pm(p)$ will be fermionic creation and annihilation operators. They take care of providing the Grassmann nature of $\Psi(x)$, so that $u_\pm(p)$ and $v_\pm(p)$ are  {\em commuting} 4-component spinors that solve \reef{DiracEqP}.

Typically the next step is to define the vacuum $|0\>$ such that 
$b_\pm(p)|0\> = d_\pm(p)|0\> = 0$. One-particle states are then defined as $|p;\pm\> \equiv d^\dagger_\pm(p) |0\>$ etc. As you have seen in your QFT course, this leads to the Feynman rules for external fermions, namely that they come equipped with wavefunctions $v_\pm(p)$ for an outgoing anti-fermion (e.g.~$e^+$) and (from the expansion of $\overline{\Psi}$) $\overline{u}_\pm(p)$ for an outgoing fermion (e.g.~$e^-$). We can choose a basis such that in the rest-frame $u_\pm$ and $v_\pm$ are eigenstates of the $z$-component of the spin-matrix; then $\pm$ denotes spin up/down along the $z$-axis. For massless fermions, $\pm$ denotes the {\bf \em heliticy}, which is the projection of the spin along the momentum of the particle. It will be our interest here to study the wavefunctions $\overline{u}_\pm(p)$ and $v_\pm(p)$ further. 

The wave function $v_\pm(p)$  solves the Dirac equation \reef{DiracEqP} and  $\overline{u}_\pm(p)$ satisfies $\overline{u}_\pm(p) (\,\pslash +m) = 0$. 
Starting with a momentum 4-vector $p^\mu = (p^0,p^i) = (E,p^i)$ with $ p^\m p_\m = - m^2$, let us use the gamma-matrix conventions \reef{gammamatrices} in  Appendix \ref{app:conv} 
to write
\be
  \label{pslashdef}
  \pslash = 
  \left( 
     \begin{array}{cc}
        0 & p_{a\dot{b}} \\
        p^{\dot{a}b} & 0 
      \end{array}
   \right) \, ,
\ee
with 
\be
  p_{a\dot{b}} ~\equiv~ p_\m\, (\sigma^\mu)_{a \dot{b}} 
  = 
    \left(
    \begin{array}{cc}
      -p^0 + p^3 & p^1 - i p^2 \\
      p^1 + i p^2 & - p^0 - p^3 \\
    \end{array}
  \right)\,,
\ee
and similarly $p^{\dot{a}b} ~\equiv~ p_\m\, (\bar{\sigma}^\mu)^{\dot{a} b}$. 
We have $\sigma^\m = (1,\sigma^i)$ and $\bar\sigma^\m = (1,-\sigma^i)$ with $\sigma^{1,2,3}$ the usual Pauli-matrices \reef{pauliM}. 
The momentum bi-spinors $p_{a\dot{b}}$ and $p^{\dot{a}b}$ can be thought of as 
2$\times$2 matrices. The determinant is Lorentz-invariant,
\be
  \det p = - p^\m p_\m = m^2 \,.
\ee
In most of this review, we study scattering processes for massless particles. You can think of this as the high-energy scattering limit in which the fermion mass can be neglected. So let us now specialize to the case of massless spinors. 

%%%%%%%
\subsection{Spinor helicity formalism}
\label{s:spinhel}
When $m=0$, the Dirac equation for the wavefunction 4-component spinors reads
\be
  \label{dirac}
  \slashed{p}\, v_\pm(p) = 0\, ,~~~~~~~~~
  \bar{u}_\pm(p)\,\slashed{p} = 0 \,.
\ee
We focus on $v_\pm(p)$ and $\overline{u}_\pm(p)$ as the wave functions associated with an {\bf \em outgoing} anti-fermion and fermions. As mentioned above, in the massless case, we can choose a basis such that the subscript $\pm$ indicates the helicity $h = \pm 1/2$. Crossing symmetry exchanges (incoming $\lra$ outgoing),  (fermions $\lra$ antifermions),  and flips the sign of the helicity, so in the massless case the wavefunctions are related as $u_\pm = v_{\mp}$ and $\overline{v}_{\pm}=\overline{u}_{\mp}$.

We write the two independent solutions to the Dirac equation \reef{dirac} as
\be \label{sps1}
  v_+(p) = 
  \left( 
  \begin{array}{c}
    |p]_a \\[1mm]  
    0
  \end{array}
  \right) ,~~~~~~~~
  v_-(p) = 
  \left( 
  \begin{array}{c}
    0 \\[1mm]  
    |p\>^{\dot{a}}
  \end{array}
  \right) ,
\ee
and 
\be
  \label{sps2}
  \overline{u}_-(p) ~=~  \big( \, 0 \, , \,\< p |_{\dot{a}} \big)\, ,
  ~~~~~~~~
  \overline{u}_+(p) ~=~
  \big(\, [ p |^{a} \, ,\, 0 \,\big) \, .
\ee
The angle and square spinors are $2$-component commuting spinors (think 2-component vectors) written in a very convenient Dirac bra-ket notation. By virtue of \reef{pslashdef} and \reef{dirac}, they satisfy the massless Weyl equation,
\be
  p^{\da b} |p]_b = 0\, , ~~~~~
  p_{a \db} |p\>^{\db} = 0 \,, ~~~~~
  [p|^b \,p_{b \da} = 0\, , ~~~~~
  \<p|_{\db} \,p^{\db a}  = 0\, . 
\ee
Raising and lowering their indices is business as usual:
\be
  \label{rlp}
  [p|^a \,=\, \eps^{ab} |p]_b \, ,~~~~~~
  |p\>^{\da} \,=\, \eps^{\da\db}  \<p|_{\db}\, .
\ee
The 2-index Levi-Civitas are defined in \reef{epsies}.

Now 
\begin{quote}
{\em The angle and square spinors are the core of what we call the \\[2mm]
\centerline{\hspace{-2.5cm}{\bf \em``spinor helicity formalism"}.} \\[2mm]
As you see, these bra-kets are nothing to be scared of: there are simply 2-component commuting spinors that solve the massless Weyl equation.}
\end{quote}

\vspace{1mm}
It is one of the powers of the spinor helicity formalism that we do not need to find explicit representations for the angle and square-spinors; we can simply work abstractly with $|p\>$ and $|p]$ and later relate the results to the momentum vectors. We'll see examples of how this works in this section. 
Let us now note a couple of properties of the spinor bra-kets.

{\bf Angle vs.~square spinors: reality conditions.}
%Let us now discuss the relation between the angle and square spinors. 
The spinor field $\overline{\Psi}$ is the Dirac conjugate of $\Psi$. Applying Dirac conjugation to the momentum space Dirac equations \reef{dirac}, we find that 
$\bar{u}_\mp = \bar{v}_\pm$ is related to $v_{\pm}$ via this conjugation \emph{provided the momentum $p^\m$ is real-valued}, i.e.~the components of $p^\mu$ are real numbers.  Thus for \emph{real} momenta 
\be
  \label{p-reality}
   p^\m \text{ real}:~~~~~~ [p|^a = (|p\>^{\da})^*~~~~~  \text{and}~~~~~ \<p|_{\da} = (|p]_a)^*. 
\ee
On the contrary, for \emph{complex-valued momenta} $p^\mu$, the angle and square spinors are independent.\footnote{With complex momenta, the angle and square spinors are independent although their little group scaling (see Section \ref{s:littlegrp}) is coupled. 
In another approach, one can keep $p^\m$ real and change the  spacetime signature to $(-,+,-,+)$; in that case, the angle and square spinors are  real and  independent.} It may not sound very physical to take $p^\m$ complex, but it is a very very very useful trick to do so. We'll see this repeatedly.

{\bf Spinor completeness relation.}
The spin-sum completeness relation with $m=0$ reads  
$u_- \overline{u}_- + u_+ \overline{u}_+ =  - \slashed{p}$. (See for example (38.23) of \cite{MSqft}.) 
With the help of crossing symmetry $\overline{u}_\mp = \overline{v}_\pm$, this can be written
in spinor helicity notation as
\be
 \label{slashp}
  - \slashed{p} ~=~ | p \> [ p |  + | p ] \< p |\, .
\ee
There is a small abuse of notation in writing \reef{slashp}: the LHS is a $4\times 4$ matrix and the RHS involves products of 2-component spinors. The relation should be read in terms of matching the appropriate L- and R-spinor indices via \reef{pslashdef}, viz.
\be
  \label{ppp}
  p_{a \db}~=~ -|p]_a\, \<p|_{\db} \, ,~~~~~~~
  p^{\da b}~=~- |p\>^{\da}\, [p|^{b}  \, .~~~~~~~
\ee
The relations \reef{ppp} may look new but they should not shock you. After all, it is taught in some algebra classes that if a $2\times 2$ matrix has vanishing determinant, it can be written as a product of two 2-component vectors, say $\lambda_a$ and $\tilde{\lambda}_{\db}$: i.e.~$\det{p}=0 ~\Leftrightarrow~ p_{a \db}~=~ -\lambda_a\, \tilde{\lambda}_{\raisebox{0pt}{{$\scriptstyle \db$}}}$. In fact, this is often the starting point of introductions to the spinor helicity formalism. 
In this presentation, we will suppress the $\lambda_a$ and $\tilde{\lambda}_{\db}$ notation in favor of the more intuitive Dirac bra-kets, $\lambda_a \to |p]_a$ and $\tilde{\lambda}_{\dot{a}} \to \<p|_{\dot{a}}$.

It is useful for keeping your feet on the ground to work out an explicit solution for $|p\>$ and $|p]$ for a given 4-momentum $p^\m$. The following exercise guides you to do just that.
\exercise{ex:one}{Consider the momentum vector
\be
  p^\m 
  = (E, ~E\, \sin\th \cos\phi , ~E\, \sin\th \sin\phi,~ E\, \cos\th) \, .
   \label{p}
\ee
Express $p_{a \db}$ and $p^{\da b}$ 
in terms of $E$, $\sin\frac{\th}{2}$, $\cos\frac{\th}{2}$ and 
$e^{\pm i \phi}$.
 
Show that the helicity spinor $ |p\>^{\dot{a}} = \sqrt{2E} 
  \left( \!\!
  \begin{array}{c}
  \cos\frac{\th}{2}\\
    \sin\frac{\th}{2} \, e^{i \phi}       
  \end{array}\!\!
  \right) $
solves the massless Weyl equation.  
Find  expressions for the spinors 
$\<p|_{\da}$, $|p]_a$, and $[p|^{a}$ and check that they satisfy $p_{a\db} = -|p]_a\<p|_{\db}$
and  
$p^{\da b} =- |p\>^{\da} [p|^{b}$.
}
You  have probably noted that the angle and square spinors are only defined up to an overall scaling that leaves $p^\mu$ invariant. This is called the {\em little group scaling} and it plays a central role which we explore much more in Section \ref{s:littlegrp}.

\vspace{1.5mm}
We are now in dire need of some examples! Before we move ahead, it is convenient to summarize the external line {\bf \em Feynman rules for outgoing massless (anti)fermions}:
\begin{itemize}
  \item Outgoing fermion with $h=+1/2$: ~$\overline{u}_+$  
  ~$\longleftrightarrow$  
   ~$\big(\, [ p |^{a} \, ,\, 0 \,\big)$\\[-2mm]
  \item Outgoing fermion with $h=-1/2$: ~$\overline{u}_-$  
  ~$\longleftrightarrow$  
 ~$\big( \, 0 \, , \,\< p |_{\dot{a}} \big)$\\[-2mm]
  \item Outgoing anti-fermion with $h=+1/2$: ~$v_+$  
  ~$\longleftrightarrow$ 
  ~$\left( 
  \begin{array}{c}
    \!\! |p]_a \!\!\! \\[1mm]  
    \!\!\!0\!\!\!
  \end{array}
  \right)$ 
  \item Outgoing anti-fermion with $h=-1/2$: ~$v_-$  
  ~$\longleftrightarrow$  
  ~$\left( 
  \begin{array}{c}
    \!\!0 \!\!\! \\[1mm]
    \!\!|p\>^{\dot{a}}\!\!\!
  \end{array}
  \right)$
\end{itemize}  
Note the useful mnemonic rule that \emph{positive helicity} of an outgoing particle is associated with \emph{square spinors} while \underline{negative helicity} comes with \underline{angle-spinors}. 
Finally, let us comment that for massless fermions we usually don't bother much to distinguish fermion-anti-fermion due to the simple crossing rules. 
In the amplitudes, we will consider all the external particles to be outgoing, so think of the rules here as the difference between the arrow on a fermion line pointing into the diagram (anti-fermion) or out of the diagram (fermion). 
\exercise{}{The helicity of a massless particle is the projection of the spin along the momentum 3-vector $\vec{p}$, so the helicity operator can be written 
$\Sigma = \mathcal{S}\cdot \vec{p}/|\vec{p}|$, where the spin 
$\mathcal{S}_i = \tfrac{1}{2} \eps_{ijk} S^{jk}$ ($i,j,k=1,2,3$) is defined by the spin matrix
$S^{\m\n} = \tfrac{i}{4}[\ga^\m,\ga^\n]$. For simplicity, you can pick a frame where $p^\m$ is along the $z$-axis. Use the results of Exercise \ref{ex:one} to show that the chiral basis \reef{sps1}-\reef{sps2} is also a helicity basis, \mbox{i.e.~show that $\Sigma v_\pm = - h_\pm v_\pm$ for 
$h_\pm = \pm \tfrac{1}{2}$.}
}

%%%%%%%%%%%%%
\subsection{Examples from Yukawa theory}
\label{s:yukawa}
Consider a Dirac fermion interacting with a real scalar $\phi$ via a Yukawa coupling:
\be
  \label{LAGyukawa}
  \lag = i \overline{\Psi} \gamma^\mu \pa_\mu\Psi
     - \tfrac{1}{2} (\pa \phi)^2 + g \phi \overline{\Psi}\Psi\,.
\ee
The interaction term gives the simple 3-vertex Feynman rule $ig$. For a diagram with two outgoing Dirac fermions connecting to the rest of the particles in the process via an internal scalar line, the usual Feynman rules give
\be
  \raisebox{-5mm}{\includegraphics[width=3.4cm]{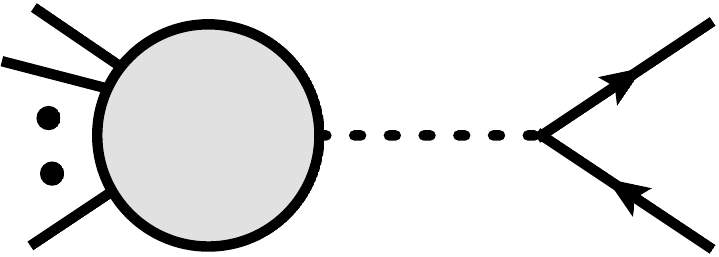}}
  ~=~ 
  ig\, \overline{u}_{h_1}(p_1) v_{h_2}(p_2)\times \frac{-i\,\,}{(p_1+p_2)^2} \times\text{(rest)}
    \label{Yukex1}
\ee
with the spinor indices contracted and the gray blob representing the rest of the diagram. We focus on the spinor product: choosing specific examples for the helicities we find
\bea 
  \overline{u}_{+}(p_1) v_{-}(p_2) 
  &=& \big(\, [ 1 |^{a} \, ,\, 0 \,\big) 
  \left( 
  \begin{array}{c}
    \!\!0 \!\!\! \\
    \!\!|2\>^{\dot{a}}\!\!\!
  \end{array}
  \right)
  ~=~ 0\, \\[1mm]
  \overline{u}_{-}(p_1) v_{-}(p_2) 
  &=& \big( \, 0 \, , \,\< 1 |_{\dot{a}} \big) 
  \left( 
  \begin{array}{c}
    \!\!0 \!\!\! \\
    \!\!|2\>^{\dot{a}}\!\!\!
  \end{array}
  \right)
  ~=~ 
  \<1|_{\dot{a}} |2\>^{\dot{a}}
  ~\equiv~   
  \<12\> \,.
\eea
Thus in the first case, the diagram vanishes. In the second case,
we introduced the {\bf \em angle spinor bracket} $\<12\>$. Together with its best friend, the {\bf  \em square spinor bracket} $[12]$, it is a key ingredient for writing amplitudes in spinor helicity formalism. So let us introduce the spinor brackets properly: for two lightlike vectors $p^\m$ and $q^\mu$, we define spinor brackets
\be
  \< p \, q \> \,=\,{\<p|}_{\da}\, |q\>^{\da} \, , \hspace{1.5cm}
  [ p \, q ]  \,=\, [p|^{a} \,|q]_a \, .
\ee
Since indices are raised/lowered with the antisymmetric Levi-Civitas \reef{epsies}, cf.~\reef{rlp}, these products are antisymmetric:
\be
   \< p \, q \> =  -\< q \, p \>\,,
   ~~~~~~
   [ p \, q ] = - [ q \, p ] \,.
\ee
All other ``bra-kets'' vanish, e.g.~  $\< p | q ] =0$. 

For real momenta, the spinor products satisfy $[p\, q]^* = \< q\, p \>$.

It is a good exercise (use \reef{Trss}) to derive the following important relation:
\be
  \label{angxsq}
  \< p \, q \>\, [ p \, q ] ~=~  2 \, p \cdot q \, ~=~ (p+q)^2 \,.
\ee
In amplitudes with momenta $p_1, p_2,\dots$ we use the short-hand notation $|1\> = |p_1\>$ etc.
Applying \reef{angxsq} to our Yukawa example above, we find that \reef{Yukex1} gives
\be 
g \<12\>\times \frac{1}{2 p_1.p_2} \times \text{(rest)}
= g \<12\>\times \frac{1}{\<12\>[12]} \times \text{(rest)}
= g  \frac{1}{[12]} \times \text{(rest)} \,.
\ee
The cancellation of the $\<12\>$-factors is the first tiny indication of  simplifications that await us in the following.

\example{Let us calculate the 4-fermion tree amplitude 
$A_4(\bar{f}^{h_1}f^{h_2} \bar{f}^{h_3}f^{h_4})$ in Yukawa theory. Our notation is that $f$ denotes an outgoing fermion and $\bar{f}$ an outgoing anti-fermion. The superscripts indicate the helicity. When specifying the helicity of each particle, we call the amplitude a {\bf \em helicity amplitude}. 

The $s$-channel diagram for the 4-fermion process is
\be
    \raisebox{-5mm}{\includegraphics[width=3cm]{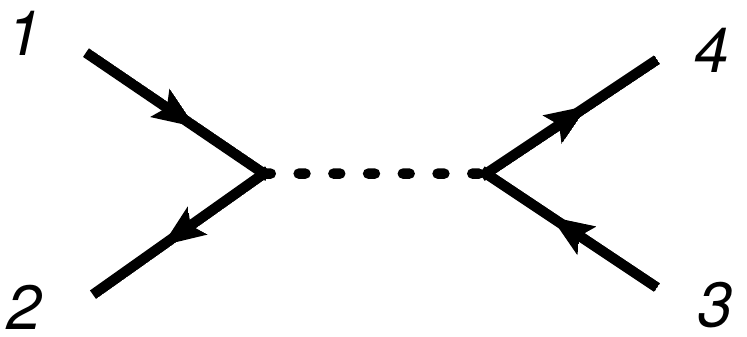}}
  ~=~ 
  ig\, \overline{u}_{4} v_3\times \frac{-i\,\,}{(p_1+p_2)^2} \times
  ig\, \overline{u}_{2} v_1 \,.
  \label{Yukex2}
\ee
Our observations in the previous example tell us that this diagram will vanish unless particles 1 and 2 have the same helicity, and 3 and 4 have the same helicity. Suppose we take particles 1 and 2 to have negative helicity and 3 and 4 positive. 
Then the $u$-channel diagram vanishes and the diagram \reef{Yukex2} is the only contribution to the 4-fermion amplitude. 
Translating the $\overline{u}v$-products to spinor brackets we find
\be
  \label{Yukex2b}
  iA_4(\bar{f}^-f^- \bar{f}^+f^+) 
  \,=\, ig^2 [43]\frac{1}{2 p_1.p_2} \<21\>
  \,=\,  ig^2 [34]\frac{1}{\<12\>[12]} \<12\>
  \,=\, i g^2 \frac{[34]}{[12]}\,.
\ee
The result is a nice simple ratio of two spinor brackets. Now it is fun to note that by momentum conservation, we have (using \reef{angxsq})
\be
  \label{12mc34}
  \<12\>[12]
  = 2 p_1.p_2 = (p_1 + p_2)^2 = 
  (p_3 + p_4)^2 = 2 p_3.p_4
  =
  \<34\>[34] \,.
\ee
Using this in the 2nd equality of \reef{Yukex2b} we get another expression for the same amplitude:
\be
  \label{Yukex2c}
  A_4(\bar{f}^-f^- \bar{f}^+f^+) 
  \,=\, g^2 \frac{\<12\>}{\<34\>}\,.
\ee
This hints at another useful lesson: there are various relationships among spinor brackets, implied for example by momentum conservation as in \reef{12mc34}, and they allow for multiple equivalent forms of the same physical amplitude.
}

\example{Next, let us see what new features appear when we calculate the 4-point tree amplitude with two scalars and two fermions. Two diagrams contribute
\bea
   \nonumber
   iA_4(\phi \,\bar{f}^{h_2} {f}^{h_3} \phi) 
   &=&~~~~
    \raisebox{-5mm}{\includegraphics[width=3cm]{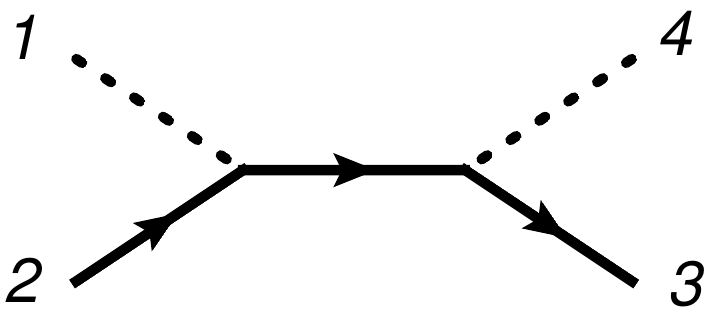}}
    ~~~~~+~~~~
    \raisebox{-5mm}{\includegraphics[width=3cm]{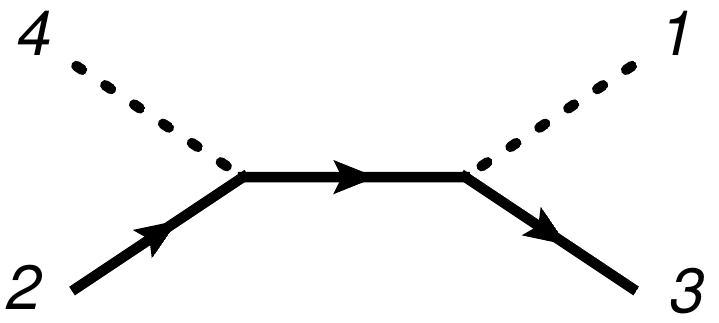}}
    \\[1mm]
    &=&
    (ig)^2 \,\overline{u}_3 \,\frac{-i(\pslash_{1}+\pslash_{2})}{(p_{1}+p_2)^2} \,v_2
    ~~~~\,+~~~~~~~~~~(1 \lra 4) \,.
  \label{Yukex3a}
\eea
If the fermions have the same helicity (say negative), then each diagram has a numerator that involves $\overline{u}_-(p_3) \gamma^\mu v_-(p_2)=0$. So they need to have opposite helicity to give a non-vanishing result: for example 
\be
   \overline{u}_-(p_3) \gamma^\m v_+(p_2) 
   ~=~
   \big( \, 0 \, , \,\< 3 |_{\dot{a}} \big) 
   \left( 
     \begin{array}{cc}
        0 & (\sigma^\mu)_{a\dot{b}} \\
        (\bar{\sigma}^\mu)^{\dot{a}b} & 0 
      \end{array}
   \right) 
   \left( 
  \begin{array}{c}
    \!\! |2]_b \!\!\! \\  
    \!\!\!0\!\!\!
  \end{array}
  \right)
  ~\equiv~
  \<3|\gamma^\m |2] \,.
\ee
Note the abuse of notation in the definition above of the {\bf \em angle-square bracket}  $\<p|\gamma^\m |k]$: it combines the 2-component spinors with the 4$\times$4 gamma-matrix. The meaning should be clear, though, in that the 2-component spinors project out the matching sigma-matrix for $\gamma^\m$.
 The spinor bracket $[p|\gamma^\m |k\>$ is defined similarly. For same-helicity fermions we have
$\<p|\gamma^\m |k\>=0=[p|\gamma^\m |k]$.
}
 Angle-square brackets appear often, so it is useful to record the following properties:
\bea
  \label{pqflip}
  [k | \g^\m | p \> 
  &=& \< p | \g^\m | k ] \, , \\[2mm]
  \label{qpconj}
  [k | \g^\m | p \>^* &=& [p | \g^\m | k \> 
  ~~~~~~~\text{(for real momenta)}
  \, 
\eea
We often use $\<p| P |k] \equiv P_\m \< p | \g^\m | k ]$. 
The notation implies that $p^\mu$ and $k^\m$ are lightlike, but no assumptions are made about $P^\m$. 
However, if $P^\m$ is also lightlike, then
\be
  \label{angsq}
  \<p| P |k] 
  ~=~ \<p|_{\da}\, P^{\da b} \,|k]_b
  ~=~ \<p|_{\da}\, ( -|P\>^{\da} [P|^b )  \,|k]_b
  ~=~ - \<pP\> [P k] \,,
  ~~~~~~~\text{($P^2=0$)}\,.
\ee
Finally, note the useful Fierz identity
\be
  \label{p1234}
 \< 1 | \g^\m | 2 ] \< 3 | \g_\m| 4 ]
   ~=~ 2 \< 1 3 \> [2 4]  \, .
\ee
\exercise{}{Prove the Fierz identity \reef{p1234}.}
\exercise{ex:kvec}{Show that $\<k|\gamma^\m |k] = 2 k^\m$ and  $\< k | P | k ] =  2 \, P \cdot k$.}
With our new tools, we return now to the tree amplitude with two scalars and two fermions. 
\example{Picking opposite helicities for the fermions in \reef{Yukex3a}, we have
\bea
\nonumber
A_4(\phi \,\bar{f}^{+} {f}^{-} \phi) 
 &=&
 -g^2 \frac{\<3| p_1 +p_2|2]}{(p_1+p_2)^2} + (1 \lra 4)\\
\nonumber
 &=&
 -g^2 \frac{\<3| p_1|2]}{(p_1+p_2)^2} + (1 \lra 4)
 ~~~~~\text{\tiny (using the Weyl eq $p_2 |2]=0$)} \\
\nonumber
 &=&
 -g^2 \frac{-\<31\> [12]}{\<12\>[12]} + (1 \lra 4) 
 ~~~~~\text{\tiny (using \reef{angsq})} \\
 &=& 
 -g^2  \frac{\<13\>}{\<12\>}+ (1 \lra 4)\,,
\eea 
so that the result is
\be
  \label{Asffs1}
 A_4(\phi \,\bar{f}^{+} {f}^{-} \phi) 
 ~=~
 -g^2 \bigg( 
 \frac{\<13\>}{\<12\>} +  \frac{\<34\>}{\<24\>}
 \bigg)\,.
\ee
Note bose-symmetry under exchange of the scalar particle momenta.
}
In amplitude calculations, {\bf \em momentum conservation} is imposed on $n$-particles as 
$\sum_{i=1}^n p_i^{\mu} = 0$ (consider all particles outgoing). This is encoded in the spinor helicity formalism as
\be
  \label{momcons}
  \sum_{i=1}^n \< q i \> [i k] ~=~0
\ee
for any lightlike vectors $q$ and $k$. 
For example, you can (and should) show that for $n=4$ momentum conservation implies $\<12\>[23] = -\<14\>[43]$. In \reef{12mc34}, we already found the  identity 
$\<12\>[12] = \<34\>[34]$ valid when $p_1+p_2+p_3+p_4 = 0$. 

With all momenta outgoing, the {\bf \em Mandelstam variables} are defined as
\be
  \label{mandelstam}
  s_{ij} = - (p_i + p_j)^2 \,,
  ~~~~
  s_{ijk} = - (p_i + p_j + p_k)^2 \,,
  ~~~~\text{etc}\,.
\ee
In particular, we have $s = s_{12}$, $t = s_{13}$, and $u = s_{14}$ for 4-particle processes.

To see some of the power of the spinor helicity formalism, let us now calculate the spin sum 
\be
\big\< |A_4(\phi \,\bar{f} {f} \phi)|^2 \big\>
= \displaystyle \sum_{h_2,h_3=\pm} \big|A_4(\phi \,\bar{f}^{h_2} {f}^{h_3} \phi)\big|^2
\ee
 for the 2-scalar 2-fermion process in the previous example. To really appreciate the difference in formalism, it is educational to first do the calculation the standard way, using the spinor-completeness relations and evaluating the gamma-matrix traces:
\exercise{}{Use standard techniques to show that 
$\big\<|A_4(\phi \,\bar{f} {f} \phi)|^2\big\> = 2 g^4 (s-t)^2/(st)$. \\[1mm]
[Hint: This is very similar to the massless limit of the example $e^- \varphi \to e^- \varphi$ in Section 48 of Srednicki \cite{MSqft}, but we include no $\tfrac{1}{2}$-factors from averages here.]}
Having resharpened your pencils after doing this exercise, let us now do the spin sum in the spinor helicity formalism. We already know that the helicity amplitudes $A_4(\phi \,\bar{f}^{h_2} {f}^{h_3} \phi)$ vanish unless the spinors have opposite helicity, so that means that 
\be
\big\< |A_4(\phi \,\bar{f}f \phi)|^2 \big\>
~=~  \big|A_4(\phi \,\bar{f}^{-} {f}^{+} \phi)\big|^2
  + \big|A_4(\phi \,\bar{f}^{+} {f}^{-} \phi)\big|^2\,.
  \label{ssEx1a}
\ee
The first term is calculated easily using the result \reef{Asffs1} for the helicity amplitude and the reality condition \reef{p-reality}:
\bea
\nonumber
 \big|A_4(\phi \,\bar{f}^{-} {f}^{+} \phi)\big|^2
 &=&
g^4 \bigg( 
 \frac{\<13\>}{\<12\>} +  \frac{\<34\>}{\<24\>}
 \bigg)
 \bigg( 
 \frac{[13]}{[12]} +  \frac{[34]}{[24]}
 \bigg) \\[1mm]
 &=&
g^4 \bigg( 
 \frac{\<13\>[13]}{\<12\>[12]} 
 +  \frac{\<34\>[34]}{\<24\>[24]}
 + \frac{\<13\>[34]}{\<12\>[24]}
 + \frac{\<34\>[13]}{\<24\>[12]}
 \bigg)\,.
 \label{ssEx1b}
\eea
In the first two terms, we can directly translate the spinor products to Mandelstam variables using \reef{angxsq}. For the last two terms, the momentum conservation identity \reef{momcons} comes in handy, giving 
$\<12\>[24] = -\<13\>[34]$ and $\<24\>[12] = - \<34\>[13]$. 
Thus \reef{ssEx1b} gives
\be
  \big|A_4(\phi \,\bar{f}^{-} {f}^{+} \phi)\big|^2
  ~=~ g^4 \bigg(
  \frac{t}{s} + \frac{s}{t} - 2 
  \bigg)
  ~=~ g^4 \, \frac{(s-t)^2}{st}\,.
\ee
The second term in \reef{ssEx1a} gives exactly the same, so $\big\<|A_4(\phi \, \bar{f} f\phi)|^2\big\> = 2 g^4 (s-t)^2/(st)$, in agreement with the result of the standard calculation (but with use of much less pencil-power).
\exercise{}{Calculate the 4-fermion `all-minus' amplitude $A_4(\bar{f}^- f^- \bar{f}^- f^-)$ in Yukawa theory.
}
\exercise{}{Calculate the spin sum 
$\big\<|A_4(\bar{f} f \bar{f} {f})|^2\big\>$
for the 4-fermion process in Yukawa theory.
}
\exercise{ex:nearSUSY}{Consider a model with a Weyl-fermion  $\psi$ and a complex scalar $\phi$:
\be
  \Lag= 
  i \psi^\dagger \bar{\sigma}^\mu \pa_\mu \psi 
   - \pa_\mu \bar\phi \,\pa^\mu \phi 
+ \tfrac{1}{2} g \, \phi\, \psi \psi + \tfrac{1}{2} g^* \, \bar\phi \, \psi^\dagger \psi^\dagger
      - \tfrac{1}{4}\lambda\,  |\phi|^4  \,.
\ee
Show that\footnote{We do not put a bar on the $f$'s here because in this model  the 4-component fermion field is a Majorana fermion so there is no distinction between $f$ and $\bar{f}$.}
\be
  A_4(\phi \phi \bar\phi \bar\phi) = - \lambda\,,
  ~~~~
  A_4(\phi \,f^- f^+ \bar\phi) =  -|g|^2 \frac{\<24\>}{\<34\>}\,,
  ~~~~
  A_4(f^- f^- f^+ f^+) = |g|^2 \frac{\<12\>}{\<34\>}\,.~~
\ee
These amplitudes serve as useful examples later in the text.
}
We end this section by discussing one more identity from the amplitudes tool-box:
the {\bf \em Schouten identity} is a  fancy name for a rather trivial fact: three vectors in a plane cannot be linearly independent. So if we have three 2-component vectors $|i\>$, $|j\>$, and $|k\>$, you can write one of them as a linear combination of the two others:
\be
  \label{lin}
  |k\> = a |i\> + b |j\>  \,~~~~~~~~
  \text{for some $a$ and $b$.}
\ee
One can dot in spinors $\<\cdot|$ and form antisymmetric angle brackets to solve for the coefficients $a$ and $b$. Then \reef{lin} can be cast in the form
\be
  |i\>\<jk\> + |j\> \<ki\> + |k\>\<ij\> ~=~ 0\,.
\ee
This is the Schouten identity. It is often written with a fourth spinor $\<r |$ ``dotted-in":
\be
   \<r i\>\<jk\> + \<r j\> \<ki\> + \<r k\>\<ij\> ~=~0\,.
\ee
A similar Schouten identity holds for the square spinors:~
$[r i][jk] + [r j][ki] + [r k][ij] = 0$. 
\exercise{}{Show that 
$A_5(f^- \bar{f}^- \phi \phi \phi) 
= g^3\, \frac{[12][34]^2}{[13][14][23][24]} + (3 \lra 5) + (4 \lra 5)$
in Yukawa theory \reef{LAGyukawa}.
}

%%%%%%%
%%%%%%%
%%%%%%%
\subsection{Massless vectors and examples from QED}
\label{s:QED}
The external line rules for outgoing spin-1 massless vectors is simply to ``dot-in" their polarization vectors. They can be written in spinor helicity notation as follows:
\be
  \label{shpolar}
   \eps_-^\m(p;q) 
   =  -\frac{\<p | \g^\m | q ] }{\sqrt{2}\,  [q \, p]} \, ,
   \hspace{1cm}
   \eps_+^\m(p;q) 
   = -\frac{ \< q | \g^\m | p ] }{\sqrt{2}\,  \<q \, p \>} \, ,
\ee
where $q \ne p$ denotes an arbitrary reference spinor. Note that the massless Weyl equation ensures that $p_\m \eps^\m_\pm(p) = 0$. It can be useful to write the polarizations as
\be
  \label{slasheps}
  \epsslash_-(p;q) ~=~ \frac{\sqrt{2}}{[q p]} 
  \Big( |p\> [q| + |q] \<p| \Big)\,,
  ~~~~~~~
  \epsslash_+(p;q) ~=~ \frac{\sqrt{2}}{\<q p\>} 
  \Big( |p] \<q| + |q\> [p| \Big)\,.
\ee
The arbitrariness in the choice of reference spinor reflects gauge invariance, namely that one is free to shift the polarization vector with any constant times the momentum vector:
$\eps_\pm^\m(p) \to \eps_\pm^\m(p) + C\,p^\m$. This does not change the on-shell amplitude $A_n$, as encoded in the familiar Ward identity $p_\m A_n^\m = 0$. For each external vector boson,  one has a free choice of the corresponding reference spinor $q_i \ne p_i$; however, one must stick with the same choice in each diagram of a given process. When summing over all diagrams, the final answer for the  amplitude is independent of the choices of $q_i$.

\exercise{}{Consider the momentum
$p^\m 
  = (E, ~E\, \sin\th \cos\phi , ~E\, \sin\th \sin\phi,~ E\, \cos\th)$.
  In Exercise \ref{ex:one}, you found the corresponding angle and square spinors $|p\>$ and $|p]$. 
In this exercise, we establish the connection between the polarization vectors \reef{shpolar} and the more familiar
polarization vectors
\be
  \tilde{\eps}^\mu_\pm(p) 
   = \pm \frac{e^{\mp i\phi}}{\sqrt{2}}
     \Big( 0,~\cos\th \, \cos\phi \pm i \sin\phi, ~
     \cos\th \, \sin\phi  \mp i \cos\phi , ~ - \sin\th \Big) \, .
  \label{pol}
\ee
Note that for $\theta=\phi=0$, we have 
$\tilde{\eps}^\mu_\pm(p) = \pm\frac{1}{\sqrt{2}} (0,1,\mp i,0)$.

(a) Show that $\tilde{\eps}_\pm(p)^2 = 0$ and $\tilde{\eps}_\pm(p) \cdot p = 0$. 

(b) Since $\tilde{\eps}^\mu_\pm(p)$ is null, 
$\big(\tilde{\eps}^\mu_\pm(p)\big)_{a\dot{b}} 
= (\sigma_\m)_{a\dot{b}} \, \tilde{\eps}^\mu_\pm(p)$ 
can be written in as a product of a square and an angle spinor. To see this specifically, first calculate $\big(\tilde{\eps}^\mu_\pm(p)\big)_{a\dot{b}}$ and then find an angle spinor $\<r|$ such that 
$\big(\tilde{\eps}^\mu_+(p)\big)_{a\dot{b}} = - |p]_a\<r|_{\dot{b}}$.\\[1mm]{}
[Hint: you should find that $\<rp\> = - \sqrt{2}$.]

(c) Next, show that it follows from \reef{shpolar} that 
$\big(\eps_+(p;q)\big)_{a\db} = \frac{\sqrt{2}}{\<qp\>}\, |p] \<q|$.

(d) Now suppose there is a constant $c_+$ such that 
$\eps^\m_+(p;q) = \tilde{\eps}^\m_+(p) + c_+ \,p^\mu$. Show that this relation requires $\<rp\> =- \sqrt{2}$ (as is consistent with the solution you found for $\<r|$ in part (b)) and then show that $c_+ = -\<r q\>/\<pq\>$.

Since $\eps^\m_+(p;q) = \tilde{\eps}^\m_+(p) + c_+ \,p^\mu$, the polarization vectors  $\eps^\m_+(p;q)$ and $\tilde{\eps}^\m_+(p)$ are equivalent. You can show the same for the negative helicity polarization. It should be clear from this exercise that the arbitrariness in the reference spinors $q$ in the polarizations \reef{shpolar} is directly related to the gauge invariance reflected in the possibility of adding any number times $p^\mu$ to the polarization vectors.
}
We now calculate some amplitudes in QED to illustrate the use of the spinor helicity formalism. 
The QED Lagrangian 
\be
  \lag = -\frac{1}{4} F_{\m\n}F^{\m\n} + i \overline{\Psi} \gamma^\m ( \pa_\m - ie A_\m) \Psi \,
\ee
describes the interaction of a massless\footnote{Think of this as the high-energy scattering limit in which we can consider electrons/positrons massless.} 
fermion with a photon via the interaction $A_\m \overline{\Psi} \gamma^\m \Psi$. The vertex rule is $i e \gamma^\mu$.

\example{
Let us consider the 3-particle QED amplitude $A_3\big(f^{h_1} \bar{f}^{h_2} \gamma^{h_3} \big)$ (here $f = e^-$ and $\bar{f} = e^+$). Choose, as an example, helicities $h_1 = -1/2$, $h_2 = +1/2$ and $h_3 = -1$. We then have
\be
  \nonumber
   iA_3\big(f^- \bar{f}^+ \gamma^- \big)
   ~=~
   \overline{u}_-(p_1) ie \gamma_\m v_+(p_2) \, \eps^\m_-(p_3;q)
   ~=~
   -ie
  \<1|\gamma_\m |2]\,  \frac{\<3 | \g^\m | q ] }{\sqrt{2}\,  [3 \, q]} 
  ~=~\sqrt{2} ie\, \frac{\<13\> [2q]}{[3 \, q]}
  \,,
\ee  
using in the last step the Fierz identity \reef{p1234}.
 We then have
\be
 A_3\big(f^- \bar{f}^+ \gamma^- \big)
 ~=~\tilde{e}\, \frac{\<13\> [2q]}{[3 \, q]}
 \,.
    \label{exQEDIIa}
\ee
We have absorbed the $\sqrt{2}$ into the definition of the coupling $e$ as 
$\tilde{e} \equiv \sqrt{2} e$.
}
Earlier, we discussed that the on-shell amplitude should be independent of the reference spinor $q$. 
Here, there are no other diagrams, and naively it appears that \reef{exQEDIIa} depends on $|q]$. However, it is in fact independent of $|q]$ --- and this brings us to discuss several important aspects: 
\begin{itemize}
\item
First, let us see how to eliminate $|q]$ from \reef{exQEDIIa}. 
Multiply \reef{exQEDIIa} by $1=\<12\>/\<12\>$. In the numerator, we then have 
$\<13\> \<12\> [2q]$. But by \reef{angsq}, 
$\<12\> [2q] = - \<1| 2 | q]$. Now use momentum conservation, 
$p_2 = - p_1 - p_3$ and the massless Weyl equation to get
\be
  \<12\> [2q] = - \<1| p_2 | q]
  = \<1|(p_1+p_3)|q]
  = \<1|3|q]
  = \<13\> [3q]\,.
\ee
The square bracket $[3q]$ cancels against the equal factor in the denominator of 
\reef{exQEDIIa}, and we are left with
\be
   \label{exQEDIIb}
   A_3\big(f^- \bar{f}^+ \gamma^- \big)
  ~=~
  \tilde{e} \,\frac{\<13\>^2}{\<12\>\,}
 \,,
\ee
which is independent of $|q]$.

\item Note that the result \reef{exQEDIIb} depends only on angle brackets, not square brackets. This is no coincidence, but a consequence of {\bf \em 3-particle special kinematics}. Note that 
if three lightlike vectors satisfy $p_1^\m +p_2^\m+p_3^\m = 0$, then
\be
   \<12\> [12] = 2 p_1.p_2 = (p_1+p_2)^2 = p_3^2 = 0
\ee
so either $\<12\>$ or $[12]$ must vanish. Suppose $\<12\>$ is non-vanishing; then 
by \reef{momcons} and the massless Weyl equation we have $\<12\>[23] = \<1|(p_1+p_3)|3] = 0$. So $[23]=0$. Similarly, $[13]=0$. This means that 
\begin{enumerate}
\item a non-vanishing on-shell 3-particle amplitude with only massless particles can only depend on either angle brackets or square brackets of the external momenta, never both. 
\item Since for real momenta, angle and square spinors are each others complex conjugates, \emph{on-shell 3-particle amplitude of only massless particles can only be non-vanishing in complex momenta.}\footnote{Or using a $(-,-,+,+)$ spacetime signature.} 
Although they do not occur in Nature, the massless complex momentum 3-point amplitudes are extremely useful for building up higher-point amplitudes recursively --- in many cases, the on-shell 3-point amplitudes  are the key building blocks.  More about this in Section \ref{s:recrels}.
\end{enumerate}
\item Finally, let us comment on the choice of $q$ in \reef{exQEDIIa}. Naively, it might seem that choosing $|q] \propto |2]$ gives zero for the amplitude; this would be inconsistent with our $q$-independent non-vanishing result \reef{exQEDIIb}. However, this choice gives  $[3q] \propto [23]$, so the denominator therefore vanishes by special kinematics. One could say that the zero [22] in the numerator is cancelled by the zero [23] in the denominator, or simply that $|q] \propto |2]$ is not a legal choice since it makes the polarization vector $\eps^\m_-(p_3;q)$ divergent. 
\end{itemize}

At this stage it is natural to ask how, then, we know if a given 3-point amplitude of massless particles should depend on angle brackets or square-brackets? This has a good answer, which we reveal in Section \ref{s:littlegrp}.
For now, let us carry on exploring QED amplitudes in the spinor-helicity formalism.

\example{Consider the QED Compton scattering process: 
$e^- \ga \to e^-  \ga$. With crossing symmetry, we can consider this as the amplitude $A_4(\bar{f}  f \ga\ga)$ with all particles outgoing and labeled by momenta 1,2,3,4:
\bea
 \nonumber
 iA_4(\bar{f}  f \ga\ga)&=&    
 ~~
 \raisebox{-5mm}{\includegraphics[width=3cm]{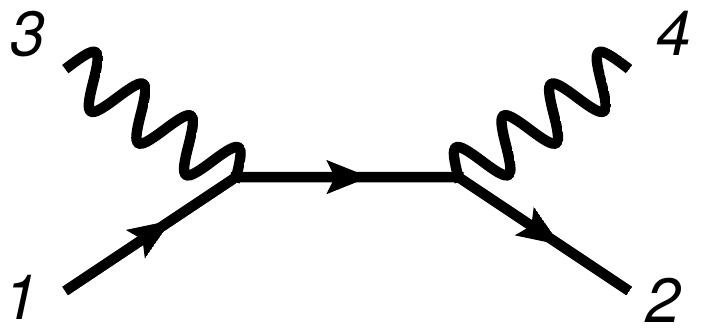}}
    ~~+~~
    \raisebox{-5mm}{\includegraphics[width=3cm]{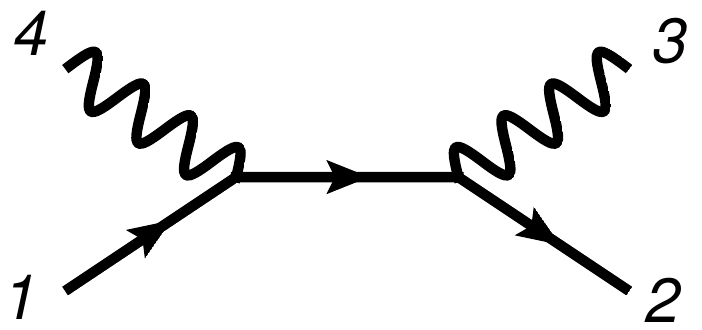}}\\[1mm]
    &=&
    (ie)^2 \,\,\overline{u}_2 \,\, \epsslash_4\,
    \frac{-i(\pslash_1+\pslash_3)}{(p_1+p_3)^2} 
    \, \epsslash_3 \, v_1
    ~+~(3 \lra 4) 
    \,.
    \label{QEDex2a}
\eea
Note that we have an odd number of gamma-matrices sandwiched between two spinors. If $\bar{f}$ and $f$ have the same helicity, then such spinor products vanish, e.g.~$\<2|\gamma^\m \gamma^\n \gamma^\r |1\> = 0$. So we need the fermions to have opposite helicity for the process to be non-vanishing. 

Suppose the photons both have negative helicity. Then the first diagram in \reef{QEDex2a} involves
$(\epsslash_{3-} \, v_{1+}) \propto |3\>[q_3 1]$ using \reef{slasheps}. By picking $|q_3] \propto |1]$ this diagram vanishes. Similarly, we can choose 
$|q_4] \propto |1]$ to make the second diagram vanish. So $A_4(\bar{f}^+  f^- \ga^- \ga^-) = 0$.
\exercise{}{As a spinor-helicity gymnastics exercise, show that $A_4(\bar{f}^+  f^- \ga^- \ga^-) = 0$ without making any special choices of the reference spinors $q_3$ and $q_4$.}
Now consider $A_4(\bar{f}^+  f^- \ga^+ \ga^-)$. We have
\bea
 \nonumber
 A_4(\bar{f}^+  f^- \ga^+ \ga^-) 
 &=&
 \frac{2e^2 \<24\> [q_4| \big( -|1]\<1| - |3]\<3| \big) |q_3\> [31]}
    {\<13\>[13] \<q_3 3\>[q_4 4]} \\
 &&+ \frac{2e^2 \<2q_3\> [3| \big( -|1]\<1| - |4]\<4| \big) |4\> [q_41]}
    {\<14\>[14] \<q_3 3\>[q_4 4]}\,.
    \label{QEDex2b}
\eea

Let us choose $q_3=q_4=p_1$. Then the second diagram in \reef{QEDex2b} vanishes and we  get 
\be
 A_4(\bar{f}^+  f^- \ga^+ \ga^-) 
 ~=~ -\tilde{e}^2\,
 \frac{ \<24\> [13] \<3 1\> [31]}{\<13\> [13] \<1 3\> [1 4]}
 ~=~-\tilde{e}^2\,
 \frac{ \<24\> [13]}{\<13\> [1 4]}\,,
\ee
where $\tilde{e} = \sqrt{2}e$. Momentum conservation lets us rewrite this using
$\<23\>[13] = - \<24\>[14]$ giving
\be
  \label{QEDex2c}
 A_4(\bar{f}^+  f^- \ga^+ \ga^-) 
 ~=~ \tilde{e}^2\,\frac{ \<24\>^2}{\<13\> \<23\>} \,.
\ee
The amplitude $A_4(\bar{f}^+  f^- \ga^- \ga^+)$ is obtained by interchanging the momentum labels 3 and 4 in \reef{QEDex2c}.
}
\exercise{}{Show that the amplitude $A_4(\bar{f}^+  f^- \ga^+ \ga^-)$ is independent of $q_3$ and $q_4$ by deriving \reef{QEDex2c} without making a special choice for the reference spinors.}
\exercise{}{Calculate the tree-level process $e^- e^+ \to e^- e^+$ using spinor helicity formalism.}
For further experience with spinor helicity formalism, consider massless {\bf \em scalar-QED}:
\bea
  \nonumber
  \lag &=& - \frac{1}{4} F_{\m\n}F^{\m\n}   - |D\varphi|^2 - \frac{1}{4} \lambda |\varphi|^4
  \\
 &=&- \frac{1}{4} F_{\m\n}F^{\m\n}  - |\pa \varphi|^2
   + i e A^\mu \big[ (\pa_\m \varphi^*) \varphi  - \varphi^* \pa_\m \varphi \big]
   - e^2 A^\m A_\m \varphi^* \varphi
   - \frac{1}{4} \lambda |\varphi|^4 \,.
   \label{scalarQED}
\eea
The Feynman rules gives a scalar-scalar-photon 3-vertex $ie(p_2-p_1)^\m$ (both momenta outgoing), a scalar${}^2$-photon${}^2$ 4-vertex $-2i e^2 \eta_{\m\n}$, and a 4-scalar vertex $-i\lambda$.

We can think of $\varphi$ and $\varphi^*$ as the spin-0 supersymmetric partners of the electron/positron and we'll loosely call them \emph{selectrons/spositrons}, though we emphasize that we are not assuming that our model is part of a supersymmetric theory.
A process like $\varphi+ \ga \to \varphi + \ga$ is then the spin-0 analogue of Compton scattering. Here we consider the extreme high-energy regime where we take the mass of the selectron/spositron to be zero. 
\exercise{ex:sQED1}{Calculate the 3-particle amplitude $A_3(\varphi \,\varphi^* \ga^{-})$. Show that it is independent of the reference spinor in the photon polarization vector and write the result in a form that only involves angle brackets. 

Use complex conjugation to write down the amplitude $A_3(\varphi \,\varphi^* \ga^{+})$.
}
\exercise{ex:sQED2}{Consider the amplitude $A_4(\varphi \,\varphi^* \ga \ga)$. Show that no matter what the photon helicities, one can always choose the reference spinors in the polarzations such that the scalar${}^2$-photon${}^2$ contact term gives a vanishing contribution to the on-shell 4-point amplitude. 
}
\exercise{ex:sQED3}{Calculate $A_4(\varphi \,\varphi^* \ga \ga)$ and massage the answer into a form that depends only on either angle or square brackets and is manifestly independent of the reference spinors.
}
\exercise{ex:sQED4}{Calculate the spin sum $\big\< |A_4(\varphi \,\varphi^* \ga \ga)|^2 \big\>$.
}
\exercise{ex:sQED5}{Calculate $A_4(\varphi\, \varphi^* \varphi\, \varphi^*)$. The answer can be expressed in terms of the Mandelstam variables, but show that you can bring it to the following form:
\be
  \label{4scalarQED}
  A_4(\varphi\, \varphi^* \varphi\, \varphi^*)
  ~=~
  -\lambda + \tilde{e}^2 
  \bigg( 
     1+ \frac{\<13\>^2\<24\>^2}{\<12\>\<23\>\<34\>\<41\>}
  \bigg) \,.
\ee
}

%%%%%%%%%%%%%%%%
%%%%%%%%%%%%%%%%
%%%%%%%%%%%%%%%%
\subsection{Yang-Mills theory, QCD, and color-ordering}
\label{s:YM}
Gluons are described by the Yang-Mills Lagrangian
\be \label{YMlag}
  \lag = -\frac{1}{4} \Tr F_{\m\n} F^{\m\n} \,, 
\ee
with 
$F_{\m\n}  
= \pa_\m A_\n - \pa_\n A_\m  -\frac{ig}{\sqrt{2}} [A_\m, A_\n]$ and $A_\m = A_\m^a T^a$. 
The gauge group is $G=SU(3)$ for QCD. We consider a case of $N$ colors and take $G=SU(N)$. The gluon fields are in the adjoint representation, so the color-indices run over $a,b,\ldots=1,2,\dots, N^2-1$. The generators $T^a$ are normalized\footnote{\label{footiefabc}A more common normalization is 
$\Tr T^a T^b = \frac{1}{2}\delta^{ab}$ and $[T^a,T^b] = i {f}^{abc} T^c$. So we have 
$\tilde{f}^{abc} = \sqrt{2} f^{abc}$, in analogue with $\tilde{e} = \sqrt{2} e$ in QED in Section \ref{s:QED}. It serves the same purpose here, namely compensating for the $\sqrt{2}$ in the polarization vectors \reef{shpolar}, so the on-shell amplitudes can be written without such factors.} such that
$\Tr T^a T^b = \delta^{ab}$ and 
$[T^a,T^b] = i \tilde{f}^{abc} T^c$. 

To extract Feynman rules from \reef{YMlag}, one unfortunately needs to fix the gauge redundancy. An amplitude-friendly choice is \emph{Gervais-Neveu gauge} for which the gauge-fixing term is $\lag_\text{gf} = - \frac{1}{2} \Tr \big(H_\m{}^\m\big)^2$ with 
$H_{\m\n}=\pa_\m A_\n - \frac{i g}{\sqrt{2}} A_\m A_\n$ \cite{MSqft}. 
In this gauge, the Lagrangian takes the  form\footnote{We ignore ghosts, since our focus here is on tree-level amplitudes.}
\be
   \lag = \Tr \Big(
      - \frac{1}{2} \,\pa_\m A_\n \pa^\m A^\n
      - i\sqrt{2} g \,\pa^\m A^\n A_\n A_\m 
      +\frac{g^2}{4} \, A^\m A^\n A_\n A_\m
   \Big)\,.
\ee
The Feynman rules then give
a gluon propagator $\delta^{ab} \frac{\eta_{\m\n}}{p^2}$. The 3- and 4-gluon vertices involve $\tilde{f}^{abc}$ and $\tilde{f}^{abx}\tilde{f}^{xcd}$ (+perms), respectively, each dressed up with kinematic factors that we'll get back to later. The amplitudes constructed from these rules can be organized into different group theory structures each dressed with a kinematic factor. For example, the color factors of the $s$-, $t$- and $u$-channel diagram of the 4-gluon tree amplitude are
\be
  \label{csctcu}
  c_s \equiv \tilde{f}^{a_1 a_2 b}\tilde{f}^{b\, a_3 a_4}\,,
  ~~~~~
  c_t \equiv \tilde{f}^{a_1 a_3 b}\tilde{f}^{b\, a_4 a_2}\,,
  ~~~~~
  c_u \equiv \tilde{f}^{a_1 a_4 b}\tilde{f}^{b\, a_2 a_3}\,,
\ee
and the 4-point contact term generically gives a sum of contributions with  $c_s$, $c_t$ and $c_u$ color-factors. The Jacobi identity relates the three color-factors:
\be
  \label{csctcuID}
   c_s + c_t + c_u = 0\,.
\ee
So there are only two independent color-structures for the tree-level 4-gluon amplitude. Let us now see this in terms of traces of the generators $T^a$. Note that 
\be
  i \tilde{f}^{abc} =  \Tr(T^a T^b T^c) - \Tr(T^b T^a T^c)\,.
\ee
The products of generator-traces in the amplitudes can be Fierz'ed using the completeness relation
\be
   (T^a)_i{}^j (T^a)_k{}^l 
   = \delta_i{}^l \delta_k{}^j  - \frac{1}{N}  \delta_i{}^j \delta_k{}^l \,.
\ee
For example, for the 4-gluon $s$-channel diagram we have
\bea
  \label{ffTT}
  &&
   \hspace{-7mm}
   \tilde{f}^{a_1 a_2 b}\tilde{f}^{b\, a_3 a_4}\\
  &&
   \hspace{-7mm}
   ~~~\propto~ 
  \Tr \big(T^{a_1} T^{a_2}  T^{a_3} T^{a_4}\big) 
  -  \Tr \big(T^{a_1} T^{a_2}  T^{a_4} T^{a_3}\big) 
  -  \Tr \big(T^{a_1} T^{a_3}  T^{a_4} T^{a_2}\big) 
  + \Tr \big(T^{a_1} T^{a_4}  T^{a_3} T^{a_2}\big)
  \,.\nonumber
\eea
Here we have also used the cyclic property of the traces to deduce the four  color-structures. 
Similarly, the 3 other diagrams contributing to the 4-gluon amplitude can also be written in terms of single-trace group theory factors. So that means that we can write the 4-gluon tree amplitude as
\be
  \label{A4full}
  A^\text{full,tree}_4 
  = g^2 \Big( A_4[1234]\, \Tr \big(T^{a_1} T^{a_2}  T^{a_3} T^{a_4}\big)
  + \text{perms of (234).} 
\Big) \,,
\ee
where the \emph{partial amplitudes} $A_4[1234]$, $A_4[1243]$ etc, are called {\bf \em color-ordered amplitudes}. Each partial amplitude is gauge invariant.\footnote{This follows from a partial orthogonality property of the single-traces \cite{Mangano:1990by}.} This color-structure generalizes to any $n$-point tree-level amplitude involving any particles that transform in the adjoint of the gauge group: we write
\be
  \label{Anfull}
   A^\text{full,tree}_n = 
   g^{n-2}
   \sum_{\text{perms } \sigma}
   A_n[1\,\sigma(2\dots n)]\,
   \Tr \big(T^{a_1} T^{\sigma(a_2} \cdots T^{a_n)}\big)\,.
\ee
where the sum is over the (overcomplete) trace-basis of $(n-1)!$ elements that takes into account the cyclic nature of the traces.
For loop-amplitudes, one also needs to consider multi-trace structures in addition to the simple single-trace -- for more about this, see \cite{dixon,Dixon:2011xs}. We have factored out the coupling constant $g$ to avoid carrying it along explicitly in all the amplitudes.

Feynman vertex rules for calculating the color-ordered amplitudes directly are
\begin{itemize}
\item 3-gluon vertex 
$V^{\m_1\m_2\m_3}(p_1,p_2,p_3)
= - \sqrt{2} \,
\big(
   \eta^{\m_1\m_2} p_1^{\m_3}
  + \eta^{\m_2\m_3} p_2^{\m_1}
  + \eta^{\m_3\m_1} p_3^{\m_2}
\big)$,
\item 4-gluon vertex $V^{\m_1\m_2\m_3\m_4}(p_1,p_2,p_3,p_4) = \eta^{\m_1\m_3}\eta^{\m_2\m_4}$\,.
\end{itemize}
The color-ordered amplitude $A_n(12\dots n)$ is calculated in terms of diagrams with no lines crossing and the ordering of the external lines fixed as given $1,2,3,\dots,n$. The polarization vectors are given in  \reef{shpolar}-\reef{slasheps}. Let us consider the simplest case, namely the 3-gluon amplitude. From the 3-vertex rule, we get
\be
  \label{gluonMMPa}
   A_3[1\, 2\, 3] =
   -\sqrt{2} 
   \Big[ 
     (\eps_1\eps_2)(\eps_3 p_1)
     +   (\eps_2\eps_3)(\eps_1 p_2)
     +   (\eps_3\eps_1)(\eps_2 p_3)   
   \Big] \,.
\ee
Let us now pick gluons 1 and 2 to have negative helicity while gluon 3 gets to have positive helicity. 
Translating to spinor helicity formalism (using the Fierz identity \reef{p1234}) we get
\be
  \label{A3calc1}
  A_3[1^- 2^- 3^+] ~=~ 
  - 
  \frac{\<12\>[q_1 q_2] \<q_3 1\> [1 3] 
   + \<2 q_3\> [q_2 3] \<12\>[2q_1] 
   + \<q_3 1\> [ 3 q_1] \<23\>[3q_2]}
  {[q_1 1][q_2 2]\<q_3 3\>} \,.
\ee
We must now consider 3-particle special kinematics (see Section  \ref{s:QED}). If $|1\> \propto |2\> \propto |3\>$  all three terms vanish in the numerator of \reef{A3calc1}. So pick  3-particle kinematics $|1] \propto |2] \propto |3]$. Then the first term vanishes and we are left with 
\be
   A_3[1^- 2^- 3^+] ~=~
   - \,\frac{\<2q_3\>[q_2 3]\<12\>[2q_1] + \<1 q_3\> [q_1 3]\<2 3\>[3q_2]}{[q_1 1] [q_2 2]\<q_3 3\>} \,.
\ee
To simplify this, first use momentum conservation to write 
$\<12\>[2q_1] = - \<13\>[3q_1]$. Then  $[q_1 3][q_2 3]$ factors and we get
\be
   A_3[1^- 2^- 3^+] ~=~
    \,\frac{[q_1 3][q_2 3] \big( \<13\>\<q_3 2\> + \<1 q_3\> \<2 3\>\big)}{[q_1 1] [q_2 2]\<q_3 3\>} \,,
\ee
which after a quick round of Schouten'ing simplifies to 
\be
   A_3[1^- 2^- 3^+] ~=~
   \,\frac{[q_1 3][q_2 3] \big( -\<12\>\<3q_3 \>\big)}{[q_1 1] [q_2 2]\<q_3 3\>}
   ~=~
    \frac{\<12\> [q_1 3][q_2 3]}{[q_1 1] [q_2 2]}
 \,.
\ee
So we got rid of $q_3$. To eliminate $q_1$ and $q_2$, use momentum conservation
$[q_1 3]\<2 3\>= - [q_1 1]\<2 1\>$
and 
$[q_2 3]\<1 3\>= - [q_1 1]\<1 2\>$.  
The result is remarkably simple:
\be
   \label{A3YMmmp}
   A_3[1^- 2^- 3^+] = \frac{\<12\>^3}{\<23\>\<31\>}\,.
\ee
The result for the `goggly' amplitude $A_3[1^+ 2^+ 3^-]$ is
\be
   \label{A3YMppm}
   A_3[1^+ 2^+ 3^-] = \frac{[12]^3}{[23][31]}\,.
\ee

\exercise{}{Fill in the details to derive the amplitude \reef{A3YMppm}.
}
\exercise{ex:polar1}{Calculate $\eps_-(p,q) \cdot \eps_-(k,q')$, ~$\eps_+(p,q) \cdot \eps_+(k,q')$, and $\eps_-(p,q) \cdot \eps_+(k,q')$. \\
Show that $\eps_\pm(p,q) \cdot \eps_\pm(k,q')$ vanishes if $q=q'$.\\
How can you make $\eps_-(p,q) \cdot \eps_+(k,q')$ vanish?
}
\exercise{ex:polar2}{Use the previous exercise to show that for any choice of  gluon helicities, it is always possible to choose the polarization vectors such that the contribution from the 4-gluon contact term to the 4-gluon amplitude vanishes.
}
\exercise{ex:polar3}{Use a well-chosen set of reference spinors to show that the entire 4-gluon amplitudes vanish if all four gluons have the same helicity.
}
\exercise{}{Calculate the color-ordered 4-gluon tree amplitude $A_4[1^- 2^- 3^+ 4^+]$ using  Feynman rules and a smart choice of reference spinors. Show that the answer can be brought to the form
\be
  \label{PT4}
  A_4[1^- 2^- 3^+ 4^+] = \frac{\<12\>^4}{\<12\> \<23\> \<34\> \<41\>}\,.
\ee
Note the cyclic structure of the numerator factor.
}
The result for the 4-gluon amplitude is an example of the famous {\bf \em Parke-Taylor $n$-gluon tree amplitude}: for the case where gluons $i$ and $j$ have helicity $-1$ and all the $n-2$ other gluons have helicity $+1$, the tree amplitude is
\be
   \label{PTn}
   A_n[1^+\ldots i^- \ldots  j^- \ldots n^+]
   = \frac{\<ij\>^4}{\<12\> \<23\> \cdots \<n1\>}\,.
\ee 
We prove this formula in Section \ref{s:recrels}.
The number of Feynman diagrams that generically contribute to an $n$-gluon tree amplitude is\footnote{This can be seen by direct counting, but see also analysis in \cite{Kampf:2013vha}.}

~~~~~~~
\begin{tabular}{ccccccccc}
$n$ = & 3 & 4 & 5 & 6 & 7 &\dots\\
\hline 
\#diagrams = & 1 & 3 & 10 & 38 & 154 &\dots
\end{tabular}

A fun little trivia point you can impress your friends with in a bar (oh, I mean at the library), is that the number of trivalent graphs that contribute to the $n$-gluon tree process is counted by the Catalan numbers. 

It should be clear that even if you have learned now some handy tricks of how  to choose the polarization vectors to reduce the complexity of the calculation, it would be no fun trying to calculate these higher-point amplitudes brute force. But despite   the complications of the many diagrams and their increased complexity, the answer is just the simple Parke-Taylor expression \reef{PTn} for the $--++\dots+$ helicity case. And that is the answer  no matter which fancy field redefinitions we might subject the Lagrangian to and no matter which ugly gauge we could imagine choosing. It is precisely the point of the modern approach to amplitudes to avoid such complications and get to an answer such as \reef{PTn} in a simple way.
\exercise{}{Rewrite the expression \reef{PT4} to show that the 4-gluon amplitude  can also be written 
\be
  A_4[1^-2^-3^+4^+] = \frac{[34]^4}{[12][23][34][41]} \,.
\ee
}
\exercise{}{Convince yourself that in general if all helicities are flipped $h_i \to -h_i$, then the resulting amplitude $A_n[1^{h_1}2^{h_2 } \ldots n^{h_n}]$ is obtained from $A_n[1^{-h_1}  2^{-h_2 } \ldots n^{-h_n}]$ by exchanging all angle  and square brackets.
}
The color-ordered amplitudes have a number of properties worth noting: 1) \emph{Cyclic:} It follows from the trace-structure that $A_n[12 \dots n] = A_n[2 \dots n\, 1]$ etc; 2) \emph{Reflection}: $A_n[12 \dots n] = (-1)^n A_n[n\dots2 \dots 1]$. Convince yourself that this is true. There is also the 3) \emph{$U(1)$ decoupling identity:} 
\be
  \label{u1dec}
  A_n[123 \dots n] + A_n[2 1 3\dots n] + A_n[2 3 1\dots n] + \cdots
  + A_n[2 3\dots 1\, n]  ~=~ 0\,.
\ee
The vanishing of this sum of $n-1$ color-ordered amplitudes is also called the photon decoupling identity. It follows from taking one of the generators $T^a$ proportional to the identity matrix.
\exercise{}{Use \reef{PTn} to show explicitly that \reef{u1dec} holds for $n=4$ for the case where gluons 1 and 2 have negative helicity and 3 and 4 have positive helicity.}
The trace-basis \reef{Anfull} is overcomplete and that implies that there are further linear relations among the partial \emph{tree-level} amplitudes: these are called the {\bf \em Kleiss-Kuijf relations} \cite{KK,DDM} and they can be written \cite{BCJ}
\be
 \label{KKn}
 A_n[1,\{\a\},n,\{\b\}] ~=~ (-1)^{|\b|} \!\!\!\!\!
 \sum_{\sigma \in \text{OP}(\{\a\},\{\b^T\})}
 A_n[1,\sigma,n] \,,
\ee
where $\{\b^T\}$ denotes the reverse ordering of the labels $\{\b\}$ and 
the sum is over ordered permutations ``OP", namely permutations of the labels 
in the joined set $\{\a\} \cup \{\b^T\}$ such that the ordering within $\{ \a \}$ and  $\{\b^T\}$ is preserved. The sign on the RHS is determined by the number of labels $|\b|$ in the set $\{\beta\}$. 

To make \reef{KKn} a little less scary, take the 5-point case as an example. Taking the LHS of \reef{KKn} to be $A_5[1,\{2\},5,\{3,4\}]$, we have 
$\{\a\} \cup \{\b^T\} = \{2\} \cup \{4,3\}$, so the sum over ordered permutations is over  
$\sigma = \{243\}, \{423\}, \{432\}$. Thus the Kleiss-Kuijf relation reads
\be
\label{KKn5a}
A_5[12534] = A_5[12435] + A_5[14235] + A_5[14325] \,. 
\ee
\exercise{}{Show that for $n=4$, the Kleiss-Kuijf relation \reef{KKn} is equivalent to the $U(1)$ decoupling relation.}
\exercise{}{Start with $A_5[1,\{2,3\},5,\{4\}]$ to show that the Kleiss-Kuijf relation gives 
\be
  \label{KKn5b}
  A_5[12345] + A_5[12354] + A_5[12435]  + A_5[14235] = 0\,.
\ee
Show then that \reef{KKn5b} together with the $U(1)$ decoupling relation implies that \reef{KKn5a}\,.   
  }
The Kleiss-Kuijf relations combine with the other identities we have mentioned to reduce the number independent $n$-gluon tree amplitudes to $(n-2)!$. However, there are further linear relationships, called the (fundamental) {\bf \em BCJ relations} --- named after Bern, Carrasco and Johansson \cite{BCJ} ---  that reduce the number of independent $n$-gluon color-ordered tree amplitudes to 
$(n-3)!$. Examples of 4-point and 5-point BCJ amplitude relations are
\bea
  \label{BCJn4}
 s_{14} A_4[1234] - s_{13} A_4[1243]  = 0\,,\\[1mm]
  \label{BCJn5}
 s_{12} A[21345]- s_{23} A[13245] - (s_{23} + s_{24}) A[13425] = 0\,.
\eea
In Section \ref{s:BCJ}, we show that the number of independent color-ordered tree amplitudes under Kleiss-Kuijf relations is $(n-2)!$ and we also discuss the origin of BCJ amplitude relations.
\exercise{}{Use the Parke-Taylor formula \reef{PTn} to verify  \reef{KKn5b}, \reef{BCJn4}, and \reef{BCJn5}.
}
\exercise{ex:earlyBCJ}{Let us get a little preview of the BCJ relations. 
Suppose we use the color-basis \reef{csctcu} to write 
the full 4-point gluon amplitude as
\be
  \label{A4ccc}
  A^\text{full,tree}_4 = \frac{n_s \, c_s}{s}
  +\frac{n_t \, c_t}{t}
  + \frac{n_u \, c_u}{u}
\ee
for some numerator factors $n_i$ that in general depend on the kinematic variables and the polarizations. Write each $c_i$ in terms of the three traces $\Tr (T^{a_1} T^{a_2} T^{a_3} T^{a_4})$ and those with orderings 1243 and 1324. (Make sure to check that the Jacobi identity \reef{csctcuID} holds.) Then use your expressions to convert \reef{A4ccc} to a basis with those three traces. 

Now use the cyclic and reflection properties of the trace and the color-ordered amplitudes to write the full amplitude $A^\text{full,tree}_4$ in \reef{A4full} in terms of the traces with the same three orderings $1234$, $1243$, and $1324$. 

Comparing the resulting expressions for $A^\text{full,tree}_4$, read off the relationship between the numerator factors $n_i$ and the color-ordered amplitudes. You should find 
\be
  A_4(1234) = 
  -\frac{n_s}{s} + \frac{n_u}{u}
\ee
and two similar expressions for $A_4(1243)$ and $A_4(1324)$.
Show that it follows directly from these expressions that the color-ordered amplitudes satisfy the $n=4$ photon decoupling relation \reef{u1dec}. 

Note that the numerator factors $n_i$ are not unique. 
Suppose that there is a choice of numerator factors $n_i$ that satisfy the same relation as the color-factors $c_i$,
\be
  \label{nsntnuID}
  n_s + n_t + n_u = 0\,.
\ee
Show that \reef{nsntnuID} implies that the color-ordered amplitudes satisfy the BCJ relation \reef{BCJn4}.

The existence of numerator factors  $n_i$  that satisfy the same identity \reef{nsntnuID} as the corresponding color-factors is called \emph{color-kinematics duality}. It 
has been of huge interest and applicability in recent studies of amplitudes in both gauge theory and gravity, and we will discuss it further in Section \ref{s:BCJ}.
}	
We end this section with a quick look at interactions between gluons and fermions. Adding
\be
  \lag ~=~ i \overline{\Psi} \ga^\m D_\m \Psi 
    ~=~ i \overline{\Psi} \ga^\m \pa_\m \Psi 
        + \frac{g}{\sqrt{2}}A_\m \overline{\Psi} \ga^\m \Psi 
\ee
to the Yang-Mills Lagrangian, we now acquire a fermion-fermion-gluon 3-vertex $i\frac{g}{\sqrt{2}} \ga^\m$. If the fermion represents a quark, $\Psi$ transforms in the fundamental of the gauge group $SU(N)$. In that case the trace-structure of the amplitudes is a little different, for example for the case of scattering 2 quarks with $n$ gluons, we get 
$(T^{a_1} T^{a_1} \dots T^{a_n})_i{}^j$.

If we want to study the interactions of gluons with their supersymmetric partners, the gluinos, then the fermion field must transform in the adjoint so we replace $\Psi$ with $\lambda = \lambda^a T^a$ and include a trace in the Lagrangian. The trace-structure for gluon-gluino scattering is exactly the same as for gluon scattering. 

We have by now seen enough examples of how to use spinor helicity formalism in the context of standard Feynman rules. It is about time that we get a little fancier. Therefore we postpone further discussion of Yang-Mills and super Yang-Mills amplitudes until we have developed a few more tools.

%%%%%%%%%%%%%%%%%%%%
%%%%%%%%%%%%%%%%%%%%
%%%%%%%%%%%%%%%%%%%%
\subsection{Little group scaling}
\label{s:littlegrp}
We have introduced $|p\>$ and $|p]$ as solutions to the massless Weyl equation, $p |p\> = 0$ and $p |p] = 0$ for $p^2=0$. Their relation to $p^\m$ was given in $\reef{ppp}$ as $p_{a \db} = - |p]_a \<p|_{\db}$. It is useful to note that these relations are invariant under the scaling
\be
  \label{lgrp}
  |p\>  \to t |p\> \,,
  ~~~~~~
  |p] \to t^{-1} |p] \,.
\ee
This is called {\bf \em little group scaling}.
Recall that the  {\em little group} is the group of transformations that leave the momentum of an on-shell particle invariant. 
For a massless particle, we can go to a frame where
$p^\m = (E,0,0,E)$. Rotations in the $xy$-plane leaves the vector invariant, so the little group representations\footnote{More precisely, the little group is $E(2)$, the group of transformations that map a 2d plane into itself. This is similar to the more familiar $SU(2)$ group, whose generators $J_+$ and $J_-$ can be identified as the two translation generators of the little group and $J_z$ can be identified with the rotation generator. Thus, just as in $SU(2)$ where representations are characterized by their $J_z$ eigenvalue, representations of the little group $E(2)$ are characterized by their spin under the 2-dimensional rotation group $SO(2)=U(1)$.} are characterized by $SO(2)=U(1)$. In the angle and square spinor representation of the momentum, the little group transformation is realized as the scaling \reef{lgrp}: for real momenta, $t$ has to be a complex phase such that $|p]^* = \<p|$ is preserved. For complex momenta, the angle and square spinors are independent so we can be more generous and let $t$ be any non-zero complex number.

Now let us consider {\em what an amplitude is made of}: each Feynman diagram consists of propagators, vertices and external line rules. When only massless particles are involved, the amplitude can always be rewritten in terms of angle and square brackets. But note that neither propagators nor vertices can possibly scale under little group transformations. Only the {\em external line rules} scale under \reef{lgrp}: 
\begin{itemize}
\item The scalar rule is a constant factor 1: it does not scale.
\item  Angle and square spinors for (Weyl) fermions: scale as $t^{-2h}$ for $h=\pm \tfrac{1}{2}$.
\item  Polarization vectors for spin-1 bosons. You can directly check \reef{shpolar} to see that under little group scaling of $|p\>$ and $|p]$, the polarization vectors $\eps_{\pm}^\m(p;q)$ scale as $t^{-2h}$ for $h=\pm 1$. They do not scale under scaling of the reference spinor.
\end{itemize}
Thus, for an amplitude of massless particles\footnote{For spin-3/2, one uses $v_\pm \eps_{\pm}^\m$, and for a spin-2 graviton, the polarization is $e^{\m\n}_\pm = \eps^\m_\pm \eps^\n_\pm$ to confirm the little group scaling.} only,
we have the following powerful result. Under little group scaling of each particle $i=1,2,\dots,n$, the on-shell amplitude transforms homogeneously with
weight $-2h_i$, where $h_i$ is the helicity of particle $i$:
\be
   \label{littlegrp}
\boxed{
   ~A_n\big( \{  |1\> , |1] ,h_1 \}, 
   \dots, \{ t_i |i\> , t_i^{-1} |i] ,h_i \} ,\dots
   \big)
   ~=~
   t_i^{-2h_i}  A_n\big( 
    \dots\{  |i\> , |i] ,h_i \}\dots \big) \, .~}
\ee

As an example, consider the QED amplitude \reef{exQEDIIb},
$A_3\big(f^- \bar{f}^+ \gamma^- \big)  ~=~
  \tilde{e}\, \,\tfrac{\<13\>^2}{\<12\>\,\,}$. 
  For the negative helicity photon (particle 3) we get $t_3^2 = t_3^{-2(-1)}$. Likewise, one confirms the scaling \reef{littlegrp} for the two fermions. 
  In fact, \emph{all massless 3-particle amplitudes are completely fixed by little group scaling!} Let's now see how.
  
\compactsubsection{3-particle amplitudes}
Recall that by 3-particle special kinematics, an on-shell 3-point amplitude with massless particles can only depend on either angle or square brackets of the external momenta. Let us suppose that it depends on angle brackets only. We can then write a general Ansatz
\be
  A_3(1^{h_1} 2^{h_2} 3^{h_3}) =
  c \<12\>^{x_{12}}  \<13\>^{x_{13}}  \<23\>^{x_{23}} \,.
\ee
Little group scaling \reef{littlegrp}  fixes
\be
  -2 h_1 = x_{12} + x_{13}\,,
  ~~~~
  -2 h_2 = x_{12} + x_{23}\,,
  ~~~~
  -2 h_3 = x_{13} + x_{23}\,.
\ee
This system is readily solved to find 
$x_{12} = h_3-h_1-h_2$ etc.~so that
\be
  \label{3pt}
  A_3(1^{h_1} 2^{h_2} 3^{h_3}) =
  c \<12\>^{h_3-h_1-h_2}  \<13\>^{h_2-h_1-h_3}  \<23\>^{h_1-h_2-h_3} \,.
\ee
This means that the helicity structure uniquely fixes the 3-particle amplitude up to an overall constant! This may remind you of a closely related fact, namely that in a conformal field theory, the 3-point correlation functions are determined uniquely (up to a multiplicative constant) by the scaling dimensions of the operators. 

We already confirmed \reef{3pt} for $A_3\big(f^- \bar{f}^+ \gamma^- \big)$. So let's do something different. Consider a 3-gluon amplitude with two negative and one positive helicity gluons. By \reef{3pt}, the kinematic structure is uniquely determined:
\be
   \label{A3YMgood}
   A_3(g_1^- g_2^- g_3^+) =  g_\text{YM} \,\frac{\<12\>^3}{\<13\> \<23\>}\,.
\ee
This matches our calculation \reef{A3YMmmp}. But --- there is perhaps a small catch in our little group scaling argument. We {\em assumed} that the amplitude depended only on angle brackets. What if it only depended on square brackets?
Then the scaling would have been the opposite, so we would have found
\be
   \label{A3YMbad}
   A_3(g_1^- g_2^- g_3^+) =  g' \, \frac{[13] [23]}{[12]^3}\,.
\ee

To distinguish between \reef{A3YMgood} and \reef{A3YMbad}, we use {\bf \em dimensional analysis}. From \reef{angxsq} we note that both angle and square brackets have mass-dimension 1. Thus the momentum dependence in \reef{A3YMgood} is (mass)$^1$; this is compatible with the fact that it comes from the $AA\pa A$-interaction in $\Tr F_{\m\n}F^{\m\n}$. However, in \reef{A3YMbad}, the momentum dependence has mass-dimension (mass)$^{-1}$, so it would somehow have to come from an interaction of the form $g' AA \tfrac{\pa}{\Box}A$. Of course, we have no such interaction term in a {\bf \em local} Lagrangian; hence we discard the expression \reef{A3YMbad} as unphysical.

The combination of {\bf \em little group scaling} and {\bf \em locality} uniquely fixes the massless 3-particle amplitudes. As we will see in the next Section,  that can be enough to determine all other tree level amplitudes in {\em some} theories. 

While we are considering dimensional analysis, it is worth making a couple of other observations. First, note that while the Yang-Mills coupling is dimensionless, the coupling $g'$ in the $g' AA \tfrac{\pa}{\Box}A$  has dimension (mass)$^2$. This means that the RHS of \reef{A3YMbad} has mass-dimension 1, just as the correct expression \reef{A3YMgood}. This is sensible since the two amplitude-expressions better have the same mass-dimension. In general,
\be 
\text{\emph{an $n$-particle amplitude in $d=4$ must have mass-dimension $4-n$.}}
\label{massdim4d}
\ee
This follows from dimensional analysis since the cross-section must have dimensions of area. You can also check it by direct inspection of the Feynman diagrams.

One more comment about \reef{A3YMgood}: you'll immediately be worried that the expression on the RHS is not Bose-symmetric in the exchange of the identical gluons 1 and 2. Fear not. The full 3-point amplitude of course comes dressed with a fully-antisymmetric group theory factor $f^{a_1a_2a_3}$: this restores Bose-symmetry. As discussed in Section \ref{s:YM}, the kinematic structure in \reef{A3YMgood} is exactly that of the color-ordered 3-point amplitude $A_3[1^- 2^- 3^+]$.

\exercise{}{Write down spinor-helicity representations of the possible color-ordered 3-point amplitudes for 
the interaction of 2 gluinos (massless spin-1/2) with a gluon in super Yang-Mills theory.}
\exercise{ex:game}{Let's play a little game. Suppose someone gives you the following amplitudes for scattering processes involving massless particles:
\bea
  \text{(a)} && A_5 = g_a\, \frac{[13]^4}{[12][23][34][45][51]}
  \,,\\[1mm]
  \text{(b)} && A_4  = g_b\, \frac{\<14\> \<24\>^2}{\<12\>\<23\>\<34\>}
  \,,\\[1mm]
  \text{(c)} && A_4  = g_c\, \frac{\<12\>^7 [12]}{\<13\>\<14\>\<23\>\<24\>\<34\>^2}
  \,.
\eea
With all particles outgoing, what are the helicities of the particles?\\
What is the dimension of the couplings $g_i$ relevant for the interactions?\\
In each case, try to figure out which theory could produce such an amplitude.
}
\example{What about a gluon amplitude with all-negative helicities? Well, let's do it. The formula \reef{3pt} immediately tells us that
\be
   \label{A3F3}
   A_3(g_1^- g_2^- g_3^-) =  a  \, \<12\> \<13\> \<23\> \,.
\ee
The mass-dimension 3 of the kinematic part reveals that ($i$) the coupling $a$ must have mass-dimension $-2$ for the whole amplitude to have mass-dimension $4-3=1$, and ($ii$) this must come from a Lagrangian interaction term with 3 derivatives, i.e.~$(\pa A)^3$. Furthermore, the kinematic terms are antisymmetric under exchanges of gluon-momenta, so Bose-symmetry tells that the couplings must be associated with antisymmetric structure constants --- as is of course the case for a non-abelian  gauge field. Thus there is a natural candidate, namely the dimension-6 operator $\Tr F^\m{}_\n F^\n{}_\l F^\l{}_\m$. Indeed this operator produces the amplitude \reef{A3F3}; but we can also conclude that in pure Yang-Mills theory or in QED, $A_3(g_1^- g_2^- g_3^-)=0$.} 
\exercise{ex:grav3pt}{Let us look at gravity scattering amplitudes. If we expand the Einstein-Hilbert action $\frac{1}{2\kappa^2} \int d^4 x \sqrt{-g}R$ around flat space $g_{\m\n} = \eta_{\m\n}+\kappa \,h_{\m\n}$, we obtain an infinite series of 2-derivative interactions involving $n$ fields $h_{\m\n}$ for any $n$. This makes it very complicated to calculate graviton scattering amplitudes using Feynman rules. (Gravitons are massless spin-2 particles; they have 2 helicity states, $h=\pm2$.) For now just focus on the 
3-point amplitude: use little group scaling to write down the result for the on-shell 3-graviton amplitudes. Check the mass-dimensions. Compare your answer with the 3-gluon amplitudes. 
}
\exercise{ex:grav3ptR3}{Consider in gravity an operator constructed from some contraction of the indices of three Riemann-tensors; we'll denote it $R^3$. If we linearize the metric around flat space, $g_{\m\n} = \eta_{\m\n}+\kappa \,h_{\m\n}$, then we can calculate graviton scattering associated with $R^3$. What is the mass-dimension of the coupling associated with $R^3$? Use little group scaling to determine $A_3(h_1^- h_2^- h_3^-)$ and $A_3(h_1^- h_2^- h_3^+)$.}
\exercise{}{Consider a the dimension-5 Higgs-gluon fusion operator $H \Tr F_{\m\n} F^{\m\n}$. Use little group scaling to determine the 3-particle amplitudes of this operator in the limit of $m_H = 0$.
(For more about on-shell methods and Higgs-gluon fusion, see \cite{alsogluonfusion}.)
} 
\example{Consider a 3-point amplitude with three scalars. We learn from \reef{3pt} that there can be no momentum dependence in the amplitude, $A_3(\phi \phi \phi)=$constant. This is of course compatible with a $\phi^3$-interaction, but what about throwing in some derivatives, as in a non-linear sigma model? --- something like  
$\phi\, \pa_\m \phi\, \pa^\m \phi$. Well, this is readily rewritten as
$\tfrac{1}{2}\pa_\m (\phi^2)\, \pa^\m \phi $, and  by partial integration this gives
$-\tfrac{1}{2} \phi^2\, \Box \phi$. On-shell this clearly vanishes for massless scalars. Now your turn: why does the 3-particle on-shell amplitude for 3 distinct massless scalars, 
e.g.~$\phi_1\, \pa_\m \phi_2\, \pa^\m \phi_3$, vanish? 
}

%%%%%%%%%%%%%%%%
%%%%%%%%%%%%%%%%
%%%%%%%%%%%%%%%%
\subsection{Fun with polarization vectors --- the MHV classification}
\label{s:MHV}
In this section we  return to the study of gluon scattering amplitudes. The Yang-Mills 
lagrangian contains two types of interaction terms, schematically
\be
  \tr F_{\m\n}F^{\m\n} 
  ~~~\longrightarrow~~~
  AA \pa A ~+~ A^4 \,.
\ee
In a typical gauge, such as Feynman gauge or Neveu-Gervais, this gives rise to Feynman rules with two types interaction vertices: the cubic vertex which depends linearly on the momenta and the quartic vertex which is independent of the momenta. Since the coupling is dimensionless, the cubic vertex is $O(\text{mass}^1)$ and the quartic is $O(\text{mass}^0)$. 

Consider tree diagrams with only cubic vertices, i.e.~trivalent tree-graphs, with $n$ external legs. If you start with a 3-point vertex ($n=3$) you can easily convince yourself that every time you add an extra external line, you have to add both a new vertex and a new propagator to keep the graph trivalent. Hence the number of vertices and propagators both grow linearly with $n$, and it takes just a few examples to see that the number of vertices is $n-2$ and the number of propagators is $n-3$. Since the cubic vertices are $O(\text{mass}^1)$ and the propagators are $O(\text{mass}^{-2})$, we find that the mass-dimension of the diagrams, and hence of the amplitude, is
\be
  \label{massdimAn}
  [ A_n ] ~\sim~ \frac{(\text{mass})^{n-2}}{(\text{mass}^2)^{n-3}} 
  ~\sim~  
  (\text{mass})^{4-n} \,.
\ee
This confirms the statement we made in the previous section. Any diagram with a mix of cubic and quartic vertices has the same mass-dimension of $ (\text{mass})^{4-n}$. But note that the number of powers of momenta in the numerator cannot exceed $n-2$; this point will be useful shortly.

Consider now the schematic form of a gluon tree amplitude:
\be
  \label{GenGlueAmp}
  A_n  \sim \sum_\text{diagrams}
   \frac{\sum \big( \prod (\eps_i.\eps_j) \big) \big( \prod (\eps_i.k_j) \big) \big( \prod (k_i.k_j) \big) }{\prod P_I^2}
\ee
i.e.~the diagrams have numerators that are some Lorentz scalar products of polarizations and momentum vectors, and in the denominators are products of momentum invariants from the propagators. 

Perhaps you know the statement that all-plus tree gluon amplitudes vanish, $A_n(1^+2^+\dots n^+)=0$? We have already seen it in exercises for $n=3,4$. Let us show it for all $n$. First recall from Exercise \ref{ex:polar1} that the polarization vector dot-products are
\be
  \eps_{i+}.\eps_{j_+} \propto \<	q_i q_j\>\,,
  ~~~~~
  \eps_{i-}.\eps_{j_-} \propto  [q_i q_j]\,,
  ~~~~~
  \eps_{i-}.\eps_{j_+} \propto  \<	i q_j\>  [j q_i]\,.
\ee
Thus, for an all-plus amplitude, we can choose all $q_i$ to be the same $q$. Then 
$\eps_{i+}.\eps_{j_+} = 0$. That means that the only way the $n$-gluon propagators can enter in the numerator of \reef{GenGlueAmp}, is as $\eps_{i+}.k_j$. We need to absorb the Lorentz indices of all $n$ polarization vectors, so that requires $n$ powers of momenta in the numerator. But as we have argued below \reef{massdimAn}, no more than $n-2$ powers of momenta is possible in any gluon tree diagram. Hence we conclude that $A_n(1^+2^+\dots n^+)=0$.

Note that if we had not known to write down a smart choice of the polarization vectors, but had worked with general expressions, we would have had to work very hard to prove that the sum of combinatorially many $n$-point tree diagrams in the all-plus amplitude add up to zero. 

Next, let us flip one of the helicities and consider an amplitude $A_n(1^-2^+\dots n^+)$. This time, choose $q_2= q_3= \dots = q_n = p_1$. This achieves $\eps_{i+}.\eps_{j_+} = 0$ and $\eps_{1-}.\eps_{j+} = 0$. So again we would need $n$ factors of $\eps_{i+}.k_j$ in the numerators of \reef{GenGlueAmp}; as before this allows us to conclude that the tree level amplitude vanishes: $A_n(1^-2^+\dots n^+)=0$.

We have shown that 
\be
  \label{allplusetc}
  \text{tree-level gluon ampl:}
  ~~~~~
  A_n(1^+2^+\dots n^+) = 0
  ~~~~\text{and}~~~~
  A_n(1^-2^+\dots n^+) = 0\, .
\ee
At loop-level, these amplitudes are actually non-vanishing in pure Yang-Mills theory (and can have a quite interesting structure). Can you see  how the argument above is changed at 1-loop level?

Let's move on and flip one more helicity: $A_n(1^-2^- 3^+\dots n^+)$. Let us try to choose the reference $q_i$'s such that as many as possible of the dot-products of polarization vectors vanish. The choice $q_1=q_2 =p_n$ and $q_3 = q_4 = \ldots = q_n = p_1$  implies that all $\eps_{i}.\eps_{j} = 0$ vanish, except $\eps_{2-}.\eps_{i+}$ for $i=3,\dots,n-1$. The polarization vector of gluon 2 can only appear once, so the terms in \reef{GenGlueAmp} can take the schematic form
\be
  A_n(1^-2^- 3^+\dots n^+) 
  ~ \sim \sum_\text{diagrams}
   \frac{\sum (\eps_{2-}.\eps_{i+})  (\eps_j.k_l)^{n-2}}{\prod P_I^2}
\ee
Since only one product of $\eps_i^\m$'s can be non-vanishing, $n-2$ factors of $(\eps_j.k_l)$ were needed, and this exactly saturates the number of momentum vectors possible by dimensional analysis \reef{massdimAn}.  
Note also that with our choice of polarization vectors, any diagram that contributes to the $A_n(1^-2^- 3^+\dots n^+)$ is trivalent. 

Thus we conclude --- based on dimensional analysis and useful choices of the polarization vectors --- that the $A_n(1^-2^- 3^+\dots n^+)$ is the ``first" gluon amplitude that can be non-vanishing, in the sense that having fewer negative helicity gluons gives a vanishing amplitude. More negative helicity states are also allowed, but one needs at least two positive helicity states to get a non-vanishing result, except for $n=3$.  

The amplitudes $A_n(1^-2^- 3^+\dots n^+)$ are called 
{\bf \em Maximally Helicity Violating} --- or simply {\bf \em MHV} for short.\footnote{The name ``Maximally Helicity Violating" comes from thinking of $2 \to (n-2)$ scattering. By crossing symmetry, an outgoing gluon with 
$\big\{ \begin{matrix} 
\text{negative} \\[-1.5mm] \text{positive}
\end{matrix}\big\}$ 
helicity is an incoming gluon with 
$\big\{ \begin{matrix} 
\text{positive} \\[-1.5mm] \text{negative}
\end{matrix}\big\}$ 
 helicity. So with all outgoing particles, the process $A_n[1^+2^+ 3^+\dots n^+]$ crosses over to $1^-2^- \to 3^+\dots n^+$ in which the outgoing states all have the opposite helicity of the incoming states; it is `helicity violating'. The process $1^-2^- \to 3^-4^+\dots n^+$ is a little less helicity violating and it crosses to $A_n[1^+2^+ 3^-4^+\dots n^+]$. We know from the above analysis that both these `helicity violating' processes vanish at tree-level in pure Yang-Mills theory. The process 
 $1^-2^- \to 3^-4^- 5^+\dots n^+$ --- equivalent to $A_n[1^+2^+ 3^- 4^- 5^+\dots n^+]$ --- is the most we can `violate' helicity and still get a non-vanishing answer at tree-level: therefore it is \emph{maximally helicity violating}.
}
The MHV gluon amplitudes are the simplest amplitudes in Yang-Mills theory. 
The next-to-simplest amplitudes are called Next-to-MHV, or {\bf NMHV}, and this refers to the class of amplitudes with $3$ negative helicity gluons and $n-3$ positive helicity gluons. This generalizes to the notation 
{\bf N$^K$MHV} amplitudes with $K+2$ negative helicity gluons and $n-K-2$ positive helicity gluons. When an amplitude has $(n-2)$ gluons of negative helicity and $2$ of positive helicity, it is called {\bf anti-MHV}. Anti-MHV is obtained from the MHV amplitude with all helicities flipped by exchanging angle brackets with square brackets. 
The result \reef{allplusetc} is actually true at any loop-order in super Yang-Mills theory. We will see why in Section \ref{s:susy} where the MHV-classification is also discussed further. 
For now, it is time for on-shell recursion relations. Go ahead to Section 3.

%%%%%%%%%%%%%%%%%%%%%%%%%%%%%%% 
%%%%%%%%%%%%%%%%%%%%%%%%%%%%%%% 
%%%%%%%%%%%%%%%%%%%%%%%%%%%%%%% 
\newpage
\setcounter{equation}{0}
\section{On-shell recursion relations at tree-level}
\label{s:recrels}
%%%%%%%%%%%%%%%%%%%%%%%%%%%%%%% 
%%%%%%%%%%%%%%%%%%%%%%%%%%%%%%% 
%%%%%%%%%%%%%%%%%%%%%%%%%%%%%%% 
Recursion relations provide a method for building higher-point amplitudes from lower-point information. In 1988, Berends-Giele developed \emph{off-shell} recursion relations \cite{BGrecrels} to construct $n$-point parton amplitudes from building blocks with one leg off-shell (see the reviews  \cite{Mangano:1990by,dixon}). This off-shell method remains useful as an algorithm for efficient numerical evaluation of scattering amplitudes. In this review, we focus on the newer (2005) recursive methods whose building blocks are themselves \emph{on-shell} amplitudes. These \emph{on-shell recursion relations} are elegant in that they  use input only from gauge-invariant objects and they have proven very powerful for elucidating the mathematical structure of on-shell scattering amplitudes. 

In the modern approaches, a key idea is to use the power of complex analysis and exploit the analytic properties of on-shell scattering amplitudes. The 
derivation of on-shell recursion relations is a great example of this, as we shall see soon. The most famous on-shell recursion relations are the  ``BCFW recursion relations" by Britto, Cachazo, Feng, and Witten \cite{bcf,bcfw}, but there are other versions based on the same idea as BCFW, namely the use of complex deformations of the external momenta. We describe this idea here, first in a very general formulation (Section \ref{s:shifts}), then specialize the results to derive the BCFW recursion relations (Section \ref{s:bcfw}). We illustrate the BCFW methods with a selection of examples, including an inductive proof of the Parke-Taylor formula \reef{PTn}. Section \ref{s:validity} contains a discussion of when to expect existence of recursion relations in general local QFTs. Finally, in Section \ref{s:csw} we outline the CSW construction (Cachazo-Svrcek-Witten \cite{cswref}), also called the MHV vertex expansion.

%%%%%%
\subsection{Complex shifts \& Cauchy's theorem}
\label{s:shifts}
An on-shell amplitude $A_n$ is characterized by the momenta of the external particles and their type (for example a helicity label $h_i$ for massless particles). We focus here on massless particles so $p_i^2 = 0$ for all $i=1,2,\dots,n$. Of course, momentum conservation  \mbox{$\sum_{i=1}^n p_i^\m = 0$} is  also imposed.

Let us now introduce $n$ complex-valued vectors $\refr^\m_i$ (some of which may be zero) such that
\begin{enumerate}
\item[(i)]~~ $\displaystyle \sum_{i=1}^n \refr_i^\m = 0$\,,
\item[(ii)]~~ $\refr_i \cdot \refr_j= 0$ for all $i,j=1,2,\dots,n$. In particular $\refr_i^2 = 0$\,, and
\item[(iii)]~~ $p_i\cdot \refr_i = 0$ for each $i$ (no sum).
\end{enumerate}
These are used to define $n$ shifted momenta
\be
   \label{pshift}
   \hat{p}_i^\m ~\equiv~ p_i^\m + z\, \refr_i^\m\
   \hspace{7mm}
   \text{with ~~$z \in \mathbb{C}$\,.}
\ee
Note that
\begin{itemize}
\item[(A)] By property (i), momentum conservation holds for the shifted momenta:
~$\displaystyle \sum_{i=1}^n\hat{p}_i^\m=0$.
\item[(B)] By (ii) and (iii), we have $\hat{p}_i^2 = 0$, so the shifted momenta are on-shell. 
\item[(C)] For a non-trivial\footnote{Non-trivial means at least two and no more than $n\!-\!2$ momenta such that $P_I^2\ne 0$.}  subset of generic momenta $\{{p}_i\}_{i \in I}$, define $P_I^\m = \sum_{i \in I} {p}_i^\m$. Then $\hat{P}_I^2$  is \emph{linear} in $z$:
\be 
\hat{P}_I^2 = \big(\sum_{i \in I} \hat{p}_i \big)^2 
= P_I^2 
  + z \,2 \,P_I \cdot R_I \,
  ~~~~~\text{with}~~~~
  R_I= \sum_{i \in I} \refr_i \,,
\ee
because the $z^2$ term vanishes by property (ii).
We can write  
\be
  \label{hatPI}
  \hat{P}_I^2
   = -\frac{P_I^2}{z_I}
      (z-z_I)\,
      ~~~~~~
      \text{with}
      ~~~~
      z_I = -\frac{P_I^2}{2 P_I \cdot R_I} \,.
\ee
\end{itemize}
As a result of (A) and (B), we can consider our amplitude $A_n$ in terms of the shifted momenta $\hat{p}_i^\m$ instead of the original momenta $p_i^\m$. In particular, it is useful  to study the shifted amplitude as a function of $z$; by construction it is holomorphic, $\hat{A}_n(z)$. The amplitude with unshifted momenta $p_i^\m$ is obtained by setting $z=0$, $A_n = \hat{A}_n(z=0)$.

We specialize to the case where $A_n$ is a {\bf \em tree-level amplitude}. In that case, the analytic structure of $\hat{A}_n(z)$ is very simple. For example, it does not have any branch cuts --- there are no log's, square-roots, etc, at tree-level. Its analytic structure is captured by its poles, and it can have only simple poles. To see this, consider the Feynman diagrams: the only places we can get poles is from the shifted propagators $1/\hat{P}_I^2$, where $\hat{P}_I$ is a sum of a nontrivial subset of the shifted momenta. By (C) above, $1/\hat{P}_I^2$ gives a simple pole at $z_I$, and for generic momenta $z_I \ne 0$. For generic momenta, no Feynman tree diagram can have more than one power of a given propagator $1/\hat{P}_I^2$; and poles of different propagators are located at different positions in the $z$-plane. Hence, for generic momenta,  $\hat{A}_n(z)$ only has simple poles and they are all located away from the origin. Note the implicit assumption of locality, i.e.~that the amplitudes can be derived from some local Lagrangian: the propagators determine the poles.

Let us then look at $\frac{\hat{A}_n(z)}{z}$ in the complex $z$-plane. Pick a contour that surrounds the simple pole at the origin. The residue at this pole is nothing but 
the unshifted amplitude, $A_n = \hat{A}_n(z=0)$. Deforming the contour to surround all the other poles, Cauchy's theorem tells us that
\be
   \label{AnRes}
   A_n = - \sum_{z_I} \text{Res}_{z=z_I} \frac{\hat{A}_n(z)}{z}  + B_n\,,
\ee
where $B_n$ is the residue of the pole at $z=\infty$. By taking $z \to 1/w$ it is easily seen that $B_n$ is the $O(z^0)$ term in the $z\to \infty$ expansion of $A_n$.

Now, then, so what? Well, at a $z_I$-pole the propagator $1/\hat{P}_I^2$ goes on-shell. In that limit, the shifted amplitude \emph{factorizes} into two on-shell parts, $\hat{A}_\text{L}$ and $\hat{A}_\text{R}$. Using \reef{hatPI}, we find
\be
  \label{resi}
   \text{Res}_{z=z_I} \frac{\hat{A}_n(z)}{z}
   ~=~ - \hat{A}_\text{L}(z_I)\, \frac{1}{P_I^2}\,\hat{A}_\text{R}(z_I) 
   ~=~
   \raisebox{-5.5mm}{\includegraphics[height=1.4cm]{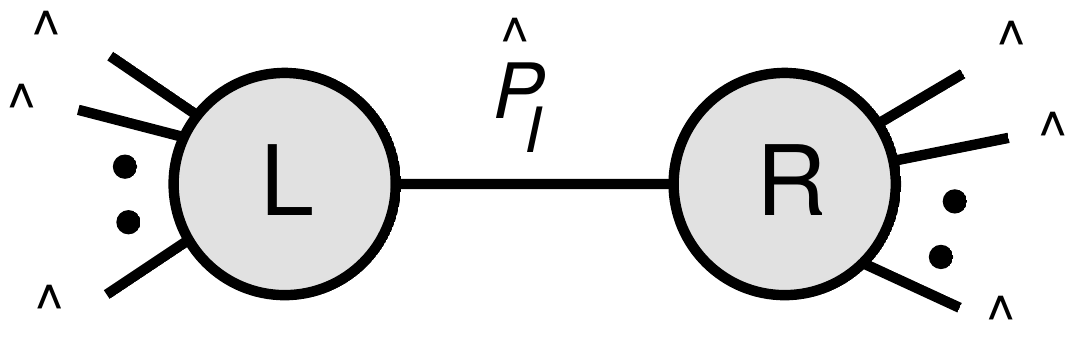}}
   \,.
\ee
Note that --- as opposed to Feynman diagrams --- the momentum of the internal line in \reef{resi} is on-shell, $\hat{P}_I^2=0$, and the vertex-blobs represent shifted \emph{on-shell amplitudes} evaluated at $z=z_I$; we call them \emph{subamplitudes}. The rule for the internal line in the diagrammatic representation \reef{resi} is to write the scalar propagator $1/P_I^2$ of the \emph{unshifted} momenta.
Each subamplitude necessarily  involves fewer than $n$ external particles, hence all the residues at finite $z$ can be determined in terms of on-shell amplitudes with less then $n$ particles. This is the basis of the recursion relations. 

The contribution $B_n$ from the pole at infinity has in general no similar expression in terms of lower-point amplitudes; there has recently been various approaches to try to compute the form of $B_n$ systematically (see for example \cite{Feng:2009ei,Conde:2012ik}), but there is currently not a general constructive method. Thus, in most applications, one assumes --- or, much preferably, proves --- that $B_n = 0$. This is most often justified by demonstrating that 
\be
 \label{largezcond}
\hat{A}_n(z) \to 0 ~~~~ \text{for} ~~~~ z \to \infty. 
\ee  
If \reef{largezcond} holds, we say that the shift \reef{pshift} 
is \emph{valid} (or \emph{good}), and in that case the $n$-point on-shell amplitude is completely determined in terms of lower-point on-shell amplitudes as
\be
   \label{recrel}
    A_n ~= \sum_{\text{diagrams}~I} \hat{A}_\text{L}(z_I) 
    \,\frac{1}{P_I^2}\,
    \hat{A}_\text{R}(z_I)
       ~~=
   \sum_{\text{diagrams}~I}
   \raisebox{-5.5mm}{\includegraphics[height=1.4cm]{recrel1}}
   \,.
\ee 
The sum is over all possible factorization channels $I$. There is also implicitly a sum over all possible on-shell particle states that can be exchanged on the internal line: for example, for a gluon we have to sum the possible helicity assignments.  
The recursive formula \reef{recrel} gives a manifestly gauge invariant construction of scattering amplitudes. This is the general form of the ``on-shell recursion relations" for tree-level amplitudes with the property \reef{largezcond}. We did not use any special properties of $d=4$ spacetime, so the recursion relations are valid in $d$ spacetime dimensions. In the following, we specialize to $d=4$ again.

%%%%%%%%%
\subsection{BCFW recursion relations}
\label{s:bcfw}
Above we shifted all external momenta democratically, but with a parenthetical remark that some of the lightlike shift-vectors $\refr_i^\m$ might be trivial, $\refr_i^\m = 0$. The BCFW shift is one in which exactly  two lines, say $i$ and $j$, are selected as the only ones with non-vanishing shift-vectors. In $d=4$ spacetime dimension, the shift is  implemented on angle and square spinors of the two chosen momenta:
\be
  \label{bcfw}
  |\hat{i}] = |i] + z\,|j]\,,   
  ~~~~~~
  |\hat{j}] = |j]\,,   
  ~~~~~~
  |\hat{i}\> = |i\>\,,
  ~~~~~~
  |\hat{j}\> = |j\> - z |i\> \,.
\ee
No other spinors are shifted. 
We call this a $[i,j\>$-shift. 
Note that $[\hat{i}k]$ and $\<\hat{j}k\>$ are linear in $z$ for $k \ne i,j$ while 
$\<\hat{i} \hat{j}\> = \<ij\>$, $[\hat{i} \hat{j}]=[ij]$, $\<\hat{i}k\>=\<ik\>$, and $[\hat{j}k]=[jk]$ remain unshifted. 
\exercise{}{Use \reef{slashp} to calculate the shift vectors $\refr_i^\m$ and $\refr_j^\m$ corresponding to the shift   \reef{bcfw}. Then show that your shift vectors satisfy the properties (i)-(iii) of the Section \ref{s:shifts}. }
Before diving into applications of the BCFW recursion relations (such as proving the Parke-Taylor amplitude), let us study the shifts a little further. As an example, consider the Parke-Taylor amplitude
\be
  \label{PT12}
   A_n[1^- 2^- 3^+ \ldots n^+]
   ~=~
   \frac{\<12\>^4}{\<12\>\<23\> \cdots \<n1\>} \, .
\ee
First check property \reef{largezcond}:\footnote{Of course, we cannot use the large-$z$ behavior of the formula \reef{PT12} itself to justify the method to prove this formula! A separate argument is needed and will be discussed shortly.}
\exercise{}{Convince yourself that for large-$z$ the amplitude \reef{PT12} falls off as $1/z$ under a $[-,-\>$-shift (i.e.~choose of $i$ and $j$ to be the two negative helicity lines.) What happens under the 3 other types of shifts? Note the difference between shifting adjacent/non-adjacent lines. 
}
\exercise{}{Consider the action of a $[1,2\>$-shift of \reef{PT12}. Identify the simple pole. Calculate the residue of $\hat{A}_n(z)/z$ at this pole. Compare with \reef{AnRes}. What happens if you try to repeat this for a $[1,3\>$-shift?
}
The validity of the BCFW recursion relations  requires that the  boundary term $B_n$ in \reef{AnRes} is absent. The typical approach is to show that 
\be
 \label{largezcond2}
\hat{A}^\text{tree}_{n \text{ gluons}} (z) \to 0 ~~~~ \text{for} ~~~~ z \to \infty. 
\ee  
In pure Yang-Mills theory, an argument \cite{ArkaniHamed:2008yf} based on the background field method establishes the following large-$z$ behavior of color-ordered gluon tree amplitudes under a BCFW shift of adjacent lines $i$ and $j$ of helicity as indicated: 
\be
   \label{largezfallof}
   \begin{array}{cccccc}
   [i,j\> & [-,-\> & [-,+\> & [+,+\> & [+,-\> &
   \\[2mm]
   \hat{A}_n(z) \sim 
   & \ds \frac{1}{z} & \ds \frac{1}{z} & \ds \frac{1}{z} & z^3
   \end{array}
\ee
If $i$ and $j$ are non-adjacent, one gains an extra power $1/z$ in each case.
Thus any one of the three types of shifts $[-,-\>$,  $[-,+\>$, $[+,+\>$ give valid recursion relations for gluon tree amplitudes. 

We are now going to use the BCFW recursion relations to construct an 
{\bf \em inductive proof of the Parke-Taylor formula} \reef{PT12}. 
The formula \reef{PT12} is certainly true for $n=3$, as we saw in Section \ref{s:YM}, and this establishes the base of the induction. For given $n$, suppose that \reef{PT12} is true for amplitudes with less than $n$ gluons. Then write down the recursion relation for $A_n[1^- 2^- 3^+ \ldots n^+]$ based on the valid $[1,2\>$-shift: adapting from \reef{recrel}, we have
\bea
  \label{PTrec1}
  A_n[1^-2^-3^+\dots n^+] &=& 
  \sum_{k=4}^{n}~~
     \raisebox{-6.5mm}{\includegraphics[height=1.6cm]{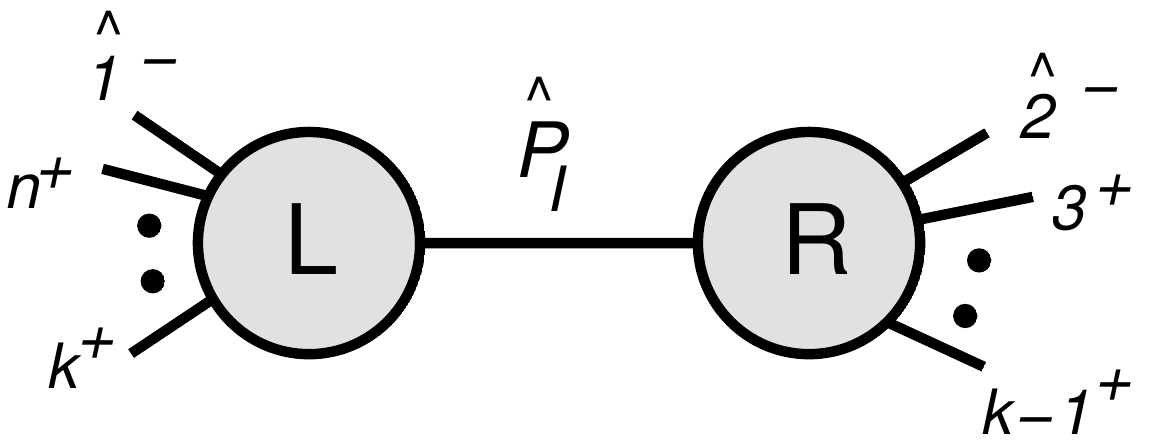}}
     \\[2mm]
   &&  \hspace{-2cm}
   =~
 \sum_{k=4}^{n} \sum_{h_I = \pm} 
  \hat{A}_{n-k+3}\big[ \hat{1}^-, \hat{P}_I^{h_I} , k^+ \ldots , n^+\big] 
  \,\frac{1}{P_I^2}\,
  \hat{A}_{k-1}\big[ -\hat{P}_I^{-h_I} ,  \hat{2}^-, 3^+ \ldots , (k-1)^+\big] \,.  
  \nonumber
\eea
The internal momentum is labelled $P_I$, meaning that for a given $k=4,\dots,n$ we have $P_I = p_2 + p_3 + \dots + p_{k-1}$ and  $\hat{P}_I = \hat{p}_2 + p_3 + \dots + p_{k-1}$. There are no diagrams where lines 1 and 2 belong to the same subamplitude, because in that case, the internal momentum would not be shifted and then there is no corresponding residue in \reef{AnRes}.
Only diagrams that preserved the color-ordering of the external states are included. 
Note that we are also explicitly including the sum over the possible helicity assignments for the particle exchanged on the on-shell internal line: if the exchanged gluon is outgoing from the L subamplitude and has negative helicity, then it will be a positive helicity outgoing gluon as seen from the R subamplitude.

Since one-minus amplitudes $A_n[-+\dots+]$ vanish except for $n=3$, \reef{PTrec1} reduces to
\bea
  \nonumber
  A_n[1^- 2^- 3^+ \ldots n^+]
   &=&
     \raisebox{-6.7mm}{\includegraphics[height=1.6cm]{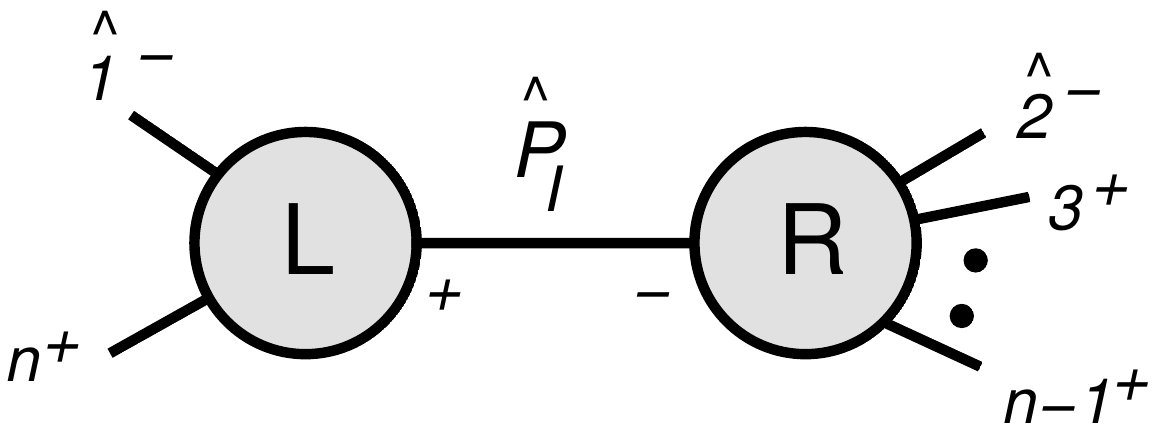}}
     ~~+~~\,
          \raisebox{-6.9mm}{\includegraphics[height=1.6cm]{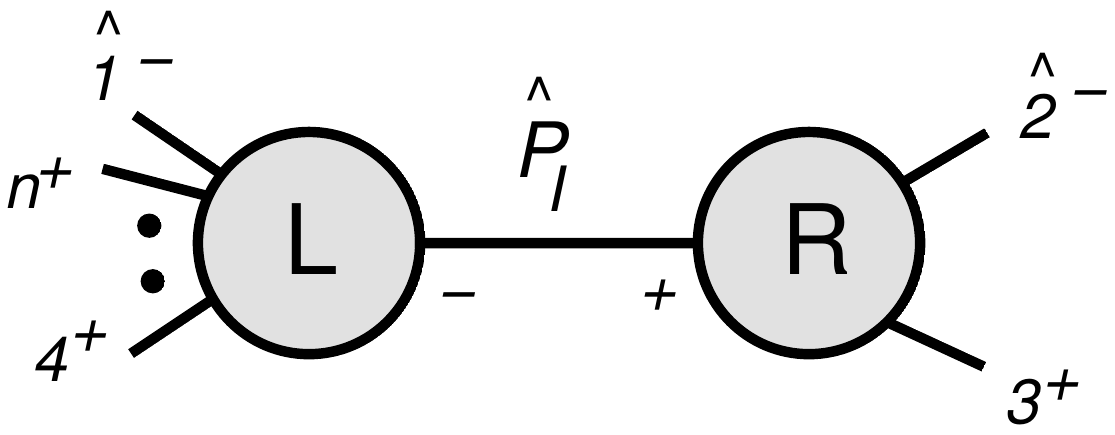}}
     \\[2mm]
     \nonumber
   &&  \hspace{-2cm}
   =~ 
  \hat{A}_{3}\big[ \hat{1}^-, -\hat{P}_{1n}^{+}\, , n^+\big] 
  \,\frac{1}{P_{1n}^2}\,
  \hat{A}_{n-1}\big[ \hat{P}_{1n}^{-}\, ,  \hat{2}^-, 3^+ \ldots (n-1)^+\big] \\[1mm]
  &&
  ~+~
  \hat{A}_{n-1}\big[ \hat{1}^-, \hat{P}_{23}^{-}\, , 4^+ \ldots , n^+\big] 
  \,\frac{1}{P_{23}^2}\,
  \hat{A}_{3}\big[ -\hat{P}_{23}^{+}\, ,  \hat{2}^-, 3^+\big] \,. 
  \label{recrelex1}
\eea
It is here understood that $\hat{P}_I$ is evaluated at the residue value of $z=z_I$ such that $\hat{P}_I^2 = 0$. The notation $P_{ij}$ means $P_{ij} = p_i +p_j$.

The next point is to implement {\bf \em special kinematics} for the 3-point subamplitudes. In the first diagram of \reef{recrelex1}, we have a 3-point anti-MHV amplitude
\be
 \label{A3SK}
   \hat{A}_{3}\big[ \hat{1}^-, -\hat{P}_{1n}^{+}\, , n^+\big]
   ~=~ \frac{[\hat{P}_{1n} \,n]^3}{[n \hat{1}][\hat{1} \hat{P}_{1n}]} \,.
\ee
Here we used the following convention for analytic continuation: 
\be
  \label{acont}
   |-p\> = - |p\> \,,~~~~~
   |-p] = +|p]\,.
\ee
Since $\hat{P}_{1n}^\m = \hat{p}_1^\m + p_n^\m$, the on-shell condition is
\be
  0 =  \hat{P}_{1n}^2 = 2  \hat{p}_1 \cdot p_n =  \<\hat{1} n\> [\hat{1} n]
  =  \<1 n\> [\hat{1} n]\,.
\ee
For generic momenta, the only way for the RHS to vanish is if $[\hat{1} n]=0$. That means that the denominator in \reef{A3SK} vanishes! But so does the numerator: from
\be
  |\hat{P}_{1n}\> [  \hat{P}_{1n}\,n] = -\hat{P}_{1n}|n] 
  \,=\, 
  -(\hat{p}_1 + p_n) |n] 
  \,=\, |1\> [\hat{1}n] \,=\, 0\, , 
\ee
we conclude that $[\hat{P}_{1n} n]=0$ since $|\hat{P}_{1n}\>$ is  not zero. Similarly, one can show that $[\hat{1}  \hat{P}_{1n}]=0$. Thus, in the limit of imposing momentum conservation, all spinor products in \reef{A3SK} vanish; with the 3 powers in the numerator versus the two in the denominator, we conclude that special 3-point kinematics force $\hat{A}_{3}\big[ \hat{1}^-, \hat{P}^{+}_{1n}\, , n^+\big] = 0$. 

The 3-point subamplitude in second diagram of \reef{recrelex1} is also anti-MHV, but it does not vanish, since the shift of line 2 is on the angle spinor, not the square spinor. This way, the big abstract recursion formula \reef{recrel} reduces --- for the case of the $[1,2\>$ BCFW shift of an MHV gluon tree amplitude --- to an expression with just a single non-vanishing diagram:
\bea
\nonumber
A_n[1^- 2^- 3^+ \ldots n^+]
   &=&
          \raisebox{-6.9mm}{\includegraphics[height=1.6cm]{recrel3b}}\\[1mm]
          &=&
   \hat{A}_{n-1}\big[ \hat{1}^-, \hat{P}_{23}^{-} \, , 4^+, \ldots , n^+\big] 
  \,\frac{1}{P_{23}^2}\,
  \hat{A}_{3}\big[ -\hat{P}_{23}^{+}\, ,  \hat{2}^-, 3^+\big]\,.
  \label{bcfwmhv1}
\eea
Our inductive assumption is that \reef{PT12} holds for $(n-1)$-point amplitudes. That, together with the result \reef{A3YMppm} for the 3-point anti-MHV amplitude, gives
\be
  A_n[1^- 2^- 3^+ \ldots n^+]
  ~=~
  \frac{\<\hat{1}\hat{P}_{23}\>^4}{\<\hat{1}\hat{P}_{23}\>\<\hat{P}_{23}\, 4\>\<45\>\cdots\<n\hat{1}\>}
  \times\frac{1}{\<23\>[23]}\times
  \frac{[3 \hat{P}_{23}]^3}{[\hat{P}_{23}\,\hat{2}][\hat{2} 3]}\,.
  \label{recrelex1b}
\ee
We could now proceed to evaluate the angle and square spinors for the shifted momenta. But it is more fun to introduce you to a nice little trick. Combine the factors from the numerator:
\be
  \label{id1}
  \<\hat{1}\hat{P}_{23}\>[3 \hat{P}_{23}]
  = - \<\hat{1}\hat{P}_{23}\>[\hat{P}_{23} \,3]
   = \<\hat{1}| \hat{P}_{23} | 3]
  = \<\hat{1}| (\hat{p}_2+p_3) |  3]
  = \< \hat{1} | \hat{p}_2 | 3]
  = -\<\hat{1} \hat{2}\> [  \hat{2} 3]
  = -\<12\> [ 2 3]\,.
\ee
In the last step we used the $\<\hat{1} \hat{2}\> = \<12\>$ and that $|\hat{2}] = |2]$. 
Playing the same game with the factors in the denominator, we find  
\be
  \<\hat{P}_{23}\, 4\> [\hat{P}_{23}\,\hat{2}]
  =
  \<4 |\hat{P}_{23} | \hat{2}] 
  =\<4 | 3 | 2] 
  =-\<4 3\> [32]
  =-\<34\> [23]\,.
   \label{id2}
\ee
Now use \reef{id1} and \reef{id2} in \reef{recrelex1b} to find
\bea
  \nonumber
  A_n[1^- 2^- 3^+ \ldots n^+]
  &=&
  -\frac{\<12\>^3 [ 2 3]^3}
  {\big(\!-\!\<34\> [23] \big)~ \<45\>\cdots\<n1\>~\<23\>[23]~[23]}
  \\[1mm]
  &=&
   \frac{\<12\>^4 }
  {\<12\>\<23\> \<34\> \<45\>\cdots\<n1\>~} \,.
  \label{recrelex1c}
\eea
This completes the inductive step. With the 3-point gluon amplitude $A_3[1^-2^-3^+]$ fixed completely by little group scaling and locality to take the form \reef{PT12}, we have then proven the Parke-Taylor formula for all $n$. This \emph{is} a lot easier than calculating Feynman diagrams!

You may at this point complain that we have only derived the Parke-Taylor formula recursively for the case where the negative helicity gluons are adjacent. Try your own hands on the proof for the non-adjacent case. In Section \ref{s:susy} we will use supersymmetry to derive a more general form of the tree-level gluon amplitudes: it will contain all MHV helicity arrangements in one compact expression.

We have now graduated from MHV to the study of {\bf \em NMHV amplitudes}. It is worthwhile to consider the 5-point example  $A_5[1^- 2^- 3^- 4^+ 5^+]$ even though this amplitude is anti-MHV: constructing it with a $[+,+\>$-shift is a calculation very similar to the MHV case --- and that would by now be boring. So, instead, we are going to use a $[-,-\>$-shift to illustrate some of the manipulations used in BCFW recursion:
\example{Consider the $[1,2\>$-shift recursion relations for 
$A_5[1^- 2^- 3^- 4^+ 5^+]$: there are two diagrams
\be
  A_5[1^- 2^- 3^- 4^+ 5^+]
  ~=~
  \raisebox{-9mm}{\includegraphics[height=1.9cm]{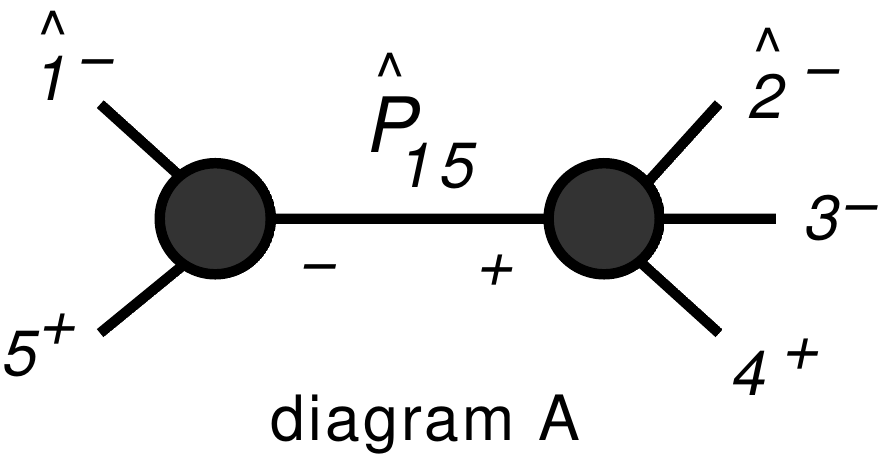}}
  ~~~+~~~
  \raisebox{-9mm}{\includegraphics[height=1.9cm]{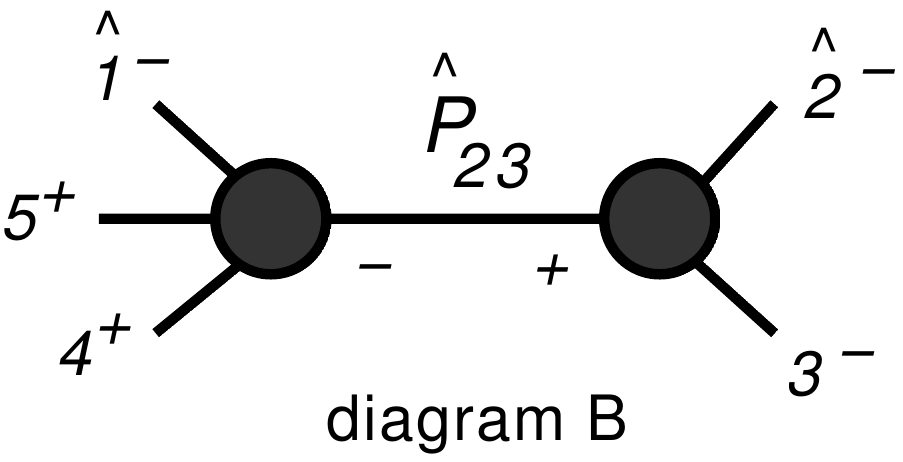}} \,.
\ee
We have indicated the required helicity for the gluon on the internal line. Had we chosen the opposite helicity option for the internal gluon in diagram A, the R subamplitude would have helicity structure $---+$, so it would vanish. Diagram B also vanishes for the opposite choice of the helicity on the internal line. For the helicity choice shown, the R subamplitude of diagram B is MHV, $A_3[-\hat{P}_{23}^+,\hat{2}^-,3^-]$, and since $|2\>$ is shifted, the special 3-particle kinematics actually makes
$A_3[-\hat{P}_{23}^+,\hat{2}^-,3^-] = 0$, just as we saw for the anti-MHV case in the discussion below \reef{A3SK}. So {\em diagram B vanishes}, and we can focus on diagram A. Using the Parke-Taylor formula for the MHV subamplitudes, we get
\be
  \label{A5bcfw1}
  A_5[1^- 2^- 3^- 4^+ 5^+]
  ~=~
  \frac{\<\hat{1} \hat{P}\>^3}{\<\hat{P} 5\>\<5 \hat{1}\>}
  \times
  \frac{1}{\<15\>[15]}
  \times
  \frac{\<\hat{2}3\>^4}{\<\hat{2}3\>\<34\>\<4\hat{P}\>\<\hat{P}\hat{2}\>}\,.
\ee
Here $\hat{P}$ stands for $\hat{P}_{15}=\hat{p}_1+p_5$.
We have three powers of $|\hat{P}\>$ in the numerator and three in the denominator. A good trick to simplify such expressions is to multiply \reef{A5bcfw1} by $[\hat{P} X]^3/[\hat{P} X]^3$ for some useful choice of $X$. It is convenient to pick $X=2$. Grouping terms conveniently together, we get:
\begin{itemize}
\item[$\bullet$] 
$\<\hat{1} \hat{P}\>[\hat{P} 2]
= - \<\hat{1} | \hat{1} \!+\! 5|  2]
= \<\hat{1}5\> [5 2]
= - \<15\> [2 5]$~~
(since $|\hat{1}\>=|1\>$).
\item[$\bullet$] 
$\<5 \hat{P}\>[\hat{P} 2]
= - \<5| \hat{1} \!+\! 5|  2]
= \<51\> [\hat{1} 2]
= \<51\> [1 2]$\,.
\item[$\bullet$] 
$\<4 \hat{P}\>[\hat{P} 2]
= - \<4| \hat{1} + 5|  2]
=  \<4| \hat{2} \!+\! 3\!+\!4|  2]
=  \<4|3|  2]
= - \<43\> [32]
=  -\<34\> [23]$.
\item[$\bullet$] 
$\<\hat{2} \hat{P}\>[\hat{P} 2]
= -2\,\hat{p}_2 \cdot\hat{P}
= 2\,\hat{p}_2 \cdot(\hat{p}_2 + p_3 +p_4)
= (\hat{p}_2 + p_3 +p_4)^2 - (p_3 + p_4)^2
\\
~\hspace{1.45cm}
= \hat{P}^2 -\<34\>[34]
=-\<34\>[34]$,\\
$~\hspace{1.45cm}$
since the amplitude is evaluated at $z$ such that $\hat{P}^2=0$.
\end{itemize}
Using these expression in \reef{A5bcfw1} gives
\be
  \label{A5bcfw2}
  A_5[1^- 2^- 3^- 4^+ 5^+]
  ~=~
 \frac{[2 5]^3\<\hat{2}3\>^3}{[1 2][23][34][15]\<34\>^3}
  \,.
\ee
Despite the simplifications, there is some unfinished business for us to deal with: \reef{A5bcfw2} depends on the shifted spinors via $\<\hat{2}3\>$. This bracket must be evaluated at the residue value of $z=z_{15}$ which is such that $\hat{P}_{15}^2 = 0$:
\be
  0 = \hat{P}_{15}^2 = \<15\>[\hat{1}5] 
  ~~~\text{i.e.}~~~
  0 = [\hat{1}5] = [15]+z_{15} [25],
  ~~~\text{i.e.}~~~
 z_{15}   = -\frac{[15]}{[25]}\,.
\ee
Use this and momentum conservation to write
\be
 \<\hat{2}3 \> = \<23\> - z_{15} \<13\>
  = \frac{\<23\>[25]+\<13\>[15]}{[25]}
  = \frac{\<34\>[45]}{[25]}
\ee
Inserting this result into \reef{A5bcfw2} we arrive at the expected anti-Parke-Taylor expression
\be
  A_5[1^- 2^- 3^- 4^+ 5^+] 
  ~=~
  \frac{[45]^4}{[12][23][34][45][51]}\,.
\ee
As noted initially, the purpose of this example was not to torture you with a difficult way to derive $A_5[1^- 2^- 3^- 4^+ 5^+]$. The purpose was to illustrate the methods needed for general cases in a simple context. 
}
You may not be overly impressed with the simplicity of the manipulations needed to simplify the output of BCFW. Admittedly it requires some work. If you are unsatisfied, go ahead and try the calculations in this section with Feynman diagrams. Good luck.

Now you have seen the basic tricks needed to manipulate the expressions generated by BCFW. So you should get some exercise.
\exercise{ex:sQED6}{Let us revisit scalar-QED from the end of Section \ref{s:QED}. Use little group scaling and locality to determine $A_3(\varphi \,\varphi^* \g^\pm)$ and compare with your result from Exercise \ref{ex:sQED1}.
Then use a $[4,3\>$-shift to show that (see Exercise \ref{ex:sQED2})
\be
  A_4(\varphi \,\varphi^* \g^+ \g^-) = g^2 \frac{\<14\> \<24\>}{\<13\> \<23\>} \,.
\ee
[Hint: this is not a color-ordered amplitude.]\\
What is the large-$z$ falloff of this amplitude under a $[4,3\>$-shift?
}
\exercise{ex:4gravitonBCFW}{Calculate the 4-graviton amplitude $M_4(1^- 2^- 3^+ 4^+)$: first recall that little group scaling \& locality to fix the 3-particle amplitudes as in Exercise \ref{ex:grav3pt}.  
Then employ the $[1,2\>$-shift BCFW recursion relations (they are valid \cite{Benincasa:2007qj,ArkaniHamed:2008yf}).

Check little group scaling and Bose-symmetry of your answer for $M_4(1^- 2^- 3^+ 4^+)$.\\{}
[Hint: your result should match one of the amplitudes in Exercise \ref{ex:game}.]

Show that $M_4(1^- 2^- 3^+ 4^+)$ obeys the 4-point ``KLT relations" \cite{Kawai:1985xq}
\be
  M_4(1234) = - s_{12}  \,A_4[1234] \,A_4[1243]\,,
\ee
where $A_4$ is your friend the Parke-Taylor amplitude and the Mandelstam variable is $s_{12}=-(p_1+p_2)^2$. 
When you are done, look up ref.~\cite{sannan} to see how difficult it is to do this calculation with Feynman diagrams.  
}

Let us now take a look at some interesting aspects of the BCFW for the 
\emph{split-helicity} NMHV amplitude $A_6[1^- 2^- 3^- 4^+ 5^+ 6^+]$. Let's first look at the recursion relations following from the $[1,2\>$-shift that we are now so familiar with. There are two non-vanishing diagrams:
\be
  \label{A6bcfw1}
  A_6[1^- 2^- 3^- 4^+ 5^+ 6^+]
  ~=~
    \raisebox{-8.5mm}{\includegraphics[height=1.8cm]{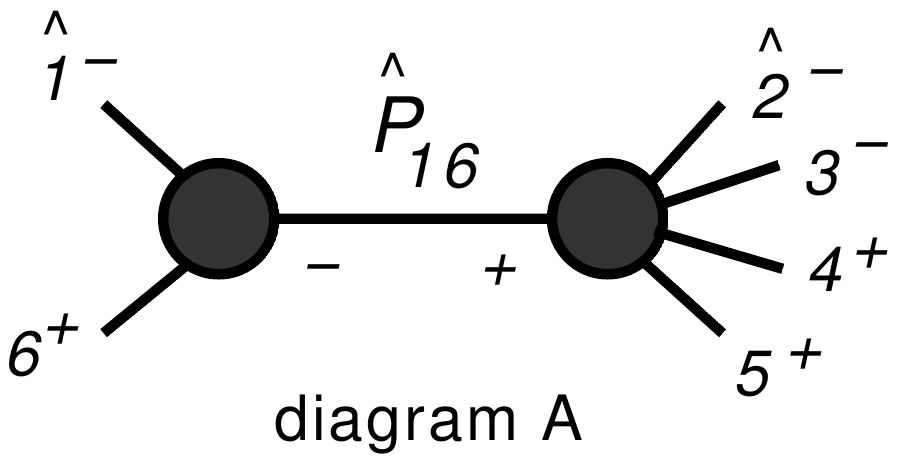}}
  ~~~+~~~
  \raisebox{-8.5mm}{\includegraphics[height=1.8cm]{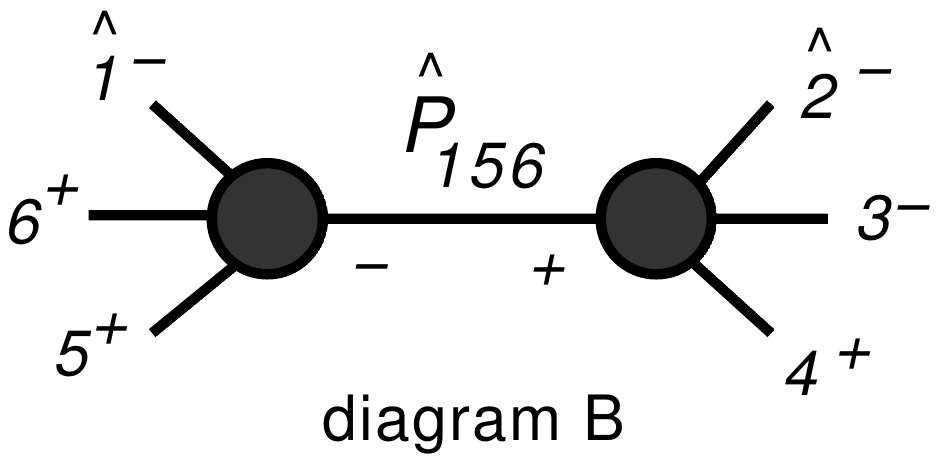}} \,.
\ee
\exercise{}{Show that the 23-channel diagram does not contribute in \reef{A6bcfw1}. 
}
The \emph{first thing} we want to discuss about the 6-gluon amplitude are the 3-particle poles in the expression \reef{A6bcfw1}. Diagram B involves a propagator $1/P_{156}^2$, so there is a 3-particle pole at $P_{156}^2=0$. By inspection of the ordering of the external states in $A_6[1^- 2^- 3^- 4^+ 5^+ 6^+]$ there should be no distinction between the $(-++)$ 3-particle channels 165 and 345, so we would expect the amplitude to have a pole also at $P_{345}^2 = P_{126}^2 = 0$.
But the $[1,2\>$-shift recursions relation \reef{A6bcfw1} does not involve any 126-channel diagram. How can it then possibly encode the correct amplitude? The answer is that it does and that the $P_{345}^2 = P_{126}^2 = 0$ pole is actually hidden in the denominator factor $\<\hat{2} \hat{P}_{16}\>$ of righthand subamplitude of diagram A in \reef{A6bcfw1}. Let us show how.

As in the 5-point example above, we multiply the numerator and denominator both with $[\hat{P}_{16}\, 3]$. Then write
\be
  \label{6ptNMHVcalc1}
  \<\hat{2} \hat{P}_{16}\>[\hat{P}_{16}\, 3]
  = \<21\> [\hat{1} 3] +   \<\hat{2} 6\>[63]\,.
\ee
It follows from $\hat{P}_{16}^2=0$ that $z_{16}=-[16]/[26]$, and this is then used to show that 
$\<\hat{2} 6\> = (\<16\>[16]+\<26\>[26])/[26]$ and 
$[\hat{1} 3] =[12][36]/[26]$. Plug these values into \reef{6ptNMHVcalc1} to find
\be
   \<\hat{2} \hat{P}_{16}\>[\hat{P}_{16}\, 3] 
   = -\frac{[36]}{[26]} \big(\<12\>[12]+\<16\>[16]+\<26\>[26]\big)
   = -\frac{[36]}{[26]}  \,P_{126}^2\,.
\ee
So there you have it: the 3-particle pole is indeed encoded in BCFW \reef{A6bcfw1}.

The \emph{second thing} we want to show you is the actual  representation for the 6-gluon NMHV tree amplitude, as it follows from \reef{A6bcfw1}:
\be
  \label{A6bcfw2}
  A_6[1^- 2^- 3^- 4^+ 5^+ 6^+]
  ~=~
  \frac{\<3|1+2|6]^3}{P_{126}^2 [21][16] \<34\> \<45\> \<5|1+6|2]}
  ~+~
  \frac{\<1|5+6|4]^3}{P_{156}^2 [23][34] \<56\> \<61\> \<5|1+6|2]}\,.
\ee
\exercise{}{Check the little group scaling of \reef{A6bcfw2}. Fill in the details for converting the two diagrams in \reef{A6bcfw1} to find \reef{A6bcfw2}.}
The expression \reef{A6bcfw2} may not look quite as delicious as the Parke-Taylor formula, but remember that it contains the same information as the sum of 38 Feynman diagrams.

The \emph{third thing} we would like to emphasize is that the $[1,2\>$-shift recursion relations is just one way to calculate $A_6[1^- 2^- 3^- 4^+ 5^+ 6^+]$. What happens if we use the $[2,1\>$-shift? Well, now there are three non-vanishing diagrams:
\be
  \label{A6bcfwAlso}
  A_6[1^- 2^- 3^- 4^+ 5^+ 6^+]
  ~=
    \raisebox{-10mm}{\includegraphics[height=1.85cm]{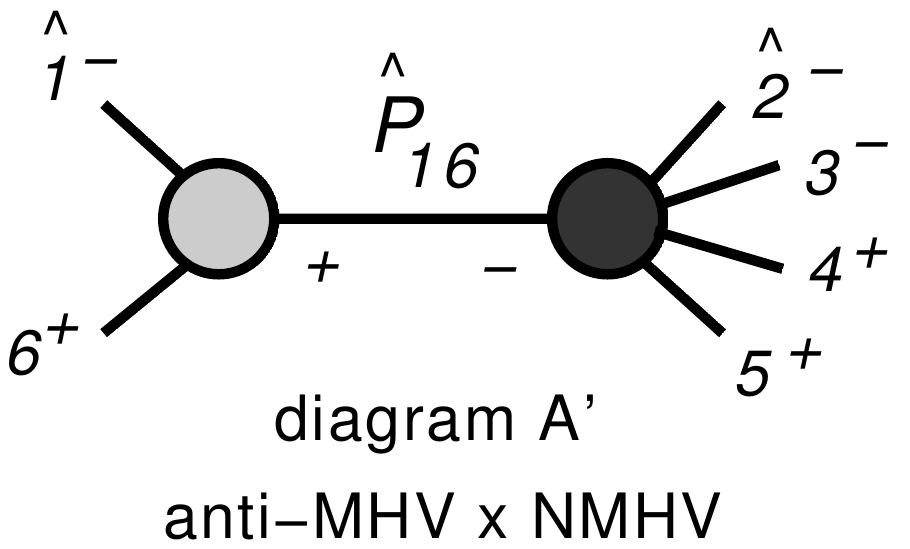}}
  ~+~
  \raisebox{-10mm}{\includegraphics[height=1.85cm]{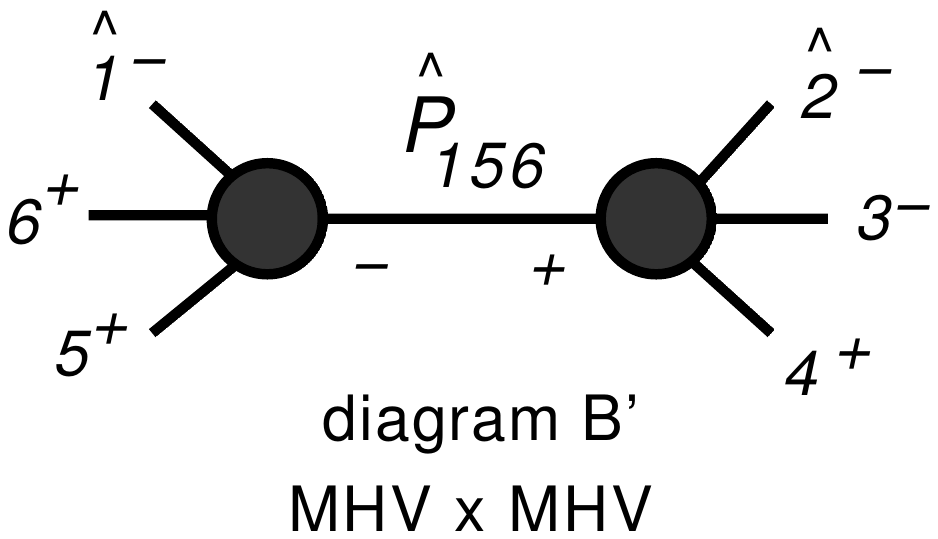}} 
  ~+~
  \raisebox{-10mm}{\includegraphics[height=1.85cm]{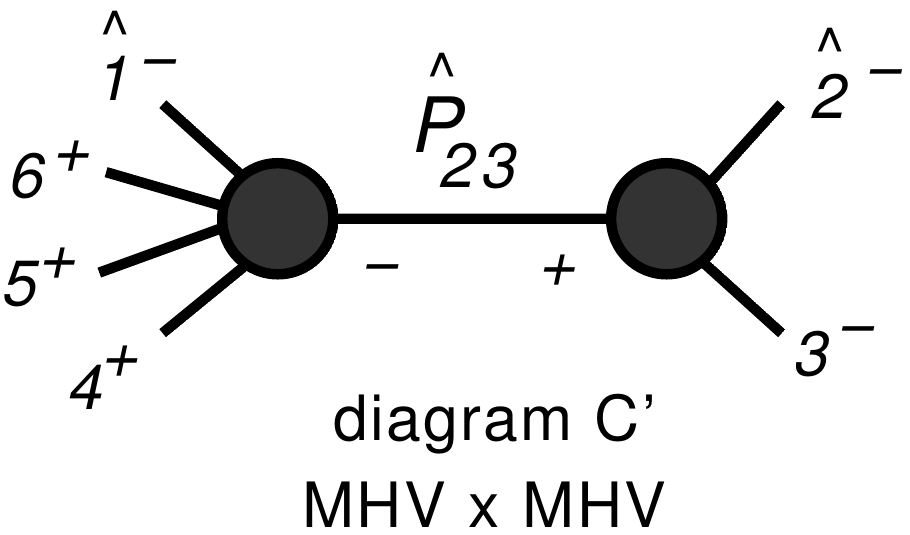}} \,.
\ee
The special 3-particle kinematics force the diagram A$'$ to have the helicity structure of anti-MHV$\times$NMHV, as opposed to the similar diagram A in \reef{A6bcfw1} which is forced to be MHV$\times$MHV.  Thus this is the first time we see a lower-point NMHV amplitude shows up in the recursion relations. This is quite generic: the BCFW relations are recursive both in particle number $n$ and in N$^K$MHV level $K$. 

The two BCFW representations \reef{A6bcfw1} and \reef{A6bcfwAlso} look quite different. In order for both to describe the same amplitude, there has to be a certain identity that ensures that diagrams A+B = A$'$ + B$'$+ C$'$. To show that this identity holds requires a nauseating trip through Schouten identities and  momentum conservation relations in order to manipulate the angle and square brackets into the right form: numerical checks can save you a lot of energy when dealing with amplitudes with more than 5 external lines. It turns out that the identities that guarantee the equivalence of  BCFW expressions such as A+B  and A$'$ + B$'$+ C$'$ 
actually originate from powerful residue theorems \cite{ArkaniHamed:2009dn} related to quite different formulations of the amplitudes. This has to do with the description of amplitudes in the Grassmannian --- we  get to that in Section \ref{s:polytopes}, but wanted to give you a hint of this curious point here.
\exercise{ex:Mi}{Show that the BCFW recursion relations based on the  $[2,3\>$-shift give the following representation of the 6-point `alternating helicity' gluon amplitude:
\be
   A_6[1^+ 2^- 3^+ 4^- 5^+ 6^-] = \{M_2\} + \{M_4\} + \{M_6\}\,,
\ee
where
\be
  \{M_i\} = \frac{\<i,i+2\>^4 [i+3,i-1]^4}
    {\tilde{P}_i^2\,\<i| \tilde{P}_i |i+3] \<i+2|\tilde{P}_i|i-1] 
     \<i,i+1\> \<i+1,i+2\> [i+3,i-2][i-2,i-1]}\,.
\ee
and $\tilde{P}_i=P_{i,i+1,i+2}$. [Hint: $\{M_4\}$ is the value of the 12-channel diagram.]\\
In Section \ref{s:polytopes} we discover that each $\{M_i\}$ can be understood as the residue associated with a very interesting contour integral (different from the one used in the BCFW argument).
}

The \emph{fourth thing} worth discussing further are the poles of  scattering amplitudes. Color-ordered tree amplitudes can have physical poles only when the momenta of adjacent external lines go collinear. We  touched this point already when we discussed the 3-particle poles. In fact, you can see from the Parke-Taylor formula that MHV amplitudes do not have multi-particle poles, only 2-particle poles. And you have seen that the 6-gluon NMHV amplitude has both 2- and 3-particle poles. But as you stare intensely at \reef{A6bcfw2}, you will also note that there is a strange denominator-factor $\<5|1+6|2]$ in the result from each BCFW diagram. This does not correspond to a physical pole of the scattering amplitude: it is a {\bf \em spurious pole}. The residue of this unphysical pole better be zero --- and it is: the spurious pole cancels in the sum of the two BCFW diagrams in \reef{A6bcfw2}. It is typical that BCFW packs the information of the amplitudes into compact expressions, but the cost is the appearance of spurious poles; this means that in the BCFW representation the  locality of the underlying field theory is not manifest. Elimination of spurious poles in the representations of amplitudes leads to interesting results \cite{Hodges} that we discuss in a later section. 

\emph{Finally}, let us for completeness note that the color-ordered amplitudes $A_6[1^- 2^- 3^+ 4^- 5^+ 6^+]$ and $A_6[1^- 2^+ 3^- 4^+ 5^- 6^+]$ with other arrangements of helicities are inequivalent to the split-helicity amplitude 
$A_6[1^- 2^- 3^- 4^+ 5^+ 6^+]$.  More about this in Section \ref{s:susy}.

\noindent {\bf Other comments:}\\
1) In our study of the recursion relations, we kept insisting on `generic' momenta. However, special limits of the external momenta place useful and interesting constraints on the amplitudes: the behavior of amplitudes under {\bf \em collinear limits} and {\bf \em soft limits} are described in the reviews \cite{dixon,Bern:2007dw}.

2) In some cases, the shifted amplitudes have ``better than needed" large-$z$ behavior. For example, this is the case for shifts of non-adjacent same-helicity lines in the color-order Yang-Mills amplitudes: $\hat{A}_n(z) \to 1/z^2$. The tree-level recursion relations can be viewed to follow from the Cauchy integral identity $\oint_\mathcal{C} \frac{\hat{A}_n(z)}{z} = 0$ with $\mathcal{C}$ a contour that surrounds all the simple poles: let's write the sum of diagrams resulting from the sum of the residues as $A_n = d_1 +\dots +d_w$. An extra power in the large-$z$ falloff $\hat{A}_n(z) \sim 1/z^2$ means that there is also a {\bf \em bonus relation}: 
$\oint_\mathcal{C} {\hat{A}_n(z)} = 0$ (with $\mathcal{C}$ as before) gives 
 $d_1\,z_1 +\dots +d_w z_w= 0$ with $z_i$ the location of the poles. The bonus relations have practical applications, for example they have been used to verify and show equivalence of different forms of MHV graviton amplitudes \cite{bonusrel}.

%%%%%%%%%%%%%

\subsection{When does it work?}
\label{s:validity}

In Section \ref{s:littlegrp} we learned that the 3-point amplitudes for massless particles are uniquely determined by little group scaling, locality and dimensional analysis. As we have just seen, with the on-shell BCFW recursion relations, we can construct all higher-point gluon tree amplitudes from the input of just the 3-point gluon amplitudes. That is a lot of information obtained from very little input! It prompts us to raise a question of suspicion: ``When can we expect on-shell recursion to work?". We will look at some examples now.

{\bf Yang-Mills theory and gluon scattering.}
From standard Feynman rules, we are familiar with the fact that the quartic term $A^4$ in the Yang-Mills Lagrangian is needed for gauge invariance. However, the recursion relations indicates that the cubic term $A^2 \pa A$ captures the information needed for the amplitudes, at least at tree-level. The key difference is that the 3-vertex is an off-shell  non-gauge invariant object, but the 3-point on-shell amplitude is gauge invariant. Since $A^4$ is fully determined from $A^2 \pa A$ by the requirement of the off-shell gauge invariance of the Lagrangian, it contains no new on-shell information. In a sense, that is why the recursion relations for on-shell gluon amplitudes even have a chance to work with input only from the on-shell 3-point amplitudes.
 
We can rephrase the information contents of $A^2 \pa A$ in a more physical way. The actual input is then this: 4d local theory with massless spin-1 particles (and no other dynamical states) and a dimensionless coupling constant. This information is enough to fix the entire gluon tree-level scattering matrix!

{\bf Scalar-QED.} As a second example, consider scalar-QED. The interaction between the photons and the scalar particles created/annihilated by a complex scalar field $\varphi$ is encoded by the covariant derivatives $D_\m = \pa_\m - i e A_\mu$ in 
\be
  \label{scalQEDlag}
 \lag  ~\supset~ - |D\varphi|^2 
  ~=~  |\pa\varphi|^2 
   +i e A^\mu \big[ (\pa_\m \varphi^*) \varphi  - \varphi^* \pa_\m \varphi \big]
   - e^2 A^\m A_\m \varphi^* \varphi \,.
\ee
In terms of Feynman diagrams, $A_4(\varphi \,\varphi^* \ga \,\ga)$ is constructed from the sum of two pole diagrams and the contact term from the quartic interaction (Exercise \ref{ex:sQED3}).  
We have seen in Exercise \ref{ex:sQED6} that this 4-point amplitude is constructible via BCFW. So it is clear that only the information in the 3-point vertices is needed, and the role of $A_\m A^\m \varphi^*\varphi$  is just to ensure off-shell gauge invariance of the Lagrangian. Thus this case is just like the Yang-Mills example above. 

Thus emboldened, let us try to compute the  4-scalar amplitude $A_4(\varphi \,\varphi^*\varphi \,\varphi^*)$ using BCFW recursion. Using a $[1,3\>$-shift, there are two diagrams and their sum simplifies to 
\be
 \label{A4bcfwsqed}
A_4^\text{BCFW}(\varphi \,\varphi^*\varphi \,\varphi^*)
~=~
\tilde{e}^2 
\frac{\<13\>^2\<24\>^2}{\<12\>\<23\>\<34\>\<41\>}
  \,.
\ee
If, on the other hand, we can calculate this amplitude using Feynman rules from the interaction terms in \reef{scalQEDlag}, we get
\be
A_4^\text{Feynman}(\varphi \,\varphi^*\varphi \,\varphi^*)
~=~
\tilde{e}^2 
  \bigg( 
     1+\frac{\<13\>^2\<24\>^2}{\<12\>\<23\>\<34\>\<41\>}
  \bigg) \,.
\ee
Ugh! So BCFW did not compute the amplitude we expected. So what did it compute? Well, let us think about the input that BCFW knows about: 4d local theory with massless spin-1 particles and charged massless spin-0 particles (and no other dynamical states) and a dimensionless coupling constant. 
Note that included in this input is the possibility of a 4-scalar interaction term $\lambda |\varphi|^4$. So more generally, we should consider the scalar-QED action from \reef{scalarQED}:
\bea
  \nonumber
  \lag &=& - \frac{1}{4} F_{\m\n}F^{\m\n}   - |D\varphi|^2 - \frac{1}{4} \lambda |\varphi|^4
  \\
 &=&- \frac{1}{4} F_{\m\n}F^{\m\n}  - |\pa \varphi|^2
   + i e A^\mu \big[ (\pa_\m \varphi^*) \varphi  - \varphi^* \pa_\m \varphi \big]
   - e^2 A^\m A_\m \varphi^* \varphi
   - \frac{1}{4} \lambda |\varphi|^4 \,.
   \label{scalarQEDsec3}
\eea
In Exercise \ref{ex:sQED5} you were asked to calculate $A_4(\varphi \,\varphi^*\varphi \,\varphi^*)$ in this model. The answer was given in
\reef{4scalarQED}: it is
\be
  \label{4scalarQEDagain}
  A_4(\varphi\, \varphi^* \varphi\, \varphi^*)
  ~=~
  -\lambda + \tilde{e}^2 
  \bigg( 
     1+\frac{\<13\>^2\<24\>^2}{\<12\>\<23\>\<34\>\<41\>}
  \bigg) \,,
\ee
So it is clear now that we have a family of scalar-QED models, labelled by $\lambda$, and that our BCFW calculation produced the very special case of $\lambda = \tilde{e}^2$. How can we understand this? Validity of the recursion relations require the absence of the boundary term $B_n$ (see Section \ref{s:shifts}). For the general family of scalar-QED models, there is a boundary term under the $[1,3\>$-shift, and its value is  $-\lambda + \tilde{e}^2$ (as can be seen from \reef{4scalarQEDagain} by direct computation). The special choice $\lambda = \tilde{e}^2$ eliminates the boundary term, and that's then what BCFW without a boundary term computes. 

The lesson is that for general $\lambda$, there is no way in which the 3-point interactions can know the contents of $\lambda |\varphi|^4$: it provides independent gauge-invariant information. That information needs to be supplied in order for recursion to work, so in this case one can at best expect recursion to work beyond 4-point amplitudes. The exception is of course if some symmetry, or other principle, determines the information in $\lambda |\varphi|^4$ in terms of the 3-field terms. This is what we find for $\lambda = \tilde{e}^2$. In fact, the expression \reef{A4bcfwsqed} actually occurs for 4-scalar amplitudes in $\mathcal{N}=2$ and $\mathcal{N}=4$ SYM theory, and in those cases the coupling of the 4-scalar contact term \emph{is} fixed by the Yang-Mills coupling by supersymmetry.

%%%%%%%% 
{\bf Scalar theory $\lambda \phi^4$.} The previous example makes us wary of  $\lambda \phi^4$-interaction in the context of recursion relations --- and rightly so. Suppose we just consider $\lambda \phi^4$-theory with no other interactions. It is clear that one piece of input must be given to start any recursive approach, namely in this case the 4-scalar amplitude $A_4 = \lambda$. In principle, one might expect on-shell recursion to determine all tree-level $A_n$ amplitudes with $n>4$ from just $A_4 = \lambda$ --- after all, what else could interfere? And this is the only interaction in the Feynman diagrams anyway. Noting that the 6-scalar amplitude is $A_6 = \lambda^2 \big( \tfrac{1}{s_{123}} + \dots \big)$, it is clear though that all BCFW shift give $O(z^0)$-behavior for large $z$ and hence there are no BCFW recursion relations without boundary term for $A_6$ in $\lambda \phi^4$-theory. Inspection of the Feynman diagrams reveals that $O(z^0)$-contributions are exactly the diagrams in which the two shifted lines belong to the same vertex. The sum of such diagrams equals the boundary term $B_n$ from \reef{AnRes}. One can in this case of $\lambda \phi^4$-theory reconstruct $B_n$ recursively.\footnote{See \cite{Feng:2009ei}. Or avoid the term at infinity by using an all-line shift, see \cite{CEK}} Thus the $A_4$ does suffice to completely determine $A_n$ for $n>4$; but it is (in many senses of the phrase) a rather trivial example.

{\bf $\mathcal{N}=4$ SYM theory.} 
This is the favorite theory of most amplitunists. The spectrum\footnote{At the origin of moduli space where all scalar vevs are zero, all the states are massless. On the Coulomb branch,  (some) scalars acquire vevs and as a result the spectrum then includes the massive $\mathcal{N}=4$ supermultiplet. Amplitudes on the Coulomb branch are discussed in \cite{Craig:2011ws,Kiermaier:2011cr}.} consists of 16 massless states: gluons $g^{\pm}$ of pos/neg helicity, 4 gluinos $\lambda^a$ and $\lambda_a$ of pos/neg helicity, and six scalars $S^{ab}$. The indices $a,b = 1,2,3,4$ are labels for the global $SU(4)$ R-symmetry. The Lagrangian contains standard gluon self-interactions, with standard couplings to the gluinos and the scalars; all fields transform in the adjoint of the $SU(N)$ gauge group. In addition, there is a scalar 4-point interaction term of a schematic form $[S,S]^2$. It contains, for example, the interaction $S^{12} S^{23} S^{34} S^{41}$. The result for the corresponding color-ordered amplitude is:
\be
  \label{N4SYM-4s-spec}
  A_4[S^{12} S^{23} S^{34} S^{41}] = 1.
\ee
Since this amplitude has no poles, it cannot be obtained via direct factorization. Actually, the amplitude \reef{N4SYM-4s-spec} and its cousin 4-scalar amplitudes with equivalent $SU(4)$ index structures are the only tree amplitudes of $\mathcal{N}=4$ SYM that cannot be obtained from  BCFW recursion relations; that may seem surprising, but it is true --- for a proof, see \cite{Elvang:2008na}.

When supersymmetry is incorporated into the BCFW recursion relations, 
\emph{all} tree amplitudes of $\mathcal{N}=4$ SYM can be determined by the 3-point gluon vertex alone. The so-called super-BCFW shift mixes the external states in such a way that even the 4-scalar amplitude \reef{N4SYM-4s-spec} can be constructed recursively. We will work with the super-BCFW shift in  Section \ref{s:N4symtrees}.

{\bf Gravity.} We have already encountered the 4-point MHV amplitude 
$M_4(1^- 2^- 3^+ 4^+)$: 
you `discovered' it from little group scaling in Exercise \ref{ex:game} and constructed it with BCFW in Exercise \ref{ex:4gravitonBCFW}.
The validity of the BCFW recursion relations for all tree-level graviton amplitudes \cite{Benincasa:2007qj,ArkaniHamed:2008yf} means that entire on-shell tree-level S-matrix for gravity is determined completely by the 3-vertex interaction of 3 gravitons. In contrast, the expansion of the Einstein-Hilbert action $\tfrac{1}{2\kappa^2}\int d^4x \,\sqrt{-g} R$ around the flatspace Minkowski metric $g_{\m\n} = \eta_{\m\n} + \kappa \, h_{\m\n}$
contains \emph{infinitely} many interaction terms. It is remarkable that all these terms are totally irrelevant from the point of view of the on-shell tree-level S-matrix; their sole purpose is to ensure diffeomorphism invariance of the off-shell Lagrangian. For on-shell (tree) amplitudes, we do not need them.

{\bf Summary.} We have discussed when to expect to have recursion relations for tree-level amplitude. The main lesson is that we do not get something for nothing: input must be given and we can only expect to recurse that input with standard BCFW when all other information in the theory is fixed by our input via gauge invariance. If another principle --- such as supersymmetry --- is needed to fix the interactions, then that principle should be incorporated into the recursion relations for a successful recursive approach. Further discussion of these ideas can be found in \cite{CEK}, mostly in the context of another recursive approach, known as CSW which we will discuss briefly next. 

%%%%%%%%%%%%%
%%%%%%%%%%%%%
%%%%%%%%%%%%%

\subsection{MHV vertex expansion (CSW)}
\label{s:csw}
We introduced recursion relations in Section \ref{s:shifts} in the context of general shifts \reef{pshift} satisfying the set of conditions (i)-(iii). Then we specialized to the BCFW shifts in Section \ref{s:bcfw}. Now we would like to show you another kind of recursive structure.

Consider a shift that is implemented via a  `holomorphic' square-spinor shift:
\be
  \label{sqshift}
  |\hat{i}] = |i] + z\, c_i |X]
  ~~~~\text{and}~~~~
   |\hat{i}\> = |i\> 
  \,.
\ee 
Here $|X]$ is an arbitrary reference spinor and the coefficients $c_i$ satisfy $\sum_{i=1}^n c_i |i\> = 0$.
\exercise{ex:cswshift}{Show that the square-spinor shift \reef{sqshift} gives shift-vectors $\refr_i$ that fulfills the requirements (i)-(iii) in Section \ref{s:shifts}.}
The choice $c_1=\<23\>$, $c_2=\<31\>$, $c_3=\<12\>$, and $c_i=0$ for $i=4,\dots,n$ implies that the shifted momenta satisfy momentum conservation. This particular realization of the square-spinor shift is called the Risager-shift \cite{Risager:2005vk}.

We consider here a situation where all $c_i \ne 0$ so that all momentum lines are shifted via \reef{sqshift} --- this is an \emph{all-line shift}. It can be shown \cite{Elvang:2008vz} that N$^K$MHV gluon tree amplitudes fall off as $1/z^K$ for large $z$ under all-line shift. So this means that all the  gluon tree-level amplitudes can be constructed with the all-line shift recursion relations; except the MHV amplitudes ($K=0$). It turns out that in this formulation of recursion relations, the tower of MHV amplitudes constitute the basic building blocks for the N$^K$MHV amplitudes. Let us see how this works for NMHV. 
The recursion relations give
\be
   \label{recrelcsw1}
    A_n^\text{NMHV}~= 
   \sum_{\text{diagrams}~I}
   \raisebox{-5.5mm}{\includegraphics[height=1.4cm]{recrel1}}
   \,.    
\ee 
If you consider the possible assignments of helicity labels on the internal line, you'll see that there are two options: either the diagram is anti-MHV${_3} \times$NMHV or MHV$\times$MHV. The former option vanishes by special kinematics of the 3-point anti-MHV vertex, just as in the case of the first diagram in \reef{recrelex1}. So all subamplitudes in \reef{recrelcsw1} are MHV. Let us write down the example of the split-helicity NMHV 6-gluon amplitude:
\be
  \begin{split}
  \hspace{-2mm}
  A_n[1^-2^-3^-4^+5^+6^+]~=~
  & 
   \raisebox{-5.5mm}{\includegraphics[height=1.3cm]{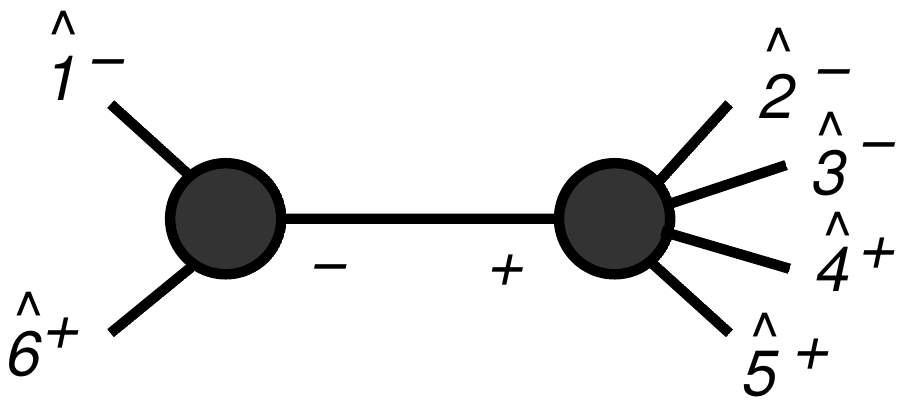}}
~+~\,
   \raisebox{-5.5mm}{\includegraphics[height=1.3cm]{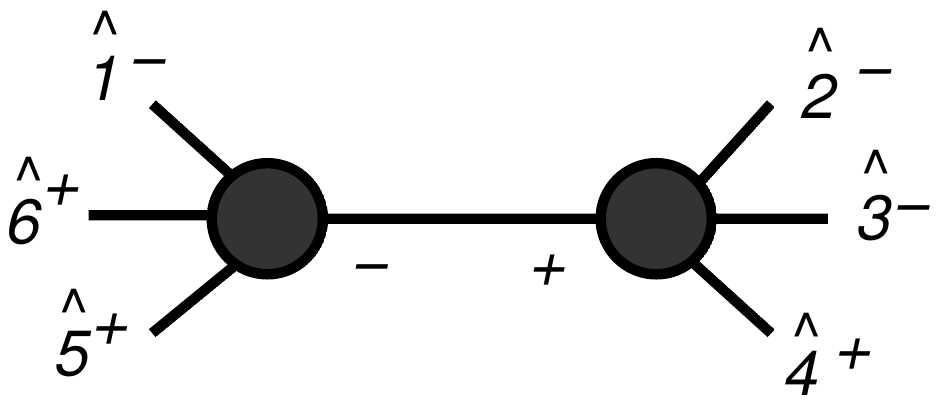}}
~+~\,
   \raisebox{-5.5mm}{\includegraphics[height=1.3cm]{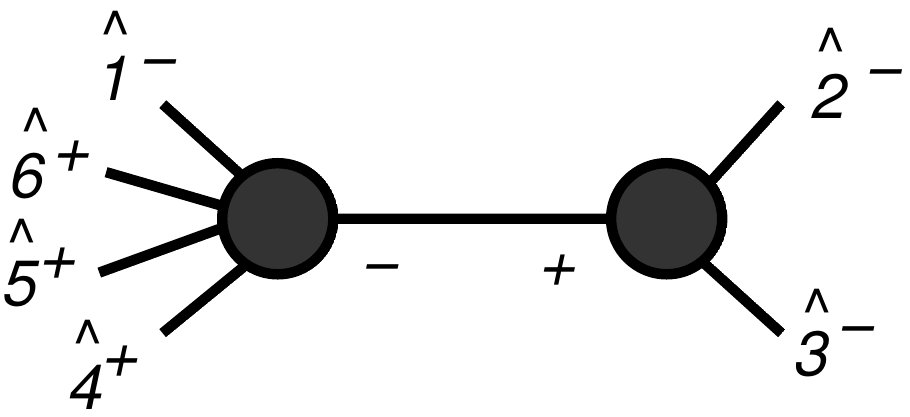}} \\
  & 
   \raisebox{-5.5mm}{\includegraphics[height=1.3cm]{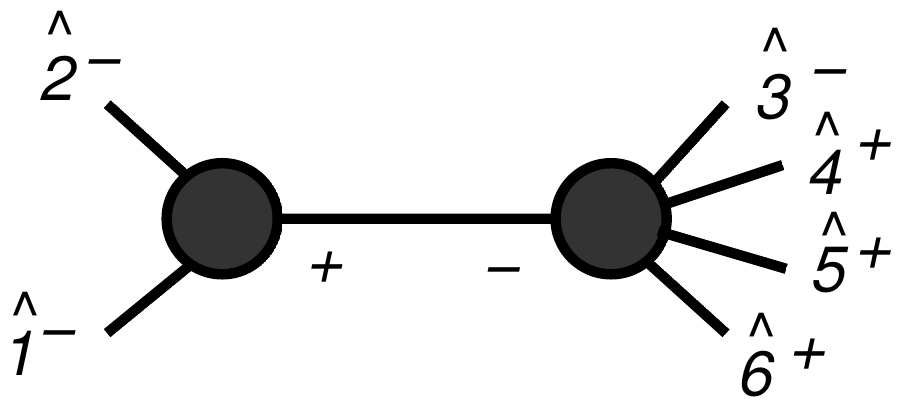}}
~+~\,
   \raisebox{-5.5mm}{\includegraphics[height=1.3cm]{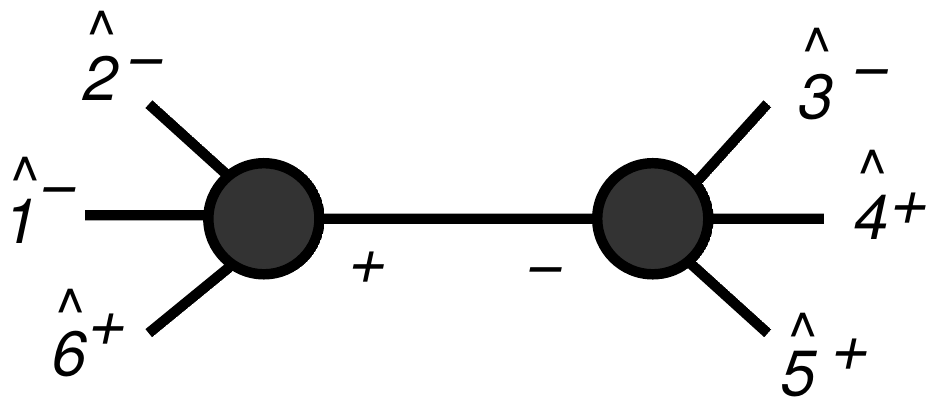}}
~+~\,
   \raisebox{-5.5mm}{\includegraphics[height=1.3cm]{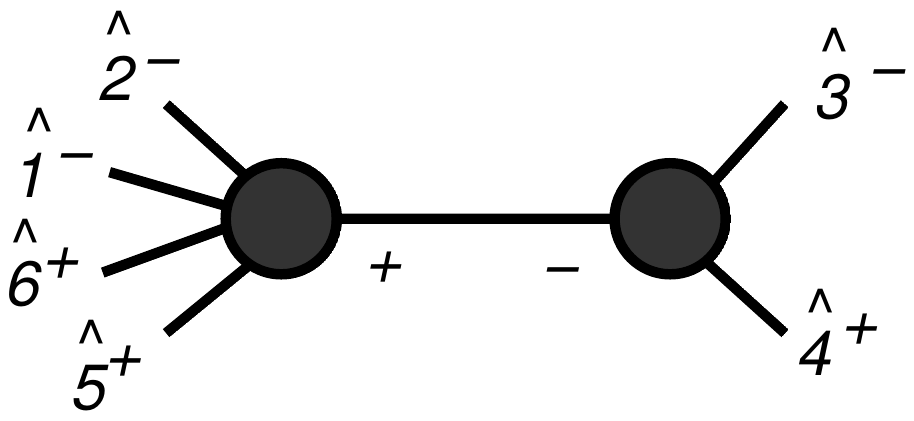}}    
     \end{split}
     \label{6ptcsw1}
\ee
All 6 diagrams are non-vanishing and this may look a little daunting, especially compared with the BCFW version where there were just 2 diagrams in the simplest version \reef{A6bcfw2}. However, the diagrams in \reef{6ptcsw1}
are easier to evaluate than the BCFW diagrams: the MHV amplitudes depend only on angle spinors, so the only way they know about the square-spinor shift is through the internal line angle spinors $|\hat{P}_I\>$, for example
\be
  \raisebox{-5.5mm}{\includegraphics[height=1.3cm]{csw-6pt-A2}}
   ~=~
   \frac{\< 1 \hat{P_I}\>^4}{\< 1 \hat{P_I}\>\< \hat{P_I} 5\>\<56\>\<61\>}
   \frac{1}{P_{156}^2} 
   \frac{\< 2 3\>^4}{\< 2 3\> \<34\> \< 4 \hat{P_I} \>\< \hat{P_I}2\>}\,.
\ee
We can write
\be
   \label{PIno1}
   |\hat{P}_I\> \frac{[\hat{P}_I X]}{[\hat{P}_I X]}
   = \hat{P}_I |X] \,\frac{1}{[\hat{P}_I X]}
   = P_I |X]\, \frac{1}{[\hat{P}_I X]} \,. 
\ee
In the last step we can drop the hat, because the shift of $(\hat{P}_I)^{\dot{a}b}$ is proportional to the reference spinor $[X|^b$ of the shift \reef{sqshift}. 
Note that the diagrams are necessarily invariant under little group scaling associated with the internal line. Therefore the factors $\frac{1}{[\hat{P}_I X]}$ in \reef{PIno1} cancel out of each diagram and we can use the prescription 
\be
   \label{cswprescription}
   |\hat{P}_I\> \to {P}_I |X]  \,.
\ee
This gives
\be
  \raisebox{-5.5mm}{\includegraphics[height=1.3cm]{csw-6pt-A2}}
   ~=~
   \frac{\< 1 |P_{156}|X]^4}{\< 1 |P_{156}|X]\< 5|P_{156}|X]\<56\>\<61\>}
   \frac{1}{P_{156}^2} 
   \frac{\< 2 3\>^4}{\< 2 3\> \<34\> \< 4 |P_{156}|X]\< 2|P_{156}|X]}
\ee
and similarly for the other `MHV vertex diagrams' in \reef{6ptcsw1}.
Note that we can drop the indication $\hat{~}$ of the shift  on the external lines in the MHV vertex diagrams diagrams since the square-spinor shift does not affect the MHV vertices and all that is needed is the CSW prescription \reef{cswprescription} for the internal lines. 

In general, each diagram depends explicitly on the reference spinor $|X]$, but of course the full tree amplitude cannot depend on an arbitrary spinor: the Cauchy theorem argument of Section \ref{s:shifts} guarantees that the sum of all the diagrams will be independent of $|X]$ and reproduce the correct tree amplitude. Numerically, it is not hard to verify  independence of $|X]$ and that the expressions \reef{6ptcsw1} and \reef{A6bcfw2} indeed produce the same scattering amplitude. 

The expansion of the amplitude in terms of MHV vertex diagrams generalizes beyond the NMHV level. In general, the  N$^K$MHV tree amplitude is written as a sum of all tree-level diagrams with precisely $K\!+\!1$ MHV vertices evaluated via the replacement rule \reef{cswprescription}. This construction of the amplitude is called the {\bf \em MHV vertex expansion}: it can be viewed as the closed-form solution to the all-line shift recursion relations.
However, it was discovered by Cachazo, Svrcek, and Witten in 2004 \cite{cswref} before the introduction of recursion relations from complex shifts. The method is therefore also known as the {\bf \em CSW expansion} and the rule \reef{cswprescription} is called the {\bf \em CSW prescription}. The first recursive derivation of the MHV vertex expansion was given by Risager \cite{Risager:2005vk} using the 3-line Risager-shift mentioned above applied to the three negative helicity line of NMHV amplitudes. The all-line shift formulation was first presented in \cite{Elvang:2008vz}. 
\exercise{}{Construct $A_5[1^-2^-3^-4^+5^+]$ from the CSW expansion. Make a choice for the reference spinor $|X]$ to simplify the calculation and show that the result agrees with anti-Parke Taylor formula.}
The MHV vertex expansion was the first construction of gluon amplitudes from on-shell building blocks. The methods is valid also in other cases, for example in super Yang-Mills theory \cite{Elvang:2008na,Elvang:2008vz} or Higgs amplitudes with gluons and partons \cite{Dixon:2004za,Badger:2004ty}. There are also applications of the MHV vertex expansion at loop-level --- for a review see \cite{Brandhuber:2011ke} and references therein. 

The MHV vertex expansion can also be derived directly from a Lagrangian \cite{MHVlagr}: a field redefinition and suitable light-cone gauge-choice brings it to a form with an interaction term for each MHV amplitude. The N$^K$MHV amplitudes are then generated from the MHV vertex Lagrangian by gluing together the MHV vertices. The reference spinor $|X]$ arises from the light-cone gauge choice. There is also an interesting twistor-action formulation of the MHV vertex expansion \cite{Boels:2007qn}.

In the case of the BCFW shift, we applied it to the gluon as well as graviton amplitudes. A version of the MHV vertex expansion was proposed for gravity in \cite{BjerrumBohr:2005jr} based on the Risager shift. However, the method fails for NMHV amplitudes for $n\ge 12$: under the Risager-shift 
 $\hat{A}_n(z) \sim z^{12-n}$ for large-$z$, so for $n \ge 12$ there is a boundary term obstructing the recursive formula \cite{Bianchi:2008pu}.
An analysis of validity of all-line shift recursion relations can be found in \cite{CEK}.

At this stage, you may wonder why tree-level gluon scattering amplitudes have so many different representations: one from the MHV vertex expansion and other forms arising from BCFW applied to various pairs of external momenta. The CSW and BCFW   representations reflect different aspects of the amplitudes, but they turn out to be closely related. We need more tools to learn more about this. So read on.

%%%%%%%%%%%%%%%%%%%%%%%%%%%%%%% 
%%%%%%%%%%%%%%%%%%%%%%%%%%%%%%% 
%%%%%%%%%%%%%%%%%%%%%%%%%%%%%%% 
\newpage
\setcounter{equation}{0}
\section{Supersymmetry}
\label{s:susy}
We begin this section with a very brief introduction to supersymmetry; it serves to give the minimal amount of information we need for our amplitude studies. We then discuss supersymmetry Ward identities for the amplitudes and introduce on-shell superspace as a tool for organizing the amplitudes into superamplitudes. This is particularly powerful in $\cn = 4$ super Yang-Mills theory where it allows us to solve super-BCFW recursion relations to find all tree-level superamplitudes. We better get started.

%%%%%%%%%%%%%%%%%%%%%%%%%%%%%%% 
%%%%%%%%%%%%%%%%%%%%%%%%%%%%%%% 
%%%%%%%%%%%%%%%%%%%%%%%%%%%%%%% 
\subsection{Introduction: $\mathcal{N}=1$ chiral supermultiplet}

Let us begin with a simple example of supersymmetry. Consider the free Lagrangian for a Weyl fermion $\psi$ and complex scalar field $\phi$:
\be
  \label{L0susy}
  \Lag_0 = 
  i \psi^\dagger \bar{\sigma}^\mu \pa_\mu \psi 
   - \pa_\mu \bar\phi \,\pa^\mu \phi \,.
\ee
The bar on $\phi$ denotes the complex conjugate.
In addition to the usual Poincar\'e symmetry, $\Lag_0$ also has a symmetry that mixes the fermions and bosons: 
\be
  \label{susytr}
  \begin{array}{rclcrcl}
  \d_\eps \phi &=& \eps \psi \,,
  &~& 
  \d_\eps \bar\phi &=& \eps^\dagger \psi^\dagger\,,\\[1mm]
  \d_\eps \psi_a &=& -i  \s^\m_{a \dot{b}} \, \eps^{\dagger \dot{b}} \pa_\m \phi\,,
  &~&
  \d_\eps \psi^\dagger_{\dot{a}} &=& 
  i  \pa_\m \bar\phi \,\eps^{b} \s^\m_{b \dot{a}} \,.
\end{array}
\ee
This is an example of a \emph{supersymmetry transformation}.
The anti-commuting constant spinor $\eps$ is the supersymmetry parameter 
(a fermionic analogue of the infinitesimal angle $\theta$ of a rotation transformation), and  $\eps \psi = \eps^a \psi_a$ and
 $ \eps^\dagger \psi^\dagger=\eps^\dagger_{\dot{a}} \psi^{\dagger \dot{a}}$ 
are the usual 2-component spinor products. 

If you have not previously seen supersymmetry, you should promptly go ahead and do these two exercises:
\exercise{}{Check that $\lag_0$ in \reef{L0susy} is invariant under the supersymmetry variation \reef{susytr} up to a total derivative. 
}
\exercise{}{Calculate $[\d_{\eps_1},\d_{\eps_2}]$ by acting with it on the fields. You should find that `the combination of two supersymmetry transformations is a spacetime translation'.
}
The 2-component spinors above can be combined into 4-component  Majorana spinors $\Psi_\text{\tiny M}$ and $\eps_\text{\tiny M}$, and the supersymmetry transformations are then defined  with suitable L- and R-projections $P_{L,R}$ from \reef{g5proj}. We write the free field expansions as 
\bea
  \nonumber
   \phi(x) &=&
   \int \widetilde{dp} 
    \,\Big[
       \,a_-(p) \,e^{ip.x}
       + a_+^\dagger(p) \,e^{-ip.x}
    \Big]
   \\[1mm]
   \psi_a(x) &=& P_L \Psi_\text{\tiny M}(x)
    ~=~
    \sum_{s=\pm} \int \widetilde{dp} 
    \,\Big[
       b_s(p) \,P_L u_s(p)\,e^{ip.x}
      +b^\dagger_s(p) \,P_L v_s(p)\,e^{-ip.x}
    \Big]\,,
\eea
and similarly for $\bar\phi(x)$ and  
$\psi^{\dagger\dot{a}}=P_R \Psi_\text{\tiny M}$.
The annihilation/creation operators for the fermion is labelled by helicity $h=\pm \tfrac{1}{2}$.  Upon canonical quantization, they satisfy the algebra of bosonic/fermionic creation-annihilation operators:
\be
  \big[ a_\pm(p) , a_\pm^\dagger(p') \big] = (2\pi)^3 \,2 E_p\, \delta^3 (\vec{p}-\vec{p}\,')
  \,,~~~~~~
  \big\{ b_\pm(p) , b_\pm^\dagger(p') \big\} = (2\pi)^3 \,2 E_p\, \delta^3 (\vec{p}-\vec{p}\,')\,,
\ee
with all other (anti)commutators vanishing.

For the fermions, the $\pm$-subscripts on the operators indicate the helicity $h=\pm \tfrac{1}{2}$. 
As a matter of later convenience, we have also labelled the two sets of annihilation/creation operators associated with the complex field $\phi$ with $\pm$-subscripts. The corresponding particles are of course scalars with $h=0$, but the label  indicates which spinor helicity state the scalar state is matched to via supersymmetry. Let us see how that works. 

The supersymmetry transformations \reef{susytr} transform the fields $\psi$ and $\phi$ into each other, and therefore the associated annihilation/creation operators are also related. The relationship is straightforward to extract from the free field expansions. Recalling from our introduction to the spinor helicity formalism that $P_L v_s(p)$ is equal to $|p]$ for $s=+$ and vanishes for $s=-$ (and similarly for $P_R$), one finds
\bea
  \label{dsusyO}
  \begin{array}{rclcrcl}
  \d_\eps a_-(p) &=& [\eps\, p]\, b_-(p)\,,
  &~&
  \d_\eps b_-(p) &=& \<\eps\, p\> \,a_-(p)\,, \\[2mm]
  \d_\eps a_+(p) &=& \<\eps\, p\> \, b_+(p)\,,
  &~&
  \d_\eps b_+(p) &=& [\eps\, p] \,a_+(p)\,.
  \end{array}
\eea 
We have introduced anti-commuting bra-kets $|\eps]_a = \eps_a$ and $\<\eps|_{\dot{a}} = \eps^\dagger_{\dot{a}}$ for the supersymmetry parameter.
Using $-|p] \<p| = p_{a\dot{b}}$, it is easy to see that $[\d_{\eps_1},\d_{\eps_2}] \mathcal{O}(p)= a^\m p_\m \mathcal{O}(p)$ for $\mathcal{O}(p)$ any one of the creation/annihilation operators. The translation parameter $a^\m$ can be written in terms of Majorana spinors as $a^\m = \eps_{\text{\tiny M},2} \ga^\m \eps_{\text{\tiny M},1}$. 

The generators 
$Q_\text{\tiny M} = 
\Big( \begin{array}{c} Q_a \\ Q^{\dagger\dot{a}} \end{array} \Big)$ 
can be found from 
$\d_\eps \mathcal{O} = \big[\bar\eps_\text{\tiny M} Q_\text{\tiny M}, \mathcal{O}\big] = \big[ [\eps Q] + \<\eps Q\>, \mathcal{O} \big]$. One finds (and you should check it) that
\be
  \begin{split}
  |Q]_a  &=~ \int \widetilde{dp}\,\, |p]_a \,
  \big( a_+(p)\, b_+^\dagger(p) - b_-(p)\, a_-^\dagger(p)\big)\,,\\
  |Q^\dagger\>^{\dot{a}}  &=~ \int \widetilde{dp}\,\,  |p\>^{\dot{a}} \,
  \big( a_-(p)\, b_-^\dagger(p) -  b_+(p)\, a_+^\dagger(p)\big)\,,
  \end{split}
  \label{QsqQang}
\ee
reproduces \reef{dsusyO}.
\exercise{}{Show that $\big\{ |Q]_a , \<Q^\dagger|_{\dot{b}} \big\}$ equals $p_{a\dot{b}}$ times the sum of the number operators.
}
The action of the supersymmetry generators \reef{QsqQang} on the annihilation operators is then 
\bea
  \label{Qact}
  \begin{array}{rclcrcl}
  [Q  \,,a_-(p) ] &=& |p] \,b_-(p)\,, 
  &~~&
  [Q^{\dagger}\,,b_-(p) ] &=& |p\> \,a_-(p)\,,   \\[2mm]{}
  [Q  \,,b_-(p) ] &=& 0\,, 
  &~~&
  [Q^{\dagger}\,,a_-(p) ] &=& 0\,,   \\[3mm]{}  
  [Q,b_+(p) ] &=& |p] \,a_+(p)\,,     
  &~~&
  [Q^\dagger,a_+(p) ] &=&  |p\>\,b_+(p)\,,  \\[2mm]{}
  [Q,a_+(p) ] &=& 0\,,     
  &~~&
  [Q^\dagger,b_+(p) ] &=&  0\,, 
  \end{array}
\eea
where  $[\,.\,,\,.\,]$ is a graded bracket that is an anti-commutator when both arguments are Grassmann and otherwise a commentator. The 2-component spinor-indices are suppressed. A similar set of relations hold for the creation operators.

It follows from our discussion that the spectrum of the model splits into a `negative helicity sector' and a `positive helicity sector'; CPT symmetry requires us to have both. In each sector, the states are related by supersymmetry --- they are said to belong to the same \emph{supermultiplet}. We note that $Q$ lowers the helicity by $\tfrac{1}{2}$ and that $Q^\dagger$ raises it by $\tfrac{1}{2}$.
The supersymmetry generators commute with the Hamiltonian and the space-translation generators, so states in the same supermultiplet must have the same mass; in our case the mass is of course zero.

\noindent {\bf Interactions.}
Next, we would like to introduce interactions --- after all, this is all about scattering amplitudes so we need something to happen! We want to study interactions that preserve supersymmetry. For our chiral model, one can introduce a `superpotential' interaction of the form
\be
  \label{LIsusy}
  \lag_I = \tfrac{1}{2} g \, \phi\, \psi \psi + \tfrac{1}{2} g^* \, \bar\phi \, \psi^\dagger \psi^\dagger
      - \tfrac{1}{4}|g|^2\,  |\phi|^4  \,.
\ee
To see that the interaction Lagrangian is supersymmetric, a small modification of the supersymmetry transformations \reef{susytr} is needed in the transformation rule of the fermion field: 
\be
  \label{susytrInt}
  \begin{array}{rclcrcl}
  \d_\eps \phi &=& \eps \psi \,,
  &~~& 
  \d_\eps \bar\phi &=& \eps^\dagger \psi^\dagger \,,\\[1mm]
  \d_\eps \psi_a &=& -i  \s^\m_{a \dot{b}} \, \eps^{\dagger \dot{b}} \pa_\m \phi
  + \tfrac{1}{2} g^* {\bar\phi}^2 \eps_a\,,
  &~~&
  \d_\eps \psi^\dagger_{\dot{a}} &=& 
  i  \pa_\m \bar\phi\,\eps^{b} \s^\m_{b \dot{a}} 
  + \tfrac{1}{2} g \phi^2\,\eps^\dagger_{\dot{a}}
  \, \,,
\end{array}
\ee
\exercise{}{Show that \reef{susytrInt} is a symmetry of $\Lag=\Lag_0 + \Lag_I$.}
Note that the coupling of 4-scalar interaction in \reef{LIsusy} is fixed in terms of the Yukawa coupling $g$ by supersymmetry. A linear version of the supersymmetry transformations can be given using an auxiliary field. Supersymmetric actions can be expressed compactly and conveniently using off-shell superspace formalism. You can find much more about this in textbooks such as Wess and Bagger \cite{Wess:1992cp}. 

\noindent {\bf Other supermultiplets.}
So far we have focussed on a very simple case of a `chiral supermultiplet' in which a spin-0 particle is partnered with spin-$\tfrac{1}{2}$ particle. One can repeat the analysis for any $\mathcal{N}=1$ supersymmetric model with particles of spin $(s,s+\tfrac{1}{2})$. For example, super QED, with a photon and a photino with helicities $\pm 1$ and $\pm \frac{1}{2}$, or $\mathcal{N}=1$ super Yang Mills with a gluon ($h=\pm1$) and a gluino ($h=\pm\tfrac{1}{2}$).  A nice feature is that the action of the supersymmetry generators on the states basically takes the same form \reef{Qact}. 

\noindent {\bf Extended supersymmetry.} The $\mathcal{N}$ counts the number of supersymmetry generators. In extended supersymmetry $\mathcal{N}>1$, there are $2^{\mathcal{N}}$ states in the massless supermultiplets (and the same in the CPT conjugates). For example, for $\mathcal{N}=2$, one supermultiplet consists of a helicity $-1$ photon/gluon, two photinos/gluinos with helicity $-\frac{1}{2}$, and a scalar with helicity 0.  Thus the multiplet has two bosonic d.o.f. and two fermionic. 
The CPT conjugate multiplet contains the same types of states but with opposite helicity.

To avoid states with spin higher than 1, the maximal amount of supersymmetry in four-dimensions is $\mathcal{N}=4$. This large symmetry-requirement places such strong constraints on the theory that it is unique (up to choice of gauge group): this is $\mathcal{N}=4$ super Yang-Mills theory (SYM). We are going to study the supersymmetry constraints on the amplitudes in much further detail in Section \ref{s:N4sym}. For now, let us content ourselves with $\mathcal{N}=1$ supersymmetry and study the consequences of it on the scattering amplitudes. 

%
%%%%%%%%%%%%%
\subsection{Amplitudes and the supersymmetry Ward identities}
\label{s:SWI}

In this section, we study the effects of supersymmetry on the amplitudes of our simple chiral model whose Lagrangian $\Lag$ is the sum of the free a Lagrangian \reef{L0susy} and \reef{LIsusy}. The 4-point tree amplitudes were essentially already presented in our earlier Yukawa theory examples of how to use spinor helicity formalism with the Feynman rules (see Exercise \ref{ex:nearSUSY}). To adapt the results from section  \ref{s:yukawa}, we just need to take the coupling of the 4-scalar interaction to be $\lambda = |g|^2$. The 4-point amplitudes are then:
\be
  \label{A4susy}
  A_4(\phi \phi \bar\phi \bar\phi) = - |g|^2\,,
  ~~~~~
  A_4(\phi \,f^- f^+ \bar\phi) =  -|g|^2 \frac{\<24\>}{\<34\>}\,,
  ~~~~~
  A_4(f^- f^- f^+ f^+) =  |g|^2 \frac{\<12\>}{\<34\>}\,.~~
\ee
By inspection of \reef{A4susy}, we see that 
\bea
 \label{srel1}
  A_4(\phi \,f^- f^+ \bar\phi) 
  &=&
  \frac{\<24\>}{\<34\>}\,
  A_4(\phi \phi \bar\phi \bar\phi)\,, \\[1mm]
 \label{srel2}
   A_4(f^- f^- f^+ f^+)
  &=&
 - \frac{\<12\>}{\<24\>}\,
  A_4(\phi \,f^- f^+ \bar\phi)\,.
\eea
These relations hold not just for the tree-level amplitudes, as we have  seen it just now;  supersymmetry ensures that \reef{srel1}-\reef{srel2} hold at all orders in the perturbation expansion. We will now see how that comes about. 

We can think of an $n$-point amplitude with all-outgoing particles as the S-matrix element 
$\< 0 | \mathcal{O}_1(p_1) \dots \mathcal{O}_n(p_n) |0\>$ in which the
$n$ annihilation operators $\mathcal{O}_i(p_i)$, $i=1,\dots, n$, act to the left on the out-vacuum.  For example, 
$A_4(\phi \,f^- f^+ \bar\phi) 
= \< 0| a_-(p_1) b_-(p_2)  b_+(p_3) a_+(p_4) |0\>$ and the tree-level result is listed in \reef{A4susy}. Suppose the vacuum is supersymmetric: $Q |0\> = 0 = Q^\dagger |0\>$. Then for any set of $n$ annihilation (or creation) operators, we have
\bea
  \nonumber
  0 &=&
  \< 0 | \big[Q^\dagger, \mathcal{O}_1(p_1) \dots \mathcal{O}_n(p_n)\big] |0\>
  \\[1mm]
  &=&
  \sum_{i=1}^n
  (-1)^{\sum_{j<i}{|\mathcal{O}_j|}}~
  \< 0 | \mathcal{O}_1(p_1) \cdots \big[Q^\dagger, \mathcal{O}_i(p_i) \big]\cdots \mathcal{O}_n(p_n) |0\> \,,
  \label{genSWI}
\eea
and similarly for $Q$. Here the sign-factor takes into account that a minus sign is  picked up from every time $Q^\dagger$ passes by a fermionic operator: so $|\mathcal{O}|$ is 0 when the operator is bosonic and 1 if fermionic. 
Now using the action of the supersymmetry generators \reef{Qact}  on the free asymptotic states, the equation \reef{genSWI} will describe a linear relation among  scattering amplitudes whose external states are related by supersymmetry. Such relations are called {\bf \em supersymmetry Ward identities}. It is easier to see how this works in an explicit example:
\bea
   \nonumber
    0
    &=&
    \< 0 | \big[Q^\dagger, a_-(p_1) b_-(p_2) a_+(p_3) a_+(p_4) \big] |0\>
    \\[2mm]    
   \nonumber
    &=&
    |2\> \,\< a_-(p_1) a_-(p_2) a_+(p_3) a_+(p_4) \>
    -|3\> \,\< a_-(p_1) b_-(p_2) b_+(p_3) a_+(p_4) \>\\
    &&
    -|4\> \,\< a_-(p_1) b_-(p_2) a_+(p_3) b_+(p_4) \>\,.
\eea
We have used that $Q^\dagger$ annihilates $a_-(p)$. Translating to amplitudes we have 
\be
  \label{SWIex1}
  0  ~=~
    |2\> \,A_4(\phi \phi \bar\phi \bar\phi)
    -|3\> \,A_4(\phi \,f^- f^+ \bar\phi) 
    -|4\> \,A_4(\phi \,f^- \bar\phi f^+ ) \,.
\ee
Note that each identity \reef{genSWI} encodes two relations, since $Q^\dagger$
($Q$) has two components. This is also visible in our example \reef{SWIex1}. We an project out the two independent relations by dotting in a suitable choice of bra-spinor $\<r|$. Picking $\<r| = \<4|$, we find
\be
  \label{exA1}
  0  ~=~
    \<42\> \,A_4(\phi \phi \bar\phi \bar\phi)
    -
    \<43\> \,A_4(\phi \,f^- f^+ \bar\phi) \,,
\ee
which is precisely the relation \reef{srel1} we found to be true at tree-level.

A second relation is extracted from \reef{SWIex1} by choosing $\<r| = \<2|$:
\be
  \label{exA2}
        A_4(\phi \,f^- \bar\phi f^+ )  ~=~
    -\frac{\<23\>}{\<24\>} \,A_4(\phi \,f^- f^+ \bar\phi) \,.
\ee
Note how the supersymmetry factor $\tfrac{\<23\>}{\<24\>}$ nicely compensates the different little group scaling of the two amplitudes.
 \exercise{}{Plug in the two SUSY Ward identities \reef{exA1} and \reef{exA2} 
 into \reef{SWIex1} to show that there are no other independent information available in \reef{SWIex1}.
 }
 \exercise{}{Derive \reef{srel2} as a SUSY Ward identity.
 }
 \exercise{}{Find a $Q$-Ward identity that shows that 
 $A_4(\phi \,f^- f^+ \bar\phi) =-
  \frac{[13]}{[12]}\,
  A_4(\phi \phi \bar\phi \bar\phi)$. Show that this relation is the equivalent to \reef{srel1}.
 }
Let us take a brief look at the SUSY Ward identities at higher points, for example for amplitudes with 6-particles. 
Starting with 
$0=\< 0 | \big[Q^\dagger, a_{1-} a_{2-} b_{3-} a_{4+} a_{5+} a_{6+} \big] |0\>$
(using a short-hand notation to indicate the momentum with a subscript) we find, after dotting in $\<r|$
\bea
   \nonumber
    0
    &=&
    \<r3\> \,A_6(\phi\phi\phi\bar\phi\bar\phi\bar\phi)
    - \<r4\> \,A_6(\phi\phi \,f^-f^+ \bar\phi\bar\phi) \\
    &&
    - \<r5\> \,A_6(\phi\phi \,f^- \bar\phi\,f^+\bar\phi)
    - \<r6\> \,A_6(\phi\phi \,f^- \bar\phi\bar\phi\,f^+)\,.
\eea
There are two pieces of information, but four `unknowns' (the amplitudes), so this time the relations do not  give simple proportionality relations among the amplitudes. Instead one gets a web of linear relations; this is typical for non-MHV type amplitudes. 

To summarize, amplitudes with external states related by supersymmetry are related to each through linear relationships called {\em supersymmetric Ward identities}. They were first studied in 1977 by Grisaru, Pendleton, and van Nieuwenhuizen \cite{SWI} and have since then had multiple applications.

In our discussion of recursion relations in Section \ref{s:validity}, we learned that in general we should not expect it to be possible to produce a 4-scalar amplitude recursively from 3-particle amplitudes because of the possibility of input from a 4-scalar contact term. However, in our supersymmetric example, the coupling of the 4-scalar interaction is determined by supersymmetry by the 3-point interactions, so one should expect that recursion relations work, in particular that all 4-point amplitudes can be determined by the 3-point ones --- but one must build supersymmetry into the recursion relations. This is done most efficiently in two steps: first one introduces \emph{on-shell superspace} and groups the amplitudes into \emph{superamplitudes}. Secondly, one incorporates a shift of the Grassmann super-parameter of the on-shell superspace into the BCFW-shift. Then one gets recursion relations for the superamplitudes. This is best illustrated for $\mathcal{N}=4$ SYM, so that is what we will turn to next.

%%%%%%%%%%%%%
\subsection{$\mathcal{N}=4$ SYM: on-shell superspace and superamplitudes}
\label{s:N4sym}

The action for $\mathcal{N}=4$ super Yang-Mills theory (SYM) 
can be written compactly as 
\be
  S = \int d^4x \, \Tr \bigg(
   -\frac{1}{4} F_{\m\n}F^{\m\n}
   -\frac{1}{2} (D\Phi_{I})^2 
   + \frac{i}{2} \overline{\Psi}\, \Dslash\,  \Psi
   +\frac{g}{2} \overline{\Psi}\, \Gamma^I [  \Phi_I, \Psi]
   +\frac{g^2}{4} [\Phi_I,\Phi_J]^2
   \bigg) \,.
   \label{N4SYMaction}
\ee
Here $D_\mu=\pa_\m - ig [A_\m, \cdot\,]$ is the covariant derivative, 
$A_\m$ is the vector potential field,  $\Phi_I$ label six real scalar fields, and $I=1,\dots, 6$ are labels of  the global $SO(6)$ R-symmetry. All fields are in the adjoint of the gauge group which we take to be $SU(N)$; the commutators in \reef{N4SYMaction} are associated with the $SU(N)$ matrix structure of the fields. 
The fermions are  represented by 10-dimensional Majorana-Weyl fields $\Psi$. The $\Gamma^I$ are gamma-matrices of the 10d Clifford algebra. This description of  $\mathcal{N}=4$ SYM follows from dimensional reduction of $\mathcal{N}=1$ SYM in 10d \cite{Brink:1976bc}. 

It is convenient to group the 6 real scalars $\Phi_I$ into 6 complex scalar fields $\varphi^{AB} = - \varphi^{BA}$, with $A,B=1,2,3,4$, that satisfy  the self-duality condition
$\overline{\varphi}_{AB} = \frac{1}{2} \eps_{ABCD}\, \varphi^{CD}$. Here  $\eps_{ABCD}$ is the Levi-Civita symbol of $SU(4)\sim SO(6)$, the scalars $\varphi^{AB}$ transform in the fully antisymmetric 2-index representation of $SU(4)$. In this language, the 10d fermion fields give 4+4  gluino states $\lambda^A$ and $\bar{\lambda}_A$ that transform in the (anti-)fundamental of $SU(4)$.

In a supersymmetric model, the value of the scalar potential $V$ at the vacuum is an order parameter of supersymmetry breaking \cite{Wess:1992cp}: in flat space, $V=0$ is necessary for preserving supersymmetry while  $V> 0$ breaks supersymmetry. In $\cn=4$ SYM, the scalar potential is $V= [\Phi_I,\Phi_J]^2$, so the theory has a moduli space of $\cn=4$ supersymmetric vacua with $[\Phi_I,\Phi_J]=0$. At the \emph{origin of  moduli space}, where all the scalar vevs vanish, $\<\varphi^{AB}\>=0$, all states are massless and the theory contains no dimensionful parameters. In fact, the theory is \emph{conformal} invariant: the trace of the stress-tensor is zero, up to the trace anomaly. In particular, 
the beta-function vanishes at all orders in perturbation theory and there is no running of the coupling. The theory is invariant under an enlarged spacetime symmetry group, namely the conformal group $SO(2,4)$.  Supersymmetry enhances the conformal symmetry to  superconformal symmetry with  $PSU(2,2|4)$; we discuss this in Section \ref{s:confsym}.

When the scalars acquire vevs in such a way that full supersymmetry is preserved, i.e.~$[\Phi_I,\Phi_J]=0$, the theory is said to be on the \emph{Coulomb branch}.\footnote{The theory also has an $\cn=2$ supersymmetric Higgs branch where the moduli are scalars of the hypermultiplet (as opposed to the Coulomb branch where the scalars are part of the vector multiplet).} We will briefly discuss scattering amplitudes on the Coulomb branch of $\cn=4$ SYM in Exercise \ref{ex:coulomb}, but otherwise focus entirely on the superconformal theory at the origin of moduli space: \emph{henceforth when we discuss amplitudes in  $\cn=4$ SYM  we are implicitly taking this to mean the theory at the origin of moduli space.}

Given that the theory is conformal, we should  clarify what we mean by the scattering-matrix in $\cn=4$ SYM. One way to think about this is to consider the theory in $4-\eps$ dimensions: then the conformal symmetry is broken and the S-matrix is well-defined. This turns out to be a little inconvenient for keeping on-shell symmetries manifest,\footnote{We discuss the symmetries of $\cn=4$ SYM amplitudes in Section \ref{s:DC}.} and it can therefore be better to consider the theory on the Coulomb branch and define the $\cn=4$ SYM S-matrix as the zero-vev limit of the Coulomb branch S-matrix. These subtleties will not affect the majority of our discussion and therefore we proceed to discuss the amplitudes of $\cn=4$ SYM without further hesitation. 

{\bf Spectrum and supersymmetry Ward identities. }
The spectrum of $\mathcal{N} = 4$ SYM consists of a CPT self-dual supermultiplet with 16 massless states: in order of descending helicity $h=1,\tfrac{1}{2},0,-\tfrac{1}{2},-1$, we list the corresponding annihilation operators as
\be \label{n4ops}
 \underbrace{a}_{1\,\text{gluon}\,g^+} , \hspace{5mm}
 \underbrace{a^A}_{4\,\text{gluinos}\,\lambda^{A}}, \hspace{5mm}
 \underbrace{\,\,a^{AB}
 }_{6\,\text{scalars}\,S^{AB}} ,\hspace{5mm}
 \underbrace{\,\,a^{ABC}}_{4\,\text{gluinos}\,\lambda^{ABC}\,\sim\,\overline{\lambda}_D} ,\hspace{5mm}
 \underbrace{\,\,a^{1234}}_{1\,\text{gluon}\,g^-}\, .
\ee
where the indices $A,B,\ldots=1,2,3,4$ are labels of the global $SU(4)$ R-symmetry\footnote{An R-symmetry is a symmetry that does not commute with supersymmetry.} that rotates the four sets of supersymmetry generators $Q^A$ and $\tilde{Q}_A \equiv Q^\dagger_A$. The helicity $h$ states transform in fully antisymmetric $2(1-h)$-index representations of $SU(4)$.

As in our $\mathcal{N}=1$ chiral supermultiplet example in the previous section, we can find the action of supersymmetry on the annihilation operators:
\bea
\begin{array}{rcl}
\big[\tQ_A,a(i)\big] &=& 0\, ,\\[2mm]
\big[\tQ_A,a^B(i)\big] &=& | i\>\,\d^B_A\,a(i)\, ,\\[2mm]
\big[\tQ_A,a^{BC}(i)\big]
&=&|i\> \, 2! \, \d^{[B}_A\,a^{\raisebox{0.7mm}{\scriptsize$C]$}}(i)\, ,\\[2mm]
\big[\tQ_A,a^{BCD}(i)\big] &=&| i\>\, 3!\, \d_A^{[B} a^{\raisebox{0.6mm}{\scriptsize$CD]$}}(i)\, ,\\[2mm]
\big[\tQ_A,a^{BCDE}(i)\big] &=& |i\>\, 4!\, \d_A^{[B} a^{\raisebox{0.6mm}{\scriptsize$CDE]$}}(i) \, ,
\end{array}
\hspace{8mm}
\begin{array}{rcl}
[Q^A,a(i)] &=& [i|\,a^A(i)\, , \\[2mm]
\big[Q^A,a^B(i)\big] &=& [i|\, a^{AB}(i)\, , \\[2mm]
\big[Q^A,a^{BC}(i)\big]
&=&[i|\, a^{ABC}(i)\, , \\[2mm]
\big[Q^A,a^{BCD}(i)\big]
&=& [i|\,a^{ABCD}(i)\, , \\[2mm]
\big[Q^A,a^{1234}(i)\big] &=& 0\, .
\end{array}
\label{n4tQ}
\eea
Note that $\tQ_A$ raises the helicity of all operators by $\tfrac{1}{2}$ and removes the index $A$ (if it is not available to be removed, then the operator is annihilated). $Q^A$ does the opposite.

To start with, we consider some examples of $\cn=4$ SYM Ward identities. Since $\tQ_A$ annihilates $a(i)$, the Ward identity 
$\<0| [\tQ_A, a^{B}_1 a_2 \dots a_n ]|0\> = 0$
gives
$\delta_A^B\, |1\> A_n[g^+ g^+ g^+ \dots g^+] = 0$. This directly says that the all-plus gluon amplitudes vanishes at all orders in perturbation theory. Similarly, one can show that the gluon amplitude with exactly one negative helicity gluon vanishes. So
\be
\label{susypm}
 \text{super Yang-Mills:}
 ~~~~~~
 A^{L\text{-loop}}_n[g^+ g^+ g^+ \dots g^+]~=~
 A^{L\text{-loop}}_n[g^- g^+ g^+ \dots g^+]~=~0\,.
\ee
We used only one supersymmetry generator for this argument, so the statement \reef{susypm} is true in any super Yang Mills theory, not just in  $\mathcal{N}=4$ SYM. In pure non-susy  Yang-Mills, we have seen  
in Section \ref{s:MHV} that \reef{susypm} holds at tree-level
\be
 \text{Yang-Mills:}
 ~~~~~~
 A^{\text{tree}}_n[g^+ g^+ g^+ \dots g^+]~=~
 A^{\text{tree}}_n[g^- g^+ g^+ \dots g^+]~=~0\,.
 \label{YMallplustree}
\ee
However, this is not true at loop-level without supersymmetry: for example in pure Yang-Mills theory, all-plus amplitudes are indeed generated at the 1-loop level. 
The reason the result \reef{YMallplustree} holds at tree-level in pure Yang-Mills theory, is that the superpartners of the gluon couple quadratically to the gluon. So an  amplitude whose external states are all gluons `sees' the superpartner states only via loops. Thus the gluon amplitudes at tree-level must obey the same Ward identity constraints as the gluon amplitudes  in super Yang-Mills. 
\exercise{}{The non-vanishing $n=3$ anti-MHV amplitude escapes the Ward identity that forces $A_n[g^- g^+ g^+ \dots g^+]=0$. Explain how.}
\exercise{}{Show that the SUSY Ward identities give the following relationships among the color-ordered amplitudes in $\cn=4$ SYM:
\be
   0 ~=~    
     -  |1\>A_n[ \lambda^{123} g^- \lambda^4 g^+\dots g^+]
     - |2\>A_n[g^- \lambda^{123} \lambda^4 g^+\dots g^+] 
     + |3\> A_n[g^- g^- g^+g^+\dots g^+]
     \label{SWIN4sym}
\ee
and
\bea
\nonumber
   A_n[g^- \lambda^{123} \lambda^4 g^+\dots g^+]
  &=&  \frac{\<13\>}{\<12\>} A_n[g^- g^- g^+g^+\dots g^+]\,,\\[1mm]
  \label{n4SWIex}
  A_n[g^- S^{12} S^{34} g^+\dots g^+]
  &=&  \frac{\<13\>^2}{\<12\>^2} A_n[g^- g^- g^+g^+\dots g^+]\,,
  \\[1mm]
\nonumber
  A_n[g^+g^+\dots g^-_i \dots g^-_j \dots g^+]
  &=&  \frac{\<ij\>^4}{\<12\>^4} A_n[g^- g^- g^+g^+\dots g^+]\,.
\eea
}
Notice the powerful 3rd identity in \reef{n4SWIex}. You have already seen this build into the Parke-Taylor formula for the tree-level MHV gluon  amplitudes. And the focus on  $A_n(1^-2^-3^+\dots n^+)$ in our recursive proof of the Parke-Taylor formula is now justified: the supersymmetry Ward identities ensure Parke-Taylor to hold for an MHV tree gluon amplitude with the two negative helicity gluons in any position.

We introduced N$^K$MHV amplitudes earlier as the gluon amplitudes with $K\!+\!2$ negative helicity gluons. Now we can expand this to define the {\bf \em N$^K$MHV sector\,} to be all amplitudes connected to the N$^K$MHV gluon amplitude via supersymmetry. For example, all the amplitudes in \reef{n4SWIex} are MHV. In fact, all MHV amplitudes are proportional to $A_n(g^- g^- g^+g^+\dots g^+)$. This requires that the supersymmetry generators connect the all  states, both positive- and negative-helicity, and that is only possible because the supermultiplet in $\mathcal{N}=4$ SYM is CPT self-conjugate.

The global $SU(4)$ R-symmetry ensures that an amplitude vanishes unless the external states combine to an $SU(4)$ singlet. This requires that the (upper) $SU(4)$ indices appear as a combination of $(K+2)$ sets of $\{1234\}$ for N$^K$MHV amplitudes. Since the supersymmetric Ward identities relate amplitudes with the same number of $SU(4)$ indices, this then provides an alternative definition of the N$^K$MHV sector.

{\bf On-shell superspace. }
It is highly convenient to introduce an on-shell\footnote{There is no (known) off-shell superspace formalism for $\mathcal{N}=4$ SYM.} superspace in order to keep track of the states and the amplitudes. We introduce four Grassmann variables $\h_A$ labeled by the $SU(4)$ index $A=1,2,3,4$.\footnote{Originally, Ferber \cite{Ferber1977qx} introduced these variables as superpartners of the bosonic twistor variables.}
This allows us to collect the 16 states into an $\cn=4$ on-shell chiral superfield
\begin{equation}\label{n4Phi}
    \Omega=g^++\h_A\l^{A}- \frac{1}{2!}\h_A\h_B S^{AB}
    -\frac{1}{3!}\h_A\h_B\h_C  {\l}^{ABC}+\h_1\h_2\h_3\h_4\, g^- \,,
\end{equation}
where the relative signs are chosen such that the Grassmann differential operators
\be
\label{n4ops2}
\cn=4~\text{SYM:}~~~~
  \begin{array}{c|c|c|c|c|c}
  \text{particle} & g^+ & \lambda^{A} & S^{AB}
  & \lambda^{ABC} &g^-=g^{1234} \\[1mm]
  \hline
  \raisebox{-1.5mm}{operator}  &
  \raisebox{-1.5mm}{1}
  & \raisebox{-1.5mm}{$\partial_{i}^A$}
  & \raisebox{-1.5mm}{$\partial_{i}^A\partial_{i}^B$}
  & \raisebox{-1.5mm}{$\partial_{i}^A\partial_{i}^B\partial_{i}^C$}
  & \raisebox{-1.5mm}{$\partial_{i}^1\partial_{i}^2\partial_{i}^3\partial_{i}^4$}
\end{array}
\ee
 select the associated state from $\Omega(p_i)$. 
 
In the on-shell formalism, the supercharges are
\be
\label{qtq}
 q^{A\,a} \equiv  [p|^a\,\frac{\pa}{\pa\h_A} \,,
~\hspace{1cm}
 q^{\dagger\dot{a}}_{A} \equiv |p\>^{\dot{a}}\,\h_A \,,
\ee
where $|p\>$ and $|p]$ are the spinors associated with the null momentum $p$ of the particle.
\exercise{}{Show that the supercharges satisfy the
standard supersymmetry anticommutation relation
$\{q^{A\, a},\tilde{q}^{\db}_B\}= \delta_{B}{}^A\,|p\>^{\db}[p|^{a}=-\delta_{B}{}^A\, p^{\db a}$. The supercharges \reef{qtq}  act on the spectrum by shifting states right or left in $\Omega$. Check that this action matches \reef{n4tQ}.
}
\example{The purpose of this example is to clarify the relation between the on-shell superspace introduced here and the usual off-shell superspace formalism described in textbooks (e.g.~\cite{Wess:1992cp}). In an off-shell $\cn=1$ formalism, the superspace is $(x^\m,\theta^a,\overline{\theta}^{\da})$ with Grassmann variables $\theta^a$ and $\overline{\theta}^{\dot{a}}$.  The algebra of the supercharges is 
\mbox{$\{ \mathcal{Q}_a,   \mathcal{Q}_b  \} = 0$,}
$\{ \overline{\mathcal{Q}}_{\da},   \overline{\mathcal{Q}}_{\db}  \}= 0$, and
$\{ \mathcal{Q}_{a},   \overline{\mathcal{Q}}_{\da}  \}=i(\s^\m)_{a\dot{a}}  \,\partial_\m$.  
As is often convenient for studies of anti-chiral superfields, we can realize the superalgebra with an anti-chiral representation 
\be
  \mathcal{Q}_a 
  =\frac{\pa}{\pa\theta^a} 
\,,~~~~~~
  \overline{\mathcal{Q}}_{\da}
  = -\frac{\pa}{\pa\overline\theta^{\da}} 
  +i {\theta}^{a}(\s^\m)_{a\dot{a}}  \,\partial_\m\,.
\ee
The algebra is represented 
faithfully also when
$\overline{\mathcal{Q}}_{\da}
\to  i {\theta}^{a}(\s^\m)_{a\dot{a}}  \,\partial_\m$.
In momentum space, the 
$\mathcal{Q}\overline{\mathcal{Q}}$-anticommutator 
can then be written 
\be
\bigg\{ 
  \frac{\pa}{\pa\theta^a} 
  ~,~  
  -{\theta}^{a}(\s^\m)_{a\dot{a}}  \,p_\m   \bigg\}
  = - (\s^\m)_{a\dot{a}}  \,p_\m \,.
  \label{QQtheta1}
\ee
Let us now go on-shell and assume that $p^\m$ is lightlike; then we can rewrite \reef{QQtheta1} in spinor helicity formalism as 
\be
\bigg\{ 
  \frac{\pa}{\pa\theta^a} 
  ~,~ 
  {\theta}^{a}|p]_a \<p|_{\dot{a}}   \bigg\}
  = |p]_a \<p|_{\dot{a}}  \,.
  \label{QQtheta2}
\ee
Introduce a new Grassmann-odd variable 
$\eta =  {\theta}^{a}|p]_a$. Then
$\frac{\pa}{\pa\theta^a} 
= |p]_a\, \frac{\pa}{\pa \eta}$, so that \reef{QQtheta2} becomes
\be
  \bigg\{ 
  |p]_a\, \frac{\pa}{\pa \eta}
  ~,~ 
  \eta \<p|_{\dot{a}}   \bigg\}
  = |p]_a \<p|_{\dot{a}} 
  ~~~~~
  \implies
  ~~~~~
  \bigg\{ 
  [p|^a\, \frac{\pa}{\pa \eta}
  ~,~ 
  |p\>^{\dot{a}} \eta    \bigg\}
  = |p\>^{\dot{a}}  [p|^a \,.
  \label{QQtheta3}
\ee
The arguments of the anticommutator, $[p|^a\, \frac{\pa}{\pa \eta}$ and $|p\>^{\dot{a}} \eta$, are recognized as 
$\cn =1$ versions of our on-shell supersymmetry generators $q^a$ and $q^{\dagger\dot{a}}$ in  \reef{qtq}.

We note that dotting  some arbitrary reference spinors $|\tilde{w}\>$ and $[w|$ (whose brackets with the $p$-spinors are non-vanishing) into \reef{QQtheta3}, we find  $\big\{ 
  \eta \,,\,  \frac{\pa}{\pa\eta}   \big\}
  =1$.
  
Consider the consequences of the above analysis. The superspace coordinates $\theta$ and $\overline{\theta}$ have mass-dimension (mass)$^{-1/2}$ and the angle and square spinors have dimension (mass)$^{1/2}$. So the on-shell superspace variables $\eta$ are dimensionless. Under little group scaling, the $\theta$ and $\overline{\theta}$ are inert, and therefore we have $\eta \to t^{-1} \eta$. Consequently, the on-shell superwavefunction \reef{n4Phi} scales homogeneously as $\Omega \to t^{-2} \Omega$ under a little group scaling and the state-operator map \reef{n4ops2} is exactly compensating this scaling when extracting component wavefunctions from $\Omega$.
}
{\bf Superamplitudes.}
We can think of $\Omega_i=\Omega(p_i)$ as a superwavefunction for the $i$'th external particle of a \emph{superamplitude} $\mathcal{A}_n(\Omega_1,\dots,\Omega_n)$. It depends on the on-shell momentum $p_i$ and a set of Grassmann variables $\eta_{iA}$ for each particle $i=1,\dots, n$. Expanding $\mathcal{A}_n(\Omega_1,\dots,\Omega_n)$ in the Grassmann variables, we note that the $SU(4)$-symmetry requires it to be a sum of polynomials in $\eta_{iA}$ of degree $4(K+2)$; an example of a legal combination is $\eta_{i 1} \eta_{i 2} \eta_{i 3} \eta_{i 4}$ corresponding to particle $i$ being a negative helicity gluon.  One can extract any amplitude from the superamplitude $\mathcal{A}_n$ by projecting out the desired external states using \reef{n4ops2}: for example
\bea
  \label{superprojex}
  A_n(1^+\ldots i^- \ldots j^- \ldots n^+)
  &=&
  \bigg( \prod_{A=1}^4 \frac{\pa}{\pa \eta_{iA}} \bigg)\,
  \bigg( \prod_{B=1}^4 \frac{\pa}{\pa \eta_{jB}} \bigg)\,
  \mathcal{A}_n(\Omega_1,\dots,\Omega_n) \bigg|_{\eta_{kC} =0}
  \\[1mm] \nonumber
  A_n(S^{12} S^{34} 3^- 4^+\ldots n^+)
  &=&
  \bigg( \frac{\pa}{\pa \eta_{11}} \frac{\pa}{\pa \eta_{12}} \bigg)\,
  \bigg( \frac{\pa}{\pa \eta_{23}} \frac{\pa}{\pa \eta_{24}} \bigg)\,
  \bigg( \prod_{A=1}^4 \frac{\pa}{\pa \eta_{3A}} \bigg)\,
  \mathcal{A}_n(\Omega_1,\dots,\Omega_n) \bigg|_{\eta_{kC} =0}.
\eea
This can of course equally well be expressed as Grassmann integrals.

The order $K$ of the Grassmann polynomial precisely corresponds to the N$^K$MHV sector. So we can organize the full tree superamplitude as
\be
  \mathcal{A}_n =
  \ca_n^\text{MHV}
  +  \ca_n^\text{NMHV}
  +  \ca_n^\text{N$^2$MHV}
  + \dots
  +   \ca_n^\text{anti-MHV} \,,
\ee
where $\ca_n^\text{MHV}$ has Grassmann degree 8, $\ca_n^\text{NMHV}$ has Grassmann degree 12 etc.

In the language of on-shell superspace, the supersymmetry Ward identities are identical to the statement that the supersymmetry generators
\be
  \label{bigQTQ}
   Q^A \equiv \sum_{i=1}^n q_i^A
   = \sum_{i=1}^n [i|\,\frac{\pa}{\pa\h_{iA}}\,,
   ~~~~\text{and}~~~~
   \tilde{Q}_A \equiv \sum_{i=1}^n q^\dagger_A
   =  \sum_{i=1}^n  |i\>\,\h_{iA}\,,~~~~~
   A=1,2,3,4,
\ee
annihilate the superamplitude:
\be
\label{SWIQtQ}
Q^A \mathcal{A}_n = 0 
~~~~~\text{and}~~~~~
\tilde{Q}_A \mathcal{A}_n = 0\,.
\ee
 Note that the requirement that $\{ Q^A,\tQ_B \}$ annihilates the superamplitude is equivalent to the statement of momentum conservation. The associated delta function $\delta^4\big(\sum_{i=1}^n p_i\big)$ has been left implicit throughout most of this review, but we will start to include it explicitly in Section \ref{s:DC}, where it plays a central role. 
\exercise{}{It may not be totally obvious to you that \reef{SWIQtQ} encodes the supersymmetry Ward identities, so the point of this  exercise is to illustrate it to you. Start by writing the (relevant terms in the) MHV superamplitude as
\be
\mathcal{A}_n^\text{MHV} = A_n[g^- g^- g^+ \dots g^+] (\eta_1)^4 (\eta_2)^4
  + A_n[g^- \lambda^{123} \lambda^{4} \dots g^+] (\eta_1)^4 (\eta_{21} \eta_{22} \eta_{23}) (\eta_{34}) + \dots\,,
  \label{AnMHVexpanded}
\ee
where $(\eta_i)^4 = \eta_{i1} \eta_{i2} \eta_{i3} \eta_{i4}$. 
(Why does the second term in \reef{AnMHVexpanded} come with a plus?)
The supersymmetry Ward identity \reef{SWIQtQ} says that the coefficient of each independent Grassmann monomial in  $\tilde{Q}_A \mathcal{A}_n^\text{MHV}$ has to vanish. Pick $A=4$ and act with $\tilde{Q}_4$ on $\mathcal{A}_n^\text{MHV}$ to extract all terms whose Grassmann structure is $(\eta_1)^4 (\eta_2)^4 (\eta_{34})$. Use that to show that the `component amplitude'  SUSY Ward identity \reef{SWIN4sym} follows from $\tilde{Q}_4 \mathcal{A}_n^\text{MHV} = 0$.  
}
Note that the action of $\tilde{Q}_A$ on the superamplitude is multiplicative. We can therefore solve it easily using a Grassmann delta function
$\d^{(8)}\big(\tilde{Q}\big)$ defined as 
\be 
 \label{delta8}
   \d^{(8)}\big(\tilde{Q}\big) 
   ~=~
   \frac{1}{2^4} \prod_{A=1}^4 \tilde{Q}_{A\dot{a}}\tilde{Q}^{\dot{a}}_A
   ~=~
   \frac{1}{2^4} \prod_{A=1}^4\sum_{i,j=1}^n \<ij\> \eta_{iA}\eta_{jA}
   \,.
\ee
\exercise{}{Show that momentum conservation ensures that $Q^A$ annihilates $\d^{(8)}\big(\tilde{Q}\big)$.}
Thus half the supersymmetry constraints, namely 
$\tilde{Q}_A \mathcal{A}_n = 0$,  are satisfied if we write the N$^K$MHV superamplitude as 
\be
  \label{susySampl}
  \ca_n^\text{N$^K$MHV} = \d^{(8)}\big(\tilde{Q}\big) \,P_{4K}\,, 
\ee
where $P_{4K}$ is a degree $4K$ polynomial in the Grassmann variables. 
If $P_{4K}$ is annihilated by each  $Q^A$, then --- by the exercise above ---  all the SUSY constraints are solved. The Grassmann delta function $\d^{(8)}\big(\tilde{Q}\big)$ can be viewed as the conservation of 
supermomentum. 

The Grassmann delta function is a degree $8$ polynomial in the $\eta_{iA}$'s. This means that for an MHV superamplitude, $\d^{(8)}\big(\tilde{Q}\big)$ fixes the $\eta_{iA}$-dependence completely and $P_0$ is in that case just a normalization constant that depends on the momenta. It is not hard to see that
\be
  \label{sA-MHV}
  \ca_n^\text{MHV}[1 2 3 \ldots n]
  ~=~
  \frac{\d^{(8)}\big(\tilde{Q}\big)}{\<12\>\<23\>\cdots \<n1\>}
\ee
produces the Parke-Taylor gluon tree amplitudes correctly. Just use the map \reef{n4ops2} to take four derivatives with respect to $\eta_{iA}$ and four with respect to $\eta_{jA}$ as in \reef{superprojex}. Then the delta function  produces the numerator-factor $\<ij\>^4$ of the \emph{`component-amplitude'} $A_n(1^+\ldots i^- \ldots j^- \ldots n^+)$. Note that {one component amplitude and supersymmetry uniquely fix the form of \emph{all} MHV amplitudes in $\mathcal{N}=4$ at each order in perturbation theory.} 
\exercise{}{Reproduce the three supersymmetry Ward identities \reef{n4SWIex} from the MHV superamplitude $\ca_n^\text{MHV}$ in \reef{sA-MHV}.
}
\exercise{}{Use the $\mathcal{N}=4$ SYM superamplitude 
$\ca_n^\text{MHV}$  \reef{sA-MHV} to calculate the 4-scalar amplitude 
$A_4[S^{12} S^{34} S^{12} S^{34}]$. Compare your answer to the 4-scalar amplitude \reef{A4bcfwsqed}. 

Calculate $A_4[S^{12} S^{23} S^{34} S^{41}]$ and compare with \reef{N4SYM-4s-spec}.}

In our discussions so far, we skipped silently over the anti-MHV 3-point amplitudes $\ca_n^\text{anti-MHV}$, whose supersymmetry orbit determines the anti-MHV sector with $K=-1$. These are encoded in degree-4 superamplitudes. We simply state the answer,\footnote{We define $ \d^{(4)}
  \big( [12] \eta_{3} +[23] \eta_{1} + [31] \eta_{2}\big) = \prod_{A=1}^4 
  \big( [12] \eta_{3A} +[23] \eta_{1A} + [31] \eta_{2A}\big)$}
\be
  \label{aMHV3pt}
  \begin{split}
  \ca_3^\text{anti-MHV} 
  &=~ \frac{1}{[12][23][31]}\, 
  \d^{(4)}
  \big( [12] \eta_{3} +[23] \eta_{1} + [31] \eta_{2}\big)\\
  &=~ \frac{1}{[12][23][31]}\, 
  \prod_{A=1}^4
  \big( [12] \eta_{3A} +[23] \eta_{1A} + [31] \eta_{2A}\big)\,,
  \end{split}
\ee
and leave it as an 
\exercise{}{to show that $Q^A \ca_n^\text{anti-MHV} = 0$ and  $\tilde{Q}_A \ca_n^\text{anti-MHV} = 0$.}

Let us now outline 3 approaches to determining the superamplitudes $\ca_n^\text{N$^K$MHV}$ beyond the MHV level.
\begin{enumerate}
\item {\bf \em Solution to the supersymmetry Ward identities.} 
The  N$^K$MHV  superamplitudes 
 in $\cn\!=\!4$ SYM must obey the supersymmetry Ward identities \reef{SWIQtQ}. 
Writing the $L$-loop superamplitude $\ca_{n,L}^\text{N$^K$MHV} = \d^{(8)}\big(\tilde{Q}\big) \,P^{(L)}_{4K}$
supersymmetry requires $Q^A P^{(L)}_{4K} = 0$. In addition, we need $P_{4K}$ to obey the Ward identities of the global $SU(4)$ R-symmetry. As we have seen, these constraints are trivially satisfied at the MHV level where one $L$-loop component-amplitude suffices to fix $P^{(L)}_{0}$ and hence the full $L$-loop MHV $n$-point superamplitude. 

For non-MHV amplitudes, how many component-amplitudes does one need to fix the N$^K$MHV superamplitude? This question is answered by analyzing the requirements $Q^A P_{4K} = 0$ and R-invariance. 
It turns out that for general $K>0$, $P_{4K}$ can be built out of the polynomials $m_{ijk,A} \equiv [ij] \eta_{kA} +[jk] \eta_{iA} + [ki] \eta_{jA}$ that also appear in $\ca_3^\text{anti-MHV}$ given in  \reef{aMHV3pt}.
The solution reveals that the number of component-amplitudes sufficient to determine 
$\ca_{n,L}^\text{N$^K$MHV}$ is the dimension of the irreducible representation of $SU(n\!-\!4)$ corresponding to a rectangular Young diagram with $K$ rows and $(\cn\!\!=\!)4$ columns. The independent component amplitudes are  labeled by the semi-standard tableaux of this Young diagram.\footnote{By manipulating the color-structure, one can improve further on this count of ``basis amplitudes" \cite{Elvang:2009wd,Elvang:2010xn}.} A given particle content (gluons, fermions, scalars) of a basis amplitude corresponds to an ordered partition of the number $4K$ into $n$ integers between 0 and 4. For each such partition, the corresponding Kostka number counts the number of independent arrangements of the $SU(4)$ R-symmetry indices. You can read more about the solution to the supersymmetry and R-symmetry Ward identities in \cite{Elvang:2009wd,Elvang:2010xn}.
\item {\bf \em The super-MHV vertex expansion.} 
The tree-level superamplitudes $\ca_{n}^\text{N$^K$MHV}$ of $\cn\!=\!4$ SYM can be constructed as  an MHV vertex expansion in which the vertices are the MHV superamplitudes  $\ca_n^\text{MHV}$ of \reef{sA-MHV}. The expansion can be derived from an all-line shift, as we discussed in Section \ref{s:csw}, whose validity was proven in \cite{Elvang:2008vz} (see also \cite{Kiermaier:2009yu}). A new feature is that one must sum over the possible intermediate states exchanged on the internal lines of the MHV vertex diagrams. This is conveniently done by integrating over all $\eta_{P_I\,A}$, $A=1,2,3,4$; this automatically carries out a super-sum \cite{Bianchi:2008pu}. More details about this aspect will be offered in the next section.  
\item {\bf \em Super-BCFW.} 
Superamplitudes can be constructed with a supersymmetric version of the BCFW-shift. This construction and its results play a central role in the latest developments, so we will treat it in detail in the following section. 
\end{enumerate}

%%%%%%%%%%%%%
\subsection{Super-BCFW and all tree-level amplitudes in $\mathcal{N}=4$ SYM}
\label{s:N4symtrees}
The BCFW shift introduced in Section \ref{s:bcfw}  preserves the on-shell conditions $p_i^2=0$ and momentum conservation $\sum_{i=1}^n p_i=0$. However, it is clear that the shift does not preserve  supermomentum conservation, $\sum_{i=1}^n |i\> \eta_{iA}=0$. As a consequence, the shifted component amplitudes have large-$z$ falloffs that depend on which types of particles are shifted, for example note the difference between the $[-,+\>$ and $[+,-\>$ BCFW shifts of gluons in \reef{largezfallof}. This can be remedied by a small modification of the BCFW shift \reef{bcfw} that allow us to conserve supermomentum. We simply  accompany the momentum shift by a shift in the Grassmann-variables \cite{nimatalk,Brandhuber:2008pf,ArkaniHamed:2008gz}: for simplicity let us write the $[1,2\>$-`supershift', 
\be
  \label{superbcfw}
  |\hat{1}] = |1] + z\,|2]\,,   ~~~~~~
  |\hat{2}\> = |2\> - z |1\> \,,~~~~~~
  \hat{\eta}_{1A} = \eta_{1A} + z\, \eta_{2A} 
  \,.
\ee
No other spinors or Grassmann variables shift.
\exercise{}{Show that the supermomentum is conserved under the supershift \reef{superbcfw} so that $\d^{(8)}\big(\tilde{Q}\big)$ is invariant.}
It follows directly from \reef{sA-MHV} that the MHV superamplitudes have a $1/z$ falloff under a supershift of any adjacent lines ($1/z^2$ for non-adjacent). In fact, this falloff behavior holds true for any tree superamplitude in $\cn=4$ SYM, as can be shown by using  supersymmetry to rotate the two shifted lines to be positive helicity gluons and using the non-supersymmetric result \reef{largezfallof} for the falloff under a $[+,+\>$-shift  \cite{ArkaniHamed:2008gz,Cheung:2008dn}.

The tree-level recursion relations that result from the super-BCFW shift \reef{superbcfw} involve diagrams with two superamplitude `vertices' connected by an internal line with on-shell  momentum $\hat{P}$. As in the non-supersymmetric case, we must sum over all possible states that can be exchanged on the internal line: in this case, this includes all 16 states of $\cn=4$ SYM. In terms of component amplitudes, the particle exchanged on the internal line depends on the external states: if they are all gluons, then the internal line is also a gluon and one must simply sum over the helicities. The superamplitude version of this helicity sum is
\be
   \bigg[ \bigg( \prod_{A=1}^4 \frac{\pa}{\pa \eta_{\hat{P}A}} \bigg)\,
   \hat{\mathcal{A}}_\text{L}  \bigg] \frac{1}{P^2} \hat{\mathcal{A}}_\text{R}
   +    
   \hat{\mathcal{A}}_\text{L} \frac{1}{P^2} 
   \bigg[ \bigg( \prod_{A=1}^4 \frac{\pa}{\pa\eta_{\hat{P}A}} \bigg)\,\hat{\mathcal{A}}_\text{R} \bigg] 
   \Bigg|_{\eta_{\hat{P}A} = 0} \,,
\ee
where $\eta_{\hat{P}A}$ is the Grassmann variable associated with the internal line. If a gluino can be exchanged\footnote{This happens when there are an odd number of external gluinos on each side of the BCFW diagram.}, then we have to move one of the four Grassmann derivatives from $\hat{\mathcal{A}}_\text{L}$ to $\hat{\mathcal{A}}_\text{R}$ in the first term --- in all four possible ways. And similarly for the second term. A scalar exchange means that two Grassmann derivatives act on $\hat{\mathcal{A}}_\text{L}$ and the two other ones on $\hat{\mathcal{A}}_\text{R}$. All in all, the entire sum over  states exchanged on the internal line can be written
\be
   \bigg( \prod_{A=1}^4 \frac{\pa}{\pa \eta_{\hat{P}A}} \bigg)\,
   \bigg[ 
   \hat{\mathcal{A}}_\text{L}  \frac{1}{P^2} \hat{\mathcal{A}}_\text{R}
   \bigg] 
   \Bigg|_{\eta_{\hat{P}A} = 0}
   ~=~
   \int d^4 \eta_{\hat{P}} 
   ~ \hat{\mathcal{A}}_\text{L}  \frac{1}{P^2} \hat{\mathcal{A}}_\text{R}\,.
   \label{supersum}
\ee
Note how the product rule distributes the Grassmann-derivatives $\pa/\pa\eta_{\hat{P}A}$ on the L and R superamplitudes in all possible ways to automatically carry out the state {\bf \em super-sum}. In \reef{supersum}, we have rewritten the Grassmann-differentiation as a Grassmann integral. Similar super-sums are used in evaluation of unitarity cuts of loop amplitudes where one includes integration over the Grassmann-variable associated with the internal line \cite{Bianchi:2008pu,Bern:2009xq}. 

The super-BCFW recursion relations can actually be solved to give closed-form expressions for \emph{all} tree-level superamplitudes in $\cn=4$ SYM \cite{Drummond:2008cr}. We are now going to show how this works. As a warm-up, we first verify that the MHV superamplitude formula \reef{sA-MHV} satisfies the super-BCFW recursion relations. Then we present the most essential details of the derivation of the tree-level NMHV superamplitude. Finally we comment briefly on the results for N$^K$MHV.

\subsubsection{MHV superamplitude from super-BCFW}
\label{s:MHVsBCFW}
Consider the super-BCFW recursion relations for the MHV superamplitude. Just as in the non-supersymmetric case \reef{bcfwmhv1}, there is just one non-vanishing diagram, but we must now include the super-sum:
\bea
\nonumber
\mathcal{A}^\text{MHV}_n[1 2 3 \ldots n]
   \!\!&=&
          \raisebox{-6.9mm}{\includegraphics[height=1.6cm]{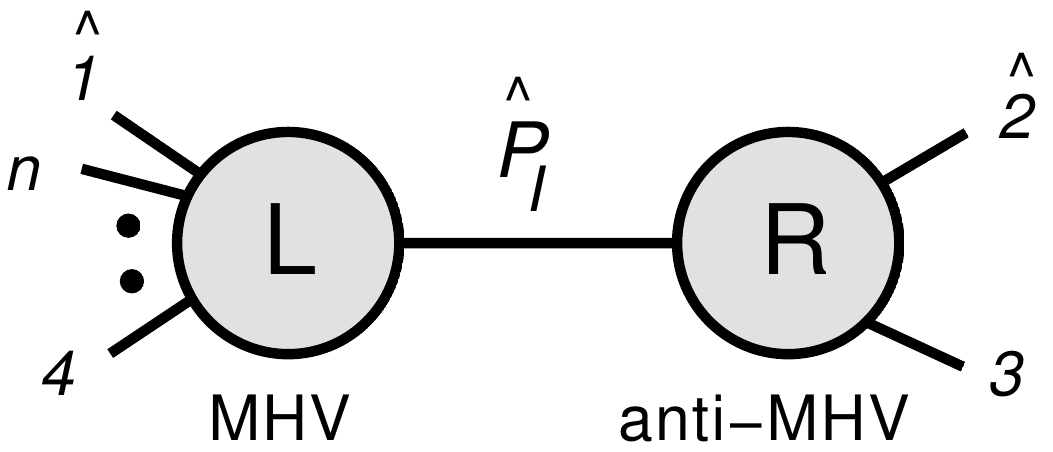}}\\[1mm]\nonumber
          \!\!&=&\!\!\!
  \int d^4 \eta_{\hat{P}}~ 
  \hat{\ca}_{n-1}\big[ \hat{1}, \hat{P} \, , 4, \ldots , n\big] 
  \,\frac{1}{P^2}\,
  \hat{\ca}_{3}\big[ -\hat{P}\, ,  \hat{2}, 3\big] 
  \\[2mm] 
  \!\!&=&\!\!\!
  \int d^4 \eta_{\hat{P}}~ 
  \frac{\d^{(8)}\big(
   \sum_{i\in \text{L}} |\hat{i}\> \hat{\eta}_i
  \big)}{\<1 \hat{P}\>\<\hat{P}\,4\>\<45\>\cdots \<n1\>}  
  \,\frac{1}{P^2}\,
  \frac{\d^{(4)}
  \big( [\hat{P}2] \eta_{3} +[23] \eta_{\hat{P}} + [3\hat{P}] \eta_{2}\big)}
  {[23][3\hat{P}][\hat{P}\,2]}
  \,,~~~~~~~~~~
  \label{sbcfwmhv1}
\eea
where $P = P_{23} = p_2 + p_3$ and we used
 \reef{sA-MHV} and \reef{aMHV3pt} for the $(n\!-\!1)$-point MHV superamplitude and 3-point anti-MHV superamplitudes. We also use the analytic continuation rule \reef{acont} for $|-P] = |P]$.

The new feature is Grassmann integral in \reef{sbcfwmhv1}.
The first delta function is
\be  
  \d^{(8)}\Big(
   \sum_{i\in \text{L}} |\hat{i}\> \hat{\eta}_i
  \Big)
  ~=~
  \d^{(8)}\Big(
  |1\> \hat{\eta}_1 
  +  |\hat{P}\> {\eta}_{\hat{P}} 
  + \sum_{i=4}^{n} |i\>\eta_i
  \Big)\,.
\ee  
On the support of the second delta function, we can set
$\eta_{\hat{P}} 
= - \big( [\hat{P}2] \eta_{3} + [3\hat{P}] \eta_{2}\big)/[23]$ to find
\bea
  \nonumber
 |1\> \hat{\eta}_1 
  +   |\hat{P}\> {\eta}_{\hat{P}} 
  &=&
 |1\> \hat{\eta}_1 
  - \frac{|\hat{P}\> }{[23]}\big( [\hat{P}2] \eta_{3} + [3\hat{P}] \eta_{2}\big)
  \\
  \nonumber
  &=&
 |1\> \hat{\eta}_1 
  +   |3\> \eta_{3}+|\hat{2}\>\eta_{2}\\[1mm]
  &=&
 |1\> {\eta}_1 
  +   |3\> \eta_{3}+ |{2}\>\eta_{2}\,.
\eea
Thus the first delta function simply becomes 
$\d^{(8)}\big(  \sum_{i=1}^{n} |i\>\eta_i \big) = 
\d^{(8)}\big( \tQ \big)$. 
Now the only $\eta_{\hat{P}}$-dependence is in
$\d^{(4)}(\dots)$ and the integral is therefore straightforward to carry out:
\be
  \nonumber
  \int d^4 \eta_{\hat{P}}~ 
  \d^{(8)}\Big(
   \sum_{i\in \text{L}} |\hat{i}\> \hat{\eta}_i \Big)
  ~\d^{(4)}
  \big( [23] \eta_{\hat{P}} +\dots \big)
  ~=~
  [23]^4 ~\d^{(8)}\big( \tQ \big)\,.
\ee
Coming back to \reef{sbcfwmhv1}, we then have 
\be
  \mathcal{A}^\text{MHV}_n[1 2 3 \ldots n]
  ~=~
  \frac{\d^{(8)}\big( \tQ \big)}{\<1 \hat{P}\>\<\hat{P}\,4\>\<45\>\cdots \<n1\>}  
  \,\frac{1}{P^2}\,
  \frac{[23]^4}
  {[3\hat{P}][\hat{P}\,2][23]}\,.
\ee
Compare this with \reef{recrelex1b} and you will see that the two expressions are the same, except that 
$\d^{(8)}\big( \tQ \big)$ has replaced $\<1 \hat{P}\>^4$ and $[23]^4$ has replaced $[3\hat{P}]^4$. Using the identities \reef{id1} and \reef{id2}, we  promptly recover the desired result
\be
  \label{sA-MHV2}
  \ca_n^\text{MHV}[1 2 3 \ldots n]
  ~=~
  \frac{\d^{(8)}\big(\tilde{Q}\big)}{\<12\>\<23\>\cdots \<n1\>} \,.
\ee
Thus we have shown that the $\cn=4$ SYM  tree-level MHV superamplitude \reef{sA-MHV2}  satisfies the super-BCFW recursion relations.
Next, we will derive an important result for NMHV superamplitudes in $\cn=4$ SYM.

%%%%%
\subsubsection{NMHV superamplitude and beyond}
\label{s:NMHV}
The super-BCFW recursion relation for the NMHV superamplitude  involves two types of diagrams
\be
 \ca^\text{NMHV}_n[12\ldots n] 
 ~~=~~
 \sum_{k=5}^n~
 \raisebox{-8.9mm}{\includegraphics[height=1.75cm]{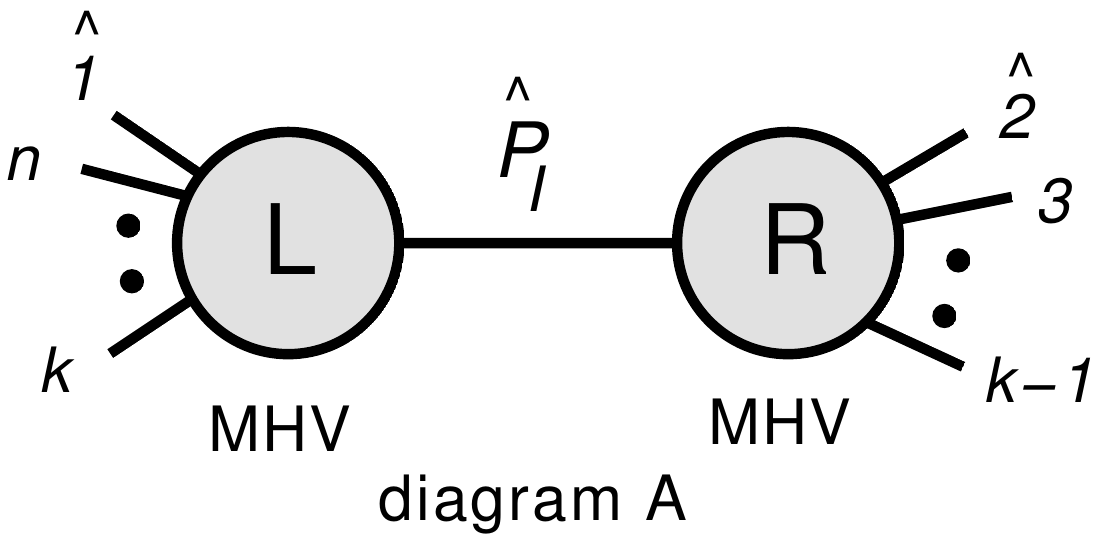}}
 ~~+~~
 \raisebox{-8.9mm}{\includegraphics[height=1.75cm]{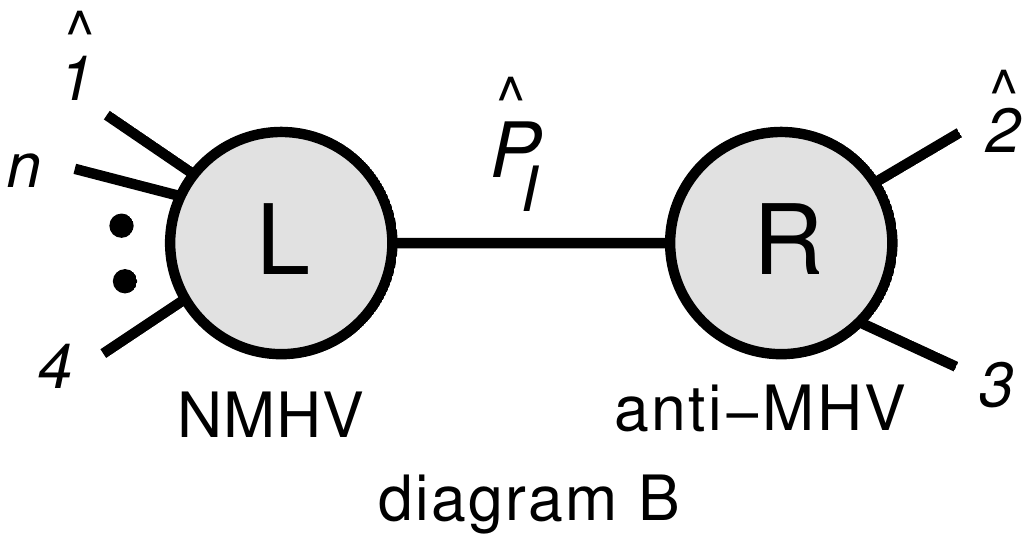}}\,.
 \label{NMHVbcfwAandB}
\ee
The diagrams of type A involve two MHV vertices. 
Diagram B is  present only for $n\ge 6$ because the L subamplitude is NMHV and therefore needs at least 5 legs; indeed, for $n=5$ we only had an MHV$\times$MHV diagram in the calculation of $A_5[1^-2^-3^-4^+5^+]$ in \reef{A5bcfw2}.

Diagram B provides the setting for an inductive proof of the NMHV superamplitude formula we are seeking, while the diagrams of type A give an `inhomogeneous' contribution. We begin with a detailed evaluation of the type A diagrams. There will be a lot of detailed calculations in this example calculation, so if you just want the result, you are free to skip ahead to the answer in \reef{NMHVdiagA4}. Right after the example, we summarize the full result for the NMHV superamplitude.
\example{Calculation of diagram A. The first step is simply to plug in the MHV superamplitudes:
{\small
\be
\label{NMHVdiagA1}
\text{Diagram A}
  ~=~
  \int d^4 \eta_{\hat{P}}~
  \frac{\d^{(8)}\big(\text{L} \big)}
     {\<1 \hat{P}\>\<\hat{P}\,k\> \<k, k+1\>\dots \<n1\>}
  ~\frac{1}{P^2}~
  \frac{\d^{(8)}\big(\text{R} \big)}
     {\<\hat{P}\,\hat{2}\> \<\hat{2}3 \> \<34\> \dots \<k-1, \hat{P}\>}  \,.
\ee
}
Here $P = p_2 + p_3 + \dots + p_{k-1}$ and (with the help of the rule \reef{acont}) we have
\be
  \begin{split}
  \d^{(8)}\big(\text{L} \big)
  ~=&~
  \d^{(8)}\Big( 
    - |\hat{P}\> \eta_{\hat{P}} 
    + |1\> \hat{\eta}_{1}
    + \sum_{r=k}^n |r\>\eta_r
    \Big),\\
  \d^{(8)}\big(\text{R} \big)
  ~=&~
  \d^{(8)}\Big( 
     |\hat{P}\> \eta_{\hat{P}} 
    + |\hat{2}\> {\eta}_{2}
    + \sum_{r=3}^{k-1} |r\>\eta_r
    \Big)\,.
  \end{split}
\ee
On the support of $\d^{(8)}\big(\text{R} \big)$, we can write $\d^{(8)}\big(\text{L} \big) = \d^{(8)}\big(\text{L} + \text{R}\big)
=  \d^{(8)}\big( \sum_{i=1}^n |i\> \eta_i\big)  
=  \d^{(8)}\big( \tQ \big)$, which is independent of $\eta_{\hat{P}}$ and expresses the conservation of supermomentum for the $n$ external states.  Now only $\d^{(8)}\big(\text{R} \big)$ remains under the state-sum integral $\int d^4 \eta_{\hat{P}}$; since $\d^{(8)}$ enforces two conditions we can project out two separate $\delta^{(4)}$'s:
\be
   \label{deltaR}
   \d^{(8)}\big(\text{R} \big)
   ~=~
   \frac{1}{\<1 \hat{P}\>^4}\,
   \d^{(4)}
   \Big( 
   \<1 \hat{P}\> \eta_{\hat{P}} 
    + \<1 \hat{2}\> {\eta}_{2}
    + \sum_{r=3}^{k-1} \<1 r\>\eta_r
   \Big)
   ~
   \d^{(4)}
   \Big( 
    \<\hat{P}\,\hat{2}\> {\eta}_{2}
    + \sum_{r=3}^{k-1} \<\hat{P} \,r\>\eta_r
   \Big)\,.
\ee
The $\eta_{\hat{P}}$-dependence is contained in just one of these two $\d^{(4)}$'s, so it is easy to perform the Grassmann-integral: it produces a factor $\<1 \hat{P}\>^4$ which cancels the normalization factor included in \reef{deltaR}. All in all, we have shown that
\be
     \label{deltaLRres}
   \int d^4 \eta_{\hat{P}}~
    \d^{(8)}\big(\text{L} \big) 
   ~\d^{(8)}\big(\text{R} \big)
   ~=~
    \d^{(8)}\big( \tQ \big)  
    ~
   \d^{(4)}
   \Big( 
    \<\hat{P}\,\hat{2}\> {\eta}_{2}
    + \sum_{r=3}^{k-1} \<\hat{P} \,r\>\eta_r
   \Big)\,.
\ee
The factor $\d^{(8)}\big( \tQ \big)$ can be used to pull out an overall factor $\ca_n^\text{MHV}$ from diagram A in \reef{NMHVdiagA1} and we can then write
\be
\text{Diagram A}
  ~=~
  \ca_n^\text{MHV} 
    ~
    \frac{\<12\>\<23\>\<k-1,k\>\,     \d^{(4)}
   \Big( 
    \<\hat{P}\,\hat{2}\> {\eta}_{2}
    + \sum_{r=3}^{k-1} \<\hat{P} \,r\>\eta_r
   \Big)}
   {\<k \hat{P}\> \<k-1, \hat{P}\>\<\hat{2}\hat{P}\>\<\hat{2}3 \>\<1 \hat{P}\>\,P^2}\,.
   \label{NMHVdiagA2}
\ee

Next, we turn our attention to the brackets in \reef{NMHVdiagA2} that involve the shifted spinors $|\hat{P}\>$ and $\<\hat{2}|$. As in the non-supersymmetric examples of Section \ref{s:bcfw}, we will manipulate these brackets by multiplying the numerator and denominator of \reef{NMHVdiagA1} by 
$[\hat{P}\, 2]^4\<21\>^4$. From $\hat{P}= - |\hat{P}\>[\hat{P}|$ we are going encounter sums of momenta that appear in the ordering fixed by color-structure, so for convenience we introduce the shorthand notation
\be
  \label{yijdef}
  y_{ij} \equiv p_i + p_{i+1} + \dots + p_{j-1} \,.
\ee
Further, we declare that $y_{ji}$ with $j>i$ equals $-y_{ij}$; when we think of the $p_i$ as cyclically ordered this simply expresses momentum conservation. We will use the variables $y_{ij}$ to write the result for diagram A in a form that may look slightly mysterious for now, but it has some very important features that are discussed in Section \ref{s:DC}.

One type of bracket is $\< r \hat{P}\>$ with $r \ne 1, \hat{2}$. We manipulate this as follows:
\bea
  \nonumber
  &&\< r \hat{P}\> [\hat{P}\, 2]\<21\>
  = - \< r | \big( \hat{2} + 3 + \dots + (k-1) \big) | 2 ] \<2 1\>
  = - \< r |y_{3k} | 2 ] \<2 1\>\\
  &&=  \< r |y_{3k}.y_{23}| 1\>
  =  \< r |y_{3k}.y_{13}| 1\>
  = -\< 1 |y_{13}.y_{3k}| r\>\,.
\eea
This result  applies to the brackets $\< r \hat{P}\>$, $\< k \hat{P}\>$ and $\< k-1, \hat{P}\>$ in \reef{NMHVdiagA2}.
\exercise{ex:Ksq}{To keep you actively engaged,  involved, and awake, here is an exercise: show that 
\be
  \label{Ksqij}
  \<i | K.K |j\> = - K^2 \<ij\>\,
\ee
for any momentum $K$ (lightlike or not) and any spinors $\<i|$ and $|j\>$.
}
The two brackets $\< 1 \hat{P}\>$ and $\< \hat{2} \hat{P}\>$ from the denominator of \reef{NMHVdiagA1} are dealt with as follows:
\be
   \< 1 \hat{P}\>[\hat{P}\, 2]
   = - \<1|y_{3k}|2]
   ~~~~~~\text{and}~~~~~~
   \<  \hat{2} \hat{P}\>[\hat{P}\, 2]
   = -2 \hat{p}_2 \cdot \hat{P}
   = - \hat{P}^2 + y_{3k}^2
   = y_{3k}^2\,.
   \label{2more}
\ee
In the last equality, we used that the BCFW diagram is evaluated on the value of $z$ such that $\hat{P}^2 = 0$. In fact, it is useful to note what this value of $z$ is:
\be
   0 = \hat{P}^2 =  \< \hat{2} | \hat{P} |2]  + y_{3k}^2
    = y_{2k}^2 - z \<1|y_{3k}|2]
    ~~~~\implies ~~~~
    z = \frac{y_{2k}^2}{\<1|y_{3k}|2]}\,.
\ee
We use this $z$ to evaluate $\<\hat{2}3\>$:
\be
  \begin{split}
   \<\hat{2}3\> 
   \!&=~
    \<23\> -  \frac{y_{2k}^2}{\<1|y_{3k}|2]} \<13\>
   =\frac{\<1|y_{3k}|2] \<23\> - y_{2k}^2\<13\>}{\<1|y_{3k}|2]}
   =\frac{\<1|y_{1k}|2] \<23\> +\<1|y_{2k}.y_{2k}|3\>}{\<1|y_{3k}|2]}
   \\  
   \!&=~
   \frac{- \<1|y_{1k}.p_2|3\> +\<1|y_{1k}.y_{2k}|3\>}{\<1|y_{3k}|2]}
   =
   \frac{\<1|y_{1k}.(-p_2+y_{2k})|3\>}{\<1|y_{3k}|2]}
   =
   \frac{\<1|y_{1k}.y_{3k}|3\>}{\<1|y_{3k}|2]}\,.
   \end{split}
   \label{long23}
\ee
In the third equality we used the result in Exercise \ref{ex:Ksq}.
Combining the result \reef{2more} and \reef{long23}, we can write
\be
    \<\hat{2}3\>  \< 1 \hat{P}\>[\hat{P}\, 2] =
   {\<1|y_{1k}.y_{k3}|3\>} \,.
\ee
We have now evaluated all four angle brackets involving $|\hat{P}\>$ and $\<\hat{2}|$ in \reef{NMHVdiagA2}. There is however, one more manipulation that we would like to do, namely one involving the propagator $1/P^2 = 1/y_{2k}^2$. It goes like this:
\be
  P^2 \<12\> = y_{2k}^2 \<12\> = - \<1| y_{2k}.y_{2k} |2\>
  = -\<1| y_{1k}.y_{3k} |2\>
  = \<1| y_{1k}.y_{k3} |2\> \,.
\ee
It is time to put everything together. We can now write diagram A from \reef{NMHVdiagA2} as
\be
\text{Diagram A}
  ~=~
 \ca_n^\text{MHV}    
    ~
    \frac{\<23\>\<k-1,k\>\,   
    [\hat{P}\, 2]^4\<21\>^4~
    \d^{(4)}
   \Big( 
    \<\hat{P}\,\hat{2}\> {\eta}_{2}
    + \sum_{r=3}^{k-1} \<\hat{P} \,r\>\eta_r
   \Big)}
   {y_{3k}^2\,\< 1 |y_{13}.y_{3k}| k\> \< 1 |y_{13}.y_{3k}| k-1\>
    \<1|y_{1k}.y_{k3}|3\> \<1| y_{1k}.y_{k3} |2\>}\,.
   \label{NMHVdiagA3}
\ee
Let us examine the $\d^{(4)}$ in the numerator. We  absorb the factors 
$[\hat{P}\, 2]^4\<21\>^4$ into the delta function whose argument then becomes (suppressing $SU(4)$-indices)
\be
  \Xi 
  ~\equiv~ 
  \<12\>[2\hat{P}]\<\hat{P}\,\hat{2}\> {\eta}_{2}
    + \sum_{r=3}^{k-1} \<12\>[2\hat{P}]\<\hat{P} \,r\>\eta_r
    = \<12\> y_{3k}^2 \,\eta_2
      +\sum_{r=3}^{k-1} \<1| y_{13}. y_{3k}|r\>\,\eta_r \,.
      \label{Xi-pre1}
\ee
We are going to do a little work on this expression in order to introduce another piece of short-hand notation, namely fermionic companions of the $y_{ij}$'s defined in \reef{yijdef}:  
\be
  |\theta_{ij,A}\> 
  ~\equiv~ 
  \sum_{r=i}^{j-1} |r\> \, \eta_{rA}\,.
  \label{thetadef}
\ee
Then $|\theta_{ji}\> = -|\theta_{ij}\>$ encodes supermomentum conservation.

We start by rewriting each of the terms in \reef{Xi-pre1}:
\be
  \begin{split}
  \<12\> \,y_{3k}^2 \,\eta_2
  &=- \<1|y_{3k}.y_{3k}|2\> \,\eta_2
  =- \<1|y_{1k}.y_{3k}|2\> \,\eta_2
  + \<1|y_{23}.y_{3k}|2\> \,\eta_2\\
  &=
  - \<1|y_{1k}.y_{3k}|\theta_{13}\>
  + \<1|y_{1k}.y_{3k}|1\> \,\eta_1
  + \<1|y_{13}.y_{3k}|2\> \,\eta_2
   \label{d4A}
\end{split}
\ee
and
\be
  \sum_{r=3}^{k-1} \<1| y_{13}. y_{3k}|r\>\,\eta_r
  = 
  \<1| y_{13}. y_{3k}|\theta_{1k}\> 
   - \<1| y_{13}. y_{3k}|1\>\,\eta_1
   - \<1| y_{13}. y_{3k}|2\>\,\eta_2 \,.
   \label{d4B}
\ee
Adding \reef{d4A} and \reef{d4B} to get $\Xi$, the two extra $\eta_2$-terms cancel and the  $\eta_1$-terms combine to
\be
   \<1|y_{1k}.y_{3k}|1\> \,\eta_1
 - \<1| y_{13}. y_{3k}|1\>\,\eta_1
  = \<1|(y_{1k} - y_{13}).y_{3k}|1\> \,\eta_1
  = \<1|y_{3k}.y_{3k}|1\> \,\eta_1  
  = 0 
\ee
by \reef{Ksqij}. Thus we have
\be
  \Xi =   
    - \<1|y_{1k}.y_{k3}|\theta_{31}\> 
  - \<1| y_{13}. y_{3k}|\theta_{k1}\> \,.
\ee
Our work brings us to the following form of a diagram of type  A:
\be
\text{Diagram A}
  =
    \ca_n^\text{MHV}
    ~
    \frac{\<23\>\<k-1,k\>\,   
   ~
    \d^{(4)}
   \Big( 
    \<1|y_{1k}.y_{k3}|\theta_{31}\> 
    + \<1| y_{13}. y_{3k}|\theta_{k1}\>
   \Big)}
   {y_{3k}^2\,\< 1 |y_{13}.y_{3k}| k\> \< 1 |y_{13}.y_{3k}| k-1\>
    \<1|y_{1k}.y_{k3}|3\> \<1| y_{1k}.y_{k3} |2\>}\,.
   \label{NMHVdiagA4}
\ee
This completes our calculation of diagram A in the super-BCFW recursion relation \reef{NMHVbcfwAandB}. Next, we discuss what the full NMHV superamplitude looks like. 
}
The result \reef{NMHVdiagA4} for diagram A is often written 
$\ca^\text{MHV}_n \, R_{13k}$, where the so-called $R$-invariants\footnote{We discuss in Section \ref{s:DC} under which symmetries $R_{ijk}$ is invariant.} are defined as
\be
  R_{1jk}
  =
  \frac{\<j-1,j\>\<k-1,k\>\,   
   ~
    \d^{(4)}
   \big( 
    \Xi_{1jk}
   \big)}
   {y_{jk}^2\,\< 1 |y_{1j}.y_{jk}| k\> \< 1 |y_{1j}.y_{jk}| k-1\>
    \<1|y_{1k}.y_{kj}|j\> \<1| y_{1k}.y_{kj} |j-1\>}\,,
    \label{Rinv1}
\ee
where
\be
\Xi_{1jk,A}=
 \<1|y_{1k}.y_{kj}|\theta_{j1,A}\> 
    + \<1| y_{1j}. y_{jk}|\theta_{k1,A}\>\,.
\ee
Note the structure: neighbor indices match up with each other. This is an important feature. For the sake of completeness, let us repeat here the definitions \reef{yijdef} and \reef{thetadef} of the variables $y_{ij}=-y_{ji}$ and 
$\theta_{ij,A} = - \theta_{ji,A}$:
\be
  \label{ythetadef}
  y_{ij} \equiv p_i + p_{i+1} + \dots + p_{j-1} 
  ~~~~~~
  \text{and}
  ~~~~~~
  |\theta_{ij,A}\> 
  ~\equiv~ 
  \sum_{r=i}^{j-1} |r\> \, \eta_{rA} \,.
\ee
For $n=5$, diagram B vanishes and diagram A with $k=5$ is the complete result:
\be
  \ca^\text{NMHV}_5 =  \ca^\text{MHV}_5\, R_{135}\,.
\ee
For $n>5$, the diagrams of type A contribute 
$\ca^\text{MHV}_n\, \sum_{k=5}^n R_{13k}$. 
Diagram B recurses this form and the result is very simple \cite{Drummond:2008cr}:
\be
  \text{Diagram B} 
  ~=~
  \ca^\text{MHV}_n \sum_{j=4}^{n-2} \sum_{k=j+2}^n R_{1jk}\,.
\ee
This means that the entire NMHV superamplitude can be expressed in terms of the $R$-invariants as
\be
 \label{superNMHV}
\ca^\text{NMHV}_n =\ca^\text{MHV}_n
~\sum_{j=3}^{n-2} \sum_{k=j+2}^n R_{1jk} \,.
\ee
For example,
\be
  \ca^\text{NMHV}_6 =\ca^\text{MHV}_6 
  \big(
    R_{135}+ R_{136} + R_{146}  
  \big)\,.
\ee
This remarkably simple result \reef{superNMHV} for the all-$n$ tree-level NMHV superamplitudes in $\cn=4$ SYM was first found in studies of loop amplitudes \cite{Drummond:2008vq}. Later the NMHV formula was constructed as we did here using super-BCFW \cite{Drummond:2008cr}. Formulas for tree-level N$^K$MHV superamplitudes was also derived --- they take the form of sums of generalized $R$-invariants \cite{Drummond:2008cr}. We refer you to the original paper \cite{Drummond:2008cr}, the discussions in \cite{Brandhuber:2008pf,ArkaniHamed:2008gz,Drummond:2009ge}, and the review \cite{Drummond:2010ep} for further details.

One final comment about the $R$-invariants \reef{Rinv1} is that we defined them here `anchored' at momentum line 1; this came about because we used a $[1,2\>$-supershift. However, by cyclic symmetry all pairs of adjacent lines in the superamplitude are on equal footing, so an $[i,i+1\>$-supershift would have resulted in an NMHV formula anchored at $i$, giving $R$-invariants $R_{ijk}$. These are defined by replacing all momentum labels $1$ in \reef{Rinv1} by $i$. Often in the literature you will find the expressions for the NMHV superamplitude given with $n$ as the anchor; 
i.e.~$\ca^\text{NMHV}_n =\ca^\text{MHV}_n
~\sum_{j=2}^{n-3} \sum_{k=j+2}^{n-1} R_{njk}$. 
\exercise{}{Write down the NMHV superamplitude formula that results from a $[2,3\>$-supershift. Then project out the gluon amplitude
$A_6[1^+ 2^- 3^+ 4^- 5^+ 6^-]$. Can you match your result to the 3-term expression in Exercise \ref{ex:Mi}? Can you guess what expression you get from a $[3,2\>$-supershift after projecting out $A_6[1^+ 2^- 3^+ 4^- 5^+ 6^-]$? You'll find the answer in Section \ref{s:polytopes}. 
}
The superamplitudes of $\cn=4$  SYM are invariant under a large symmetry group, which is enhanced in the planar limit. In particular, both the  obvious and hidden symmetries are realized for the tree-level superamplitudes and the form of the $R$-invariants is essential for this. 
We will discuss these symmetries next in Section \ref{s:DC}.

%%%%%%%%%%%%%%%%%%%%%%%%%%%%%%% 
%%%%%%%%%%%%%%%%%%%%%%%%%%%%%%% 
%%%%%%%%%%%%%%%%%%%%%%%%%%%%%%% 
\newpage 
\setcounter{equation}{0}
\section{Symmetries of $\cn=4$ SYM}
\label{s:DC}
%%%%%%%%%%%%%%%%%%%%%%%%%%%%%%% 
%%%%%%%%%%%%%%%%%%%%%%%%%%%%%%% 
%%%%%%%%%%%%%%%%%%%%%%%%%%%%%%% 
We learned in the previous section that all tree-level superamplitudes in 
$\cn=4$ SYM can be constructed by solving the super-BCFW recursion relations. This section is dedicated to a detailed description of the symmetries of superamplitudes in $\cn=4$ SYM.

%First we introduce \emph{dual space} and discover a `hidden' symmetry called \emph{dual superconformal symmetry}. Then we introduce new variables, the \emph{momentum (super)twistors}, in order to find a manifestly dual superconformal invariant expression for the $\cn=4$ SYM tree-level superamplitudes.

%%%%%%%%%%%%%%%%%%%%%%%%%%%%%%%%%%%%
\subsection{Superconformal symmetry of $\mathcal{N}=4$ SYM}
\label{s:confsym}
%%%%%%%%%%%%%%%%%%%%%%%%%%%%%%%%%%%%
All the theories we study here are Poincar\'e  invariant. The ten symmetry generators $P^\m$ and $M_{\m\n}$ can be written in spinor-index notation by contracting the Lorentz indices with $(\bar\sigma^\m)^{\da b}$ and   
$({\s}^{\m}\bar{\s}^{\n}-{\s}^{\n}\bar{\s}^{\m})_{a}{}^{b}$ and its conjugate. 
The action of the Poincar\'e  generators on the scattering amplitudes can then be realized by the operators
\be
   P^{\dot{a}b}= - \sum_{i} 
      |i\>^{\da}[i|^b \,,
   ~~~~~~
    M_{ab}
    =\sum_{i}|i]_{(a}\,
     \partial_{[i|^{\scriptstyle b)}}\,,
   ~~~~~~
   M_{\dot{a}\dot{b}}
   =\sum_{i} \<i|_{(\dot{a}}\,
   \partial_{|i\>^{\scriptstyle \dot{b})}}\,,
   \label{PMM}
\ee
where the sum is over external particle labels $i=1,2,\dots,n$ and $(\dots)$ indicate symmetrization of the enclosed indices. 
It is important to note that the operators satisfy the usual Poincar\'e  commutator algebra without imposing momentum conservation on the $n$ momenta. 
\exercise{MMonA}{
Show that 
\be
  \sum_{i} \<i|_{\dot{a}}\,
   \partial_{|i\>^{\scriptstyle \dot{b}}} \<jk\> 
   = 
   \eps_{\dot{a}\db}
   \<jk\> \,.
   \label{Monang}
\ee
and hence $M_{\dot{a}\dot{b}}$ annihilates angle brackets.  $M_{ab}$ trivially gives zero on any angle brackets. The equivalent conclusions hold for square brackets. \\{}
[Hint: use $A^{ab}-A^{ba}=-A^c{}_c \,\epsilon^{ab}$ which is valid for any $2\times2$ matrix.]
}
When discussing the symmetry properties, we are going to include the momentum conservation delta function $\delta^{4}\big(\sum_i p_i\big)$ explicitly in the amplitudes.
Then $P^{\da a}$ annihilates the amplitude in the distributional sense \mbox{$P^{\da a}\, \delta^4(P) = 0$}. The action of the rotations/boosts follow from the following useful identity.
\example{Calculate
\be 
  \sum_{i} |i\>^{\dot{a}}\,
   \partial_{|i\>^{\scriptstyle \dot{b}}} ~
   \delta^{4}\big(P\big)
   \,=\,
   \sum_{i} |i\>^{\dot{a}}\,
   \frac{\partial P^{\dot{c}d}}{\partial {|i\>^{\scriptstyle \dot{b}}}} ~
   \frac{\pa}{\partial P^{\dot{c}d}}~
   \delta^{4}\big(P\big)
   \,=\,
   P^{\dot{a}d} \,
   \frac{\pa}{\partial P^{\dot{b}d}}~
   \delta^{4}\big(P\big)
   \,=\,
   -2 \d^{\dot{a}}{}_{\dot{b}} \,
   \delta^{4}\big(P\big) \,,
   \label{diffdelP}
\ee
where the last equality holds as a distribution since
$\int x \,f(x)\, \partial_x \,\delta(x) = -\int f(x)\,\delta(x)$   
and
$\frac{\pa P^{\dot{a}d}}{\partial P^{\dot{c}d}} = 2 \d^{\dot{a}}{}_{\dot{b}}$.
It follows that 
$M_{\dot{a}\dot{b}}\, \delta^{4}\big(P\big) = 0$.
}
 We conclude from the above that the Poincar\'e  generators annihilate the amplitudes.

For a supersymmetric theory, the Poincar\'e  generators \reef{PMM} are supplemented by the supersymmetry generators $Q$ and $\tQ$; for $\cn=4$ SYM, the supersymmetry generators were given in  \reef{bigQTQ} in on-shell superspace.  We have already discussed that the annihilation of the superamplitudes by the supersymmetry generators encode the supersymmetry Ward identities. In particular, the supermomentum conserving Grassmann-delta function 
$\delta^{(8)}\big(\sum_i |i\> \eta_i\big)$ implies that $\tQ_A$ annihilates the superamplitude.

The spacetime symmetry of $\mathcal{N}=4$ SYM is enlarged to the superconformal group. This means that in addition to the super-Poincar\'e  generators,  
\begin{itemize}
\item[\footnotesize $\Diamond$] 4 translations and 6 boosts \& rotations in \reef{PMM},
\item[\footnotesize $\Diamond$] 16 fermionic supersymmetry generators $Q^{Aa}$ and $\tQ_{A}^{\da}$ in \reef{bigQTQ},
\end{itemize}
the superconformal algebra also has 
\begin{itemize}
\item[\footnotesize $\Diamond$] 4 conformal boosts $K_{a\dot{a}}$, 
\item[\footnotesize $\Diamond$] 1 dilatation $D$, 
\item[\footnotesize $\Diamond$] 15 $SU(4)$ R-symmetry generators $R^{A}\,_B$, satisfying the traceless condition $R^C{}_C=0$,
\item[\footnotesize $\Diamond$] 16 fermionic conformal supersymmetry generators  
$\tilde{S}^A_{\dot{a}}$ and $S_{aA}$\,. 
\end{itemize}
Together, these 16+16=32 fermonic and 4+6+4+1+15=30 bosonic generators form the graded Lie algebra $su(2,2|4)$ of the superconformal group.\footnote{$U(2,2|4)$ has 32 fermonic and 32 bosonic generators, i.e.~it has two more $U(1)$'s than $SU(2,2|4)$.} 
Introducing a collective index $\mathsf{A}=(a\,,\dot{a},\,A)$
we can write the superconformal generators
 $G^{\mathsf{A}}{}_{\mathsf{B}}\in su(2,2|4)$. 

We are going to realize the generators  $G^{\mathsf{A}}{}_{\mathsf{B}}\in su(2,2|4)$ in the following form, organized here according to their mass dimensions:
\eq
\begin{array}{ccc}\; 
& P^{\dot{a}b}= - \sum_{i} 
      |i\>^{\da}[i|^b \,, 
&  \\[1mm] 
\displaystyle 
\tilde{Q}^{\dot{a}}_A= \sum_{i}|i\>^{\dot{a}}\,\eta_{iA} & \; & 
Q^{aA}=\sum_{i}[i|^a_{i}\,\partial_{\eta_{iA}}\\[1mm] 
\displaystyle 
   ~~~~M_{\dot{a}\dot{b}}
   =\sum_{i} \<i|_{(\dot{a}}\,
   \partial_{|i\>^{\scriptstyle \dot{b})}}
& 
\raisebox{-4pt}{$
\begin{array}{c} \quad\; 
\displaystyle 
D=\sum_i \Big(\tfrac{1}{2}
   |i\>^{\dot{a}}\,\partial_{|i\>^{\scriptstyle \dot{a}}}
+\tfrac{1}{2}|i]_{a}\, \partial_{|i]_{\scriptstyle a}}
+1\Big) \\[3mm] 
\displaystyle 
R_A\,^B=\sum_i\Big(
  \eta_{iA}\,\partial_{\eta_{iB}}
    -\tfrac{1}{4}\delta_A{}^B\eta_{iC}\,\partial_{\eta_{iC}}\Big)
    \end{array} 
$}    &
\displaystyle 
~~M_{ab}
    =\sum_{i}|i]_{(a}\,
     \partial_{[i|^{\scriptstyle b)}}
      \\[1mm]
\displaystyle 
~~~\tilde{S}^{A}_{\dot{a}}=\sum_i \partial_{|i\>^{\scriptstyle \dot{a}}}\,\partial_{\eta_{iA}}& \; & 
\displaystyle 
~S_{aA}=\sum_i \partial_{[i|^{\scriptstyle a}} \, \eta_{iA} \\[1mm]
 \;  & 
\displaystyle 
K_{a\dot{a}}=
-\partial\raisebox{2pt}{{}$_{\scriptstyle |i\>^{\scriptstyle \dot{a}}}$}\,
\partial_{[i|^{\scriptstyle a}}
\,.
& \;
\end{array}
\label{SCGen}
\eqe
These generators are given as a sum of an operator $G_i^{\mathsf{A}}\,_{\mathsf{B}}$ that is defined on \emph{one} leg, i.e.~$G^{\mathsf{A}}{}_{\mathsf{B}}  = \sum^n_{i=1} G_i^{\mathsf{A}}\,_{\mathsf{B}}$; this reflects the local nature of the symmetry. In Section \ref{s:DC}, we will encounter  symmetries whose  generators are ``non-local" in that they involve products of operators that act on different legs. 
\exercise{ex:dilatation}{Show that the action of first two terms in $D$  extracts the mass dimension from an expression constructed from angle and square brackets. 
The show that $D$ annihilates amplitudes when including $\delta^{4}(P)$.
}

For on-shell kinematics we define the helicity generator $H$ as
\be
H=\sum_i 
\left[ 
 \, \,|i\>_{\dot{a}}\,\partial_{|i\>_{\scriptstyle \dot{a}}}
  -|i]_{a}\, \partial_{|i]_{\scriptstyle a}}
  -\eta_{iA}\,\partial_{\eta_{iA}}+2\,\,
  \right] .
\ee
Indeed one finds $H\,|i\>_{\dot{a}}=\,|i\>_{\dot{a}}$, $H \, |i]_{a}=-|i]_{a}$ and $H\,\eta_A=-\eta_A$, i.e.~it gives the correct little group weight of each on-shell variable. Thus when $H$ acts on a component amplitude it extracts the sum of the helicity weights plus $2n$: in the N$^K$MHV sector this is $\sum_i (-2h_i) + 2n = 4K + 8$. But such a component amplitude appears in the superamplitude multiplied by a factor of $4(K+2)$ Grassmann-variables $\eta_i$'s. Therefore $H \, \ca^\text{N${}^K$MHV}_n = 0$.

Together with the helicity generator, the generators in (\ref{SCGen}) forms a closed algebra.  For example, we encourage you to check that:
\eq
 \label{SQacom}
\{S_{aA}, Q^{bB}\}=
\frac{1}{2}\delta_A{}^B M_a{}^b
+\delta_a{}^b R_A{}^B
+\frac{1}{2}\delta_A{}^B\delta_a{}^b\big(D-\frac{1}{2}H\big)\,.
\eqe
Since $H$ vanishes on the amplitude, the generators in (\ref{SCGen}) close into the superconformal group when acting on the on-shell amplitude.

We are now ready to study the action of the superconformal symmetry on the amplitude of $\mathcal{N}=4$ SYM: 
the superamplitudes should be invariant under the full superconformal symmetry group, so one should find
$G^{\mathsf{A}}{}_{\mathsf{B}}\,\mathcal{A}_n=0\,.$
We restrict our analysis to the MHV superamplitude
\be
\label{MHVPQ}
\mathcal{A}_n^{\rm MHV}=\frac{\delta^4(P)\delta^{(8)}(\tilde{Q})}{\prod_{i=1}^n\langle i,i+1\rangle}\,.
\ee

We have already discussed that this superamplitude is super-Poincar\'e invariant. To prove that it enjoys the full $SU(2,2|4)$ symmetry, the superconformal algebra ensures that it is sufficient to check that the amplitude vanishes under the conformal supersymmetries $S_{aA}$ and $\tilde{S}^{A}_{\dot{a}}$. For example, the anticommutator of $S_{aA}$ and $\tilde{S}^{A}_{\dot{a}}$ gives the conformal boost $K_{a\dot{a}}$. The anticommutator \reef{SQacom} is a another example. The following two examples show that $\mathcal{A}_n^{\rm MHV}$ is annihilated by $S_{aA}$ and  $\tilde{S}^{A}_{\dot{a}}$.
\example{
Let us show that 
$S_{aA}=\sum_i \partial_{[i|^{\scriptstyle a}} \, \eta_{iA}$ annihilates $\ca_n^\text{MHV}$.
Note that 
\be
   S_{aA} \,\d^4\big(P\big)
   = - \sum_i |i\>^{\da} \, \eta_{iA} \,  \pa_{P^{\da a}}\, \d^4\big(P\big)   = - \tilde{Q}_{A}^{\da} ~\pa_{P^{\da a}}\, \d^4\big(P\big)\,
\ee
 vanishes on the support of $\delta^{(8)} \big(\tilde{Q}\big)$. Since $\ca_n^\text{MHV}$ does not have any further dependence on square spinors, we conclude 
$S_{aA} \,\ca_n^\text{MHV} = 0$.
}
\example{To start with, consider  the action of $\tilde{S}^{A}_{\dot{a}}=\sum_i \partial_{|i\>^{\scriptstyle \dot{a}}}\,\partial_{\eta_{iA}}$    on $\delta^{(8)} \big(\tilde{Q}\big)$: it follows from direct calculation that 
\be 
  \partial_{\eta_{iA}} \delta^{(8)} \big(\tilde{Q}\big)
   =  |i\>^{\da}\, \partial_{\tilde{Q}_{A}^{\da}} \delta^{(8)} \big(\tilde{Q}\big)
  ~~~~~\text{and}~~~~~
  \tilde{S}^{A}_{\dot{a}}~\delta^{(8)} \big(\tilde{Q}\big)
  = (n-1+3)\, \pa_{\tilde{Q}_A^{\da}} \delta^{(8)} \big(\tilde{Q}\big)\,.
\ee
To show the second identity, you need
$\tQ_{C\da} \tQ_C^{\db} = \frac{1}{2} \delta_{\da}{}^{\db}\, \tQ_{C\dot{c}} \tQ_C^{\dot{c}}$ (no sum on $C$). We use this result when we write
\be
  \tilde{S}^{A}_{\dot{a}}\,\ca_n^\text{MHV}
  = \Big( \tilde{S}^{A}_{\dot{a}}\,\delta^{(8)} \big(\tilde{Q}\big) \Big) \,
     \frac{\delta^4(P)}{\prod_{i=1}^n\langle i,i+1\rangle}
     + \Big(\pa_{\tilde{Q}_A^{\db}} \delta^{(8)} \big(\tilde{Q}\big) \Big)
     \Big( \sum_{i} |i\>^{\da} \partial_{|i\>^{\scriptstyle \dot{b}}}~
     \frac{\delta^4(P)}{\prod_{i=1}^n\langle i,i+1\rangle} \Big)\,.
\ee
We can evaluate the second term using \reef{Monang} and \reef{diffdelP}, and we then have
\be
  \tilde{S}^{A}_{\dot{a}}\,\ca_n^\text{MHV}
  ~=~ \Big[
  (n-1+3)\, \pa_{\tilde{Q}_A^{\da}} \delta^{(8)} \big(\tilde{Q}\big) 
  ~+~ 
  \Big(\pa_{\tilde{Q}_A^{\db}} \delta^{(8)} \big(\tilde{Q}\big) \Big)
  (-2 - n) \delta_{\db}{}^{\da}
  \Big]\frac{\delta^4(P)}{\prod_{i=1}^n\langle i,i+1\rangle}
  ~=~0\,.~
\ee
This completes the proof that $\ca_n^\text{MHV}$ respects the superconformal symmetry $SU(2,2|4)$ of $\mathcal{N}=4$ SYM.\footnote{We have ignored here subtle points of non-generic momenta and anomalies.}
}

We end this section with a remark about the operation of {\bf \em inversion}, which will be useful for us later.
 Inversion acts on the spacetime coordinates as
\be
    \mathcal{I}(x^\m) = \frac{x^\m}{x^2}\,.
\ee
The inversion operation generates the (super)conformal symmetry group from the (super)Poincar\'e group, for example $K^\mu =  \mathcal{I}\, P^\mu\,  \mathcal{I}$. 
\exercise{ex:confboost}{In this exercise, we derive the form of the conformal boost generator in position space; this will be useful for us later. Write the momentum generator in position space as 
$P^{\da a} = \frac{\pa}{\pa x_{a \da}}$. Show that 
$\mathcal{I} \,P^{\da a} \,\mathcal{I}$ is equivalent to
\be
  \mathcal{K}^{\da a} 
  = -x^{\da c} x^{\dot{c} a} \frac{\pa}{\pa x^{\dot{c} c}}
\ee
by demonstrating that $\mathcal{I} \,P^{\da a} \,\mathcal{I}$ and 
$\mathcal{K}^{\da a}$ give the same result when acting on  
$x^{\db b}$.\\{}
[Hint: you'll need the same type of identity given in the hint of Exercise \ref{MMonA}.]

}
%

%%%%%%%%%%%%%%%%%%%%%%%%%%%%%%%%%%%%
\subsection{Twistors}
\label{s:twist}
%%%%%%%%%%%%%%%%%%%%%%%%%%%%%%%%%%%%
The representation of the superconformal generators given \reef{SCGen} is unusual in that the generators appear with various degrees of derivatives. For example, the bosonic $SU(2,2)$ subgroup has a 2-derivative generator $K_{a\dot{a}}$ as well as a multiplicative generator $P^{\dot{a}a}$. Since the form of the generators depend on the choice of variables --- here spinor helicity variables --- we can hope to find a set of variables such that all generators are linearized. 
This is actually simple and can be achieved by performing a `Fourier transformation' on the angle spinor variables: 
\be
\langle i|_{\dot{a}}\rightarrow i\frac{\partial}{\partial |\tilde\mu_i\rangle^{\dot{a}}},
\hspace{1cm}
 \frac{\partial}{\partial |i\rangle^{\dot{a}}}\rightarrow -i\langle\tilde\mu_i|_{\dot{a}}\,.
\ee
We are assuming here that our spacetime metric signature is $(--++)$ so that the angle and square spinors are all real. 
\exercise{}{Show that 
$\displaystyle
| i\rangle^{\dot{a}}\rightarrow -i \frac{\partial}{\partial \langle\tilde\mu_i|_{\dot{a}}}$.
}
For example, the translation and conformal boost generator each becomes linearized after the Fourier transform:
\be
 P^{\dot{a}b}\rightarrow i\sum_{i} 
      [i|^b\partial_{\langle\tilde\mu_i|_{\dot{a}}},\quad~~~ K_{a\dot{a}}=i\sum_i 
\langle\tilde\mu_i|_{\dot{a}} \partial_{[i|^{\scriptstyle a}}\,,
\ee
\exercise{}{
Show that the dilation generator then becomes 
\be
D=\sum_i \Big(\tfrac{1}{2}
   |i\>^{\dot{a}}\,\partial_{|i\>^{\scriptstyle \dot{a}}}+\tfrac{1}{2}|i]_{a}\, \partial_{|i]_{\scriptstyle a}}+1\Big)
   ~\rightarrow~
   \sum_i \Big(\tfrac{1}{2}|i]_{a}\, \partial_{|i]_{\scriptstyle a}}-\tfrac{1}{2}\langle\tilde\mu_i|_{\dot{a}}\partial_{ \langle\tilde\mu_i|_{\dot{a}}}\Big)\,.
\ee
}
The new set of parameters 
\be
  \mathcal{W}_i^{\mathsf{A}}
  =
  \big(\,[i|^a,\, |\tilde\mu_i\rangle^{\dot{a}},\,\eta_{iA}\,\big)\,,
\ee  
with a collective index $\mathsf{A} = (\dot{a},a,A)$ are called {\bf \em supertwistors}. In these variables the generators of the superconformal algebra $su(2,2|4)$ can be written compactly as
\be
G^{\mathsf{A}}{}_{\mathsf{B}}  = \sum^n_{i=1} G_i^{\mathsf{A}}\,_{\mathsf{B}}=
 \sum^n_{i=1} \Big(\mathcal{W}_i^{\mathsf{A}}\partial_{\mathcal{W}_i^\mathsf{B}}-\frac{1}{4}\delta^{\mathsf{A}}\,_{\mathsf{B}}\mathcal{W}_i^{\mathsf{C}}\partial_{\mathcal{W}_i^\mathsf{C}}\Big)\,.
\ee
The  $\delta^{\mathsf{A}}\,_{\mathsf{B}}$-term is necessary for the bosonic subgroups, $SU(2,2)$ and $SU(4)$, to be traceless. However, as the term proportional to $\delta^{\mathsf{A}}\,_{\mathsf{B}}$ simply counts the degree of $\mathcal{W}_i^{\mathsf{A}}$, in practice if the function one is interested in has manifestly vanishing weight in each 
$\mathcal{W}_i^{\mathsf{A}}$, the generators simplify to:
\be
G^{\mathsf{A}}{}_{\mathsf{B}}  = \sum^n_{i=1} \mathcal{W}_i^{\mathsf{A}}\partial_{\mathcal{W}_i^\mathsf{B}}\,.
\ee  

The bosonic components 
${W}_i^{I}  =
  \big(\,[i|^a,\, |\tilde\mu_i\rangle^{\dot{a}}\big)$ 
  of the supertwistors are simply called {\bf \em twistors}. They were first introduced by Penrose \cite{Penrose} in the context of describing flat Minkowski spacetime. Later they were supersymmetrized by Ferber \cite{Ferber1977qx} and used to form representations of the superconformal group. Note that under little group scaling, the supertwistors scale homogeneously, 
  $\mathcal{W}_i \to t_i \mathcal{W}_i$. That means that we can define them projectively:  the bosonic twistors are points in $\mathbb{CP}^3$ while the supertwistors live in $\mathbb{CP}^{3|4}$. We give a very brief introduction to twistors in appendix \ref{app:twistor}. 

As we will often see, a well-motivated introduction of new variables sometimes leads to the realization of hidden structures in the amplitude. Lets us consider how the $n$-point anti-MHV amplitude looks like in bosonic twistor space, $W^I_i= \big(\,[i|^a,\, |\tilde\mu_i\rangle^{\dot{a}}\big)$. To obtain anti-MHV amplitudes,  simply take the $|i\rangle$ of the MHV amplitude, without the momentum delta function, and change it to $|i]$, so one can straightforwardly conclude that the only $|i\rangle$ dependence in the anti-MHV amplitude is via $\d^4(P)$. Thus Fourier transforming $|j\rangle$, one has:
\eq
\int \left(\prod_{j=1}^nd^2|j\rangle e^{i\langle j\,\mu_j\rangle}\right) \mathcal{A}^\text{anti-MHV}_n
=\left[\int \left(\prod_{j=1}^nd^2|j\rangle e^{i\langle j\,\mu_j\rangle}\right) \delta^4(P)\right]f(|i])\,.
\eqe
Here $f(|i])$ is a function that only depends on $|i]$. To ease the  integration, write the delta function itself as a Fourier integration (ignoring factors of $2\pi$),
\be
\delta^4(P) = \int d^4x\,
e^{-i \,x_{a\dot{a}}\, \sum_{j}|j\rangle^{\dot{a}}[j|^a}\,.
\ee
The integration over $|j\rangle$ can now be carried out and we find 
\eq
\int \left(\prod_{j=1}^nd^2|j\rangle e^{i\langle j\,\tilde\mu_j\rangle}\right) \mathcal{A}^\text{anti-MHV}_n=
\int d^4x\bigg(\prod_{j=1}^n\delta^{2}\big(\langle \tilde\mu_j|_{\dot{a}}+[j|^ax_{a\dot{a}}\big)\bigg)\,f(|i])\,.
\eqe
This mean that in twistor space, the twistor variables ${W}_i^{I}=\big(\,[i|^a,\, |\tilde\mu_i\rangle^{\dot{a}}\big)$ are localized by the delta functions. The delta function enforces for each $i$, 
\eq
\langle \tilde\mu_i|_{\dot{a}}+[i|^a x_{a\dot{a}}=0\,.
\label{curves}
\eqe
This equation says that ${W}_i^{I}$ is determined by just the input of $[i|$. This naively has two components, but the projective nature of ${W}_i^{I}$ reduces this to just 1 degree of freedom. 
Thus the solution to \reef{curves} is parameterized by a 1-dimensional variable, say $[i|^1$, and that means the solution is described as a degree 1 curve in $\mathbb{CP}^3$ (it is defined as the zeroes of a degree 1 polynomial in $W_i$).

Since each $x_{a\dot{a}}$ defines a different curve, the integration $\int d^4x$ can be understood as integrating over all possible curves, 
i.e.~it is an integration over the moduli of degree 1 curves. Thus amazingly, the anti-MHV amplitude in twistor space corresponds to $n$ twistors $W_i$ living on a degree 1 curve!

This amazing observation was due to Witten~\cite{Witten:2003nn}, and it provided an important ingredient as well as inspiration for the modern development of scattering amplitudes. For a tree-level amplitude with $q$ number of plus helicities, the amplitudes live on a degree $(q\!-\!1)$-curve.\footnote{In~\cite{Witten:2003nn}, the twistor was defined with $\mathcal{W}_i^{\mathsf{A}}
  =
  \big(\,|i\rangle^{\dot{a}},\, [\mu_i|^{a},\,\eta_{iA}\,\big)$, where $[\mu_i|^{a}$ is the Fourier conjugate to $[i|^a$. Thus, instead, one had the MHV amplitude to be of degree 1 in twistor space, while N$^K$MHV corresponds to degree $K$-curves in twistor space. Note that in Section \ref{s:momtwist} we will introduce a variable $[\mu_i|^{a}$ which is \emph{not}  the same as the one in Witten's supertwistor $\mathcal{W}_i$. But it will be part of a different type of supertwistor $\mathcal{Z}_i$. Please don't be too confused.} This geometric interpretation was found to be given by a twistor string theory, whose tree-level amplitudes are precisely those of the 4-dimensional $\mathcal{N}=4$ SYM. More details can be found in the original paper \cite{Witten:2003nn}. We  leave this story for now, but it will sneak back into the limelight later on when supertwistors $\mathcal{W}_i^{\mathsf{A}}$  make another appearance in Section \ref{s:polytopes}.
  
Next, we explore the symmetries of the $\cn=4$ SYM superamplitudes further (Section \ref{s:DC}) and also take a tour at loop-level (Section \ref{s:loops}).

%%%%%%%%%%%%%%%%%%%%%%%%%%%%%%%%%%%%
%%%%%%%%%%%%%%%%%%%%%%%%%%%%%%%%%%%%
%%%%%%%%%%%%%%%%%%%%%%%%%%%%%%%%%%%%
\subsection{Emergence of dual conformal symmetry}
\label{s:emDCS}
%%%%%%%%%%%%%%%%%%%%%%%%%%%%%%%%%%%%%
In the previous sections we have experienced the advantage of using spinor helicity formalism for scattering amplitudes, both in terms of the restrictive power of consistency conditions, such as little group scaling, and in terms of the simplicity of the final result. Many of these properties stem from two facts: (1) these spinor variables trivialize the on-shell condition $p^2=0$, and (2) at the same time they transform linearly under Lorentz symmetries, so that we get manifestly Lorentz-invariant expressions for the amplitudes. In contrast, the ordinary way of representing the amplitudes using polarization vectors realizes Lorentz invariance by introducing a redundancy, namely gauge invariance; but this makes the  amplitudes  overly complicated. One can also consider working with only on-shell degrees of freedom by using light-cone (or space-cone) gauge. However, as these gauges are not Lorentz invariant,  the symmetry generators act non-linearly on the kinematic variables. Thus using the spinor helicity formalism essentially allows us to ``have our cake and eat it". It allows us to work with only on-shell degrees of freedom, yet the global symmetries are linearly realized. 

At this point you may have noticed that there is a glaring hole in the above story: there is an important part of the Poincar\'e  symmetry that does not act linearly on the spinor variables: the translations. In momentum space, translation invariance  corresponds to momentum conservation. This symmetry, as well as its supersymmetric partner, is respected by the scattering amplitudes in a rather {\em ad hoc} fashion, namely by being enforced through the presence of the delta functions:
\eq
\delta^{4}\Big(\sum_{i=1}^np_i^{\dot{a}a}\Big)
~~~~~~\text{and}~~~~~~
\delta^{(2\mathcal{N})}\Big(\sum_{i=1}^n|i\rangle\eta_{i}^A\Big)\,.
\eqe  
Here we have indicated the $\cn$-fold supersymmetric case, though in this section we are going to study only $\cn=4$ SYM.
The point here is that 
if we follow the spirit of what the spinor helicity formalism brought us, we should try to find new variables that either simplify, or at least encode, the information of momentum and supermomentum conservation.

As an inspiration, let us visualize momentum conservation geometrically. The fact that $n$ momenta $p_i^\mu$ add to zero implies that the vectors close into a closed contour, e.g.~for \mbox{$n=5$}:
\be
  \includegraphics[width=1.8in]{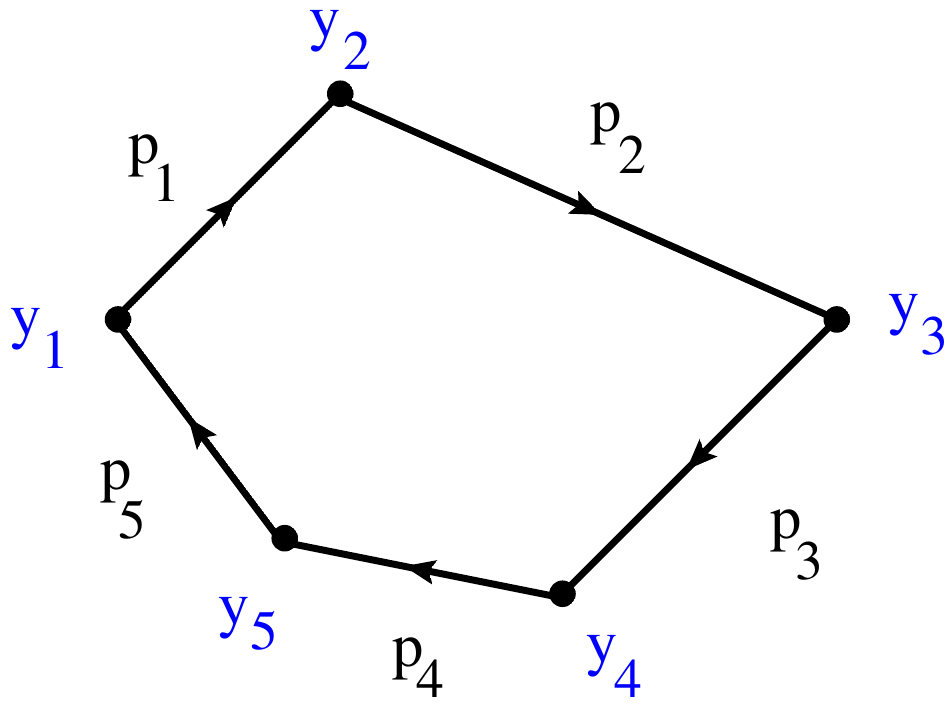}
\ee
 Now there are two different ways to define the contour: it can be defined by the edges or by the cusps. The former is just the usual momentum representation. For the latter, we take the cusps to be located at points $y_i^\mu$ in a {\bf \em dual-space} \cite{Drummond}. They are defined by their relation to the momentum vectors:
\eq
y^{\dot{a}a}_i-y^{\dot{a}a}_{i+1}=p_i^{\dot{a}a}\,.
\label{DualDef}
\eqe
The {\bf \em dual coordinates} $y_i$ (sometimes called  {\bf \em zone variables}) are \emph{not} spacetime coordinates; they are dual momentum variables defined by \reef{DualDef}. In particular, they have mass-dimension 1. In dual space, $n$-point momentum conservation simply corresponds to the periodicity condition that $y_{n+1} = y_1$. For massless particles, the edges of the $n$-edge polygon are lightlike.

Since the ordering of the external lines is crucial for the definition of the polygon, we need a well-defined notion of ordering. For (super) Yang-Mills theory, this is simply the color-ordering. The ordering lets us define
\be
  y_{ij} ~\equiv~ y_i - y_j ~=~ p_i + p_{i+1} +\dots + p_{j-1}.
\ee
The resulting variables $y_{ij}$ precisely match the $y_{ij}$'s introduced in  \reef{ythetadef} when we calculated the NMHV tree-level superamplitudes in $\cn=4$ SYM; this is of course no coincidence. In \reef{ythetadef} we also defined fermionic variables  $\theta_{ijA}^a$; those now arise as differences 
$|\theta_{ij,A}\> \equiv |\theta_{iA}\>-|\theta_{jA}\>$ 
of  dual space fermionic coordinates $|\theta_{iA}\>$ defined as
\eq
|\theta_{iA}\>-|\theta_{i+1,A}\>~=~|i\> \,\eta_{iA}\,,
\label{DualFDef}
\eqe
where the $\eta_{iA}$'s are the on-shell superspace Grassmann variables introduced in Section \ref{s:N4sym} and $A=1,2,3,4$ are the $SU(4)$ R-symmetry labels.  

In the dual coordinates,  the $n$-point tree-level MHV superamplitude of 
$\mathcal{N}=4$ SYM is
\eq
\mathcal{A}_n^\text{MHV}=\frac{\delta^{4}\left(y_1-y_{n+1}\right)\delta^{(8)}\left(\theta_1-\theta_{n+1}\right)}{\prod_{i=1}^n\langle i, i+1\rangle}\,,
\label{DualMHV}
\eqe
and the tree-level NMHV superamplitude takes the form
\be
\mathcal{A}_n^\text{NMHV}
=
\mathcal{A}_n^\text{MHV}\,
\sum_{j=2}^{n-3}\sum_{k=i+2}^{n-1} R_{njk}\,,
\label{DualNMHV}
\ee
where 
\be
  \begin{split}
  R_{njk}
  ~=~&
  \frac{\<j-1,j\>\<k-1,k\>\,   
   ~
    \d^{(4)}
   \big( 
    \Xi_{njk}
   \big)}
   {y_{jk}^2\,\< n |y_{nj}.y_{jk}| k\> \< n |y_{nj}.y_{jk}| k-1\>
    \<n|y_{nk}.y_{kj}|j\> \<n| y_{nk}.y_{kj} |j-1\>}\,,
\\[2mm]
\Xi_{njk,A}~=~&
 \<n|y_{nk}.y_{kj}|\theta_{jn,A}\> 
    + \<n| y_{nj}. y_{jk}|\theta_{kn,A}\>\,.
    \end{split}
    \label{Mess}
\ee
The NMHV expression is exactly the same as the super-BCFW result in \reef{NMHVdiagA4}-\reef{Rinv1}; here we have chosen to anchor the expressions on line $n$.

The representations in \reef{DualMHV} and \reef{DualNMHV} may appear somewhat disappointing since the amplitudes in dual space are basically identical to the original expressions. However, the dual space description allow us to study a new symmetry, namely (super)conformal symmetry in the dual coordinates $y$. This is called {\bf \em dual (super)conformal symmetry!} To describe it, first note that since the defining relation \reef{DualDef} of the 
$y_i$'s  is invariant under translations, the amplitude is guaranteed to be translational invariant in the $y$-space. Next, using the fact that the conformal boost generator is ${\mathcal{K}}^\mu=I \mathcal{P}^\mu I$, the dual superconformal  property of the amplitude can be extracted by simply studying how the amplitude transforms under \emph{dual inversion}:
\eq
I\big[y_i^\mu\big]
  =\frac{y_i^\mu}{y_i^2}\,, \quad~
I\big[\,|\theta_{iA}\>^{\da}\big]
  =  \<\theta_{iA}|_{\db}\, \frac{y_i^{\dot{b}a}}{y_i^2} \, ,\quad~
I\big[ \,[i|^a\big]
  =\frac{y_i^{\dot{a}b}}{y_i^2}\,|i]_b\,, \quad~ 
I\big[\,|i\>^{\dot{a}}\big]
  = \< i|_{\db}\,\frac{y_{i+1}^{\dot{b}a}}{y_{i+1}^2}\,.
\label{InvertRule}
\eqe 
These rules are well-defined only when we have a notion of ordering. Note that the inversion rules for $|i\>$ and $[i|$ are defined only up to a relative scaling.
\exercise{ex:dci}{The inversion rules for $y_i^\m$ and $|\theta_{iA}\>$ are standard. Verify the consistency of the rules for $|i\>$ and $[i|$ using \reef{DualDef} and $y^2_{i,i+2}=\langle i,i+1\rangle[i, i+1]$. \\[1.4mm]
{}[Hint: Show first that the definition \reef{DualDef} implies that  
$\<i|_{\da}\,y_i^{\dot{a}a}=\<i|_{\da}\,y_{i+1}^{\dot{a}a}$. It follows that 
 $I[\langle i,i+1\rangle]=\langle i,i+1\rangle/ y_{i+2}^2$.]
}
The momentum and supermomentum delta function transform under dual inversion as
\be
I\big[\delta^{4}(y_1-y_{n+1})\big]
=y_1^{8}\,\delta^{4}(y_1-y_{n+1})\,,
~~~~~~~
I\big[\delta^{(8)}(\theta_1-\theta_{n+1})\big]
=y_1^{-8}\,\delta^{(8)}(\theta_1-\theta_{n+1})\,.
\label{dualinvdeltas}
\ee
 Thus we see that the inversion weight of the bosonic delta function exactly cancels\footnote{Note that in $D=4$ this cancellation only happens for $\mathcal{N}=4$. 
 In $D=3$, a similar cancellation happens for $\mathcal{N}=6$; indeed a theory with $\mathcal{N}=6$ supersymmetry exists, namely ABJM theory, and its amplitudes also respect dual superconformal symmetry. 
 We discuss this theory in more detail in Section \ref{s:ABJM}. 
A similar result in $D=6$ would require a supersymmetric theory with 24 supercharges. It is not clear if an interacting theory exists that can realize the symmetry in $D=6$.}  that of the Grassmann delta function. So for the $\mathcal{N}=4$ SYM MHV superamplitude one obtains
\eq
I\big[\mathcal{A}_n^{\rm MHV}\big]
~=~
\Big(\prod_{i=1}^n y_i^2\Big)\,\mathcal{A}_n^{\rm MHV}\,.
\label{AmpInvert}
\eqe
Hence, under dual superconformal inversion, the MHV superamplitude transforms \emph{covariantly} with equal weights on all legs. 

At this point, you may wonder if this new dual conformal symmetry is secretly just another incarnation of  the conventional conformal symmetry. It is straightforward to see that this is not the case. Consider the pure Yang-Mills tree-amplitude, which is conformal invariant (because it takes the same form as in $\cn=4$ SYM). Under dual inversion, the split-helicity amplitude transforms as:
\eq
I\big[A_n[1^-2^-3^+\cdots n^+]\big]
~=~\Big(\prod_{i=1}^n y_i^2\Big)\big(y^2_1\big)^4 \, 
A_n[1^-2^-3^+\cdots n^+]\,.
\eqe
Clearly, this amplitude does not have homogeneous inversion properties. 
The situation is worse for a gluon amplitude without the split-helicity arrangement, for example
$A_n[1^-2^+3^-\cdots n^+]$. The result of $I\big[\langle13\rangle\big]$ is not  proportional to $\langle13\rangle$, so $A_n[1^-2^+3^-\cdots n^+]$ does not even transform covariantly under dual inversion. This shows that one can have a conformal invariant amplitude that is not dual conformal covariant. Hence the two symmetries are  inequivalent.

What about the NMHV superamplitude? Well, remarkably, the complicated mess in \reef{Mess} is invariant under dual inversion, 
i.e.~$I\big[R_{njk}\big]=R_{njk}$. 
Thus $\mathcal{A}_n^\text{NMHV}$ has the same homogeneous dual inversion weight as the MHV superamplitude. 
In fact, using super-BCFW recursion relations it can be shown   \cite{Drummond:2008cr,Brandhuber:2008pf} that \emph{all} tree superamplitudes of $\mathcal{N}=4$ SYM transform covariantly under dual inversion,
\be
I\big[\mathcal{A}^{\rm tree}_n\big]~=~\Big(\prod_{i=1}^n y_i^2\Big)\,\mathcal{A}^{\rm tree}_n\,.
\label{AmpInvert1}
\ee 
We prove this statement using recursion relations in Section \ref{s:bcfwMTtree}.
\exercise{ex:Risinv}{Use the inversion rules \reef{InvertRule} to show that $R_{njk}$ is invariant. Note that it was crucial that the spinor-products in $R_{njk}$ could be arranged to involve adjacent lines.  \\{}
[Hint: The identity 
  $(y_{nj}y_{jk}+y_{nk}y_{kj})_a{}^{b}+y^2_{jk}\,\delta_a{}^{b}=0$ is useful for calculating  $I[\Xi_{njk}]$.]}

Due to the non-trivial weights in \reef{AmpInvert1}, the dual conformal boost generator does not vanish on the amplitudes. Rather, it generates an `anomaly' term, 
\eq
\mathcal{K}^\mu\mathcal{A}_n^{\rm tree}=\Big(-\sum_{i=1}^n y_i^\mu \Big)\mathcal{A}_n^{\rm tree}\,.
\label{DCAnom}
\eqe
If we bring this term to the LHS and redefine $\tilde{\mathcal{K}}^\mu\equiv \mathcal{K}^\mu+\sum_{i=1}^n y_i^\mu$, then the new generator $\tilde{\mathcal{K}}^\mu$  annihilates  the amplitudes.

The dual conformal symmetry can be enlarged into an $SU(2,2|4)$ dual superconformal symmetry. Recall that $\mathcal{N}=4$ SYM is also superconformal invariant, with the same $SU(2,2|4)$ group. If we combine the two sets of generators, we obtain an infinite dimensional algebra called a {\bf \em Yangian}~\cite{Drummond3}. The generators of this algebra are organized by levels. For the $SU(2,2|4)$ Yangian, level $0$ consists of the ordinary superconformal generators 
$G^{\mathsf{A}}{}_{\mathsf{B}}  = \sum^n_{i=1} G_i^{\mathsf{A}}\,_{\mathsf{B}}$, where $\mathsf{A}=(I,A)$ and $I = (\dot{a},a)$ is the index of conformal symmetry $SU(2,2)$ and $A$ is the $SU(4)$ $R$-symmetry index. At level 1, the generators are bi-local in their particle index:
\eqa
\nonumber \text{level 0:}&&\sum^n_{i=1}G_i^{\mathsf{A}}\,_{\mathsf{B}}\\
\nonumber \text{level 1:}&&\sum^n_{i<j}(-1)^{|\mathsf{C}|}[G_i^{\mathsf{A}}\,_{\mathsf{C}}\,G_j^{\mathsf{C}}\,_{\mathsf{B}}-(i\leftrightarrow j)]\\
&&\vdots
\eqae
where $|\mathsf{C}|$ is 0 for $\mathsf{C}=I$ and 1 for $\mathsf{C}=A$.
It turns out that the shifted dual conformal boost generator $\tilde{\mathcal{K}}^\mu$ (not the unshifted one, $\mathcal{K}^\mu$) belongs to level 1. So the `anomaly' in \reef{DCAnom} was not a nuisance, but rather it was needed in order for the tree-level superamplitude of $\mathcal{N}=4$ SYM to be Yangian invariant! Beyond level 1, the new generators can be obtained simply by repeated (anti)-commutation of level 1 and level 0 generators. For further information about Yangian symmetry we refer to the original work \cite{Drummond3}.  The message here is that superamplitudes of 
$\mathcal{N}=4$ SYM are Yangian invariant.

%%%%%%%%%%%%%%%%%%%%%%%%%%%%%%%%%%%%%%%%%%%%%%%%%%%%%%
\subsection{Momentum twistors}
\label{s:momtwist}
%%%%%%%%%%%%%%%%%%%%%%%%%%%%%%%%%%%%%%%%%%%%%%%%%%%%%%%
Now that we have seen that the $\cn=4$ SYM tree superamplitudes have dual superconformal symmetry (in fact even Yangian symmetry), we are again set on the path to find new variables that transform covariantly under the new symmetry. This is especially justified given that $R_{njk}$ is very unwieldy in its current form: we would like to write it as an expression that is manifestly invariant under the dual superconformal symmetry. Also, the presence of both helicity spinors and the vectors $y_i^\m$ in $R_{njk}$ is a further redundancy of variables that we would like to eliminate. 

As a first step, we redefine the dual-space coordinate $y_i^\m$ in terms of spinor variables. Recall that we introduced the $y_i^\m$ coordinates  by their relation to the momenta: $p_i = y_i^\m - y_{i+1}^\m$. This relation implies that 
$\<i|_{\da}\,y_i^{\dot{a}a}=\<i|_{\da}\,y_{i+1}^{\dot{a}a}$. Instead of referring to the momentum, we can take this relation to be the defining relation for the dual coordinates $y_i^\m$: these are called the 
{\bf \em incidence relations} and take the form
\eq
[\mu_i|^a=
 \<i|_{\da}\,y_i^{\dot{a}a}=
\<i|_{\da}\,y_{i+1}^{\dot{a}a} \,.
\label{incidence}
\eqe   
The incidence relations define the new variable $[\mu_i|^a$.
The statement of the incidence relations is that 
for a given pair of spinors $Z_i^I=\big( |i\>, [\mu_i|\big)$, with $I=(\da,a)$ being an $SU(2,2)$ index, any two points, $y_i^\m$ and $y_{i+1}^\m$, in $y$-space that satisfies \reef{incidence} must be null-separated by the vector $y_i^\m - y_{i+1}^\m=p_i^{\dot{a}a}$. Thus the line in $y$-space determined by the two points, say $y_i^\m$ and $y_{i+1}^\m$, corresponds to a point $Z_i^I=\big( |i\>, [\mu_i|\big)$ in $Z$-space (which we are going to discuss further in the following). 

\begin{figure}[t]
\begin{center}
\includegraphics[width=6cm]{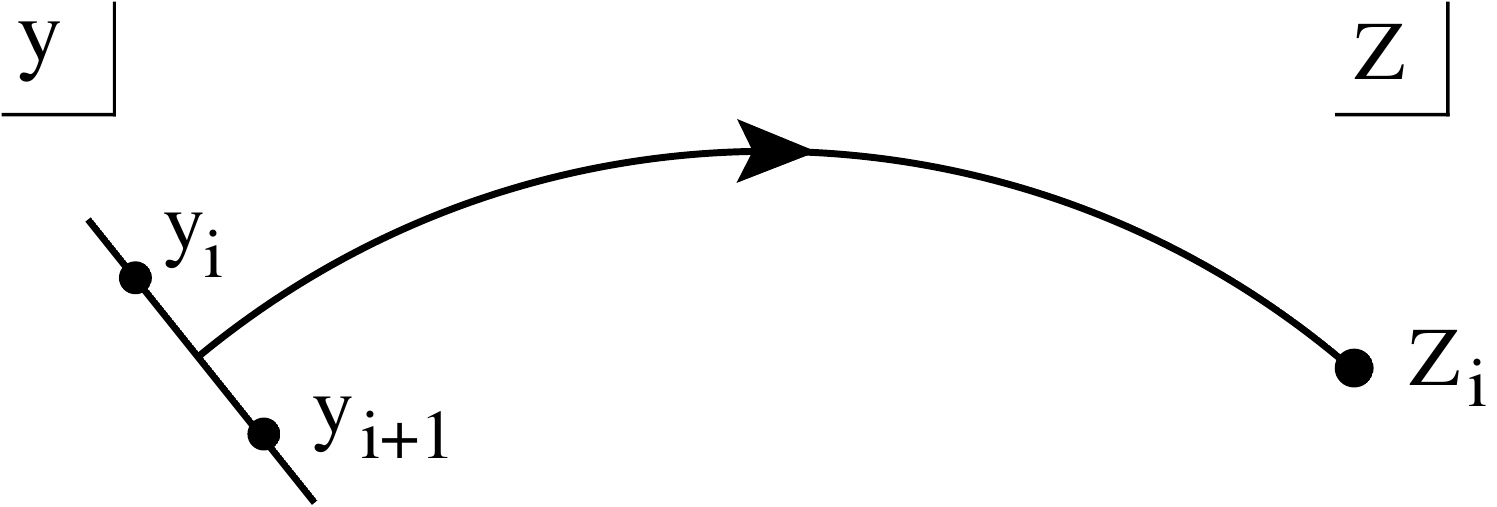}
\hspace{1.6cm}
\includegraphics[width=6cm]{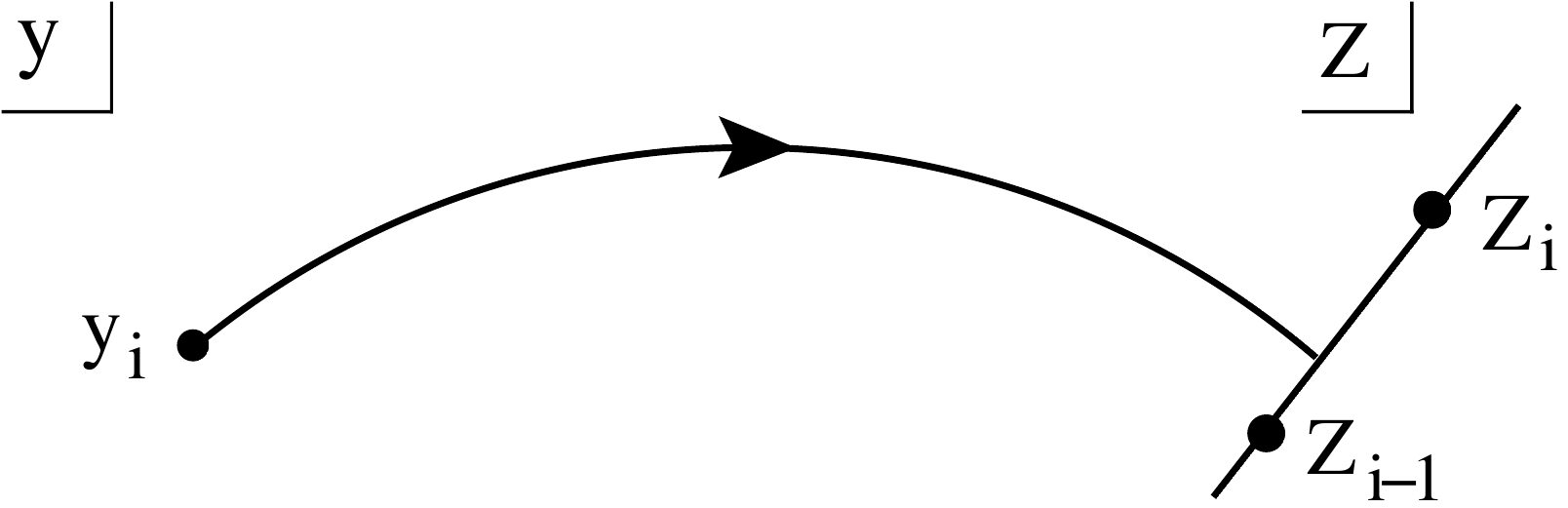}
\caption{\small A graphical representation of the map between dual space $y^\mu$ and momentum twistor space 
$Z_i^I=\big(|i\>, [\mu_i|\big)$. The lefthand figure illustrate the incidence relations \reef{incidence}: a null line in dual space, defined by the two points $y_i$ and $y_{i+1}$, corresponds to a point 
$Z_i^I=\big(|i\>, [\mu_i| \big)$ in momentum twistor space. 
 The righthand figure shows how a point $y_i$ in dual space maps to a line in momentum twistor space via the relation \reef{inversemap}.}
\label{coordinates}
\end{center}
\end{figure}

On the other hand, any point in $y$-space is determined by a line in 
 $Z$-space. To see how this comes about, note that the point
  $y_i^{\da a}$ is involved in two incidence relations:
$[\m_i| = \<i|y_i$ and $[\m_{i-1}| = \<i-1|y_i$. Combining these leads to
\be
  |i\>^{\db} \,[\m_{i-1}|^a - |i-1\>^{\db}\, [\m_i|^a
  ~=~
  \Big(
     |i\>^{\db}\<i-1|_{\da} - |i-1\>^{\db} \<i|_{\da} 
  \Big)\, y_i^{\da a}
  ~=~
  \< i-1,i\> \,y_i^{\db a}
\ee
so that
\be
   \label{inversemap}
   y_i^{\da a}
   ~=~
   \frac{|i\>^{\da} \,[\m_{i-1}|^a - |i-1\>^{\da}\, [\m_i|^a}{\< i-1,i\>} \,.
\ee
This means that $y_i$ is determined by $Z_{i-1}^{I}$ and $Z_i^{I}$: these two points define a unique line in $Z$-space.
The relationship between $y$-space and $Z$-space is illustrated in 
Figure \ref{coordinates}.
\exercise{}{Use the identities from Appendix \ref{app:conv} to show that
$|\mu_i]_a = -(y_i)_{a\dot{b}} |i\>^{\db}$.
}
We have translated the dual coordinates $y_i$ to $Z_i^{I}\equiv(|i\>^{\dot{a}}, [\mu_i|^a)$. The new four-component spinor variables $Z_i^I$ are called {\bf \em momentum-twistors}~\cite{Hodges}. The name stems from the analogy with spacetime twistors: a point in position space maps to a line in twistor space, and vice versa. The relationship between a flat four-dimensional space and the twistor variables can be stated in an $SU(2,2)$ covariant fashion, which we review briefly in 
Appendix \ref{app:twistor}. 

Note that the defining incidence equations \reef{incidence} imply that 
$|\mu_i] \to t_i |\mu_i]$ under little group scaling \reef{lgrp}: this means that the momentum twistors undergo a uniform rescaling ${Z}_i^I\rightarrow t_i{Z}_i^I$. Hence the momentum twistors are defined projectively.

So what have we achieved by going from $y_i$ to $Z_i^{I}\equiv(|i\>^{\dot{a}}, [\mu_i|^a)$? Well, the new variables transform linearly under the dual conformal transformations.\footnote{For simplicity we will consider $U(2,2)$ which includes
the $SU(2,2)$. 
}
The generators $\mathcal{G}^{I}{}_{J}$ of the dual conformal group can now be written compactly together with the group algebra as 
\eq
\mathcal{G}^{I}{}_{J}\equiv \sum_{i}Z_i^I\frac{\partial}{\partial Z_i^{J}},
\hspace{1.7cm}
 \big[\mathcal{G}^I{}_J,\mathcal{G}^K{}_L\big]
 =\delta_{J}{}^K \,\mathcal{G}^{I}{}_L
 -\delta^I{}_{L}\,\mathcal{G}^{K}{}_J\,,
\label{DCgen}
\eqe
with $I,J,\dots=(\dot{a},a)$. $\mathcal{G}^{I}{}_{J}$ can be thought of as a $4\times 4$ matrix operator with a  block diagonal $2\times 2$ structure. 
To make the generators more concrete, consider the $2\times 2$ block 
 with $I=a$ and  $J=\dot{a}$: it is  
$\mathcal{G}^{a}{}_{\dot{a}}=\sum_i [\mu_i|^a\frac{\partial}{\partial |i\>^{\dot{a}}}$.  Its index structure, and the fact that has mass-dimension 1, indicates that this should be the dual conformal boost $\mathcal{K}^{a}{}_{\dot{a}}$. 
In analogue with the regular conformal boost, given in Exercise \ref{ex:confboost},  the dual conformal boost generator can be written  in dual $y$-space as
$\mathcal{K}^{a}{}_{\dot{a}}
= - \sum_i \eps_{\dot{a}\dot{c}}\, y_i^{\dot{b}a}\,y_{i}{}^{\dot{c}b}\,\frac{\partial}{\partial y_i^{\dot{b}b}}$. Comparing this expression to $\mathcal{G}^{a}{}_{\dot{a}}$, it is not obvious that 
$\mathcal{G}^{a}{}_{\dot{a}} = \mathcal{K}^{a}{}_{\dot{a}}$, but the following exercises shows you how it works.
\exercise{}{ 
Show that this $\mathcal{K}^{a}{}_{\dot{a}}$ and $\mathcal{G}^{a}{}_{\dot{a}}$ are equivalent by demonstrating that they give the same result when acting on $y_i^{\dot{c}c}$ in \reef{inversemap}.
  }
Since the $y_i$'s and the momenta $p_i = -|i\> [i|$ are related, the variable change from $(|i\>, y_i)$ to $Z_i=(|i\>, [\m_i|)$ implies that we should be able to express $[i|$ in terms of $|i\>$ and $[\m_i|$. Indeed one finds
\eq
  [i|
  ~=~
  \frac{\langle  i+1, i\rangle[\mu_{i-1}|+\langle  i, i-1\rangle[\mu_{i+1}|+\langle i-1, i+1\rangle [\mu_{i}|}{\langle i-1, i\rangle\langle i, i+1\rangle}
\,.
\label{bosonicmap}
\eqe
\exercise{}{Derive \reef{bosonicmap} using 
the incidence relations \reef{incidence} and Schouten identities.
}
Since the momentum twistors $Z^I$ carry the dual conformal $SU(2,2)$ index $I$ we can form a dual conformal invariant by contracting four $Z^I$'s with the Levi-Civita $\eps_{ABCD}$ of $SU(2,2)$: we  use a 
{\bf \em 4-bracket} to denote this invariant:
\eqa
\nonumber\langle i, j, k, l\rangle
~\equiv~ 
\epsilon_{IJKL}Z_i^IZ_j^JZ_k^KZ_l^L &\,=\,&
\langle i j\rangle[\mu_k\mu_l]
+\langle i k\rangle[\mu_l\mu_j]
+\langle il\rangle[\mu_j\mu_k]\\  
&&+~\langle kl\rangle[\mu_i\mu_j]
+\langle lj \rangle[\mu_i\mu_k]
+\langle jk \rangle[\mu_i\mu_l]\,.
\label{FourBracket}
\eqae 
 On the RHS we have expanded out the product in terms of $SL(2,\mathbb{C})$ invariants, with $[\mu_i \mu_j]\equiv [\mu_i|^a |\mu_j]_{a}$. 
 
We can get some intuition for the new 4-bracket by evaluating them in special cases. For example
\be
\langle k,j-1,j,r\rangle 
~=~ 
\langle j-1,j\rangle \,
\langle k|y_{kj}y_{jr}|r\rangle\,.
\label{ID2}
\ee
\exercise{}{Prove \reef{ID2} by first using \reef{FourBracket} to rewrite the LHS as a sum of $\langle ij\rangle[\mu_k \mu_l]$'s. Then apply \reef{incidence} and Schouten away to pull out an overall factor $\langle j-1,j\rangle$.}
 A special case of \reef{ID2} is
\be
\langle j-1,j,k-1,k\rangle
= 
 \langle j-1,j\rangle \,
\langle k-1|y_{k-1, j}\,y_{jk}|k\rangle
= 
 \langle j-1,j\rangle \,
\langle k-1,k\rangle \,y_{jk}^2\,,
\ee
i.e.
\be
y_{jk}^2~=~  \frac{\langle j-1,j,k-1,k\rangle}{\langle j-1,j\rangle \,
\langle k-1,k\rangle}\,.
\label{ID3}
\ee
Since $1/y_{ij}^2$ are propagators, the relation \reef{ID3} will appear repeatedly in our discussions. 

Looking at \reef{ID2} and \reef{ID3} makes us realize that these are exactly the type of objects that appear in the denominators of the $R$-invariants \reef{Mess} of the NMHV tree-amplitudes, so we can write it
\be
  R_{njk}
  ~=~
  {
  \frac{\<j-1,j\>^4\<k-1,k\>^4\,   
   ~
    \d^{(4)}
   ( 
    \Xi_{njk}
   )}
   {\langle n,j-1,j,k-1\rangle\langle j-1,j,k-1,k\rangle\langle j,k-1,k,n\rangle\langle k-1,k,n,j-1\rangle\langle k,n,j-1,j\rangle}
   }\,.
\label{Rinv3}
\ee
Note that is was possible to arrange the five denominator factors such that the input of the 4-brackets go cyclically through the set of five labels 
$(n,j-1,j,k,k-1)$. Now the denominator is manifestly dual conformal invariant since it is composed entirely out of the $SU(2,2)$-invariant 4-brackets.
However, the 4-brackets transform under little group scaling with weight 1 for each line in the argument. Thus the denominator is not really an invariant since the $Z^I$'s are defined only projectively. So let us take a closer look at the numerator; for this purpose we need Grassmann-companions for the $Z^I$'s. 
 
Similarly to the bosonic incidence relations \reef{incidence}, we use  
the spinor-fermionic coordinate $\theta_{iA}^{\dot{a}}$ to introduce a Grassmann-odd (spacetime-)scalar coordinate $\chi_{iA}$:
\eq
\chi_i^A= \<i \,\theta_{iA}\>= \<i \,\theta_{i+1,A}\> \,.
\label{Fermiincidence}
\eqe
With these new fermionic twistor variables, we have extended the $SU(2,2)$ momentum twistors $Z^I$ to $SU(2,2|4)$ {\bf \em momentum super-twistors}
\be
 \mathcal{Z}_i^{\mathsf{A}}
 \equiv \big(|i\>^{\dot{a}}, [\mu_i|^a\, \big|\,\chi_{iA}\big)\,, 
 ~~\text{ where $\mathsf{A}=(\dot{a},a,A)$.}
 \label{smomtwis}
\ee 
 Under the little group scaling,  $\mathcal{Z}_i^{\mathsf{A}}\rightarrow t_i\mathcal{Z}_i^{\mathsf{A}}$, so the momentum super-twistors are defined projectively.
\exercise{}{Derive the fermionic versions of \reef{inversemap} and \reef{bosonicmap}.
}
To rewrite $\delta^{(4)}(\Xi_{njk})$ in the $R$-invariant \reef{Rinv3} with 
$\Xi_{njk,A}=\<n|y_{nk}.y_{kj}|\theta_{jn,A}\> + \<n| y_{nj}. y_{jk}|\theta_{kn,A}\>$, we need the identity
\eq
\langle k|y_{kr}y_{rj}|\theta_{jA}\>=-\frac{\langle k,r-1,r,j-1\rangle\chi_{jA}-\langle k,r-1,r,j\rangle\chi_{j-1,A}}{\langle r-1,r\rangle\langle j-1,j\rangle}
\,.
\label{Dist}
\eqe  
\exercise{}{Derive \reef{Dist} by manipulating the RHS using \reef{ID2} and employing the Schouten identity.
}
\exercise{}{Use  \reef{Dist} to show that
\be
  \label{Xinew}
  \Xi_{njk,A} =- \frac{\big[ \langle j-1,j,k-1,k\rangle \,\chi_{nA}+ \text{cyclic}\big]}{\langle j-1,j\rangle\langle k-1,k\rangle}  \,.
\ee
[Hint: use the hint in Exercise \ref{ex:Risinv}.]
}
Plugging \reef{Xinew} into \reef{Rinv3}, we then have
\be
  R_{njk}
  ~=~
  {
  \frac{  
    \d^{(4)}
   \big( 
    \langle j-1,j,k-1,k\rangle \,\chi_{n}+ \text{cyclic}
   \big)}
   {\langle n,j-1,j,k-1\rangle\langle j-1,j,k-1,k\rangle\langle j,k-1,k,n\rangle\langle k-1,k,n,j-1\rangle\langle k,n,j-1,j\rangle}
   }\,.
\label{Rtwistor}
\ee
The ``$+ $cyclic" is the instruction to sum 
cyclically over the labels $(n,j-1,j,k,k-1)$,  similarly to the product structure in the denominator.
Now $R_{njk}$ is manifestly invariant under both the little group scaling and dual $SU(2,2)$. Together with the overall factor of $\ca_n^\text{MHV}$, 
we then have the building blocks of the NMHV superamplitude in a form that makes it manifestly dual superconformal invariant. 

The expression \reef{Rtwistor} for $R_{njk}$ is cyclic in the labels $(n,j-1,j,k-1,k)$. This motivates us to define the {\bf \em 5-bracket} notation
\eq
R_{njk}=\big[n,j-1,j,k-1,k\big]\,.
\label{fivebrackets}
\eqe
The 5-bracket is cyclic in its five arguments, so for example
$[6,1,2,3,4] = [1,2,3,4,6]$. 

In terms of the 5-bracket, the $n$-point NMHV amplitude is simply given by 
\eq
\mathcal{A}_n^{\rm NMHV}=
\mathcal{A}_n^{\rm MHV}~
\sum_{j=2}^{n-3}\sum_{k=j+2}^{n-1} \big[n,j-1,j,k-1,k\big]\;.
\label{Final}
\eqe
Let us review what we have accomplished so far. Starting with the simple observation that momentum conservation is imposed in a rather {\em ad hoc} fashion, we introduced the auxiliary variables $y_i$ such that the constraint is encoded in a geometric fashion. This led us to the realization of a new symmetry of the tree amplitude for $\mathcal{N}=4$ SYM, a conformal symmetry in the dual space $y_i$. The new symmetry put us on the journey to search for new variables that linearize their transformation rules, culminating in the simple symmetric form of the $n$-point NMHV superamplitude in \reef{Final}. For N$^{K}$MHV, equation \reef{Final} generalizes to a sum involving products of $K$ 5-brackets. 

The 5-bracket in \reef{Final} correspond to the terms in the super-BCFW expansion of the super\-amplitude; specifically we have seen in Section \ref{s:NMHV} how each $R_{n2k}$ arises from an MHV$\times$MHV BCFW diagram while the remaining $R_{njk}$'s with $j>2$ appear via recursion from the BCFW diagram with NMHV$\times$anti-MHV subamplitudes. As we have discussed, this means that the representation \reef{Final} is not be unique, since there are many equivalent BCFW expansions for a given amplitude, depending on the choice of lines in the BCFW shift. This implies that the dual conformal invariants  \reef{fivebrackets} are linearly dependent. For example, compare for $n=6$  the result of the recursions relations based on the BCFW supershifts $[6,1\>$ and $[1,2\>$: they have to give the same result, so
\be
  [6,1,2,3,4]+[6,1,2,4,5]+[6,2,3,4,5] = 
  [1,2,3,4,5]+[1,2,3,5,6]+[1,3,4,5,6]\,.
\ee
Using the cyclic property of the 5-bracket, we can write this 
\eq
 [2,3,4,6,1]+[2,3,4,5,6]+ [2,4,5,6,1]
 =
 [3,4,5,6,1]+[3,5,6,1,2]+[3,4,5,1,2]
 \,.
\eqe
Now you see that the LHS looks like the result of a $[2,3\>$ supershift, while the RHS comes from a $[3,4\>$ supershift. In fact, you'll note that the LHS and independently the RHS are invariant under $i \to i+2$. 
We can also reverse the labels in the 5-brackets at no cost to get
\eq
 [2,3,4,6,1]+[2,3,4,5,6]+ [2,4,5,6,1]
 =
 [3,1,6,5,4]+[3,2,1,6,5]+[3,2,1,5,4]
 \,.
 \label{susy6termID}
\eqe
This states that the `parity conjugate' supershifts $[2,3\>$ and $[3,2\>$ give identical results. In fact, we can now conclude that any adjacent supershifts are equivalent.
The identity \reef{susy6termID} shows up again in Sections \ref{s:grassmannia} and \ref{s:polytopes} where we will understand its origins better.
\exercise{ex:preloop}{Use cyclicity of the 5-brackets to show that the tree-level 6-point NMHV superamplitude can be written in the form
\be
  \mathcal{A}_6^{\rm NMHV}
  = \mathcal{A}_6^{\rm MHV}\times
     \frac{1}{2} \Big(
         R_{146} + \text{cyclic}
      \Big)\,,
\ee
where ``cyclic" means the sum over advancing the labels cyclically, 
i.e.~$R_{146} + R_{251} + $ 4 more terms.
}
The presence of these equivalence-relations between the dual conformal invariants may strike you as rather peculiar and you may wonder if it has a deeper meaning. Furthermore, while the expressions in \reef{Rtwistor}  and \reef{Final} are extremely simple, they lack one key aspect when compared to the Parke-Taylor superamplitude: cyclic invariance. The presence of dual conformal symmetry relies heavily on the cyclic ordering of the amplitude, and hence it is somewhat surprising that the  manifestly dual conformal invariant form of the superamplitude \reef{Final} breaks manifest cyclic invariance. One might say that we are  asking too much of the amplitude, but considering the payoff we have reaped from the innocent chase of manifest momentum conservation, we will boldly push ahead with our pursuit of ``having cakes and eating them"   in  Sections \ref{s:grassmannia} and \ref{s:polytopes}. 

{\bf Momentum twistors.} 
For our further studies, it is worth making a few observations about the momentum twistors. We introduced the dual $y_i$'s in order to make momentum conservation manifest; but the $y_i$'s could not be chosen freely since they are subject to the constraint $(y_i - y_{i+1})^2= 0$ that ensures the corresponding momenta $p_i$ to be on-shell. On the other hand, the momentum twistors $Z_i$ are free variables: they are subject to the scaling equivalence $Z_i \sim t Z_i$, so they live in projective space $\mathbb{CP}^{3}$. (The momentum supertwistors $\mathcal{Z}_i$ are elements of $\mathbb{CP}^{3|4}$.) We can choose $n$-points $Z_i$ in $\mathbb{CP}^{3}$, subject to no constraints, then study the $n$ lines defined by consecutive points $(Z_i,Z_{i+1})$, with the understanding that the $n$'th line is $(Z_n,Z_{1})$.  Equation \reef{inversemap} maps each line $(Z_i,Z_{i+1})$ to $y_i$ and the incidence relation \reef{incidence} guarantees that the points $y_i$ and $y_{i+1}$ are null-separated; thus the corresponding momenta $p_i = y_i - y_{i+1}$ are on-shell. 
Since the lines 
$(Z_i,Z_{i+1})$ per definition close into a closed contour ensures momentum conservation $y_{n+1} = y_1$. So all in all, the map to momentum twistors geometrizes the kinematic constraints of momentum conservation and on-shellness by simply stating these requirements as the intersection of $n$ lines $(i,i+1)\equiv (Z_i,Z_{i+1})$ at the points $(i) \equiv Z_i$ in the $\mathbb{CP}^{3}$ momentum twistor space. The momentum supertwistors similarly make conservation of supermomentum automatic.

Pursuing the geometric picture a little further, consider intersections of lines and planes. In $\mathbb{CP}^{3}$, the point $(p) \equiv Z_p$ that corresponds to the intersection of line $(i,j)\equiv (Z_i,Z_j)$ with a plane defined by $(k,l,m)\equiv (Z_k,Z_l, Z_m)$ is given by 
\eq
(p)=(i,j)\bigcap(k,l,m)=Z_{i}\,\langle j,k,l,m\rangle-Z_{j}\,\langle i,k,l,m\rangle\,.
\label{TwistorPoint}
\eqe
The symbol $\bigcap$ indicates the intersection of the two objects.
Similarly, the line $(p,q)$ that corresponds to the intersection of plane $(Z_i,Z_j,Z_k)$ and $(Z_l, Z_m, Z_n)$ is given by 
\eq
 (p,q)=(i,j,k)\bigcap(l,m,n)=(i,j)\,\langle k,l,m,n\rangle+(j,k)\,\langle i,l,m,n\rangle+(k,i)\,\langle j,l,m,n\rangle\,.
\label{TwistorLine}
\eqe
A more detailed discussion of twistor geometry can be found in \cite{Mason}. 

Propagators $1/y_{ij}^2$ are expressed in terms of momentum twistors via \reef{ID3}. This means that the on-shell condition $y_{ij}^2 = 0$ becomes the requirement
$\langle i-1,i,j-1,j\rangle = 0$. This is the statement that the four momentum twistors labelled by $i-1$, $i$, $j-1$, $j$ are linearly dependent. Geometrically, it means that they lie in the same plane in
$\mathbb{CP}^{3}$. 
In Section \ref{s:bcfwMTtree} we will see that in the momentum twistor space a  pole $\hat{y}_{ij}^2 = 0$ in the BCFW-shifted amplitude is characterized as the intersection between the line $(i-1,i)$ and the plane $(i-1,j-1,j)$; this motivates why we are interested in formulas such as  \reef{TwistorPoint}. 

It is now time to venture beyond tree-level and wrestle with loops: 
in the next three sections, we discuss various approaches to loop-amplitudes.

%%%%%%%%%%%%%%%%%%%%%%%%%%%%%%% 
%%%%%%%%%%%%%%%%%%%%%%%%%%%%%%% 
%%%%%%%%%%%%%%%%%%%%%%%%%%%%%%% 
\newpage
\setcounter{equation}{0}
\section{Loops I: Unitarity methods}
\label{s:loops}
%%%%%%%%%%%%%%%%%%%%%%%%%%%%%%% 
%%%%%%%%%%%%%%%%%%%%%%%%%%%%%%% 
%%%%%%%%%%%%%%%%%%%%%%%%%%%%%%% 

Up to now, we have focused exclusively on tree-level amplitudes. The 
loop-corrections are of course highly relevant, both in particle physics applications and for our understanding of the mathematical structure of the S-matrix. 

An $L$-loop amplitude can be written schematically as 
\be
  {\cal A}^{L\text{-loop}}_n  =  i^L\,
\sum_{j}{\int{
 \Big( \prod_{l = 1}^L \frac{d^D \ell_l}{ (2 \pi)^D} \Big)
  \frac{1}{S_j}  
 \frac {n_j \,c_j}{\prod_{\alpha_j}{p^2_{\alpha_j}}}}}\,, 
 \label{genform}
\ee
where $j$ labels all possible $L$-loop Feynman diagrams. For each diagram, 
  $\ell_l$ are the $L$ loop momenta, $\alpha_j$ label the propagators, and $S_j$ is the symmetry factor. 
The kinematic numerator factors $n_j$ are polynomials of Lorentz-invariant 
contractions of external- and loop-momenta and polarization vectors (or other external wavefunctions). The constants  $c_j$ capture the information about couplings and gauge group factors. 

At loop-level, we discuss three distinct objects:
\begin{enumerate}
\item The {\em loop-integrand} is the rational function inside the loop momentum integration.
\item The {\em loop-integral} is the combination of the integrand and the loop-momentum integration measure: this is a formal object, since we have not specified the integration region of the loop momentum or addressed divergences. 
\item  The {\em  loop-amplitude} is the result of carrying out the loop-integrations in the loop-integral.
If we integrate over  physical momentum space $\mathbb{R}^{1,3}$, the integral  may have infrared (IR) and ultraviolet (UV) divergences. We need to regulate such divergences in order to make the integrated result, the amplitude, well-defined.
\end{enumerate}

The analytic structure of loop-amplitudes is more complicated than for tree amplitudes. Where tree-amplitudes are simple rational functions, the loop-integrations typically give rise to various generalized logarithms and special functions. Thus  loop-amplitudes have branch cuts in addition to poles. 
The well-understood pole structure of tree amplitudes was instrumental for developing the on-shell recursion relations (Section \ref{s:recrels}), so at first sight it looks challenging  to develop a similar approach for loop-amplitudes. 
Nonetheless, the analytic structure of the loop-integrands can be exploited to reconstruct the amplitude from lower-order on-shell data. The purpose of this and the following two sections is to show you how.  Our focus in this section is on the widely used and very successful unitarity method. 
Next, in Section \ref{s:loops2}, we present  BCFW recursion relations for the loop-integrands, and finally we discuss  Leading Singularities and on-shell diagrams in Section \ref{s:loops3}.

The generalized unitarity method \cite{UnitarityMethod} is a subject that deserve much more attention than we are able to offer here. Our introduction to the unitarity method covers just the minimum needed for you to see the idea and appreciates is power. 
The method of generalized unitarity has been reviewed extensively and you can learn more about it and its applications to  supersymmetric as well as non-supersymmetric theories in the reviews \cite{Review1, Review2, Review3,Review4}.

%%%%%%%%%%%%%%%%%%%%%%%%%%%%%%%%%%%%
\subsection{Unitarity and the generalized unitarity method}
\label{s:genunit}
\begin{figure}
\begin{center}
\includegraphics[scale=0.85]{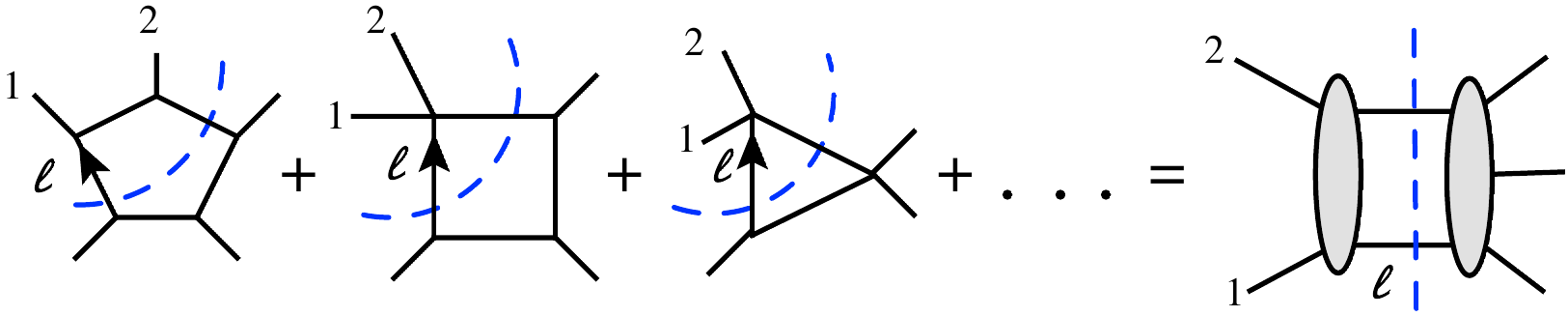}
\caption{\small The sum of residues from all Feynman diagrams with propagators $\ell^2$ and $(\ell-p_1-p_2)^2$ on-shell must give the product of two tree-amplitudes.}
\label{GenU}
\end{center}
\end{figure}

%%%%%%%%%%%%%%%%%%%%%%%%%%%%%%%%%%%%%
We begin with a concrete example: the color-ordered planar 5-point 1-loop gluon amplitude in pure Yang-Mills theory.  Suppose we identify\footnote{More about this choice in Section \ref{s:bcfwloop}.} the loop-momentum such that in each Feynman diagram, $\ell$ is the momentum that flows between legs $1$ and $5$, as indicated in Figure \ref{GenU}.
Then we can collect all the distinct Feynman diagrams under one integral,
\be
\int d^D\ell \,\sum_{j}J_j\,.
\ee   
The integrands $J_j$ take the form indicated in \reef{genform}.
To compute the full amplitude we need to integrate $\ell$ over $\mathbb{R}^4$ (after Wick rotation from  $\mathbb{R}^{1,3}$), but let us focus on the subplane where the loop-momentum satisfies the two {\em cut conditions}
\eq
\ell^2~=~(\ell-p_1-p_2)^2~=~0\,.
\label{LoopConstraint}
\eqe
On this subplane, integrands of the form 
\be
 J_i=\frac{1}{S_i}\frac{c_i n_i}{\cdots(\ell^2)\cdots(\ell-p_1-p_2)^2\cdots}
\ee
become singular. The singularity corresponds to a kinematic configuration where two propagators go on-shell. So the sum of the corresponding residues from all such integrands must be equivalent to the product of two {\em on-shell} tree amplitudes, as shown schematically in Figure \ref{GenU}. 
In other words, if the enemy gives us an integrand and claims that it corresponds to the 1-loop amplitude of some (unitary) theory, we can test the claim by checking if the integrand factorizes correctly into products of tree amplitudes. This way, our knowledge of tree amplitudes can be recycled into information about the loop-integrand! 
The operation of taking loop propagators on-shell is called a 
{\bf \em unitarity cut}. It originates from the unitary constraint of the $S$-matrix. To see how, recall that unitarity requires $S^{\dagger}S=1$. Writing $S=1+iT$, where $T$ represents the interacting part of the $S$-matrix, unitarity requires $-i(T-T^{\dagger})=T^{\dagger}T$. If we examine this constraint order by order in perturbation theory, it tells us that the imaginary part of the $T$-matrix at a given order is related to the product of lower-order results. In particular, the imaginary part of the 1-loop amplitude is given by a product of two tree amplitudes. This is illustrated by the diagram
\be
  \vcenter{\hbox{\includegraphics[scale=0.5]{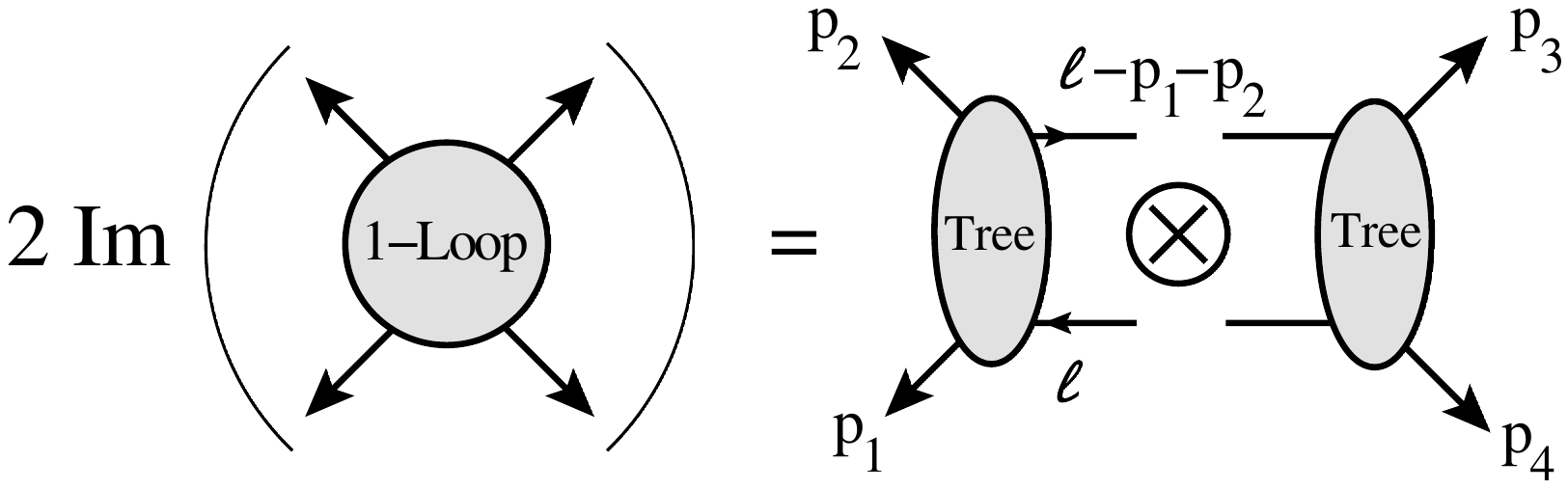}}} \,.
  \label{fig:CuttingRules}
\ee
The product of two tree amplitudes on the RHS  involves a sum over all possible on-shell states that can `cross' the cut. Only states from the physical spectrum of the theory are included in this sum. 
In gauge theory, Feynman diagram calculations of loop-amplitudes require  Feynman diagrams with ghosts in the loops: the purpose of the ghosts is to cancel unphysical modes in the loops. In contrast, in the unitarity cut \reef{fig:CuttingRules} we restrict the loop-momenta to be on-shell and only physical modes are included in the two on-shell amplitudes on the RHS of \reef{fig:CuttingRules}.

The cut rules also include integrals of any remaining freedom in the loop momentum after imposing the cut constraints, such as \reef{LoopConstraint}, and momentum conservation. 
The integral over all allowed kinematic configurations, with respect to the  diagram \reef{fig:CuttingRules}, can be written as
\eq
\int d^D\ell~ \delta_+\big(\ell^2\big)\,\delta_+\big((\ell-p_1-p_2)^2\big)
\,.
\label{ellmeasure}
\eqe
The subscript ${}_+$ means that we are choosing the solution to the on-shell condition that has a positive time component, $\ell^0>0$, i.e.~it is associated with a particle (as opposed to an anti-particle) crossing the cut. 
Note that \reef{ellmeasure} just replaces the two cut propagators with their on-shell conditions, exactly as we did in Figure \ref{GenU}.

The imaginary part of the amplitude probes the branch cut structure, hence the unitarity cut allows us to relate the ``pole structure" of the integrand with the ``branch cut structure" of the loop-integral. 
One can reconstruct the integrand by analyzing different sets of unitarity cuts.
The unitarity cuts can also involve several `cut' lines, i.e.~several internal lines taken on-shell, such that the 1-loop amplitude factorizes into multiple on-shell tree amplitudes. (A higher-loop amplitude would factorize into on-shell lower-loop amplitudes.) An $N$-line cut simply means that $N$ internal lines are taken on-shell. 
Reconstructing the full loop-amplitude  from systematic application of  unitarity cuts is called the  {\bf \em generalized unitarity method} \cite{UnitarityMethod}. It has been applied to a wide range of scattering problems, from next-to-leading order precision QCD predictions to the ultraviolet behavior of perturbative supergravity theories. We discuss its implementation at 1-loop level in the following.

%%%%%%%%%%%%%%%%%%%%%%%%%%%%%%%%%%%%
\subsection{One-loop amplitudes from unitarity}
\label{s:1loopUni}
%%%%%%%%%%%%%%%%%%%%%%%%%%%%%%%%%%%%%
The information of unitarity cuts can be utilized most efficiently if we know, {\em a priori}, a complete basis of integrals that can appear in the scattering amplitudes. As an example, consider a 1-loop amplitude in $D$-dimensions. It can be shown 
 \cite{Scalar1, Scalar2, Scalar3, Scalar4} 
 that all 1-loop amplitudes can be rewritten as a sum of $m$-gon 1-loop scalar integrals $I_m$ for $m=2,3,\cdots,D$:
\be
A^{\rm 1-loop}=\sum_{i}C_D^{(i)}I_{D}^{(i)}
+\sum_{j}C_{D-1}^{(j)}I^{(j)}_{D-1}+\cdots+\sum_{k}C^{(k)}_{2}I^{(k)}_2+\text{rational terms}\,,
\label{IntBasis}
\ee
where $C_m^{(i)}$ are kinematic-dependent coefficients for the $m$-gon scalar integrals $I_m^{(i)}$. {\bf\em Scalar integrals} are the Feynman diagrams that appear in $\phi^n$-theory; the dependence on loop- and external momenta is contained solely in the propagators. 
As an example, a {\bf\em box integral} $I_4^{(i)}$ takes the form
\be
I_4^{(i)} 
~~=\!\!\!\!\!
\raisebox{-15mm}{
\includegraphics[scale=0.45]{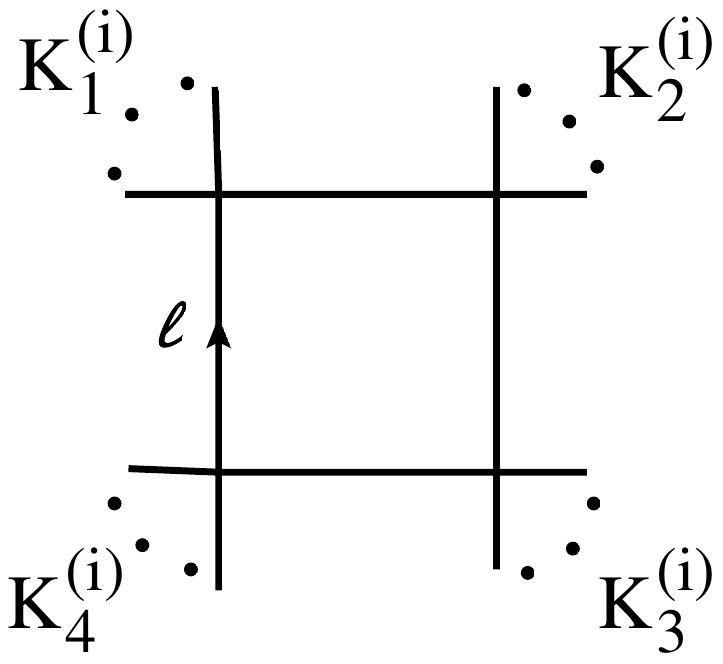}
}
\!\!\!\!\!\!=
\int \frac{d^D \ell}{ (2 \pi)^D}
\,\frac{1}{\ell^2\big(\ell-K^{(i)}_1\big)^2\big(\ell-K^{(i)}_1-K^{(i)}_2\big)^2\big(\ell+K^{(i)}_4\big)^2}\,,
\ee
where $\big(K^{(i)}_1,K^{(i)}_2,K^{(i)}_3,K^{(i)}_4\big)$ are sums of the external momenta at each of the four subamplitudes. The label $i$ indicates a particular choice of distributing the external lines on the four subamplitudes, i.e.~different 4-line cuts.

The origin of the integral basis \reef{IntBasis} is that  one can use the external momenta to form a basis for any vectors in the integrals. Since there are only $D$ independent vectors in $D$-dimensions, the set of needed integrals can be reduced to the set of scalar integrals shown in \reef{IntBasis}. 
A more detailed discussion of ``integral reductions" can be found in  Section 4.2 of \cite{dixon}. 

The expression \reef{IntBasis} makes the task of computing the 1-loop amplitude  a matter of determining the coefficients $C_{m}^{(i)}$. Since the scalar integrals have distinct propagator structures, only a subset contribute to a given unitarity cut. By applying multiple unitarity cuts, one obtains a set of linear equations that relate the  $C_{m}^{(i)}$'s to the results of the cuts. Each unitarity cut is computed as a product of tree amplitudes. Solving these linear equations gives us the coefficients $C_{m}^{(i)}$'s as a combination of products of tree amplitudes. By \reef{IntBasis}, this determines the 1-loop amplitude, up to the possibility of rational terms that we discuss below. 

\begin{figure}
\begin{center}
\includegraphics[scale=0.6]{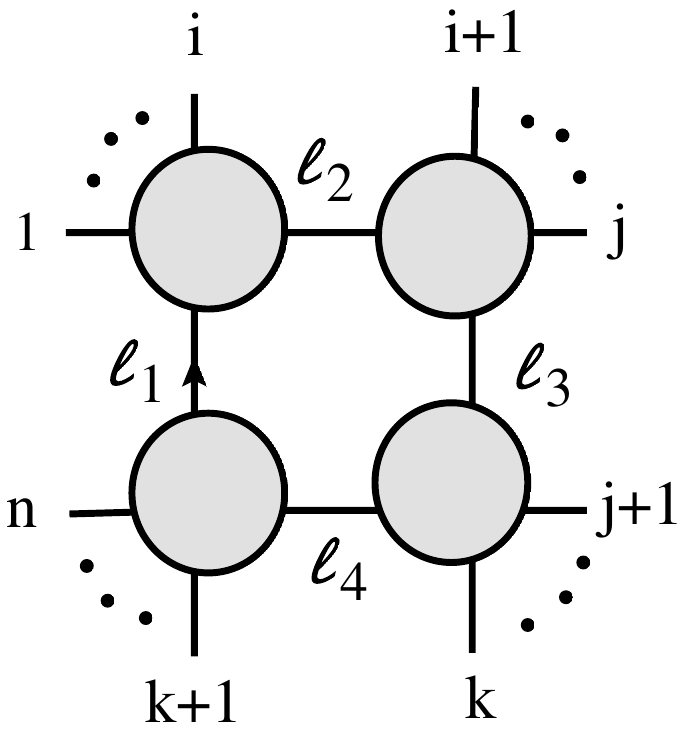}
\caption{\small 1-loop box diagram with $K^{(i)}_1=p_1+\cdots+p_i$, $K^{(i)}_2=p_{i+1}+\cdots+p_{j}$, $K^{(i)}_3=p_{j+1}+\cdots+p_{k}$ and $K^{(i)}_2=p_{k+1}+\cdots+p_{n}$. The corresponding box coefficient $C_4^{(i)}$  in \reef{BoxCoef} is the product of the four tree amplitudes at each corner.}
\label{QuadCut}
\end{center}
\end{figure}

Solving for the $C_{m}^{(i)}$'s can be organized according to the number of propagators present in the scalar integrals. In $D$-dimensions, $\ell$ has $D$ components, so one can find isolated solutions for $\ell$, labelled $\ell^*$, such that all propagators in the scalar integral $I_{D}^{(i)}$ are on-shell: this corresponds to a $D$-line cut. Since the cut constraints are quadratic in loop-momentum, there are two solutions, denoted $\ell^{*(1)}$ and $\ell^{*(2)}$.
The corresponding coefficient $C_D^{(i)}$ is completely determined by the product of $D$ tree amplitudes:
\be
  C_D^{(i)} = \frac{1}{2} \sum_{\ell=\ell^{*(1)},\ell^{*(2)}} 
  A_{n_1}^\text{tree} \cdots A_{n_D}^\text{tree}\,.
  \label{CDexpr}
\ee
Note that one averages over the two solutions, $\ell^{*(1)}$ and $\ell^{*(2)}$. 
At 1-loop, the relative weight between the two solutions can be determined by considering special integrands that integrate to zero. The associated maximal cut must also vanish and this fixes the above prescription. See \cite{HenrikLoopsLegs} for a concise discussion.
\example{Let us make \reef{CDexpr} concrete. 
For an $n$-point 1-loop amplitude in $D=4$, the coefficient of the box integral shown in Figure \ref{QuadCut} is given by
\be
\begin{split}
 C_{4}^\text{(fig\,\ref{QuadCut})}&=\, \frac{1}{2}
\sum_{\ell=\ell^{*(1)},\ell^{*(2)}}
\bigg[~ \sum_\text{states}  A_{n_1}\big[-\ell_1,1,\cdots,i,\ell_2\big]
\times
A_{n_2}\big[-\ell_2,i+1,\cdots,j,\ell_3\big]\\[-3mm]
&
\hspace{3.7cm}
\times A_{n_3}\big[-\ell_3,j+1,\cdots,k,\ell_4\big]
\times A_{n_4}\big[-\ell_4,k+1,\cdots,n,\ell_1\big]\bigg]\,,
\end{split}
\label{BoxCoef}
\ee
where $\sum_\text{states}$ indicates a state sum for each internal line $\ell_1=\ell$, $\ell_2=\ell-(p_1+\ldots+p_i)$, $\ell_3=\ell-(p_1+\ldots+p_j)$, and 
$\ell_4=\ell+(p_{k+1}+\ldots+p_{n})$. The vectors $\ell^{*(1)}$ and $\ell^{*(2)}$  solve the on-shell conditions $\ell_1^2=\ell_2^2=\ell_3^2=\ell_4^2=0$ of the 4-line cut. 
}

Coefficients $C_m^{(i)}$ with $m<D$ are not quite as simple to calculate, but they can be obtained systematically. 
After determining all $D$-gon coefficients, we treat $(D\!-\!1)$-cuts. Both $I_D^{(i)}$ and $I_{D-1}^{(j)}$ integrals can potentially contribute to $(D\!-\!1)$-cuts, but since we have already determined all the $C_D^{(i)}$'s, we can unambiguously determine all $C_{D-1}^{(j)}$'s. Similarly, all integral coefficients
can be determined iteratively. This way, the generalized unitarity method offers a systematic way to determine the 1-loop amplitude in terms of tree-amplitudes. Detailed discussions of extracting 1-loop integral coefficients in $D=4$ can be found in~\cite{OneLoopMethods,ArkaniHamed:2008gz}.  

Of course, there is a big elephant in room  --- you met it already in \reef{IntBasis}:  it is the {\bf \em rational terms}. Rational terms are rational functions that do not possess branch cuts, so they are undetectable by unitarity cuts. The rational terms arise from the need to regularize the loop integrals. 
In dimensional regularization, the loop momentum $\ell$ is really $(D\!-\!2\epsilon)$-dimensional. If we separate the loop momentum into a $D$-dimensional part $\ell^{(D)}$ and an $(-2\epsilon)$-dimensional part $\mu^{-2\epsilon}$, there can also be contributions from the $\mu$-integrals. An example of an integrand that gives a branch-cut-free contribution is the $(D/2+1)$-gon scalar integral with $\mu^2$ numerator: it integrates to a finite value
\eq
\int \frac{d\ell^{(D)}d\mu^{-2\epsilon}}{(2\pi)^{D-2\epsilon}}\; I_{D/2+1}[\mu^2]=-\frac{1}{(4\pi)^{D/2}}
\frac{1}{(D/2)!}
+\mathcal{O}(\epsilon)\,.
\label{rationals}
\eqe 
One cannot capture this from the ordinary unitarity cut since it is just a rational function (here, a constant). The unitarity cut forces the  loop momentum to be on-shell in $D$-dimensions and this implies $\mu^2=0$, so the contribution from the above integrand vanishes. On the other hand, if one considers unitarity cuts where the internal lines become massless in $D-2\epsilon$ dimensions $(\ell^{(D)})^2+\mu^2=0$, or equivalently \emph{massive} in $D$-dimensions with mass $m^2=\mu^2$, such terms \emph{are} detectable. Thus rational terms can be reconstructed from unitarity cuts if we allow the states crossing the cut to be massive. 

Rational terms are \emph{absent} for supersymmetric Yang-Mills theories because supersymmetry cancellations ensure that  the powers of loop momentum in the integrals do not lead to rational terms after integral reduction. For non-supersymmetric theories --- for example $\lambda\phi^4$, QED, QCD --- rational terms are present and they are the most time-consuming ones to compute. We will not discuss this important issue in further detail, but simply refer you to \cite{DDimUnitarity, Rational} and references therein. Instead we  illustrate  the unitarity method by working out an explicit example.
\example{In this example, we calculate the {\bf 4-point 1-loop amplitude in $\mathcal{N}=4$ SYM} using the generalized unitarity method. We will do so by first considering a 2-line cut and then infer from it which terms contribute in the integral basis expansion \reef{IntBasis}. 

The $s$-channel unitarity cut is
\be
\text{Cut}_s~=
\raisebox{-10mm}{\includegraphics[width=4cm]{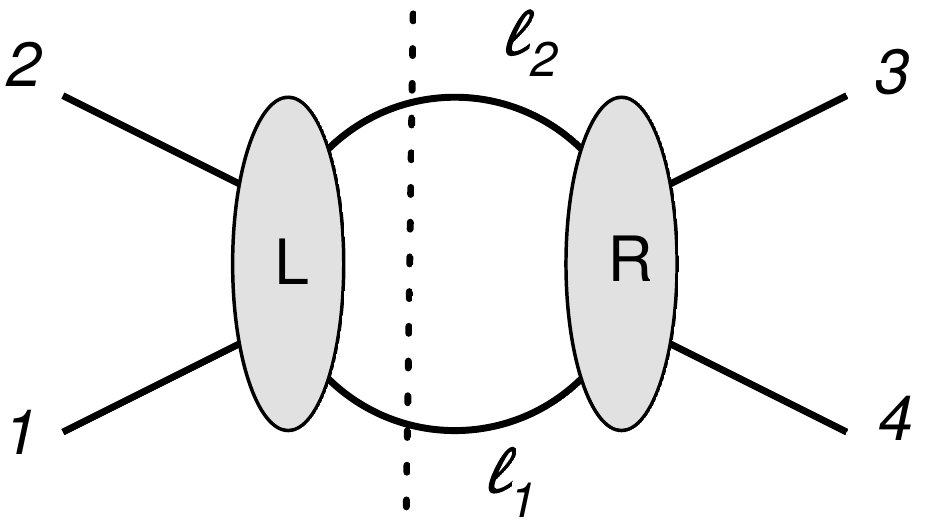}}
=
\sum_\text{states}
\mathcal{A}_4[-\ell_1,1,2,\ell_2]
\times \mathcal{A}_4[-\ell_2,3,4,\ell_1]\,.
\ee
Using the analytic continuation \reef{acont},  the MHV superamplitudes are
\be
\mathcal{A}_4[-\ell_1,1,2,\ell_2]=
\frac{\delta^{(8)}(L)}{\langle \ell_1 1\rangle\langle1 2\rangle\langle2 \ell_2\rangle\langle\ell_2 \ell_1\rangle}\,,
\hspace{7mm}
\mathcal{A}_4[-\ell_2,3,4,\ell_1]=
\frac{\delta^{(8)}(R)}{\langle \ell_2 3\rangle\langle3 4\rangle\langle4 \ell_1\rangle\langle\ell_1 \ell_2\rangle}\,.
\ee
The arguments of the Grassmann delta functions are 
$L=-|\ell_1\rangle\eta_{\ell_1}+|1\rangle\eta_1+|2\rangle\eta_2+|\ell_2\rangle\eta_{\ell_2}$, and 
$R=-|\ell_2\rangle\eta_{\ell_2}+|3\rangle\eta_3+|4\rangle\eta_4+|\ell_1\rangle\eta_{\ell_1}$. As in the case of tree-level recursion (see the discussion around \reef{supersum}), the intermediate state sum is performed as an integration of the on-shell Grassmann variable $\eta_{\ell_i}$ associated with each internal line \cite{Bianchi:2008pu}. 
These integrals are easy to perform when we use $\delta^{(8)}(L) \,\delta^{(8)}(R) = \delta^{(8)}(L+R)\,\delta^{(8)}(R) =\delta^{(8)}(\tQ)\,\delta^{(8)}(R)$. 
We find
\eqa
\nonumber 
\text{Cut}_s
&=&
\frac{\delta^{(8)}\big(\tQ\big)}{\langle12\rangle\langle34\rangle}
\int d^4\eta_{\ell_1}d^4\eta_{\ell_2}~
\frac{\delta^{(8)}(R)}{\langle \ell_1 1\rangle\langle2 \ell_2\rangle\langle\ell_2 \ell_1\rangle\langle \ell_2 3\rangle\langle4 \ell_1\rangle\langle\ell_1 \ell_2\rangle}\\[1mm]
&=&
-\frac{\delta^{(8)}\big(\tQ\big)}{\langle12\rangle\langle34\rangle}\frac{\langle\ell_1\ell_2\rangle^2}{\langle \ell_1 1\rangle\langle2 \ell_2\rangle\langle \ell_23\rangle\langle4\ell_1\rangle}\,.
\label{2ptCut1}
\eqae
On the unitarity cut, one can convert the loop-momentum part of denominator in the above expression into propagators:
\be
\text{Cut}_s \,=\,
\mathcal{A}_4^{\rm tree}[1234]\times
\frac{-su}{(\ell_2+p_2)^2(\ell_1+p_4)^2}\bigg|_{\ell_1^2=\ell_2^2=0}\,.\label{2ptCut2}
\ee
\exercise{}{Show that (\ref{2ptCut1}) is indeed equivalent to \reef{2ptCut2}. 
}

Let us now consider the possible integrals from \reef{IntBasis} that contribute to the $s$-channel cut: the integrals that contain the propagators with $\ell_1^2 = \ell^2$ and $\ell_2^2=(\ell-p_1-p_2)^2$ are the \emph{box-integral} $I_4(p_1,p_2,p_3,p_4)$, and the \emph{triangle-integrals}  $I_3(p_1,p_2,p_{3}+p_{4})$ and $I_3(p_3,p_4,p_{1}+p_2)$, and the 
\emph{bubble-integral} $I_2(p_1+p_2,p_3+p_4)$. In each case, we have indicated  the distribution of the external lines. 
The result \reef{2ptCut2} for $\text{Cut}_s$ shows that there are two uncut propagators left after cutting $\ell_1$ and 
$\ell_2$, so this excludes the triangle- and bubble-integrals. Thus we conclude that only the box integral is present, i.e.~$\mathcal{A}_4^\text{1-loop}[1234]  =
C_{4} \,I_4(p_1,p_2,p_3,p_4)$. The box coefficient $C_{4}$ is readily determined from \reef{2ptCut2}, giving\footnote{The minus sign compared with \reef{2ptCut2} comes from the $(-i)^2$ in the cut propagators.}
\eq
\mathcal{A}_4^\text{1-loop}[1234] 
~=~ 
su \,\mathcal{A}_4^{\rm tree}[1234]\,I_4(p_1,p_2,p_3,p_4)\,
\label{N4Answ}
\eqe 
To make sure that \reef{N4Answ} is the correct result for the amplitude, we  examine other distinct cuts to see if there could be terms that vanish in the $s$-channel cut and were therefore not captured in our analysis. The only other available cut is the $u$-channel cut. (Color-ordering excludes the $t$-channel cut.) But since the RHS of \reef{N4Answ} is invariant under cyclic permutation, it is guaranteed to produce the correct $u$-channel cut. Hence \reef{N4Answ} is indeed the correct 1-loop 4-point amplitude for 
$\mathcal{N}=4$ SYM.  
We discuss the evaluation of $I_4$ in dimensional regularization in Section \ref{s:loopN4}.
  }

Working through the details in the example above, you will notice that the unitary method does take some work and you may wonder how it compares with a brute-force 1-loop Feynman diagram calculation. The answer is that the unitarity method is superior, since it heavily reduces the number of diagrams needed 
and it avoids gauge obscurities. For the unitarity method, the input is gauge-invariant on-shell amplitudes. You might find it curious that the first computation of the 1-loop 4-gluon amplitude $A_4^\text{1-loop}$ in $\cn=4$ SYM was not done  in QFT, but in string theory: in 1982, Green, Schwarz, and Brink \cite{Green:1982sw} obtained $A_4^\text{1-loop}$ as the low-energy limit of the superstring scattering amplitude for four gluon states. 

We close this section with some general comments on the 1-loop integral expansion \reef{IntBasis}. The representation of 1-loop amplitudes in terms of scalar integrals provides an interesting categorization scheme in terms of whether or not particular classes of integrals --- in 4d: boxes, triangles, bubbles, and rationals ---  appear or not. For example, we mentioned earlier that rational terms are absent in $\cn>0$ super Yang-Mills theory. 
One can ask which theories involve only box-integrals, i.e.~no triangle- or bubble-integrals and no rational terms. In 4d, such ``no-triangle" theories include $\mathcal{N}=4$ SYM \cite{UnitarityMethod}, $\mathcal{N}=8$ supergravity \cite{NoTria2,Bern:2007xj}, and $\mathcal{N}=2$ SYM coupled to specific tensor matter fields \cite{NoTria3}. For pure $\mathcal{N}=6$ supergravity, only box and triangle integrals appear~\cite{NoTria4}, while for pure $\mathcal{N}\leq4$ supergravity all integrals in \reef{IntBasis}, including rational terms, appear. 

One relevant aspect of the above analysis is that in 4d only the bubble integrals $I_2^{(i)}$ contain ultraviolet divergences. In dimensional regularization, all bubble integrals have a common leading $1/\eps$-term, and hence contribute $\tfrac{1}{\eps}\sum_{i}C^{(i)}_2$ to the amplitude. As a result, the beta function for a given theory vanishes at 1-loop order precisely when $\sum_{i}C^{(i)}_2=0$. In fact, in a renormalizable theory, one must have $\sum_{i}C^{(i)}_2\sim A^{\rm tree}_n$ and the proportionality constant is related to the 1-loop beta function \cite{ArkaniHamed:2008gz,NoTria3,Elvang:2011fx,Huang:2012aq}. Note that even though bubble coefficients are non-trivial for pure $\mathcal{N}\leq4$ supergravity theories, their sum $\sum_{i}C^{(i)}_2$ must vanish since the theory is known to be free of ultraviolet divergences at 1-loop order.

 Next, we offer  a quick survey of results for  loop amplitudes in planar $\cn=4$ SYM.

%%%%%%%%%%%%%%%%%%%%%%%%%%%%%%%%%%%%
\subsection{1-loop amplitudes in planar $\cn=4$ SYM}
\label{s:loopN4}
%%%%%%%%%%%%%%%%%%%%%%%%%%%%%%%%%%%%%
When we introduced the $\cn=4$ SYM theory in Section \ref{s:N4sym}, we mentioned that this is a conformal theory, there is no running of the coupling. This means all ultraviolet (UV) divergences cancel in the on-shell scattering amplitudes, order by order in perturbation theory.  

The loop-amplitudes in $\cn=4$ SYM  theory do have infrared (IR) divergences, though, as is typical in a theory with massless states. The IR divergences are well-understood: in dimensional regularization, $D=4-2\eps$, 
the $L$-loop $\cn=4$ SYM amplitude behaves as $1/\eps^{2L}$ for small $\eps$. Each $1/\eps$ can be understood as a loop-momentum going {\em collinear} with an external momentum or becoming {\em soft}. When this happens simultaneously for each of the $L$ loop-momenta, one gets the leading behavior, $1/\eps^{2L}$. Soft and collinear limits for amplitudes in massless gauge theories have been studied since the late 1970s and is an entire subject on its own; we refer you to the very brief outline in \cite{Roiban:2010kk} and references therein. Here, we focus on recent work on loop amplitudes in planar $\cn=4$ SYM. 

In the example in Section \ref{s:1loopUni}, we used the unitarity method to construct the 1-loop 4-point superamplitude in planar $\cn=4$ SYM. We found (see \reef{N4Answ}) that it could be written in terms of a single scalar box integral:
\eq
\mathcal{A}_4^\text{1-loop}[1234] 
~=~ 
su \,\mathcal{A}_4^{\rm tree}[1234]\,I_4(p_1,p_2,p_3,p_4)\,.
\label{N4AnswAgain}
\eqe 
Evaluating the scalar box integral $I_4$ in dimensional regularization  $D=4-2\eps$, one finds 
\be
  \mathcal{A}_4^\text{1-loop}[1234]  
  ~=~ 
  \mathcal{A}_4^{\rm tree}[1234]\, 
  \bigg\{ \frac{2}{\eps^2} 
  \Big[ 
  \big(-\m^{-2} y_{13}^2 \big)^{-\eps} 
  + \big(-\m^{-2} y_{24}^2 \big)^{-\eps} 
  \Big]
  -
   \ln^2\Big( \frac{y_{13}^2}{y_{24}^2} \Big) 
  - \pi^2
  + O(\eps)
 \bigg\}\,,
 \label{4pt1loopA}
\ee
where $\mu$ is the regularization scale and $y_{ij} = y_i-y_j$ are the zone-variables defined in \reef{DualDef}. In terms of Mandalstam variables, we have $s = - y_{13}^2$ and $u = -y_{24}^2$. 

Let us now use the notation
\be
  \ca_{n;L}^\text{N$^K$MHV} ~~=~~
  \text{$n$-point $L$-loop N$^K$MHV superamplitude of planar $\cn=4$ SYM}\,,
\ee 
with the color-ordering $12\dots n$ of external particles implicit. 
Since supersymmetry and $SU(4)$ R-symmetry\footnote{The $SU(4)$ R-symmetry is non-anomalous \cite{Marcus:1985yy,diVecchia:1984jh}.} Ward identities hold at each loop-order, this decomposition of the loop-amplitude is sensible and $\ca_{n;L}^\text{N$^K$MHV}$ has Grassmann degree $4(K+2)$.

It is convenient to factor out an MHV tree-level superamplitude and write the loop-expansion as
\be
  \ca_{n;L}^\text{N$^K$MHV}(\eps) 
  ~=~
  \ca_{n;0}^\text{MHV} 
  \Big(
     \mathcal{P}_{n;0}^\text{N$^K$MHV} 
     + \lambda\,  \mathcal{P}_{n;1}^\text{N$^K$MHV}(\eps) 
     + \dots
  \Big)\,,
  \label{stripMHV}
\ee 
where $\lambda \sim g^2 N$ is the t'Hooft coupling written in terms of the gauge coupling $g$ and the rank of the gauge group $SU(N)$.
We include the dependence on the $\eps$-regulator explicitly in the loop  amplitudes. 
At tree-level, we have
\be
\mathcal{P}_{n;0}^\text{MHV} = 1 
~~~~\text{and}~~~~ 
\mathcal{P}_{n;0}^\text{NMHV} 
=
\sum_{j=3}^{n-2} \sum_{k=j+2}^n R_{1jk}\,.
\ee
The NMHV result is given in terms of the dual superconformal invariants $R_{1jk}$ defined in \reef{Rinv1} and discussed further in Section \ref{s:momtwist}.
\exercise{}{Why is it possible to factor out $\ca_{n;0}^\text{MHV}$ even at loop-level?}
The $\eps$-regulator explicitly breaks the conformal and dual conformal symmetry. You can see that explicitly in the expression \reef{4pt1loopA} for the 1-loop 4-point superamplitude: not even the finite part $O(\eps^0)$ respects dual conformal inversion \reef{InvertRule}:
\be
I(y_{ij}^2) = \frac{y_{ij}^2}{y_i^2 y_j^2}\,. 
\label{dualInvSQy}
\ee
So the raw output of the loop-amplitudes does not entertain the ordinary or dual conformal symmetries of the $\cn=4$ SYM theory. However, the IR divergences take a universal form that  facilitate construction of IR-finite quantities that turn out to respect the symmetries. This will be discussed below. Let us begin at 1-loop with the structure of the IR divergences. 

At 1-loop order, the IR divergent part of $\mathcal{A}_{n;1}^\text{N$^K$MHV}$
 is captured entirely by the MHV superamplitude in the sense that 
\be
  \mathcal{A}_{n;1}^\text{N$^K$MHV}(\eps)
  =   \mathcal{A}_{n;0}^\text{N$^K$MHV}
      \times \text{IRdiv}\big[ \mathcal{P}_{n;1}^\text{MHV}(\eps) \big]
      + O(\eps^0)\,,
   \label{IR1Luniv}
\ee
where 
\be
 \text{IRdiv}\big[ \mathcal{P}_{n;1}^\text{MHV}(\eps) \big]
 ~=~
 \frac{1}{\eps^2} \sum_{i=1}^n (- \mu^{-2} \,y_{i,i+2}^2 )^{-\eps} \,.
\ee
Note that for $n=4$, this reproduces the IR divergent terms in \reef{4pt1loopA}. The \emph{finite} part of the 4-point MHV superamplitude is 
\be
  \mathcal{F}_{4;1}^\text{MHV}(\eps)
  ~\equiv~
  \mathcal{P}_{4;1}^\text{MHV}(\eps)
  -\text{IRdiv}\big[ \mathcal{P}_{4;1}^\text{MHV}(\eps) \big]
  ~=~ 
  - \ln^2\Big( \frac{y_{13}^2}{y_{24}^2} \Big) 
  - \pi^2
  + O(\eps)\,.
  \label{F41}
\ee
The universal form \reef{IR1Luniv} of the 1-loop IR divergences implies that the {\bf \em ratio functions}\footnote{The RHS of \reef{ratiofct} can be viewed as the $O(\lambda)$ term in the small $\lambda$ expansion of the ratio $\ca_{n}^\text{N$^K$MHV}/\ca_{n}^\text{MHV}$.}
\be
  \mathcal{R}_{n;1}^\text{N$^K$MHV}(\eps)
  ~\equiv~
  \mathcal{P}_{n;1}^\text{N$^K$MHV}(\eps)
  - \mathcal{P}_{n;0}^\text{N$^K$MHV}\, 
     \mathcal{P}_{n;1}^\text{MHV}(\eps)\,
  \label{ratiofct}
\ee
are IR finite. 
Moreover, it has been proposed  \cite{Drummond:2008vq,Drummond:2008bq}  that $\mathcal{R}_{n;1}^\text{N$^K$MHV}$(0)'s are actually dual conformal invariant. This was shown at NMHV level for $n\le 9$ in \cite{Drummond:2008vq,Drummond:2008bq} and for general $n$ in \cite{Brandhuber:2009xz,Elvang:2009ya} using generalized unitarity. 
 To give you a sense of the expressions, 
 we present the result \cite{Drummond:2008bq} for the ratio function for the 6-point 1-loop NMHV superamplitude. It is 
\be
  \label{ratio6pt}
   \mathcal{R}_{6;1}^\text{NMHV}(0) 
   =  \frac{1}{2} \Big( R_{146} \,V_{146}+ \text{cyclic} \Big)\,,
\ee 
where 
\be
  V_{146} = 
   - \ln u_1 \, \ln u_2 
   + \frac{1}{2} \sum_{k=1}^3 
         \Big[ \ln u_k \, \ln u_{k+1}  + \text{Li}_2 (1-u_k)\Big]
    - \frac{\pi^2}{6}\,.
    \label{V146}
\ee
The $u_i$'s are dual conformal cross-ratios,
\be
   u_1 = \frac{y_{13}^2 y_{46}^2}{y_{14}^2 y_{36}^2}\,,
   ~~~~
   u_2 = \frac{y_{24}^2 y_{51}^2}{y_{25}^2 y_{41}^2}\,,
   ~~~~
   u_3 = \frac{y_{35}^2 y_{62}^2}{y_{36}^2 y_{52}^2}\,,
   \label{u123}
\ee
so each $V_{ijk}$ is a dual conformal invariant, as you can see by applying dual inversion \reef{dualInvSQy}. The ``+ cyclic'' in \reef{ratio6pt} is the instruction to sum over the cyclic sum of  the external state labels; note that $V_{251}$ is just $V_{146}$ with $u_1 \to u_2 \to u_3 \to u_1$. 

The dilogarithm Li$_2$ in \reef{V146} is the $q\!=\!2$ case of the polylogarithm Li$_q$. Starting with the familiar logarithm $\text{Li}_{1}(x) = - \ln(1-x)$, the polylogarithms are    defined iteratively as 
\be
  \text{Li}_{q}(x) = \int_0^x dt \, \frac{\text{Li}_{q-1}(t)}{t} \,.
  \label{polylog}
\ee
Recalling that the BCFW recursion relations for the tree-level 6-point NMHV superamplitude only has three terms, you might be surprised to see six terms in the 1-loop result \reef{ratio6pt}. However,  in Exercise \ref{ex:preloop}  we used the cyclically invariant 5-brackets $[i,j-1,j,k-1,k] = R_{ijk}$ to rewrite the  tree-level  superamplitude as  
$\ca_{6;0}^\text{NMHV}/\ca_{6;0}^\text{MHV}
=\mathcal{P}_{6;0}^\text{NMHV} = \frac{1}{2} \big( R_{146} + \text{cyclic} \big)$. This was done in anticipation of the 1-loop ratio function \reef{ratio6pt}, and now you see that \reef{ratio6pt} is just like the tree-level result but with  each $R_{ijk}$ dressed with a dual conformal invariant $V_{ijk}$. Adding loop-orders 0 and 1, we can therefore write 6-point ratio function
\be
  \label{ratio6pt2}
   \mathcal{R}_{6}^\text{NMHV}(0) 
   =  \frac{1}{2} \Big( R_{146} \big( 1+ \lambda \, V_{146}  \big) + \text{cyclic} \Big)
     + O(\lambda^2)\,.
\ee 

There are two properties worth noting about the 1-loop ratio function $\mathcal{R}_{6;1}^\text{NMHV}(0)$:
\begin{itemize}
   \item It is dual conformal invariant, but not dual \emph{super}conformal invariant. For a discussion of this, see \cite{Korchemsky:2009hm}.
   \item $V_{146}$ --- and hence $\mathcal{R}_{6;1}^\text{NMHV}(0)$ ---
   has uniform transcendentality 2. This can be extended to the $\eps$-dependent terms if $\eps$ is assigned transcendentality $-1$.
\end{itemize}
Both of these properties carry over to all $\mathcal{R}_{n;1}^\text{NMHV}(0)$. At higher-loop order in \emph{planar} $\cn=4$ SYM, the degree of transcendentality is expected to be uniformly $2L$. 

At higher-point, there are more dual conformal invariant cross-ratios available than just the three $u_i$'s for $n=6$. Consequently, the NMHV 1-loop ratio functions $\mathcal{R}_{n;1}^\text{NMHV}(0)$ are more involved; however, they are all known explicitly and they take a similar form as \reef{ratio6pt}. You can find the results for $\mathcal{R}_{n;1}^\text{NMHV}(0)$ in \cite{Elvang:2009ya}.

%%%%%%%%%%%%%%%%%%%%%%%%%%%%%%%%%%%%
\subsection{Higher-loop amplitudes in planar $\cn=4$ SYM}
\label{s:multiloop}
%%%%%%%%%%%%%%%%%%%%%%%%%%%%%%%%%%%%%
The generalized unitarity method can be applied successfully to higher-loop amplitudes, both at the planar and non-planar level; for a recent review see \cite{Review1}. The application of unitarity is most efficient when a complete integral basis is available; for 1-loop amplitudes in 4d, the basis consists of  the scalar box-, triangle-, and bubble-integrals in \reef{IntBasis}. 

Beyond 1-loop, there is not a complete understanding of the basis integrals for amplitudes in generic quantum field theories, although partial results have been achieved at 2-loops in the planar limit, see \cite{2LoopBasis,HenrikLoopsLegs,2LoopBasisCut} and \cite{Badger:2012dp,Zhang:2012ce,Sogaard:2013yga}. One thing worth noting is that the integral basis is finite \cite{Smirnov:2010hn}. 

\begin{figure}
\begin{center}
\begin{eqnarray*}
I_4^\text{1-loop} \!\!&=&\! y_{13}^2\, y_{24}^2~ \times ~
\raisebox{-0.9cm}{\includegraphics[scale=0.45]{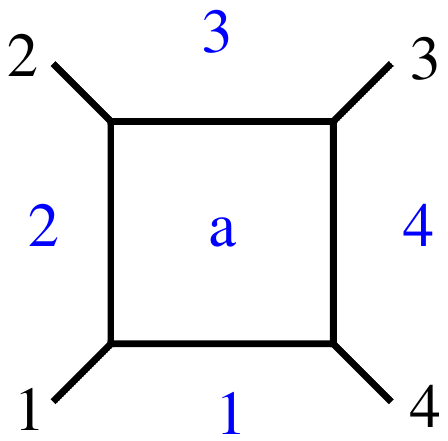}}\,,
~~~~~~
I_4^\text{2-loop}
~=~(y_{13}^2)^2\, y_{24}^2~ \times ~
\raisebox{-0.85cm}{\includegraphics[scale=0.45]{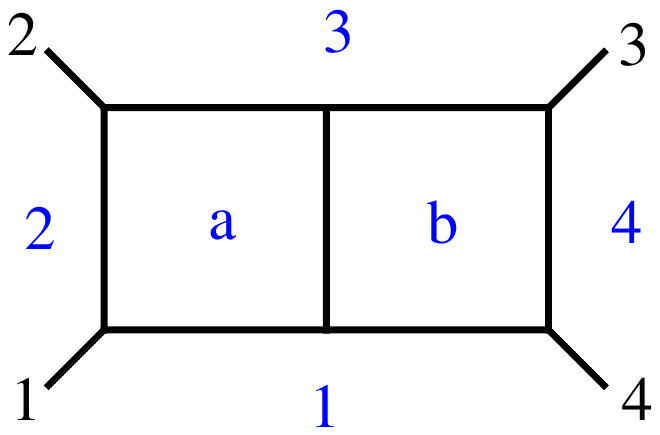}}
~+~\text{cyclic}\,,
\\[2mm]
I_4^\text{3-loop} \!\!&=&\! 
(y_{13}^2)^3\, y_{24}^2\, \times 
\raisebox{-0.85cm}{\includegraphics[scale=0.45]{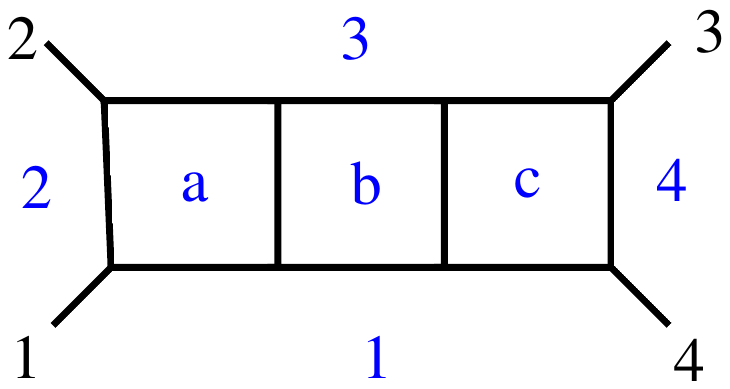}}
~+~
(y_{13}^2)^2\, y_{24}^2 \,y_{a4}^2\,\times
\raisebox{-1.12cm}{\includegraphics[scale=0.45]{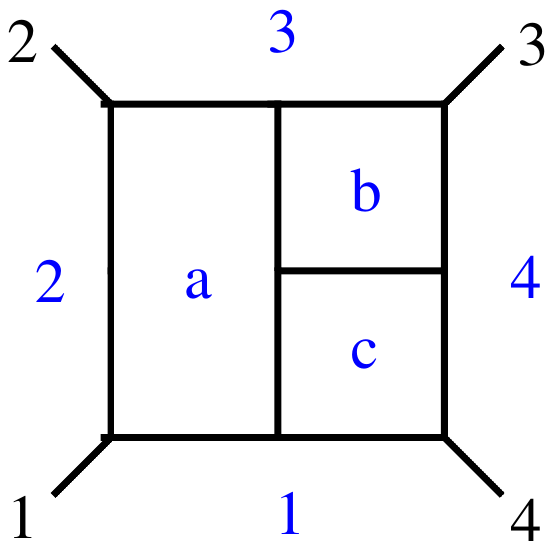}}
~+~\text{cyclic}\,.\\[-7mm]
\end{eqnarray*}
\caption{\small The integrands of $\mathcal{N}=4$ SYM 4-point amplitude to 3-loop order. These are the unique scalar integrands that are dual conformal invariant.}
\label{4ptintegrands}
\end{center}
\end{figure}

Without a given basis of integrals, one strategy is to construct the most general integral Ansatz that satisfies certain criteria, such as dimension-counting, and then use various integral identities to recast the Ansatz into a basis of independent integrals. Further symmetries, such as dual conformal invariance in planar $\mathcal{N}=4$ SYM,  can be a strong handle on finding a complete integral basis. As an example, the diagrams in Figure \ref{4ptintegrands} correspond to the only dual conformal invariant scalar integrals for the 
 4-point 1-, 2- and 3-loop integrands of planar $\mathcal{N}=4$ SYM. 
 The coefficients of each integral is fixed by applying unitarity cuts \cite{N42Loop,N43Loop}, so that the LHS of the equations in Figure \ref{4ptintegrands} are the full integrands for the  4-point 1-, 2- and 3-loop amplitudes in  planar $\mathcal{N}=4$ SYM. 
 The evaluation of these integrals leads to interesting results that we discuss next.

The analytical result \cite{N=42Loop1} for the {\bf 2-loop 4-point amplitude} in planar $\cn=4$ SYM was shown by 
Anastasiou, Bern, Dixon and Kosower (ABDK) \cite{N42Loop} to be expressible  in terms of the 1-loop amplitude as
\be
   \mathcal{P}_{4;2}^\text{MHV}(\eps)
   ~=~
   \frac{1}{2}
   \Big[\mathcal{P}_{4;1}^\text{MHV}(\eps)\Big]^2
   +\mathcal{P}_{4;1}^\text{MHV}(2\eps) \, f^{(2)}(\eps)
   + C^{(2)} + O(\eps) \,,
   \label{P42}
\ee
where the MHV factor is stripped off as in \reef{stripMHV}, 
 $f^{(2)}(\eps)=-\zeta_2 - \zeta_3\, \eps - \zeta_4 \,\eps^2$ 
and $C^{(2)} =-\zeta_2^2/2$. Here $\zeta_s = \sum_{k=1}^\infty k^{-s}$ is the Riemann zeta function; note
$\zeta_2=\tfrac{\pi^2}{6}$, $\zeta_3 \approx 1.202$, and $\zeta_4 = \tfrac{\pi^4}{90}$.

It is interesting that  the 2-loop 4-point amplitude in planar $\cn=4$ SYM can be written in terms of the 1-loop result. But at 3-loops, the plot thickens! 
By explicit calculation of the {\bf 3-loop 4-point amplitude} in planar $\cn=4$ SYM, 
Bern, Dixon, and Smirnov (BDS) \cite{N43Loop} found that the iterative structure continues:
\be
  \mathcal{P}_{4;3}^\text{MHV}(\eps)
  ~=~ 
  - \frac{1}{3} \Big[ \mathcal{P}_{4;1}^\text{MHV}(\eps) \Big]^3
  + \mathcal{P}_{4;1}^\text{MHV}(\eps)  \, \mathcal{P}_{4;2}^\text{MHV}(\eps)
  +f^{(3)}(\eps)\, \mathcal{P}_{4;1}^\text{MHV}(3\eps) 
  + C^{(3)} + O(\eps)
  \,.
  \label{P43}
\ee
Here $f^{(3)}(\eps) = \tfrac{11}{2} \zeta_4 + O(\eps)$ and $ C^{(3)}$ is a constant.
 
The 2- and 3-loop results indicate an exponentiation structure. This motivates the 
{\bf \em ABDK/BDS Ansatz for the full MHV superamplitude in $\cn=4$ SYM}:
\be
  \mathcal{P}_{n}^\text{MHV(BDS)}(\eps)
  ~=~
  \exp\bigg[
    \sum_{L=1}^\infty  \lambda^L  
    \, \Big( f^{(L)}(\eps) \, 
     \mathcal{P}_{n;1}^\text{MHV}(L \eps)
     + C^{(L)}  + O(\eps) \Big)
  \bigg]\,.
  \label{BDSansatz}
\ee
This Ansatz is almost correct: keep reading! 
In the ABDK/BDS Ansatz, the functions $ g^{(L)}$ are of the form 
$f^{(L)}(\eps) = f_0^{(L)}+\eps\, f_1^{(L)} + \eps^2 f_2^{(L)}$, and the constants $C^{(L)}$ and $g_{0,1,2}^{(L)}(\eps)$ are independent of the number of external legs $n$. In particular,  at 1-loop order $f^{(1)}(\eps) =1$ and $C^{(1)}=0$, and at 2-loops the results for $f^{(2)}(\eps)$ and $C^{(2)}$ were given below \reef{P42}.
\exercise{}{Show that \reef{BDSansatz} reproduces the 4-point 2- and 3-loop expressions \reef{P42}  and \reef{P43}.}
Of course, the way one would go about testing the ABDK/BDS exponentiation Ansatz \reef{BDSansatz} is by direct calculation of the $n$-point $L$-loop amplitudes at $L=2,3,\dots$. 
But how many 2-loop amplitudes have you ever calculated? Yeah, it is not an easy task, nonetheless  progress has been made. It has been shown numerically in \cite{Bern:2006vw,Cachazo:2006tj} that the exponentiation Ansatz correctly produces the {\bf 5-point 2-loop amplitude}.
It is very interesting that something new happens at 6- and higher-point: while {\em the ABDK/BDS Ansatz matches the IR divergent structure, it does not fully produce the correct finite part. The ABDK/BDS Ansatz determines the finite part of the amplitude only up to a function of dual conformal cross-ratios of the external momenta.} This function is called the {\bf \em remainder function} and it is defined as
\be
  \mathbf{r}_{n;L}(\eps) \,\equiv\,
  \mathcal{P}_{n;L}^\text{MHV}(\eps)-\mathcal{P}_{n;L}^\text{MHV(BDS)}(\eps) \,,
\ee
where $\mathcal{P}_{n;L}^\text{MHV}(\eps)$ is the actual MHV $L$-loop amplitude and
$\mathcal{P}_{n;L}^\text{MHV(BDS)}(\eps)$ is the $O(\lambda^L)$ terms in the expansion of the exponential Ansatz \reef{BDSansatz}.
The remainder function does not show up for $n\!=\!4,5$ because in those cases there are no available conformal cross-ratios.

The first indication of the remainder function came from a strong coupling calculation by Alday and Maldacana \cite{Alday:2007he} who proposed \cite{Alday:2007hr} to use the AdS/CFT correspondence to calculate $\mathcal{P}_{n}^\text{MHV}$. Subsequently, it was verified numerically that a remainder function is needed for the parity-even part of the {\bf 6-point 2-loop MHV amplitude} \cite{Bern:2008ap}, whereas ABDK/BDS successfully determines the parity-odd part \cite{Cachazo:2008hp}. 
The analytic form of the remainder function $\mathbf{r}_{6;2}$ for the 6-point 2-loop MHV amplitude was calculated (as a hexagonal Wilson-loop) by Del Duca, Duhr, and Smirnov 
\cite{DelDuca:2009au,DelDuca:2010zg}. The result, written in terms of the three dual conformal cross-ratios $u_{1,2,3}$ in \reef{u123}, is a respect-inducing 17-page long sum of generalized polylogarithms; all terms have transcendentality 4. 
In an impressive application of a mathematical tool known as  {\bf \em the Symbol}, Goncharov, Spradlin, Vergu, and Volovich \cite{Goncharov:2010jf} managed to simplify this complicated result for $\mathbf{r}_{6;2}$ to an expression that involves only regular polylogs --- $\text{Li}_s$ and $\ln$ --- and fits in just a few lines of \LaTeX. 

The simple answer \cite{Goncharov:2010jf}  for $\mathbf{r}_{6;2}$ is an important step towards a better understanding  of  loop-amplitudes in planer $\cn=4$ SYM. The Symbol is now being used to understand higher-loop amplitudes, 
however, there will be amplitudes in planer $\cn=4$ SYM involving integrals that the Symbol does not help with. 
Thus techniques are eventually needed beyond the Symbol. 

\vspace{3mm}
We have reviewed the unitarity method and shown you how it allows us to construct $L$-loop amplitudes from {\em on-shell} lower-loop input. While the approach explores the analytic structure of the loop-integrands, it is somewhat different from the recursive techniques you know from tree-level amplitudes. BCFW is available at the level of loop-integrands, and that is the subject of the next section.

%%%%%%%%%%%%%%%%%%%%%%%%%%%%%%% 
%%%%%%%%%%%%%%%%%%%%%%%%%%%%%%% 
%%%%%%%%%%%%%%%%%%%%%%%%%%%%%%% 
\newpage
\setcounter{equation}{0}
\section{Loops II: BCFW recursion} 
\label{s:loops2}
%%%%%%%%%%%%%%%%%%%%%%%%%%%%%%% 
%%%%%%%%%%%%%%%%%%%%%%%%%%%%%%% 
%%%%%%%%%%%%%%%%%%%%%%%%%%%%%%% 

It is a curious aspect of our discussion of unitarity cuts in the previous section that we have always cut at least two propagators. This contrasts the tree-level recursion relations where the amplitude is constructed from the 
factorization-structure of a single propagator going on-shell.  It is  tempting to ask if loop amplitudes can be reconstructed from the singularities associated with taking  a single propagator on-shell? The answer leads to a recursive approach to constructing loop-integrands.

%%%%%%%%%%%%%%%%%%%%%%%%%%%%%%%%%%%%
\subsection{Loop-integrands}
\label{s:bcfwloop}
%%%%%%%%%%%%%%%%%%%%%%%%%%%%%%%%%%%%%
As we have discussed previously, the loop amplitudes have complicated analytic structure, so we focus on the \emph{loop integrand} which is just a rational function with poles at the location of the propagators, much similar to the tree amplitudes. Suppose we do a BCFW-shift on the external legs, for example  $p_1^\m\rightarrow p_1^\m+z\,q^\m$ and 
$p_n^\m\rightarrow p_n^\m-z\,q^\m$, with $q^2=0$ as usually. We can deduce from the Feynman diagrams that the shifted loop integrand possess two types of poles in $z$: (1) poles in loop-independent propagators and (2) poles in propagators involving loop-momentum. The residue of a type 1 pole corresponds to  factorization of the integrand into a product of two lower-loop integrands. The residue of a type 2 pole in an $L$-loop $n$-point integrand is an $(n\!+\!2)$-point $(L\!-\!1)$-integrand with two adjacent legs evaluated in the {\bf \em forward limit} 
\be
  p_i^\m=r^\m\,,
  \hspace{7mm}
  p_{i+1}^\m=-r^\m\,,
  ~~~
  \text{with}~~~r^2=0\,.
  \label{fwlimit}
\ee  
   This is illustrated for the example of a 4-point 3-loop amplitude in Figure \ref{BCFWInt}.  
The poles of type 2 are precisely what we would call single-line cuts in the unitarity method~\cite{SingleCut1,SingleCut2}. 

Thus --- provided that the large-$z$ behavior is well-understood --- it appears that one can straightforwardly set up a recursion relation for loop integrands. However, there are subtleties we have to resolve:
\begin{itemize}
\item  The first issue has to do with the identification of the loop-momenta in the loop-integrand. In the amplitude, we have to integrate the loop-momenta, so $\ell_i$ are just dummy variables that can be redefined while still giving the same integrated answer. 
But the integrand itself can have different pole structures depending on how the $\ell_i$ are identified. As an example, consider the 1-loop 4-point box-integral and compare the equivalent parameterizations $I_4^{(a)}$ and $I_4^{(b)} = I_4^{(a)}(\ell \to \ell + p_1)$. 
 BCFW-shifting legs $1$ and $2$ yields two distinct analytic functions in $z$: 
\be
\begin{split}
 I^{(a)}_4(\hat{1},\hat{2},3,4)~=~&\frac{1}{\ell^2(\ell-p_1-zq)^2(\ell-p_1-p_2)^2(\ell+p_4)^2}\\
I^{(b)}_4(\hat{1},\hat{2},3,4)~=~&\frac{1}{(\ell+p_1+zq)^2\ell^2(\ell-p_2+zq)^2(\ell-p_2-p_3+zq)^2}\,.
\end{split}
\label{IaIb}
\ee     
In general there is no canonical way to identify how the loop momentum is parameterized, so that is the first subtlety that needs to be resolved. It basically comes down to the definition of what we mean by \emph{the} `un-integrated integrand'.
\begin{figure}
\begin{center}
\includegraphics[scale=0.75]{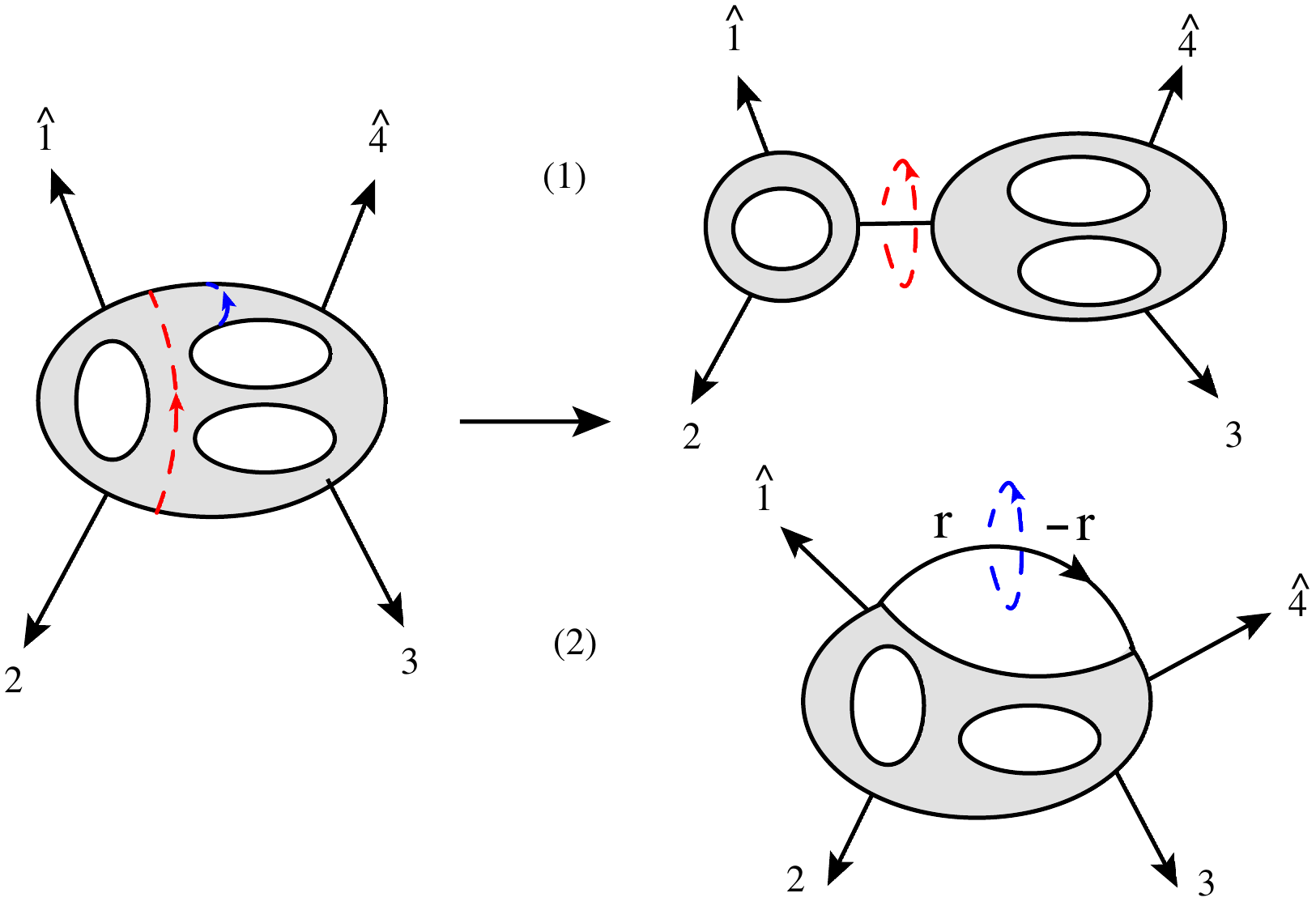}
\caption{\small Schematic representation of the two types of poles occurring in a BCFW-shift of a 4-point 3-loop integrand. Type 1 are poles associated with loop-momentum \emph{independent} propagators and type 2 poles are loop-momentum \emph{dependent} propagators. The former factorizes into a product of a 2-loop and a 1-loop integrand while the latter gives a forward limit of a 2-loop amplitude with two extra legs.
}
\label{BCFWInt}
\end{center}
\end{figure}
%

%%%%%%%%%%

\item
The second subtlety has to do with the forward limit. When the loop-momentum dependent propagators go on-shell, there is a residue corresponding to a lower-loop $(n\!+\!2)$-point integrand in the forward limit, but that limit suffers from singularities. For example, from the explicit Feynman diagrams one sees that if the forward legs are attached to the same external line, then  due to momentum conservation there is a $1/p^2$ singularity as $p^2 \to 0$.  Such diagrams can be identified with cuts of bubbles on external legs or tadpole diagrams. This is illustrated in Figure \ref{ForwardTrouble}. In massless theories, these integrate to zero in dimensional regularization and do not contribute to the loop amplitude. However, prior to integration, they are part of the integrand and will contribute to the single cuts. Therefore an important, but difficult, task is to identify these contributions in the forward limit such that one can consistently remove them.
\begin{figure}
\begin{center}
\includegraphics[scale=0.7]{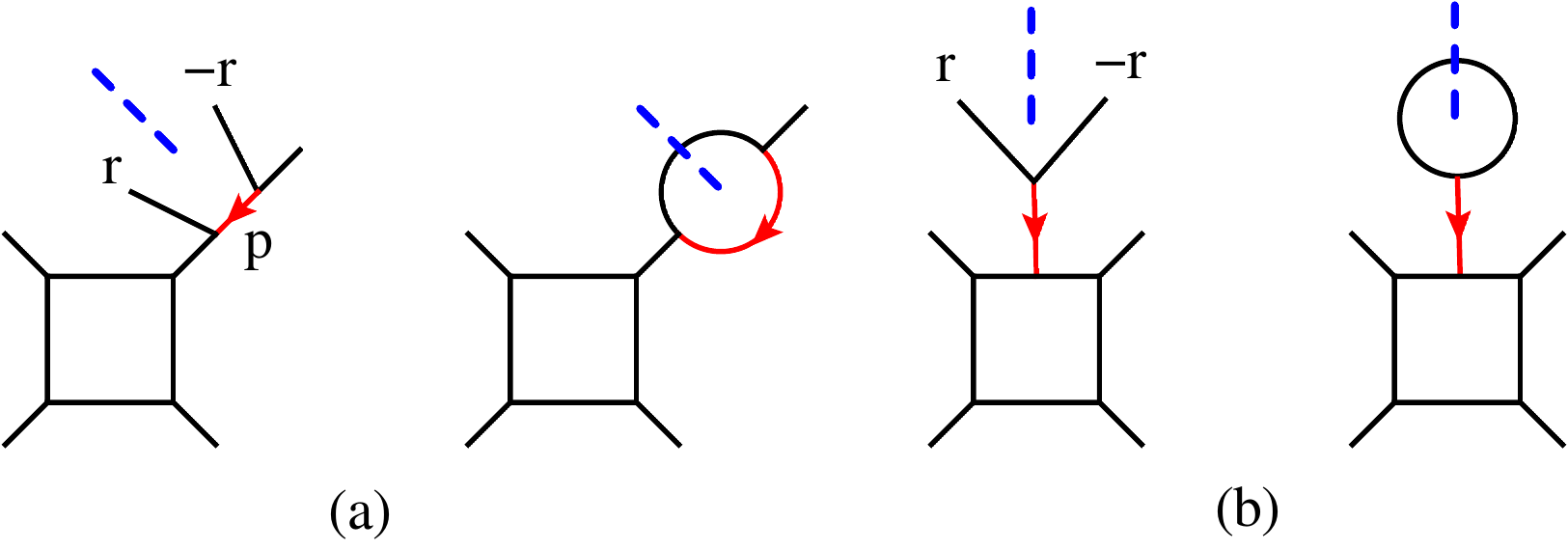}
\caption{\small Examples of  diagrams problematic for the forward limit. In the lefthand diagram of (a), the propagator between the forward legs diverges due to momentum conservation. The righthand part of diagram (a) illustrates that such a diagram corresponds to the single-cut of a bubble on an external leg. Similar remarks apply to the diagrams in (b), where the limit corresponds to cutting a tadpole.}
\label{ForwardTrouble}
\end{center}
\end{figure}
\end{itemize}

Resolution of the above subtleties have been partially achieved for non-supersymmetric theories \cite{SingleCut1} and \emph{completely resolved in supersymmetric theories in the planar limit}\footnote{Here \emph{planar} means the partial amplitudes associated with a single-trace structure in the  color-trace decomposition we discussed in Section \ref{s:YM}.} \cite{SingleCut2, SingleCut3}. 
In particular, it was shown that for supersymmetric theories the problematic terms associated with the tadpole and external bubbles cancel in the state sum over the supermultiplet, and thus one has a perfectly well-defined residue. Furthermore, in the planar limit, the loop momenta in the integrand can be defined unambiguously. This is done by defining the $\ell_i$'s with a specific relation to the ordering of the external momenta. For example at 1-loop, one choice is to declare that $\ell$ is the momentum associated with the internal line immediately before line 1. For the 4-point 1-loop box integral, this selects integrand $I_4^{(a)}$ in \reef{IaIb}.

The identification of the loop-momentum is naturally done in dual space $y$, that we defined in \reef{DualDef} in order to make momentum conservation manifest.  We noted there that the $y_i$'s are also sometimes called zone variables; that is because we can think of them as labeling the `zones', or regions, that the external lines of the amplitude separate the plane into. 
This assumes a well-defined ordering of the external lines based on the color-ordering, and to do something similar at loop-level further requires the graphs to be planar. Let us illustrate this for a 6-point tree-graph and the 4-point 1-loop box diagram:
\be
\begin{split}
&
\includegraphics[width=4cm]{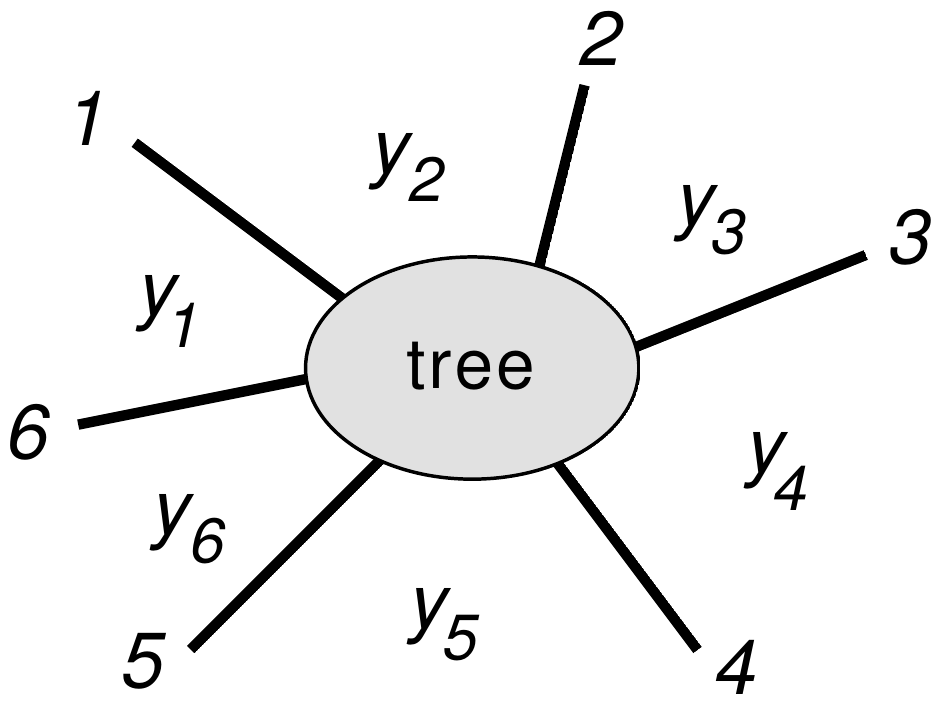}
\hspace{2cm}
\includegraphics[width=3.3cm]{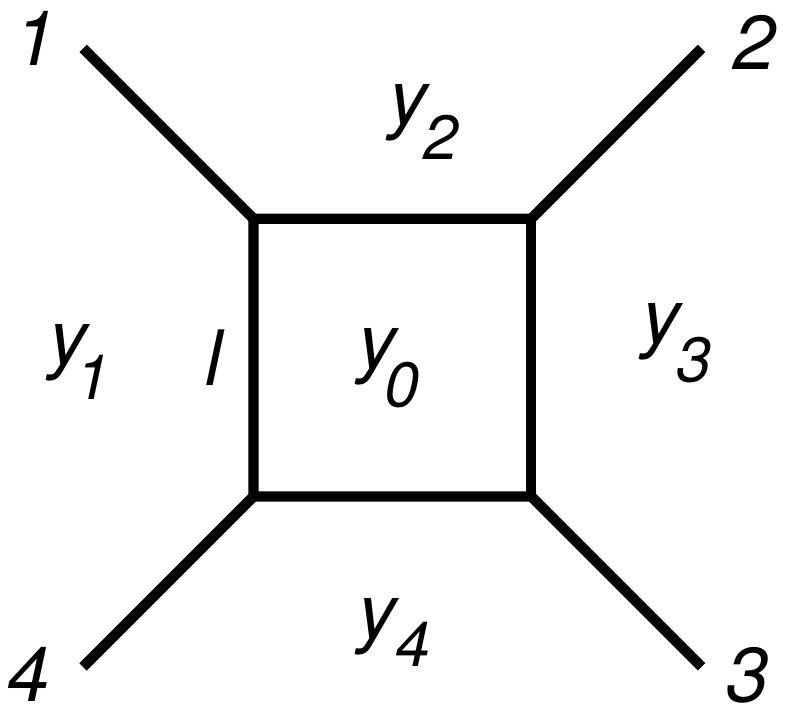}
\end{split}
\label{zones}
\ee
Obviously, at loop-level there are ``internal" zones, one for each loop. This offers the opportunity to switch the integration variable from $\ell_i$ to the new internal zone variables. For example in the  4-point 1-loop box graph, we can use $y_0$ as the loop-parameter instead of $\ell$. They are related by $\ell = y_1 - y_0$, similarly to the relationship $p_i = y_i - y_{i+1}$. The dual variables therefore give an unambiguous definition of the loop-momentum and this facilitates the loop-level recursion relations for planar integrands.   

As an example, let us express the integrand of the box-diagram in \reef{zones} in dual variables. Following the rules for identifying the momentum on each internal line in terms of the zone-variables of the two adjacent zones, we have $\ell_i^2 = (y_0 - y_i)^2 = y_{0i}^2$. The box-integrand is therefore simply
\eq
I_4(p_1,p_2,p_3,p_4)\rightarrow \frac{1}{y^2_{01}\,y^2_{02}\,y^2_{03} \, y^2_{04}}\,.
\label{I4}
\eqe      
The loop integral performed over $\int d^4 y_0$.

To reiterate,  for planar supersymmetric theories, we have overcome all subtleties. In the following, we review a BCFW recursion relation that generates planar loop-integrands for $\mathcal{N}=4$ SYM; it was developed in \cite{SingleCut3} and also considered in \cite{RutgerBCFW}.

%%%%%%%%%%%%%%%%%%%%%%%%%%%%%%%%%%%%%%%%%%%%%%%%%%%%%%%%%%%%%%%%%%%
\subsection{BCFW shift in momentum twistor space}
\label{s:bcfwMTshift}
%%%%%%%%%%%%%%%%%%%%%%%%%%%%%%%%%%%%%%%%%%%%%%%%%%%%%%%%%%%%%%%%%%%

The planar integrand is well-defined in the dual coordinates $y_i$, so we would like to formulate the BCFW shifts in the dual representation. Actually, it is even more natural to use the momentum supertwistors $\mathcal{Z}_i^\mathsf{A}$ that we introduced in Section \ref{s:momtwist}. This is because the $\mathcal{Z}_i^\mathsf{A}$'s can be chosen freely in $\mathbb{CP}^{3|4}$, giving  momentum conservation and the on-shell conditions automatically. We can therefore set up the BCFW shift without worrying about these constraints. The simplest possibility is to write
\eq
\hat{\mathcal{Z}}_i = \mathcal{Z}_i+w\mathcal{Z}_{i+1}\,.
\label{ZDeform}
\eqe
and leave all other $\mathcal{Z}_i$'s unshifted. The shift parameter $w$ is a complex variable, $w \in \mathbb{C}$.
Geometrically, \reef{ZDeform} is the statement that the point $\hat{\mathcal{Z}}_i$ lies on the line $(i,i+1)=(\mathcal{Z}_i,\mathcal{Z}_{i+1})$.\footnote{As in Section \ref{s:momtwist}, we denote the line in momentum twistor space defined by two points $(Z_j,Z_k)$ as $(j,k)$, and the plane defined by three points $(Z_j,Z_k,Z_l)$ as $(j,k,l)$.}

Let us translate \reef{ZDeform} back to the spinor helicity formalism. 
In components \reef{ZDeform} says
\be
  |\hat{i}\> = |i\> + w \, |i+1\>\,,~~~~~
  |\hat{\m}_i] = |{\m}_i] + w \, |{\m}_{i+1}]\,,~~~~~
  \hat{\chi}_{iA} = \chi_{iA} + w\,\chi_{i+1,A}\,.
\ee
Using \reef{inversemap} and the incidence relations \reef{incidence}, one finds that 
\be
  \hat{y}_i =  y_i
  +
  z\,
  |i-1\>[i|\,,
\label{xDeform}
\ee
where
\be
 z = \frac{w\langle i,i+1\rangle}
  {\langle i-1, i\rangle+w\langle i-1,i+1\rangle} \,.
  \label{zwmap}
\ee
All other $y_j$'s are unshifted. 
\exercise{}{Use \reef{inversemap} and  \reef{incidence} to show that $y_{i+1}$ and $y_{i-1}$ are unshifted. Then derive \reef{xDeform}.
}
The shift in $y$-space makes sense geometrically, because by Figure \ref{coordinates} the point $\hat{y}_{i+1}$ is determined by the line $(\hat{i},i+1)$ which is equivalent to the line 
$(i,i+1)$ since \reef{ZDeform} exactly tells us that the point $\hat{\mathcal{Z}}_{i}$ lies on $(i,i+1)$. So we conclude $\hat{y}_{i+1} = y_{i+1}$. On the other hand, $\hat{y}_{i}$ is determined by the line
$(i-1,\hat{i})$ which is different from $(i-1,i)$, so $\hat{y}_{i} \ne y_i$ for $w \ne 0$. 

Translating from dual $y$-space to momentum space, we have
\bea
 \hat{p}_i &=&  \hat{y}_i - y_{i+1} 
 ~=~ - \big(|i\> - z |i-1\>\big) [i| \,,\\[1mm]
 \hat{p}_{i-1} &=&  
 y_{i-1}  -  \hat{y}_{i}  
 ~=~ - |i-1\> \big([i-1| + z[i|\big) \,.
\eea
No other momenta shift.
We immediately read off that this is a $[i-1, i\>$ BCFW-shift
\be
  |\hat{i}\> \,=\, |i\> - z |i-1\>\,,~~~~~~
  |\widehat{i-1}] \,=\, |i-1] + z|i]\,.
  \label{zwmap0}
\ee
Since there is also a shift of the Grassmann-components of the momentum supertwistors, \reef{ZDeform} actually induces a  BCFW $[i-1, i\>$-supershift.
\exercise{}{Show that the Grassmann-part of the shift in \reef{ZDeform}
is $\hat{\eta}_{i-1} = \eta_{i-1} + z \, \eta_{i}$. }
It may seem surprising that the shift $\hat{\mathcal{Z}}_i = \mathcal{Z}_i+w\mathcal{Z}_{i+1}$ is equivalent to a $[i-1, i\>$-supershift; one might have expected a shift involving lines $i$ and $i+1$ instead. Actually, the
shift \reef{ZDeform} is also equivalent to a $[i+1, i\>$ shift: this is  because the momentum twistors are defined projectively, so we could supplement \reef{ZDeform} with an overall scaling. For example, one finds that the angle spinor shift in \reef{zwmap0} is equivalent to 
\be
  |\hat{i}\>=
  \frac{\langle i-1,i\rangle}{\langle i-1,i\rangle+w\langle i-1,i+1\rangle}\,
  \big(|i\>+w|i+1\>\big)\,.
  \label{alsozwmap0}
\ee
\exercise{}{Manipulate $|\hat{i}\>$ in  \reef{zwmap0} to find \reef{alsozwmap0}.
}

\begin{figure}
\begin{center}
\includegraphics[scale=0.65]{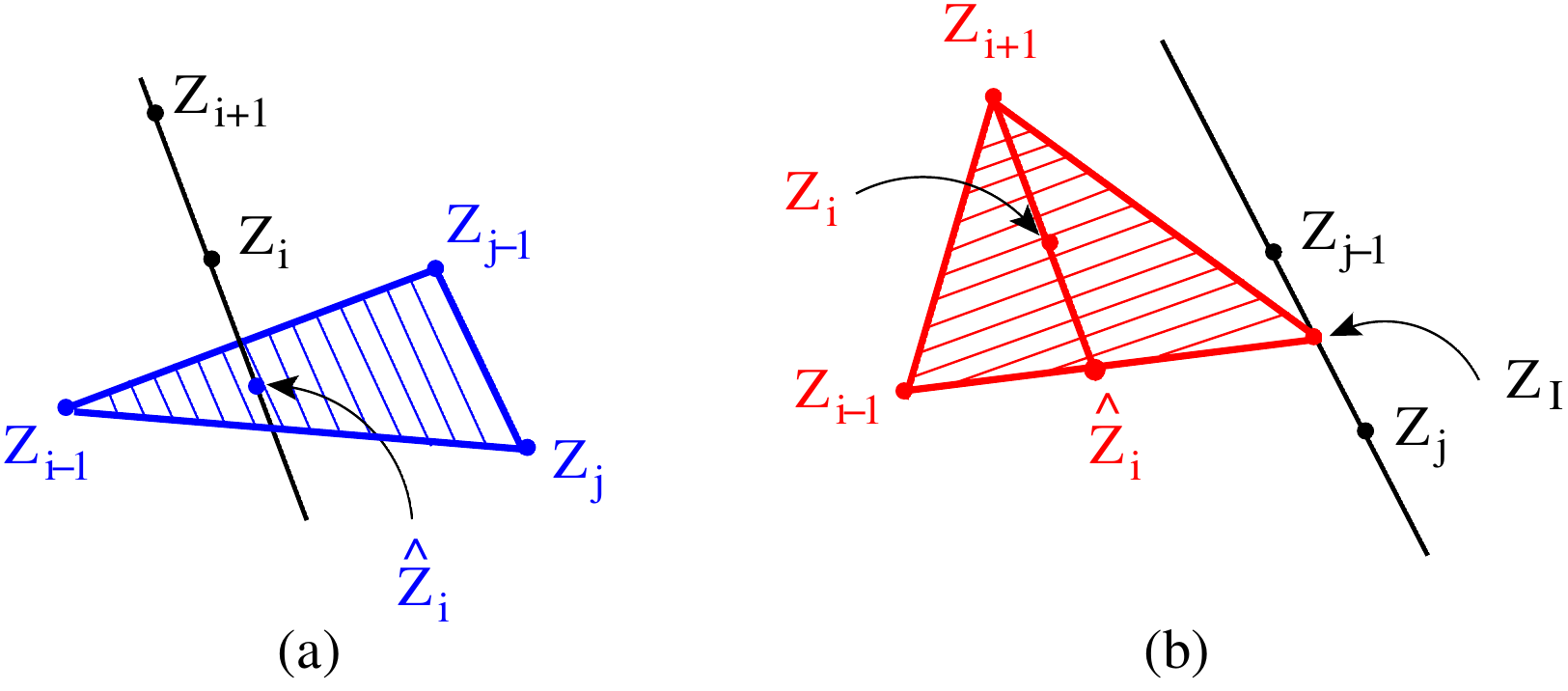}
\caption{\small The geometry  of \reef{ihat} and \reef{ZI}: 
(a) shows that the on-shell condition $\hat{P}_I^2=0$ fixes the shifted momentum twistor $\hat{Z}_i$ to be at the intersection of the line $(i,i+1)$ with the plane $(i-1,j,j-1)$. In (b) the momentum twistor $Z_I$ is located at the intersection of line $(j, j-1)$ and the plane  $(i-1,i,i+1)$. }
\label{Geometry}
\end{center}
\end{figure}

Our next  task  is to describe the kinematics associated with the internal lines in the BCFW diagrams --- it turns out to have a nice geometric description in momentum twistor space. Consider a typical BCFW diagram associated with a factorization channel $P_I$:
\be
\raisebox{-1.8cm}{\includegraphics[height=4cm]{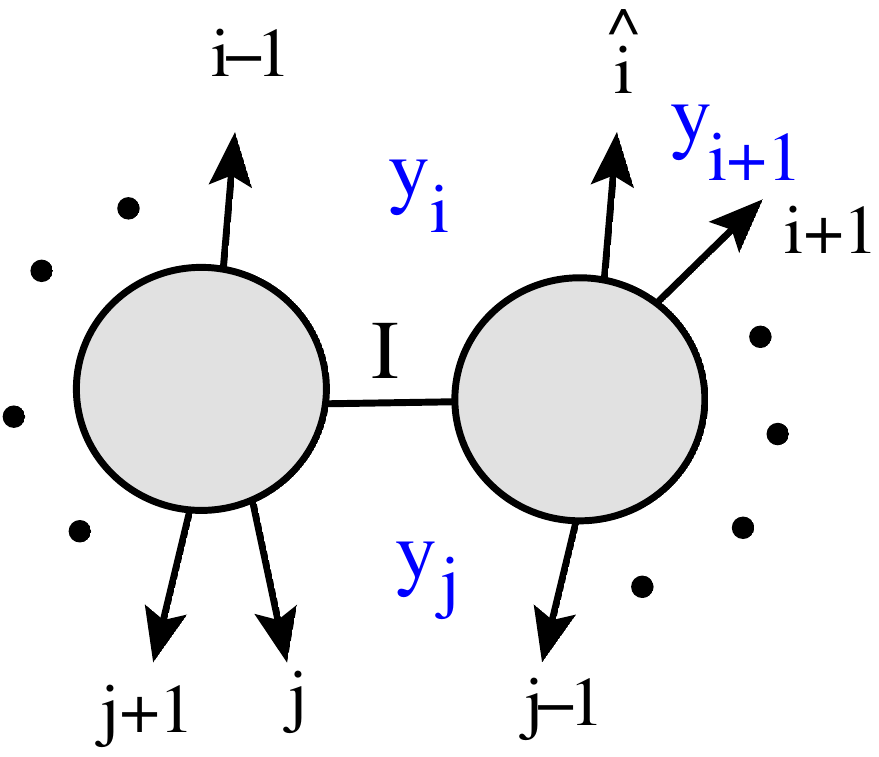}}
\label{BCFWchannel}
\ee
For simplicity, let us for now suppose that there are no loop-momenta in $P_I$.  
The shifted momentum on the internal line is 
\be 
  \hat{P}^2_I= (\hat{p}_i+p_{i+1}+\cdots+p_{j-1})^2
  ~=~
  (\hat{y}_i - y_j)^2
  ~=~
  \hat{y}^2_{ij}
  ~=~
  \frac{\langle i-1,\hat{i},j-1,j\rangle}{\langle i-1,i\rangle\langle j-1,j\rangle}
  \,.
\ee
We have used that  $\<i-1,\hat{i}\>=\<i-1,i\>$. 
The shift $\hat{{Z}}_i =  {Z}_i + w \, {Z}_{i+1}$ says that the point 
$\hat{{Z}}_i$ lies on the line $(i,i+1)$ and its position on that line is parameterized by $w$.  The condition  $\hat{P}_I^2=0$ is the statement that $w$ is chosen such that $\langle i-1,\hat{i},j-1,j\rangle=0$, so this value $w_*$ is such that the point $\hat{{Z}}_i$ lies in the plane $(i-1,j-1,j)$. In other words, $\hat{{Z}}_i$ can be characterized as the point of intersection between the line $(i,i+1)$ and the plane  $(i-1,j-1,j)$, viz.
\eq
(\,\hat{i}\,)=(i,i+1) \bigcap(i-1,j-1,j)\,.
\label{ihat}
\eqe
The intersection formula was given in \reef{TwistorPoint} in terms of the 4-brackets. The geometry is illustrated in Figure \ref{Geometry}(a).

We now determine the momentum twistor $Z_I$ associated with the internal line $\hat{P}_I$. Take a look at the BCFW diagram in \reef{BCFWchannel}. The point $y_j$ in dual space can  be determined by the line $(j-1,j)$ in momentum twistor space. But by inspection of \reef{BCFWchannel}, $y_j$ can also be determined by the line $(I,j)$. This means that the three points $Z_I$, $Z_{j-1}$, and $Z_{j}$ lie on the same line. Similarly, the point $\hat{y}_i$ can be determined by the line $(i-1,\hat{i})$ or by the line $(I,\hat{i})$, so $Z_I$, $Z_{i-1}$, and $\hat{Z_{i}}$ lie on the same line. Since $Z_I$ lie on both the two lines, we conclude that $Z_I$ can be characterized as the intersection point of the lines $(i-1,\hat{i})$ and 
$(j-1,j)$. We previously learned that $\hat{Z_{i}}$ lies in the plane $(i-1,j-1,j)$, and therefore plane contains the line $(i-1,\hat{i})$. Thus we conclude that $Z_I$ is the point where the line $(j-1,j)$ intersects the plane $(i-1,i,i+1)$:
\eq
(\,I\,)=(j,j-1) \bigcap(i-1,i,i+1)\,.
\label{ZI}
\eqe
The geometry is sketched in Figure \ref{Geometry}(b). 
These results will be useful in the following.

Now we are ready to study the BCFW recursion relations in momentum twistor space. Consider the BCFW shift  \reef{ZDeform} of a $n$-point $L$-loop integrand ${\mathcal{I}}_n^L$ (for tree-level you can translate `integrand' to `superamplitude' in your head). The recursion relations are based on the usual contour argument for $\int \frac{dw}{w}\,\hat{\mathcal{I}}^L_n(w)$. 
Poles at finite values of $w$ are equivalent, via \reef{zwmap}, to poles at finite $z$: they arise from  propagators with momentum $\hat{y}_{ij}$ going on-shell and the corresponding BCFW diagrams are those in \reef{BCFWchannel}.
However, in completing the contour integral argument we also need to consider the large-$w$ limit. It is clear from the relation \reef{zwmap} that $z$ goes to a finite value $z_*$ as $w \to \infty$. Specifically, 
\be
  z~\xrightarrow{w \to \infty}~ z_* ~\equiv~ \frac{\langle i, i+1 \rangle}{\langle i-1, i+1\rangle} \,.
  \label{zstar}
\ee
Thus the pole at infinity in the $w$-plane maps to a finite point in the $z$-plane and we will have to consider this contribution too. 
 At $w=\infty$, the shifted angle spinors are
\eq
  |\hat{i}\>\Big|_{w\to \infty}
  =
  \frac{\langle i -1,i \rangle}{\langle i-1, i+1\rangle}\,|i+1\>\,.
\eqe 
Thus, in the limit $w \to \infty$, the spinors $|\hat{i}\>$ and 
$|i+1\>$ become proportional, and that implies that $\hat{P}_{i,i+1}= \hat{p}_i+p_{i+1}$ is on-shell: 
$\hat{P}_{i,i+1}^2 = \<\hat{i},i+1\>[i,i+1] \to 0$ for $w \to \infty$. Or equivalently, we may note that $z_*$ is exactly the solution to $\<\hat{i},i+1\>=0$. 
Hence the pole at $w = \infty$ corresponds to a factorization channel of an  integrand into a 3-point anti-MHV part --- the only possibility that can support the special kinematics $|\hat{i}\> \propto |i+1\> \propto |\hat{P}_{i,i+1}\>$ --- and the remainder $L$-loop integrand:
\be
\raisebox{-18mm}{\includegraphics[width=9.5cm]{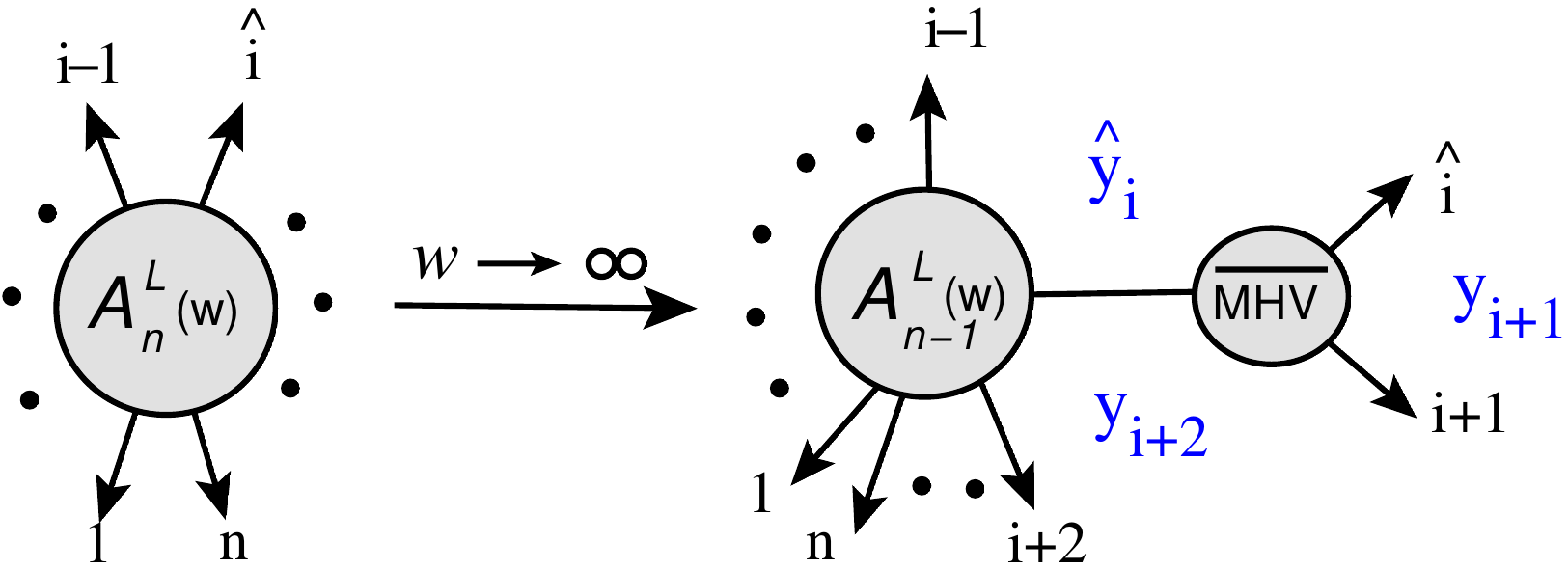}}
\label{Largew}
\ee
Note that the lines in this figure are labeled by momentum supertwistors, so we place a hat just on line $i$  since that is the only shifted momentum twistor. 
\exercise{}{Show that for $w\rightarrow\infty$, the kinematics give
\eqa
\nonumber
|\widehat{i-1}\>=|i-1\>\,,
&&
|\widehat{i-1}]
=|i-1]-\frac{\langle i,i+1\rangle}{\langle i+1,i-1\rangle}\,|i]\,,
\\
|{\hat{P}_I}\>=|{i+1}\>\,,&&\;\;
~\,|\hat{P}_I]=-
\left(\frac{\langle i,i-1\rangle}{\langle i+1,i-1\rangle}|i]+|i+1]\right)\,.
\label{Shifted}
\eqae
}

To summarize, BCFW recursion relations in momentum twistor space express the integrand (superamplitude) as a boundary contribution 
$\mathcal{B}_\infty$ from $w=\infty$ plus a sum of residues at finite $w$. Schematically, we have
\be
\mathcal{I}^L_n 
~=~ \hat{\mathcal{I}}^L_n(w=0) 
~=~ \mathcal{B}^L_\infty 
- \sum_{w_* \ne 0} 
\Big( \text{Residues of $\frac{\hat{\mathcal{I}}^L_n(w)}{w}$ at finite $w_*$} \Big)\,.
\label{genBCFW}
\ee
The boundary term $\mathcal{B}_\infty$ is computable and is given by the diagram \reef{Largew}. The rest of the residues come from diagrams such as \reef{BCFWchannel}.

To become familiar with how this works in practice, we first apply the  recursion relations to tree-level superamplitudes before moving on to loop-integrands in Section \ref{s:bcfwMTloop}.

%%%%%%%%%%%%%%%%%%%%%%%%%%%%%%%%%%%%%%%%
\subsection{Momentum twistor BCFW at tree-level}
\label{s:bcfwMTtree}
%%%%%%%%%%%%%%%%%%%%%%%%%%%%%%%%%%%%%%%%
In Section \ref{s:NMHV}, we used the super-BCFW recursion relations to show that the NMHV tree-level superamplitude of $\mathcal{N}=4$ SYM can be written as the MHV superamplitude times a sum of the dual superconformal invariants $R_{ijk}$\,; see \reef{superNMHV}. We then rewrote the NMHV formula in terms of momentum twistors in Section \ref{s:momtwist} and found \reef{Final}
\be
\mathcal{A}_n^{\rm NMHV}~=~
\mathcal{A}_n^{\rm MHV}~
\sum_{j=2}^{n-3}\sum_{k=j+2}^{n-1} \big[n,j-1,j,k-1,k\big]\,,
\label{Final2}
\ee
where the 5-brackets $\big[n,j-1,j,k-1,k\big] = R_{njk}$ are invariant under cyclic permutations of the five labels. It was claimed then that the N$^K$MHV tree superamplitudes took a similar form but with the sum involving products of $K$ 5-brackets. We now prove this statement using the momentum twistor formulation of super-BCFW. This also serves to prove that the tree superamplitudes of $\mathcal{N}=4$ SYM are dual superconformal covariant. 

Adapted to tree-level, the  $\hat{Z}_i$-shift BCFW relation \reef{genBCFW} reads
\be
\mathcal{A}_n = \mathcal{B}_\infty - \sum_{j=i-3}^{i+2} 
\Big( \text{Residues of $\frac{\hat{\mathcal{A}}_n(w)}{w}$ at } \hat{y}_{ij}^2 = 0\Big)
\ee
We begin with a detailed evaluation of the boundary term.

{\bf The boundary term $\mathcal{B}_\infty$.}
Per definition, the boundary contribution is the residue of the pole at infinity,\footnote{We ignore the $2\pi i$ of the Cauchy theorem since all such factors drop out at the end.}
\be
  \mathcal{B}_\infty =  -\oint_{\mathcal{C}_\infty} \frac{dw}{w} \,\hat{\mathcal{A}}_n(w)\,.
\ee
Here $\mathcal{C}_\infty$ is a contour that surrounds $w=\infty$ counterclockwise. Since we are more familiar with the shift in momentum space, let us change variables from $w$ to $z$. With the help of \reef{zwmap} we find
\be
 \mathcal{B}_\infty = \oint_{\mathcal{C}_{z_*}} dz\, \frac{z_*}{z(z-z_*)}\,
 \,\hat{\mathcal{A}}_n(z)\,,
\ee
where $z_*=\frac{\langle i, i+1 \rangle}{\langle i-1, i+1\rangle}$ is the value of $z$ at $w=\infty$.
Now we need to find out how the shifted $n$-point amplitude behaves for $z$ near $z_*$. We already established in \reef{Largew} that the N$^K$MHV superamplitude factorizes as  N$^K$MHV$_{n-1}$$\times$anti-MHV$_3$ at $z=z_*$. Let us focus on the {\bf MHV case} ($K=0$) first to see explicitly how this comes about. Under the $[i\!-\!1,i\>$-supershift, the Grassmann delta function in the MHV superamplitude is inert, and the only part of the amplitude affected by the shift is the denominator factor $\<\hat{i},i+1\>$. This exactly is the factorization pole for $z \to z_*$. Therefore, near $z_*$ we can write
\be
  \hat{\mathcal{A}}_n^\text{MHV}(z)
  ~\xrightarrow{z\to z_*}~ 
  \hat{\mathcal{A}}_{n-1}^\text{MHV}(z_*)
     \,  \frac{1}{\hat{P}_I^2}\,
     \hat{\mathcal{A}}_{3}^\text{anti-MHV}(z_*)
   ~=~
   \frac{P_I^2}{\hat{P}_I^2}~
     \Big[ \hat{\mathcal{A}}_{n-1}^\text{MHV}(z_*)
       \,\frac{1}{P_I^2}\,
     \hat{\mathcal{A}}_{3}^\text{anti-MHV}(z_*) \Big] \,. 
     \label{Ahatfactor}
\ee
We know from super-BCFW'ing the MHV superamplitude in Section \ref{s:MHVsBCFW} that the factor $\big[ \ldots \big]$ in \reef{Ahatfactor} equals $\mathcal{A}_{n}^\text{MHV}$ (remember, for MHV only one diagram contributed in the recursion relations based on a BCFW shift of adjacent lines). The prefactor 
is $\frac{P_I^2}{\hat{P}_I^2} = - z_*/(z-z_*)$. Thus  
\be
  \mathcal{B}_\infty^\text{MHV} ~=~
  -\mathcal{A}_{n}^\text{MHV} 
  \oint_{\mathcal{C}_{z_*}} dz\, \frac{z_*^2}{z(z-z_*)^2}
   ~=~ \mathcal{A}_{n}^\text{MHV} \,.
\ee
In the second equality, we evaluated the double pole integral using 
\be
  \oint dz\, \frac{f(z)}{(z-z_*)^2} = \frac{d}{d z_*} 
\oint dz\, \frac{f(z)}{(z-z_*)}
  = f'(z_*)\,
\ee
for the case $f(z) = 1/z$.

What we have achieved for the MHV case here is a verification of the  simple statement that the MHV tree-level superamplitude satisfies the super-BCFW recursion relation which for MHV only include one term, namely the MHV$_{n-1}$$\times$anti-MHV$_3$ diagram. We knew that already more than 35 pages ago (Section \ref{s:MHVsBCFW}), but the point is that here we have set up the calculation in a way that facilitates the generalization to N$^K$MHV level. And that is what we do next.

{\bf  N$^K$MHV case.} Assume inductively that the $(n\!-\!1)$-point tree-level N$^K$MHV  superamplitude can be written as an MHV prefactor times a dual superconformal invariant that we call $Y_{n-1}^{(K)}$; we already know this to be true for all $n$ when $K=1$, since $Y_{n-1}^{(1)}$ is the sum of 5-brackets given in \reef{Final2}. The calculation of the contribution from $w=\infty$ follows the same steps as the MHV case, expect that the factorization \reef{Ahatfactor} is now replaced by
\be
 \hat{\mathcal{A}}_n^\text{MHV}(z)
  ~\xrightarrow{z\to z_*}~ 
  \hat{\mathcal{A}}_{n-1}^\text{MHV}(z_*)
  \, \widehat{Y}^{(K)}_{n-1}(z_*)
       \,\frac{1}{\hat{P}_I^2}\,
     \hat{\mathcal{A}}_{3}^\text{anti-MHV}(z_*)
\ee
Let us take a closer look at the $Y$-factor. It is naturally a function of momentum supertwistors
\be
\widehat{Y}^{(K)}_{n-1}(z_*)
~=~
\widehat{Y}^{(K)}_{n-1}(\ldots,\mathcal{Z}_{i-1},\mathcal{Z}_I,\mathcal{Z}_{i+2},\ldots)\,.
\ee
Now in our analysis of the kinematics, we learned that the momentum twistor $\mathcal{Z}_I$ is characterized as the intersection \reef{ZI} between the line $(j-1,j)$ and the plane $(i-1,i,i+1)$. In our case here, we have $j=i+2$, so \reef{ZI} says that 
$\mathcal{Z}_I$ is the point of intersection between the line $(i+1,i+2)$ and the plane $(i-1,i,i+1)$. Obviously this intersection point is $\mathcal{Z}_{i+1}$:
\eq
\text{For $w=\infty$ case:}~~~~
(\,I\,)=(i+1,i+2) \bigcap(i-1,i,i+1)
= (i+1)\,.
\eqe
So we can freely substitute $\mathcal{Z}_I \to \mathcal{Z}_{i+1}$ to find
\be
\widehat{Y}_{n-1}^{(K)}(z_*)
~=~
{Y}_{n-1}^{(K)}(\ldots,\mathcal{Z}_{i-1},\mathcal{Z}_{i+1},\mathcal{Z}_{i+2},\ldots )\,.
\ee  
This factor is independent of $z$ and we can therefore repeat our argument from the MHV case to find
\be
  \mathcal{B}_\infty^\text{N$^K$MHV} ~=~
  \oint_{\mathcal{C}_\infty} \frac{dw}{w} \,\hat{\mathcal{A}}^\text{N$^K$MHV}_n(w)
  ~=~
   \mathcal{A}_{n}^\text{MHV}~
   {Y}^{(K)}_{n-1}(\mathcal{Z}_1,\ldots,\mathcal{Z}_{i-1},\mathcal{Z}_{i+1},\ldots \mathcal{Z}_n)
  \,.
\ee
This completes the calculation of the boundary term.

{\bf Residues at finite $w$.}
Now we extract the residues of the finite poles in the $w$-plane. They arises from propagators $1/\hat{y}^2_{ij}$ going on-shell. Writing the shifted propagator in terms of momentum twistor, we find 
\be
  \frac{1}{\hat{y}^2_{ij}} 
  ~=~ 
  \frac{\langle \hat{i},i-1\rangle\langle j j-1\rangle}
  {\langle   \hat{i},i-1,j,j-1\rangle}
  ~=~
  \frac{\langle i,i-1\rangle\langle j j-1\rangle}
  {\langle i,i-1,j,j-1\rangle + w \langle i+1,i-1,j,j-1\rangle}
  ~=~
  \frac{w_*}{y^2_{ij}} \,\frac{1}{w-w_*}
\ee
where $w_* = - \frac{\langle i,i-1,j,j-1\rangle}{\langle i+1,i-1,j,j-1\rangle}$.
This means that
\be
   -\int_{\mathcal{C}(w_*)} \frac{dw}{w} \,  \frac{1}{\hat{y}^2_{ij}} \, f(w)
   ~=~
   -\int_{\mathcal{C}(w_*)} \frac{dw}{w} \, \frac{w_*}{y^2_{ij}} \,\frac{1}{w-w_*} \, f(w)
   ~=~\frac{1}{y^2_{ij}}~f(w_*)\,.
\ee
Hence the contribution from the shifted propagator is simply the unshifted propagator, exactly the same as the usual BCFW rules. In a given factorization channel, we can write the left and the right subamplitudes as:
\eq
\mathcal{A}_{n_L}=\mathcal{A}_{n_L}^\text{MHV}\,
Y_{n_L}^{(K_{L})}(\mathcal{Z}_I,\mathcal{Z}_{j-1},\cdots,\mathcal{Z}_{i+1},\mathcal{Z}_{\hat{i}}),
~~~~
\mathcal{A}_{n_R}=\mathcal{A}_{n_R}^\text{MHV}\,
Y_{n_R}^{(K_{R})}(\mathcal{Z}_I,\mathcal{Z}_{i-1},\cdots,\mathcal{Z}_{j+1},\mathcal{Z}_{j})\,.
\eqe
Here the Grassmann degrees obey $K_R + K_L +1 =K$; in particular this is why there were no such diagrams for the MHV case.
Then the contribution of the BCFW channel is simply 
\eq
\left(\sum_\text{states}
\frac{\hat{\mathcal{A}}_{n_L}^\text{MHV}\,
\hat{\mathcal{A}}_{n_R}^\text{MHV}}{P^2_I}\right)\,
\widehat{Y}^{(K_{L})}_{n_L}(\mathcal{Z}_I,\mathcal{Z}_{j-1},\cdots,\mathcal{Z}_{i+1},\hat{\mathcal{Z}}_{i})~
\widehat{Y}^{(K_{R})}_{n_{R}}(\mathcal{Z}_I,\mathcal{Z}_{i-1},\cdots,\mathcal{Z}_{j+1},\mathcal{Z}_{j})\,.
\eqe
The $\mathcal{Z}_I$ appearing in the dual superconformal invariants $Y_\text{L,R}$ can be written in terms of the external line supermomentum twistors using $\mathcal{Z}_k$ the characterization of $\mathcal{Z}_I$ as an intersection point \reef{ZI} and the formula \reef{TwistorPoint}. Similarly for $\hat{\mathcal{Z}}_{i}$, via the intersection rule \reef{ihat}.

As a consequence, the state sum --- which is an integration over the $\eta_I$ variables --- acts solely on the MHV prefactors.\footnote{In \cite{Drummond:2008cr}, this was achieved by cleverly using cyclic symmetry to ensure that the $\eta_I$'s appear only in the MHV prefactors.} Furthermore, the factor in the parenthesis is simply the BCFW term of the NMHV amplitude arising from the $P_I$ factorization channel. We have calculated this in Section \ref{s:NMHV}, and later learned that in the momentum twistor language the answer is written in terms of the 5-bracket:
\be
\left(\sum_\text{states}
\frac{\hat{\mathcal{A}}_{n_L}^\text{MHV}\,
\hat{\mathcal{A}}_{n_R}^\text{MHV}}{P^2_I}\right)
~=~
{\mathcal{A}}^\text{MHV}_{n}\times
\big[i-1 ,i ,i+1,j-1, j\big]\,.
\ee
Thus we have finally arrived at the BCFW recursion relation for tree-level amplitudes in $\mathcal{N}=4$ SYM, written in momentum twistor space: 
\bea
\nonumber
\mathcal{A}^\text{N$^K$MHV}_n
&=&
\mathcal{A}^\text{MHV}_{n}\bigg\{Y^{(K)}_{n-1}(\ldots,\mathcal{Z}_{i-1},\mathcal{Z}_{i+1},\mathcal{Z}_{i+2},\ldots )\\
\label{TreeBCFW}
~~&&
+\sum_{j=i+3}^{i-2}
\big[i-1, i, i+1,j-1, j\big]
   \\[-3mm] \nonumber
 ~~&&
   \hspace{1.65cm}
   \times 
\widehat{Y}^{(K_{L})}_{n_{L}}(\mathcal{Z}_I,\mathcal{Z}_{j},\mathcal{Z}_{j+1},\ldots,\mathcal{Z}_{i-1})\times
\widehat{Y}^{(K_{R})}_{n_{R}}(\mathcal{Z}_I,\hat{\mathcal{Z}}_{i},\mathcal{Z}_{i+1},\ldots,\mathcal{Z}_{j-1})\bigg\}\,.
\eea
The above relation corresponds to the shift defined in \reef{ZDeform}, and the momentum twistors $\hat{\mathcal{Z}}_{i}$ and $\mathcal{Z}_I$ are given by \reef{ihat} and \reef{ZI} respectively. Also, $K_R + K_L +1 =K$ and $n_L + n_R = n+2$.

The result \reef{TreeBCFW} verifies the claim that all tree-level amplitudes of $\mathcal{N}=4$ SYM can be written as an MHV prefactor times polynomials of 5-brackets: given that this is true for the NMHV amplitudes, \reef{TreeBCFW} ensures that the 5-brackets are recycled into the higher-$K$ results. Since the 5-brackets are manifestly dual superconformal invariant, so are all tree-level superamplitudes of $\mathcal{N}=4$ SYM.

%%%%%%%%%%%%%%%%%%%%%%%%%%%%%%%%%%%
\subsection{Momentum twistor BCFW for planar loop integrands}
\label{s:bcfwMTloop}
%%%%%%%%%%%%%%%%%%%%%%%%%%%%%%%%%%%%%%%
To initiate the discussion of BCFW recursion for planar loop-integrands of $\cn=4$ SYM, let us examine a specific example to get some intuition for the good looks and behaviors of integrands. In other words, we start with the answer and let that guide our discussion.

In Section \ref{s:genunit}, we used the generalized unitarity method to construct the 1-loop $\cn=4$ SYM superamplitude.  We found
\eq
\mathcal{A}_4^\text{1-loop}[1234] 
~=~ 
su \,\mathcal{A}_4^{\rm tree}[1234]\,I_4(p_1,p_2,p_3,p_4)\,,
\label{N4Answ2}
\eqe 
where $I_4$ is the 1-loop box-integral which we wrote in dual $y$-space in \reef{I4} as 
\eq
I_4(p_1,p_2,p_3,p_4)= \int d^4 y_0~ \frac{1}{y^2_{01}\,y^2_{02}\,y^2_{03} \, y^2_{04}}\,.
\label{I4b}
\eqe      
Here the propagator-terms $y_{0i}^2 = (y_0 - y_i)^2$ involve the zone-variable $y_0$ associated with the loop momentum, as indicated in \reef{zones}.

The expressions \reef{N4Answ2}-\reef{I4b} determine the loop-integrand for the 4-point 1-loop $\cn=4$ SYM superamplitude to be (using $-s = y_{13}^2$ and $-u = y_{24}^2$)
\be
 \mathcal{I}_4^\text{1-loop}[1234]
 ~=~
 \mathcal{A}_4^{\rm tree}[1234]~ 
 \frac{y_{13}^2\, y_{24}^2}{y^2_{01}\,y^2_{02}\,y^2_{03} \, y^2_{04}}\,.
 \label{4ptIntResPre}
\ee
Now, to translate this to momentum twistor space, recall that a point $y$ in dual space maps to a line in momentum twistor space. So let us take $y_0$ to be mapped to some line $(A,B)$ determined by two points  $\mathcal{Z}_A$ and $\mathcal{Z}_B$; the loop-integral $\int d^4 y_0$  maps to an integral over all inequivalent lines $(A,B)$. There is a story here of how to define the integration measure appropriately --- we  postpone this until later in this section in order to first discuss the structure of the loop-integrands. 

Using \reef{ID3} to rewrite all the dual variables $y$ in the integrand in terms of 4-brackets,  in particular
$y_{0i}^2 =  \frac{\<A,B,i-1,i\>}{\<AB\>\<i-1,i\>}$,
we arrive at the expression 
\be
 \mathcal{I}_4^\text{1-loop}[1234]
 ~=~
 - \mathcal{A}_4^{\rm tree}[1234]~ 
 \frac{\langle1234\rangle^2 \<AB\>^4}
 {\langle AB12\rangle \langle AB23\rangle \langle AB34\rangle \langle AB41\rangle}\,.
 \label{int}
\ee
Note that all $\<i-1,i\>$'s dropped out. The factor $\<AB\>^4$ will eventually be absorbed in the integration measure and what remains is manifestly dual conformal invariant. 

The expression \reef{int} is an example of what a 1-loop integrand looks like in momentum twistor space. 
Under a shift $\hat{\mathcal{Z}}_4 =\mathcal{Z}_4 + w \mathcal{Z}_3$, the integrand has a pole that  involves the loop-momentum: it comes from $\langle AB\hat{4}1\rangle = 0$. The residue of such a pole is the new input we need for the loop-level recursion relations. 
\exercise{}{What type of super-BCFW shift is induced in momentum space by the momentum twistor shift $\hat{\mathcal{Z}}_4 =\mathcal{Z}_4 + w \mathcal{Z}_3$?}
Next, we outline the form of the recursion relations for general $L$-loop integrands.

%%%%%%%%%%%%%%%%%%%%%%%%%%%%
\compactsubsection{Structure of the BCFW recursion for loop integrands.}
Without loss of generality, we consider the recursion relations derived from a shift of the $n$'th momentum twistor,
\eq
\hat{\mathcal{Z}}_n=\mathcal{Z}_{n}+w\mathcal{Z}_{n-1}\,.
\eqe
For an $n$-point $L$-loop integrand, there will be three distinct contributions to the recursion relations:
\begin{enumerate}
\item
 The boundary contribution from $w\to\infty$. This contribution is calculated just as in the tree-level case of the previous section, so we simply just state the result (suppressing the $K$ of the N$^K$MHV classification)
\be
   \text{term at $w\to\infty$:}~~~~~~
   \mathcal{A}^{\rm tree}_{n,\text{MHV}}~Y^{L}_{n-1}\big(\mathcal{Z}_1,\cdots,\mathcal{Z}_{n-1}\big) \,.
\ee
Here $Y^{L}_{n-1}$ is, by the inductive assumption, an $L$-loop dual superconformal invariant.
\item
Residues of factorization channels from propagators that do \emph{not} involve loop-momenta correspond to poles in $\hat{y}_{1j}^2 \propto \langle\hat{n},1,j-1,j\rangle=0$. 
The results for these also follow the same steps as the tree-level case, and one finds 
\be
\mathcal{A}^{\rm tree}_{n,\text{MHV}}
\sum_{j=3}^{n-2} \big[j-1,j,n-1,n,1\big]
\,
Y^{L_1}_\text{L}\big(\mathcal{Z}_{I_j},\mathcal{Z}_{j},\mathcal{Z}_{j+1},\cdots,\hat{\mathcal{Z}}_{{n}_j}\big)
\,
Y^{L_2}_\text{R}\big(\mathcal{Z}_{I_j},\mathcal{Z}_{1},\mathcal{Z}_{2},\cdots,\mathcal{Z}_{j-1}\big) \,.
\label{poles2}
\ee
where $\hat{\mathcal{Z}}_{{n}_j}=(n-1,n)\bigcap(1,j-1,j)$ and 
 $\mathcal{Z}_{I_{j}}=(j,j-1)\bigcap(n-1,n,1)$. 
This includes an implicit sum over loop-orders $L_1$ and $L_2$ in the sub-integrands such that $L_1+L_2=L$. Also, 
the N$^K$MHV level was suppressed so one must sum over the Grassmann degrees associated with the sub-integrands such that $K_\text{L} + K_\text{R} = K-1$.

\item Residues of factorization channels from propagators that \emph{do} involve loop-momenta; they correspond to 
\be
\langle A\,B\,\hat{n}\,1\rangle = 0\,. 
\label{ForwardConstraint}
\ee   
These are the new contributions at loop-level, so we will take a closer look at  them now.

\end{enumerate}

%%%%%%%%%%%%%%%%%%%%%%%%%%%%
\compactsubsection{Forward limit contributions.}
%%%%%%%%%%%%%%%%%%%%%%%%%%%%
For an $L$-loop $n$-point integrand, the residue of the pole  \reef{ForwardConstraint} is an $(L\!-\!1)$-loop $(n\!+\!2)$-point integrand whose two extra legs are evaluated in the forward limit \reef{fwlimit}, as shown in Figure \ref{BCFWInt}. The example of $n=4$ will illustrate the idea of how to do this.
\begin{figure}
\begin{center}
\includegraphics[scale=0.8]{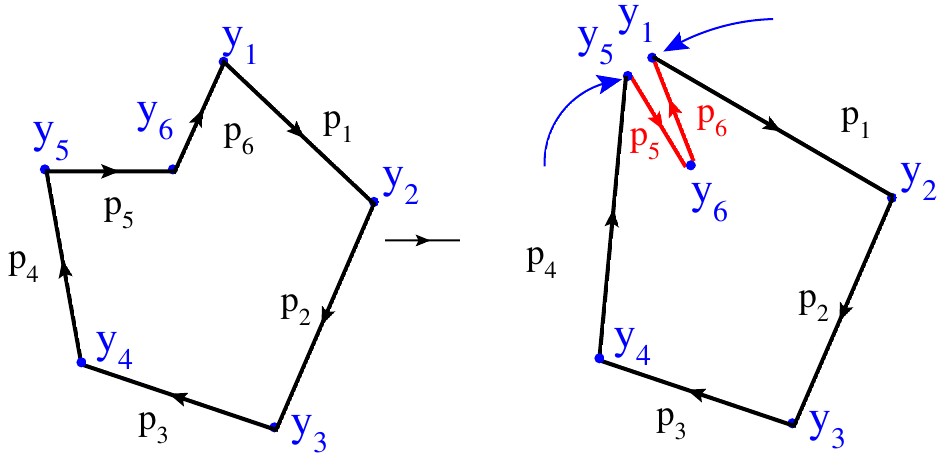}
\caption{\small The forward limit illustrated in dual coordinates. The limit corresponds to $y_5$ and $y_1$ approaching a point while satisfying $y^2_{14}=0$.}
\label{Forward}
\end{center}
\end{figure}%
\example{Start with a 6-point integrand. Translated to dual coordinates, the forward limit of $p_5$ and $p_6$ approaching $p_5=-p_6=r$ is the limit of taking $y_1\to y_5$ while $y_6$ remains fixed. This is illustrated in Figure \ref{Forward}. 
In momentum twistor space, the point $y_1$ is determined by the line $(1,6)$ and $y_5$ by $(4,5)$, so $y_1$ and $y_5$ can be identified only when $(Z_1,Z_6,Z_5,Z_4)$ lie on the same line. It is easy to achieve this configuration if the line $(1,4)$ intersects line $(6,5)$, because then we can send $Z_5$ and $Z_6$ to the intersection point $(6,5)\bigcap(1,4)$. Note that this does not change $y_6$, but the result is $y_1 \to y_5$.

However, momentum twistors live in $\mathbb{CP}^3$ where two lines generically do not intersect. So we cannot take the limit as naively as above. Instead,  we modify the  momentum twistor $Z_4\rightarrow \hat{Z}_{4}=Z_{4}+wZ_3$, and tune $w$ such that the new line $(1,\hat{4})$ intersects $(5, 6)$: let $Z_{\hat{B}}$ be the  point of intersection. Since $\hat{Z}_{4}$ per construction lies on the line $(3,4)$, we can characterize $Z_{\hat{B}}$ as the intersection point between the line $(5,6)$ and the plane $(3,4,1)$ (see Figure \ref{Geometry2}):
\be
 (\hat{B})=(5,6)\bigcap\,(3,4,1)\,.
 \label{hatA}
\ee
Since the lines $(1,\hat{4})$ and $(5, 6)$ are arranged to intersect, it follows that $\hat{Z}_{4}$ lies in the plane $(5,6,1)$; see Figure \ref{Geometry2}. But $\hat{Z}_{4}$ is also on the line $(3,4)$, so the shifted momentum twistor can be identified in terms of the unshifted lines as 
\be
 \hat{Z}_{4}=(3,4)\bigcap\,(5,6,1)\,.
 \label{hatZ4}
\ee
The setup with \reef{hatA} and \reef{hatZ4} allows us to take the forward limit directly by sending $Z_5, Z_6$ to the intersection point $Z_{\hat{B}}$. We can summarize the deformation and forward limit $p_5=-p_6=r$ as
\be
(Z_1,Z_2,Z_3,Z_4,Z_5,Z_6)\rightarrow (Z_1,Z_2,Z_3,\hat{Z}_{{4}},Z_5,Z_6)\Big|_{Z_5,Z_6\rightarrow Z_{\hat{B}}}\,.
\ee
It is important to note that $\hat{Z}_{4}$ satisfies
\be
\langle5,6,\hat{4},1\rangle=0\,.
\label{56hat41}
\ee
Comparing \reef{56hat41} with \reef{ForwardConstraint}, we recognize  the single-cut condition (or equivalently, momentum dependent pole in BCFW) provided that $Z_5$ and $Z_6$ are identified as the loop-momentum twistors $Z_A$ and $Z_B$. This is also the statement that $y_6$ has been identified as our loop integration region, as Figure \ref{Forward} indicates that it should be.
}

\begin{figure}
\begin{center}
\includegraphics[scale=0.65]{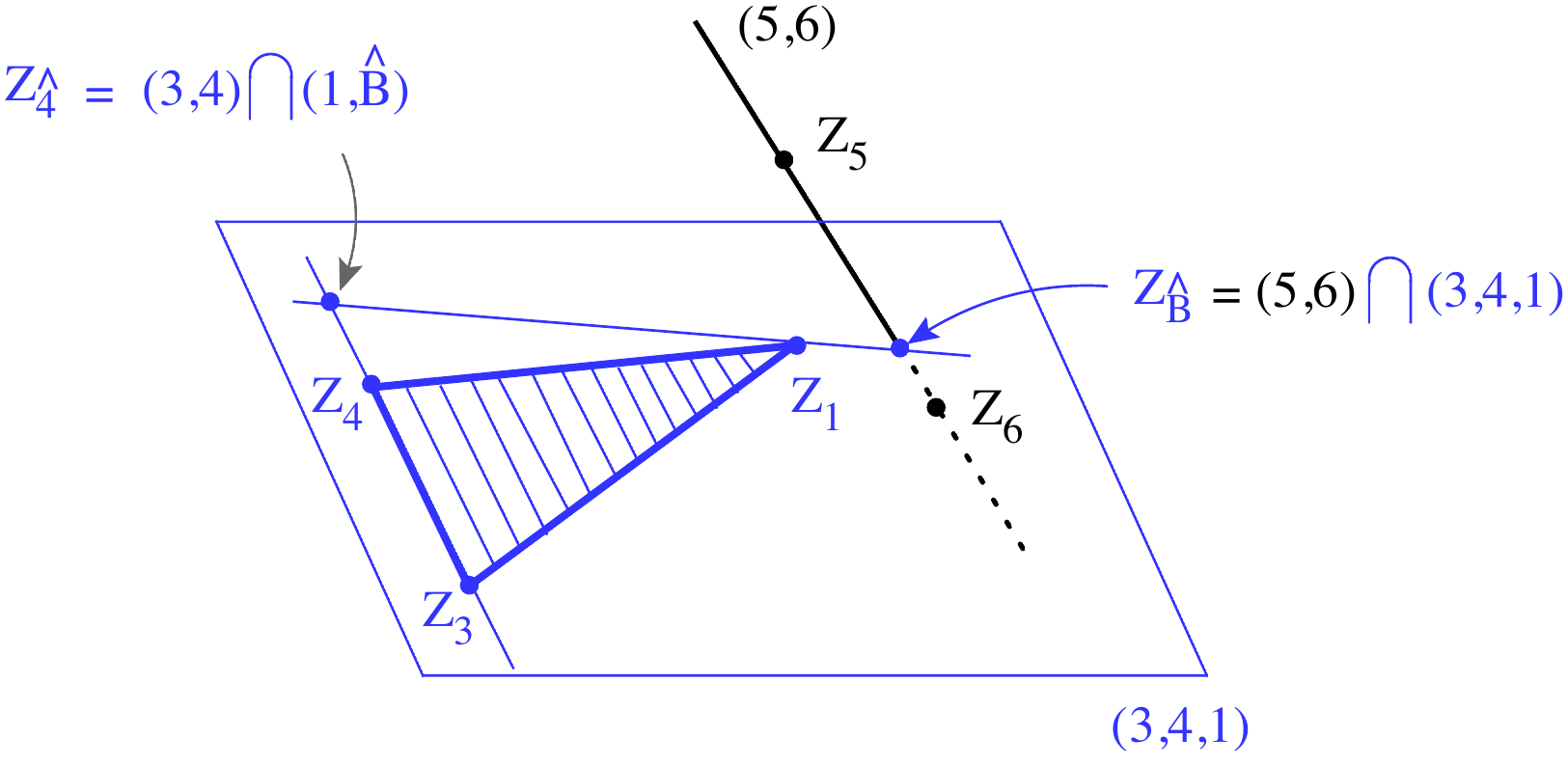}
\caption{\small The geometry of the forward limit Figure \ref{Forward} illustrated here in momentum twistor space. The point $Z_{\hat{B}}$ is defined as the intersection of line $(Z_5, Z_6)$ with plane $(Z_1,Z_3,Z_4)$. The BCFW-deformed $\hat{Z}_{{4}}$ is fixed at the interaction of line $(Z_3,Z_4)$ and $(Z_{\hat{B}},Z_{1})$. Again the blue points lie in the same plane.}
\label{Geometry2}
\end{center}
\end{figure}
Let us return to the general case.  The forward limit is taken for   
$(L\!-\!1)$-loop $(n\!+\!2)$-point integrands by sending the momentum twistors $(\hat{Z}_{{n}},Z_{A},Z_{B})$ to the forward configuration discussed in the example. 
One must multiply by an overall MHV factor as well as the result for the cut propagator. The result (which we discuss further below) is 
\be
\mathcal{A}^{\rm tree}_{n,\text{MHV}} 
\times 
 f(A,B,n-1,n,1)
\times
\bigg( 
Y_{n+2}^{L-1}[\mathcal{Z}_{1},\mathcal{Z}_{2},\cdots,
\hat{\mathcal{Z}}_{{n}_{AB}},\mathcal{Z}_A,\mathcal{Z}_B]\bigg|_{A,B \to \hat{B}}
\,\bigg) 
\,,
\label{ForwardPrelim}
\ee   
where adapting \reef{hatA} and \reef{hatZ4} to the $n$-point case with $5 \to A$ and $6 \to B$ identifies
\eq
\hat{\mathcal{Z}}_{{n}_{AB}}=(n-1,n)\bigcap(A,B,1)
~~~~\text{and}~~~~
\hat{B}=(A,B)\bigcap (n-1,n,1)
\,.
\eqe
Here, $f(A,B,n-1,n,1)$ represents the kinematic function which includes the 
cut  propagator as well as possible Jacobian factors that arise from solving the single cut constraint, $\langle A,B,\hat{n},{1}\rangle=0$. We will determine this function shortly, but first we will address one important missing piece: what to do about the $Z_A$ and $Z_B$ momentum twistors and how the extra loop momenta integral emerges from the forward limit. That is the next step.

%%%%%%%%%%%%%%%%%%%%%%%%%%%%
\compactsubsection{The integration measure.}
Since the forward limit is taken of a higher point amplitude/integrand, we need to devise a way to remove the information of the two extra external legs. The most naive proposal is to apply the following integration:
\eq
\int {d^{4|4}\mathcal{Z}_{A} d^{4|4}\mathcal{Z}_{B}}\,.
\label{Measure1}
\eqe
Surprisingly, this is the correct answer! The reason it is correct is rather non-trivial. Let us first consider the bosonic part of the integration. The integration over $Z_A$ and $Z_B$ can be decomposed into two pieces, one is the integration over all possible lines $(A,B)$, and the other is the movement of $Z_A$ and $Z_B$ along a particular line $(A,B)$. To aid this separation, consider the following $GL(2)$ transformation on $(Z_A,Z_B)$,
\be
\bigg(\begin{array}{c}Z_{A'} \\Z_{B'}\end{array}\bigg)
~=~
\bigg(\begin{array}{cc}c_{A'}{}^{A} & c_{A'}{}^{B} \\ c_{B'}{}^{A} & c_{B'}{}^{B}\end{array}\bigg)\,
\bigg(\begin{array}{c}Z_{A} \\Z_{B}\end{array}\bigg)\,.
\label{GL2}
\ee
The above $2\times2$ matrix exactly parametrize the movement along the line $(A,B)$, because the new pair $(A',B')$ defines the same line as $(A,B)$. In the forward limit we are sending $Z_A$ and $Z_B$ on a given line to the intersection of $(A,B)$ with the plane $(n-1,n,1)$, so this limit  corresponds to a particular solution for the $GL(2)$ matrix. In light of this, it will be convenient to separate the bosonic integral as
\eq
\int {d^4Z_{A} d^4Z_{B}} = \int \frac{d^4Z_{A} d^4Z_{B}}{{\rm Vol}[GL(2)]}\int_{GL(2)}\,.
\eqe
We have separated the $GL(2)$ part of the integration $\int_{GL(2)}$ from the $Z_A,Z_B$ integration. The $\text{Vol}[GL(2)]$ in the denominator indicates that as one integrates over the $4\times2$ dimensional space of $Z_A$ and $Z_B$, one needs to mod out the $2\times2$ $c$-matrix in \reef{GL2} that parameterizes an arbitrary $GL(2)$ transformation. The explicit integration measure for $\int_{GL(2)}$ can be fixed by requiring it to be $SL(2)$ invariant and having $GL(1)$ weight $4$ in both $A$ and $B$.\footnote{This follows from the fact that there are four-components in $Z_A$ and $Z_B$.} This fixes the form to be
\eq
\int_{GL(2)} =\int \langle c_{A'}dc_{A'}\rangle\langle c_{B'}dc_{B'}\rangle\langle c_{A'}c_{B'}\rangle^2\,,
\label{GL2measure}
\eqe 
where $\langle c_{A'}c_{B'}\rangle=
c_{A'}{}^{A}c_{B'}{}^B-c_{A'}{}^{B}c_{B'}{}^A$. After one has separated out the $GL(2)$ integral, the remaining integration measure can be naturally related to the $y_0$ measure. To see this note that after stripping off the $GL(2)$ part, the remaining measure is purely integrating over all distinct lines $(A,B)$. Recall that distinct lines in twistor space define distinct points, this tells us that this measure is precisely proportional to $\int d^4y_0$. The precise momentum twistor integral that is equivalent to the loop-integral over $y_0$ is
\eq
 \int d^4y_0
~=~
 \int \frac{d^4Z_{A} d^4Z_{B}}{{\rm Vol}[GL(2)]\langle AB\rangle^4}
\,,
\eqe
where the four extra factors of $\langle AB\rangle$ in the denominator are necessary for the measure to be projective. The angle bracket $\langle AB\rangle$ breaks the $SL(4)$ invariance because it picks only the angle spinor piece of the momentum twistors. This breaks dual conformal invariance --- but that is expected because the  $d^4y_0$ inverts non-trivially under dual conformal inversion. From our example \reef{int}, we see that the $\langle AB\rangle^4$ factor in the measure is exactly canceled but the same factor appearing when we rewrote the box-integral in momentum twistor space. This is a general feature which follows from (or, if you prefer, is necessary for) the dual conformal invariance of the loop-integrand. Henceforth, we simply implicitly assume the cancellation of the $\langle AB\rangle^4$'s. Let us for later reference write what the 1-loop 4-point superamplitude looks like when dressed in full momentum twistor regalia:
\be
 \mathcal{A}_4^\text{1-loop}[1234]
 ~=~
 - \mathcal{A}_4^{\rm tree}[1234]~ 
 \int \frac{d^4Z_{A} d^4Z_{B}}{{\rm Vol}[GL(2)]}~
 \frac{\langle1234\rangle^2}
 {\langle AB12\rangle \langle AB23\rangle \langle AB34\rangle \langle AB41\rangle}\,.
 \label{intfinal}
\ee

Back to the forward-limit discussion. 
To integrate over all possible configuration of the forward limit, then one should only integrate over all distinct lines $(A,B)$. This requires us to remove the $GL(2)$-part of the integration in \reef{Measure1}. Thus we have two problems to solve, how to put the higher-point amplitude on the forward limit and how to remove the $GL(2)$ redundancy. Fortunately, we can scare two birds with one stone!\footnote{No need to be aggressive and hurt any birds.} We begin by simply presenting the resolution: the correct prescription for the computation of the forward limit is
\be
\mathcal{A}^{\rm tree}_{n,\text{MHV}} \int \frac{d^{4|4}\mathcal{Z}_{A} d^{4|4}\mathcal{Z}_{B}}{{\rm Vol}[GL(2)]}\int_{\rm GL(2)}
[A,B,n-1,n,1]
\times
Y_{n+2}^{L-1}[\mathcal{Z}_{1},\mathcal{Z}_{2},\cdots,
\hat{\mathcal{Z}}_{{n}_{AB}},\mathcal{Z}_A,\mathcal{Z}_{\hat{B}}]\,,
\label{ForwardPrelim2}
\ee
where $\mathcal{Z}_B$ is sent to the intersection $\hat{B}=(A,B)\bigcap (n-1,n,1)$. Notice the appearance of the factor $[A,B,n-1,n,1]$. The role this factor plays is two-fold: 
\begin{itemize}
  \item It contains the invariants $\langle A, n-1,n,1\rangle$ and $\langle B, n-1,n,1\rangle$ in the denominator. The vanishing of these invariants is precisely the forward limit, and therefore these poles can be used to localize the $GL(2)$ integral on to the forward limit (the two birds fly). 
  \item It also contains the factor $1/\langle A,B, n,1\rangle$, which is precisely the cut propagator.
\end{itemize}
Hence the $GL(2)$ integration is understood to encircle poles that correspond to the forward limit. One may ask if $[A,B,n-1,n,1]$ is the unique function that satisfies the above two points? The answer is no, however, it can be easily justified by dual conformal invariance. The recursion better preserve this symmetry. With $Y_{n+2}^{L-1}$ already an invariant, $[A,B,n-1,n,1]$ is the unique invariant that satisfies the above two properties. Thus, using symmetry arguments we did not need to know {\em a priori} what the function $f(A,B,n-1,n,1)$ in \reef{ForwardPrelim} should be; it is whatever $[A,B,n-1,n,1]$ evaluates to once the $GL(2)$ integral is localized. This is admittedly rather abstract, but we are going to realize the contents of the discussion here explicitly when we compute the 4-point 1-loop amplitude in Section \ref{s:bcfwMTloop4pt}. 

Finally, we need to sum over all $\cn=4$ SYM states that can run in the forward limit loop. In \reef{ForwardPrelim2} this is naturally achieved in a way that preserves the dual superconformal symmetry by simply extending the bosonic momentum twistor integration to include the Grassmann-components, $\chi_A$ and $\chi_B$. We are now ready to put everything together.

%%%%%%%%%%%%%%%%%%%%%%%%%%%%
\compactsubsection{Result of the BCFW recursion for $L$-loop integrands.}
Summarizing the preceding discussion, the loop-level BCFW recursion relation is given by
\eqa
\label{LoopBCFW}
\mathcal{A}^{L\text{-loop}}_n \!\!&=&\!\!
\mathcal{A}^{\rm tree}_{n,\text{MHV}}\bigg\{
Y^{L}_{n-1}\big(\mathcal{Z}_1,\cdots,\mathcal{Z}_{n-1}\big)
\\
&&\nonumber
+\sum_{j=3}^{n-2} \big[j-1,j,n-1,n,1\big]
\,
Y^{L_1}_\text{L}\big(\mathcal{Z}_{I_j},\mathcal{Z}_{j},\mathcal{Z}_{j+1},\cdots,\hat{\mathcal{Z}}_{{n}_j}\big)
\,
Y^{L_2}_\text{R}\big(\mathcal{Z}_{I_j},\mathcal{Z}_{1},\mathcal{Z}_{2},\cdots,\mathcal{Z}_{j-1}\big) 
\\
&&
\nonumber
+
\int \frac{d^{4|4}\mathcal{Z}_{A} d^{4|4}\mathcal{Z}_{B}}{{\rm Vol}[GL(2)]}\int_{GL(2)}
[A,B,n-1,n,1]\,
Y_{n+2}^{L-1}[\mathcal{Z}_{1},\mathcal{Z}_{2},\ldots,
\hat{\mathcal{Z}}_{{n}_{AB}},\mathcal{Z}_A,\mathcal{Z}_{\hat{B}}]
\bigg\}\,.
\eqae   
In the second line, $L_1$ and $L_2$ are summed over subject to $L_1+L_2=L$, as are the Grassmann degrees $K_1+K_2 = K-1$, 
and we have 
\eq
\hat{\mathcal{Z}}_{{n}_j}=(n-1,n)\bigcap(1,j-1,j)\,,
~~~~
 \mathcal{Z}_{I_{j}}=(j,j-1)\bigcap(n-1,n,1)\,,
~~~~
\hat{\mathcal{Z}}_{{n}_{AB}}=(n-1,n)\bigcap(A,B,1)\,.
\eqe

Before moving on to an explicit application of the loop-integrand recursion relations, it is important to note that \reef{LoopBCFW} provides us with the tool to prove dual conformal properties of loop amplitudes, just as what was done with the tree-level recursion. Assuming the $n$-point $L'<L$-loop as well as the $(n\!+\!2)$-point  $(L\!-\!1)$-loop amplitude is given by a MHV tree-amplitude times a dual conformal invariant function, then through \reef{LoopBCFW} the $n$-point $L$-loop amplitude will have the same property.

We will now apply the recursion relations developed in this section to show how the  4-point 1-loop integrand \reef{int}, 
 can be derived recursively from the recursion relation with the input of just a tree-amplitudes. Sharpen your pencils and keep your eraser close at hand.
 
%%%%%%%%%%%%%%%%%%%%%%%%%%%%%%%%%%%%
\subsection{Example: 4-point 1-loop amplitude from recursion}
\label{s:bcfwMTloop4pt}
%%%%%%%%%%%%%%%%%%%%%%%%%%%%%%%%%%%%%
The 4-point 1-loop amplitude is the simplest example that can illustrate all the novel details in the loop-recursion. Let us examine the potential terms in the recursion formula \reef{LoopBCFW}:

The \emph{first term} with  $Y^{L=1}_{3}$ is absent. This is because there are no 3-point 1-loop amplitudes. Another way to understand this is that this contribution came from the pole at $w \to \infty$. If we sneak-peak at the answer for the 4-point 1-loop amplitude \reef{int}, we realize that while the shifted MHV prefactor does have a   $w \to \infty$ pole (as we saw and used in Section \ref{s:bcfwMTtree}), its residue is actually zero for the 1-loop integrand because the $1/\langle AB\hat{4}1\rangle \to 0$ as $w\to \infty$. In other words, this is a consistent picture for the absence of the first term $Y^{L=1}_{3}$ in \reef{LoopBCFW}.

The \emph{second term} in \reef{LoopBCFW} is absent because $Y^{L=1}_{3}=0$. This is consistent with \reef{int} not having any momentum-independent poles at finite $w$.

The \emph{third term} in \reef{LoopBCFW} is
\be
\mathcal{A}^\text{1-loop}_4
=
\mathcal{A}^{\rm tree}_{4,\text{MHV}}
\int \frac{d^{4|4}\mathcal{Z}_{A} d^{4|4}\mathcal{Z}_{B}}{{\rm Vol}[GL(2)]}\int_{GL(2)}[A,B,3,4,1]
\times
Y_{6}[\mathcal{Z}_{1},\mathcal{Z}_{2},\mathcal{Z}_{3},\hat{\mathcal{Z}}_{{4}_{AB}},\mathcal{Z}_A,\mathcal{Z}_{\hat{B}}]\,.~~
\label{rec4pt1loop}
\ee  
Unfortunately, we now have to evaluate this thing. 

$Y_{6}$ is the tree-level NMHV dual conformal invariant for $n=6$, discussed previously in \reef{Final}:
\be
Y_{6}[\mathcal{Z}_{1},\mathcal{Z}_{2},\mathcal{Z}_{3},\hat{\mathcal{Z}}_{{4}_{AB}},\mathcal{Z}_A,\mathcal{Z}_{\hat{B}}]
~=~
[\hat{B},1,2,3,\hat{4}]+[\hat{B},1,2,\hat{4},A]+[\hat{B},2,3,\hat{4},A]\,.
\label{Y6ex}
\ee
The hatted momentum twistors can be found explicitly using  the intersection formulas \reef{TwistorPoint}. Since the twistors are defined projectively, one can freely include a scaling-factor:
\bea
(\hat{4})=
(3,4)\bigcap(A,B,1)&\implies&\;
\mathcal{Z}_{\hat{4}}=
\frac{1}{\langle3AB1\rangle}\big(\mathcal{Z}_{4}\langle3AB1\rangle-\mathcal{Z}_{3}\langle4AB1\rangle\big)\,, 
\label{hattedMT4}\\
(\hat{B})=(A,B)\bigcap(3,4,1)&\implies&\;
\mathcal{Z}_{\hat{B}}=\frac{1}{\langle A341\rangle}\big( - \mathcal{Z}_{A}\langle B341\rangle+\mathcal{Z}_{B}\langle A341\rangle\big)\,.~~~~~~~
\label{hattedMT}
\eea
For convenience, we picked overall factors such that the `hatted' twistors have the same projective weights as the un-hatted ones. 
Note that some 4-brackets remain unshifted: $\<3,\hat{4},.\,,.\> = \<3,4,.\,,.\>$
and $\<A,\hat{B},.\,,.\> = \<A,B,.\,,.\>$. 

Let us first do the fermionic integrals in $d^{4|4}\mathcal{Z}_{A} d^{4|4}\mathcal{Z}_{B}$, i.e.~$d^4\chi_Ad^4\chi_B$. 
We have to saturate the Grassmann integrals, so it is only relevant to look at the $\chi_A$- and $\chi_B$-terms in the Grassmann delta functions. Begin with the 5-bracket $[A,B,3,4,1]$ that multiplies each of the three terms in $Y_6$. Its Grassmann delta function involves
\be
  [A,B,3,4,1] 
  ~\propto~
  \d^{(4)}\big( \chi_A \< B 341\> - \chi_B \< A 341\> + \dots \big)\,.
  \label{AB341}
\ee
It follows from \reef{hattedMT} that
$\chi_{\hat{B}} \propto 
\chi_A\langle B 3 4 1\rangle-\chi_B\langle A 3 4 1\rangle$, so any appearance of 
$\chi_{\hat{B}}$ in the three 5-brackets in \reef{Y6ex} vanishes on the support of the $\d^{(4)}$ in \reef{AB341} under the $\int d^4\chi_Ad^4\chi_B$-integral. In particular, the only contribution from $\chi_A$, $\chi_B$ in $[\hat{B},1,2,3,\hat{4}]$ is through $\chi_{\hat{B}}$, so we immediately conclude that
\be
  \int d^4\chi_Ad^4\chi_B ~[A,B,3,4,1] \times [\hat{B},1,2,3,\hat{4}] = 0\,.
\ee
In the next case, $[\hat{B},1,2,\hat{4},A]$, we have 
$\d^{(4)}(\chi_{\hat{B}}\<12\hat{4}A \>+ \chi_A\, \<\hat{B}12\hat{4}\> + \dots)$. 
As before the $\chi_{\hat{B}}$-term can be dropped. Moreover, one can show that $\<\hat{B}12\hat{4}\>$ vanishes (see Exercise \ref{ex:5schouten} below), so we conclude
\be
  \int d^4\chi_Ad^4\chi_B ~[A,B,3,4,1] \times [\hat{B},1,2,\hat{4},A] = 0\,.
\ee
\exercise{ex:5schouten}{The 3-term Schouten identity for angle and square spinors is the statement that 3 vectors in a plane are linearly dependent. As 4-component objects, the momentum twistors $Z^I$, $I=(\dot{a},a)$, similarly satisfy a 5-term Schouten identity
\be
\langle i,j,k,l\rangle Z_m+\langle j,k,l,m\rangle Z_i+\langle k,l,m,i\rangle Z_j+\langle l,m,i,j\rangle Z_k+\langle m,i,j,k\rangle Z_l
~=~0\,.
\label{FiveSchouten}
\ee
Use \reef{FiveSchouten} to derive the two identities
\be
 \<\hat{B}12\hat{4}\> ~=~ 0 \,,
 ~~~~~~~~~~
 \<234 \hat{B}\>   ~=~ - \frac{\<1234\>\<34AB\>}{\<A341\>}\,.
 \label{5tsch}
\ee
Use $\<AB \hat{4}1\> = 0$, it follows from \reef{hattedMT4}.
}
With the help of the second identity in \reef{5tsch}, one finds that the result of integrating the $\delta^{(4)}$'s in $[A,B,3,4,1] \times [\hat{B},2,3,\hat{4},A]$ gives 
$\<1234\>^4\<34AB\>^4$. 

In conclusion, after Grassmann integration, only the third 5-bracket in \reef{Y6ex} contributes. After some simplifications one finds
\be
   \int d^4\chi_Ad^4\chi_B ~[A,B,3,4,1] \times [\hat{B},2,3,\hat{4},A] 
   ~=~
   \frac{\<1234\> \<AB 34\>}{\<A234\> \<B341\>} \times I_4(A,B)\,,
   \label{grassdone}
\ee
where
\be
  I_4(A,B) = \frac{\langle1234\rangle^2}
 {\langle AB12\rangle \langle AB23\rangle \langle AB34\rangle \langle AB41\rangle} 
\ee
is the answer we expect, cf.~\reef{intfinal}.
\exercise{}{Derive \reef{grassdone}.
}
Now the recursion relations \reef{rec4pt1loop} instruct us to finish the forward limit by performing the $GL(2)$ integral:
\be
\mathcal{A}^\text{1-loop}_4
=
\mathcal{A}^{\rm tree}_{4,\text{MHV}}
\int \frac{d^{4}{Z}_{A} d^{4}{Z}_{B}}{{\rm Vol}[GL(2)]}~
\int_{GL(2)} 
I_4(A,B)  \times \frac{\<1234\> \<AB 34\>}{\<A234\> \<B341\>}
\label{rec4pt1loop2}\,.
\ee  
 This can be done by first doing a $GL(2)$ rotation \reef{GL2} of $Z_A, Z_B$ and then integrating over the $GL(2)$ parameters. Since the integral is $GL(1)$ invariant, we can fix the scale in the $GL(2)$ matrix and  set $c_{A'}{}^{A}=c_{B'}{}^{B}=1$. With this  `gauge fixing' we have 
\be
\bigg(\begin{array}{c}Z_{A} \\Z_{B}\end{array}\bigg)
~\to~
\bigg(\begin{array}{cc}1 & c_{A'} \\ c_{B'} & 1\end{array}\bigg)\,
\bigg(\begin{array}{c}Z_{A} \\Z_{B}\end{array}\bigg)\,.
\label{GL2gf}
\ee
The result of this transformation on the 4-brackets is
\be 
\langle ABij\rangle
~\rightarrow~ \langle ABij\rangle\langle c_{A'}c_{B'}\rangle\,,
~~~~~~~
\begin{array}{rcl}
\langle Aijk\rangle&\rightarrow& 
 \langle Aijk\rangle  + c_{A'}\,\langle Bijk\rangle\,,
  \\[1mm]
\langle Bijk\rangle&\rightarrow& c_{B'} \, \langle Aijk\rangle
  + \langle Bijk\rangle \,,
\end{array}  
\label{GL2Change}
\ee
and the gauge fixing means that $\langle c_{A'}c_{B'}\rangle = 1 - c_{A'} c_{B'}$. Also, $\langle c_{A'}dc_{A'}\rangle = dc_{A'}$ and
$\langle c_{B'}dc_{B'}\rangle = dc_{B'}$. So including the appropriate $GL(2)$ measure \reef{GL2measure}, we then have
\bea
\nonumber
\mathcal{A}^\text{1-loop}_4
&=&
\mathcal{A}^{\rm tree}_{4,\text{MHV}}
\int \frac{d^{4}{Z}_{A} d^{4}{Z}_{B}}{{\rm Vol}[GL(2)]}~I_4(A,B)~ \<1234\> \<AB 34\> 
\\
&&\hspace{1cm}
\times
\int 
 \frac{ dc_{A'}\,  dc_{B'}}
  {\Big(1 - c_{A'} c_{B'}\Big)\,
    \Big( \langle A234\rangle  + c_{A'}\,\langle B234\rangle \Big) \,
    \Big( c_{B'} \, \langle A341\rangle + \langle B341\rangle \Big) }
    \,.~~~~~~~~
\label{rec4pt1loop3}
\eea
Now the plan all along was the $GL(2)$ integration was supposed to localize us on the forward limit. So consider the denominator factor 
$\big( c_{B'} \, \langle A341\rangle + \langle B341\rangle \big)$. The vanishing of this expression is the statement that $\hat{Z}_{B} = Z_B + c_{B'} Z_A$ is sent to the intersection point of the line $(A,B)$ and the plane $(3,4,1)$: but this is exactly part of the forward limit  
$\hat{Z}_{B} \to Z_{\hat{B}} = (A,B)\bigcap (3,4,1)$. So to realize this, we take the contour \reef{rec4pt1loop3} in the $c_{B'}$-plane to surround the pole 
$c_{B'*} = -\langle B341\rangle/\langle A341\rangle$. 
Now we also want to send $A$ to the intersection point $\hat{B}$, but the integral \reef{rec4pt1loop3} appears not to have a pole that achieves this. However, when we evaluate the $c_{B'}$-integral to localize $B \to \hat{B}$, the factor 
$\big(1 - c_{A'} c_{B'}\big)$ actually develops the desired pole, namely
$c_{A'*} = -\langle A341\rangle/\langle B341\rangle$.  Let's just do it:
\bea
\nonumber
  &&
  \int_{\mathcal{C}(c_{A'*})} dc_{A'}\int_{\mathcal{C}(c_{B'*})} dc_{B'}~
  \frac{1}{\Big(1 - c_{A'} c_{B'}\Big)\,
    \Big( \langle A234\rangle  + c_{A'}\,\langle B234\rangle \Big) \,
    \Big( c_{B'} \, \langle A341\rangle + \langle B341\rangle \Big)}\\
\nonumber
    &&~~~=~
      \int_{\mathcal{C}(c_{A'*})} 
  \frac{dc_{A'}}{\Big( \langle A341\rangle  + c_{A'} \langle B341\rangle\Big)\,
    \Big( \langle A234\rangle  + c_{A'}\,\langle B234\rangle \Big)}\\
\nonumber
    &&~~~=~
  \frac{1}{
   \langle A234\rangle \langle B341\rangle + \langle A341\rangle\,\langle B234\rangle}\\
    &&~~~=~ - 
  \frac{1}{\langle AB34\rangle \langle 1234\rangle}   \,.
\eea
In the last line we used the 5-term Schouten identity \reef{FiveSchouten}.
Plugging this result into \reef{rec4pt1loop3}, the factors 
$\langle AB34\rangle \langle 1234\rangle$ cancel and we are left with
\be
\mathcal{A}^\text{1-loop}_4
~=~
-\mathcal{A}^{\rm tree}_{4,\text{MHV}}
\int \frac{d^{4}{Z}_{A} d^{4}{Z}_{B}}{{\rm Vol}[GL(2)]}~I_4(A,B)\,.
\label{rec4pt1loop4}
\ee
This is the correct result, as we discussed in Section \ref{s:bcfwMTloop}.

Note that in this derivation, the $GL(2)$ integral ended up localizing the integrand on the forward limit, where the poles that were used in this localization was given by the extra $[A,B,3,4,1]$. This precisely realizes the idea we described around \reef{ForwardPrelim2}. Now you might be a little concerned that we could have chosen to localized on ``non-forward" poles in the $GL(2)$ integral instead, but the answer would have been the same, as guaranteed by the large-$c_{A',B'}$ falloff of the integrand.

Finally, you may find it discouraging that it takes much more work and sophistication to calculate even the simplest of all $\cn = 4$ SYM amplitudes with BCFW than it did with the generalized unitarity method, as we showed in Section \ref{s:genunit}. However, while it is not directly practical, it is morally encouraging --- and perhaps even fascinating --- that all information about the 4-point MHV 1-loop amplitude is encoded already in the 6-point NMHV tree-level amplitude. This is a realization of an interesting connection between amplitudes with different number of particles
$n$, different N$^K$MHV levels, and different loop-orders $L$. 

%%%%%%%%%%%%%%%
\subsection{Higher loops}
The planar loop-integrand recursion relations studied above can also be directly applied to higher-loop order in the planar limit of $\cn=4$ SYM. This was already shown in \reef{LoopBCFW}. However, in contrast to the unitarity approach, the integrands obtained from recursion generally contain spurious poles. Local poles (non-spurious) refer to propagator-like poles, these take the form of $1/\langle i,i-1,j,j-1\rangle$ or $1/\langle A,B,i-1,i\rangle$ in momentum twistor space. Spurious poles, on the other hand, could take the form $1/\langle A,B,\hat{4},2\rangle$; this is non-local in that it does not arise from a propagator in the loop-diagrams.
At 4-point, since there is only one term in the recursion, such spurious poles must vanish by themselves, and indeed the final result is free of spurious poles. However, at higher-points the spurious poles cancels between various terms in the recursion relations, and this makes it difficult to carry out the integration of the loop-integrand to obtain the actual amplitude. Spurious poles are a hallmark of BCFW recursion relations --- we already discussed this for tree-level BCFW at the end of Section \ref{s:bcfw}. BCFW builds in unitarity and gauge-invariance at the expense of manifest locality. While it provides us with a method to compute  loop-integrands, it leads to complications as one eventually has to integrate these non-local functions in momentum space. 

Given the large amount of symmetry enjoyed by $\mathcal{N}=4$ SYM --- superconformal symmetry and dual conformal symmetry as well as their enhancement to the Yangian --- it is tempting to be ambitious and ask if it is possible to manifest \textit{both} locality and dual conformal invariance at the same time. Certainly the unitary method discussed previously would suffice for this purpose, since the dual conformal invariant scalar integrals are local. However, when extended beyond 4-point, the number of dual conformal invariant integrals becomes large and not all of them may contribute to a given amplitude.

At 1-loop level, the 1-loop box integral in \reef{intfinal} is the only available dual conformal invariant local 4-point integral. In other words, dual conformal symmetry forces the triangle and bubble-contributions to be absent in $\cn=4$ SYM. Could it be that all amplitudes $\cn=4$ SYM are fixed by similar considerations? To study this involves \emph{maximal cuts} and \emph{Leading Singularities} --- and some of the principles involved also extend beyond the planar limit and to SYM with less supersymmetry. This is currently an active area of research and we will discuss the basic setup in detail in the following section. 

%%%%%%%%%%%%%%%%%%%%%%%%%%%%%%% 
%%%%%%%%%%%%%%%%%%%%%%%%%%%%%%% 
%%%%%%%%%%%%%%%%%%%%%%%%%%%%%%% 
\newpage
\setcounter{equation}{0}
\section{Loops III: Leading Singularities and on-shell diagrams}
\label{s:loops3}
%%%%%%%%%%%%%%%%%%%%%%%%%%%%%%% 
%%%%%%%%%%%%%%%%%%%%%%%%%%%%%%% 
%%%%%%%%%%%%%%%%%%%%%%%%%%%%%%% 

Unitarity cuts in $D$-dimensions (see Section \ref{s:loops}) can involve at most
 $D$ cut propagators per loop since each loop-momentum only has $D$ components. When the maximum number of propagators, $D\times L$, are cut, the unitarity cut is called a {\bf \em maximal cut} \cite{MaximalCut,LS}. The maximal cuts are very useful for determining the integrand, in particular in 4d planar $\cn = 4$ SYM.
\example{As an example of a maximal cut, consider the quadruple cut of the 1-loop $n$-point amplitude in 4d:
\be
\raisebox{-17mm}{\includegraphics[scale=0.5]{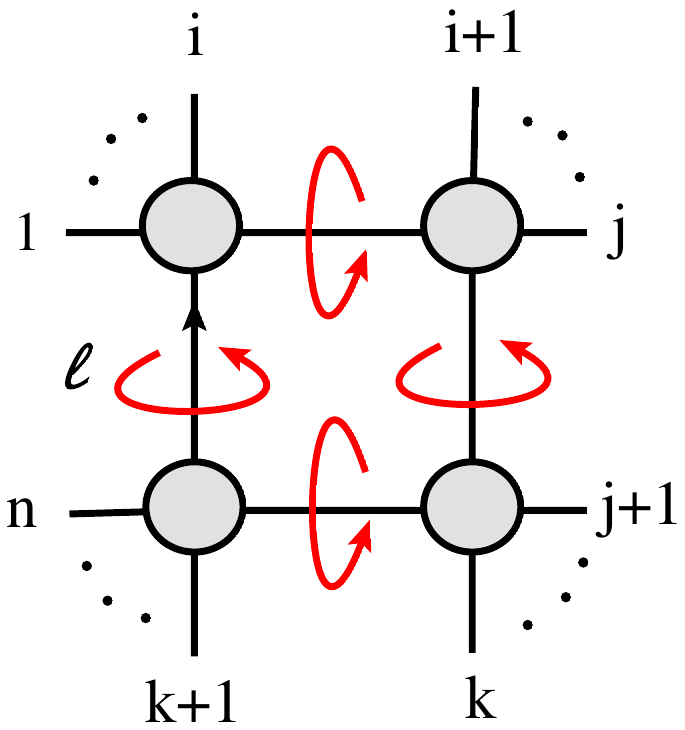}}
~~~~
\begin{array}{l}
\text{Four propagators are put on-shell:}\\[3mm]
\ell^2=\big(\ell-K_1^{(i)}\big)^2=\big(\ell-K_2^{(i)}\big)^2=\big(\ell-K_3^{(i)}\big)^2=0\,.
\end{array}
\label{LSconstraint}
\ee
Here $K^{(i)}_1=p_1+\cdots+p_i$, $K^{(i)}_2=p_{i+1}+\cdots+p_{j}$, $K^{(i)}_3=p_{j+1}+\cdots+p_{k}$ and $K^{(i)}_4=p_{k+1}+\cdots+p_{n}$. No more propagators can be put on-shell in 4d since the loop-momentum only has four components. 
}
The result of a maximal cut is a product of on-shell tree amplitudes, $A_{n_1}^\text{tree} \cdots A_{n_j}^\text{tree}$, appropriately summed over all possible intermediate states, with the loop-momenta evaluated on the solutions to the cut constraints. For example, for the 1-loop box in 4d $\cn=4$ SYM, the value of the maximal cut \reef{LSconstraint} is 
$\int d^4\eta_I\,A_{n_1}^\text{tree}A_{n_2}^\text{tree}A_{n_3}^\text{tree}  A_{n_4}^\text{tree}$ evaluated on the 2 solutions to the quadratic loop-momentum constraints. At $L>1$, the cut constraints generically have 2$^L$ distinct solutions, however, there are situations where there are not enough propagators to cut; a simple example in 4d is the following  2-loop double-box integral\label{panic}
\eq
\raisebox{-10mm}{
\includegraphics[scale=0.4]{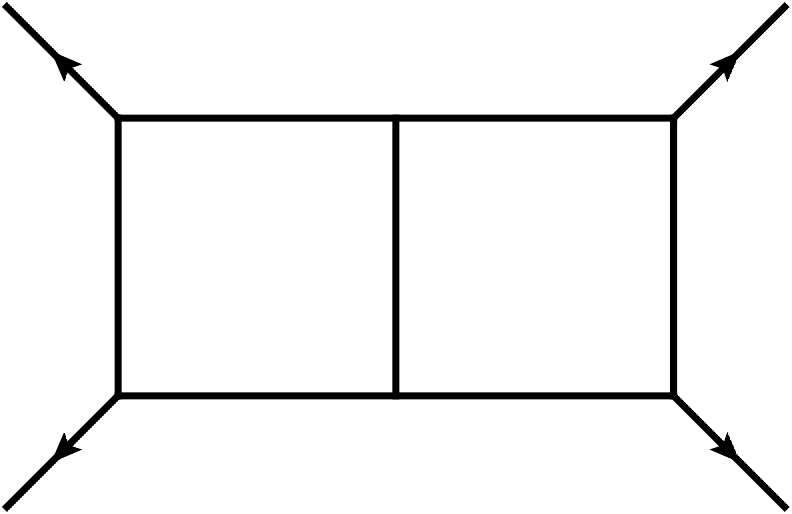}}\,.
\label{2Lbox}
\eqe 
It has only 7 propagators, but we need to take $2\times 4=8$ propagators on-shell for a maximal cut. For such cases, the solution space for the loop-momenta is not a set of isolated points but rather a continuous multi-dimensional manifold. If we choose to impose the cut constraint on one loop-momentum at a time, new poles appear and they can be used to fix the remaining degrees of freedom, again leaving us with a set of isolated solutions for the loop momenta. We demonstrate this explicitly for the double-box \reef{2Lbox} in Section \ref{s:compLS}.

The method of generalized unitarity is to find an integrand that reproduces all the unitarity cuts, including of course all the maximal cuts. But how exactly do we treat the distinct solutions to the maximal cut constraints? There are two ways to proceed:\footnote{The maximal cut was formulated in \cite{LS} for 1-loop amplitudes of $\mathcal{N}=4$ SYM and generalized to multi-loop amplitudes in \cite{MaximalCut}. A more detailed review is offered in \cite{Review1}. 
}
\begin{enumerate}
  \item {\em Appropriate sampling over all solutions}. We require that the correct integrand matches the maximal cut evaluated on a sampling of all $2^L$ solutions, with each solution given an appropriate weight. 
  At 1-loop there are just 2 solutions and the proper weight is $1/2$ for both, thus in effect averaging over the 2 solutions, as in \reef{CDexpr} and \reef{BoxCoef}. For higher-loops, one starts with a set of integrals that integrate to zero. The appropriate weight for each solution is determined by the requirement that their contributions to the vanishing integrals need to sum to zero. Explicit examples and further discussions at 2-loop order can be found in \cite{2LoopBasisCut}.
  \item {\em Match each solution.} We require  the integrand to  reproduce \emph{each} of the cut solutions individually. In this approach, the individual cut solutions are treated as independent entities and the resulting value for the maximal cut evaluated on each solution is called a {\bf \em Leading Singularity (LS)}.  The name reflects that these objects are the residues of the most singular configuration of the loop-integrand (for generic external kinematics). Note that, despite the name, these contributions are finite.
\end{enumerate}

We focus here on the second approach. A major motivation is that all 
planar loop amplitudes of $\mathcal{N}=4$ SYM can be written as a  
linear combination of dual conformal invariant ``unit Leading Singularity integrands" (to be introduced below)  \cite{UnitLS}.  The characterization of Leading Singularities turns out to be quite interesting mathematical problem; it has been studied in the recent paper \cite{ArkaniHamed:2012nw}. The Leading Singularities offer insight into the structure of planar  $\mathcal{N}=4$ SYM amplitudes at all-loop orders, but to  obtain the actual amplitudes, one still needs to perform the loop-integrations; this is an area of active research.  
While the notion of dual conformal invariance is only well-defined in the planar limit, it is a well-defined  question whether the full non-planar loop-amplitudes of  $\mathcal{N}=4$ SYM  can also be determined by the Leading Singularities. This is another current area of investigation. 

We begin our study of the Leading Singularities at 1-loop order. All amplitudes in this section are in 4d planar $\cn=4$ SYM.

%%%%%%%%%%%%%%%%%%%%%%%%%%%%%%%%%%%%
\subsection{1-loop Leading Singularities}
\label{s:LS1loop}
%%%%%%%%%%%%%%%%%%%%%%%%%%%%%%%%%%%%%

To build intuition for the Leading Singularities at 1-loop order, we start with the simplest case of 4-point, then consider the new features  at 5-point, and finally generalize to $n$-point. 

%%%%%%%%%%%%%%%%%%%%%%%%%%%%%%%%%%%%
\compactsubsection{4-point.}
%%%%%%%%%%%%%%%%%%%%%%%%%%%%%%%%%%%%%
For $n=4$, the maximal cut conditions \reef{LSconstraint} are simply
\be
\raisebox{-14mm}{\includegraphics[scale=0.5]{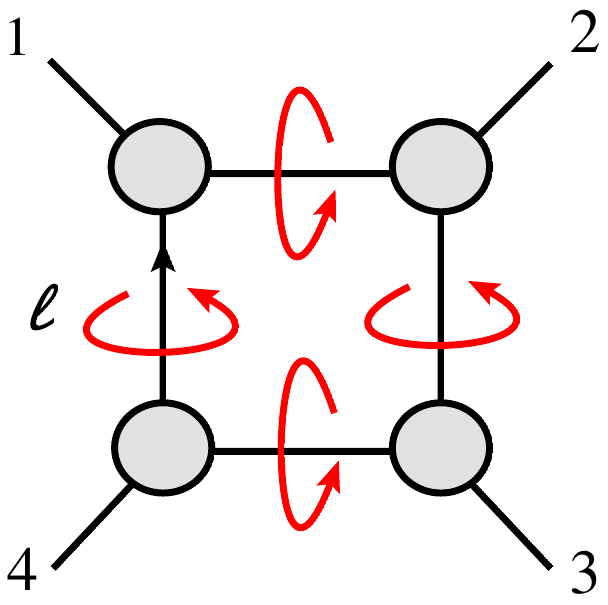}}
~~~~
  \ell^2=(\ell-p_{1})^2=(\ell-p_1-p_2)^2=(\ell-p_1-p_2-p_3)^2=0\,.
\label{LSconstraint4pt}
\ee 
In dual variables, this is simply $y^2_{01}=y^2_{02}=y^2_{03}=y^2_{04}=0$ (see Section \ref{s:bcfwloop}). 
And translating that to momentum twistor space (as in the early part of Section \ref{s:bcfwMTloop}), we have
\eq
\langle AB12\rangle=\langle AB23\rangle=\langle AB34\rangle=\langle AB41\rangle=0\,.
\label{AB1234}
\eqe 
The geometric interpretation of the cut constraints \reef{AB1234} is that $(A,B)$ is a line in $\mathbb{CP}^3$ that intersects each of the four lines $(1,2)$, $(2,3)$, $(3,4)$, and $(4,1)$. As anticipated from the quadratic nature of the constraint \reef{LSconstraint4pt}, there are two independent solutions. This is rather obvious geometrically:
\be
\text{(a)}\;~(A,B)=(1,3)
\vcenter{\hbox{\includegraphics[scale=0.5]{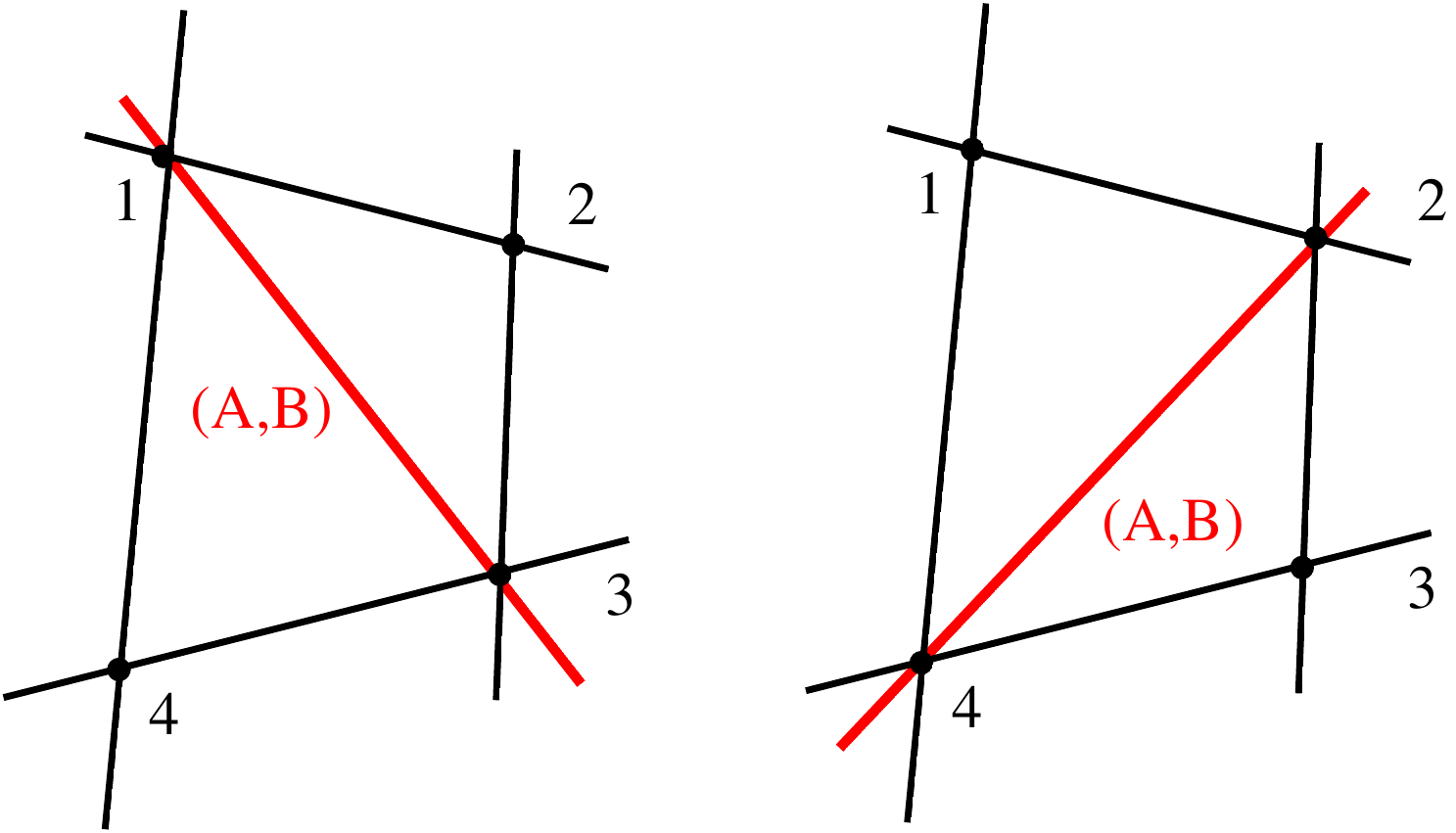}}}
~~
\text{(b)}\;~(A,B)=(2,4)\,.~~
\label{4ptLSSol}
\ee
Each of the four 3-point tree-amplitudes in the quadruple cut can be either MHV or anti-MHV. Recall that special kinematics apply to the 3-point amplitudes --- we summarize it here:
\be
  \text{MHV}\!\!\!\!
  \raisebox{-0.95cm}{\includegraphics[width=1.8cm]{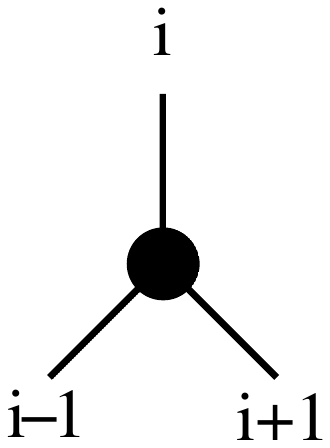}}
  |i-1] \propto |i] \propto |i+1]\,,
  ~~~~~~~~~
  \overline{\text{MHV}}\!\!\!\!
  \raisebox{-0.95cm}{\includegraphics[width=1.8cm]{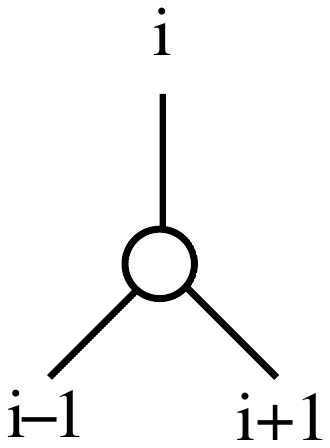}}
  |i-1\> \propto |i\> \propto |i+1\>\,.
  \label{specblob}
\ee
We use a black blob to indicate an MHV 3-point subamplitude (or vertex), and  a white blob for 3-point $\overline{\rm MHV}$ = anti-MHV.

Consider a configuration where two MHV subamplitudes are  adjacent, for example
\be
  \raisebox{-10mm}{\includegraphics[scale=0.4]{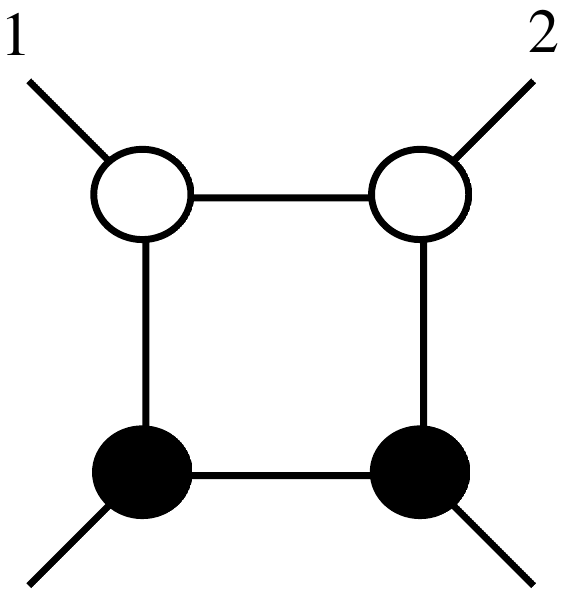}}.
  \label{4ptLSCutNo}
\ee
By the special kinematics \reef{specblob}, we must have $|1\> \propto |2\>$ which implies $s_{12} = -(p_1 + p_2)^2 = 0$. This is of course not true for generic  momenta $p_1$ and $p_2$. Hence we conclude that for generic external momenta, we are not allowed to have helicity configurations such as \reef{4ptLSCutNo} where two MHV or two anti-MHV subamplitudes are adjacent.

The only helicity options for the 4-point quadruple cut are therefore
\be
   \raisebox{-11mm}{\includegraphics[width=6cm]{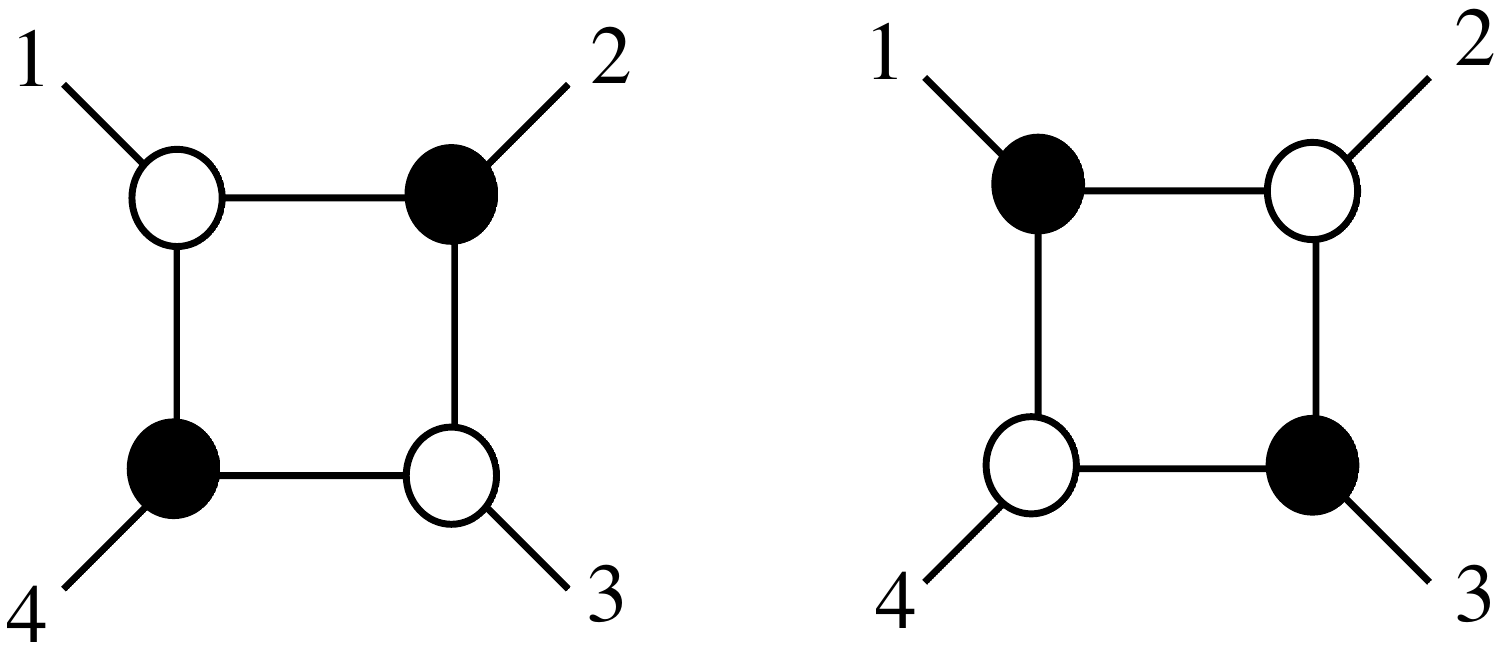}}.
   \label{heloptions}
\ee
Now these have to be evaluated on the kinematic solutions \reef{4ptLSSol}.  For the solution  (a) where the line $(A,B)$ is $(1,3)$, we can simply pick the momentum twistor of the loop line $\ell$ to be $Z_1 = ( \,|1\> , [\mu_1| \,)$. This means that  $|\ell\> \propto |1\>$ and that selects the kinematics where the vertex that line 1 attaches to is anti-MHV, i.e.~this picks the first helicity configuration in \reef{heloptions}. Likewise, solution (b) with $(A,B)=(2,4)$ selects the opposite helicity configuration. So we conclude that the maximal cuts have the two  solutions:
\be
\text{(a)}~~(A,B)=(1,3)\!\!\!
\raisebox{-10.5mm}{\includegraphics[width=6cm]{4ptLSSolHel}}
\text{(b)}~~(A,B)=(2,4)\,.
\label{4ptLSSolHel}
\ee
These two diagrams encode the {\bf \em Leading Singularities} for the 4-point 1-loop amplitude. 
The Leading Singularities, $\text{LS}_\text{(a)}$ and $\text{LS}_\text{(b)}$, are calculated as the product of the four subamplitudes summed over all intermediate states and evaluated on the respective solutions (a) and (b), \emph{times a Jacobian factor}. The Jacobian factor $1/J$ comes from a change of variables that converts the associated loop-integral over $\mathbb{R}^4$ to a contour integration with four contours encircling each of the four propagator poles in the quadruple cut.\footnote{Since the solutions to the cut conditions may be complex-valued, we should really consider the loop-integral as an integral over the real section in $\mathbb{C}^4$; that makes it more natural to convert to a contour integral encircling the poles corresponding to the on-shell propagators.}  
The conversion of the integral can be done via a change of variables 
$u_i=y^2_{0i}$, for $i=1,2,3,4$, giving
\be
\label{y0-ui}
\int \frac{d^4y_0}{y^2_{01}y^2_{02}y^2_{03}y^2_{04}}
~=~\int \frac{du_1}{u_1}\frac{du_2}{u_2}\frac{du_3}{u_3}\frac{du_4}{u_4}\,
J \,,
\ee 
where $J=\det(\partial y^\mu_0/\partial u_i)$ is the Jacobian.
As we show explicitly in the example below, the Jacobian is
\be
  J
  ~=~
  \frac{1}{y_{13}^2y_{24}^2}
  ~=~
  -\frac{\<12\>\<23\>\<34\>\<41\>}{\langle 1234\rangle^2}\,.
\label{theJ}
\ee
The Leading Singularity for the 4-point 1-loop amplitude is then
\bea
\nonumber 
\text{LS}_\text{(a)}
&=&
J
\int \left[\prod_{i=1}^4d^4\eta_{\ell_i}\right]
\bigg(
\mathcal{A}_3^{\rm \overline{MHV}}(-\ell_4,p_1,\ell_1)
\,\mathcal{A}_3^{\rm MHV}(-\ell_1,p_2,\ell_2)\\
&&\hspace{3.2cm}
\times\,\mathcal{A}_3^{\rm \overline{MHV}}(-\ell_2,p_3,\ell_3)
\,\mathcal{A}_3^{\rm MHV}(-\ell_3,p_4,\ell_4)
\bigg)\bigg|_{\ell = \ell^{(a)}}
\,.
\label{LS4AAAA}
\eea
A similar expression is found for $\text{LS}_\text{(b)}$. 
Evaluating the $\text{LS}_\text{(a)}$ and $\text{LS}_\text{(b)}$, one finds
\be
\text{LS}_\text{(a)}\,=\,
\text{LS}_\text{(b)}\,=\,
\mathcal{A}^{\rm tree}_4\,.
\ee 
\exercise{}{Evaluate the RHS of \reef{LS4AAAA} to show that $\text{LS}_\text{(a)} = \mathcal{A}^{\rm tree}_4$.
}
Now before exploring the Leading Singularities further, let us illustrate how the Jacobian is obtained. It can of course be calculated brute-force, but in the example below we carry out the calculation via a tour to momentum twistors.
\example{We calculate $J$ in \reef{y0-ui} via the momentum twistor formulation. From \reef{N4Answ2}-\reef{int}, we read off
\be
 \int \frac{d^4y_0}{y^2_{01}y^2_{02}y^2_{03}y^2_{04}}
 = 
 \int \frac{d^4Z_Ad^4Z_B}{\text{vol}(GL(2))}\,\frac{\<12\>\<23\>\<34\>\<41\>}{\langle AB12\rangle\langle AB23\rangle\langle AB34\rangle\langle AB41\rangle}\,.
\ee 
The loop momentum twistors $Z_A$ and $Z_B$ can be expanded on the basis of the four external line momentum twistors as
\be
Z_A=a_1Z_1+a_2Z_2+a_3Z_3+a_4Z_4\,,\;\quad Z_B=b_1Z_1+b_2Z_2+b_3Z_3+b_4Z_4\,.
\ee
This linear transformation gives 
$d^4 Z_A\, d^4 Z_B = \langle1234\rangle^2 \, d^4a_i d^4b_i$. The 4-brackets
$\langle A,B,i-1,i\rangle = \<1234\> M_{i+1}$ where $M_j$ is the $j$th minor of the $2\times 4$ matrix
\be
   \Big(
     \begin{array}{cccc}
       a_1 & a_2 & a_3 & a_4 \\
       b_1 & b_2 & b_3 & b_4 
     \end{array}
   \Big)\,.
\ee
For example, $\langle AB34\rangle = \<1234\> M_{1} =  \<1234\> (a_1 b_2 - a_2 b_1)$. Now, consider a
$GL(2)$-transformation \reef{GL2} of $Z_A, Z_B$. We can use it to set $a_4=b_2=0$ and $a_2=b_4=1$.
\exercise{}{Convince yourself that a $GL(2)$ rotation of  $Z_A, Z_B$ allows you to make the above choice of parameters, but that setting $a_4=b_4=0$ would be illegal.
}
In this gauge, we have $\<AB12\> = \<1234\>\, a_3$ etc, and the integrand then has no dependence (obviously) on $a_4,b_2,a_2,b_4$. This means that the $GL(2)$-volume factor cancels and we are then left with
\be
 \int \frac{d^4y_0}{y^2_{01}y^2_{02}y^2_{03}y^2_{04}}
 = 
 \int \frac{da_1}{a_1}
  \frac{db_1}{b_1}\frac{da_3}{a_3}\frac{db_3}{b_3} 
 \frac{\<12\>\<23\>\<34\>\<41\>}{-\langle1234\rangle^2}\,.
 \label{a1a3b1b3now}
\ee 
This way we have brought the loop-integral to the form on the LHS of \reef{y0-ui} and we see that the Jacobian is indeed \reef{theJ}.

Now our integration variables $a_i$ and $b_i$ in \reef{a1a3b1b3now} are not exactly the $u_i = y_{0i}^2$ that we introduced above \reef{y0-ui}: for example 
$u_1 = y_{01}^2 = \frac{\<41AB\>}{\<41\>\<AB\>} 
= -\frac{\<1234\>}{\<41\>\<AB\>}\,b_3$. So the $u_i$'s are proportional to the $a_{1,3}$ and $b_{1,3}$, but the factors of proportionality drop out of the $du_i/u_i$ measure.
}

Recall that we are studying the Leading Singularities in order to find an integrand that faithfully reproduces both $\text{LS}_\text{(a)}$ and $\text{LS}_\text{(b)}$. The integrand that we already know for the 4-point 1-loop amplitude does the job --- let us see how. We have previously found (see \reef{N4Answ}) that  
\be
 \ca_4^\text{1-loop} 
 ~=~  
 \ca_4^\text{tree}  \,y_{13}^2 y_{24}^2\, \int \frac{d^4y_0}{y^2_{01}y^2_{02}y^2_{03}y^2_{04}}
 \,.
\ee
When we convert this to the contour integral, the prefactor $y_{13}^2 y_{24}^2$ exactly cancels the Jacobian \reef{theJ}, so we are left with
\be
  \ca_4^\text{1-loop} 
 ~=~  
 \ca_4^\text{tree}  \,  
 \int \frac{da_1}{a_1}
  \frac{db_1}{b_1}\frac{da_3}{a_3}\frac{db_3}{b_3} \,.
\ee 
From this, we can directly read off the residue at the propagator poles 
$a_1=b_1=a_3=b_3 = 0$. The result is independent of which of the two solutions (a) or (b) we use to localize the loop-integral, so the quadruple cuts of the integral matches exactly with the Leading Singularities, 
$\text{LS}_\text{(a)}=
\text{LS}_\text{(b)}=
\mathcal{A}^{\rm tree}_4\,.$
This may not shock you, but once we venture beyond 4-point amplitudes, things are not so simple.

The result $\text{LS}_\text{(a)}=\text{LS}_\text{(b)}$  is special for the 4-point case. It can be represented diagrammatically as
\be
\raisebox{-12mm}{\includegraphics[scale=0.45]{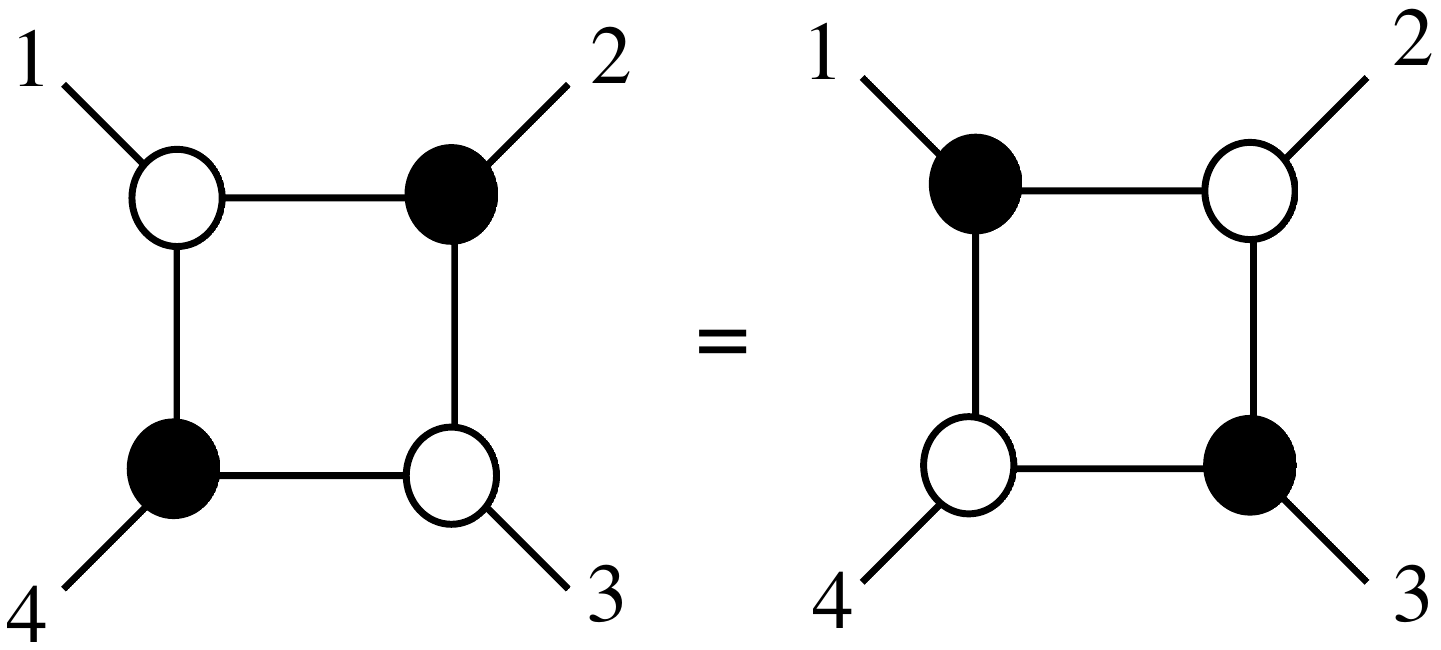}}\,.
\label{sqmove}
\ee
This identity is called the {\bf \em square move} and it will show up later in our discussions of higher-loop Leading Singularities and on-shell diagrams.

\vspace{7mm}
 
%%%%%%%%%%%%%%%%%%%%%%%%%%%%%%%%%%%%
\compactsubsection{5-point.}
%%%%%%%%%%%%%%%%%%%%%%%%%%%%%%%%%%%%%
At this point, we have constructed the 4-point 1-loop amplitude of $\cn=4$ SYM in three different ways: generalized unitarity, loop-level BCFW, and Leading Singularities. It is time to move ahead. 

We consider a specific maximal cut of the 5-point 1-loop amplitude:
\be
\raisebox{-17mm}{\includegraphics[scale=0.5]{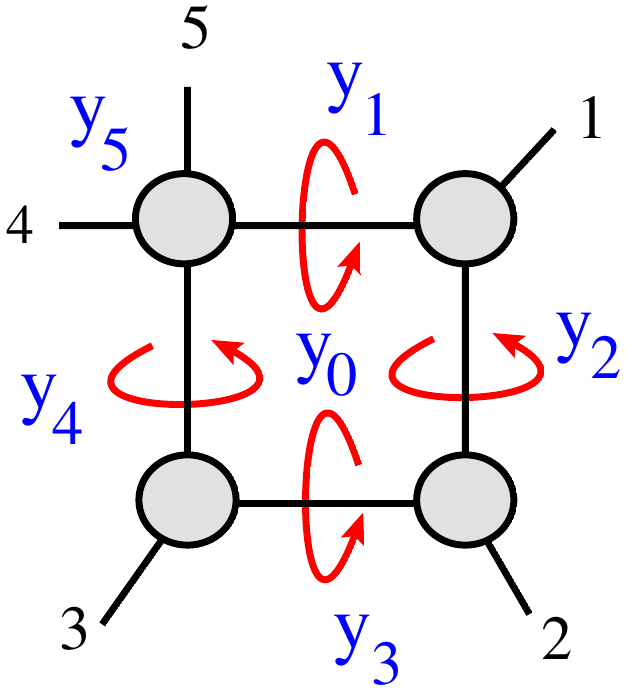}}
~~~
\hspace{1cm}
y^2_{01}=y^2_{02}=y^2_{03}=y^2_{04}=0\,.
\label{5ptcutcond}
\ee
The cut constraints for this maximal cut 
can be written in momentum twistor space as  
\eq
\langle AB12\rangle=\langle AB23\rangle=\langle AB34\rangle=\langle AB51\rangle=0\,.
\label{5ptConst}
\eqe
There are two solutions: 
\bea
\nonumber
&&\raisebox{-1cm}{\includegraphics[scale=0.55]{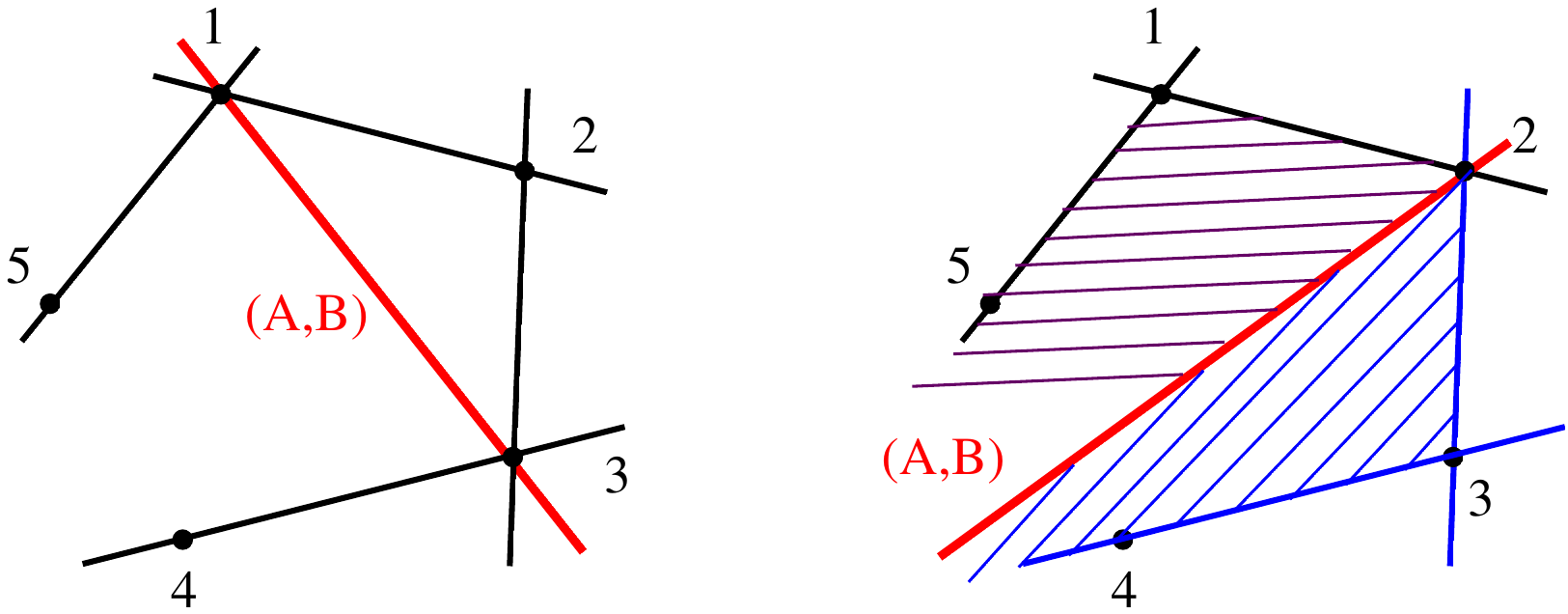}}\\[1mm]
~~
&&~(a)~~(A,B)=(1,3)\quad\quad
 \hspace{0.5cm}
 (b)~~(A,B)=(5,1,2)\bigcap(2,3,4)\,.
\label{MomTwistSol}
\eea
It is straightforward to see that $(a)$ is a solution to \reef{5ptConst}.
As for $(b)$, note that generically two planes in $\mathbb{CP}^3$ intersect in a line. Any points $A$ and $B$ on the intersection of the two planes in $(b)$ will be linearly dependent with any two points in either plane.
This establishes that $(b)$ is a solution to \reef{5ptConst}.

Since intersections of planes in $\mathbb{CP}^3$ may not feel as natural to you as brushing your teeth (hopefully),  let us make the solutions \reef{MomTwistSol}
 explicit in momentum space.  With $\ell = y_{10}$, the constraints are $\ell^2=(\ell-p_1)^2=(\ell-p_1-p_2)^2=(\ell-p_1-p_2-p_3)^2=0$ and it is not hard to verify that the two solutions for the loop-momentum $\ell$ can be written
\eq
\ell^{(1)}
=-|1\>\Big([1|+\frac{\langle23 \rangle}{\langle13\rangle}[2|\Big)\,,
\hspace{8mm}
\ell^{(2)}=-\Big(|1\>+\frac{[23]}{[13]}|2\>\Big)[1|\,.
\label{5ptmomSol}
\eqe
Note that even though $\ell^{(1)}$ is formally the complex conjugate of $\ell^{(2)}$, their geometric interpretations in momentum twistor space are quite different. This is because momentum twistors are chiral objects (only $|i\>$ appears,  not $|i]$).
\exercise{}{Show that $\ell^{(1)}$ and $\ell^{(2)}$ in \reef{5ptmomSol} solve the cut constraints. Check little group scaling. Then show that two solutions, $\ell^{(1)}$ and $\ell^{(2)}$, correspond to the two geometric solutions $(a)$ and $(b)$  of \reef{MomTwistSol}, respectively, in momentum twistor space.
}
The solution $\ell^{(1)}$ has $|\ell\> \propto |1\>$ and by momentum conservation these are also proportional to the angle spinor of $(\ell-p_1)$. This means that the special 3-point kinematics forces the vertex where line 1 attaches to be anti-MHV. Likewise, the solution $\ell^{(2)}$ forces the same vertex to be anti-MHV. By \reef{4ptLSCutNo}, the rest of the helicity structure is fixed, and we see that the two solutions (a) and (b) correspond to the two options
\be
  \raisebox{-20mm}{\includegraphics[scale=0.5]{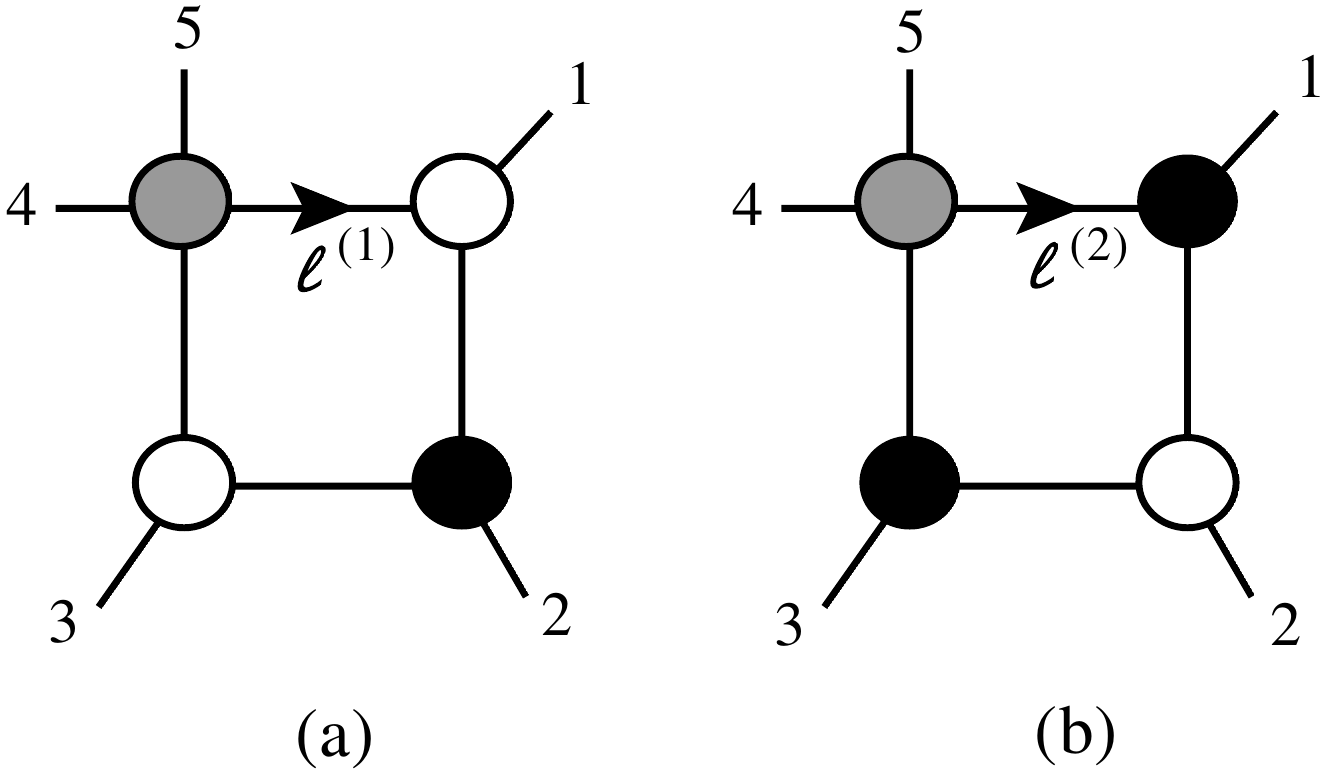}} 
  \,.
  \label{5ptLSCutab}
\ee
The gray blob for the 4-point vertex does not have specific helicity designation because  4-point on-shell amplitude is simultaneously MHV and anti-MHV. 

Now, let us count the number of Grassmann $\eta$'s of these blob-diagrams. MHV has 8 $\eta$'s and anti-MHV 4, and for each of the 4 internal lines we have to do an $d^4\eta$-integral. For diagram (a), this then gives $8+4+4+8 - 4\times 4 = 8$ corresponding to the 5-point MHV sector. For diagram (b): $8+8+8+4 - 4\times 4 = 12$ which identifies it as belonging to the NMHV sector.  
Including the appropriate Jacobians, the two diagrams \reef{5ptLSCutab}  evaluate to the respective MHV or NMHV 5-point tree-level amplitudes.

The above discussion tells us that for a given MHV or NMHV 1-loop 5-point amplitude, only one of these solutions to the maximal cut conditions \reef{5ptcutcond} is relevant. If we focus on the MHV sector, only  diagram $(a)$ matters and equals  $\ca_5^\text{tree}$ for solution $\ell^{(1)}$ and is zero when evaluated on solution $\ell^{(2)}$. This leads us to the crux of problem we mentioned in the beginning of this section: whether the integral basis Ansatz we write for the integrand faithfully reproduces all Leading Singularities. Let us illustrate this explicitly. Consider the scalar box-integral whose propagators are those considered in the maximal cut \reef{5ptcutcond}:
\be
 I_{5,\text{box}}(1,2,3,4)~=~
 \frac{\langle 5123\rangle\langle 1234\rangle}{\langle AB51\rangle\langle AB12\rangle\langle AB23\rangle\langle AB34\rangle} \,.
 \label{5ptbox1234}
\ee
We use the labels on the $n$-point box-integral $I_{n,\text{box}}(i,j,k,l)$ to specify the first external leg on each of the vertices. For example, for the arrangement in \reef{LSconstraint}, the corresponding scalar box-integral would be labeled $I_n(1,i+1,j+1,k+1)$.

When evaluating the quadruple cut for the integral  $I_{5,\text{box}}(1,2,3,4)$, the Jacobian cancels the numerator factor $\langle 5123\rangle\langle 1234\rangle$, and since there is no other dependence on the loop momenta than the 4 propagators we are cutting, this integral produces the same answer, namely 1, no matter if we evaluate it on solution $\ell^{(1)}$ or  $\ell^{(2)}$: i.e.~
\be
 I_{5,\text{box}}\big|_{(1)}~=~I_{5,\text{box}}\big|_{(2)}~=~1\,.
\ee
On the other hand, we now know that the corresponding Leading Singularities of diagram $(a)$ are 
\be
  \text{LS}_{(1)} =\ca_{5,\text{MHV}}^\text{tree} \,,
  ~~~~~~
  \text{LS}_{(2)} = 0 \,.
\ee
This means that the Ansatz
\be
  \ca_{5,\text{MHV}}^\text{1-loop}
  = \ca_{5,\text{MHV}}^\text{tree}
     \times \Big( I_{5,\text{box}}(1,2,3,4) + \text{other box-integrals}\Big)
  \label{ansatz5pt1L1}
\ee
does \emph{not} produce the Leading Singularities faithfully. However, it does produce the average of the two maximal cuts correctly because 
\be
  \frac{1}{2} 
  \Big( I_{5,\text{box}}\big|_{(1)} + I_{5,\text{box}}\big|_{(2)}\Big) 
  = \frac{1}{2} (1+1) =    1 
\ee
equals the sum of the Leading Singularities $\text{LS}_{(1)} +\text{LS}_{(2)} = 1+0 =1$. The message is that the integrand Ansatz \reef{ansatz5pt1L1} can produce the correct maximal cut when one averages over the two constraints (as is usually done in applications of the generalized unitarity method), but it does not produce each Leading Singularity honestly. If you just want an answer for the amplitude, you don't have to care. But let us try to be caring people and see where it takes us.

We have learned now that we need something else in the Ansatz \reef{ansatz5pt1L1} in order to match the Leading Singularities. That something else turns out to be the pentagon integral
\be
  I_{5,\text{pentagon}} = \frac{\langle A,B|(1,2,3)\bigcap(3,4,5)\rangle\langle2451\rangle}{\langle AB12\rangle\langle AB23\rangle\langle AB34\rangle\langle AB45\rangle\langle AB51\rangle}\,.
  \label{defpent5}
\ee
The numerator includes the bi-twistor $(1,2,3)\bigcap(3,4,5)$ that characterizes the line of intersection between the planes $(1,2,3)$ and $(3,4,5)$; the intersection formula was given in \reef{TwistorLine}. 
When we evaluate the maximal cut \reef{5ptcutcond} of the pentagon, the residue depends on the loop-momentum and hence on which solution \reef{5ptmomSol} we evaluate it. Including the Jacobian $J=\langle 5123\rangle\langle 1234\rangle$, one finds
\eq
I_{5, \text{pentagon}}\big|_{(a)}=0,~~~~~
 \quad I_{5, \text{pentagon}}\big|_{(b)}=-1\,.
\label{5ptLSResult}
\eqe
This is good news, because now the improved Ansatz
\be
  \ca_{5,\text{MHV}}^\text{1-loop}
  = \ca_{5,\text{MHV}}^\text{tree}
     \times \Big( I_{5,\text{box}}(1,2,3,4) 
       +  I_{5,\text{pentagon}}
       +
     \text{other}\Big)
  \label{ansatz5pt1L2}
\ee
(where ``other" is assumed to not contribute to our cut)
has the following maximal cut \reef{5ptcutcond}: on two solutions,  $\ell^{(1)}$ and $\ell^{(2)}$, it gives
\be
  \begin{split}
  I_{5,\text{box}}\big|_{(1)} + I_{5,\text{pentagon}}\big|_{(1)}
   &= \ca_{5,\text{MHV}}^\text{tree} \times( 1 + 0 ) 
  = \ca_{5,\text{MHV}}^\text{tree}
  = \text{LS}_{(1)}\,,
  \\[1mm]
  I_{5,\text{box}}\big|_{(2)} + I_{5,\text{pentagon}}\big|_{(2)}
   &= \ca_{5,\text{MHV}}^\text{tree} \times( 1 -1 ) 
  = 0
  = \text{LS}_{(2)}\,.
  \end{split}
\ee
So it produces the correct Leading Singularities for the cut \reef{5ptcutcond}!

Now, unfortunately we are not done yet, because we have to worry about all the other cuts: there are a total of $2 \times 5=10$ Leading Singularities for the 5-point 1-loop amplitude. With just the box diagram and the pentagon diagram in \reef{ansatz5pt1L2}, there is no chance that this can be the full answer: the reason is simply that the sum of those two integrals is not cyclically invariant. It takes just one more integral to achieve cyclic invariance, namely the box integral 
$ I_{5,\text{box}}(3,4,5,1)$. 
Diagrammatically we can express  the final answer as
\bea
  \nonumber
  \ca_{5,\text{MHV}}^\text{1-loop}
  &=& \ca_{5,\text{MHV}}^\text{tree}
     \times \Big( I_{5,\text{box}}(1,2,3,4) 
       +  I_{5,\text{pentagon}}
       +
     I_{5,\text{box}}(3,4,5,1)\Big)\\[1.5mm]
     &=&
     \ca_{5,\text{MHV}}^\text{tree}
     \left(
     \raisebox{-11mm}{\includegraphics[scale=0.54]{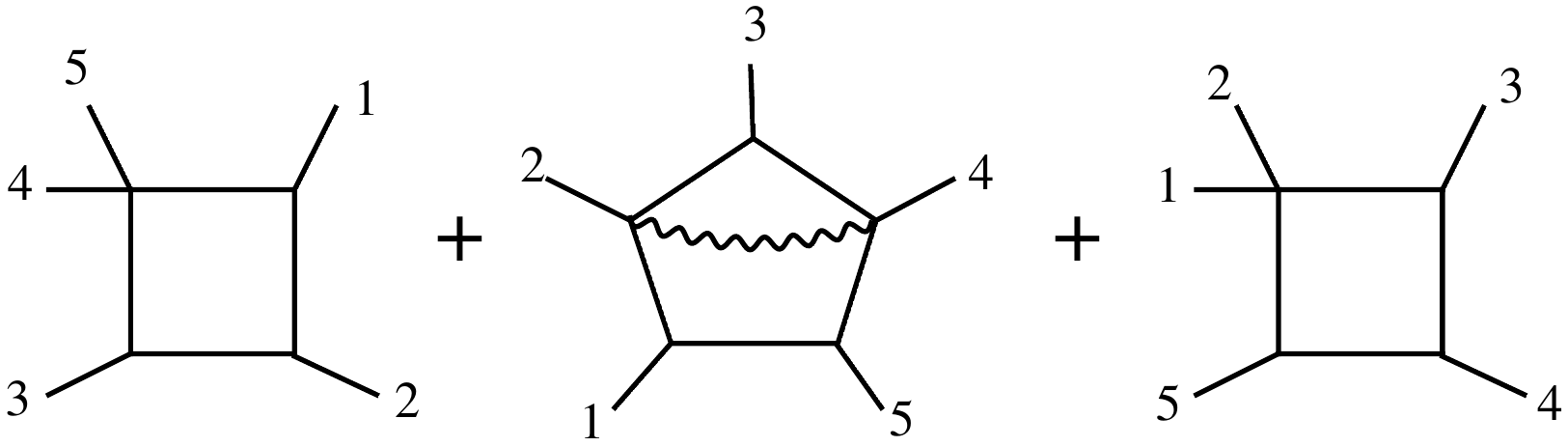}}
     \right)
     \,.~~~~~~~
  \label{5ptResult}
\eea
The diagrammatic notation for the pentagon integral has a wavy line indicating that the bi-twistor $(1,2,3)\bigcap(3,4,5)$ goes in the numerator in \reef{defpent5}. 
Our previous results plus cyclic invariance then guarantee that \reef{5ptResult} produces all 10 Leading Singularities correctly. 
\exercise{}{Show that \reef{5ptLSResult} is true.
}
\exercise{}{Show that \reef{5ptResult} is invariant under cyclic permutations of the external lines.
}
We have introduced here box and pentagon 1-loop integrands, \reef{5ptbox1234} and \reef{defpent5}, whose quadruple cuts evaluate to either $+1$, $-1$ or $0$. Such integrands are called {\bf \em unit Leading Singularity integrands}.

It has been shown \cite{UnitLS} that all 
planar loop amplitudes of $\mathcal{N}=4$ SYM can be obtained as a  
linear combination of unit Leading Singularity integrands (times a tree amplitude).
The coefficients in front of the unit Leading Singularity integrands are determined by the Leading Singularity, thus knowing them is sufficient to determined the entire amplitude.  
For planar amplitudes, we need unit Leading Singularity integrands that are also dual conformal invariant and local, and this is a rather restrictive class of integrands. 
We have seen the Leading Singularity method at work for 4- and 5-point 1-loop amplitudes. The structure generalizes to higher points, as we now outline.

%%%%%%%%%%%%%%%%%%%%%%%%%%%%%%%%%%%%
\compactsubsection{6-point and beyond.}
%%%%%%%%%%%%%%%%%%%%%%%%%%%%%%%%%%%%%
The 1-loop $n$-point MHV amplitude is given by a simple generalization of the 5-point result: 
\bea
\nonumber
\mathcal{A}^\text{1-loop}_{n,\text{MHV}}
&=&
\mathcal{A}^\text{tree}_{n,\text{MHV}}\left(\;\sum_{1<i<j<n}\;\vcenter{\hbox{\includegraphics[scale=0.54]{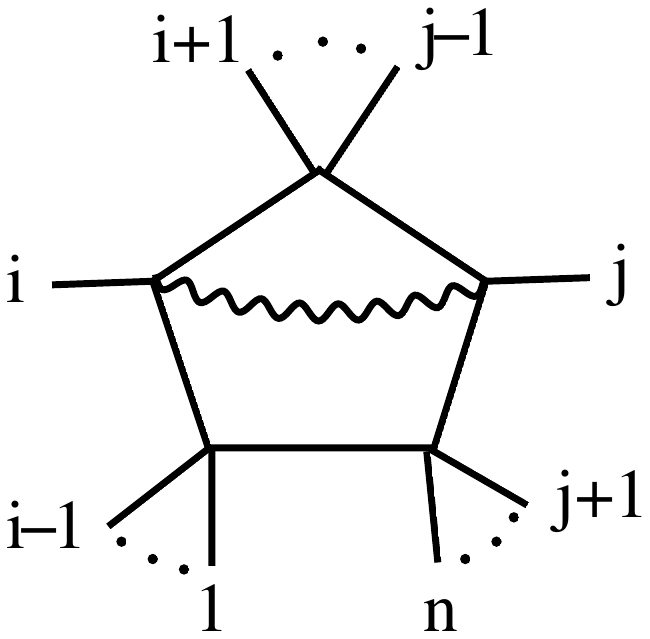}}} \;\right)
\\[3mm]
&=&\!\!
\int _{A,B}\frac{\langle A,B|(i-1,i,i+1)\bigcap(j-1,j,j+1)\rangle\langle i,j,n,1\rangle}{\langle A,B i,i-1\rangle\langle AB,i,i+1\rangle\langle A,B,j-1,j\rangle\langle A,Bj,j+1\rangle\langle A,B,n,1\rangle}\,.~~~~~~~~~~
\label{AllnAnsw}
\eea
In the sum, there are two boundary cases: $i=2$, $j=3$ and $i=n-2$, $j=n-1$. These correspond to box integrals whose numerators are simply the Jacobian coming from cutting all four propagators. More precisely, we have 
\be
\vcenter{\hbox{\includegraphics[scale=0.4]{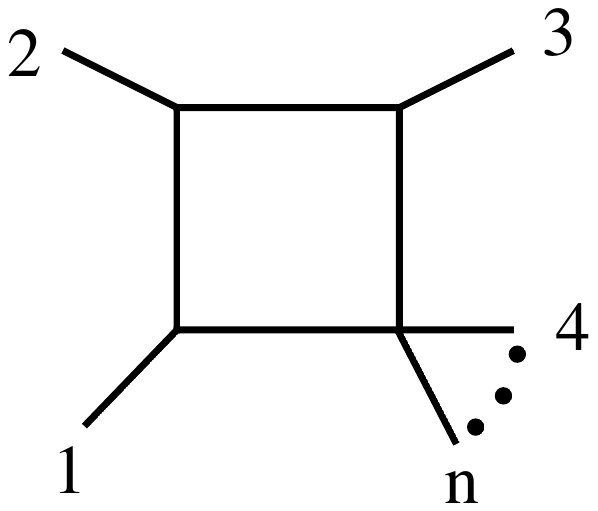}}}
~=~
\displaystyle
\int _{A,B}\frac{\langle n123\rangle\langle 1234\rangle}{\langle AB 12\rangle\langle AB23\rangle\langle AB34\rangle\langle ABn1\rangle}
\ee
and
\be
\vcenter{\hbox{\includegraphics[scale=0.4]{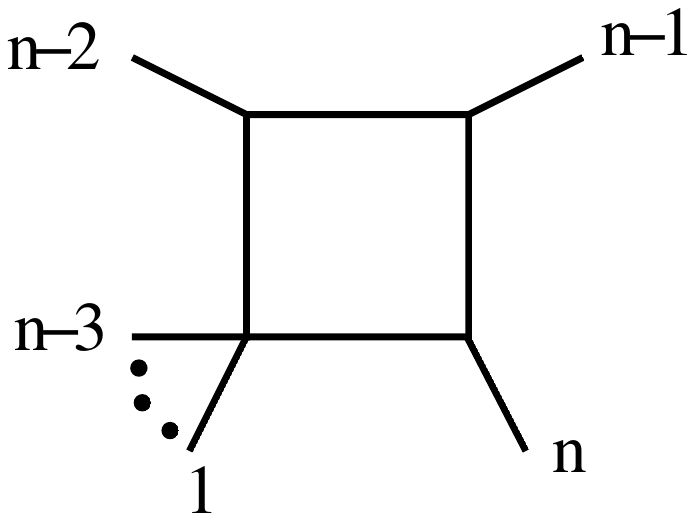}}}
\!\!\!=
\displaystyle
\int _{A,B}\frac{\langle n{\rm-3},n{\rm-}2,n{\rm-}1,n\rangle\langle n{\rm-}2,n{\rm-}1,n,1\rangle}{\langle A,B, n{\rm-}3,n{\rm-}2\rangle\langle A,B,n{\rm-}2,n{\rm-}1\rangle\langle A,B,n{\rm-}1,n\rangle\langle A,B,n,1\rangle}\,.
\ee
In conclusion, the two Leading Singularities of arbitrary 1-loop MHV amplitudes can be reproduced by including the simple combination of tensorial (due to the loop momentum dependence in the numerator) pentagon integrals. These are all  local unit Leading Singularity integrands. Note that these integrands can be used as part of the basis for 1-loop amplitudes in any massless quantum field theory. 
The special situation for $\mathcal{N}=4$ SYM is that these integrals provide the entire answer, whereas for a generic QFT, one needs in addition the various lower-gon integrals that are not captured by the maximal cuts. 

You may (and should) be puzzled that in the beginning of Section \ref{s:genunit}, we stated that the 1-loop amplitudes in a unitary 4d quantum field theory can be expanded on a basis of scalar box-, triangle-, and bubble-integrals with possible additional input from rational terms. This was summarized in  equation \reef{IntBasis}, and we noted that in $\cn=4$ SYM, the only non-vanishing contributions were the box-integrals. There were no pentagons in that story! So what is the deal? The point of the pentagon integrals in the present section is that they allow us to write the  $\cn=4$ SYM 1-loop integrand in a form that reproduces  each Leading Singularity faithfully. On the other hand, \reef{IntBasis} determines the 1-loop $\cn=4$ SYM amplitudes a sum of box-integrals whose coefficients are evaluated by quadruple cuts, evaluated as the \emph{average} of the two loop-constraint solutions. More box-diagrams contribute in \reef{IntBasis} than in \reef{AllnAnsw}. So what \emph{is} the deal?
Well, the two representations of the integrand must yield the same answer for the amplitude. The integrals have to be regulated, and if one uses dimensional regularization $4-2\eps$, the difference between the two representations is only in the $O(\epsilon)$-terms. Specifically, the pentagon integrals contain the information about the `missing' boxes plus $O(\epsilon)$ \cite{Sharpening}. Thus the two procedures yield the same integrated answer. 

%%%%%%%%%%%%%%%%%%%%%%%%%%%%%%%%%%%%
\subsection{2-loop Leading Singularities}
\label{s:compLS}
%%%%%%%%%%%%%%%%%%%%%%%%%%%%%%%%%%%%%
Back on page  \pageref{panic}, we noted  that not all loop-diagrams appear to have enough propagators available for a maximal cut of $4L$-lines. 
A representative example is the double-box diagram of the 
2-loop 4-point amplitude:
\be
\raisebox{-11mm}{\includegraphics[scale=0.4]{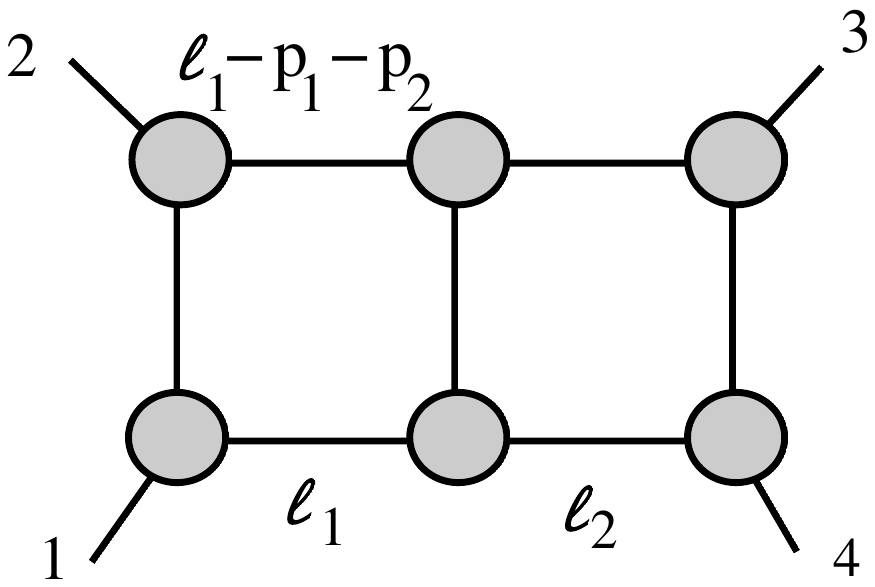}}\,.
\label{dbox}
\ee
With 7 propagators, we can only localize 7  of the 8 components of the two 
loop-momenta, leaving behind a 1-dimensional loop integral. 
However, when the 7 propagators are on-shell, the 4-point 1-loop analysis tells us that the lefthand box in \reef{dbox} is a Leading Singularity
that equals the 4-point tree amplitude, $A_4^\text{tree}[\ell_2,p_1,p_2,\ell_2-p_1-p_2]$, where $\ell_2$ parameterizes the loop-momentum in the righthand box. But this tree amplitude has a propagator $1/(\ell_2-p_1)^2$ that can now be used to localize the final component of the loop-momenta, thus providing a maximal cut. Moreover, on this pole, the 4-point tree amplitude factorizes into two 3-point amplitudes, and therefore the result is simply an on-shell 4-point 1-loop box diagram.  The procedure is illustrated here:
\be
 \raisebox{-10mm}{\includegraphics[scale=0.8]{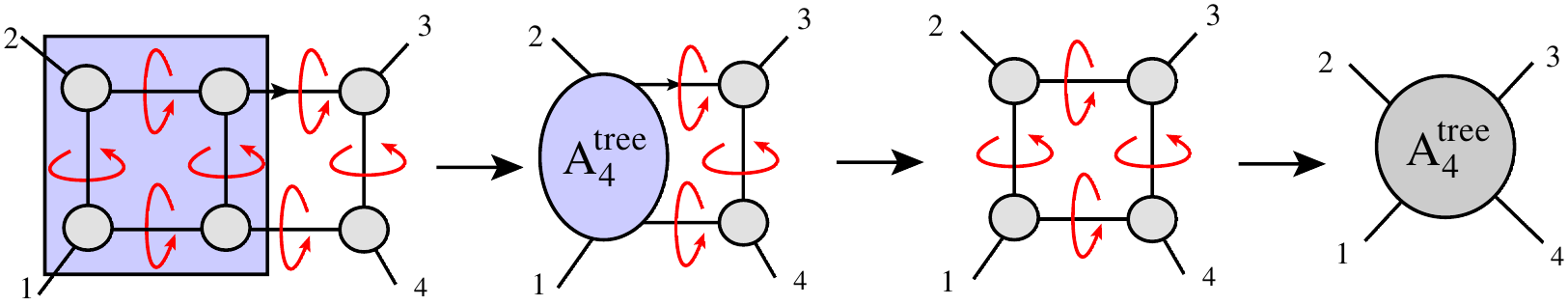}}
\ee
The last step uses that the 1-loop 4-point Leading Singularity is equal to the tree amplitude $A_4^\text{tree}[1234]$. In conclusion: 
\begin{itemize}
\item[(i)] the Leading Singularities are well-defined for the 2-loop 4-point amplitude, even though the double-box only has 7 loop propagators, and
\item[(ii)]  the 4-point `double-box Leading Singularity'  equals the 4-point  tree amplitude.
\end{itemize}

A Leading Singularity that involves a ``pole-under-a-pole" is called a {\bf \em composite Leading Singularity}. A proto-type of such a composite object is the 3-variable contour integral
\eq
\oint dx\,dy\,dz\,\frac{1}{x(x+yz)}\,.
\eqe
This integrand appears to have only two poles, insufficient to localize the 3d integral. However, if the $x$-contour circles the pole at $x=0$, then an additional pole emerges in the form $1/(yz)$ and this can then localize the remaining two integrals, giving the residue $1$ (ignoring $2\pi i$'s).

The idea of composite Leading Singularities resolves the subtlety about defining maximal cuts and Leading Singularities for higher-loop amplitudes. Henceforth we work with the understanding that the  Leading Singularities of multi-loop amplitudes are always well-defined.

%%%%%%%%%%%%%%%%%%%%%%%%%%%%%%%%%%%%
\subsection{On-shell diagrams}
\label{s:onshelldiag}
%%%%%%%%%%%%%%%%%%%%%%%%%%%%%%%%%%%%%
We have found in Section \ref{s:LS1loop} that the 4-point 1-loop  Leading Singularity is equal to the 4-point tree amplitude:
\be
  \raisebox{-11mm}{\includegraphics[scale=0.4]{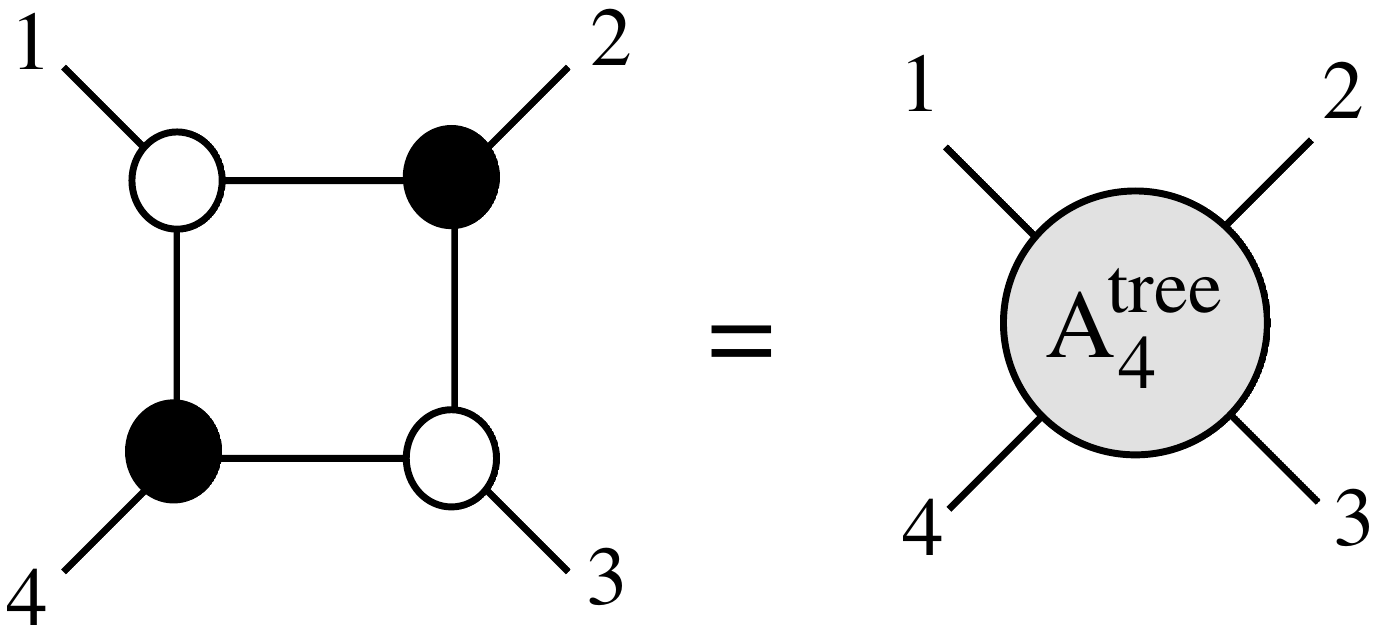}}\,.
  \label{boxistree}
\ee
This looks rather peculiar since the LHS is a 1-loop diagram while the RHS is a tree-amplitude. It actually turns out that the LHS can be interpreted as a super-BCFW diagram! We now show how.

Consider the top two vertices in the Leading Singularity diagram 
\be
  \raisebox{-11mm}{\includegraphics[scale=0.4]{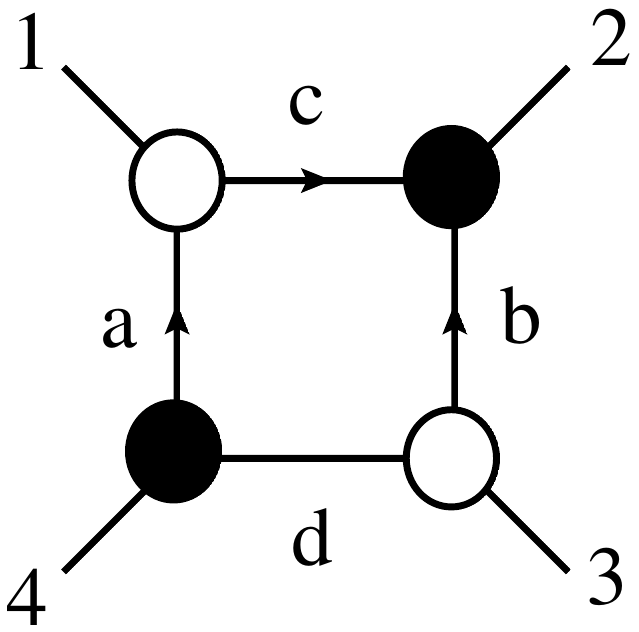}}
  ~~~~~~~~
  \begin{array}{l}
   |a\> \propto |c\>  \propto |1\> \,,
  \\[3mm]
  \,|b] \propto |c]\,\propto  |2]\,.
  \end{array}
  \label{bcfwBRabcd}
\ee
The MHV and anti-MHV designations imply the indicated special 3-particle kinematics. Up to an overall factor $z$, this determines $p_c$ to be $p_c = - z |1\> [2|$. Momentum conservation then fixes $p_a$ and $p_b$ to be 
\be
  p_a=-|1\>\big([1|+z[2|\big)\,
  ~~~~\text{and}~~~~
  p_b=-\big(|2\> - z|1\>\big)[2|\,.
\label{LSBCFW}
\ee
We recognize $p_a$ and $p_b$ as BCFW $[1,2\>$-shifted momenta
$\hat{p}_1$ and $\hat{p}_2$! 

What about the Grassmann variables? Let us carry out the $\eta_c$-integral in the product of the Grassmann delta functions of the first two vertices
\eq
 \begin{split}
 &\int d^4\eta_a\,d^4\eta_b\,d^4\eta_c\,
\delta^{(4)}\big([1c]\eta_a+[ca]\eta_1+[a1]\eta_c\big)
\,\delta^{(8)}\big(|2\rangle\eta_2-|b\rangle\eta_b-|c\rangle\eta_c\big)\\[1mm]
  &~~\propto \int d^4\eta_a\,d^4\eta_b\,
  \delta^{(8)}\Big(
  |1\> \big(\eta_a - (\eta_1 - z\eta_b)\big)
  - |2\> \big(\eta_b - \eta_2\big)\Big)\,.
  \end{split}
\eqe 
The last integral localizes $\eta_a$ and $\eta_b$ to be 
\eq
\eta_a=\eta_1-z\eta_2
  ~~~~\text{and}~~~~
\eta_b=\eta_2\,.
\eqe
This is exactly the shift of the Grassmann variables associated with the supersymmetrization of the BCFW shift \reef{LSBCFW}. 

Finally, let us see how the internal line $d$ in \reef{bcfwBRabcd} fixes $z$. The on-shell condition is $0=p_d^2 = (p_3 + p_b)^2 = (\<23\> - z \<13\>)[23]$ i.e.~$z=\<23\>/\<13\>$. This   corresponds exactly to the pole where the propagator $1/\hat{p}_{23}$ in the $[1,2\>$-shifted 4-point tree amplitude goes on-shell. As we know from the super-BCFW calculation \reef{sbcfwmhv1}, this is exactly the factorization pole that allows us determine the full  4-point tree-amplitude in $\cn=4$ SYM from the MHV$_3 \times$ anti-MHV$_3$ super-BCFW diagram. 

We have established the connection between the Leading Singularity diagram on the LHS of \reef{boxistree} and the 4-point super-BCFW diagram \reef{sbcfwmhv1}, and this allow us to understand why the 4-point 1-loop Leading Singularity is just the 4-point tree superamplitude. The connection is summarized diagrammatically as the {\bf \em BCFW-bridge}  
\be
  \raisebox{-11mm}{\includegraphics[scale=0.5]{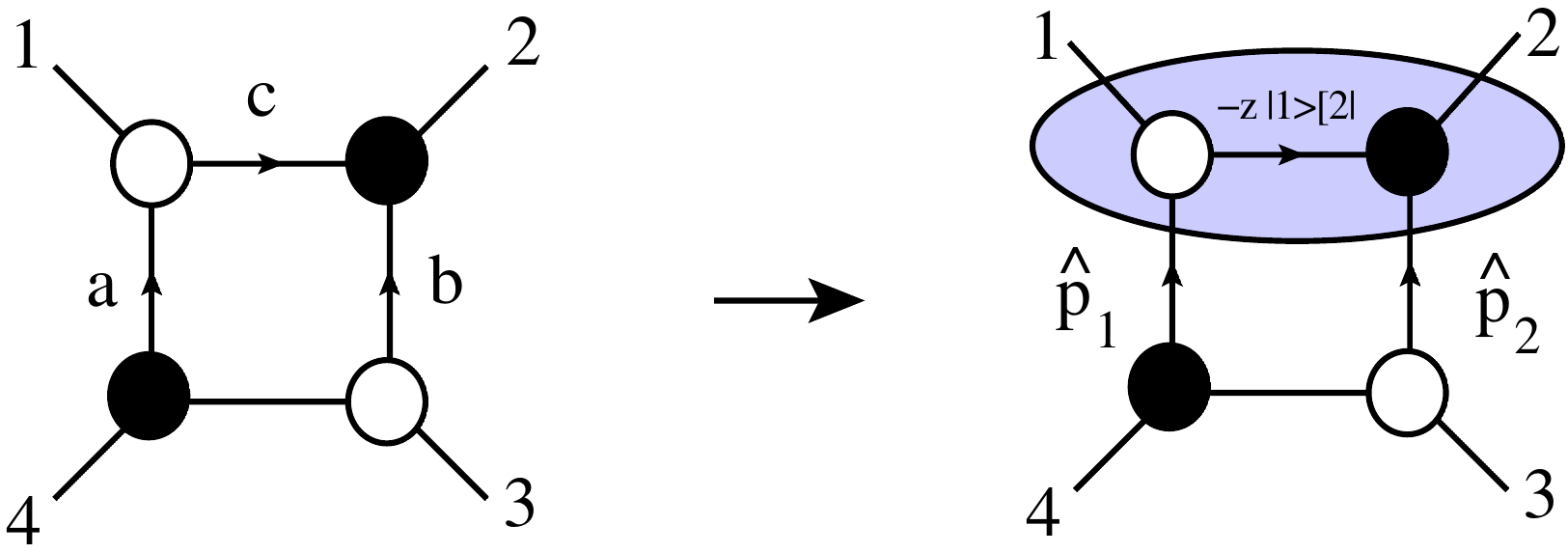}}\,,
\ee
where the upper two vertices, surrounded by the blue region, is the `bridge'. The bridge provides the BCFW super-shift.

Exchanging black and white dots in the BCFW bridge, simply corresponds to the conjugate BCFW shift. This also gives another meaning to the {\bf \em square move} \reef{sqmove}:   
\be
\raisebox{-12mm}{\includegraphics[scale=0.4]{SquareMove}}\,.
\label{sqmove2}
\ee
It simply says that the two BCFW super-shifts $[1,2\>$ and $[2,1\>$ give the same 4-point amplitude.

With the square move and the BCFW shift, it becomes fun to calculate Leading Singularities. Starting from the fundamental 3-point vertices, we can build {\bf \em on-shell diagrams} that contain information about the higher-loop amplitudes. Each 3-point vertex represents the MHV or anti-MHV 3-point amplitude, along with the implication that the square or angle spinors of its legs are proportional. The vertices are glued together by `on-shell propagators' whose rules can be written
\eq
\raisebox{1mm}{\includegraphics[scale=0.5]{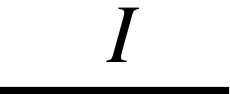}}
~~=~\int \frac{d^2|I\> d^2|I] d^4\eta_I}{U(1)}\,.
\label{internalLine}
\eqe 
The $\eta_I$ integral is the usual state sum. The integration over the momentum variables will be localized by the momentum conservation delta function on both sides of the propagator. 

In addition to the square move \reef{sqmove2}, there are two rules that help us  simplify complicated on-shell diagrams. 
The first rule follows from the observation that each MHV 3-vertex imposes that the square spinors of the associated lines are proportional, so two consecutive MHV vertices imply that all four square spinors are proportional. This gives
the rule 
\eq 
\raisebox{-8mm}{\includegraphics[scale=0.5]{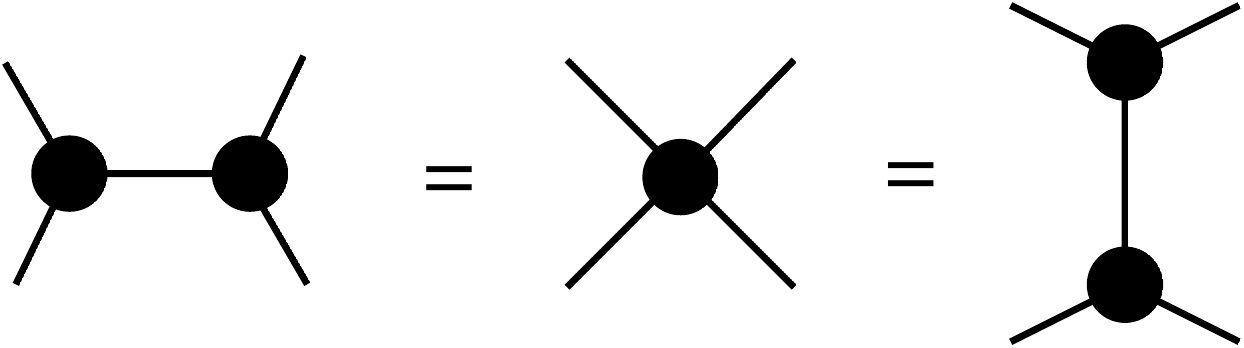}}\,.
\label{STEquiv}
\eqe
There is of course an equivalent rule for anti-MHV. 
The black 4-vertex blob imposes, per definition, that the four lines have proportional square spinors. This blob does not represent a 4-point MHV tree amplitude; it is just a short-hand notation for the double-blob diagrams.

The second rule is
\eq
\raisebox{-4mm}{\includegraphics[scale=0.5]{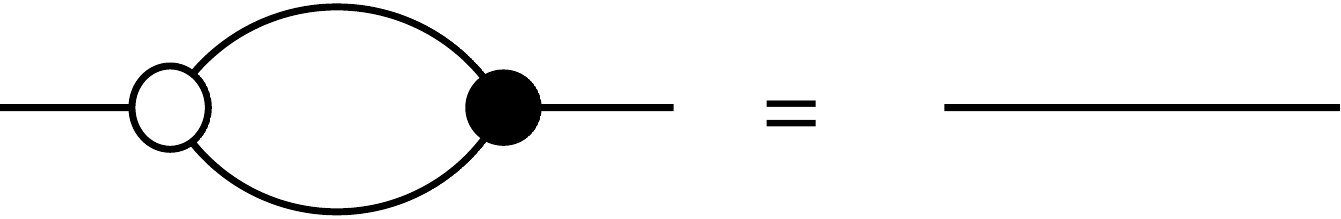}}\,.
\label{CollapseProp0}
\eqe
The 3-particle kinematics forces the internal lines in the bubble to be collinear, and this collapses the bubble. This formally eliminates a loop-integral.

Combining the two rules \reef{STEquiv} and \reef{CollapseProp0} gives 
\eq
\raisebox{-7mm}{\includegraphics[scale=0.5]{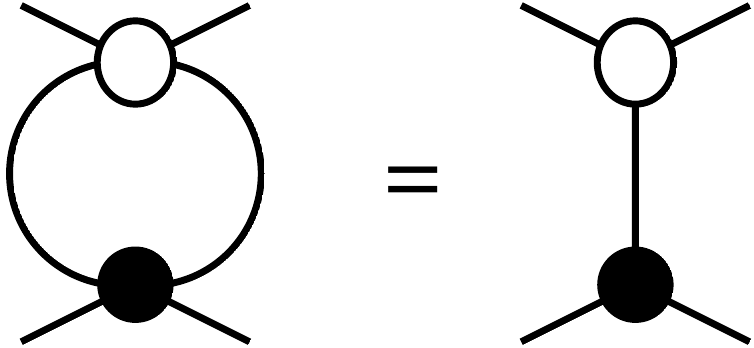}}\,.
\label{CollapseProp}
\eqe
\exercise{}{Show that the internal lines in the bubble \reef{CollapseProp0} are collinear and that \reef{CollapseProp} follows from \reef{STEquiv} and \reef{CollapseProp0}.
}

The point of these rules is to simplify the evaluation of on-shell diagrams. 
For the on-shell diagram of the 2-loop 4-point Leading Singularity, 
 we first apply the square move \reef{sqmove2} and then the collapse-moves \reef{STEquiv} and \reef{CollapseProp} to get
\be
  \raisebox{-8mm}{\includegraphics[scale=0.8]{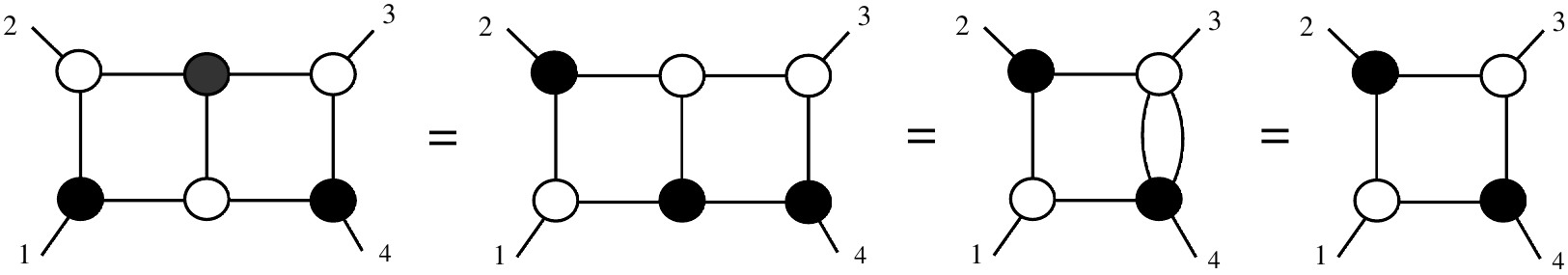}}\,.
\ee
This shows that the 4-point 2-loop Leading Singularity equals  the 4-point 1-loop Leading Singularity, which in turn is just the 4-point tree amplitude. We had already found this result in Section \ref{s:compLS} by evaluating the   composite Leading Singularity. The rules for the on-shell diagrams offer a simpler diagrammatically derivation.
\exercise{}{Write down 3-loop on-shell diagrams for the 4-point MHV amplitude and show that they reduce to 1-loop result.} 
Using the BCFW bridge, we can begin to build up more complicated on-shell diagrams. For example, we can use the BCFW bridge to interpret the on-shell diagram:
\be
  \raisebox{-13mm}{\includegraphics[scale=0.5]{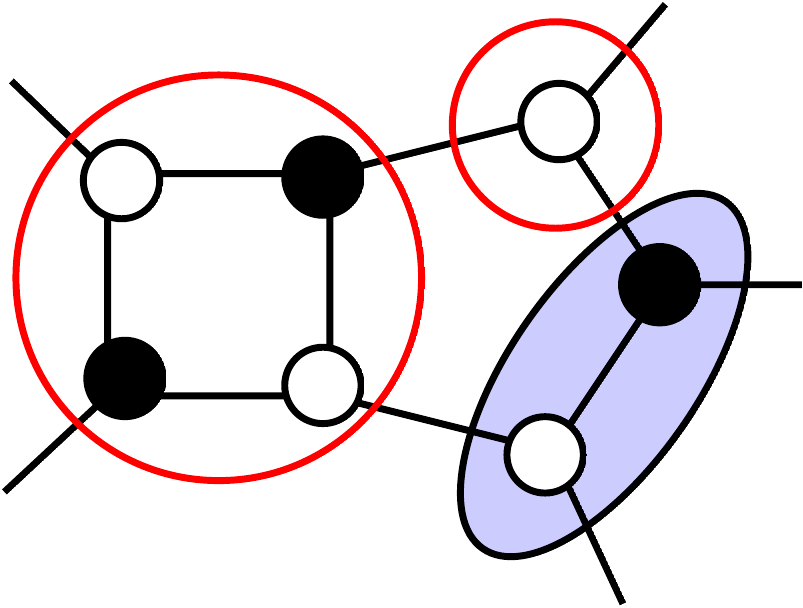}}.
  \label{5ptBCFWLS}
\ee
The red circles highlight the 4-point MHV tree amplitude and 3-point anti-MHV tree subamplitudes. The  BCFW bridge, indicated with the blue-shaded area, induces the BCFW shift on the two affected lines shows and this shows that this on-shell diagram represents the BCFW diagram for the 5-point MHV tree-level superamplitude.

The super-BCFW recursion relations for the 6-point NMHV tree amplitude can be represented with on-shell diagrams as \\[-6mm]
\be
    \raisebox{-20mm}{\includegraphics[scale=0.83]{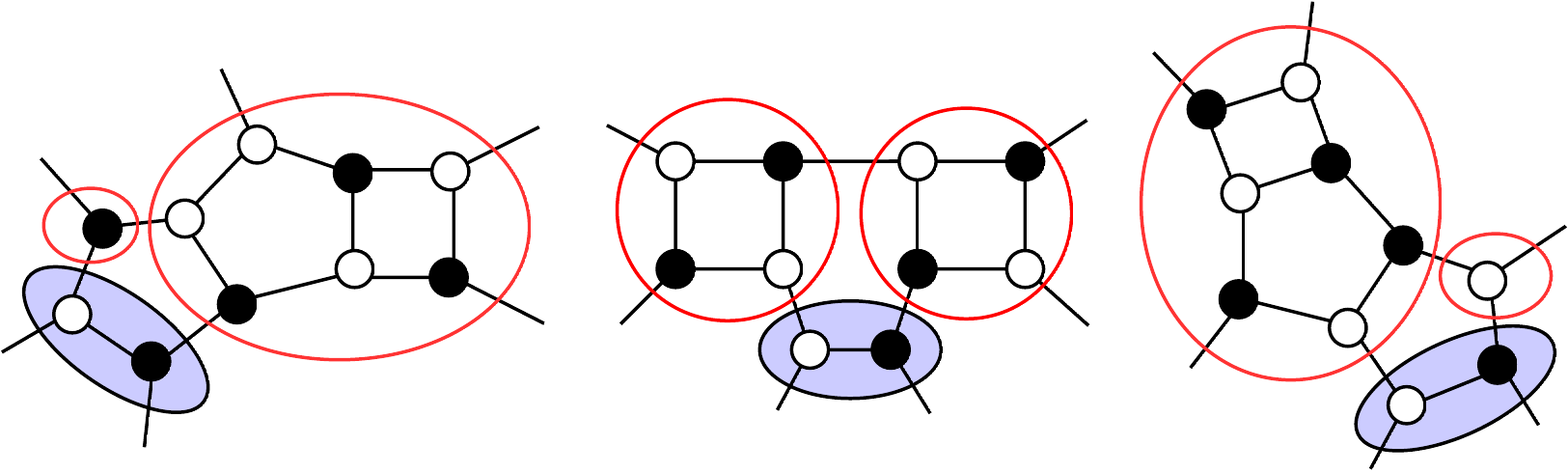}}.
    \label{6ptBCFWLS2}
\ee
The lefthand diagram is the MHV$_3 \times$ anti-MHV$_5$ BCFW diagram, the middle diagram is the BCFW diagram with two  4-point  MHV tree subamplitudes and the righthand diagram is the MHV$_5 \times$ anti-MHV$_3$ BCFW diagram.
\exercise{}{Do the $\eta$-counting to show that \reef{5ptBCFWLS} represents an on-shell diagram for an MHV amplitude and \reef{6ptBCFWLS2} an NMHV amplitude.}
\exercise{}{Interpret the effect of the rule \reef{STEquiv} on the on-shell diagram \reef{5ptBCFWLS}. Show that the 3-loop diagram 
\be
  \raisebox{-13mm}{\includegraphics[scale=0.4]{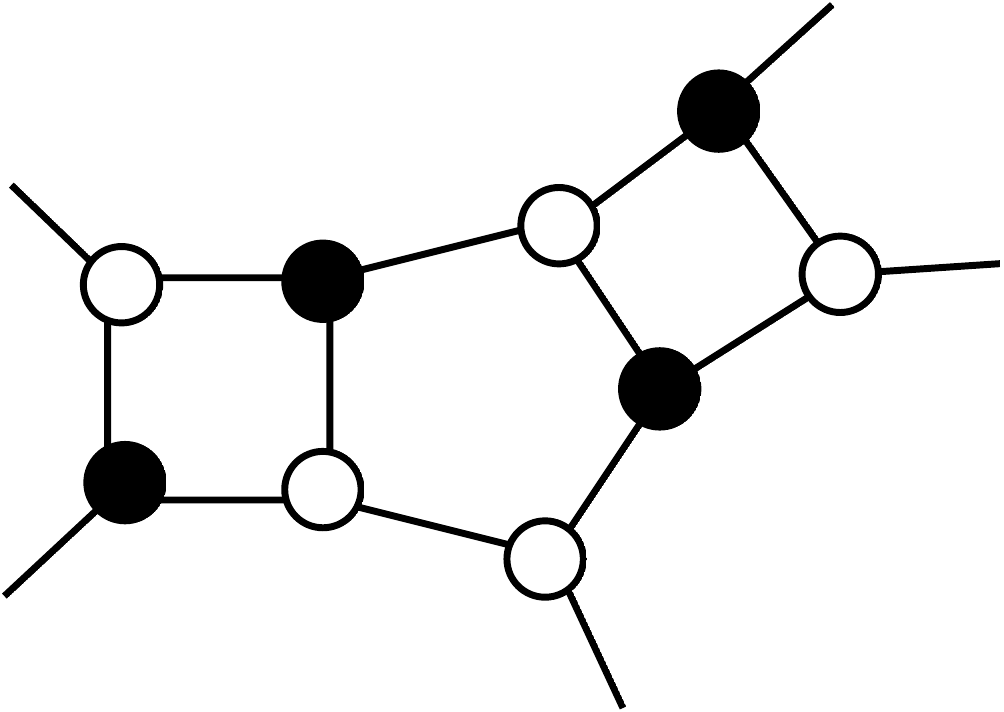}}.
  \label{5ptBCFWLS3L}
\ee
is equivalent to \reef{5ptBCFWLS}. What does that tell you about Leading Singularities?
}
The concept of the Leading Singularity is well-defined for lower-$\cn$ SYM and for non-super\-symmetric theories, and so are the on-shell diagrams. The only distinction is that the edges need arrows because the on-shell states split into two CPT conjugate multiplets, one with the positive helicity gluon and the other with the negative helicity gluon. ($\cn=4$ SYM is special in that its supermultiplet is CPT self-conjugate.) Opposite helicity multiplets must sit at different ends of the propagators, so it is dressed with an arrow that indicates the assignment.

In the next section, we discuss a formula that reproduces all Leading Singularities in planar $\cn=4$ SYM. This  formula will tell us that the number of distinct Leading Singularities for a given $n$ and given N$^K$MHV level is fixed. For example, for $K=0$ (MHV) there is only one Leading singularity: that is why both the 1- and 2-loop Leading Singularities for the 4-point amplitude evaluated to the same value, namely the 4-point tree-amplitude. That this pattern continues it rather remarkable, since it says that the on-shell diagrams with 4 external lines all evaluate to the 4-point tree amplitude, no matter how many hundreds of loops we add in. 

 The problem of determining and classifying all distinct on-shell diagrams, under the equivalence-moves, turns out to be an interesting mathematical problem that has become an exciting research topic \cite{ArkaniHamed:2012nw}.

%%%%%%%%%%%%%%%%%%%%%%%%%%%%%%% 
%%%%%%%%%%%%%%%%%%%%%%%%%%%%%%% 
%%%%%%%%%%%%%%%%%%%%%%%%%%%%%%% 
\newpage
\setcounter{equation}{0}
\section{Grassmannia}
\label{s:grassmannia}
%%%%%%%%%%%%%%%%%%%%%%%%%%%%%%% 
%%%%%%%%%%%%%%%%%%%%%%%%%%%%%%% 
%%%%%%%%%%%%%%%%%%%%%%%%%%%%%%%
The on-shell BCFW recursion formulas \reef{TreeBCFW} and \reef{LoopBCFW} have taught us that tree superamplitudes and loop-integrands of \emph{planar} $\mathcal{N}=4$ SYM 
can be  written as
\eq
\mathcal{A}_n^{L\text{-loop}}=\mathcal{A}_{n, \text{MHV}}^{\rm tree} \times
Y_n^{\text{$L$-loop}}\,,
\label{resN4amp}
\eqe
where $Y_n^{\text{$L$-loop}}$ is a  dual conformal invariant. In the N$^K$MHV sector, $Y_n^{\text{tree}}$ is a sum of $K$ products of 5-brackets (or $R$-invariants in Section \ref{s:N4symtrees}), as found in Sections \ref{s:emDCS} and \ref{s:momtwist}. At loop-level, $Y_n^{\text{$L$-loop}}$ is a linear combination of dual conformal invariant integrands. 

The MHV tree amplitude prefactor in \reef{resN4amp} serves an  important purpose:  since it is dual conformal {\em covariant} with homogenous dual conformal inversion-weight of the external particles --- as given in \reef{AmpInvert} ---  it generates the necessary dual conformal ``anomaly" that modifies the dual conformal  generators in such a way that they become part of the level 1 generators of a Yangian symmetry. Hence the MHV factor in \reef{resN4amp} is essential for Yangian symmetry.

There are several interesting points to consider:
\begin{enumerate}
  \item The color-ordered planar $\cn=4$ SYM superamplitudes have {\bf \em cyclic symmetry} in the labels of the external states. However, in the dual conformal  representations \reef{TreeBCFW} and \reef{LoopBCFW}, the cyclic symmetry is completely obscured. This is not surprising, because these recursion formulas are based on shifts of two adjacent external lines: making two lines special breaks the cyclic symmetry.  In the pursuit of happiness and manifest symmetries, we may ask if there is a formalism for the planar $\cn=4$ SYM superamplitudes in which both the (dual) conformal symmetry and the cyclic symmetry are manifest? This suggestion will guide us in Section \ref{s:YangInv}.
   \item In Section \ref{s:onshelldiag}, we gave examples of how the individual Leading Singularity diagrams can be understood as the values of BCFW diagrams of  tree amplitudes. Since each BCFW diagram is Yangian invariant, as indicated in the recursion formula \reef{TreeBCFW}, this implies that the Leading Singularities are also Yangian invariant. 
Thus, understanding the {\bf \em most general Yangian invariants} is a step towards gaining control of the planar superamplitudes in $\cn=4$ SYM at \emph{any} loop order. Of course, one still needs to understand how to put the Yangian invariants together to obtain a given superamplitude; read on.
  \item  BCFW recursion can be based on any choice of two shifted external momenta.  Different choices can give drastically different representations of the same amplitude, in particular with distinct spurious poles. For the amplitude to be \emph{local}, i.e.~free of spurious poles, the residues of the spurious poles must cancel in the sum of BCFW diagrams. Thus, we can view the {\bf \em equivalence of  two different BCFW representations} as intimately related with {\bf \em locality}. Each BCFW diagram is Yangian invariant, so by understanding how to enforce locality in a Yangian invariant way, it turns out that the equivalence between the different BCFW representations can be trivialized. 
\end{enumerate}

It may seem surprising, but the above three  points  can be addressed jointly. 
The strategy is to find a way to generate the most general Yangian invariant rational function from a formula in which cyclic symmetry is manifest. That's our job now, so let's get to work.

%%%%%%%%%%%%%%%%%%%%%%%%%%%%%%%%%%%%
\subsection{Yangian invariance and cyclic symmetry}
\label{s:YangInv}
%%%%%%%%%%%%%%%%%%%%%%%%%%%%%%%%%%%%%  
The level 0 Yangian generators are the superconformal generators studied in Section \ref{s:confsym}. In Section \ref{s:twist}, we introduced the supertwistors 
$\mathcal{W}_i^{\mathsf{A}} =
  \big(\,[i|^a,\, |\tilde\mu_i\rangle^{\dot{a}},\,\eta_{iA}\,\big)$
in order to linearize the action of the superconformal generators. 
 The Grassmann components of the supertwistor are simply the on-shell superspace coordinates $\eta_{iA}$, and 
  $|\tilde{\mu}_i\>^{\dot{a}}$ is the Fourier conjugate coordinate of 
  $|i\>^{\dot{a}}$. A function $f$ of on-shell momentum space spinor-helicity variables is Fourier transformed to the (super)twistor space as
\be
\int \bigg[\prod_{i=1}^n d^2|i\>\bigg]
\,f\big([i|,|i\>,\eta_i\big)
~e^{i \sum_{j=1}^n\! \<j \tilde\mu_j\>}
~\equiv~ 
\tilde{f}(\mathcal{W}^{\mathsf{A}}_i) \,.
\label{FourierW}
\ee
In supertwistor space, the superconformal generators are 
\be
G^{\mathsf{A}}\,_{\mathsf{B}}=\sum_{i=1}^n G_i^{\mathsf{A}}\,_{\mathsf{B}}=\sum_{i=1}^n\mathcal{W}^{\mathsf{A}}_i\frac{\partial}{\partial \mathcal{W}^{\mathsf{B}}_i}\,.
\label{WGen}
\ee
The level 1 generators can be written in bi-local form as (Section \ref{s:emDCS})
\be
\sum^n_{i<j}(-1)^{|\mathsf{C}|}
 \big[G_i^{\mathsf{A}}\,_{\mathsf{C}}\,G_j^{\mathsf{C}}\,_{\mathsf{B}}-(i\leftrightarrow j)\big]\,.
\label{Level1}
\ee

Our aim is a cyclic invariant formula that generates Yangian invariant rational functions; these are the building blocks for N$^K$MHV superamplitudes of planar $\cn=4$ SYM. Let us try to motivate the construction, step by step. To start with, note that the level 0 generators \reef{WGen} act on the supertwistor variables as
$SL(2,2|4)$ linear transformations. 
Any $\delta^{4|4}$ delta function whose argument is a linear combination of the supertwistors  is  invariant under the linear $SL(2,2|4)$ transformation: for example 
\be
\delta^{4|4}\big(\sum_{i=1}^n C_{i}\,\mathcal{W}_i^{\mathsf{A}}\big)
~\equiv ~
\delta^2\big(\sum_{i=1}^n C_{i} [i|^a \,\big) \,
\delta^2\big(\sum_{i=1}^n |i\>^{\dot{a}} C_{i}  \,\big) \,
\delta^{(4)}\big(\sum_{i=1}^n C_{i} \,\eta_{iA} \,\big) \,
\,,
\label{d44pre}
\ee
with some arbitrary auxiliary coefficients $C_i \in \mathbb{C}$. 
This is because a level 0 generator transforms the argument of one of the delta functions to that of another delta function;  schematically 
\eq
x\frac{\partial}{\partial y}\delta(x)\delta(y)=x\,\delta(x)\,\delta'(y)=0\,.
\eqe   
For an N$^K$MHV superamplitude, we need Yangian invariants that are  Grassmann polynomials of degree $4(K+2)$; so it is natural to take 
\be
k\equiv K+2
\ee
 products of \reef{d44pre}. 
 Note that $k$ counts the number of negative helicity gluons in the pure gluon amplitude. 
 To avoid having $k$ identical delta functions, we introduce $k$ sets of the auxiliary variables $C_{\mathsf{a}i}$  labelled  by  an index $\mathsf{a}=1,2,\dots,k$.
So now we have a Grassmann degree $4(K+2)$ object 
\be
  \prod_{\mathsf{a}=1}^k 
  \delta^{4|4}\big(\sum_{i=1}^n 
  C_{\mathsf{a}i}\mathcal{W}_i^{\mathsf{A}}\big)
\label{d44}
\ee 
that is $SL(2,2|4)$ invariant. The parameters $C_{\mathsf{a}i}$ are sometimes called {\bf \em link variables} \cite{ArkaniHamed:2009dn}.

The $n\times k$ parameters $C_{\mathsf{a}i}$ are arbitrary so to remove the dependence on them, let us integrate \reef{d44} over all $C_{\mathsf{a}i}$. This has the further benefit of making the integrated result cyclically invariant: a permutation of the 
$\mathcal{W}_i$'s is compensated by a permutation of the integration variables $C_{\mathsf{a}i}$'s (and such a transformation has unit Jacobian).
However, it is not clear what measure we should use when integrating over the 
$C_{\mathsf{a}i}$'s. So let us allow for a general cyclically invariant function $f(C)$ and write our candidate `generating function' as
\be
\int d^{k\times n}C~f(C)~\prod_{\mathsf{a}=1}^{k}\delta^{4|4}\bigg(\sum_{i=1}^n C_{\mathsf{a}i}\mathcal{W}_i^{\mathsf{A}}\bigg)\,.
\label{PreGrass}
\ee
The integral of the $k\times n$ complex parameters is intended to be carried out as a contour integral. The choice of contour is a very important and physically relevant aspect that will be discussed in Section \ref{s:GrassRes}.

When the level 1 generators are considered, it turns out that there is a unique choice of $f(C)$ such that \reef{PreGrass} is Yangian invariant. We will not repeat the argument here, but refer you to \cite{DrummondTDual,YangianFix}. The unique function that gives  \reef{PreGrass} full Yangian symmetry is
\be
  f(C)=\frac{1}{M_1M_2\cdots M_n}\,,
\label{Minor}
\ee
where $M_i$ is the $i$th ordered minor of the $k\times n$ matrix $C_{\mathsf{a}i}$: this is the determinant of the $k\times k$ submatrix whose first column is the $i$th column of $C_{\mathsf{a}i}$, specifically
\eq
M_i\equiv \epsilon^{\mathsf{a}_1\,\mathsf{a}_2\,\ldots\, \mathsf{a}_{k}}C_{\mathsf{a}_1i}C_{\mathsf{a}_2,i+1}\cdots C_{\mathsf{a}_{k},i+k-1}\,,
\eqe
with $i=1,2,\dots, n$. One goes around cyclically when reaching the end of the $C$-matrix. 
For example, the $n=5$ and $k=2$ matrix 
\be
C=
\bigg(
\begin{array}{cccccc}
C_{11} & C_{12}&C_{13}& C_{14} &C_{15} \\
C_{21} & C_{22}&C_{23}& C_{24} &C_{25}
\end{array}
\bigg)
\ee 
gives $M_1 = C_{11} C_{22} - C_{12} C_{21}$ and 
$M_5 = C_{15} C_{21} - C_{11} C_{25}$.

Thus we have learned that the integral
\eq
\int \frac{d^{k\times n}C}{M_1M_2\cdots M_n}\,\prod_{\mathsf{a}=1}^{k}\delta^{4|4}\bigg(\sum_{l=1}^n C_{\mathsf{a}l}\mathcal{W}_l^{\mathsf{A}}\bigg)\,
\label{PreGrass2}
\eqe
 is invariant under the  Yangian generators and has cyclic symmetry.

Before declaring victory, there are loose ends that we must comment on. \emph{First}, a minor issue (yes, a pun) is that if $M_i$ contains columns that are not strictly increasing due to cyclicity (for example $M_{n-1}=\cdots C_{\mathsf{a}n}C_{\mathsf{a}1}\cdots$) then the proof of Yangian invariance goes through only on the support of the bosonic delta functions. \emph{Second}, a major issue is that the integral we so proudly wrote down in (\ref{PreGrass2}) is not at all well-defined --- it is divergent. To see this, note that the product of delta functions is invariant under a $GL(k)$ rotation of the $k$ $\mathsf{a}$-indices. The minors only respect $SL(k)$ transformations: $GL(1)$ takes $C_{\mathsf{a}i}\rightarrow tC_{\mathsf{a}i}$, hence 
$M_i\rightarrow t^{k}M_i$, but this excess weight is canceled by the Jacobian of $d^{k\times n}C$. Thus the integral  has $GL(k)$ symmetry.
To define a proper integral we need to ``gauge fix" the $GL(k)$ redundancy. We indicate the need to gauge fix $GL(k)$ by writing
\eq
\mathcal{L}_{n,k}\big(\mathcal{W}_i\big)
~=~\int \frac{d^{n\times k}C_{\mathsf{a}i}}{GL(k)\,\prod_{j=1}^n M_j}\,\prod_{\mathsf{a}=1}^{k}\delta^{4|4}\bigg(\sum_{l=1}^n C_{\mathsf{a}l}\mathcal{W}_l^{\mathsf{A}}\bigg)\,.
\label{TheGrass}
\eqe
It turns out \cite{DrummondTDual,YangianFix} that for given $n$ and $k$, \emph{$\mathcal{L}_{n,k}$ is the unique cyclically invariant integral-expression that generates all Yangian invariants!} We are going to give examples in the following sections. 
The formula \reef{TheGrass} was first introduced by Arkani-Hamed, Cachazo, Cheung and Kaplan~\cite{ArkaniHamed:2009dn}, who at the time conjectured that it produces all Leading Singularities of planar $\mathcal{N}=4$ SYM. A similar integral formula was presented by Mason and Skinner~\cite{Mason} based on   momentum supertwistors $\mathcal{Z}$, as opposed to the `regular' supertwistors $\mathcal{W}$, thus interchanging the role of the ordinary superconformal and dual superconformal symmetries. 

We used supertwistors 
$\mathcal{W}_i^{\mathsf{A}} =
  \big(\,[i|^a,\, |\tilde\mu_i\rangle^{\dot{a}},\,\eta_{iA}\,\big)$ to emphasize superconformal and Yangian symmetry in the construction above. However, since we are more familiar with scattering amplitudes in momentum space 
 $\big(\,[i|^a,\, |i\rangle^{\dot{a}},\,\eta_{iA}\,\big)$, 
we are going to  inverse-Fourier transform all $|\tilde{\mu}_i\>$ in \reef{TheGrass} back to $|i\>$. This is conveniently done in a gauge-fixing of  the $GL(k)$ symmetry where the first $k\times k$ block of $C_{\mathsf{a}i}$ is  the unit matrix. For example for $n=7$ and $k=3$, we have
\eq
C=
\left(\begin{array}{ccccccc}
1 & 0 & 0 & c_{14} & c_{15} & c_{16}& c_{17} \\
0 & 1 & 0 & c_{24} & c_{25} & c_{26}& c_{27} \\
0 & 0 & 1 & c_{34} & c_{35} & c_{36}& c_{37}
\end{array}\right)\,.
\label{C37}
\eqe 
In this gauge, \reef{TheGrass} becomes
\eq
  \int \frac{d^{(n-k)\times k}c}{\prod_{j=1}^n M_j}\,
  \prod_{\mathsf{a}=1}^{k}
   \delta^{2}\bigg([{\mathsf{a}}|+\sum_{l=k+1}^n c_{\mathsf{a}l}[l|\bigg)\,
   \delta^{2}\bigg( |\tilde{\mu}_{\mathsf{a}}\>+\sum_{l=k+1}^n 
   |\tilde{\mu}_l\>\,c_{\mathsf{a}l}\bigg)\,
   \delta^{(4)}\bigg(\eta_{\mathsf{a}}+\sum_{l=k+1}^n c_{\mathsf{a}l}\,\eta_l \bigg)\,.
\eqe
Performing the inverse-Fourier transform 
$\int d^2|\tilde{\mu}_j\> \, e^{-i \<j \,\m_j\>}$ 
for each $j=1,\dots,n$ gives
\eqa
  \nonumber
  \mathcal{L}_{n,k}\big([i|,|i\>,\eta_i\big)
  &=&
  \int \frac{d^{(n-k)\times k}c}{\prod_{j=1}^n M_j}
  \left[\,\prod_{\mathsf{a}=1}^{k}
  \delta^{2}\bigg([{\mathsf{a}}|+\sum_{l=k+1}^n c_{\mathsf{a}l}[l|\bigg)\,
  \delta^{(4)}\bigg(\eta_{\mathsf{a}}+\sum_{l=k+1}^n c_{\mathsf{a}l}\,\eta_l \bigg)   
  \right]\\ 
&&
  \hspace{2.4cm}
  \times
  \left[\,
  \prod_{i=k+1}^n\delta^{2}
  \bigg(|i\>-\sum_{\mathsf{a}=1}^{k} |\mathsf{a}\> c_{\mathsf{a}i}
  \bigg)
  \right]\,.
\label{TheGrassMom}
\eqae
\exercise{}{Fill out the details of the inverse Fourier transformation to derive 
\reef{TheGrassMom}.
}
The representation \reef{TheGrassMom} is central in the next section where we study the geometric interpretation  of $\mathcal{L}_{n,k}$. In Section \ref{s:GrassRes}, we show that familiar amplitude expressions can be derived from  $\mathcal{L}_{n,k}$.

%%%%%%%%%%%%%%%%%%%%%%%%%%%%%%%%%%%%%%
\subsection{The Grassmannian}
\label{s:Grassmannia}
%%%%%%%%%%%%%%%%%%%%%%%%%%%%%%%%%%%%%%
It is very convenient to view the $n\times k$ matrices $C_{\mathsf{a}l}$ in \reef{TheGrass} as $k$ $n$-component vectors that define a $k$-plane in $\mathbb{C}^n$. The space of all $k$-planes in an $n$-dimensional space
 is called the {\bf \em Grassmannian} Gr($k,n$). The formula \reef{TheGrass} for  
$\mathcal{L}_{n,k}$ is therefore naturally viewed as a cyclic invariant integral over all $k$-planes in the Grassmannian. Since any non-degenerate linear transformation of the $k$ $n$-vectors gives the same plane, there is a natural $GL(k)$ invariance. 
It is precisely the same $GL(k)$ redundancy  we encountered  previously in the discussion of the integral \reef{TheGrass}: the Grassmannian integral \reef{TheGrass} is well-defined only when `gauge fixing' the $GL(k)$ redundancy.
Because of the $GL(k)$ redundancy, the dimensions of the Grassmannian  Gr($k,n$) is  $k\times n-k^2=k(n-k)$.

With this geometric picture in mind, let us now examine the  bosonic delta functions in the gauge-fixed expression \reef{TheGrassMom} for  
$\mathcal{L}_{n,k}$. They enforce the constraints
\be
\sum_{i=1}^n
C_{\mathsf{a}i}\, [i|^a=0\,,
~~~~~~~
\sum_{i=1}^n
\tilde{C}_{\mathsf{a}' i} \<i| =0\,,
\label{MomConstraint}
\ee  
where $C$ and $\tilde{C}$ are
$k\times n$ and $(n-k)\times n$ matrices respectively, so $\mathsf{a}'=k+1,\cdots,n$. They are explicitly given as
{\small
\be
C=\left(\begin{array}{ccccccc}
1 & 0&\cdots & 0 & c_{1,k+1} & \cdots & c_{1n} \\
0 & 1&\cdots & 0 & c_{2,k+1} & \cdots & c_{2n} \\ 
\vdots &\vdots & \vdots & \vdots & \vdots & \vdots & \vdots \\
0 & \cdots &0& 1 & c_{kk+1} & \cdots & c_{kn}
\end{array}\right),
~~~
\tilde{C}=\left(\begin{array}{ccccccc}
-c_{1,k+1} & \cdots& -c_{k,k+1}  & 1 &0 & \cdots &0 \\
-c_{1,k+2} & \cdots &-c_{k,k+2} & 0 & 1 & \cdots & 0 \\
 \vdots &\vdots & \vdots & \vdots & \vdots & \vdots & \vdots \\
 -c_{1n} & \cdots &-c_{kn} & 0 & 0 & \cdots & 1\end{array}\right)\,.
 \label{GeoPic}
\ee}%
An important feature is that 
\be
  C \, \tilde{C}^T 
  = 
  \sum_{i=1}^nC_{\mathsf{a}i}\,\tilde{C}_{\mathsf{a}'i}=0 \,.
  \label{CCtzero}
\ee
\exercise{}{Construct $\tilde{C}$ associated with \reef{C37} and check that \reef{CCtzero} holds.
}
We can view  $\tilde{C}$ as $(n-k)$ $n$-vectors spanning an  $(n-k)$-plane in $n$ dimensions.  The condition \reef{CCtzero} states that the $(n-k)$-plane defined by $\tilde{C}$ is the orthogonal complement of the $k$-plane defined by $C$. 

In this notation, we can reinstate the $GL(k)$ redundancy and write our momentum space Grassmannian integral \reef{TheGrassMom} as
\be
\mathcal{L}_{n,k}
=
\int \frac{d^{n\times k}C}{{GL}(k)\prod_{j=1}^n M_j}
\left[\,\prod_{\mathsf{a}=1}^{k}
\delta^{2}\Big({\textstyle \sum_{i}}C_{\mathsf{a}i}\, [i|\Big)
\delta^{(4)}\Big({\textstyle \sum_{i}} C_{\mathsf{a}i}\,\eta_{iA}\Big)\right]
\left[\,\prod_{\mathsf{a}'=k+1}^n
\delta^{2}\Big({\textstyle \sum_{i}}
\tilde{C}_{\mathsf{a}' i} \<i|\Big)\right]\,,\\
\label{TheBetterGrassMom}
\ee
with the understanding that  $\tilde{C}$ is defined as the complement to $C$ in the sense of \reef{CCtzero}. 

Now a geometric picture is emerging of the meaning of the constraints \reef{MomConstraint}. We can consider the collection of the $n$ $|i\>$'s as defining a 2-plane in an $n$-dimensional space,
\be
  \left(\begin{array}{cccc}
  |1\>^{\dot{1}} & 
  |2\>^{\dot{1}} &
  \cdots&
  |n\>^{\dot{1}} 
  \\
  |1\>^{\dot{2}} & 
  |2\>^{\dot{2}} &
  \cdots&
  |n\>^{\dot{2}} 
\end{array}\right)\,.
\ee
Similarly the $[i|$'s define a 2-plane in an $n$-dimensional space.

\begin{figure}
\begin{center}
\includegraphics[scale=0.4]{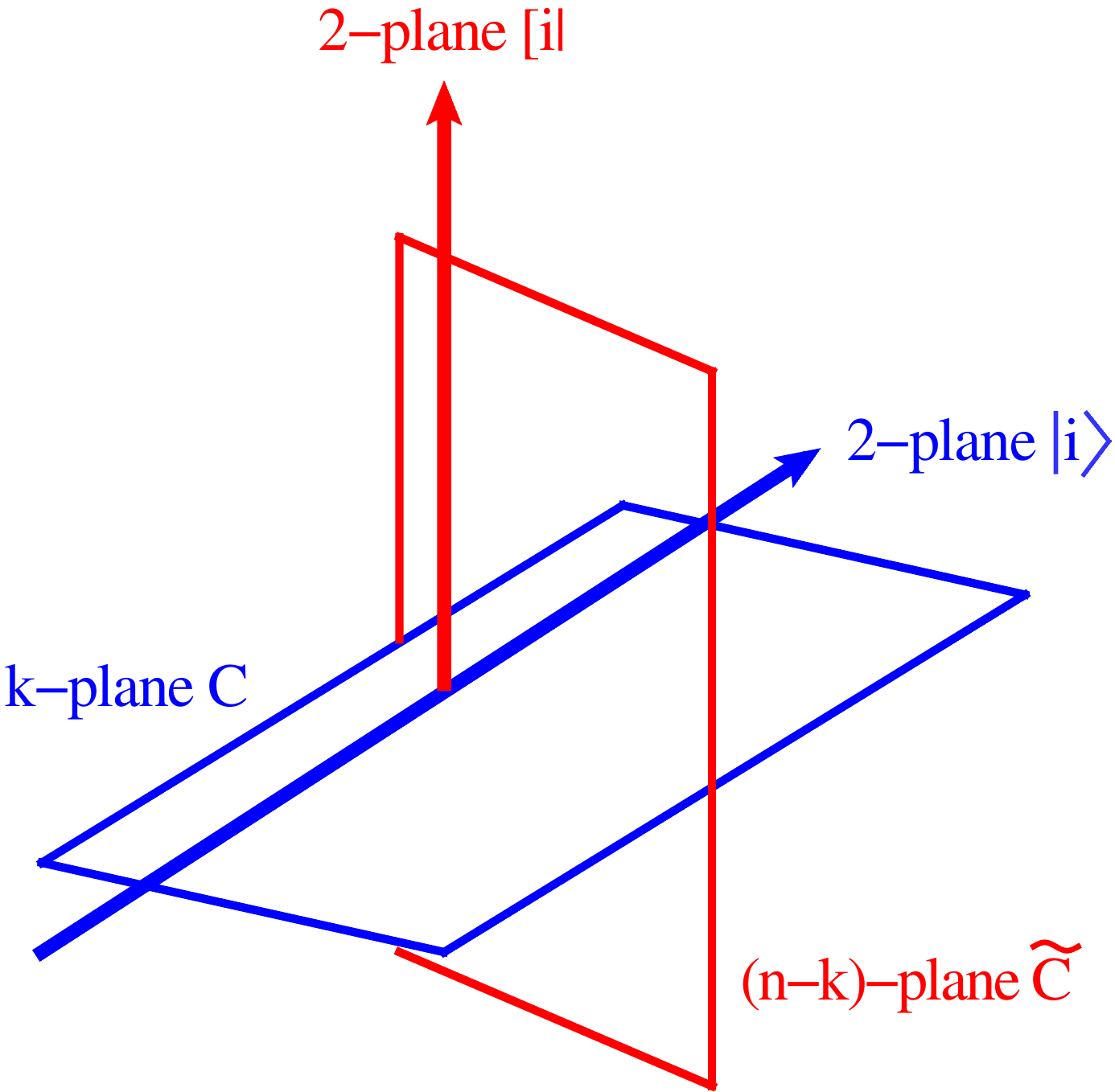}
\caption{\small The geometry of planes in the Grassmannian. The $k$-plane $C$ and $(n-k)$-plane $\tilde{C}$ are orthogonal complements, so the constraint $\sum_i C_{\mathsf{a}i} [i|=0$ in  \reef{MomConstraint} means that the 2-plane spanned by the $n$ $[i|$'s  is orthogonal to $C$ and hence must be contained in $\tilde{C}$. Similarly, $C$ contains the 2-plane spanned by the $|i\>$'s. It follows from the geometry that the 2-planes spanned by $[i|$ and $|i\>$, respectively, are orthogonal, but that is exactly the statement of momentum conservation $\sum_{i=1}^n |i\> [i| = 0$.}
\label{CPlanes}
\end{center}
\end{figure}

The constraints \reef{MomConstraint} say that the 2-plane spanned by the $[i|$'s  is orthogonal to the $k$-plane $C$ and the 2-plane defined by $|i\>$'s is orthogonal to the $(n-k)$-plane $\tilde{C}$. 
This is illustrated in Figure \ref{CPlanes}. 
Since $\tilde{C}$ and $C$ are  orthogonal complements, we immediately conclude that $\tilde{C}$ must contain the 2-plane $[i|$ while $C$ must contain $|i\>$. This in turn tells us that the 2-plane of $[i|$ must be orthogonal to the 2-plane of $|i\>$, i.e. 
\be
  \label{momconsGrass}
  \sum_{i=1}^n | i\rangle[i|=0\,.
\ee
This is just the statement that the external momenta satisfy momentum conservation. Thus, seemingly out of nowhere, the cyclic- and Yangian-invariant generating function $\mathcal{L}_{n,k}$ `knows' about momentum conservation.

As we have just shown, the bosonic delta functions in \reef{TheBetterGrassMom} give non-vanishing results only on the support of momentum conservation 
$\d^4(P)$. With this in mind, let us  count the number of `free' integration variables in \reef{TheBetterGrassMom}, i.e.~the number of $c_{\mathsf{a}i}$'s not localized by the bosonic delta functions.
After gauge-fixing $GL(k)$, we have a total of $k\times (n-k)$ $c_{\mathsf{a}i}$-variables. There are $[2k + 2(n-k)] = 2n$ bosonic delta functions, but this  includes momentum conservation $\d^4(P)$, so 4 of the delta functions do not localize any $c_{\mathsf{a}i}$-variables. Therefore a total of 
\be
\#\text{(integration variables)} = [ k\times (n-k)-(2n-4) ]=(k-2)(n-k-2)
\label{intvarno}
\ee
$c_{\mathsf{a}i}$-variables are left to be integrated. 

In the MHV sector, $k=2$ so we learn from \reef{intvarno} that the integral \reef{TheBetterGrassMom} is fully localized by the bosonic delta functions. For generic $k$ and $n$, $\mathcal{L}_{n,k}$ is a  multi-dimensional integral that localizes on the poles in the minors; we work out an explicit example in the next section. For $k=0$, the integral vanishes since it is  proportional to $\delta^2( |i\>)$ which does not have support for generic momenta. This is simply the statement that the `all-plus' gluon amplitude  vanishes in $\cn=4$ SYM. 
For $k=1$, the last delta function in \reef{TheBetterGrassMom} forces all of the $|i\>$'s to be proportional to each other, but this lacks support for    generic momenta, with the exception of special kinematics for $n\!=\!3$. Not surprisingly, this says that the all-plus-and-one-minus gluon amplitudes vanish in $\mathcal{N}=4$ SYM for $n>3$. 

The counting of integration variables in \reef{intvarno} is invariant under $k \to (n-k-2)$. This corresponds to a flip of what we identify as positive and negative helicity, i.e.~which states are associated with highest/lowest Grassmann weight. Indeed, for $k=n-2$, the (super)amplitude is anti-MHV, so it makes sense that the Grassmann integral is localized completely by the bosonic delta functions, just as it is for the MHV sector $k=2$.

Perhaps you have noticed that the dimension $(k-2)(n-k-2)$ of the $\mathcal{L}_{n,k}$-integral is also the dimension of the Grassmannian Gr($k-2,n-4$). This is not a coincidence. Since the bosonic delta functions enforce that $C_{\mathsf{a}i}$  contains the 2-plane $|i\>$, this reduces the matrix down to $(k-2)\times n$. We can then use part of the $GL(k)$ redundancy to remove $4(k-2)$ components of the remaining $C_{\mathsf{a}i}$, leaving behind a $(k-2)\times (n-4)$ matrix with a residual $GL(k-2)$ redundancy. This matrix lives in Gr$(k-2,n-4)$. 

So far we have constructed a cyclic and Yangian invariant integral $\mathcal{L}_{n,k}$ in the Grassmannian, but we have not really done anything with it. In fact, other than showing that it captures momentum conservation, we have given you little reason to believe that there is any connection to scattering amplitudes in planar $\cn=4$ SYM. 
So now we better show you how it works.

%%%%%%%%%%%%%%%%%%%%%%%%%%%%%%%%%%%%%%
\subsection{Yangian invariants as residues in the Grassmannian }
\label{s:GrassRes}
%%%%%%%%%%%%%%%%%%%%%%%%%%%%%%%%%%%%%%
We carry out the Grassmannian integral \reef{TheBetterGrassMom} in the simplest cases to illustrate how the familiar MHV and NMHV superamplitudes appear. 

%%%%%%%%%%%%%%%%%%%%%%%%%%%%%%%%%%%%%%%%
\subsubsection{MHV amplitudes}
%%%%%%%%%%%%%%%%%%%%%%%%%%%%%%%%%%%%%%%%
As the counting \reef{intvarno} shows,  the bosonic delta functions completely localize the integral \reef{TheBetterGrassMom} for the MHV sector ($k=2$). In fact, the geometric description in the previous section tells us that for MHV, the bosonic delta functions exactly encode conservation of 4-momentum on the $n$ external states: $\delta^{4}(P)$. What about the Grassmann delta function? Well, when $k=2$, $C$ defines a 2-plane, and since the 2-plane $|i\>$ must be contained in $C$, we can simply identity to the two 2-planes; up to a $GL(2)$ transformation we therefore have
\eq
\left(\begin{array}{cccc}C_{11} & C_{12}& \cdots &C_{1n} \\C_{21} & C_{22}& \cdots &C_{2n}\end{array}\right)=
  \left(\begin{array}{cccc}
  |1\>^{\dot{1}} & 
  |2\>^{\dot{1}} &
  \cdots&
  |n\>^{\dot{1}} 
  \\
  |1\>^{\dot{2}} & 
  |2\>^{\dot{2}} &
  \cdots&
  |n\>^{\dot{2}} 
\end{array}\right)\,.
\eqe 
Using this explicit representation of the $C_{\mathsf{a}i}$'s, the Grassmann delta function becomes the familiar statement of supermomentum conservation,
\eq
\prod_{\mathsf{a}=1}^{2}
\delta^{(4)}\Big( {\textstyle\sum_i} C_{\mathsf{a}i}\,\eta_{iA}\Big)
=\prod_{\dot{a}=1}^{2}\delta^{(4)}\Big({\textstyle\sum_i}  |i\>^{\dot{a}}\,\eta_{iA}\Big)
=\delta^{(8)}\big(\tilde{Q}\big)\,.
\eqe
The minors are
\eq
M_i=\epsilon^{\mathsf{a}\mathsf{b}}\,C_{\mathsf{a}i}\,C_{\mathsf{b},i+1}
=-\epsilon_{\dot{a}\dot{b}}\,|i\>^{\dot{a}}\,|i+1\>^{\dot{b}}
=-\langle i,i+1\rangle\,.
\eqe
Putting everything together, we find  
\be
  \mathcal{L}_{n,2}
~=~
  (-1)^n\frac{\delta^{(8)}(\tilde{Q})\,\delta^4(P)}{\prod_{i=1}^n\langle i ,i+1\rangle}
~=~ (-1)^n \mathcal{A}_{n,\text{tree}}^\text{MHV}
\,.
\label{k=2Result}
\ee
So for $k=2$, the cyclic invariant integral $\mathcal{L}_{n,2}$ nicely produces the MHV tree-amplitude (up to an overall convention-dependent sign).
\example{Did that go a little fast? Fair enough, let us evaluate $\mathcal{L}_{n,2}$ in full detail, starting with the gauged-fixed expression \reef{TheGrassMom}, which for $k=2$ gives
\be
\begin{split}
  \mathcal{L}_{n,2} &=~ \int \frac{d^{(n-2)\times 2}c}{M_1 \cdots M_n}\,
  \left[\,\prod_{\mathsf{a}=1}^{2}
  \delta^{2}\bigg([{\mathsf{a}}|+\sum_{l=3}^n c_{\mathsf{a}l}[l|\bigg)\,
  \delta^{(4)}\bigg(\eta_{\mathsf{a}}+\sum_{l=3}^n c_{\mathsf{a}l}\,\eta_l \bigg)   
  \right]\\
  &\hspace{3.2cm}
  \times
  \left[\,
  \prod_{i=3}^n\delta^{2}
   \bigg(|i\>
  -  |1\> c_{1i}
  -  |2\> c_{2i}
  \bigg)
  \right]\,.
\end{split}
\label{L2n}
\ee
We rewrite the last set of delta functions with $i = 3,4,\dots, n$
\eq
  \delta^{2}
  \bigg(|i\>
  -  |1\> c_{1i}
  -  |2\> c_{2i}
  \bigg)
=
\frac{1}{\<12\>}\,
\delta\left(c_{1i}-\frac{\langle i2\rangle}{\langle12\rangle}\right)\,
\delta\left(c_{2i}-\frac{\langle i1\rangle}{\langle21\rangle}\right)
  \,,
\label{DeltaString} 
\eqe
to show how they localize the $2(n-2)$ components $c_{1i}$ and $c_{2i}$. Thus, on the support of these delta functions, the first four bosonic delta functions in \reef{L2n} give 
\eq
  \delta^{2}\bigg([1|+\sum_{l=3}^n c_{1l}[l|\bigg)\,
  \delta^{2}\bigg([2|+\sum_{l=3}^n c_{2l}[l|\bigg)\,
~=~
\langle 12\rangle^2~
\delta^4(P)\,;
\eqe
 this is how the momentum conservation delta function appears. 

Likewise for the Grassmann delta function: on the support of \reef{DeltaString} it gives
\be
  \prod_{\mathsf{a}=1}^{2}
  \delta^{(4)}\bigg(\eta_{\mathsf{a}}+\sum_{l=3}^n c_{\mathsf{a}l}\,\eta_l \bigg)   
    ~=~
    \frac{1}{\<12\>^4}~\delta^{(8)}\big(\tilde{Q}\big)\,.
\ee

Finally, we evaluate the minors $M_i$. With the help of the Schouten identity we find
\be
  M_1 = 1\,,~~~
  M_2 = \frac{\<23\>}{\<12\>}\,,~~~
  M_3 = -\frac{\<34\>}{\<12\>}\,,~~~
  M_4 = -\frac{\<45\>}{\<12\>}\,,~~
  \dots~~\,,
  M_n = -\frac{\<n1\>}{\<12\>}\,,~~
  \label{Ms}
\ee
and hence
\eq
\prod_{i=1}^nM_i =\frac{(-1)^n}{\langle 12\rangle^n}\left(\prod_{i=1}^n\langle i,i+1\rangle\right)\,.
\label{prodMs}
\eqe
Inserting everything into \reef{L2n} we indeed obtain $(-1)^n$ times the MHV tree superamplitude, as in \reef{k=2Result}.}
\exercise{}{Derive \reef{DeltaString}-\reef{prodMs}. Then plug the results into \reef{L2n} to verify that all powers of $\<12\>$ cancel.}
The Grassmannian integral $\mathcal{L}_{n,k}$ has given a unique result, 
$\mathcal{A}_{n,\text{tree}}^\text{MHV}$, for $k=2$. Given that the Grassmannian integral produces all possible Yangian invariants \cite{DrummondTDual,YangianFix}, this means that all MHV superamplitudes have the same Leading Singularities in planar $\cn=4$ SYM, up to a sign, \emph{to all loop-orders}. We have already seen a non-trivial manifestation of this fact at 2-loops in Section \ref{s:compLS} (and again in Section \ref{s:onshelldiag}), where the Leading Singularities of the 2-loop 4-point superamplitude was found to be the MHV tree superamplitude. Thus, if we were to take the 234-loop MHV superamplitude and solve the on-shell constraints that localize the $234\times4=936$ loop momenta, the result of the Leading Singularities will again be  MHV tree superamplitudes! No other Yangian invariants are available at MHV order.

%%%%%%%%%%%%%%%%%%%%%%%%%%%%%%%%%%%%%%%%%%%%%%%%%%%%%%%%%%%% 
\subsubsection{6-point NMHV amplitudes}
\label{s:6ptGrassmannian}
%%%%%%%%%%%%%%%%%%%%%%%%%%%%%%%%%%%%%%%%%%%%%%%%%%%%%%%%%%%%
Let us now move on to a slightly more complicated --- hence more exciting --- example, the 6-point NMHV amplitude. With $k=3$ and $n=6$, the counting formula \reef{intvarno} reveals that the Grassmannian integral $\mathcal{L}_{6,3}$ involves just one non-trivial integration. To evaluate it, we choose the gauge
\eq
\left(\begin{array}{cccccc}c_{21} & 1 & c_{23} & 0 & c_{25} & 0 \\ c_{41} & 0 & c_{43} & 1 & c_{45} & 0 \\c_{61} & 0 & c_{63} & 0 & c_{65} & 1\end{array}\right)
\,.
\label{36gauge}
\eqe 
The $c$-variables are labeled  such that the bosonic delta functions in \reef{TheBetterGrassMom} can be written as:
\be
\delta^2\Big([\,\bar{i}\,|+\sum_{j}c_{\,\bar{i}j}\,[j|\Big)\,,
\,\quad 
\delta^2\Big(|j\>-\sum_{\bar{i}} |\,\bar{i}\,\>\,
  c_{\,\bar{i}j}\Big)\,,
\label{6ptconstraint}
\ee
where  $\bar{i}=2,4,6$ and $j=1,3,5$. Since the integral is 1-dimensional,  there must be a 1-parameter family of solutions that solve the delta function constraints \reef{6ptconstraint}. Indeed, if $c^*_{\,\bar{i}j}$ is a solution, then 
\eq
\hat{c}_{\,\bar{i}j}(\tau)=c^*_{\,\bar{i}j}
+\tfrac{1}{4}\,
\tau\, \,
\epsilon_{\bar{i}\bar{j}\bar{k}}\,\langle \bar{j}\bar{k}\rangle \, \,
\epsilon_{jkl}[k l]\,,
\label{ctau}
\eqe 
is also a solution for any $\tau$. 
Here $\epsilon_{\bar{i}\bar{j}\bar{k}}$ is a Levi-Civita symbol  for the indices $\bar{i}=2,4,6$, and similarly for $\epsilon_{jkl}$. There are implicit sums over repeated labels in \reef{ctau}.  That $\hat{c}_{\,\bar{i}j}(\tau)$ is a solution can be seen from the result that the $\tau$ dependence drops out from the constraints in \reef{6ptconstraint} due to the Schouten identity:
\eqa
\nonumber
[\,\bar{i}\,|^a+\sum_{j} \hat{c}_{\,\bar{i}j}(\tau)[j|^a
&=&
\tfrac{1}{4}\, \tau\,\,\epsilon_{\bar{i}\bar{j}\bar{k}}\langle \bar{j}\bar{k}\rangle
\sum_{j}\epsilon_{jkl}[{j}|^a[kl]\\
&=&\tfrac{1}{2}\, \tau\,\,\epsilon_{\bar{i}\bar{j}\bar{k}}\langle \bar{j}\bar{k}\rangle 
\big([1|^a[35]+[{3}|^a[51]+[{5}|^a[13]\big)
~=~0\,.
\eqae 
We can now remove the bosonic delta functions by localizing the integral on the solution to the constraints \reef{6ptconstraint} such that the remaining integral is over the 1-dimensional parameter $\tau$. That gives
\eqa
\nonumber
\mathcal{L}_{6,3}
&=&
\int \frac{d^{9}c_{\,\bar{i}j}}{M_1 \cdots M_n}
\bigg[
\prod_{j}
\delta^2\Big(|j\>-\sum_{\bar{i}} |\,\bar{i}\,\>\,
  c_{\,\bar{i}j}\Big)
\bigg]
\bigg[
\prod_{\bar{i}}
\delta^2\Big([\,\bar{i}\,|+\sum_{j}c_{\,\bar{i}j}\,[j|\Big)
\delta^{(4)}\Big(\eta_{\,\bar{i}}+\sum_{j}c_{\,\bar{i}j}\eta_{j}\Big)
\bigg]\\
\nonumber
&=&
\delta^4(P)
\int \frac{d^{9}c_{\,\bar{i}j}d\tau}{M_1 \cdots M_n}~
\delta^9\Big(c_{\,\bar{i}j}-\hat{c}_{\,\bar{i}j}(\tau)\Big)~
\prod_{\bar{i}}
\delta^{(4)}\Big(\eta_{\,\bar{i}}+\sum_{j}c_{\,\bar{i}j}\eta_{j}\Big)\,\\
&=&
\delta^4(P)\int \frac{d\tau}{\hat{M}_1 \cdots \hat{M}_n}\prod_{\bar{i}}\delta^{(4)}\Big(\eta_{\,\bar{i}}+\sum_{j}\hat{c}_{\,\bar{i}j}\eta_{j}\Big)\,.
\label{6ptInt}
\eqae
The `hat' indicates dependence on $\tau$ via \reef{ctau}. 
In the gauge \reef{36gauge}, the minors are  
\be
\hspace{-0.4mm}
\begin{array}{rclrclrcl}
 \hat{M}_1&=&\hat{c}_{43}\hat{c}_{61}-\hat{c}_{41}\hat{c}_{63}\,,~~
 &\hat{M}_3&=&\hat{c}_{23}\hat{c}_{65}-\hat{c}_{25}\hat{c}_{63}\,,~~
 &\hat{M}_5&=&\hat{c}_{21}\hat{c}_{45}-\hat{c}_{25}\hat{c}_{41}\,,~~\\[1mm]
\hat{M}_2&=&-\hat{c}_{63}\,,&
\hat{M}_4&=&-\hat{c}_{25}\,,&
\hat{M}_6&=&-\hat{c}_{41}\,.
\end{array}
\label{mihats}
\ee

At this stage, there appears to be no \emph{a priori} prescription of which contour to pick in the $\tau$-plane. Each minor $M_i$ has a simple pole in $\tau$, so there are six different residues that we denote $\{M_i\}$.
 Let us focus on the pole in $M_4$. This means that $\tau$ is evaluated at  
 $\tau_*$ such that $\hat{c}_{25}(\tau_*)=0$. We can make the calculation simpler by choosing the origin for $\tau$ such that $\hat{M}_4 = 0$ for $\tau=0$; in other words, we choose  
 $\hat{c}_{25}^* = 0$. Let us use the constraints \reef{6ptconstraint} to solve for the 8 other $\hat{c}_{\,\bar{i}j}^*$'s.  From  
\eqa
|5\> - |4\>c^*_{45}-|6\>c^*_{65}=0,\quad ~~~~
[2|+c^*_{21}[1|+c^*_{23}[3|=0
\eqae 
we deduce
\eqa
c^*_{45}=\frac{\langle 56\rangle}{\langle 46\rangle},\;~~~
c^*_{65}=\frac{\langle 45\rangle}{\langle 46\rangle},\;~~~
c^*_{21}=-\frac{[23]}{[13]},\;~~~
c^*_{23}=-\frac{[12]}{[13]}\,.
\eqae
And this in turn allow us to solve
\eqa
[4|+c^*_{41}[1|+c^*_{43}[3|+c^*_{45}[5|=0,\quad~~~~
[6|+c^*_{61}[1|+c^*_{63}[3|+c^*_{65}[5|=0\,,
\eqae
to find
\eq
c^*_{41}=-\frac{\langle 6|4\!+\!5|3]}{\langle46\rangle[13]},~~~
\,c^*_{43}=\frac{\langle 6|4\!+\!5|1]}{\langle46\rangle[13]},~~~
\,c^*_{61}=\frac{\langle 4|5\!+\!6|3]}{\langle46\rangle[13]},~~~
\,c^*_{63}=-\frac{\langle 4|5\!+\!6|1]}{\langle46\rangle[13]}\,.
\eqe
\exercise{}{Use the above results for $c^*_{\,\bar{i}j}$ to show that the unused constraints in \reef{6ptconstraint} give $\d^4(P)$.}
We can now substitute the solutions $c^*_{\hat{i}j}$ into the minors \reef{mihats} and the Grassmann delta functions to obtain the residue of the integral \reef{6ptInt} of the pole $1/M_4$, denoted by $\{M_4\}$. For simplicity, consider a particular component amplitude, namely the gluon amplitude with helicity assignments $(+,-,+,-,+,-)$. For this amplitude, the coefficient from the Grassmann delta functions is just $1$. 
Taking into account the extra factor of $\langle 46\rangle[13]$ coming from 
$\hat{c}_{25}=-\tau \langle 46\rangle[13]$, 
we find
\eq
\{M_4\}~=~
\frac{\langle 46\rangle^4[13]^4}{\langle 4|5\!+\!6|1]\langle 6|4\!+\!5|3][21][23]\langle 54\rangle\langle 56\rangle P_{456}^2}\,.
\label{M4}
\eqe
\exercise{}{Show that $\hat{M}_1 \big|_{\tau=0} = \tfrac{P_{456}^2}{\<46\>[13]}$.
Evaluate the other minors at $\tau = 0$ and use them to derive the result\reef{M4} for the residue at $\tau=0$.}
We could calculate the residues associated with each of the other minors similarly. $\{M_6\}$ and $\{M_2\}$ are just cyclic permutations of 
$\{M_4\}$ by two sites, so we have
\be
\begin{split}
\{M_6\}&=~\frac{\langle 62\rangle^4[35]^4}
{\langle 6|1\!+\!2|3]\langle 2|6\!+\!1|5][43][45]\langle 16\rangle\langle 12\rangle P_{612}^2}\,,\\
\{M_2\}&=~\frac{\langle 24\rangle^4[51]^4}
{\langle 2|3\!+\!4|5]\langle 4|2\!+\!3|1][65][61]\langle 32\rangle\langle 34\rangle P_{234}^2}\,.
\end{split}
\ee
For the residues $\{M_1\},\,\{M_3\},\,\{M_5\}$, it is convenient to choose the  gauge 
\eq
\left(\begin{array}{cccccc}1 & c_{12} & 0 & c_{14} & 0 & c_{16} \\ 0 & c_{32} & 1 & c_{34} & 0 & c_{36} \\ 0 & c_{52} & 0 & c_{54} & 1 & c_{56}\end{array}\right)\,.
\eqe 
Then following the same steps as before we find that the $\{M_1\}$ residue for the $(+,-,+,-,+,-)$ amplitude is
\eq
\{M_1\}~=~\frac{-\langle 6|2\!+\!4|3]^4}
{\langle 1|5\!+\!6|4]\langle 5|6\!+\!1|2][23][34]\langle 56\rangle\langle 61\rangle P_{561}^2}\,.
\label{M1}
\eqe
The other residues, $\{M_3\}$ and $\{M_5\}$, are obtained by relabeling the external states in \reef{M1}. 

We have now extracted six residues $\{M_i\}$ from \reef{6ptInt} for a projection that corresponds to the helicity configuration $(+,-,+,-,+,-)$ of a gluon amplitude. But it is not yet clear what the residues have to do with the amplitude. 
Each of the $\{M_i\}$'s contains spurious poles, such as $\langle 4|5\!+\!6|1]$  in $\{M_2\}$ and $\{M_4\}$. However, in the sum $\{M_2\}+\{M_4\}$, this spurious pole cancels. In fact, in the sum $\{M_2\}+\{M_4\}+\{M_6\}$ all three spurious poles ---  $\langle 4|5\!+\!6|1]$, $\langle 6|4\!+\!5|3]$, and $\langle 2|6\!+\!1|5]$ --- cancel, so this is a local object. Your brain may even be tingling with the sensation that you have seen this combination before\dots Go back to look at Exercise \ref{ex:Mi}: there we calculated the 6-point tree-amplitude
$A_6[1^+ 2^- 3^+ 4^- 5^+ 6^-]$ from a $[2,3\>$-BCFW shift and found that it was exactly
\be
  A_6[1^+ 2^- 3^+ 4^- 5^+ 6^-] = \{M_2\}+\{M_4\}+\{M_6\}\,.
  \label{m2m4m6}
\ee
In the BCFW construction, each of the three terms in \reef{m2m4m6} corresponds exactly to a BCFW diagram. Now we have also seen that each term can be understood as the residue of a pole associated with the minor $M_i$ in the cyclically invariant Grassmannian integral. So for $\mathcal{N}=4$ SYM, individual BCFW diagrams are in one-to-one correspondence with the residues of the Grassmannian integral. Since the Grassmannian integral was constructed to produce Yangian invariants, we now understand that each super-BCFW diagram is a Yangian invariant.

\begin{figure}
\begin{center}
\includegraphics[scale=0.65]{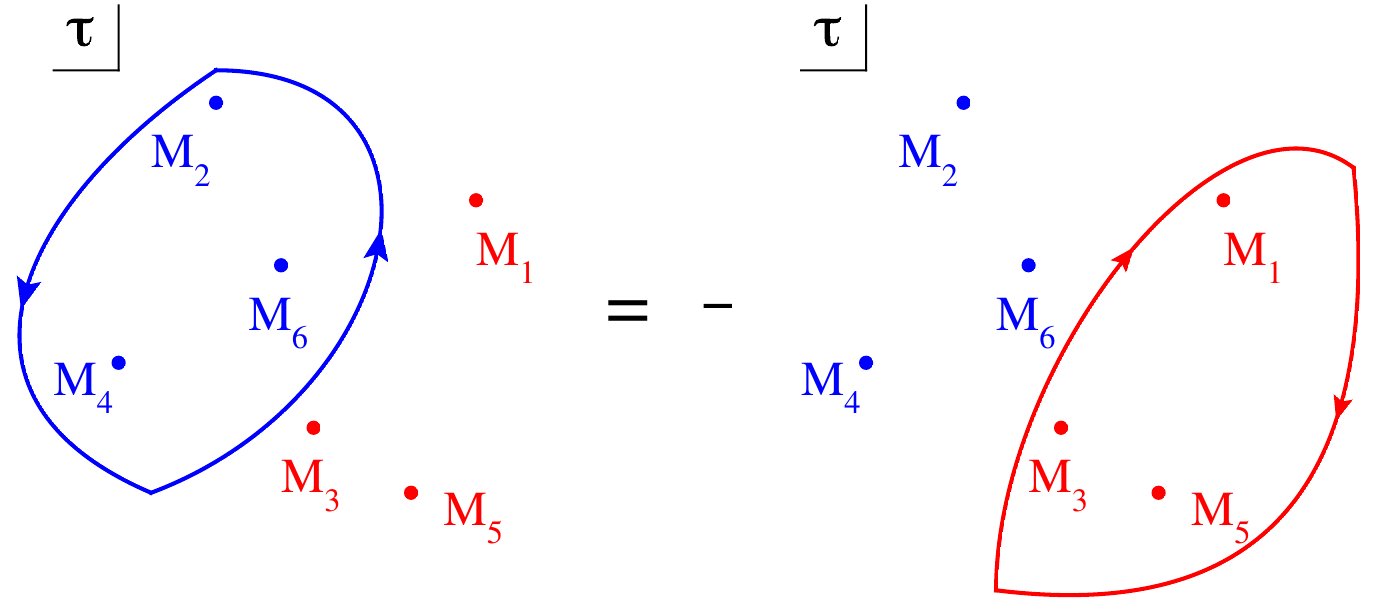}
\caption{\small The ``tree-contour" in the Grassmannian. It circles the residues of the poles $\{M_2\},\,\{M_4\},\,\{M_6\}$. Through contour deformation, the result is equivalent to minus the sum of  $\{M_1\},\,\{M_3\},\,\{M_5\}$.}
\label{Deform}
\end{center}
\end{figure}%

Consider the contour that encircles  the minors $\{M_2\},\,\{M_4\},\,\{M_6\}$. It is this contour that gives a Yangian invariant rational function that is local and free of spurious singularities. The statement of locality has become a choice of contour.

Through contour deformation, illustrated schematically in Figure \ref{Deform}, we have
\eq
\{M_2\}+\{M_4\}+\{M_6\}=-\{M_1\}-\{M_3\}-\{M_5\}\,.
\label{6termid}
\eqe
This means that the tree amplitude $A_6[1^+ 2^- 3^+ 4^- 5^+ 6^-]$ can also be represented by (minus) the sum of $\{M_1\}$, $\{M_3\}$, and $\{M_5\}$. Indeed, this is the representation that one obtains from the BCFW-shift $[3,2\>$, the `parity conjugate' of the shift $[2,3\>$ that produced the $\{M_2\},\,\{M_4\},\,\{M_6\}$ representation. Actually, this is a little too quick, because for the component-amplitude $A_6[1^+ 2^- 3^+ 4^- 5^+ 6^-]$, the shift $[3,2\>$ would be an illegal $[+,-\>$-shift; the precise statement is that the $\{M_1\},\,\{M_3\},\,\{M_5\}$ representation is the result of a $[3,2\>$ BCFW \emph{super}-shift recursion relation with a projection to the $(+,-,+,-,+,-)$ gluon helicity states. 

The insight gained here is that the mysterious six-term identity \reef{6termid} that arises from the equivalence of the two conjugate BCFW super-shifts $[2,3\>$  and $[3,2\>$ is  simply a consequence of the residue theorem of the Grassmannian integral $\mathcal{L}_{n,k}$! Actually, the identity \reef{6termid} is the 5-bracket six-term identity \reef{susy6termID} projected to the $(+,-,+,-,+,-)$ gluon helicity states. In section  \ref{s:polytopes} we expose an underlying geometric interpretation of such identities.

\compactsubsection{The Grassmannian, the tree contour, and the twistor string}
In Witten's \emph{twistor string} \cite{Witten:2003nn}, mentioned briefly at the end of Section \ref{s:twist}, the 
N$^{K}$MHV superamplitudes in $\cn=4$ SYM are calculated as open string current algebra correlators integrated over the moduli space of degree 
$(K\!+\!1)$ curves in supertwistor space. It turns out that the \emph{RSV connected prescription} \cite{Roiban:2004yf} for the twistor string has a direct relation to the BCFW recursion relations. Moreover, different BCFW representations are related via (higher-dimensional versions of) Cauchy's theorem \cite{Spradlin:2009qr}. Sounds similar to the properties of the Grassmannian integral, right? In fact, it can be shown that the tree-contour in the Grassmannian precisely gives the RSV connected prescription for the twistor string! More precisely, instead of first solving the bosonic delta functions as we did above, one can consider first localizing on the zeroes of the minors. 
Then the tree contour reduces the Gr$(k,n)$ Grassmannian integral to a Gr$(2,n)$ integral which, after a Fourier transform, becomes precisely the twistor string formula \cite{ArkaniHamed:2009dg}. Thus the underlying property that makes the tree-contour special is that it localizes the Gr$(k,n)$ Grassmannian integral  to Gr$(2,n)$ in a particular fashion that is intimately tied to locality. We will see this story repeat itself when we consider the Grassmannian formula for the 3-dimensional ABJM theory in Section \ref{s:3D}. 
There are several papers in the literature on the relationship between amplitudes and the twistor string, and you may like to consult for example \cite{Witten:2003nn,ArkaniHamed:2009dg,Bourjaily:2010kw,Spradlin:2009qr,Roiban:2004yf,Bullimore:2009cb,Dolan:2011za}.

The Grassmannian picture is interesting and has given us insight about locality, but there is perhaps a small stone in our shoe: we have yet not been able to see how the cancelation of spurious poles takes place within the tree superamplitude in a manifest fashion. There is a geometric story about how that happens and we draw it in Section \ref{s:polytopes}.

%%%%%%%%%%%%%%%%%%%%%%%%%%%%%%%%%%%%%%%%%%%%%%%%%%%%%%%%%%%%% 
\subsection{From on-shell diagrams to the Grassmannian}
%%%%%%%%%%%%%%%%%%%%%%%%%%%%%%%%%%%%%%%%%%%%%%%%%%%%%%%%%%%%%
The terms appearing in the BCFW expansion of the 6-point NMHV amplitude have now appeared in two distinct entities. We have seen in this Section that they are given as residues of an integral over a Gr(3,6) Grassmannian manifold. In Section \ref{s:onshelldiag}, they were the result of gluing on-shell cubic vertices together into on-shell diagrams. So can we make a connection between the Grassmannian and the on-shell diagrams? Yes, we can!

In Section \ref{s:onshelldiag}, we did not explicitly compute any of the on-shell diagrams beyond  the simplest box-diagram. The reason is simple: explicitly solving all momentum conservation constraints at each vertex is a complicated task because it is quadratic in spinor variables. In this section, we have seen that momentum conservation can be converted into a linear constraint with the aid of the Grassmannian. This means that if we convert all the 3-point vertices in the on-shell diagram into  Grassmannian integrals, then the momentum conservation constraints are just a set of linear equations. Let us see how this is done in practice. For the MHV 3-point amplitude, the Grassmannian integral is simply Gr(2,3):
\eq
\mathcal{A}_3^{\rm MHV}=\int 
\frac{d^{2\times3}C}{M_1M_2M_3}~
\delta^{2\times 2}\big(C_{i} [i| \big)
~\delta^{(4\times 2)}\big(C_{i} \eta_i \big)
~\delta^{2\times1}\big(\tilde{C}_{i}\langle i|\big)\,,
\eqe  
where we have used the momentum space representation introduced in Section \ref{s:Grassmannia}: $C_{\mathsf{a}i}$ is a $2\times 3$ matrix, and $\tilde{C}_{i}$ is its $1\times3$-dimensional orthogonal complement. For the anti-MHV 3-point amplitude, the analogue Grassmannian integral in  Gr(1,3) is
\eq
\mathcal{A}_3^\text{anti-MHV}=\int \frac{d^{1\times3}C}{M_1M_2M_3}~
  \delta^{2\times 1}\big(C_{i} [i| \big)
  ~\delta^{(4\times 1)}\big(C_{i} \eta_i \big)
  ~\delta^{2\times2}\big(\tilde{C}_{i}\langle i|\big)\,,
\label{anti-MHV3pt}
\eqe  
where now $C_{i}$ is an $1\times3$-dimensional matrix and $\tilde{C}_{\mathsf{a}i}$ is its $2\times3$-dimensional orthogonal complement. To see that this indeed gives the correct 3-point amplitude, note that the first bosonic delta function requires that $C_{i}$ is orthogonal to the 2-plane $[i|$, so this localizes the integral (up to an irrelevant overall rescaling) to 
\eq
C_{i}=\Big(\;[23]\;,\;[31]\;,\;[12]\;\Big)\,.
\label{G13sol}
\eqe
Substituting this into \reef{anti-MHV3pt}, one indeed recovers the anti-MHV 3-point amplitude.
\exercise{}{Given \reef{G13sol}, determine a representation of $\tilde{C}_{\mathsf{a}i}$. Substitute the result into \reef{anti-MHV3pt} to recover the 3-point amplitude. Note that all potential Jacobian factors can be fixed by dimension-counting and symmetry analysis.}
In summary, the MHV and anti-MHV 3-point amplitudes can be viewed as providing $2\times 2$ and $1\times2$ linear constraints for the $[i|$'s respectively. For later convenience, we parametrize the $Gr(2,3)$ and $Gr(1,3)$ Grassmannians as follows:
\eqa
\nonumber &&\quad\; b\quad c \quad a\\
&&\left(\begin{array}{ccc}1 & 0 & \alpha_{b} \\0 & 1 & \alpha_{c}\end{array}\right)\quad \vcenter{\hbox{\includegraphics[scale=0.55]{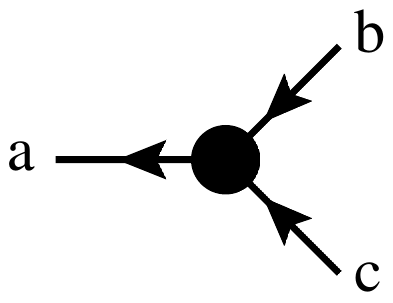}}}
\label{Bvert}
\\[4mm]
\nonumber &&\quad\; a\quad b \quad c\\[-3.5mm]
&&\left(\begin{array}{ccc}1 & \beta_b & \beta_{c} \end{array}\right)\quad \vcenter{\hbox{\includegraphics[scale=0.6]{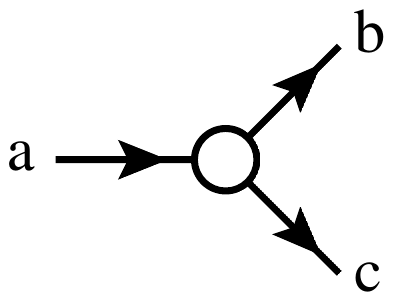}}}
\label{Wvert}
\eqae
This provides a graphical representation for the particular gauge that we have chosen for each Grassmannian. Incoming lines on the 3-point vertex indicates that the corresponding columns in the Grassmannian are $GL(k)$-gauge-fixed to be the identity matrix. Outgoing lines correspond to unfixed columns. In this gauge, the $[i|$ part of the bosonic delta functions are:
\eq
\text{MHV:}\quad 
\delta^2\big([b|+\alpha_{b}[a|\big)~\delta^2\big([c|+\alpha_{c}[a|\big)\,,
\hspace{1cm} 
\text{anti-MHV:}\quad \delta^2\big([a|+\beta_{b}[b|+\beta_{c}[c|\big)\,.
\eqe
For each vertex, the spinors of the incoming lines are expressed as a linear combination of those of the outgoing lines. One can perform a similar analysis for the fermionic delta functions and the bosonic delta functions of $\langle i|$. The analysis is exactly parallel, so we leave them implicit.
\exercise{}{What does the bosonic delta function for $\langle i|$ look like? What constraints do they impose?}
We are now ready to start gluing! Recall from \reef{internalLine} that each internal line in the on-shell diagram corresponds to an integral over the set of internal variables
\eq
\int \frac{d^2|I\> \,d^2|I] \,d^4\eta_I}{U(1)}\,.
\label{internalLine2}
\eqe
Since the spinors $\big(|I\>,\, |I]\big)$ also appear in the vertices on each end of the line, the bosonic delta functions of these vertices can be used to localize the  integral \reef{internalLine2}.  This can be made manifest in a graphical way. For each on-shell diagram, we decorate the lines with arrows following the rule that for each black-vertex, there should be two incoming lines and one outgoing line, while for each white-vertex there should be one incoming line and two outgoing lines, just as in \reef{Bvert}-\reef{Wvert}. One might wonder if it is always possible to find such decoration consistent throughout the on-shell diagram. For diagrams of {\em physical relevance} the answer is yes, since one can interpret the outgoing lines as $+$ helicity, incoming lines as $-$ helicity, and a consistent decoration is equivalent to consistent helicity assignments. For example, consider gluing six vertices together to form a double box diagram. We can have consistent decoration if there is at least one different color vertex, but not if they are all the same: 
\eq
\raisebox{-8mm}{\includegraphics[scale=0.5]{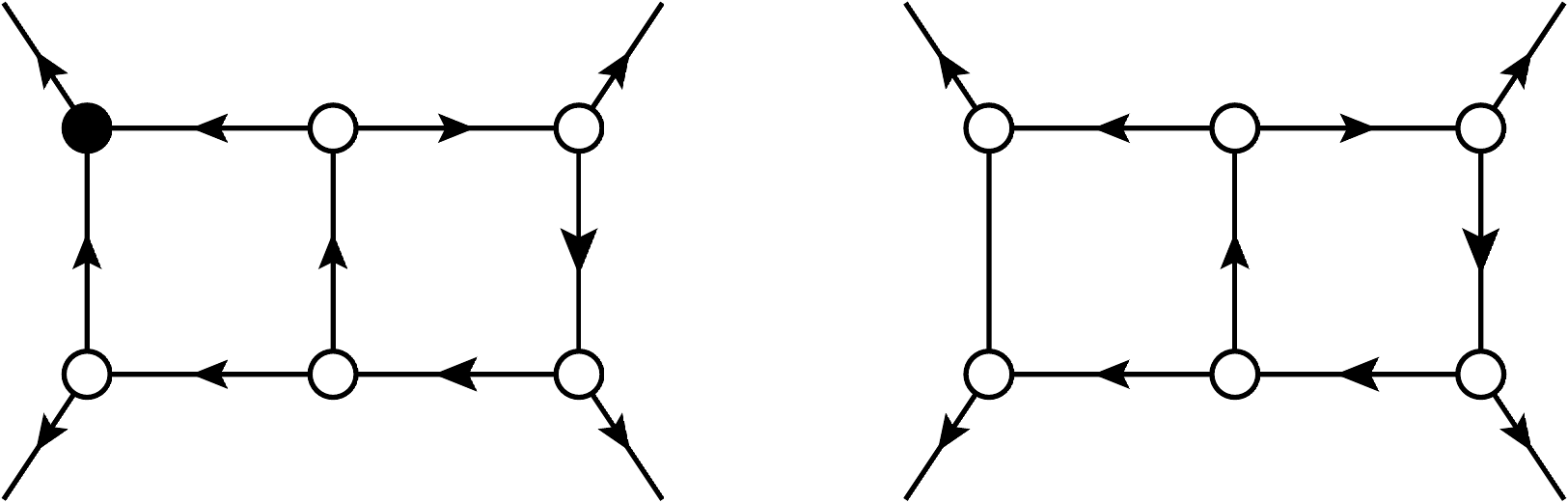}}\,.
\eqe
The lines of the second diagram cannot be consistently oriented. It follows from the previous discussion that the spinors of the internal line are  completely determined by the outgoing lines of one of the vertices. For example,  decorations of  the diagram
\be
\raisebox{-9mm}{\includegraphics[scale=0.5]{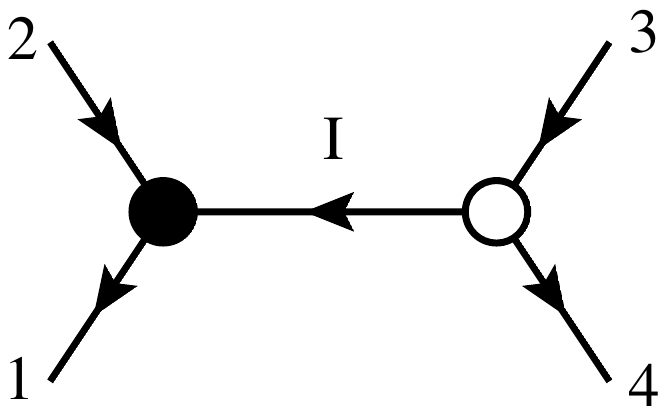}} 
\ee
dictate the bosonic delta functions for the square spinors to be
\eq
\delta^2\big([2|+\alpha_{2}[1|\big)\,,\quad 
\delta^2\big([I|+\alpha_{I}[1|\big)\,,\quad  
\delta^2\big([3|+\beta_{4}[4|+\beta_{I}[I|\big)\,.
\eqe
The second bosonic delta function localizes the $\int  d^2|I] $ integral, while the remaining two delta functions become
\eq
\delta^2\big([2|+\alpha_{2}[1|\big)\,,\quad 
\delta^2\big([3|+\beta_{4}[4|-\beta_{I}\alpha_{I}[1|\big)\,.
\label{G24Delta}
\eqe
The delta functions in \reef{G24Delta} can be combined to the form $\prod_{\mathsf{a}=1}^2\delta^2(C_{\mathsf{a}i}[i|)$ with the Gr(2,4) Grassmannian is given as 
\eqa
\nonumber&&\quad\quad\; 1 \quad\quad  ~\,2\quad ~~~3 \quad ~~~\,4\\
\quad C_{\mathsf{a}i}&=&
\left(\begin{array}{cccc}\alpha_2 &~ 1 &~~~ 0 & ~~~0 \\ 
-\beta_{I}\alpha_{I} & ~0 & ~~~1 & ~~~\beta_{4}\end{array}\right)\,.
\label{Gr(24)sol}
\eqae
So gluing the two 3-point vertices  together now gives a new Grassmannian integral
\eq
\int \frac{d\alpha_2}{\alpha_2}\frac{d\alpha_I}{\alpha_I}\frac{d\beta_4}{\beta_4}\frac{d\beta_I}{\beta_I}\frac{1}{U(1)}\;\;
\delta^{2\times 2}\big(C_{i} [i| \big)~
\delta^{(4\times 2)}\big(C_{i} \eta_i \big)~
\delta^{2\times2}\big(\tilde{C}_{i}\langle i|\big)\,,
\label{GrCex}
\eqe
where the $C_{\mathsf{a}i}$ is identified in \reef{Gr(24)sol}. 

Notice the leftover $1/U(1)$ in \reef{GrCex}. We have been treating $(|I\>,|I])$ as independent variables, each being fixed by the bosonic delta functions. However, there remains a gauge-fixing functional that is present to remove the little-group redundancy. This functional is represented by this $1/U(1)$ factor.  We do not need its explicit form, just remember that it can be used to localize an additional degree of freedom. 

The above simple example generalizes to arbitrary decorated on-shell diagram. A diagram with $n_b$ black vertices and $n_w$ white vertices contains $2\times (2 n_b+n_w)$  constraints on the $|i]$'s. If it has $n_I$ internal lines, then the $2\times n_I$ integrations over the internal $|I]$'s can be localized by these bosonic delta functions. At the end of the day, there will be $2\times (2 n_b+n_w-n_I)$ constraints left and they can be conveniently grouped into a degree $2\times k $ delta function 
$\prod_{\mathsf{a}=1}^k\delta^2\big(C_{\mathsf{a}i}[i|\big)$ where $k=(2 n_b+n_w-n_I)$ and $n=3(n_w+n_b)-2n_I$. Note that by counting the Grassmann degrees of the black ($2 \times 4$) and white blobs ($1 \times 4$) minus the $n_I$ internal Grassmann integrations ($n_I \times 4$), this $k$ is exactly the same $k$ as in the N$^{k+2}$MHV classification. 
Thus each on-shell diagram corresponds to the following Grassmannian integral:
\eq
\int\left( \prod_{i=1}^{n_b}\frac{d\alpha_{i1}}{\alpha_{i1}}\frac{d\alpha_{i2}}{\alpha_{i2}}\right)
\left( \prod_{i=1}^{n_w}\frac{d\beta_{i1}}{\beta_{i1}}\frac{d\beta_{i2}}{\beta_{i2}}\right)
\left(\prod_{i=1}^{n_I}\frac{1}{U(1)_i}\right)~
\delta^{2\times k}\big(C_{i} [i| \big)~
\delta^{(4\times k)}\big(C_{i} \eta_i \big)~
\delta^{2\times (n-k)}\big(\tilde{C}_{i}\langle i|\big)\,.
\label{Gint}
\eqe
As we have seen, this Gr($k,n$) Grassmannian integral is parametrized by an   on-shell diagram decorated with arrows consistently thoughout the diagram. The  Grassmannian integral \reef{Gint} is in a $GL(k)$-gauge-fixed form. Each vertex contains two degrees of freedom, but the $n_I$  internal lines each leave a $1/U(1)$ gauge-fixing function, so the dimension of this Grassmannian is
\eq
\text{dim}(C)=2\times (n_v)-n_I
\eqe 
where the number of vertices is $n_v=n_b+n_w$. Using Euler's formula for a planar diagram $(n_f-n)-n_I+n_v=1$, where $n_f$ is the number of faces in a diagram, we find that the dimension of the Grassmannian corresponding to a particular on-shell diagram is 
\eq
\text{dim}(C)=n_f-1\,.
\eqe  
The total number of bosonic delta functions are $2\times (k+n-k)-4=2n-4$. If $\text{dim}(C)=2n-4$, then all the degrees of freedom in the integral are completely localized by the bosonic delta functions. This is the case for the on-shell diagrams that correspond to the BCFW terms in Section \ref{s:onshelldiag}. 
We saw this in Section \ref{s:6ptGrassmannian} for Gr(3,6): 
prior to solving the bosonic delta functions, 
each BCFW term is obtained by localizing on the zeroes of one of the minors. This is precisely the $9-1=8$-dimensional Grassmannian manifold indicated by the on-shell diagrams.  

If $\text{dim}(C)<2n-4$, then the bosonic delta functions over-constrain the external data and can only be satisfied in special kinematics. This is precisely the scenario for the example that led to \reef{Gr(24)sol}, where  $\text{dim}(C)=3<4$. From \reef{Gr(24)sol} one can readily read off what the special kinematics is: $[1|\sim [2|$. 

What happens if $\text{dim}(C)>2n-4$? It corresponds to a term in the BCFW representation of a loop amplitude, where the remaining integrations can be translated into the loop momentum integration! For example, consider attaching a BCFW bridge on the forward limit of \reef{6ptBCFWLS2}. Recall that in our ``brief" calculation of the loop-recursion in Section \reef{s:bcfwMTloop4pt}, only the middle term of \reef{6ptBCFWLS2} has a non-trivial forward limit contribution. Attaching a BCFW bridge to the middle term and perform a series of equivalence moves one finds:
\eq
\raisebox{-12mm}{\includegraphics[scale=0.7]{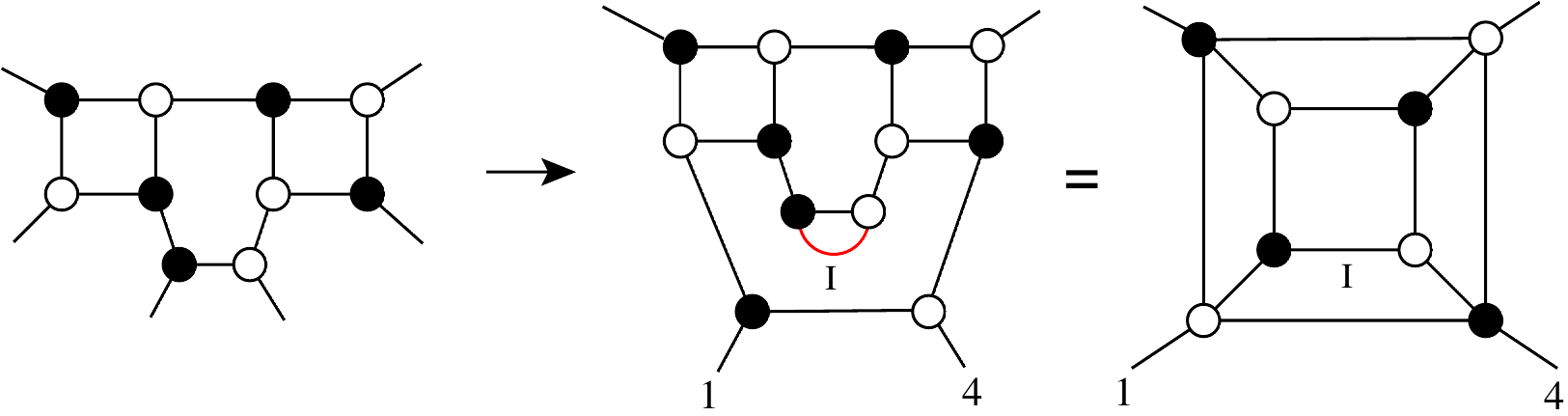}}\,.
\label{6to4}
\eqe 
The final diagram is precisely the 1-loop 4-point amplitude. One can count that with $n=4$, $\text{dim}(C)=9-1=4+(2n-4)$, i.e.~four  integrals remain after solving all bosonic delta functions. 
These four extra integrals correspond to the integration over the four components of the loop momentum. We can readily identify these extra components in the on-shell diagram: since the original tree-digram contains no extra integration variable, the new degrees of freedom must arise from the procedure of taking the forward limit. This introduces a factor of 
$\int d^2|I\> \,d^2|I] \,d^4\eta_I/U(1)$ in \reef{6to4}; that is 3 integrations because the $U(1)$ is mod'ed out. The presence of the BCFW bridge introduces the 4th integration $\int \frac{dz}{z}$. The loop-momentum $\ell$ can then be written 
\eq
\ell=|I\>[I|+ z|1\>[4|\,.
\eqe
It is quite remarkable that starting out with a fully on-shell construction of ``on-shell diagrams" leads to a loop-integrand construction in which the loop momentum is off-shell. 
\exercise{}{Use equivalence moves to prove the last two diagrams of \reef{6to4} are equivalent to each other.} 
There is much more information in the connection between the Grassmannian and the on-shell diagrams, but we also have other fish to fry (and birds to scare). If we have awoken your appetite for blob diagrams and Grassmannians and you are interested in learning more about their relations to permutations, stratifications, amalgamation, dimers, bipartite graphs, and quivers, you should take a look at the exciting paper \cite{ArkaniHamed:2012nw} for more details.

%%%%%%%%%%%%%%%%%%%%%%%%%%%%%%% 
%%%%%%%%%%%%%%%%%%%%%%%%%%%%%%% 
%%%%%%%%%%%%%%%%%%%%%%%%%%%%%%% 
\newpage
\setcounter{equation}{0}
\section{Polytopes}
\label{s:polytopes}
%%%%%%%%%%%%%%%%%%%%%%%%%%%%%%% 
%%%%%%%%%%%%%%%%%%%%%%%%%%%%%%% 
%%%%%%%%%%%%%%%%%%%%%%%%%%%%%%% 

In Section \ref{s:grassmannia}, we learned that the individual terms in a BCFW expansion of an N$^{k-2}$MHV $n$-point superamplitude are residues of a cyclically invariant integral-formula in the Grassmannian Gr$(k,n)$. In this language, we found that the different BCFW representations of the 6-point NMHV amplitude $A_6[1^+ 2^- 3^+ 4^- 5^+ 6^-]$ are related by a simple contour deformation. This was manifested in the six-term identity \reef{6termid}. 
In {\em momentum supertwistor space}~\cite{Mason}, the six-term identity is promoted to the relation
\eq
 [2,3,4,6,1]+[2,3,4,5,6]+ [2,4,5,6,1]
 =
 [3,1,6,5,4]+[3,2,1,6,5]+[3,2,1,5,4]
 \,.
\label{Tempting}
\eqe
We already encountered this version of the six-term identity in \reef{susy6termID} when we discussed the equivalence of the BCFW recursion relations based on $[2,3\>$ and $[3,2\>$ super-shifts. The 5-brackets were defined in Section~\ref{s:momtwist} as
\be
[i,j,k,l,m]\equiv \frac{\delta^4\big(\chi_{iA}\langle jklm\rangle+{\rm cyclic}\big)}{\langle ijkl\rangle\langle jklm\rangle\langle klmi\rangle\langle lmij\rangle\langle mijk\rangle}\,,
\label{5bracket}
\ee
with 4-brackets 
$\nonumber\langle ijkl\rangle
\equiv 
\epsilon_{IJKL}Z_i^IZ_j^JZ_k^KZ_l^L$ involving the bosonic components 
$Z_i^I=(|i\>, [\m_i|)$ of the $SU(2,2|4)$ momentum supertwistors  
$\mathcal{Z}_i^{\mathsf{A}} \equiv \big(|i\>^{\dot{a}}\,,\, [\mu_i|^a\,\, \big|\,\,\chi_{iA}\big)$, $\mathsf{A}=(\dot{a},a,A)$. The $[\mu_i|$ are defined by the incidence relations \reef{incidence}.

Recall that momentum conservation is automatic for momentum twistors, so the six-term identity \reef{Tempting} must hold for any six momentum twistors; it is not specific to the NMHV 6-point amplitude but is an intrinsic property of the 5-brackets \reef{5bracket}. Hence it seems worthwhile to  try to understand the structure of \reef{Tempting} better. 
Using cyclic and reflection symmetry of the 5-brackets, we can rewrite \reef{Tempting} as
\be
 [1,2,3,4,5]-[2,3,4,5,6]+[3,4,5,6,1]-[4,5,6,1,2]+[5,6,1,2,3]-[6,1,2,3,4]
 ~=~0
 \,.~~~
\label{alsoTempting}
\ee
Pretend for a moment that we do not know what the 5-brackets are. Consider a fully antisymmetric 5-bracket $\langle i,j,k,l,m\rangle$ defined as the contraction of five 5-component vectors $\mathsf{Z}_i^\mathcal{I}$ with a 5-index Levi-Civita tensor. Such an object would satisfy the 
 `Schouten identity'
\eqa
\nonumber&&\langle1,2,3,4,5\rangle\,\mathsf{Z}_6^\mathcal{I}
-\langle2,3,4,5,6\rangle\,\mathsf{Z}_1^\mathcal{I}
+\langle3,4,5,6,1\rangle\,\mathsf{Z}_2^\mathcal{I}\\
&& - \langle 4,5,6,1,2\rangle\,\mathsf{Z}_3^\mathcal{I}
+\langle 5,6,1,2,3\rangle\,\mathsf{Z}_4^\mathcal{I}
-\langle 6,1,2,3,4\rangle\,\mathsf{Z}_5^\mathcal{I}~=~0\,.
\label{5schout}
\eqae
Since this looks quite similar to \reef{alsoTempting}, including relative signs, we might be tempted to think that \reef{alsoTempting} somehow arises as a Schouten identity. This is of course too speculative:  the 5-brackets $[i,j,k,l,m]$ really represent rational functions of the 4-component momentum twistors $Z_i$, not some 5-index objects contracted with a 5-index Levi-Civita tensor.
 However, we can entertain the idea a little further. Could the 5-bracket \reef{5bracket} be written in terms of some new 5-vectors? Clearly, the fermionic variables $\chi_{iA}$ appear in \reef{5bracket} on different footing than their bosonic counterparts. In the name of democracy, let us define the following purely bosonic 5-component vector 
\eq
\mathsf{Z}^{\mathcal{I}}_i=\left(\begin{array}{c} Z^I_i \\ \chi_i\cdot \psi\end{array}\right)\,,~~~~~
\mathcal{I}=1,\cdots,5\,,
\label{5vec}
\eqe
where $\chi_i\cdot \psi=\chi_i^A\psi_{A}$ and $\psi_{A}$ is an $SU(4)$ auxiliary Grassmann variable common for all external particles $i=1,2,\dots, n$. If we define $\langle i,j,k,l,m\rangle$ as the contraction of five of these 5-vectors with a 5-indexed Levi-Civita tensor, then they will satisfy the Schouten identity \reef{5schout} --- but that is not what we are after, so read on.

To write the 5-bracket $[i,j,k,l,m]$ in terms of the 5-vectors \reef{5vec}, we must  remove the auxiliary variable $\psi_{A}$. Since it is fermionic, this can be  done via a Grassmann-integration: 
one finds that the 5-bracket \reef{5bracket} can be written as
\eq
[i,j,k,l,m]=\frac{1}{4!}\int d^4\psi\,
\frac{\langle i,j,k,l,m\rangle^4}{\langle0, i,j,k,l\rangle\langle0, j,k,l,m\rangle\langle 0,k,l,m,i\rangle\langle 0,l,m,i,j\rangle\langle 0,m,i,j,k\rangle}\,,
\label{FiveVec}
\eqe
where we have introduced the auxiliary reference 5-vector
\eq
\mathsf{Z}^{\mathcal{I}}_0=\left(\begin{array}{c}0 \\0 \\0 \\0 \\1\end{array}\right)\,.
\eqe
The representation \reef{FiveVec} is certainly not just contractions of five 5-vectors with a Levi-Civita tensor, so the origin of the identity in \reef{alsoTempting} is not a Schouten identity. But let us not give up just yet, for it will be worthwhile to  examine \reef{FiveVec} further. 

Since the integral $\int d^4\psi$ is universal for all 5-brackets, we ignore it for the time being and focus on the integrand of \reef{FiveVec}. Each $\mathsf{Z}_i^{\mathcal{I}}$ appears an equal number of times in the numerator and the denominator, so the integrand is invariant under $\mathsf{Z}_i^{\mathcal{I}} \to t_i\, \mathsf{Z}_i^{\mathcal{I}}$ for each $i=1,2,\dots,n$.
In other words, the 5-vectors $\mathsf{Z}_i^{\mathcal{I}}$ appear projectively in \reef{FiveVec},  and therefore we can think of the $\mathsf{Z}_i^{\mathcal{I}}$ as homogeneous coordinates of points in projective space $\mathbb{CP}^4$. 
The presence of the reference vector $\mathsf{Z}_0^\mathcal{I}$ in the denominator breaks projective invariance, but only at this particular point. 

There is an analogous case where we have encountered something similar. 
The momentum twistors $Z^{I}_i$ in Section \ref{s:momtwist} are defined projectively and are elements in $\mathbb{CP}^3$. The map in Figure \ref{coordinates} shows how to relate momentum twistors $Z^{I}_i$ with points $y_i$ in dual space. Specifically, the distance between two points $y_i$ and $y_j$ in dual space is
\eq
y^2_{ij}
~=~
\frac{\langle i-1,i,j-1,j\rangle}{\langle i-1,i\rangle\langle j-1,j\rangle}
~=~
\frac{\langle i-1,i,j-1,j\rangle}{\langle I_0, i-1,i\rangle\langle I_0, j-1,j\rangle}
\,.
\label{Distance}
\eqe
The first equality is simply \reef{ID3}: it has a momentum twistor 4-bracket in the numerator and regular angle brackets in the denominator. In the second equality we have rewritten the denominator  in a more suggestive form involving only 4-brackets, at the cost of introducing a reference bi-twistor $I_0^{IJ}$ defined as
\eq
I_0^{IJ}=\left(\begin{array}{cc}0 & 0 \\0 & \epsilon_{\dot{a}\dot{b}}\end{array}\right)\,.
\label{Infty}
\eqe   
In the literature, $I_0$ is often referred to as the {\bf \em infinity twistor},\footnote{The infinity twistor in \reef{Infty} corresponds to a flat space metric. For AdS$_4$ it is given as $I_0^{IJ}=\bigg(\begin{array}{cc}\epsilon_{ab}\Lambda & 0 \\0 & \epsilon_{\dot{a}\dot{b}}\end{array}\bigg)$, where $\Lambda$ is the cosmological constant.}  and its role is to break $SL(4)$ conformal invariance and provide a preferred metric for the definition of distance.

The expression \reef{Distance} is similar to the integrand of \eqref{FiveVec}: both are projectively defined, except for the reference bi-twistor/vector. The reference bi-twistor appears twice in the denominator of \eqref{Distance}, reflecting that this expression gives the distance between the two points $i$ and $j$. An analogous expression involving three points and a reference vector appearing thrice defines the area of a triangle. And so on. 
The appearance of the reference vector $\mathsf{Z}^{\mathcal{I}}_0$ five times in the denominator of \reef{FiveVec} gives us a hint that \textit{the rational integrand in \reef{FiveVec} is the volume geometric figure defined by five points in $\mathbb{CP}^4$!} In the following, we pursue the interpretation of 5-brackets as volumes of simplices and their sum --- the superamplitudes --- as volumes of {\bf \em polytopes}. Definitions and explanations follow next.

%%%%%%%%%%%%%%%%%%%%%%%%%%%%%%%%%%%%%%%%%%%%%%%%%%%%%%%%%%%%%%
\subsection{Volume of an $n$-simplex in $\mathbb{CP}^n$}
%%%%%%%%%%%%%%%%%%%%%%%%%%%%%%%%%%%%%%%%%%%%%%%%%%%%%%%%%%%%%%
%%%%%%%%
Let us begin with the concepts of polytopes and simplices before reconnecting with the motivation above.

\compactsubsection{Polytopes and simplices: definitions and examples}
 We are all familiar with polygons: triangles, squares (or more generally quadrilaterals), pentagons, hexagons, chiliagons, star-shapes etc. These are figures in the plane bounded by a finite number of straight line-segments. Their 3-dimensional analogues --- tetrahedrons, cubes, prisms, dodecahedrons etc --- are solids whose faces are polygons. The $n$-dimensional versions of polygons and polyhedrons are called {\bf \em polytopes} or  {\bf \em $n$-polytopes}. A $2$-polytope is a polygon and a $3$-polytope is a polyhedron. 

A simplex is in a sense the simplest example of an polytope. To define it, recall first that a {\bf \em convex set} $C$ (in, for example, $\mathbb{R}^n$ or $\mathbb{CP}^n$) has the property that the line segment between any two points in $C$ lies entirely in $C$. In the plane, triangles are convex, but star-shaped polygons are not. Given a set of points $S$, the {\bf \em convex hull} of $S$ is the intersection of all convex sets containing $S$. Examples from the plane: 1) the convex hull of a circle is the closed disk bounded by the circle. 2) The convex hull of three points is a triangle; adding a fourth point that lies inside the triangle, the convex hull of the four points is the same triangle. But for a fourth point outside the triangle (but in the same plane), the convex hull is now a convex quadrilateral:
\eq
\vcenter{\hbox{\includegraphics[scale=0.4]{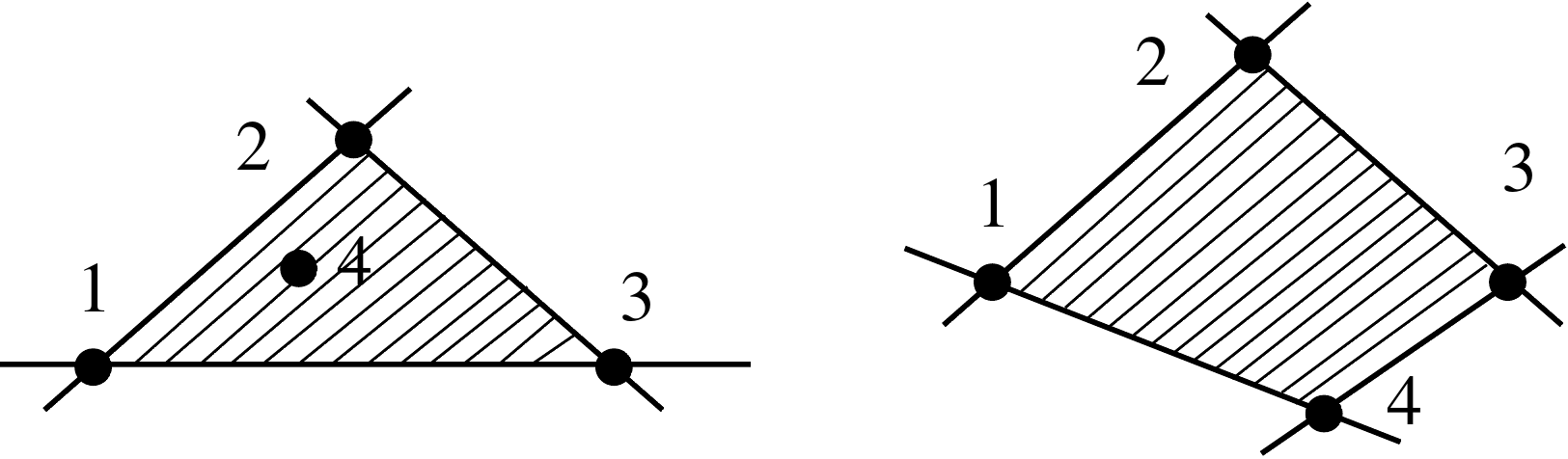}}}\,\,.
\eqe
An {\bf \em $n$-simplex} is the convex hull of a set of $n+1$ points. Examples:
\be
\begin{array}{l}
0\text{-simplex = a point}\\
1\text{-simplex = line segment}\\
2\text{-simplex = triangle}\\
3\text{-simplex = tedrahedron}.
\end{array}
\ee
An $n$-simplex is bounded by $n\!+\!1$ $(n\!-\!1)$-simplices who intersect each other in $n+1 \choose 2$ $(n\!-\!2)$-simplices.
 For $n\!+\!1$ generic points in $\mathbb{R}^n$, an $n$-simplex has an $n$-dimensional volume. (For $\mathbb{CP}^n$ it will be $n$-complex dimensional.)  The volume of a polytope can be calculated by `tessellating' it into simplices, whose volumes are easier to calculate. 

Now that we know what simplices and polytopes are, let us progress towards understanding how the integrand in \reef{FiveVec} represents the volume of a 4-simplex in $\mathbb{CP}^4$, as claimed. As a warm-up, we begin in 2 dimensions with a 2-simplex (a triangle). 

%%%%%%%%
\compactsubsection{Area of a $2$-simplex in $\mathbb{CP}^2$}
The area of a triangle in a 2-dimensional plane can be computed as
\eq
{\rm Area}\left[\;\vcenter{\hbox{\includegraphics[scale=0.4]{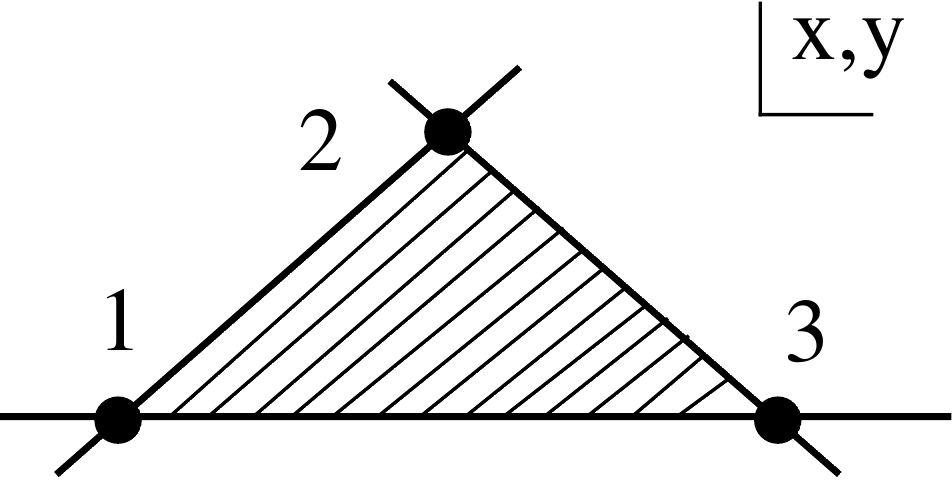}}}\;\right]\;
=\;\frac{1}{2}\left|\begin{array}{ccc}x_1 & x_2 & x_3 \\ y_1 & y_2 & y_3 \\1 & 1 & 1\end{array}\right|\,,
\label{triarea}
\eqe
where the $(x_i,y_i)$ are the coordinates of the three vertices.
\exercise{}{If the area formula \reef{triarea} is not familiar, you should derive it by showing that it is equivalent to the ``$\tfrac{1}{2}bh$"-formula that was imprinted on your brain in elementary school.}
The $1$'s in the last row of \reef{triarea} are redundant as we can write the same formula as a sum of the $2\times2$ minors. In physics, when faced with a redundancy we can choose to eliminate it or promote it to a feature. Choosing the latter, we define three 3-vectors along with a reference vector:
\eq
\mathsf{W}_{iI}=\left(\begin{array}{c} x_i \\ y_i \\ 1 \end{array}\right)\,,
\quad~~ 
\mathsf{Z}_{0}^I=\left(\begin{array}{c} 0 \\ 0 \\ 1 \end{array}\right)\,,
~~~~~I=1,2,3 \,.
\label{canonical}
\eqe
The area can now be written  
\eq
{\rm Area}\left[\;\vcenter{\hbox{\includegraphics[scale=0.4]{CP2Area}}}\;\right]~=~\frac{1}{2}\,\frac{\langle 1,2,3\rangle }{(\mathsf{Z}_0\cdot \mathsf{W}_1)(\mathsf{Z}_0\cdot \mathsf{W}_2)(\mathsf{Z}_0\cdot \mathsf{W}_3)}\,,
\label{Points}
\eqe
where the 3-bracket is the contraction of a 3-index Levi-Civita tensor with the three $\mathsf{W}_i$ vectors: 
$\langle 1,2,3\rangle = \eps^{IJK} \mathsf{W}_{1I} \mathsf{W}_{2J} \mathsf{W}_{3K}$. Using \reef{canonical}, the $\langle 1,2,3\rangle$-numerator exactly equals the $3\times 3$-determinant in \reef{triarea}, so you might consider the trivial dot-products $\mathsf{Z}_0\cdot \mathsf{W}_i = \mathsf{Z}_0^I \mathsf{W}_{iI} = 1$ in the denominator a provocation of your sense of humor. However, written in this form, the redundancy has been promoted to projective symmetry: the new area-formula \reef{Points} is invariant under scalings $\mathsf{W}_i\rightarrow t_i\,\mathsf{W}_i$. When the 3-vectors $\mathsf{W}_i$ are ``gauge fixed" to the canonical form in \reef{canonical}, we immediately recover the original area formula. Since the triangle vertices are specified in terms of projectively defined 3-vectors, we can think of the triangle as an object in $\mathbb{CP}^2$ and the $\mathsf{W}_i$'s as the homogenous coordinates of the vertices.

The area formula in \reef{Points} involves an antisymmetric 3-bracket as well as the inner product of 3-vectors. To make contact with \reef{FiveVec} and \eqref{Distance}, we would like to have a representation that is given solely in terms of 3-brackets. To achieve this, it is useful to characterize the triangle by its edges instead of its vertices. We define a `dual space' whose points $\mathsf{Z}^I_a$ are associated with lines in $\mathsf{W}$-space: a given line is defined as the set of points $\mathsf{W}_{I}$ satisfying the incidence relations 
\be
  \mathsf{Z}^{I}\mathsf{W}_{I}=0 \,.
\ee
Since $\mathsf{Z}^I$ is a vector in the 2-dimensional space $\mathbb{CP}^2$, the constraint indeed defines a 1-dimensional subspace, i.e.~a line.

Now, to define the triangle in terms of three lines in dual space, note that each $W_{iI}$ is characterized by lying simultaneously on two lines. Labeling the three edges of the triangle as $a$, $b$ and $c$, the vertex $W_{1I}$ is the intersection of lines $a$ and $c$, so that
$\mathsf{Z}_a^{I}\mathsf{W}_{1I}=\mathsf{Z}_c^{I}\mathsf{W}_{1I}=0$. These two constraints are easily solved and we have
\eq
\vcenter{\hbox{\includegraphics[scale=0.4]{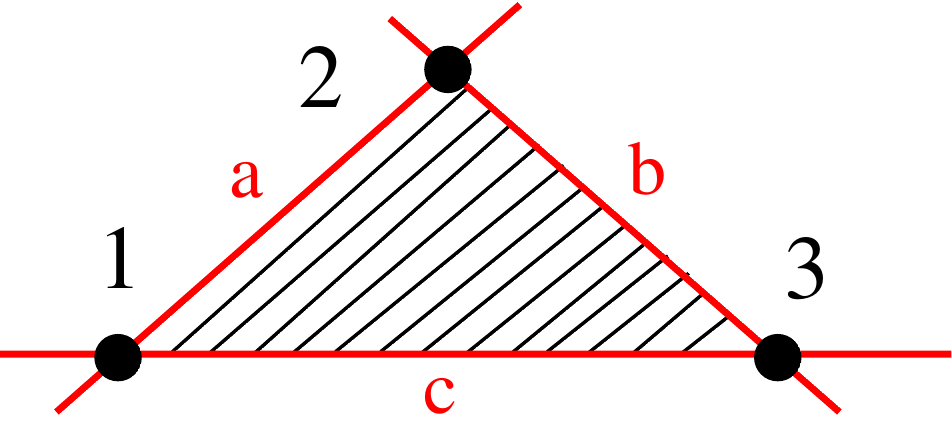}}}
~\rightarrow~
\begin{array}{l} 
\mathsf{W}_1=\langle *, \mathsf{Z}_c, \mathsf{Z}_a\rangle \\ 
\mathsf{W}_2=\langle *, \mathsf{Z}_a, \mathsf{Z}_b\rangle \\ 
\mathsf{W}_3=\langle *, \mathsf{Z}_b, \mathsf{Z}_c\rangle\,,
\label{WfromZ}
\end{array}
\eqe
where the $*$ indicates the free index, 
e.g.~$\mathsf{W}_{1I}=
\langle *, \mathsf{Z}_c, \mathsf{Z}_a\rangle
= \eps_{IJK}\mathsf{Z}_c^J \mathsf{Z}_a^K$.

Plugging the map \reef{WfromZ} into \eqref{Points}, we find that the area is now given as
\eq
{\rm Area}\left[\;\vcenter{\hbox{\includegraphics[scale=0.4]{CP2Area2}}}\;\right]
~=~
\frac{1}{2}\,\frac{\langle a,b,c\rangle^2 }{\langle 0,b,c\rangle\langle 0,a,b\rangle\langle 0,c,a\rangle}~\equiv~\big[a,b,c\big]\,,
\label{2simplex}
\eqe
where 
$\langle a,b,c\rangle = \eps_{IJK}\mathsf{Z}_a^I\mathsf{Z}_b^J \mathsf{Z}_c^K$.
\exercise{}{Show that \reef{2simplex} follows from  \reef{Points}.}
As advertised earlier, we now see  that the ``volume" (i.e.~area) of a 2-simplex is given by a rational function whose denominator is the product of all 3-brackets involving two of the edge variables $\mathsf{Z}^I_{i}$, $i=a,b,c$, and a reference vector $\mathsf{Z}^I_{0}$. Requiring projective invariance for each $\mathsf{Z}^I_i$ with $i=a,b,c$ uniquely fixes the numerator. For later convenience, we have introduced the notation $\big[a,b,c\big]$ to denote the volume \reef{2simplex}. Note that the area-formula comes with an ``orientation" in the sense that $\big[a,b,c\big]$ is fully antisymmetric in $a,b,c$.

%%%%%%%%%%%
\compactsubsection{Volume of an $n$-simplex in $\mathbb{CP}^n$}The expression in  \reef{2simplex} can be  generalized to the volume of $n$-simplex in $\mathbb{CP}^n$: we denote it by an antisymmetric $(n\!+\!1)$-bracket
\eq
\big[\mathsf{Z}_{i_1},\ldots \mathsf{Z}_{i_{n+1}}\big]
~=~
\frac{1}{n!}\,\frac{\langle i_1, i_2,\ldots, i_{n+1}\rangle^n}
{\langle 0,i_1,\ldots,i_ n\rangle 
\langle 0,i_2,\ldots,i_ {n+1}\rangle
\cdots\langle 0,i_{n+1},i_1,\ldots,i_{n-1}\rangle}\,.
\label{nvolume}
\eqe
where the angle-brackets are the contractions of the $n\!+\!1$ listed $\mathbb{CP}^n$ vectors with an $(n\!+\!1)$-index Levi-Civita. 
The $n\!+\!1$ variables   $\mathsf{Z}_{i_1},\cdots, \mathsf{Z}_{i_{n+1}} \in \mathbb{CP}^n$
carry the information about the $n\!+\!1$ boundaries of the $n$-simplex: for a given vector $\mathsf{Z}_{i}^I$, the set of $\mathsf{W}_{I}$'s satisfying the 
 incidence relation $\mathsf{Z}_{i}^I\mathsf{W}_{I}=0$ span of subspace $\mathbb{CP}^n$ of dimension $n\!-\!1$. These contain the 
 $n\!+\!1$ $(n\!-\!1)$-simplex faces of the $n$-simplex. 
 
 In the example of the 2-simplex in $\mathbb{CP}^2$, the 3 $\mathsf{Z}_i$'s label the 1-dimensional lines $a$, $b$, $c$ bounding the triangle. Each pair of lines intersect in a point that is a vertex of the triangle: we can label the vertices $(a,b)$, $(b,c)$, and $(c,a)$. They are defined in terms of the $\mathsf{Z}_i$'s in \reef{WfromZ} and the denominator of the volume formula \reef{2simplex} is the dot-product of all vertex point vectors with the reference vector. 

As a second example, consider a 3-simplex (tetrahedron) in $\mathbb{CP}^3$.
The four 4-component homogeneous coordinates of (dual) $\mathbb{CP}^3$ --- $\mathsf{Z}_{a}^I$, $\mathsf{Z}_{b}^I$, $\mathsf{Z}_{c}^I$, $\mathsf{Z}_{d}^I$ --- have 2-dimensional orthogonal complements spanned by the $\mathsf{W}_I$'s satisfying $\mathsf{Z}_{i}^I\mathsf{W}_{I}=0$. Pairwise, these generic 2-planes intersect in a line: this gives 6 lines, $(a,b)$, $(b,c)$, etc,  that define the 1-simplex edges of the tetrahedron. Three generic 2-planes in  $\mathbb{CP}^3$ intersect in a point: this defines the four vertices of the tetrahedron and we label them $(a,b,c)$, $(b,c,d)$, $(c,a,d)$, and $(d,a,b)$, as illustrated here:
 \eq
 \vcenter{\hbox{\includegraphics[scale=0.4]{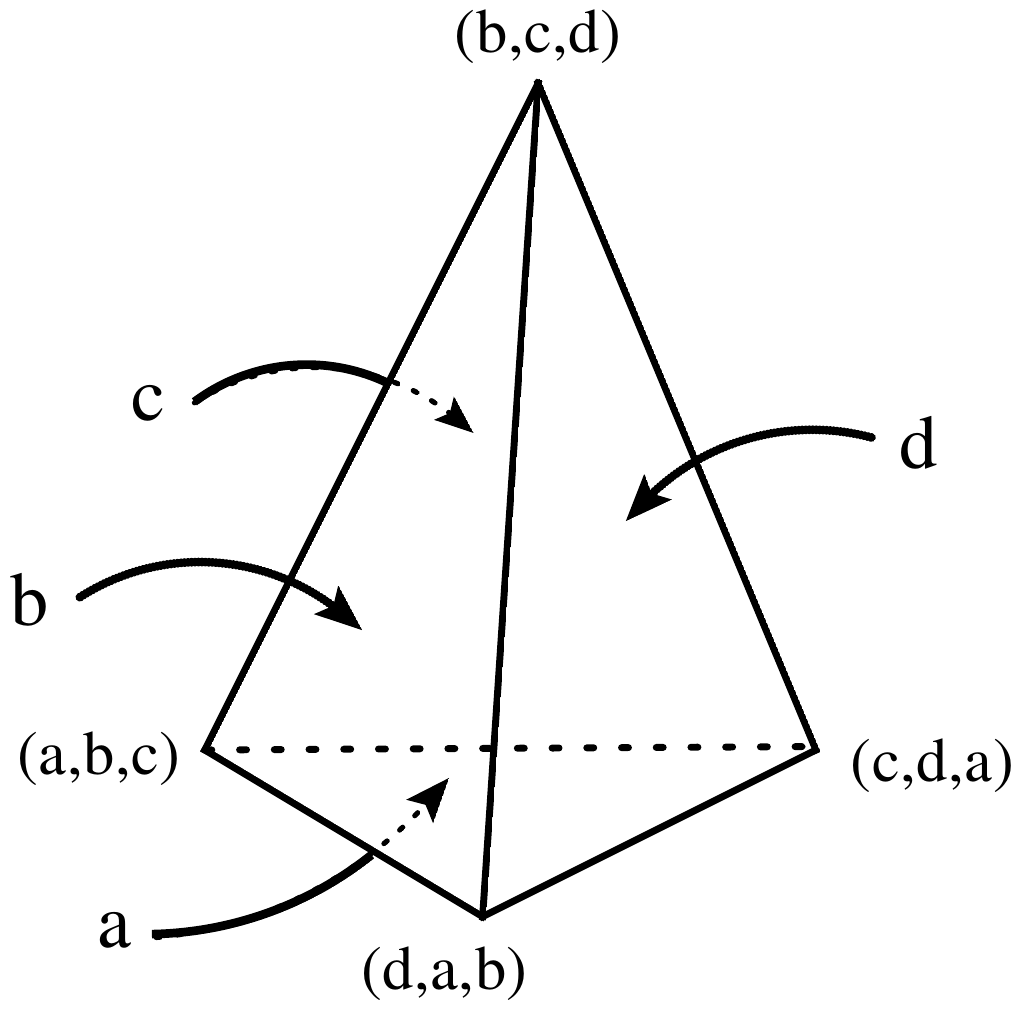}}}
 \eqe
Just as in the case of the triangle, the denominator of the volume formula \reef{nvolume} is the product of each vertex coordinate dotted into the reference vector $\mathsf{Z}_0$. The numerator compensates the scaling of each `face'-variable 
$\mathsf{Z}_{a}^I$, $\mathsf{Z}_{b}^I$, $\mathsf{Z}_{c}^I$, $\mathsf{Z}_{d}^I$ to make the volume formula projective.
 
Now, for $n=4$ the  5-bracket volume expression \reef{nvolume} for a 4-simplex is identical to the integrand in the amplitude 5-bracket expression \eqref{FiveVec}! This verifies our statement at the beginning of the section that the rational function in \eqref{FiveVec} is indeed the volume of a 4-simplex in $\mathbb{CP}^4$. The denominator factors in \reef{nvolume} involve the five vertices of the 4-simplex. Since an NMHV tree superamplitude is a sum of 5-brackets, we are led to view the amplitude as a volume of a polytope in $\mathbb{CP}^4$. We realize this expectation in the next section and discuss its consequences.

%%%%%%%%%%%%%%%%%%%%%%%%%%%%%%%%%%%%%%%%%%%%%%%%%%%%%%%%%%%%%%
\subsection{NMHV tree superamplitude as the volume of a polytope\label{VolSec}}
%%%%%%%%%%%%%%%%%%%%%%%%%%%%%%%%%%%%%%%%%%%%%%%%%%%%%%%%%%%%%%
The simplest NMHV case is the 5-point (anti-MHV)superamplitude
\be
A^\text{NMHV}_5[1,2,3,4,5] ~=~ A^\text{MHV}_5 \times  \big[1,2,3,4,5\big]
\ee
Thus, up to the MHV factor, $A^\text{NMHV}_5$ is the volume of a 4-simplex in $\mathbb{CP}^4$.

Next, for the NMHV 6-point superamplitude, 
consider the $[2,3\>$ super-BCFW representation on the LHS of \reef{Tempting}:
\be
A^\text{NMHV}_6[1,2,3,4,5,6] ~\propto~
\underbrace{\big[2,3,4,6,1\big]}_{\langle 4,6,1,2\rangle,\; \langle2,3,4,6\rangle}
+\underbrace{\big[2,3,4,5,6\big]}_{\langle 6,2,3,4\rangle,\; \langle4,5,6,2\rangle}
+ \underbrace{\big[2,4,5,6,1\big]}_{\langle2,4,5,6\rangle, \;\langle 6,1,2,4\rangle}
\,.
\label{Tempting2}
\ee
Apart from the overall MHV factor, the 6-point NMHV superamplitude is the sum of the volumes of three 4-simplices in $\mathbb{CP}^4$; we expect this to be the volume of a polytope obtained by somehow gluing the three simplices together. But how exactly does this work? To address this question, it is useful to examine the poles in the 5-brackets.

Recall from  \reef{ID2} and \reef{ID3} that momentum twistor 4-brackets $\langle i-1,i,j-1,j\rangle$ in the denominator gives local poles, whereas 4-brackets like 
$\langle k,i-1,i,j\rangle$ give spurious `non-local' poles. 
Examining the denominator terms of the 5-brackets in \reef{Tempting2}, we find that each of them has 2 spurious poles; they are listed under each 5-bracket. The spurious poles come in pairs 
--- for example $\langle 4,6,1,2\rangle$ and $\langle 6,1,2,4\rangle$ in the first and third 5-brackets ---
and cancel in the sum \reef{Tempting2}, as required by locality of the physical amplitude. In the geometric description of a 5-bracket as a 4-simplex in $\mathbb{CP}^4$, each of the five factors in the denominator of the volume-expression (i.e.~the 5-bracket) is determined by a vertex of the associated 4-simplex. In particular, spurious poles must be associated with vertices in $\mathbb{CP}^4$ that somehow `disappear' from the polytope whose volume equals the sum of the simplex-volumes in \reef{Tempting2}. 
We now discuss how the `spurious' vertices disappear in the sum of simplices. Let us start in $\mathbb{CP}^2$ where the polytopes are easier to draw.

%%%%%%%
\compactsubsection{Polytopes in $\mathbb{CP}^2$}
In 2 dimensions, consider the 4-edge polytope
\eq
\includegraphics[scale=0.4]{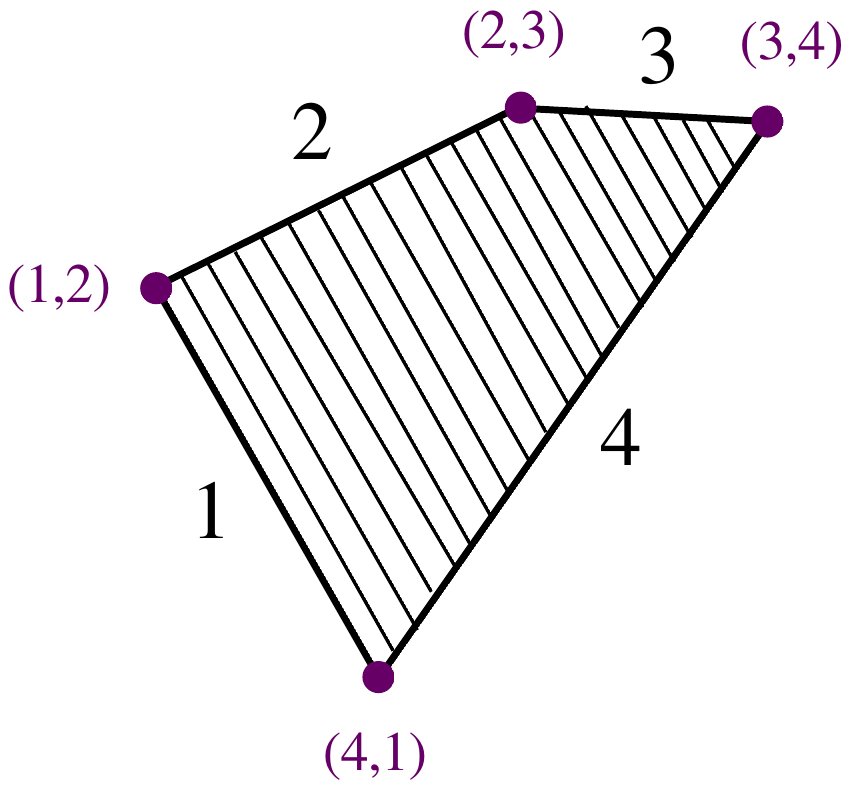}
\label{4pt2poly}
\eqe
All vertices for this $\mathbb{CP}^2$ ``amplitude" are defined by adjacent edges and is in this sense  local. We would like to compute the area of the 2-polytope \reef{4pt2poly} using 2-simplex volumes $[a,b,c]$. There are several different ways to do this, corresponding to different triangulations of the polytope. As an example, introduce a `non-local' point $(1,3)$ as the intersection of lines 1 and 3. The resulting triangulation is
\eqa
\nonumber\vcenter{\hbox{\includegraphics[scale=0.4]{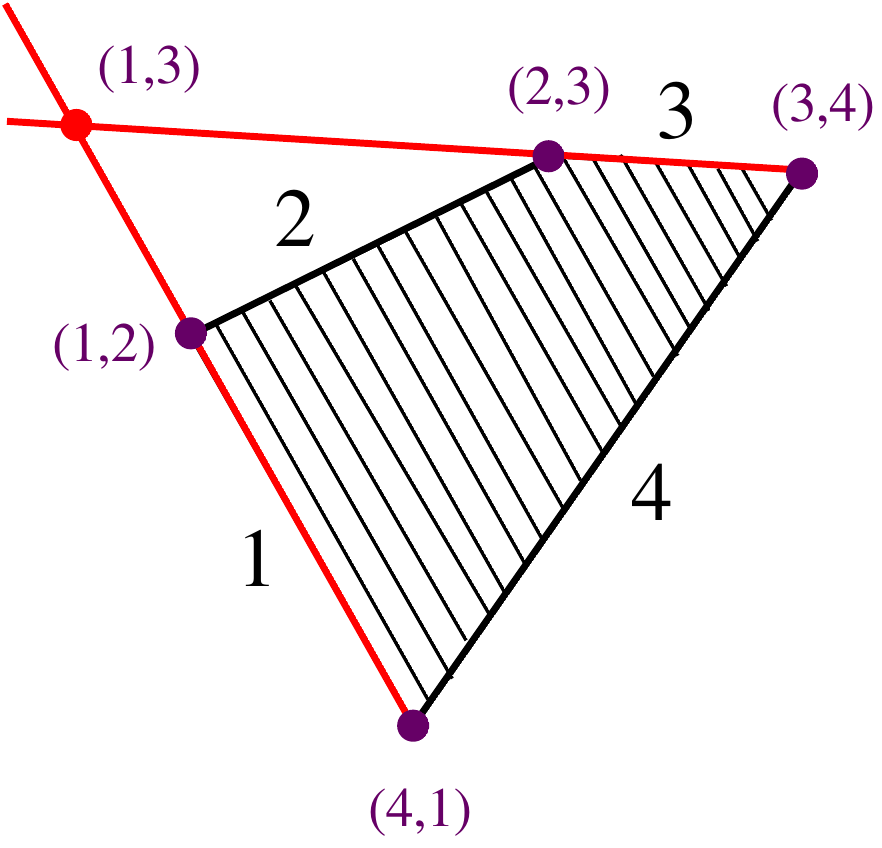}}}&=&\vcenter{\hbox{\includegraphics[scale=0.4]{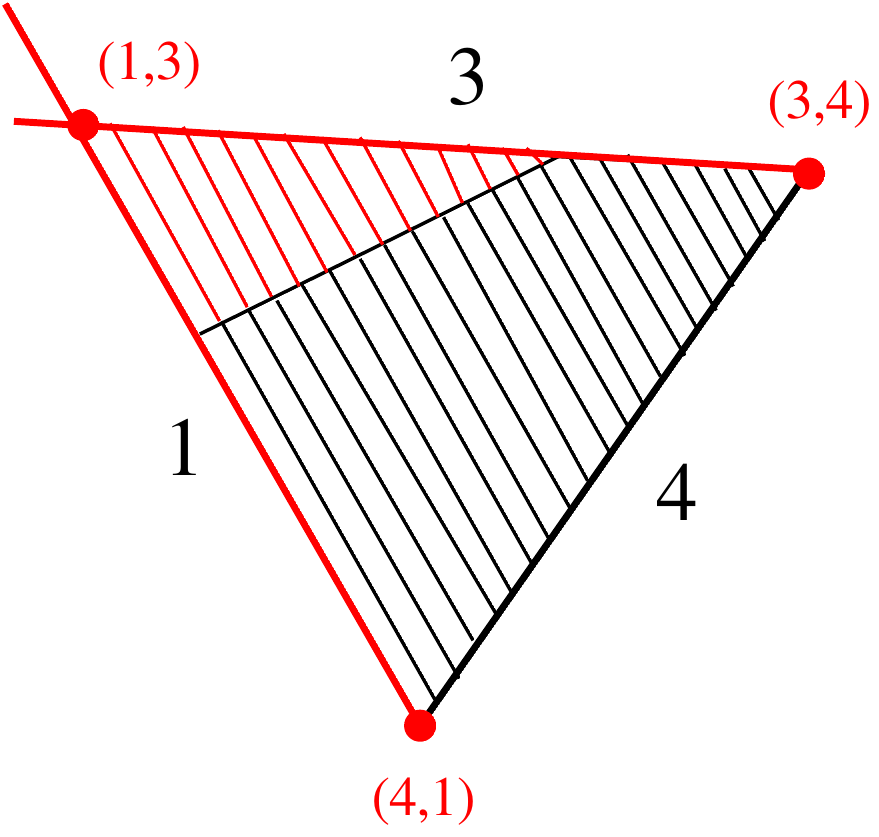}}}\quad-\quad\vcenter{\hbox{\includegraphics[scale=0.4]{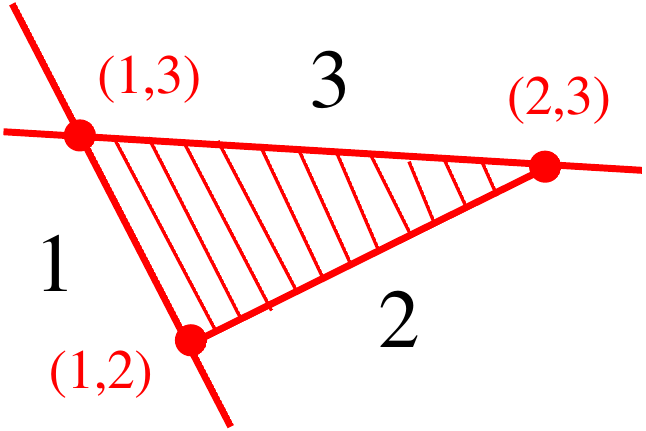}}}\,\\
\nonumber&=&\big[4,1,3\big]-\big[2,1,3\big]\\
&=&\big[4,1,3\big]+\big[1,2,3\big]\,.
\label{cp2example0}
\eqae 
The area of the 4-edge polytope is given by the difference of two triangular areas. The non-local vertex $(1,3)$ appears in both triangles. Comparing the last two lines, the sign of the 3-bracket indicates the orientation of the triangle with respect to a particular predetermined ordering of all edges (or, in higher dimensions, boundaries). 

It is useful to also consider another triangulation, so introduce the point $(2,4)$:
\bea
\nonumber\vcenter{\hbox{\includegraphics[scale=0.4]{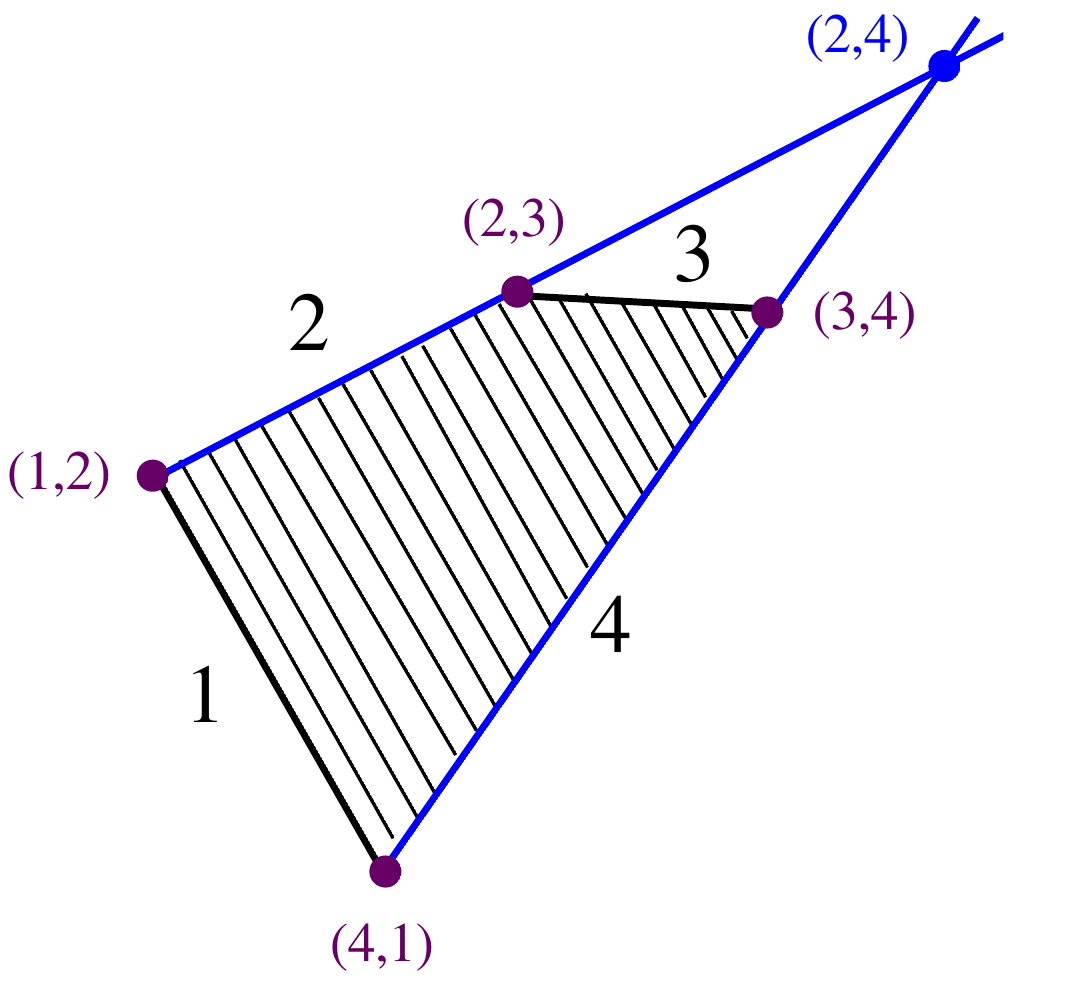}}}&=&\vcenter{\hbox{\includegraphics[scale=0.4]{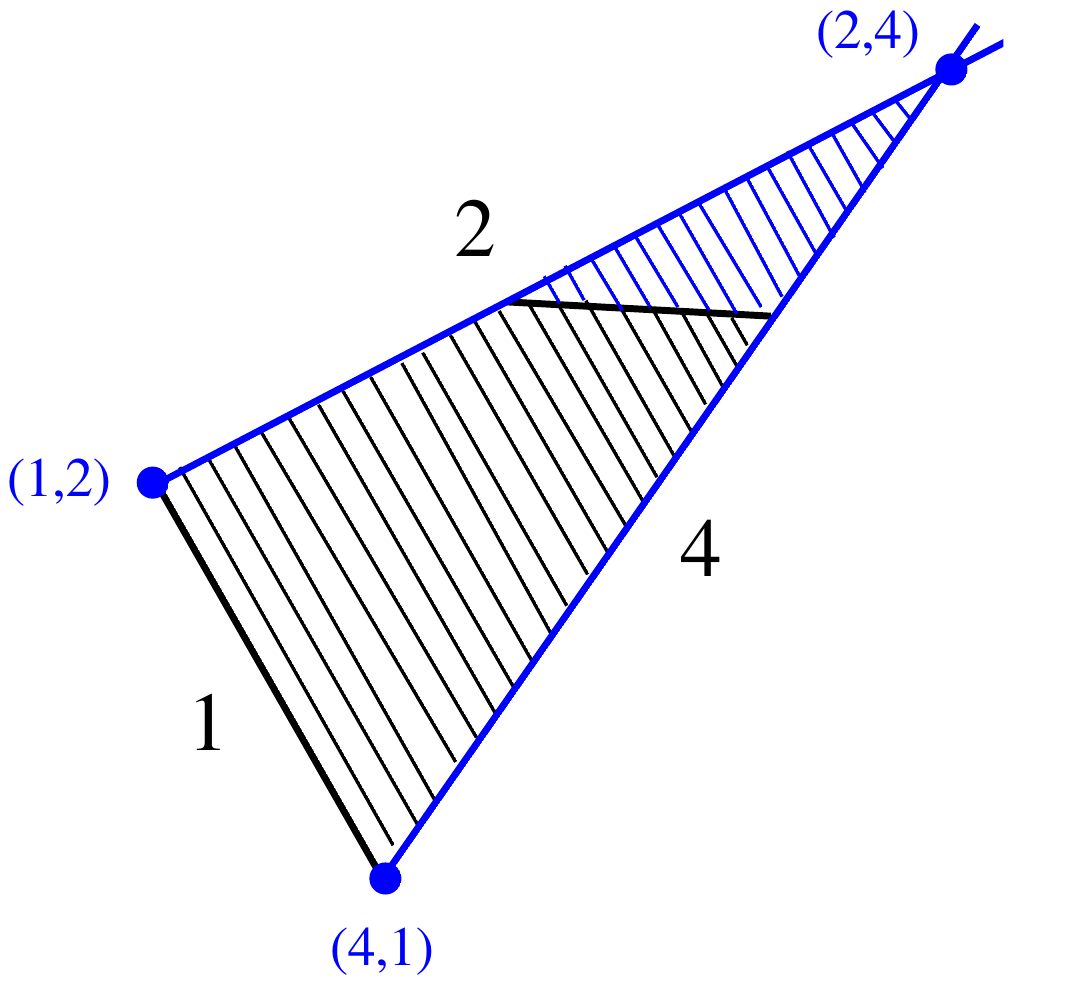}}}-\quad\vcenter{\hbox{\includegraphics[scale=0.4]{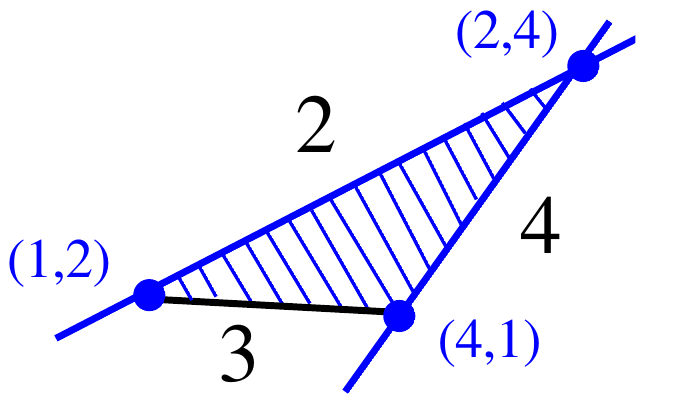}}}\,\\
\nonumber
&=&\big[1,2,4\big]-\big[2,4,3\big]\\ 
&=&\big[1,2,4\big]+\big[2,3,4\big]\,.
\label{cp2example}
\eea
The two triangulations \reef{cp2example0} and \reef{cp2example} compute the same area (``amplitude"), so we have a $\mathbb{CP}^2$ version of the identity \reef{Tempting}, namely $[4,1,3]+[1,2,3]=[1,2,4]+[2,3,4]$ which can also be written
\eq
\big[2,3,4\big] - \big[1,3,4\big] + \big[1,2,4\big]-\big[1,2,3\big]
~=~0\,.
\label{ThreeSimplex}
\eqe
\exercise{}{Suppose the 4-vertex polytope in the example above was not convex as drawn in \reef{cp2example}: show that the volume of a non-convex 4-vertex polytope can also be written $[1,2,4]+[2,3,4]$. 
}
%
%%%%%%%
\compactsubsection{Polytopes in $\mathbb{CP}^4$}
Extending the simple $\mathbb{CP}^2$ example to $\mathbb{CP}^4$, one finds that the BCFW representation of a 6-point NMHV tree superamplitude corresponds to a triangulation of the associated polytope by introduction of three new auxiliary vertices. This allows one to use the given external data, the boundaries of the polytope, to efficiently construct the corresponding triangulation. Efficiency here means using a minimum number of 4-simplices; we  come back to this point in Section \ref{s:geomath}. Different BCFW constructions simply correspond to different choices of auxiliary vertices. As an example, the auxiliary vertices for the BCFW representation in \eqref{Tempting2} are $(2,4,5,6)$, $(6,1,2,4)$ and $(2,3,4,6)$. Note that these exactly label the spurious poles in \reef{Tempting2}.
 
Let us now see how the removal of an auxiliary vertex works in 
$\mathbb{CP}^4$. Since it can be slightly challenging to draw a 4-dimensional object on paper, we go to the 3d boundary of the 
4-dimensional polytope. Specifically, at the 3d boundary defined by $\mathsf{Z}_1\cdot \mathsf{W}=0$, only the two simplices
$\big[2,3,4,6,1\big]$ and $\big[2,4,5,6,1\big]$ in \reef{Tempting2} contribute, and their projections to the boundary are the tetrahedrons defined by the faces 
$\mathsf{Z}_2$, $\mathsf{Z}_3$, $\mathsf{Z}_4$ and $\mathsf{Z}_6$ and, respectively,  by 
$\mathsf{Z}_2$, $\mathsf{Z}_4$, $\mathsf{Z}_5$ and $\mathsf{Z}_6$.
The two boundary tetrahedrons share the non-local vertex $(1,2,4,6)$, which we simply label $(2,4,6)$ on the boundary. Since this vertex does not appear in other terms of \reef{Tempting2}, we should be able to visualize its cancellation geometrically on the boundary defined by $\mathsf{Z}_1$. On the 3d subspace, the superamplitude contains the combination 
 \eq
 \big[2,3,4,6,1\big]+\big[2,4,5,6,1\big]
 ~\xrightarrow{\text{$\mathsf{Z}_1$  bdr}}~
 \big[6,2,4,5\big]-\big[6,2,4,3\big] 
 ~=~\text{vol(bdr polytope)} \,.
 \label{FinalStraw}
 \eqe 
 On the RHS we have arranged the common faces, $\mathsf{Z}_6$, $\mathsf{Z}_2$ and $\mathsf{Z}_4$, to appear in the same order to facilitate the geometrical interpretation. The pictorial representation of \reef{FinalStraw} is
 \eq
 \vcenter{\hbox{\includegraphics[scale=0.8]{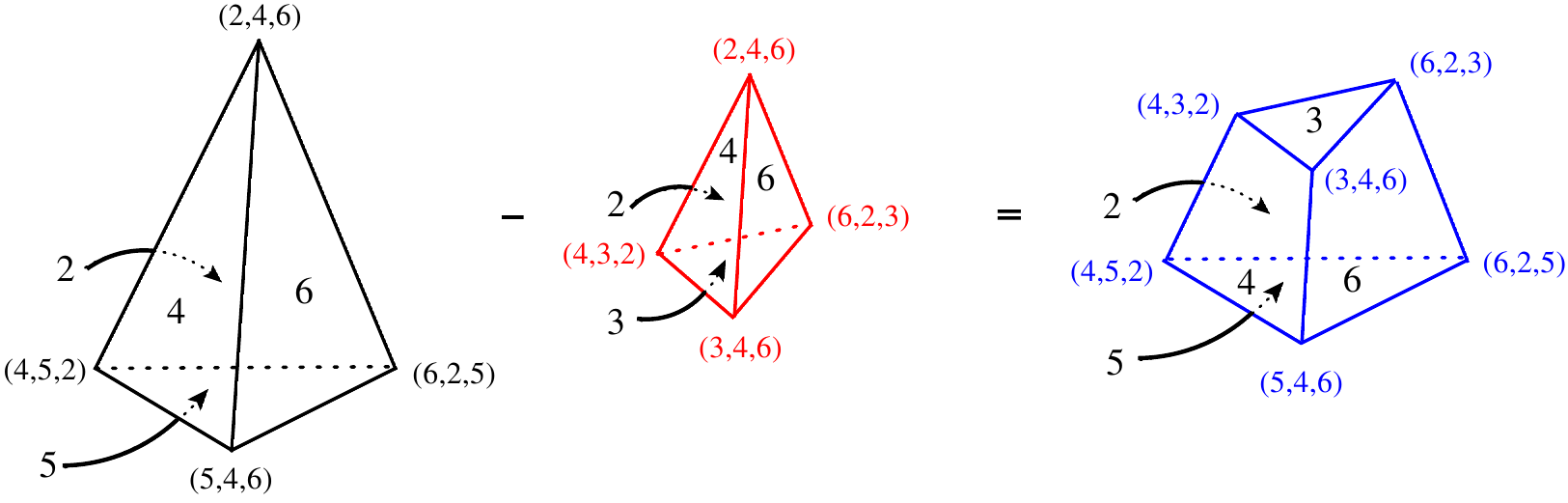}}}
 \label{bdrcp3}
 \eqe
The `non-local' auxiliary vertex, $(2,4,6)$ indeed `cancels' in the sum leaving behind the volume of the 3-dimensional polytope with five faces and six \emph{local}
 vertices! To see that the remaining vertices are local, remember that we are in the subspace $\mathsf{Z}_1\cdot W=0$, so each vertex is really represented as $(1,*,*,*)$ in $\mathbb{CP}^4$. Thus each of the six vertices, 
\be
 (1,2,3,4),\;~~(1,2,3,6),\;~~(1,3,4,6),\;~~(1,4,5,2),\;~~(1,4,5,6),\;~~(1,2,5,6)\,,
\label{locptsZ1}
\ee
involve two pairs of adjacent labels and by \reef{ID3} therefore they  correspond to local poles.  
 Thus we conclude that on the subspace $\mathsf{Z}_1\cdot W=0$, which involves only two of the simplices in \reef{Tempting2},  the amplitude is free of non-local vertices. One can similarly understand the cancellation of the two other spurious poles in \reef{Tempting2}.

We have found that the 6-point NMHV tree superamplitude is given by the volume of a polytope in $\mathbb{CP}^4$. It is defined as the sum of the three 4-simplices in \reef{Tempting2} and its six boundaries are in 1-1 correspondence with the momentum supertwistors $\mathsf{Z}^I_{i}$, $i=1,\ldots,6$. Different BCFW representations correspond to the different tessellations of the polytope into 4-simplices; each representation requires introduction of `spurious' vertices and the associated spurious poles cancel because they are absent in the original polytope. The vertices of the polytopes are all local and can be characterized as the nine quadruple intersections $(i,i+1,j,j+1)$ of the six boundaries determined by $\mathsf{Z}^I_{i}$.

The polytope interpretation of the amplitudes was first presented by Hodges \cite{Hodges} with the goal of geometrizing the cancellation of spurious poles in the BCFW expansion. Building on Hodges' work, the authors of \cite{NimaPoly} constructed the representation of the NMHV superamplitude where both dual superconformal symmetry and locality are manifest. 

%%%%%%%%%%%%%%%%%%%%%%%%%%%%%%%%%%%%%%%%%%%%%%%%%%%%%%%%%%%%%%
\subsection{The boundary of simplices and polytopes}
%%%%%%%%%%%%%%%%%%%%%%%%%%%%%%%%%%%%%%%%%%%%%%%%%%%%%%%%%%%%%%
We have studied the volumes of the simplices and polytopes; can we also learn something from studying their boundaries? Let us again start with a simple triangle in $\mathbb{CP}^2$,
\be
 \vcenter{\hbox{\includegraphics[scale=0.4]{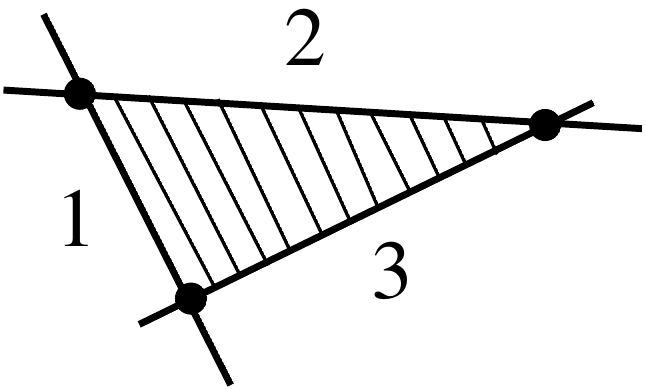}}}
\ee
The subspace defined by $\mathsf{Z}_1\cdot \mathsf{W} = 0$ contains part of the boundary of the triangle, namely the line segment bounded by the intersections of lines 2 and 3 with line 1. The length of the line segment is just the projection to the subspace defined by $\mathsf{Z}_1$, namely  $[2,3]$.
Note that since the `volumes' (i.e.~lengths) of the line segments are defined with a choice of sign, we have to pick an orientation for each on: here and in the following, we pick the orientation of the faces to point into the volume of the polytope that they are bounding. 
With this choice of orientation, the circumference is $[12]+[23]+[31]$.

To see a little more structure, consider the tetrahedron in $\mathbb{CP}^3$,
\be
 \vcenter{\hbox{\includegraphics[scale=0.4]{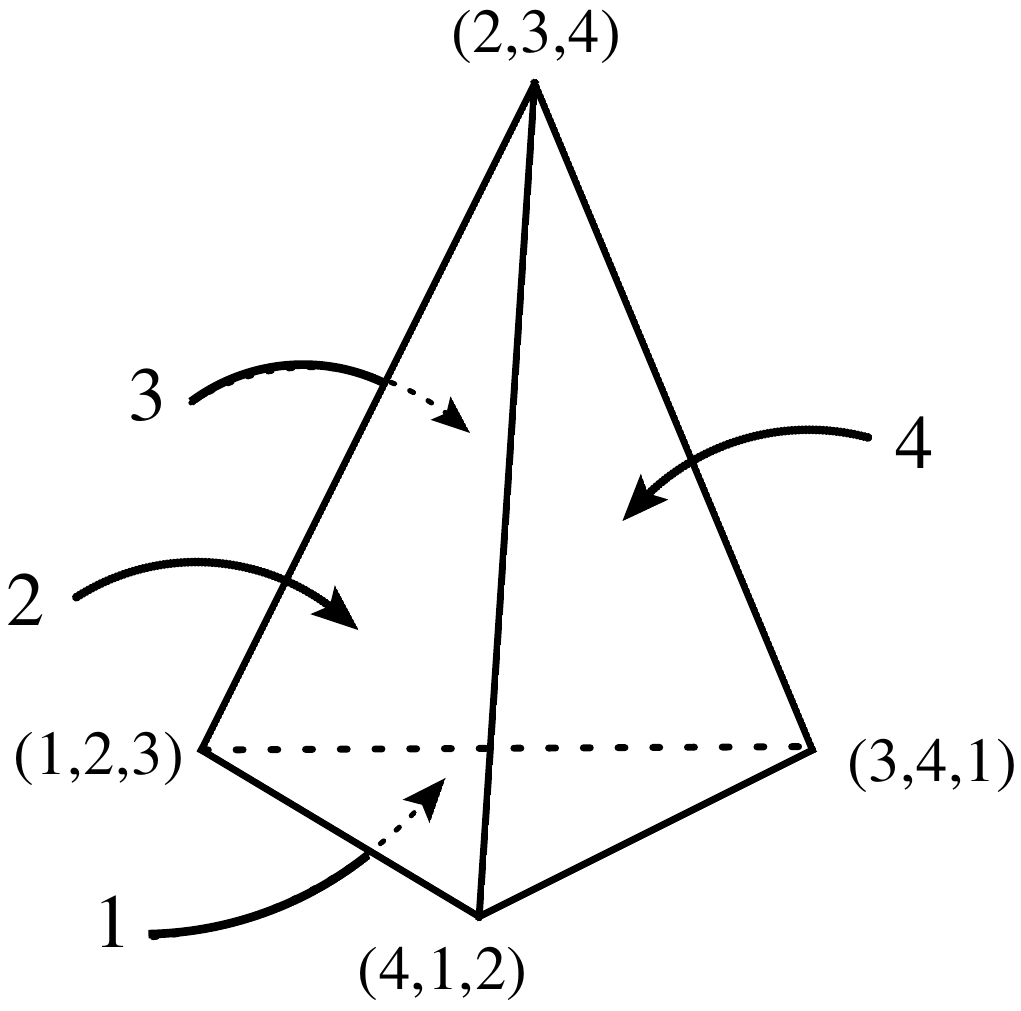}}}
\ee
The volume is $[1,2,3,4]$. The 2-plane defined by $\mathsf{Z}_1$  contains the face bounded by the intersections of the plane 1 with the planes 2, 3, and 4. Therefore the area of this face is $[2,3,4]$. Keeping careful track of the orientations of the faces, we find that the area of boundary of the tetrahedron is $[2,3,4]+[1,4,3]+[2,4,1]+[2,1,3]$. 

We can summarize the results for the boundary `volumes' so far as
\be
\begin{split}
\text{bdr of } \mathbb{CP}^2 \text{ triangle}:~~
\pa  \big[ 1,2,3 \big] &=~ \big[2,3\big]-\big[1,3\big]+\big[1,2\big]\,,\\
\text{bdr of } \mathbb{CP}^3 \text{ tetrahedron}:~~
\pa  \big[ 1,2,3,4 \big]&=~ \big[2,3,4\big]-\big[1,3,4\big]+\big[1,2,4\big]-\big[1,2,3\big]\,.\\
\end{split}
\label{cp23bdrs}
\ee
This motivates us to define the {\bf \em boundary operation} for any $(n\!-\!1)$-simplex:
\be
\pa  \big[ 123 \dots n \big] ~=~
\sum_{i=1}^n (-1)^{i+1}\, \big[1,2,\dots,i-1,i+1,\dots,n\big]\,.
\label{defbdrop}
\ee
\exercise{}{Since the boundary of a boundary vanishes, our definition \reef{defbdrop} better have the property that $\pa^2 = 0$. 
Show that the action of $\partial^2$ on any $n$-simplex is zero .
}
At this stage you might have noticed the similarity between the RHS of the 
tetrahedron boundary identity
\reef{cp23bdrs} and the $\mathbb{CP}^2$ vanishing identity \reef{ThreeSimplex}.
This is easy to understand: in $\mathbb{CP}^2$, we cannot construct a 3-simplex  $\big[ 1,2,3,4 \big]$ with a 3d volume, so in particular the boundary of such a formal object must vanish: 
\be
\mathbb{CP}^2\!:~~~~~  
0~=~\pa  \big[ 1,2,3,4 \big]~=~ \big[2,3,4\big]-\big[1,3,4\big]+\big[1,2,4\big]-\big[1,2,3\big] \,.
\ee
This gives another geometric interpretation of the $\mathbb{CP}^2$ BCFW identity \reef{ThreeSimplex}.

Similarly, a 5-simplex $\big[1,2,3,4,5,6\big]$ in $\mathbb{CP}^4$ must have vanishing boundary:
\be
\begin{split}
\mathbb{CP}^4\!:~~~~~
 0 ~=~ \partial \big[1,2,3,4,5,6\big]
 ~\equiv~&
\big[2,3,4,5,6\big]-\big[3,4,5,6,1\big]+\big[4,5,6,1,2\big]\\
&
-\big[5,6,1,2,3\big]+\big[6,1,2,3,4\big]-\big[1,2,3,4,5\big]\,.
\end{split}
\label{bdrop}
\ee
The RHS is exactly the six-term identity \reef{alsoTempting} which originated from the equivalence of different super-BCFW shifts \reef{Tempting}. This was the identity that motivated our study at the beginning of the section: we understand of course now that it is not a Schouten identity, but here it is interpreted as the vanishing boundary of a formal 5-simplex in $\mathbb{CP}^4$.

Let us now take a look at the action of the boundary operation on a superamplitude. As per usual, we start with  $\mathbb{CP}^2$ to get intuition for the problem. Consider the triangulation used in \reef{cp2example0} to calculate the volume of a 4-sided polygon in $\mathbb{CP}^2$ 
\eqa
\label{cp2bdrex}
\vcenter{\hbox{\includegraphics[scale=0.4]{2Area}}}&=&\vcenter{\hbox{\includegraphics[scale=0.4]{2Area3}}}\quad-\quad\vcenter{\hbox{\includegraphics[scale=0.4]{2Area4}}}\,\\
\nonumber
&=&\big[4,1,3\big]-\big[2,1,3\big] 
\,.
\eqae 
We apply the boundary operator to each 2-simplex and find:
\be
\begin{split}
 \pa \big[4,1,3\big] &=~  \big[1,3\big]-\big[4,3\big]+\big[4,1\big]\,,\\
 \pa \big[2,1,3\big] &=~  \big[1,3\big]-\big[2,3\big]+\big[2,1\big]\,.
\end{split}
\label{cp2bdrex2}
\ee
The RHS of each equation is the circumference of the respective triangles.
Let us try to interpret $\pa\big( [4,1,3]-[2,1,3] \big)$.
It contains $\big([4,1] - [2,1]\big)$: this is a difference of lengths of two line segments in the subspace defined by $\mathsf{Z}_3$ and thus it is the length of the side labeled 3 in the 4-sided polygon on the LHS of \reef{cp2bdrex}.
Similarly, $\big([4,3] - [2,3]\big)$ is the length of side 1 of the polygon. Now we are left with two terms are both labeled $[1,3]$ in \reef{cp2bdrex2}. It is tempting to cancel these two terms, but this is not quite correct:  $[1,3]$ in $\pa [4,1,3]$ lives in the subspace defined by $\mathsf{Z}_4$ while in $\pa [2,1,3]$ is in the $\mathsf{Z}_2$-subspace. So the two $[1,3]$'s are the lengths of the sides 2 and 4 in the polygon \reef{cp2bdrex}. Why does their difference show up in $\pa\big( [4,1,3]-[2,1,3] \big)$ instead of their sum? Easy: that is because they have the opposite orientations: in our conventions, side 4 in the big triangle in \reef{cp2bdrex} is oriented to point into the polygon, but side 2 in small triangle  points out of the 4-sided polygon. Flipping the orientation and labeling the 2-brackets by the subspace $\mathsf{Z}_i$ they live on, we see that the difference of the two terms in \reef{cp2bdrex2} exactly calculate the circumference of the 4-sided polygon on the LHS of \reef{cp2bdrex}:
\be
   \pa\big( [4,1,3]-[2,1,3] \big) =
   \big([4,3] - [2,3]\big)_{\mathsf{Z}_1}
   + \big[3,1\big]_{\mathsf{Z}_2}
    + \big([4,1] - [2,1]\big)_{\mathsf{Z}_3}
    + \big[1,3\big]_{\mathsf{Z}_4} \,.
\ee

The boundary operator $\pa$ was introduced \cite{NimaPoly} as a formal operation useful for studying the cancellation of spurious poles in the BCFW expansion, without emphasis on the interpretation as the `boundary volume' we have presented here. Let us now comment on the application of $\pa$ in \cite{NimaPoly}. Note that for a simplex, there is a unique point `opposite' each face: in particular in the triangles in \reef{cp2bdrex} the point labeled $(1,3)$ is the non-local `spurious' point that sits across from the line segments $\big[1,3\big]_{\mathsf{Z}_2}$ and $\big[3,1\big]_{\mathsf{Z}_4}$, respectively. So since the two $\big[1,3\big]$ define the same point $(1,3)$, one can in  a \emph{vertex-interpretion} of the boundary operation cancel them in $\pa\big( [4,1,3]-[2,1,3] \big)$: one can think of this as the cancelation of the spurious point in the $(1,3)$ in this particular triangulation. To distinguish the vertex-interpretation from the boundary volume, we include a $V$ (for vertex) with each term; then we write
\be
   \pa\big( [4,1,3]-[2,1,3] \big) =
  V\big[3,4\big]+V\big[4,1\big]
  +V\big[2,3\big]+V\big[1,2\big]\,.
\ee
Note how each term on the RHS is of the from $V\big[i,i+1\big]$ indicating that the polytope has only local vertices; the non-local vertex $V\big[1,3\big]$ cancelled. This is the interpretation of the boundary operation given in \cite{NimaPoly}.
\exercise{}{As an example in $\mathbb{CP}^3$, consider the dissection of the 5-faced polytope into two tetrahedrons in \reef{bdrcp3}. Keep careful track of the orientations of the boundaries to show that 
$\pa\big( \big[6,2,4,5\big]-\big[6,2,4,3\big]  \big)$ calculates the surface area of the 5-faced 3-polytope on the RHS of \reef{bdrcp3}. Next use the vertex-interpretation discussed above to show that the spurious poles are cancelled. Lift the example back to $\mathbb{CP}^4$ (remember that \reef{bdrcp3} was the projection on the subspace defined by $\mathsf{Z}_1$) to see that each boundary vertex term is of the from $V[i,i+1,j,j+1]$ as in \reef{locptsZ1}.}
Enough of toy-examples! Let us compute the boundary of the NMHV 6-point superamplitude in the BCFW representation
\be
  A_6^\text{NMHV}[1,2,3,4,5,6]= 
  A_6^\text{MHV} \times\
  \Big(
  \big[1,3,4,5,6\big]+\big[3,5,6,1,2\big]+\big[5,1,2,3,4\big]
  \Big)\,.
  \label{A6poly}
\ee
Using the vertex-interpretation, $\pa$ acts on the first two 4-simplices to give
\be
\begin{split}
\partial\Big(\big[1,3,4,5,6\big]+\big[3,5,6,1,2\big]\Big)
~=~& V\big[3,4,5,6\big]+V\big[4,5,6,1\big]+V\big[6,1,3,4\big]
+V\big[5,6,1,2\big] \\
&
+V\big[2,3,6,1\big]
+V\big[2,3,5,6\big]+V\big[1,3,4,5\big]+V\big[1,2,3,5\big]\,.
\end{split}
\ee
All vertices on the RHS are local except that last two.
Including the third 5-bracket from \reef{A6poly}, the non-local vertices cancel and we have
\be
\partial \Big(\big[1,3,4,5,6\big]+\big[3,5,6,1,2\big]+\big[5,1,2,3,4\big]\Big)=
\sum_{i=1}^6V\big[i,i+1,i+2,i+3\big]+\sum_{i=1}^3V\big[i,i+1,i+3,i+4\big]\,,
\ee
where the arguments are understood cyclically.
This shows that, indeed, the boundary of the polytope that corresponds to our amplitude $A_6^\text{NMHV}$ contains only local vertices. The power of the boundary operation is that it makes the cancellation of spurious points clear without a need to draw any polyhedrals. 
\exercise{}{
How many local vertices are there in the polytope corresponding to the 7-point NMHV superamplitude? 
}
%answer: 14. For general n, you can find a recursive formula. I find the sequence:
%{1, 5, 9, 14, 20, 27, 35, 44, 54, 65, 77, 90,...}
%the case 1 is fake. 5 is the for the 5-pt ampl, 9 for the 6-pt amp, etc.

%%%%%%%%%%%%%%%%%%%%%%%%%%%%%%%%%%%%%%%%%%%%%%%%%%%%%%%%%%%%%%
\subsection{Geometric aftermath}
\label{s:geomath}
%%%%%%%%%%%%%%%%%%%%%%%%%%%%%%%%%%%%%%%%%%%%%%%%%%%%%%%%%%%%%%

The BCFW triangulation is an efficient representation of the tree-level NMHV superamplitudes in the sense that it involves only relatively few terms. 
However, we can imagine other triangulations. For example, consider the 4-sided polygon in $\mathbb{CP}^2$. Introducing an auxiliary point $\mathsf{W}^*$ inside the polygon, we can triangulate is as
\be
\vcenter{\hbox{\includegraphics[scale=0.4]{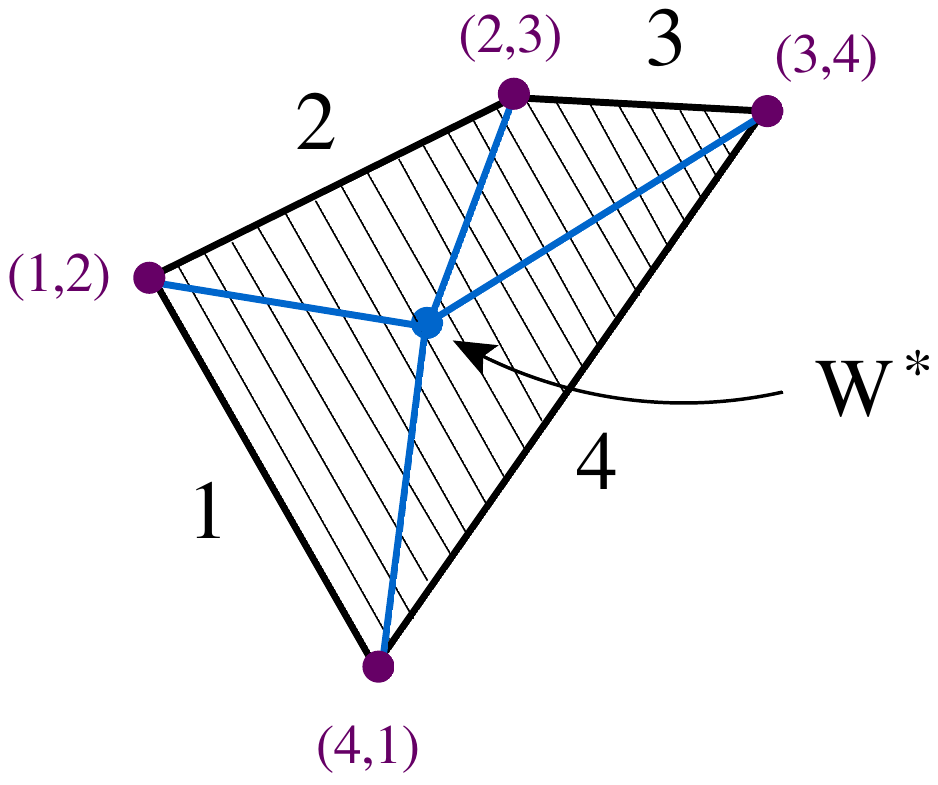}}}
\quad =\quad
\sum_{i=1}^4\,\frac{1}{2}\,
\frac{\langle \mathsf{W}^*,\mathsf{W}_{(i-1,i)},\mathsf{W}_{(i,i+1)}\rangle}{(\mathsf{Z}_0\cdot \mathsf{W}^*)\langle 0,i-1,i\rangle\langle0,i,i+1\rangle}\,.
\label{cswpolytope}
\ee
This gives a 4-term expression for the volume of the polygon, as opposed to the 2-term BCFW triangulations in \reef{cp2example0} or \reef{cp2example}. In this sense, BCFW is more efficient. 
The representation \reef{cswpolytope} may remind you of another representation of scattering amplitudes, namely the CSW expansion (or MHV vertex expansion) of Section \ref{s:csw}. It is actually not quite the same; CSW in momentum twistor space involves a reference supertwistor $\mathcal{Z}^* = (0,|X],0)$ instead of $\mathsf{W}^*$; see \cite{Bullimore:2010pj} for details.

As yet another way to calculate the amplitudes, we might ask if there is a triangulation that does not give spurious poles? A prescription for such  a representation was given in \cite{NimaPoly} for the tree-level NMHV superamplitudes.
To give a hint of how it works, consider the boundary $\mathsf{Z}_1\cdot \mathsf{W}=0$ that we also analyzed in \reef{bdrcp3}. Instead of the 2-term triangulation applied in  \reef{bdrcp3} at the cost of a non-local vertex, we can triangulate the 5-sided polytope as
\eqa
\nonumber\vcenter{\hbox{\includegraphics[scale=0.4]{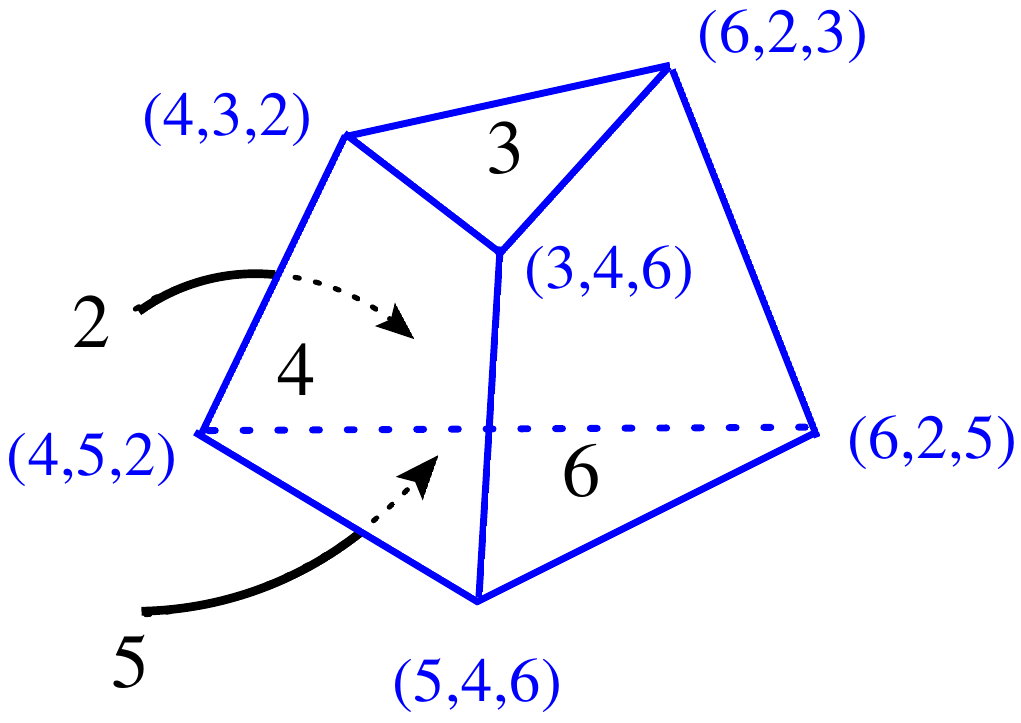}}}&&=
~~~
\vcenter{\hbox{\includegraphics[scale=0.4]{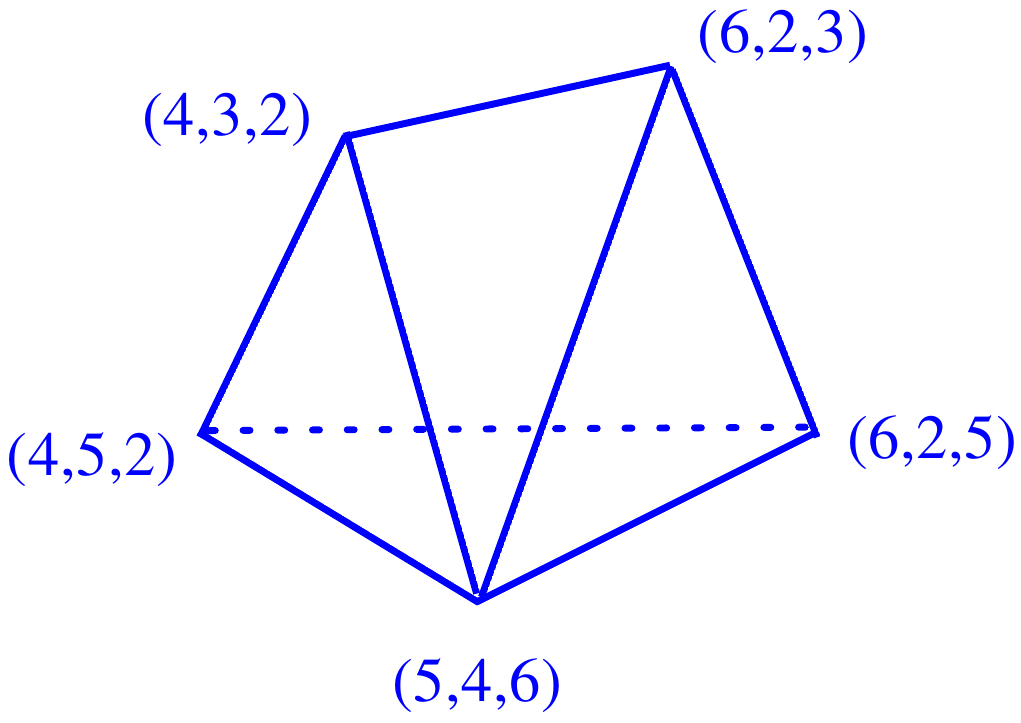}}}+\vcenter{\hbox{\includegraphics[scale=0.4]{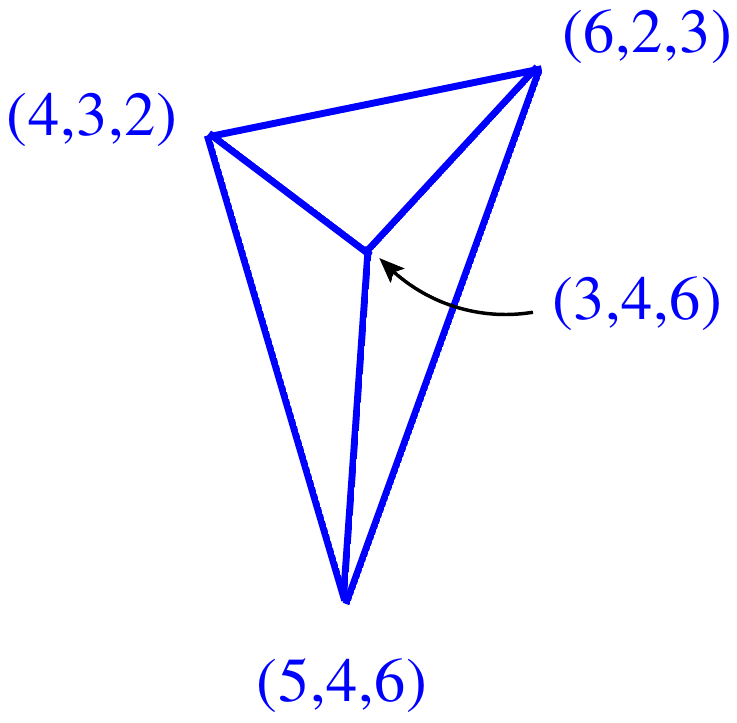}}}\\
&&~\hspace{-4cm}=~
\vcenter{\hbox{\includegraphics[scale=0.4]{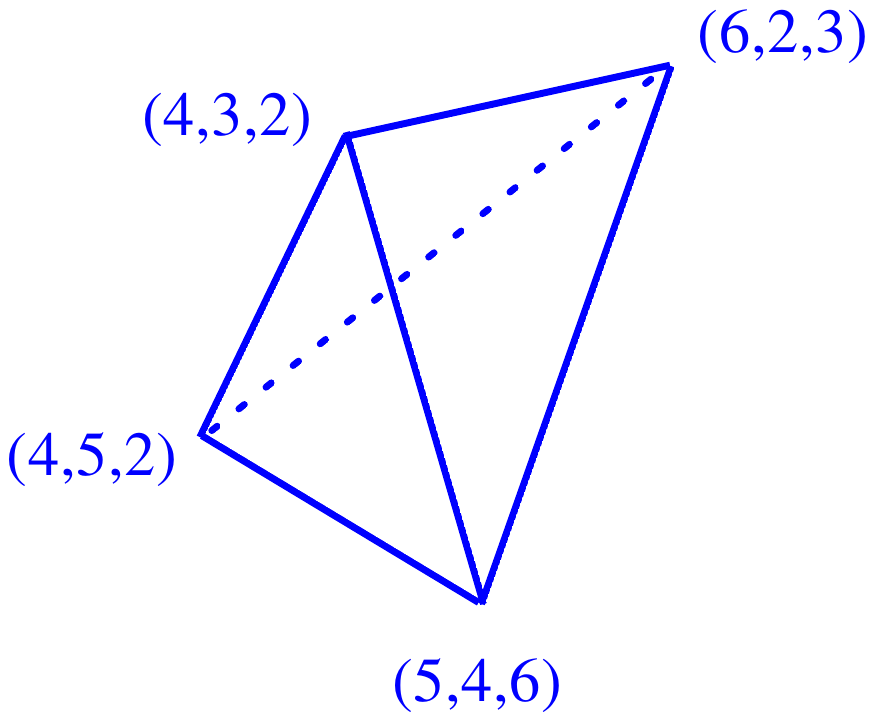}}}\!\!\!+\vcenter{\hbox{\includegraphics[scale=0.4]{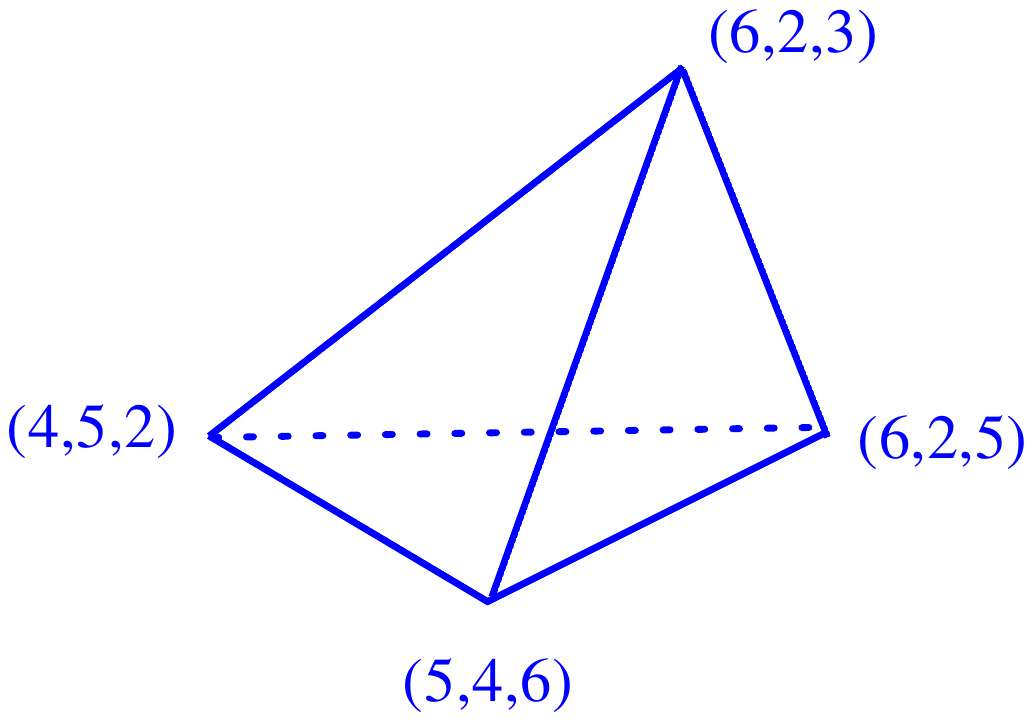}}}+\!\!\vcenter{\hbox{\includegraphics[scale=0.4]{FinalArea4}}}\,.
\eqae
The manifestly local tessellation of the polytope gives more terms than the BCFW representation. 
You can find the general expression in \cite{NimaPoly}.

We have argued that each $n$-point tree-level NMHV superamplitude of $\cn=4$ SYM can be interpreted as the volume of a polytope in $\mathbb{CP}^4$. It should be rather obvious by now that the reverse is not true: not all polytopes in $\mathbb{CP}^4$ correspond to superamplitudes in $\cn=4$ SYM. An example is the polytope obtained by gluing together two of the three simplices in the BCFW representation \reef{A6poly}: that is a perfectly fine polytope, but it has non-local poles so does not correspond to a physical amplitude. The color-ordering plays a key role in interpreting the superamplitudes polytopes. It is of course an interesting questions if this geometric picture can be extended beyond the leading color level --- or if other polytopes might have interpretations in terms of scattering processes.

The current discussion of tree superamplitudes utilizes the dual superconformal invariance of planar $\mathcal{N}=4$ SYM, focusing on the 5-brackets invariants.
Since the tree amplitudes of pure Yang-Mills theory can be projected out from $\mathcal{N}=4$ SYM superamplitudes, a similar analysis can be applied to pure YM as well. In fact, it was in pure Yang-Mills theory that Hodges realized the polytope picture \cite{Hodges}. 

The polytope interpretation described here is valid for NMHV $n$-point tree superamplitudes as well as 1-loop $n$-point MHV integrands in planar $\cn=4$ SYM \cite{NimaPoly}. The generalization is not obvious. Tree-level N$^K$MHV superamplitudes involve sums of products of $K$ 5-brackets, so a geometric interpretation in terms of simplex-volumes is not straightforward. There is nonetheless a geometrization of all N$^K$MHV tree superamplitudes and loop-integrands in planar $\cn=4$ SYM: it goes under the name of the {\bf \em amplituhedron}\index{amplituhedron} \cite{NT}. 
This is a polytope defined in a space whose coordinates are a union of  momentum supertwistors and Grassmannian coordinates that extend the $\chi_i \cdot \psi$-construction in \reef{5vec}. For tree-level NMHV, the amplituhedron reduces to a dual of the polytope discussed here, with  vertices and faces interchanged. The amplituhedron description makes locality manifest, while unitarity is an emergent property. BCFW arises as a particular triangulation. Loops appear from integrating out pairs of `hidden' points, in a somewhat similar way to the description of the loop-integrands in Chapter \ref{s:loops2}.

It is curious that in connection to amplitudes, polytopes can appear
in different guises. An example, different from our discussion so far,
is the observation \cite{MasonSkinnerPoly} that a 1-loop box integral
can be interpreted as the volume of a tetrahedron in
AdS$_5$. The vertices of the tetrahedron are the four dual region
variables $y_i$ (in the embedding formalism) and the edges are
geodesics in AdS$_5$. Since all $\mathcal{N}=4$ SYM 1-loop amplitudes
are given by an expansion in box integrals, the amplitudes can be interpreted as sums of volumes of such AdS$_5$  tetrahedrons, weighted by the appropriate box-coefficients. Non-planar 1-loop amplitudes 
can be given as linear combination of planar ones, so the same
conclusion extends to non-planar amplitude as
well~\cite{NonPlanarPoly}.

We have encountered many different representations of the tree-amplitude and now seen a unifying geometric interpretation. But there is yet another representation of amplitudes that we had a glimpse of in Section \ref{s:YM}, namely the `BCJ representation'  in which the color- and kinematic-structures enter on a dual basis. For such representations the amplitude can be manifestly local. We discuss BCJ further in Section~\ref{s:BCJ}, but note here that the color-ordering is crucial for the relation between polytopes and amplitudes: it allows us to relate the polytope to the momentum space representation of an amplitude and this is key for the statements about locality. If you ask about polytope interpretations for the planar BCJ representation, non-planar, or non-color-ordered amplitudes, then you have found yourself a bunch of research projects.

%%%%%%%%%%%%%%%%%%%%%%%%%%%%%%% 
%%%%%%%%%%%%%%%%%%%%%%%%%%%%%%% 
%%%%%%%%%%%%%%%%%%%%%%%%%%%%%%% 
\newpage
\setcounter{equation}{0}
\section{Amplitudes in dimensions $D\ne4$} 
\label{s:Dne4}
%%%%%%%%%%%%%%%%%%%%%%%%%%%%%%% 
%%%%%%%%%%%%%%%%%%%%%%%%%%%%%%% 
%%%%%%%%%%%%%%%%%%%%%%%%%%%%%%% 
Just in case it slipped your mind, 
our discussion up to now has focused on scattering amplitudes in $D\!=\!4$ spacetime dimensions. There is a good reason for this: for one,  this review was written in 3+1 dimensions (as far as we know) and this is where our particle physics experiments take place. And secondly, the power of the $D\!=\!4$ spinor helicity formalism and its extensions to twistors and momentum twistors allowed us to explore the rich and exciting mathematical structure of 4d scattering amplitudes, especially those in planar $\cn=4$ SYM. However,  there are quantum field theories worthwhile studying in other dimensions too;  in this section we take a look at their  scattering amplitudes. We discuss $D\!=\!6$ briefly, but otherwise our eyes are on $D\!=\!3$, particularly  on the interesting $\cn=8$ and $\cn=6$ superconformal theories BLG and ABJM.

%%%%%%%%%%%%%%%%%%%%%%%%%%%%%%
\subsection{Helicity formalism in $D\ne 4$}
\label{s:spDnot4}
%%%%%%%%%%%%%%%%%%%%%%%%%%%%%%
We have often emphasized in this review that the modern on-shell approach relies heavily on having a `good' set of variables that parameterize the on-shell degrees of freedom: `good'  means that the variables trivialize (part of) the kinematic constraints and transform linearly under the global symmetries of the theory. This is realized strikingly by the supertwistors and momentum supertwistors of planar $\cn=4$ SYM, but the trivialization of the massless on-shell condition $p_i^2=0$ in the spinor helicity formalism with $|i\>$ and $|i]$  was our starting point. So this is also where we begin for $D \ne 4$.

To parametrize massless kinematics in $D$-dimensions, consider bosonic spinors that carry  a spinor index $A$ of the Lorentz group {Spin}($1,D-1$) and a fundamental index $\mathsf{a}$ of the little group $SO(D-2)$:
\eq
\lambda_{i\mathsf{a}~\leftarrow~\text{little grp}}^{A ~\leftarrow~\text{Lorentz}}\;.
\label{GenSpin}
\eqe
As per usual, $i=1,2,\dots,n$ is a particle label.
The spinor-type (Weyl, Majorana etc) will be specified when we specialize to a given dimension $D$.  If the spinors are complex, there will be a conjugate spinor $\tilde\lambda$ whose $A$ and $\mathsf{a}$ indices are in the appropriate conjugate representations. 

The 4d Lorentz group is  {Spin}(1,3)= $SL(2,\mathbb{C})$ and the little group is $SO(2)=U(1)$. The spinors 
\eq
   \text{$D=4$:}\qquad  
   \tilde{\lambda}_{i-}^{\dot{a}} = |i\>^{\dot{a}}\,,
   \hspace{6mm}
   \lambda_{i+}^a = [i|^a\,.
   \label{D4spinhel}
\eqe 
are Weyl-spinors, so the index $A$ is the familiar $SL(2,\mathbb{C})$ indices $a,\dot{a}$. The little group index $\mathsf{a}$ is $+$ or $-$ depending on how the spinors transform under the $U(1)$ little group transformations. 
The $D=4$ lightlike momentum is written as the bi-spinor as the familiar relation
\eq
   \text{$D=4$:}\qquad  
   p_i^{\dot{a}a}=-\tilde{\lambda}_{i-}^{a}\,{\lambda}_{i+}^{\dot{a}}
    =- |i\>^{\dot{a}} [i|^a
   \,.
\label{D=4Result}
\eqe
As discussed in Section \ref{s:sh}, $p_i^\mu$ is real when the spinors \reef{D4spinhel} are conjugate.

The relation \reef{D=4Result} implies that the 2$\times$2 matrix $p_i^{\dot{a}a}$ has rank 1 and therefore solves the $D=4$ massless constraint $p_i^2=-\det(p_i)=0$. To see if a similar construction could be available in $D$ dimensions, we simply count degrees of freedom. A real lightlike vector has $D\!-\!1$ degrees of freedom with the $-1$ from the condition $p_i^2=0$.\footnote{An on-shell massive momentum has $p_i^2=-m_i^2$, but we view the constraint as imposed on $D+1$ degrees of freedom, ${p}_i^\mu$ and  $m_i^2$.} So the strategy is to find a {Spin}(1,$D-$1) spinor representation that allows forming a little group invariant bi-spinor with $D\!-\!1$ degrees of freedom. 
Here is how the counting works in $D=4$. 
The complex 2-component spinor $\lambda_{i+}^a = [i|^a$ has four real degrees of freedom, and when combined with its complex conjugate $\tilde{\lambda}_{i-}^{\dot{a}}=|i\>^{\dot{a}}$, the resulting bi-spinor \reef{D=4Result} is invariant under the $U(1)$ little group rotation. Thus subtracting out the $U(1)$ redundancy, we indeed have $4-1=3$ degrees of freedom, matching that of a real lightlike vector in 4d. Now let us look at how the counting works in other dimensions.

For $D=3$, the Lorentz group is {Spin}(1,2)= $SL(2,\mathbb{R})$ and the minimal spinor representation is a 2-component Majorana spinor $\lambda_i^a$, where $a=1,2$ an $SL(2,\mathbb{R})$ index. The null momentum is given by 
\eq
  \text{$D=3$:}\qquad 
  p_i^{ab}=\lambda_i^a\lambda_i^b \,.
\label{D=3LDef}
\eqe
For real momentum, the spinors $\lambda_i^a$ may be either real or purely imaginary. Either way, they encode 2 real degrees of freedom. So the RHS of \reef{D=3LDef} has $2$ degrees of freedom, the correct count for a 3d lightlike vector.
Note that no little group index was included on the spinors $\lambda_i^a$ because the little group $\mathbb{Z}_2$ is discrete. It acts as $\lambda_{i}^a\rightarrow -\lambda_{i}^a$, indeed leaving the momentum \reef{D=3LDef} invariant.

For $D=6$, we have {Spin}(1,5)=$SU^*(4)$ and the little group is $SO(4)=SU(2)\times SU(2)$. The $*$ on the $SU^*(4)$ indicates it is pseudo-real.\footnote{Pseudo real means that for each group element $g$, the complex conjugate $g^*$ is related to $g$ via a similarity transformation $g=\Omega g^*\Omega ^{-1}$, where $\Omega$ is an antisymmetric matrix. (If $\Omega$ is symmetric, then the representation is a real.)} We pick a chiral spinor $\lambda_{i\mathsf{a}}^A$ in the fundamental of $SU^*(4)$, so $A=1,2,3,4$. The spinor is chiral, as opposed to anti-chiral, because it is in the fundamental, not anti-fundamental, representation of $SU^*(4)$. The two $SU(2)$-factors of the little group belong to the chiral and anti-chiral spinors, respectively, so $\lambda_{i\mathsf{a}}^A$ carries a little group index $\mathsf{a}=1,2$ of the chiral $SU(2)$ factor.  
A candidate for the lightlike momentum can now be formed as the little group invariant bi-spinor
\be
  \text{$D=6$:}\qquad 
p_i^{AB}=\lambda_{i}^{A\mathsf{a}}\lambda_{i\mathsf{a}}^B \,.
\label{D=6LDef}
\ee
This works to give the right number of degrees of freedom, namely 5, for a massless momentum in 6d: the spinor $\lambda_{i}^{A\mathsf{a}}$ has $4 \times 2$ degrees of freedom, but we have to mod out by the little group $SU(2)$-factor, giving $4\times2-3=5$.

The results for $D=3,4,6$ can be summarized as follows:
\be
  \begin{array}{|c|c|c|c|}
  \hline
   & \text{Spin}(1,D-1) & \text{little group} & p^2=0\\ 
   \hline
   D=3  & SL(2,\mathbb{R}) & \mathbb{Z}_2 &  
    p_i^{ab}=\lambda_i^a\lambda_i^b \\ 
    \hline
   D=4 & SL(2,\mathbb{C}) & SO(2) = U(1) & 
   p_i^{\dot{a}a}=-\lambda_{i}^{a}\tilde{\lambda}_{i}^{\dot{a}}   \\ 
   \hline
   D=6 & SU^*(4) & SO(4) = SU(2) \times SU(2) &  
   p_i^{AB}=\lambda_{i}^{A\mathsf{a} }\lambda_{i\mathsf{a}}^B\\
\hline
  \end{array}
\ee

\vspace{1mm}
How about general $D$ dimensions? The strategy is to introduce a bosonic spinor $\lambda_{i\mathsf{a}}^{A}$ (where $\mathsf{a}$ transforms under the little group, or a subgroup as in the 6d example) and use it (and possibly its conjugate spinor) to form a (real) lightlike vector as a little group invariant bi-spinor, e.g.~$\lambda_{i\mathsf{a}}^{A} \tilde\lambda_{i}^{\mathsf{a}B}$. However, for this to encode a null momentum, the number of real degrees of freedom of the bi-spinor, modulo the number of little group generators, has to match that of a lightlike vector:
\eq
{\rm DOF}\big[\lambda_{i\mathsf{a}}^{A} \tilde\lambda_{i}^{\mathsf{a}B}\big]-\#(\text{little group generators)}~=~D-1\,.
\label{OnShellConstraint}
\eqe
 This is a non-trivial constraint because the bi-spinor typically has more than $D-1$ degrees of freedom. One has to find a minimal spinor representation with maximal little group redundancy; this was particularly clear in the 6d example above. Indeed, we know solutions to these constraints only for $D=3,4,6$.\footnote{$D=3,4,6$ are precisely the dimensions in which twistor constructions that describe conformal symmetry are known; see Section II.C.5 of Siegel's ``Fields"~\cite{Fields}.} 
\exercise{}{What is the smallest possible number of degrees of freedom for a little group invariant bispinor in $D=10$? ~
[Hint: In $D=10$, the minimum spinor representation is a Majorana-Weyl spinor; it has 16 real components.]}
It is possible to reduce the number of independent spinor degrees of freedom further by imposing the equations of motion, i.e.~the zero-mass Dirac equation.
Now you may be puzzled, because back in Section \ref{s:sh} we set up the $D=4$ spinor helicity formalism by requiring at the starting point that the spinors $|i\>$ and $|i]$ satisfied the Dirac/Weyl equation. For $D=3,4,6$, this approach is equivalent: the Lorentz contraction of the $(D\!-\!1)$-component bi-spinor with one of its spinors is zero, so the momentum space form of the massless Dirac equation is automatic.  

For $D\ne 3,4,6$, setting up a spinor helicity formalism is possible but the resulting spinors are constrained in the sense that the Dirac equation is imposed as a non-trivial condition \cite{Boels:2009bv,CaronHuot:2010rj,Boels:2012ie}.  
Constrained spinors are more difficult to work with, especially if one wants to construct symmetry generators in order to study  symmetries of the $D$-dimensional amplitudes. For this reason, we focus on $D=3,6$ in this section: we describe $D=6$ briefly, then offer more details about the interesting structure  of $D=3$ amplitudes.

%%%%%%%%%%%%%%%%%%%%%%%%%%%%%%%%%%%%%%%%%%%%%%%%%%%
\subsection{Scattering amplitudes in $D=6$}
\label{s:6d}
%%%%%%%%%%%%%%%%%%%%%%%%%%%%%%%%%%%%%%%%%%%%%%%%%%%
Oh, who cares about 6d scattering amplitudes!! Don't we live in 4d? Well, the 6d massless condition 
\be
-p_0^2 +p_1^2+p_2^2+p_3^2+p_4^2+p_5^3 = 0
\label{6dmassless}
\ee
 can be viewed from 4d spacetime as the on-shell condition for a massive 4d momentum vector: take $p_4^2 +p_5^2 = m^2$ (or $=m \tilde{m}$ if you are willing to accept complex masses). Then $p_\text{4d}^2 = -m^2$ follows from \reef{6dmassless} with $p_\text{4d}$ denoting the first four components of the 6d momentum. This makes the 6d formalism very useful for studies of 4d amplitudes with massive particles. For such uses, see for example~\cite{6DSYM1,Craig:2011ws,Elvang:2011fx} as well as \cite{Scott} for explicit applications to Higgs production processes.

The 6d spinor helicity formalism was first developed by Cheung and O'Connell~\cite{CheungO'Connell} and its supersymmetrization was carried out in~\cite{Dennen:2009vk}. It has been applied to tree- and loop-level scattering amplitudes in maximal super Yang-Mills theory in 6d \cite{6DSYM1,6DSYM2,6DSYMDC} and also used in other 6d theories~\cite{6DSC, M5,M52,M5a}.

In 4d, we used $(\sigma^\m)_{a\db}$ and $(\bar{\sigma}^\m)^{\da b}$ to define the $2 \times 2$ matrices $p_{a\db} = p_\mu (\sigma^\m)_{a\db}$ and $p^{\da b} =p_\m (\bar{\sigma}^\m)^{\da b}$. 
Similarly, the 6d Lorentz group $SO(1,5) \sim SU^*(4)$ has antisymmetric 4$\times$4 matrices $(\sigma^\mu)_{AB}$ and $(\tilde \sigma^{\mu})^{AB}$, $A,B = 1,2,3,4$, that allow us to define
\begin{equation}
  p_{AB} =p_{\mu}\,(\sigma^\mu)_{AB}\,, \hspace{1cm} 
  p^{AB} =p_{\mu}\,(\tilde \sigma^{\mu})^{AB}\,.
\end{equation}
The explicit form of the $(\sigma,\tilde\sigma)$ matrices as well as their relation to the 6d $8\times8$ $\gamma$-matrices can be found in  Appendix A of \cite{CheungO'Connell}.

The pseudo-real property of $SU^*(4)$, implies that $p^{AB}$ and $p_{AB}$ are related as
\eq
p^{AB}=\frac{1}{2}\epsilon^{ABCD}p_{CD}\,.
\eqe 
In this notation, the $SO(1,5)$ invariant product $p^\mu p_\mu$ can be written as the manifestly $SU^*(4)$-invariant contraction
\eq
p^\mu p_\mu=-\frac{1}{4}p^{AB}p_{AB}=-\frac{1}{8}\epsilon_{ABCD}p^{AB}p^{CD}\,.
\eqe

Now, in momentum space, the 6d Dirac equation for massless spinors is 
\begin{equation}
	p_{AB} \lambda^{B\mathsf{a}}_i = 0\,, \hskip 2 cm 
	p^{AB}\, \tilde\lambda_{iB\dot{\mathsf{a}}} = 0\,,
	\label{Dirac6d}
\end{equation}
where $\lambda^{B\mathsf{a}}$ and $\tilde\lambda_{B\dot{\mathsf{a}}}$ are chiral and anti-chiral spinors, and $\mathsf{a}=1,2$ and $\dot{\mathsf{a}}=1,2$ are fundamental indices  of the two $SU(2)$'s of the little group $SO(4)=SU(2) \times SU(2)$. The two pairs of Weyl spinors
\be
 \lambda_i^{A\mathsf{a}}=\langle i^{\mathsf{a}} |^A=\,^A|i^{\mathsf{a}}\rangle\qquad\text{and}\qquad
  \tilde\lambda_{iB\dot{\mathsf{a}}}=[ i_{\dot{\mathsf{a}}}|_B=\,_B| i_{\dot{\mathsf{a}}} ]
 \label{6dbra-kets}
 \ee
are the building blocks of the 6d spinor-helicity formalism. There is no distinction between bras and kets because there is no raising or lowering of the $SU^*(4)$ indices. 

The little group indices can be raised/lowered using the $SU(2)$ Levi-Civita symbol as 
$\lambda_{\mathsf{a}}
=\e_{\mathsf{a}\mathsf{b}}\lambda^{\mathsf{b}}$ and
$\tilde\lambda^{\dot{\mathsf{a}}}
=\e^{\dot{\mathsf{a}}\dot{\mathsf{b}}}\tilde\lambda_{\dot{\mathsf{b}}}$. This allows us to form little group invariants, as in the bi-spinor construction \reef{D=6LDef}. 
Indeed,  the massless momentum is given as 
\begin{eqnarray} 
 p_i^{AB}=\lambda_i^{A \mathsf{a}}\lambda_{i\mathsf{a}}^{B} \,, 
 \hspace{1cm} 
  p_{iAB}=\tilde\lambda_{iA\dot{\mathsf{a}}}\tilde\lambda_{iB}\,^{\dot{\mathsf{a}}}\,,
  \label{DefP6}
\end{eqnarray}
Due to the antisymmetric contraction of the $SU(2)$ indices, the bi-spinors in \reef{DefP6} are automatically antisymmetric in the $SU^*(4)$ indices $A$ and $B$. By \reef{DefP6}, the $4 \times 4$ matrix $p_{i}^{AB}$ has rank 2, so $p_i^2\sim \epsilon_{ABCD}\,p_i^{AB}p_i^{CD}$ is  zero. Hence the massless on-shell condition $p_i^2=0$ is satisfied. Thus this realizes the construction \reef{D=6LDef}.

Reverting the momentum in  \reef{DefP6} from matrix form to vector form, we have
\eq
p_i^\mu=-\frac{1}{4}\langle i^a|\sigma^\mu |i_{a}\rangle=-\frac{1}{4}[ i_{\dot{\mathsf{a}}}|\tilde{\sigma}^\mu |i^{\dot{\mathsf{a}}}]\,.
\eqe
These expressions are the 6d versions of the 4d relation 
$k^\m = \tfrac{1}{2}\<k|\gamma^\m |k]$ that you derived in Exercise \ref{ex:kvec}. 

The Dirac equation \reef{Dirac6d} implies that 
$\lambda_i^{A \mathsf{a}}\,\tilde\lambda_{iA\dot{\mathsf{a}}}=0$, so the chiral and anti-chiral spinors are related. Construction of  symmetry generators using these variables must take these constraints  into account. However, if only chiral spinors are needed, we can still work with unconstrained variables.

To get a better feeling for the 6d $4\times2$ spinors --- and to facilitate reduction to 4d --- consider the embedding of our  good old 4d spinors in the new 6d spinors. Choosing $\mu=0,1,2,3$ to be the 4d subspace and setting $p_4=p_5=0$, the 4d spinors appear in the 6d ones as
\eq
\lambda^A_{i\mathsf{a}}=\left(\begin{array}{cc}
			0 & \langle i|_{\dot{a}}\\ 
			\;[ i|^{a} & 0
		\end{array}\right)\,, \hskip 1.5 cm 
	\tilde{\lambda}_{iA\dot{\mathsf{a}}}=\left(\begin{array}{cc}
		0  & |i\rangle^{\dot{a}}\\ 
		-| i ]_{a} & 0
		\end{array}\right)\,.
		\label{6D4D}
		\eqe
Thus the constraint $\lambda_i^{A \mathsf{a}}\tilde\lambda_{iA\dot{\mathsf{a}}}=0$ becomes nothing but the familiar $\langle i\, i\rangle=[i\,i]=0$. 

In 6d  massless kinematics, the basic Lorentz invariant spinor products  are:
\begin{itemize}
  \item $\langle i^{\mathsf{a}} | j_{\dot{\mathsf{b}}} ] 
  \,\equiv\, 
  \lambda_i^{A \mathsf{a}} \tilde{\lambda}_{j A \dot{\mathsf{b}}} \,=\, 
 [ j_{\dot{\mathsf{b}}} | i^{\mathsf{a}}\rangle\,,$
  \item $\langle i^\mathsf{a} j^\mathsf{b} k^\mathsf{c} l^\mathsf{d} \rangle
  \,\equiv\, 
  \epsilon_{ABCD} \lambda_i^{A\mathsf{a}}\lambda_j^{B\mathsf{b}} \lambda_k^{C\mathsf{c}}
   \lambda_l^{D\mathsf{d}} \,,$
  \item 
  $[ i_{\dot{\mathsf{a}}} j_{\dot{\mathsf{b}}} k_{\dot{\mathsf{c}}} l_{\dot{\mathsf{d}}} ]
  \,\equiv\,
 \epsilon^{ABCD} \tilde{\lambda}_{iA\dot{\mathsf{a}}}\tilde{\lambda}_{jB\dot{\mathsf{b}}}\tilde{\lambda}_{kC\dot{\mathsf{c}}}
   \tilde{\lambda}_{lD\dot{\mathsf{d}}}\;.$
\end{itemize}
In particular, the Mandalstam variable $s_{ij} = -(p_i+p_j)^2$ is 
\eq
s_{ij}
=-\frac{1}{2}\epsilon^{\mathsf{a}\mathsf{b}}\epsilon^{\dot{\mathsf{a}}\dot{\mathsf{b}}}\langle i_\mathsf{a}|j_{\dot{\mathsf{a}}}]\langle i_\mathsf{b}|j_{\dot{\mathsf{b}}}]
=-\det\langle i_\mathsf{a}|j_{\dot{\mathsf{a}}}]\,.
\label{sijDef}
\eqe

We have outlined the 6d spinor helicity formalism, so now it is time to apply it to amplitudes. 
Let us begin with 3-point amplitudes; this involves {\bf \em special 3-particle kinematics} because all $s_{ij}$ vanish. In 4d, we got around this by working with complex kinematics such that $\langle ji \rangle \neq ([ij])^*$ and that allowed us to choose either all the angle- or the square-brackets to be non-vanishing, but not both. 
In 6d 3-particle kinematics, the only Lorentz invariants available are the brackets $\langle i_{\mathsf{a}} | j_{\dot{\mathsf{a}}} ]$. But since $0=s_{ij}=-\det\langle i_\mathsf{a}|j_{\dot{\mathsf{a}}}]$, the $2\times2$ matrix $\langle i_{\mathsf{a}} |j_{\dot{\mathsf{a}}} ]$ must be rank 1. We have encountered $2\times2$ matrices of rank 1 before, namely the 4d massless $p_{a\da}$, and by now it should be a simple reflex to introduce two 2-component spinors, $u_{ia}$ and $\tilde{u}_{j\dot{a}}$ such that $\langle i_a|j_{\dot{a}}]=u_{ia}\tilde{u}_{j\dot{a}}$ \cite{CheungO'Connell}. So the 3-point amplitudes in 6d are written in terms of  
these `auxiliary' 2-component spinors.

Just as in 4d, the 6d {\bf \em 3-point amplitudes} are highly constrained by little group and Lorentz invariance. For example, one finds that the 3-vector amplitudes only come in two types, one is generated by the $AA\partial A$ vertex of the Yang-Mills action while the other is generated by the operator $F_\mu\,^\nu F_\nu\,^\rho F_\rho\,^\mu$~\cite{CheungO'Connell}. A wide class of possible 3-point interactions was categorized in~\cite{M5}. In particular, for 6d self-dual antisymmetric tensors --- which are part of the (2,0) supermultiplet that describes the degrees of freedom of M5-membranes in M-theory --- one can demonstrate \cite{M5} that a 3-point amplitude cannot be both Lorentz invariant and carry the correct little group indices to describe scattering of 3 self-dual tensors; so it does not exist.

The 6d {\bf \em  4-point Yang-Mills amplitude} is given by:
\eq
A_4(1,2,3,4)=\frac{\langle 1^\mathsf{a}2^{\mathsf{b}}3^{\mathsf{c}}4^{\mathsf{d}}\rangle[1_{\dot{\mathsf{a}}}2_{\dot{\mathsf{b}}}3_{\dot{\mathsf{c}}}4_{\dot{\mathsf{d}}}]}{s\,u}\,.
\label{YM4pt}
\eqe
The 6d gluons are not labelled by the 4d concept of helicity: instead, a 6d massless spin-1 particle has 4 physical states labeled by the little group indices 
${}^\mathsf{a}{}_{\dot{\mathsf{a}}}$.
\exercise{}{Use the map in \reef{6D4D} to reduce the 6d amplitude \reef{YM4pt} to 4d. You should find the usual suspect, the MHV gluon amplitude. But that is not all: identify the other possibilities and describe their origin.}
In maximal SYM in 6d, the 4-point superamplitude takes the simple form
\eq
\mathcal{A}_{4}(1,2,3,4)\,=\,
\delta^6\big(P\big)\,
\delta^{(4)}\big(Q\big)\,\delta^{(4)}\big(\tilde{Q}\big)\,\frac{1}{y_{13}^2y_{24}^2}\,.
\label{6d4pt}
\eqe
We have used dual space to write $s=-y_{13}^2$ and $u=-y_{24}^2$.
The supermomentum delta functions are defined in~\cite{Dennen:2009vk}. If we write the $n$-point superamplitude as $\mathcal{A}_n=\delta^6\big(P\big)\,
\delta^{(4)}\big(Q\big)\,\delta^{(4)}\big(\tilde{Q}\big)\,f_n$, we note from \reef{6d4pt} that 
$I[f_4]=y^2_1y^2_2y_3^2y_4^2\,f_4$
  under dual conformal inversion \reef{InvertRule}, i.e.~$f_4$ inverts in exactly the same way as the 4d 4-point superamplitude of $\mathcal{N}=4$ SYM. Using a 6d version super-BCFW recursion, it was proven \cite{6DSYMDC} for all $n$ that
\eq
I[f_n]=\left[\prod_{i=1}^ny^2_i\right]f_n\,.
\eqe
In 4d, it was essential for dual superconformal symmetry of planar superamplitudes in $\cn=4$ SYM that the inversion weights of the bosonic and fermonic delta functions cancelled, as shown in \reef{dualinvdeltas}. This, however, does not happen in 6d maximal SYM: $\d^6(P)$ inverts with weight $6$, while the bosonic delta function has weight $-(4+4)/2$. Therefore, the planar superamplitudes of 6d maximal SYM do not have uniform inversion weight. Nonetheless, as is often the case with scattering amplitudes, even if a symmetry is not exact, it is still useful if it is broken in a predetermined fashion, as is the case here.  Remarkably, using generalized unitarity methods it has been shown \cite{6DSYMDC} that the planar $L$-loop integrands of the 6d maximal SYM theory have the same dual conformal inversion weight as in 4d. (A similar result was found for 10d SYM \cite{CaronHuot:2010rj}.) While, the origin of this form of dual superconformal symmetry is not clear (and the 6d and 10d SYM theories are not (super)conformal), it has non-trivial implications in 4d for the structure of (super)amplitudes on the Coulomb branch of $\mathcal{N}=4$ SYM~\cite{Alday:2009zm,Craig:2011ws}.

%%%%%%%%%%%%%%%%%%%%%%%%%%%%%%%%%%%%%%%%%%%%%%%%%%%
\subsection{Scattering amplitudes in $D=3$}
\label{s:3D}
%%%%%%%%%%%%%%%%%%%%%%%%%%%%%%%%%%%%%%%%%%%%%%%%%%%
Scattering amplitudes in $D=3$ turn out to have very interesting properties. After introducing the nessacery kinematic tools and basic examples of amplitudes, we focus on scattering processes in the 3d $\cn=8$ and $\cn=6$ superconformal theories called BLG and ABJM.

%%%%%%%%%%%%%%%%%%%%%%%%%%%%%%%%%%%%%%%%%%%%%%%%%%%
\subsubsection{$D=3$ kinematics}
\label{s:3dkin}
%%%%%%%%%%%%%%%%%%%%%%%%%%%%%%%%%%%%%%%%%%%%%%%%%%%
We construct 3d kinematics by reduction from 4d using that the 4d massless condition, $-p_0^2+p_1^2+p_2^2+p_3^2=0$, is equivalent to a 3d massive constraint. It is convenient to identity the $p_2$-component with the 3d mass as $p_2^2=m^2$ so that the 3d momentum $p^\mu$ with $\mu=0,1,3$  satisfies $p^\mu p_\mu = - m^2$.

Recall that in 4d, the momentum can be given as
\be
\text{$D=4$:}\qquad   p_{a\dot{b}}   = 
    \left(
    \begin{array}{cc}
      -p^0 + p^3 & p^1 - i p^2 \\
      p^1 + i p^2 & - p^0 - p^3 \\
    \end{array}
  \right)\,.
\ee
We restrict this to 3d by removing  $p_2$ and writing
\be
\text{$D=3$:}\qquad   p_{ab} 
  = 
    \left(
    \begin{array}{cc}
      -p^0 + p^3 & p^1 \\
      p^1 & - p^0 - p^3 \\
    \end{array}
  \right)\,.
\ee
Then $\det p_{ab} = -(-p_0^2+p_1^2+p_3^2) =  m^2$.

The $2\times2$ matrix $p_{ab}$ is symmetric. If the 3d momentum $p^\mu$, $\mu=0,1,3$, is real, $p_{ab}$ is also real.\footnote{This contrasts the 4d case, where $p_{a\db}$ is complex valued, and it reflects the different  Lorentz groups,  $SO(1,2)=SL(2,\mathbb{R})$ in 3d and  $SL(2,\mathbb{C})$ in 4d.} A generic real symmetric $2\times2$ matrix can be written as~\cite{TristanBeisert}\footnote{We could also have written 
$p_{ab}=\lambda_a\lambda_b+\mu_a\mu_b$, but
 this is equivalent to \reef{3Dmassive} by a linear redefinition.} 
\eq
\text{$m\neq0$:}
\qquad p_{ab}=\lambda_a\bar\lambda_b+\lambda_b\bar\lambda_a\,,
\label{3Dmassive}
\eqe
where $\bar{\lambda}_a=(\lambda_a)^*$ when $p^\mu$ is real. 
\exercise{}{If $p^\mu$ is complex, we take $\lambda_a$ and $\bar{\lambda}_a$ to be independent. For each case, $p^\mu$ real or complex, count the number of degrees of freedom on each side of \reef{3Dmassive}.}
By direct calculation of the determinant of \reef{3Dmassive}, we find that 
$m^2 = \det p_{ab}=-\langle\lambda\bar\lambda\rangle^2$, where $\langle \lambda\bar{\lambda}\rangle=\lambda^a\bar{\lambda}_a$ and spinor indices are raised and lowered with the 2-index Levi-Civita of the $SL(2,\mathbb{R})$ Lorentz group. 
For 3d massless kinematics, $m=0$, we must therefore have $\langle \lambda\bar{\lambda}\rangle=0$, implying that $\bar{\lambda} \propto \lambda$.
Thus, we can write
\be
  \text{$m=0$:}
  \qquad 
  p_{ab}\,=\,\lambda_a\lambda_b 
  \,=\,
  \<p|_a\, \<p|_b
  \,,
  \label{massless3dp}
\ee
where  $\lambda = \<p|$ was rescaled such that the prefactor is just 1. Note that 
$\<p|$ must be either purely real or purely imaginary for $p_{ab}$ to be real.

It follows from \reef{massless3dp} that in 
\emph{3d massless kinematics, all Lorentz invariants are built out of one kind of angle brackets, namely $\langle ij \rangle=\lambda_i^a{\lambda_j}_a$.}
For example, since $2 p_i . p_j = - \<ij\>^2$, the Mandelstams $s_{ij}$  are 
\eq
  \text{$D=3$:}\qquad 
  s_{ij}= - (p_i+p_j)^2 = \langle ij\rangle^2\,.
\eqe
Momentum conservation $\sum_{i=1}^n p_i^\mu = 0$ can be written
\be
     \text{$D=3$:}\qquad  
     \sum_{i=1}^n |i\> \<i| = 0\,.
\ee

Our 3d kinematics is ready, so let us see some amplitudes. As it is our style, we start with {\bf \em 3-particle amplitudes}. These is particularly easy in 3d, because 3-particle kinematics requires all $s_{ij} =  \<ij\>^2 = 0$ and hence there are no Lorentz invariants available for a massless 3-point amplitude. {\em Thus for massless kinematics, there are no 3-point on-shell  amplitudes in 3d.}
 
 The {\bf \em little group} for massless kinematics in 3d is the discrete group $\mathbb{Z}_2$; it acts on the spinor variables as $|i\> \rightarrow-|i\>$. The homogeneous scaling of the scattering amplitudes distinguishes only two types of particles in 3d: {\em scalar particles} scale with $+1$ and {\em fermions} scale with $-1$.
And spin-1 vector particles? A massless vector in $D$-dimensions has $D-2$ degrees of freedom, so in $D=3$ this is just 1, the same as a scalar.   

Tree-level scattering amplitudes of 3d super Yang-Mills theory can be obtained directly from 4d ones using dimensional reduction. 
For example, the  
dimensional reduction of the 4-point gluon amplitude $A_4[1^- 2^+ 3^- 4^+]$ of 4d Yang-Mills theory gives
\eq
A_4[1^- 2^+ 3^- 4^+]
~=~
\frac{\langle13\rangle^4}{\langle12\rangle\langle23\rangle\langle34\rangle\langle41\rangle}
~~\xrightarrow{\text{4d} \,\to\, \text{3d}}~~
-\frac{\langle13\rangle^4}{\langle12\rangle^2\langle23\rangle^2}\,.
\label{A43dYM}
\eqe
We have used 3d momentum conservation 
$\langle 34\rangle\langle 41\rangle=-\langle 32\rangle\langle 21\rangle$ to simplify the result.
The two helicity states of the 4d gluon become 2 degrees of freedom in 3d that we can organize  as a 3d ``gauge boson" and a scalar.

\subsubsection{3d SYM and Chern-Simons theory}
\label{s:3dYMCS}
The 3-dimensional Yang-Mills action 
\be
  \mathcal{L}_\text{YM} =\frac{1}{g^2}\int d^3x\,\Tr F_{\mu\nu} F^{\mu\nu}
\ee  
has a coupling $g^2$ of mass dimension $(\text{mass})^{1}$. We are particularly interested in theories with extra symmetry (after all, we keep getting milage out of $\cn=4$ SYM), but a  superconformal theory needs dimensionless couplings. 

In 3d, the gauge field can be introduced with a dimensionless coupling via the {\bf \em Chern-Simons Lagrangian} 
\eq
\mathcal{L}_\text{CS}
~=~
\frac{\kappa}{4\pi}\,\epsilon^{\mu\nu\rho}\,
\Tr\bigg(A_\mu\partial_\nu A_{\rho}+\frac{2i}{3}A_{\mu}A_{\nu}A_{\rho}\bigg)\,.
\label{lagCS}
\eqe
The coupling $\kappa$ is an integer and is called the {\bf \em Chern-Simons level}. 
 
The equation of motion derived from varying $\mathcal{L}_\text{CS}$ with respect to the gauge field is
\eq
\partial_{[\mu} A_{\nu]}+i[A_{\mu}, A_{\nu}]=F_{\mu\nu}=0\,.
\eqe 
The solution to this equation is simply $A_{\mu}=g\partial_\mu g^{-1}$, where $g$ is an arbitrary element in the gauge group. This means that the gauge field is  pure gauge, or a flat connection. For us, the relevant implication is that the Chern-Simons gauge field does not carry any physical degrees of freedom, since one can always choose a gauge such that $A_{\mu}=0$. This is an important difference between a gauge field whose dynamics is governed by $\mathcal{L}_\text{CS}$ versus the usual Yang-Mills Lagrangian $\mathcal{L}_\text{YM}$: the gauge boson scattering amplitude of 3d Yang-Mills theory are non-trivial, but for a theory with just a Chern-Simons term the scattering amplitudes are trivially zero because there are no physical states to scatter. 

There can be non-trivial scattering amplitudes for Chern-Simons theory provided matter fields are introduced. 
The  {\bf \em Chern-Simons matter Lagrangian} is typically written 
\eq
\mathcal{L}=\mathcal{L}_\text{CS}+\mathcal{L}_{\phi\psi}\, ,
\label{CSGen}
\eqe 
where the  matter Lagrangian $\mathcal{L}_{\phi\psi}$ encodes the interactions of the scalar(s) $\phi$ and fermion(s) $\psi$  with the gauge field as well as their mutual interactions. In 3d, the (complex) scalar- and fermion-interactions with dimensionless couplings are of the form $\phi^3 \bar\phi^3$ and $\bar\psi \psi \bar\phi \phi$. Thus for superconformal theories, $\mathcal{L}_{\phi\psi}$  takes the form 
\eq
\mathcal{L}_{\phi\psi}=-D^\mu \bar\phi D_\mu{\phi}+i\bar{\psi}\displaystyle{\not}D\psi+V_{\psi\bar{\psi}\phi\bar{\phi}}+V_{\phi^3\bar{\phi}^3}\,,
\eqe
where $V_{\psi\bar{\psi}\phi\bar{\phi}}$ and $V_{\phi^3\bar{\phi}^3}$ are quartic and sextic interaction terms. The explicit form of these terms depends on the  theory; we will show you two examples, namely the $\cn=8$ and $\cn=6$ superconformal 3d theories (Sections \ref{s:BLG} and \ref{s:ABJM}). But let us first explore the properties of  amplitudes in 3d a little further.

%%%%%%%%%%%%%%%%%%%%%%%%%%%%%%%%%%%%%%%%%%%%%%%%%%%
\subsubsection{Special kinematics and poles in amplitudes}
\label{s:spk3dfac}
%%%%%%%%%%%%%%%%%%%%%%%%%%%%%%%%%%%%%%%%%%%%%%%%%%%

The are 3-particle interaction terms in the Lagrangians discussed in Section \ref{s:3dYMCS}, but we have learned in Section \ref{s:3dkin} that all on-shell 3-point amplitudes vanish in 3d. Nonetheless, the 3-particle vertices still make their presence felt by hiding in special kinematic limits of higher-point amplitudes.  
As an example of this, consider the limit 
$s_{12}=\langle 12\rangle^2\rightarrow0$ of a 4-point amplitude. In this limit, $|1\>$ becomes proportional to $|2\>$, so $|1\>=\alpha|2\>$ for some $\alpha$. Further, we must have $(1+\alpha^2)s_{23}=0$, since 
\eq
0=p_4^2=(p_1+p_2+p_3)^2~~~\xrightarrow{s_{12}\to0}~~~
0=s_{13}+s_{23}=(1+\alpha^2)s_{23}\,.
\label{3ptConst}
\eqe 
There are two types of solutions to this constraint. For generic $\alpha$, $s_{23}$ must be zero and one can conclude that all Lorentz invariants vanishes, which is in line with our previous discussion that there are no Lorentz invariants for on-shell 3-point kinematics. However, the constraint 
$(1+\alpha^2)s_{23}=0$ also admits a solution that allows non-trivial Lorentz invariants: $\alpha=\pm i$. For $\alpha=\pm i$, we have $p_1=-p_2$ and similarly $p_3=-p_4$. Thus this corresponds to the kinematic configuration where two particles are traveling in straight lines:
\be
  \includegraphics[scale=0.75]{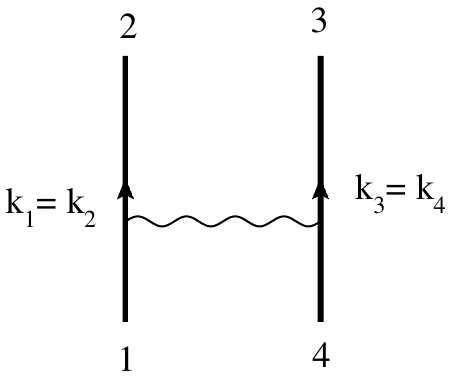}\,
\ee
From momentum conservation, any exchange between the two particle lines must have zero momentum, so when we approach the $s_{12} \to 0$ limit the amplitude should develop a singularity associated with the propagator of an exchanged soft particle. 

To see this in an explicit example, consider the 3d YM gluon amplitude \reef{A43dYM}
\be
A_4[1 2 3 4]
~=~
-\frac{\langle13\rangle^4}{\langle12\rangle^2\langle23\rangle^2}\,.
\label{A43dYM2}
\ee
Taking the limit $|1\> \to i\,|2\>$, the amplitude indeed develops a non-trivial $1/s_{12}$ singularity with a non-vanishing residue:
\eq
A_4[1^- 2^+ 3^- 4^+]\Big|_{|1\>\, \to \,i\,|2\>}=-\frac{s_{23}}{s_{12}}\,.
\label{softA43d}
\eqe
The  $1/s_{12}$ singularity reflects the $1/p^2$ behavior of the gluon propagator.

We have seen that although there are no massless 3-point amplitudes in 3d, the 4-point amplitude still develops a non-trivial ``soft" pole. The origin of this singularity comes from the exchange of a soft particle between two particles going in straight lines.  Note that the exchanged particle has momentum  $p^\mu\rightarrow0$, so it is not strictly going on-shell. This is also reflected in the observation that amplitude \reef{softA43d} does not factorize into two 3-point amplitudes. 

Importantly, the precise behavior of the singularity is dictated by the propagator of the intermediate particle. If the exchanged particle is an ordinary Yang-Mills gluon, then we should observe a $1/p^2$ singularity. That is what happened in the example \reef{softA43d}. However, if it is a fermion or a Chern-Simons gauge boson, one should find a $1/\sqrt{p^2}$ singularity.\footnote{Our reasoning here is valid only for $n=4$. For example, the $n$-point Parke-Taylor amplitude with  $n>4$ has only  $1/\langle i,i+1\rangle$ poles that do not exhibit the $1/p^2$ of the Yang-Mills gluon propagator. A careful inspection of the 3-point gluon vertex reveals that the only non-vanishing term is proportional to $k_2^\mu (\epsilon_1\cdot\epsilon_2)$ in the limit where legs $1,2$ are the two gluons propagating in a straight line. This is dotted into the remaining Feynman diagram which for $n=4$ is simply another 3-point vertex that in this limit  contributes just one term proportional to $k_3^\mu (\epsilon_3\cdot\epsilon_4)$. Hence, on this soft pole, the residue is simply given by the product of the two 3-point vertices. For $n>4$, the remaining Feynman diagram has multiple contributions, and thus the residue of this soft pole contains several terms and it is possible that they might cancel, leaving behind a milder singularity. Indeed this is the case. Thus our discussion of soft-pole structure is only valid for $n=4$.} In the case of a Chern-Simons boson, it  comes from the propagator of the gauge field in the Lagrangian $\mathcal{L}_\text{CS}$ of \reef{lagCS}; in Landau gauge it is
\eq
\langle A^\mu(p)A^\nu(-p)\rangle=\frac{\epsilon^{\mu\rho\nu}p_{\rho}}{p^2}
\,.
\eqe
We are going to use  information about poles in this special kinematic limit to constrain the possible 4-point amplitudes in Section \ref{s:BLG}.

%%%%%%%%%%%%%%%%%%%%%%%%%%%%%%%%%%%%%%%%%%%%%%%%%%%
\subsubsection{$D=3$ superconformal algebra}
\label{s:SCFT3d}
%%%%%%%%%%%%%%%%%%%%%%%%%%%%%%%%%%%%%%%%%%%%%%%%%%%
We stated in Section \ref{s:spDnot4} that the minimal spinors in 3d are 2-component Majorana spinors. They satisfy the Majorana reality condition, and so do the supersymmetry charges. Thus for $\mathcal{N}$-fold supersymmetry in 3d,  we have $\mathcal{N}$ real supercharges and the R-symmetry group is $SO(\mathcal{N})$. Since the 3d theories discussed in this review have $\mathcal{N}=8$ or $\mathcal{N}=6$ supersymmetry, we focus on $\mathcal{N}$=\,even in the following. 
When $\mathcal{N}=2M$, the real supercharges can be grouped into $M=\mathcal{N}/2$ complex spinors  $Q^{aA}$  and their complex conjugate $\widetilde{Q}^{a}_A$. Here $A=1,\ldots,M$ is the index of the reduced $SU(M)$ R-symmetry. 

We introduce $M$ on-shell superspace coordinates $\eta_i^{A}$ for each external leg. The supercharge can now be written as 
\eq
\widetilde{Q}^{a}_{A}=\sum_i |i\>^a\eta_{iA}\,,
\hspace{1cm}
 Q^{aA}=|i\>^a\frac{\partial}{\partial \eta_{iA}}\,.
 \label{3dQQt}
\eqe
You  can quickly see that 
$\{\widetilde{Q}^{a}_{A},Q^{bB}\}=\delta_A{}^B P^{ab}$. 
The generators \reef{3dQQt} are part of a larger symmetry group: the $OSp(\mathcal{N}|4)$ superconformal group. The notation $OSp(\mathcal{N}|4)$ means that the bosonic generators include the $SO(\mathcal{N})$ R-symmetry as well as the $Sp(4)$ conformal symmetry generators. More precisely, the generators are:
\eqa 
\begin{array}{lcl}
&P^{ab}=\sum_{i}|i\>^a\,|i\>^b&\\
\widetilde{Q}^{a}_{A}=\sum_{i}|i\>^a\eta_{iA}&\;&Q^{aA}=\sum_{i}|i\>^a\partial_{\eta_{iA}}\\[2mm]
M_{ab}=\sum_i \<i|_{(a}\partial_{|i\>^{b)}}&\;& 
D=\sum_i\big(\frac{1}{2}|i\>^a\partial_{|i\>^a}+\frac{1}{2}\big)\\[2mm]
R_{AB}=\sum_i \eta_{iA}\eta_{iB}\qquad& 
R_A{}^B=\sum_i\big(\eta_{iA}\partial_{\eta_{iB}}-\frac{1}{2}\delta^A{}_B\big)~~~&
R^{AB}=\sum_i \partial_{\eta_{iA}}\partial_{\eta_{iB}}\\[2mm]
\tilde{S}_{aA}=\sum_{i}\partial_{|i\>^a}\eta_{iA}
&\;&
S_{a}^{A}=\sum_{i}\partial_{|i\>^a}\partial_{\eta_{iA}}\\[2mm]
 &K_{ab}=\sum_{i}\partial_{|i\>^a}\partial_{|i\>^b}\,.&
\end{array}
\label{3DGen}
\eqae
The $SO(\mathcal{N})$ R-symmetry generators are separated into  $U(\mathcal{N}/2)$ generators $R_A{}^B$ and coset generators $R_{AB}$ and $R^{AB}$ of $SO(\mathcal{N})/U(\mathcal{N}/2)$. 

As an important application for these generators, let us explore what kind of constraint the $U(1)$ piece of the $U(\mathcal{N}/2)$ R-symmetry imposes on the superamplitudes in a 3d $\cn=2M$ superconformal theory. The $U(1)$ piece is given by
\eq
R_C{}^C=\sum_i\Big(\eta_{iC}\partial_{\eta_{iC}}-\frac{M}{2}\Big)\,.
\eqe 
The $R_C{}^C$ generator annihilates the superamplitude, $R_C{}^C\,\mathcal{A}_n=0$, if
\eq
 \sum_i \eta_{iC}\partial_{\eta_{iC}}\,\mathcal{A}_n\,=\,n\frac{M}{2}\mathcal{A}_n
 \,.
\label{U1Constraint}
\eqe
The LHS simply counts the Grassmann degree of $\eta_A$'s in $\mathcal{A}_n$. Since one cannot have fractional degree of $\eta$ in $\mathcal{A}_n$, the equation \reef{U1Constraint} can only hold for odd $M$ if $n$=\,even. So we learn that {\em only even-multiplicity scattering amplitudes can be non-vanishing for a superconformal theory with $M$ odd}. 

The {\em same is actually  also true for $M$ even}. This is because superconformal theories generally require the presence of a gauge field whose self-interaction is described by the Chern-Simons action \reef{lagCS}. As discussed previously, the Chern-Simons gauge field does not carry any physical degrees of freedom. It follows from the Lagrangian \reef{CSGen} that any odd-multiplicity Feynman diagram has at least one external leg associated with  gauge field. Since it carries no degrees of freedom, the scattering amplitude vanishes. 

So we learn that in a 3d superconformal theory, the Grassmann degree of the superamplitudes, i.e.~the N$^K$MHV-level, is rigidly tied to the number of external particles, contrary to its freer life in 4d. For example for $\cn=8$, an MHV superamplitude has Grassmann degree 8 and by \reef{U1Constraint} it  exists only for $n=4$ external particles in a 3d superconformal theory; a 6-point superamplitude on the other hand must have Grassmann degree 12, so it has to be NMHV. Thus in a 3d  $\cn=8$ superconformal theory, there is no tower of MHV superamplitudes, no equivalent of the $n$-gluon Parke-Taylor amplitude. 
Similarly, in a 3d  $\cn=6$ superconformal theory, the 4-point superamplitude must have Grassmann degree 6.

%%%%%%%%%%%%%%%%%%%%%%%%%%%%%%%%%%%%%%%%%%%%%%%%%%%
\subsubsection{$\cn=8$ superconformal theory: BLG}
\label{s:BLG}
%%%%%%%%%%%%%%%%%%%%%%%%%%%%%%%%%%%%%%%%%%%%%%%%%%%
A 3d superconformal theory with $\mathcal{N}\!=\!8$-fold supersymmetry has  an on-shell spectrum with 8 scalars $(\phi, \phi^{AB},\bar{\phi})$ and 8 fermions $(\psi^A,\bar{\psi}_A)$. Just as in $\cn=4$ SYM in 4d, it is convenient to encode the degrees of freedom in an  on-shell superfield 
\eq
\Phi=\phi+\eta_A\,\psi^A
-\frac{1}{2}\eta_A\eta_B\,\phi^{AB}
-\frac{1}{3!}\epsilon^{ABCD}\eta_A\eta_B\eta_C\,\bar{\psi}_{D}+
\eta_1\eta_2\eta_3\eta_4\,\bar{\phi}\,.
\label{SF3d}
\eqe
We have $\eta_A \to - \eta_A$
under little group transformations, so the superfield $\Phi$ is inert. This means that the superamplitudes are also invariant under little group transformations.

Since there are no massless 3-point amplitudes, let us consider the most general 4-point tree superamplitude that enjoys $\mathcal{N}=8$ superconformal symmetry. 
To start with, invariance under $\mathcal{N}=8$ supersymmetry implies that the $n$-point superamplitude takes the form
\eq
  \mathcal{A}_n
  \,=\,\delta^3\big(P\big)\,
  \delta^{(8)}\big(\widetilde{Q}\big)\,\,
  f_n\big(|i\>,\eta_i\big)\,,
\eqe
where 
$\delta^{(8)}\big(\widetilde{Q}\big)
=\prod_{A=1}^4
\big(\frac{1}{2}\widetilde{Q}^{a}_{A}\widetilde{Q}_{aA}\big)$. The function $f_n$ is  constrained further by the superconformal generators  \reef{3DGen}. 

As noted at the end Section \ref{s:SCFT3d}, we know that the $U(1)$ generator  \reef{U1Constraint} requires the 4-point superamplitude to have degree 8. Since the supermomentum delta function is already degree 8 in the $\eta_i$'s, we infer that $f_4$ can only depend on the bosonic variables $|i\>$. 

Next, annihilation of the superamplitude by the dilatation operator $D$ in \reef{3DGen}   implies
\be
D\mathcal{A}_n=0
~~~~~\rightarrow ~~~~~
\sum_i\Big(\tfrac{1}{2}|i\>^a\partial_{|i\>^a}\Big)\mathcal{A}_n=-\frac{n}{2}\mathcal{A}_n\,.
\label{3dDonAn}
\ee
As in 4d (see Exercise \ref{ex:dilatation}), the operator $\sum_i\big(\frac{1}{2}|i\>^a\partial_{|i\>^a}\big)$ counts the mass dimension when acting on a function of spinor brackets. It also acts on the delta functions in $\ca_n$, giving a factor of $-3$ on $\delta^3(P)$ and $4$ on  $\delta^{(8)}(\widetilde{Q})$. Thus, by \reef{3dDonAn}, dilatation invariance requires the mass-dimension of 
$f_4$ to be $-\tfrac{4}{2}-(-3+4) = -3$. 
\exercise{}{Use the example around  equation \reef{diffdelP} to show that 
$\sum_i\big(\frac{1}{2}|i\>^a\partial_{|i\>^a}\big) \,
\delta^3(P) = -3\delta^3(P)$.}
To summarize, from the $U(1)$ R-symmetry and dilatation invariance, we conclude that $f_4$ is a purely bosonic function of mass-dimension $-3$. Finally, taking into account that the superamplitude must be little group invariant,  we can write the
4-point superamplitude \cite{M5a} as
\eq
\mathcal{A}_4\,=\,\delta^3\big(P\big)\,
  \delta^{(8)}\big(\widetilde{Q}\big)\,\,
  \frac{1}{\langle12\rangle\langle23\rangle\langle31\rangle}\,.
\label{BLG4pt}
\eqe
For example, we can project out the 4-scalar  amplitude $A_4(\phi\phi\bar{\phi}\bar{\phi})$ using \reef{SF3d}: the Grassmann delta function produces a factor of $\<34\>^4$, so we get (with the help of momentum conservation)
\be
 A_4(\phi\phi\bar{\phi}\bar{\phi}) 
 = \frac{\<34\>^4}{\<12\>\<23\>\<13\>} 
 = -\frac{\<34\>^3}{\<24\>\<23\>} \,,
 \label{A4N8-3d}
\ee

The astute reader should object:  multiplying the solution \reef{BLG4pt} by an arbitrary function of 
$\frac{\langle 13\rangle \langle 24\rangle}{\langle 14\rangle \langle 23\rangle}$  
still satisfies all previous criteria. This is indeed a valid objection; however, such a function would change the pole structure of component amplitudes, such as \reef{A4N8-3d}, generated by $\ca_4$.
We have imposed in \reef{BLG4pt} that the amplitudes only have $\frac{1}{\sqrt{p^2}}$ poles. Why? Well, since the only scalar-fermion interactions are  of the form $\phi^3 \bar\phi^3$ and $\bar\psi \psi \bar\phi \phi$, poles in 
the tree-level amplitude $A_4(\phi\phi\bar{\phi}\bar{\phi})$ cannot arise from scalar or fermion propagators. Hence the only option is that they come from gauge boson exchanges. Since we are considering a 3d superconformal theory, the gauge boson self-coupling must be dimensionless; this rules out 3d Yang-Mills theory and rules in Chern-Simons gauge theory. Hence all poles in $A_4(\phi\phi\bar{\phi}\bar{\phi})$ must be $\frac{1}{\sqrt{p^2}}$ and this fixes the 4-point tree superamplitude in a $\cn=8$ superconformal 3d theory to be \reef{BLG4pt}.

The result \reef{BLG4pt} for the superamplitude has a very important property: it is antisymmetric under the exchange of any two external particles. This property is inherited by the component amplitude $A_4(\phi\phi\bar{\phi}\bar{\phi})$ in \reef{A4N8-3d},  contradicting with the expected Bose symmetry. We encountered something similar in Section \ref{s:littlegrp} when we wrote down the 3-point gluon amplitudes in 4d from just little group scaling and dimensional analysis. The resolution was to include the antisymmetric structure constants $f^{abc}$ of the Yang-Mills gauge group.

At the superamplitude level, the same issue arises: the physical degrees of freedom are contained in the bosonic  superfield $\Phi$, so 
$\ca_4(\Phi_1\Phi_2\Phi_3\Phi_4)$ should be Bose symmetric under the exchange of any two external legs. But --- as you see from \reef{BLG4pt} --- it is fully antisymmetric. We could avoid this contradiction if the amplitudes are  a color-ordered. However, the presence of the $1/\langle 24\rangle$ pole in for example \reef{A4N8-3d} invalidates this interpretation. Instead, the contradiction can be resolved if we include more than one supermultiplet, giving each one a label $a_i$. Then we can introduce a new 4-index ``coupling constant" $f^{a_1a_2a_3a_4}$ that is completely antisymmetric in all four indices. Using this we write
\eq
\ca_4\big(\Phi^{a_1}_1\Phi^{a_2}_2\Phi^{a_3}_3\Phi^{a_4}_4\big)
\,=\,\delta^3\big(P\big)\,\delta^{(8)}\big(\widetilde{Q}\big)\,\frac{f^{a_1a_2a_3a_4}}{\langle12\rangle\langle23\rangle\langle31\rangle}\,.
\label{BLG4pt2}
\eqe 
Now Bose symmetry is respected. Thus {\em by requiring $\mathcal{N}=8$ superconformal symmetry in 3d, the 4-point superamplitude forces us to  introduce a completely {\bf \em antisymmetric 4-index coupling constant}.} 

This new coupling constant looks similar to the totally antisymmetric 3-index structure constant of Yang-Mills theory $f^{a_1a_2a_3}$. This resemblance is not a coincidence. In the search for a $\mathcal{N}=8$ super Chern-Simons matter theory, Bagger, Lambert, and Gustavsson (BLG)~\cite{BLG1,BLG2} found a Lagrangian whose gauge symmetry is built on a {\bf \em Lie 3-algebra}. This algebra is defined through a triple product
\be
[T^a,T^b,T^c]=f^{abc}{}_d\,T^d\,.
\ee
The gauge indices are raised/lowered with $h^{ab} = \Tr T^a T^b$ and its inverse. The 
structure constants $f^{abcd} =  f^{abc}{}_e h^{ed}$ are totally antisymmetric.  Much like the structure constants of the usual gauge Lie 2-algebra  satisfy the Jacobi identity \reef{csctcuID}, the 3-algebra structure constants are required to satisfy a four-term ``fundamental identity":
\be
  f^{fgd}{}_ef^{abce}-f^{fga}{}_ef^{bcde}
  +f^{fgb}{}_ef^{cdae}-f^{fgc}{}_ef^{dabe}
  \,=\,0\,.
\label{FundamentalId}
\ee
The fields in the {\bf \em BLG theory} \cite{BLG1,BLG2} consist of 8 scalars $X^{I_v}_{a}$ with $I_v=1,\ldots,8$ transforming as a vector of $SO(8)$, 8 real spinors $\Psi^{I_c}_a$ with $I_c=1,\ldots,8$ transforming as a chiral spinor of $SO(8)$, and a Chern-Simons gauge field $A_\mu^{ab}$. The BLG Lagrangian is \cite{Schwarz}
\bea
\nonumber
 \frac{1}{\kappa}\mathcal{L}_\text{BLG}
 &=&\frac{1}{48}\epsilon^{\mu\nu\rho}\Big(\frac{1}{2}f^{abcd}A_{\mu ab}\partial_\nu A_{\rho cd}+\frac{1}{3}f^{cda}\,_gf^{efgb}A_{\mu ab}A_{\nu cd}A_{\rho ef}\Big)-\frac{1}{2}D^\mu X^{I_v}_a D_\mu X^{I_v}_a\\
&&+\frac{i}{2}\bar{\Psi}^{I_c}_a\displaystyle{\not}D\Psi^{I_c}_a
 +i3f^{abcd}\bar{\Psi}_a\Gamma^{I_vJ_v}\Psi_bX^{I_v}_cX^{J_v}_d
  \label{LcsN8}
 \\
 &&
  -12f^{abcd}f_a\,^{efg}(X^{I_v}_bX^{J_v}_cX^{K_v}_d)(X^{I_v}_eX^{J_v}_fX^{K_v}_g)\,.
\nonumber  
\eea
In the Lagrangian construction \cite{BLG1,BLG2}, the need for the antisymmetric 4-index structure constant comes from the requirement that the supersymmetry transformations on the fields close into the correct algebra. Linear combinations of the eight scalars and fermions can be identified as the $(\phi, \phi_{AB},\bar{\phi})$ and $(\psi_A,\bar{\psi}^A)$ components of our superfield \reef{SF3d}. Indeed, the 4-point amplitudes computed from the Lagrangian \reef{LcsN8} match \cite{M5a} the component amplitudes of the 4-point superamplitude \reef{BLG4pt2}.

It is quite non-trivial for an antisymmetric $f^{abcd}$ to satisfy \reef{FundamentalId} and currently the only known example is if $a$ is an index of $SO(4)$ and $f^{abcd}\sim\epsilon^{abcd}$. In search for other examples, there were many attempts to relax the symmetry properties of the 4-index structure constant. However, as we have shown from the on-shell analysis, $\mathcal{N}=8$ superconformal symmetry only allows for a totally antisymmetric structure constant. Indeed all known examples of Lie 3-algebras with $f^{abcd}$ 
 not totally antisymmetric correspond to Chern-Simons matter theories with $\mathcal{N}<8$ supersymmetries.

%%%%%%%%%%%%%%%%%%%%%%%%%%%%%%%%%%%%%%%%%%%%%%%%%%%
\subsubsection{$\cn=6$ superconformal theory:  ABJM}
\label{s:ABJM}
%%%%%%%%%%%%%%%%%%%%%%%%%%%%%%%%%%%%%%%%%%%%%%%%%%%
Let us now consider a 3d superconformal theory with $\mathcal{N}=6$ supersymmetry.  The R-symmetry is $SO(6)=SU(4)$ and the physical degrees of freedom are 4 complex scalars $X_{\mathsf{A}}$ and 4 complex fermions $\psi^{\mathsf{A}a}$ as well as their complex conjugates $\bar{X}^{\mathsf{A}}$ and $\bar{\psi}_{\mathsf{A}a}$. They transform in the fundamental or anti-fundamental of $SU(4)$ and $\mathsf{A}=1,2,3,4$. To arrange these states in  on-shell superspace, 
we introduce three anticommuting variables $\eta_A$ and write
\be
\begin{split}
  \Phi
  ~=&~
  X_4+\eta_A\,\psi^A
  -\frac{1}{2}\epsilon^{ABC}\,\eta_A\eta_B\,X_C
  -\eta_1\eta_2\eta_3\,\psi^4\,,\\
  \bar{\Psi}
  ~=&~\bar{\psi}_4+\eta_A\bar{X}^A
  -\frac{1}{2}\epsilon^{ABC}\,\eta_A\eta_B\,\bar{\psi}_C
  -\eta_1\eta_2\eta_3\,\bar{X}^4\,.
\end{split}
\label{ABJMmap}
\ee
We have split the fields as $X_{\mathsf{A}}\rightarrow(X_4, X_{A})$ and $\psi^{\mathsf{A}}\rightarrow(\psi^4, \psi^A)$, and similarly for $\bar{X}^{\mathsf{A}}$ and $\bar{\psi}_{\mathsf{A}}$. So only an $SU(3)$ subgroup of the $SU(4)$ is manifest in this on-shell superspace formalism.

The on-shell superspace representation \reef{ABJMmap} involves a bosonic superfield $\Phi$ and a fermionic superfield $\bar{\Psi}$. Having two superfields is standard for superamplitudes in theories with less-than-maximal supersymmetry. For example in 4d $\cn<4$ SYM, the spectrum is not CPT self-conjugate and therefore a superfield is needed for each of the CPT conjugate supermultiplets;  details and applications of the formalism can be found in \cite{Elvang:2011fx}. In 3d, the need for two superfields comes from R-symmetry. Just as in 4d $\cn<4$ SYM, where the two superfields contain states that are parity-conjugate with respect to each other, in 3d the two superfields contain states that are conjugate to each other under R-symmetry. 

Since fermions transform with a minus under 3d little group transformations, the superamplitude must by odd under $|i\>\rightarrow-|i\>$ and $\eta_i\rightarrow-\eta_i$ if $i$ is a $\bar{\Psi}$ state. Following the same steps as for the $\mathcal{N}=8$ BLG theory in Section \ref{s:BLG}, we then find that the 4-point superamplitude in a 3d $\cn=6$ superconformal theory is fixed up to a multiplicative constant to be~\cite{Bargheer}
\eq
\mathcal{A}_4\big[\bar{\Psi}_1\Phi_2\bar{\Psi}_3\Phi_4\big]=
\delta^3\big(P\big)\,\delta^{(6)}\big(\widetilde{Q}\big)\,\frac{1}{\langle14\rangle\langle43\rangle}\,.
\label{ABJM4pt}
\eqe
The 4-point superamplitude \reef{ABJM4pt}  precisely encodes the color-ordered 4-point amplitudes of an $\mathcal{N}=6$ Chern-Simons matter theory that was constructed by Aharony, Bergman, Jafferis and Maldacena (ABJM)~\cite{ABJM}. The theory, known as {\bf \em ABJM theory}, contains two gauge fields $A^{\mathsf{a}}\,_{\mathsf{b}}$ and $\hat{A}^{\dot{\mathsf{a}}}\,_{\dot{\mathsf{b}}}$ with gauge group $U(N)\times U(N)$. The matter fields are bi-fundamental, meaning that they transform in the fundamental of one $U(N)$ gauge group and the anti-fundamental of the other $U(N)$. More precisely the index structure of the matter fields are  $(X_{\mathsf{A}})^{\dot{\mathsf{a}}}\,_{\mathsf{a}}$, $(\bar{X}^{\mathsf{A}})^{\mathsf{a}}\,_{\dot{\mathsf{a}}}$, $(\psi^{\mathsf{A}})^{\dot{\mathsf{a}}}\,_{\mathsf{a}}$ and $(\bar{\psi}_{\mathsf{A}})^{\mathsf{a}}\,_{\dot{\mathsf{a}}}$.  The Lagrangian is \cite{ABJML, ABJML2}
\be
\begin{split}
  \mathcal{L}_\text{ABJM}
  &=~
  \frac{k}{2\pi}
  \bigg[\frac{1}{2} \epsilon^{\mu\nu\rho}\,
  \Tr\left(A_\mu\partial_\nu A_{\rho}
    +\frac{2i}{3}A_\mu A_\nu A_\rho-\hat{A}_\mu\partial_\nu \hat{A}_{\rho}-\frac{2i}{3}\hat{A}_\mu \hat{A}_\nu \hat{A}_\rho\right)\\
  &\hspace{1.7cm}
  -(D^\mu X_{\mathsf{A}})^\dagger D_{\mu}X_{\mathsf{A}}
  +i\bar{\psi}_{\mathsf{A}}\displaystyle{\not}D\psi^{\mathsf{A}}
  +\mathcal{L}_4+\mathcal{L}_6\bigg]\,,
\end{split}
\label{ABJML}
\ee
where the covariant derivatives for the bi-fundamental fields are 
\eqa
\nonumber 
  D^\mu X_{\mathsf{A}}
  &\equiv&
  \partial_{\mu}X_{\mathsf{A}}
  +i\hat{A}_\mu X_{\mathsf{A}}
  -i X_{\mathsf{A}}A_\mu\,,\\
  (D^\mu X_{\mathsf{A}})^\dagger
  &\equiv&
  \partial_{\mu}\bar{X}^{\mathsf{A}}
  +iA_\mu \bar{X}^{\mathsf{A}}
  -i\bar{X}^{\mathsf{A}}\hat{A}_\mu \,,
\eqae
with the same definitions for $\psi^{\mathsf{A}}$ and $\bar{\psi}_{\mathsf{A}}$. 

The quartic and sextic interaction terms in \reef{ABJML} are
\bea
  \nonumber
  \mathcal{L}_4
  &=&
  i\Tr\Big(
    \bar{X}^{\mathsf{B}}X_{\mathsf{B}}\bar{\psi}_{\mathsf{A}}\psi^{\mathsf{A}}
    -X_{\mathsf{B}}\bar{X}^{\mathsf{B}}\psi^{\mathsf{A}}\bar{\psi}_{\mathsf{A}}
    +2X_{\mathsf{A}}\bar{X}^{\mathsf{B}}\psi^{\mathsf{A}}\bar{\psi}_{\mathsf{B}}
    -2\bar{X}^{\mathsf{A}}X_{\mathsf{B}}\bar{\psi}_{\mathsf{A}}\psi^{\mathsf{B}}
    \\
    &&
    \hspace{1.3cm}
    -\epsilon_{\mathsf{A}\mathsf{B}\mathsf{C}\mathsf{D}}\bar{X}^{\mathsf{A}}\psi^{\mathsf{B}}\bar{X}^{\mathsf{C}}\psi^{\mathsf{D}} 
    +\epsilon^{\mathsf{A}\mathsf{B}\mathsf{C}\mathsf{D}}X_{\mathsf{A}}\bar{\psi}_{\mathsf{B}}X_{\mathsf{C}}\bar{\psi}_{\mathsf{D}} \Big)\,,
    \\[1mm]
    \nonumber
    \mathcal{L}_6
    &=&
    \frac{1}{3}\Tr\Big( 
    X_{\mathsf{A}}\bar{X}^{\mathsf{A}}X_{\mathsf{B}}\bar{X}^{\mathsf{B}}X_{\mathsf{C}}\bar{X}^{\mathsf{C}}
    +\bar{X}^{\mathsf{A}}X_{\mathsf{A}}\bar{X}^{\mathsf{B}}X_{\mathsf{B}}\bar{X}^{\mathsf{C}}X_{\mathsf{C}}
    +4\bar{X}^{\mathsf{A}}X_{\mathsf{B}}\bar{X}^{\mathsf{C}}X_{\mathsf{A}}\bar{X}^{\mathsf{B}}X_{\mathsf{C}}\\
     && 
     \hspace{1.3cm}
     -6X_{\mathsf{A}}\bar{X}^{\mathsf{B}}X_{\mathsf{B}}\bar{X}^{\mathsf{A}}X_{\mathsf{C}}\bar{X}^{\mathsf{C}}\Big)\,.
\label{QuarticV}
\eea

For theories whose external states are bi-fundamental matter fields, the color structure of the amplitude is given in terms of a product of Kronecker deltas. In particular, 
with $n=2m$ the full color-dressed amplitude is  \cite{Bargheer}
\eq
\sum_{\sigma \in S_{m},~ \bar \sigma \in \bar S_{m-1}}
\hskip-0.5cm \mathcal{A}_n(\bar1,\sigma_1, \bar \sigma_1, \ldots, \bar \sigma_{m-1},\sigma_m) \, 
\delta_{\dot{\mathsf{a}}_{\bar{1}}}^{\dot{\mathsf{a}}_{\sigma_1}}
\cdots
\delta^{\dot{\mathsf{a}}_{\sigma_m}}_{\dot{\mathsf{a}}_{\bar \sigma_{m-1}}}\,
\delta^{\mathsf{a}_{\bar \sigma_1}}_{\mathsf{a}_{\sigma_1}}
\cdots 
\delta^{\mathsf{a}_{\bar{1}}}_{\mathsf{a}_{\sigma_m}}\,,
\label{BiFunOrder}
\eqe
where the sums are 
over all distinct permutations of  $m$ even sites and $m-1$ odd sites.  Each partial amplitude $\mathcal{A}_n$ is multiplied by a product of Kronecker deltas, and this naturally defines an ordering, very similar to Yang-Mills amplitudes. However, since the on-shell degrees of freedom are contained in two distinct supermultiplets, the color-ordered superamplitude is not cyclically invariant, but  invariant up to a sign under cyclic rotation of two sites:
\eq
\mathcal{A}_{n=2m}\big[\bar{\Psi}_1\Phi_2\ldots\Phi_{2m}\big]=(-1)^{m-1}\mathcal{A}_{n=2m}\big[\bar{\Psi}_3\Phi_4\ldots\Phi_{2m}\bar{\Psi}_1\Phi_{2}\big]\,;
\label{ABJMCycle}
\eqe
the  minus signs come from the exchanges of $\bar{\Psi}$'s. 
For the superamplitude \reef{ABJM4pt}, the 2-site cyclic property \reef{ABJMCycle} is ensured by momentum conservation. 

After having seen a Lie 3-algebra appear in the $\cn=8$ superconformal BLG theory in Section \ref{s:BLG}, you may wonder if the above Lagrangian can also be rewritten in terms of a 3-algebra. Indeed it can! 
In fact, we can read off the properties of the 4-index structure constants from the 4-point superamplitude \reef{ABJM4pt}. It is symmetric under the exchange of the legs that correspond to the fermionic supermultiplet $\bar{\Psi}$, while it is antisymmetric under the exchange of the bosonic multiplets $\Phi$. This is opposite from the expected symmetry properties of 
$\mathcal{A}_4(\bar{\Psi}_1\Phi_2\bar{\Psi}_3\Phi_4)$, and therefore one can consider dressing the superamplitude with a 4-index structure constant $f^{a_2a_4\bar{a}_1 \bar{a}_3}$ that is antisymmetric with respect to the exchange of barred or unbarred indices, respectively. The color-dressed superamplitude is then\footnote{You might wonder why this issue did not come up when we stated that the amplitude in \reef{ABJM4pt} matched that derived from the Lagrangian \reef{ABJML}. The reason is that it matched in the context of a color ordered amplitude where the exchange of external lines is not a symmetry. In contrast, here we are considering  a fully color-dressed amplitude. In other words, we are asking what properties should the color factor have such that the amplitude can be considered as a color-dressed amplitude.}
\be
\mathcal{A}_4\big(\bar{\Psi}_1^{\bar{a}_1}\Phi_2^{a_2}\bar{\Psi}_3^{\bar{a}_3}\Phi_4^{a_4}\big)=
\delta^3\big(P\big)\,\delta^{(6)}\big(\widetilde{Q}\big)\,
\frac{f^{a_2a_4\bar{a}_1 \bar{a}_3}}{\langle14\rangle\langle43\rangle}\,.
\label{ABJM4pt2}
\ee
It has been shown \cite{ABJM3} that the Lagrangian \reef{ABJML} is completely equivalent to an alternative one where the matter fields carry the 3-algebra indices indicated in \reef{ABJM4pt2}.

%%%%%%%%%%%%%%%%%%%%%%%%%%%%%%%%%%%%%%%%%%%%%%%%%%%
\subsubsection{BCFW recursion}
%%%%%%%%%%%%%%%%%%%%%%%%%%%%%%%%%%%%%%%%%%%%%%%%%%%
We argued in Section \ref{s:SCFT3d} that only even-point amplitudes are non-vanishing in 3d superconformal theories. This means that the 4-point superamplitudes are the building blocks of higher-point amplitudes in these theories. Conveniently, we found that the 4-point tree-level superamplitudes in $\cn=8$ and $\cn=6$ theories in Sections \ref{s:BLG} and \ref{s:ABJM} are completely determined by the requirements of symmetries and pole structure. Now is time to go to higher-point and of course our favorite tool is BCFW recursion. 

To get started, we have to set up a BCFW recursion relation in 3d. And 3d is different from all other $D>3$ in terms of defining a BCFW deformation. To see this, recall from Section \ref{s:recrels} that we shift two external momenta $i$ and $j$ linearly 
\be
p_i\rightarrow p_i+zq,\;\; p_j\rightarrow p_j-zq\,,
\ee
with a vector $q$ that satisfies
\be
q\cdot p_i=q\cdot p_j=q^2=0\,.
\label{Ddimshift}
\ee
This ensures that the shifted momenta remain on-shell and that invariants $\hat{P}_{ij\dots k}^2$ are linear in $z$, so that each propagator going on-shell corresponds to a unique pole in the $z$-plane.

Unfortunately (or, very interestingly, if that is how you like it), in 3d the only $q$ that satisfies these constraints is $q=0$. The reason is this. A 3d vector $q$ with  $q^2=0$ can be written as a bi-spinor $q= |q\>^a |q\>^b$. The $|q\>$ is a  
2-component spinor so it cannot be linearly independent from the $|i\>$ and $|j\>$ of the two lightlike momenta we are shifting. Hence
\be
|q\>=\alpha|i\>+\beta|j\>\,
\ee
for some numbers $\alpha$ and $\beta$. Solving for $\alpha$ and $\beta$ subject to the constraints $q\cdot p_i=q\cdot p_j=0$ in  \reef{Ddimshift} gives 
$\alpha=\beta=0$ and hence $q=0$. 

So in order to make progress, we need to relax some of the constraints imposed on the shifted momenta. We cannot give up on momentum conservation and on-shellness for the shifted momenta.  Instead, we can either shift 3 or more external momenta or give up on the property that the momenta shift linearly in $z$. The former is similar to the shift associated with CSW (Section \ref{s:csw}) and comes at the price of involving many diagrams and less compact expressions for the superamplitudes. Opting for the solution with fewer diagrams, we  choose the latter and consider the following general 2-line ``deformation" \cite{Gang}
\begin{align}
\begin{pmatrix} |\hat{i}\>  \\[1mm] |\hat{j}\> \end{pmatrix} 
= R(z)  \begin{pmatrix} |i\> \\[1mm]  |j\> \end{pmatrix} \,,
\label{R}
\end{align}
where $R(z)$ is a $2\times2$ matrix that depends on $z$. Since we want the shift to respect momentum conservation, the matrix $R$ must satisfies:
\be
R(z)^T\,R(z)=I \,.
\label{Rcondition}
\ee
Since $R(z)$ an orthogonal matrix, we can parametrize it as 
\eq
R(z)=\left(\begin{array}{cc} \cos\theta & -\sin\theta \\ \sin\theta & \cos\theta \end{array}\right)=
\left(\begin{array}{cc} 
\frac{z+z^{-1}}{2} & -\frac{z-z^{-1}}{2i} \\[1.5mm] 
\frac{z-z^{-1}}{2i} & \frac{z+z^{-1}}{2} 
\end{array}\right)\,.
\label{3DBCFW}
\eqe
If we define the deformation on the fermionic variables $\eta_i$ and $\eta_j$ in the same fashion,  supermomentum conservation is also preserved by the shift:
\eq
\hat{\tilde{q}}_{iA}+\hat{\tilde{q}}_{jA}
\,=\,
\big(\,|\hat{i}\> \,,\, |\hat{j}\>\,\big)
\left(\begin{array}{c}\!\hat{\eta}_{iA}\! \\[1mm] \!\hat{\eta}_{jA}\!\end{array}\right)
\,=\,\big(\,|i\>\,, \,|j\>\,\big)\,
R^T(z)R(z)
\left(\begin{array}{c}\!\eta_{iA}\! \\[1mm] \!\eta_{jA}\!\end{array}\right)
\,=\,\tilde{q}_{iA}+\tilde{q}_{jA}\,.
\eqe 
The deformation matrix \reef{3DBCFW} becomes the identity when $z=1$, so the unshifted kinematics correspond to $z=1$ and not $0$. This leads to the following contour integral representation of the unshifted tree-level amplitude
\eq
\mathcal{A}_n =  \frac{1}{ 2 \pi i } \oint_{z=1} \frac{\hat{\mathcal{A}}_n (z)}{z-1}
\,,
\label{zDeform}
\eqe
where the contour wraps just the pole at $z=1$.
If the deformed superamplitude $\hat{\mathcal{A}}_n (z)$ vanishes as $z\rightarrow \infty$,\footnote{One should also make sure that there are no poles at $z=0$. Exchanging $1/z\leftrightarrow z$ in \reef{3DBCFW} can be compensated by extra sign factors in the kinematics of the shifted legs, so if $\mathcal{A}_n(z)$ vanishes as $z\rightarrow\infty$ for generic kinematics, then it  also vanishes at $z=0$.} one can  perform a contour-deformation and evaluate the amplitude as a sum of the residues at finite $z \ne 0,1$.

Just as in 4d, the poles at finite $z \ne 0,1$ correspond to propagators going on-shell. Let us take a closer look at what the singularities look like. Without loss of generality, we choose $1$ and $n$ as the deformed momenta:
\begin{align}
&\hat p_1^{ab}  =  \frac{1}{2} ( p^{ab}_1 + p^{ab}_{n} ) +  z^2 q^{ab} +  z^{-2} \tilde q^{ab}\,,
\nonumber \\
&\hat p^{ab}_{n}  =  \frac{1}{2} ( p^{ab}_1 + p^{ab}_{ n} ) -   z^2 q^{ab} -  z^{-2} \tilde q^{ab}\,.
\end{align}
Here $q$ and $ \tilde q$ are given by
\begin{align}
q^{ab} = \frac{1}{4} ( |1\> + i  |n\> )^{a} ( |1\> + i  |n\> )^{b} , 
\hspace{1cm} 
\tilde{q}^{ab} = \frac{1}{4} ( |1\> - i  |n\> )^{a} ( |1\> - i  |n\> )^{b} \,.
\end{align}
Defining $P^{ab}_{12\dots i}=p^{ab}_1+p^{ab}_2+\cdots+p^{ab}_i$, the on-shell condition for the shifted propagator $\hat{P}_{12\dots i}^2$ takes the form 
\eq
\hat{P}_{12\dots i}^2\,=\,
\langle\tilde{q}|P_{23\dots i}|\tilde{q}\rangle z^{-2}+\langle q|P_{23 \dots i}|q\rangle z^{2}-(P_{23\dots i}\cdot P_{i+1\dots n-1})
\,=\,0\,,
\label{P12iSQ}
\eqe 
where $(p_i\cdot p_j)=p_i^\mu p_{j\mu}$ and $\langle i|P|j\rangle\equiv \lambda_i^a P_{a}\,^b\lambda_{jb}$. One can explicitly write down the values of $z$ that correspond to the propagator going on-shell  
\be
\big\{ (z_{1,i}^\ast)^2, (z_{2,i}^\ast)^2 \big\} =    \frac{ (P_{2\dots i}\cdot P_{i+1\dots n-1}) \pm \sqrt{ ( P_{2\dots i})^2 (P_{i+1 \dots n-1} )^2 } }{ 2 \langle q|P_{2 \dots i}|q\rangle}   \, \label{zeros}\,.
\ee
\exercise{}{Prove the following useful identity
\eq
\left((z_{1,i}^\ast)^2-1\right)\left((z_{2,i}^\ast)^2-1\right)=\frac{P_{12\dots i}^2}{\langle q|P_{2 \dots i}|q\rangle}\,.
\label{BCFWKeys}
\eqe
}
As the propagator goes on-shell, the amplitude factorizes into two lower-point amplitudes. This allows us to write the sum of residues at $z \ne 1$  as a sum over distinct single propagator diagrams where legs $1$ and $n$ are on opposite sides of the propagator. For Chern-Simons matter theories, we also require that only even multiplicity subamplitudes appear on each side of the propagator. For each propagator, one needs to sum over the four solutions, 
$(z_{1,i}^\ast,\, -z_{1,i}^\ast, \,z_{2,i}^\ast, \,-z_{2,i}^\ast)$ to the on-shell constraint \reef{P12iSQ}. The final result is then \cite{Gang}
\begin{align}
\mathcal{A}_n =
&  \sum_i    \int d^3 \eta_I   \bigg(   
\hat{\mathcal{A}}_L \big( z_{1,f}^*; \,\eta_I \big)
\frac{H(z_{1,f}^{ \ast} , z_{2,f}^{ \ast} ) }{ P_{12\dots i}^2 } 
 \hat{\mathcal{A}}_R \big(z_{1,f}^* ;\, i\eta_I\big)
 +\big(z_{1,f}^* \leftrightarrow z_{2,f}^* \big)  \bigg) \,,\label{fact-limit}
\end{align}
where the function $H(a,b)$ is 
\eq
H(a,b)\equiv\frac{a (b^2 -1)}{a^2 - b^2}
\label{Hfunction}
\eqe
and the Grassmann integral takes care of the intermediate state sum. Did you notice the $i$ in $\hat{\mathcal{A}}_R$? That comes from the analytic continuation of the incoming $\to$ outgoing internal line. In 3d massless kinematics, we only have one type of spinor, namely $|p\>$, so with $p^{ab} = -|p\>^a |p\>^b$ we must have
\be
  |-p\> = i \, |p\>\,.
\ee
Hence we must also have $\eta_{-p} = i\, \eta_{p}$, since --- as you can check --- this ensures that the arguments of the L and R Grassmann delta functions add up to the overall supermomentum $\widetilde{Q}$.
\exercise{}{Show that a contour deformation of \reef{zDeform} gives the representation \reef{fact-limit}. \\{} 
[Hints: The identity in \reef{BCFWKeys} will be useful. Furthermore, since one of the shifted legs, $1$ or $n$,  necessarily corresponds to the fermionic multiplet, we have 
$\mathcal{A}_L (-z) \mathcal{A}_R(-z) = -\mathcal{A}_L (z) \mathcal{A}_R (z)$.]  }

The validity of \reef{fact-limit} relies on whether or not the super-shifted superamplitude vanishes as $z\rightarrow\infty$. It was shown in \cite{Gang} that this criteria is satisfied for ABJM and BLG theories.
\exercise{}{Recall in Section \ref{s:validity} that we discussed when a BCFW recursion is valid: we showed that the presence of contact terms, for example $\phi^4$, in the action tend to spoil the recursion since such terms go to a constant as $z\rightarrow\infty$. This issue can be avoided in supersymmetric theories since amplitudes where such terms are present are related via supersymmetry to those where it is absent. This is accomplished via the super-BCFW shifts. One can illustrate the idea by carefully choosing the external states such that contact terms do not contribute to a particular component amplitude; then (loosely speaking) supersymmetry ensures that the superamplitude which contains this well-behaved component amplitude, also goes to zero for  $z\rightarrow\infty$. Let us test whether such a component amplitude can be found for ABJM theory at 6-point. Consider the 6-point contact term in \reef{QuarticV}. Show that if we choose all R-symmetry indices to be the same, say 1, then the sextic interaction terms vanish. Thus the 6-point scalar amplitude with all scalars having the same $SU(4)$ indices has good large-$z$ behavior. }

One thing is deriving the recursion relations, another thing is using them! So let us now apply the 3d recursion relations to compute the 6-point amplitude $A_6(\bar{X}^4 X_4 \bar{X}^4 X_4 \bar{\psi}_4 \psi^4)$ in ABJM theory.  For simplicity, we drop the $SU(4)$ indices on the component-fields, i.e.~$\bar{X}^4\rightarrow \bar{X}$. Choosing lines 1 and 6 for the shift, the only factorization channel is (123$|$456), so there is only one diagram, namely 
\be
\raisebox{-5mm}{\includegraphics[width=3.5cm]{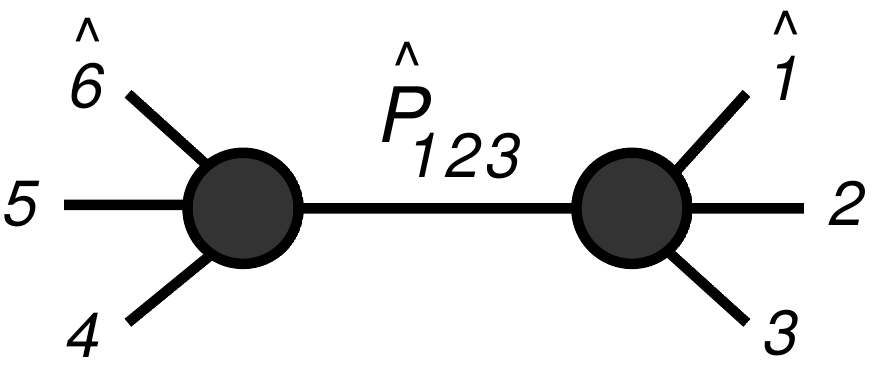}}
\,.
\ee
The recursion relation \eqref{fact-limit} then reads
\bea
\label{SuperBCFW6}
&&
\hspace{-5mm}\mathcal{A}_6\big(\bar{\Psi} \Phi \bar{\Psi}\Phi\bar{\Psi}\Phi\big) 
\\ \nonumber
&&
=  \int d^3 \eta  \bigg[ 
\hat{\mathcal{A}}_{4}\big(\bar{\Psi}_{\hat{1}} \Phi_2 \bar{\Psi}_3 \Phi_{\hat{P}_{123}}\big)\Big|_{z=z^*_1} 
\frac{H( z^*_1 , z^*_{2}  ) }{P_{123}^2}~
\mathcal{A}_{4}\big(\bar{\Psi}_{-\hat{P}_{123}}\Phi_4\bar{\Psi}_5\Phi_{\hat{6}} \big)\Big|_{z=z^*_1} 
~+~( z^*_1 \leftrightarrow z^*_2 ) \bigg] \,. ~~~~~~
\eea
To project out the amplitude $A_6(\bar{X} X \bar{X} X\bar{\psi} \psi)$ from the superamplitude \reef{SuperBCFW6} we need the coefficient of the  $(\eta_1)^3(\eta_3)^3(\eta_6)^3$ monomial, where $(\eta_i)^3=\eta_{i1}\eta_{i2}\eta_{i3}$. This follows from \reef{ABJMmap}.
After manipulation of the Grassmann delta functions and using $\hat\eta_1 (z) \hat\eta_6 (z) = \eta_1 \eta_6$, 
we find
\be
A_6\big(\bar{X} X \bar{X} X\bar{\psi} \psi\big) \\
 =
\hat{A}_{4}\big( \bar{X}_{\hat{1}} X_2 \bar{X}_3 X_{\hat{P}_{123}} \big)\Big|_{z=z^*_1} 
\frac{H( z^*_1 , z^*_{2}  ) }{P_{123}^2}
\,
\hat{A}_{4}\big(\bar{X}_{\hat{P}_{123}} X_4 \,\bar{ \psi}_5\, \psi_{\hat{6}} \big)\Big|_{z=z^*_1} 
 + ( z^*_1  \leftrightarrow z^*_2)   \,,
\label{bcfw analytic}
\ee
where the 4-point amplitudes, obtained from the superamplitude (\ref{ABJM4pt}), are
\be
  \hat{A}_{4}\big( \bar{X}_{\hat{1}} X_2 \bar{X}_3 X_{\hat{P}_{123}} \big)
   =-\frac{\<\hat{1}3\>^3}{\<\hat{1} \hat{P}_{123}\> \<\hat{P}_{123}3\> }
   ~~~~\text{and}~~~~
  \hat{A}_{4}\big(\bar{X}_{\hat{P}_{123}} X_4 \,\bar{ \psi}_5 \,\psi_{\hat{6}} \big)
  = \frac{\<\hat{P}_{123} \hat{6}\>^2}{\<\hat{6}5\>}\,.
  \label{3dintrees}
\ee
By \reef{zeros}, the poles in the $z$-plane are located at
\be
{z^*_1}^2
= \frac{\langle 16\rangle^2 - \big(\langle 23 \rangle - \langle 45\rangle\big)^2}{
\big(\langle 1| + i \langle 6|\big) P_{45}\big(|1\rangle + i | 6 \rangle\big) } \,, 
\qquad
{z^*_2}^2 =  \frac{\langle 16\rangle^2 - \big(\langle 23 \rangle + \langle 45\rangle\big)^2}{
\big(\langle 1| + i \langle 6|\big) P_{45}\big(|1\rangle + i | 6 \rangle\big) } \,.
\ee
After repeated use of momentum conservation and Schouten's identity, we find that the 6-point amplitude is 
\be
\begin{split}
&A_6\big(\bar{X} X \bar{X} X\bar{\psi} \psi\big)
= -\frac{1}{2 P_{123}^2} 
\left[ 
\frac{ \big(\langle 2 \vert P_{123} \vert 6 \rangle + i \langle 31 \rangle \langle 45 \rangle \big)^3}
{\big(\langle 1 \vert P_{123} \vert  4 \rangle +i \langle 23 \rangle \langle 56 \rangle \big)\big(  \langle 3 \vert P_{123} \vert 6 \rangle  + i \langle 12 \rangle \langle 45 \rangle \big)}
 \right. 
\\
&\qquad \qquad \qquad  \qquad \;\;\;
\left.
~~~~~~~~~~
- \frac{\big( \langle 2 \vert P_{123} \vert 6 \rangle- i \langle 31 \rangle \langle 45 \rangle \big)^3}
{\big(\langle 1 \vert P_{123} \vert 4 \rangle - i \langle 23 \rangle \langle 56 \rangle \big)\big( \langle 3 \vert P_{123} \vert 6 \rangle - i \langle 12 \rangle \langle 45 \rangle \big)}\right].
\end{split}
\label{re:6pt-4b2f}
\ee
Here the first term is the result of evaluating the first term in \reef{SuperBCFW6} while the second term is the $( z^*_1  \leftrightarrow z^*_2)$ contribution.
\exercise{}{Let us derive \reef{re:6pt-4b2f} from \reef{bcfw analytic}. First prove 
\eq
\left((z_{1}^\ast)^2+1\right)\left((z_{2}^\ast)^2-1\right)=\frac{-i\langle 1|P_{23}|6\rangle+\langle23\rangle\langle45\rangle}{\langle q|P_{2 \dots i}|q\rangle}\,.
\label{BCFWKey2}
\eqe
Next, use \reef{BCFWKey2}  to show that 
\eq
\<\hat{1}3\>\,z_{1}^\ast\left((z_{2}^\ast)^2-1\right)=\frac{i\langle23\rangle\big(\langle 2 \vert P_{123} \vert 6 \rangle + i \langle 31 \rangle \langle 45 \rangle \big)}{2\langle q|P_{2 3}|q\rangle}\,.
\eqe
Now continue to manipulate the tree-amplitudes \reef{3dintrees} to derive the first line in  \reef{re:6pt-4b2f}.
}

You may worry about the apparently spurious poles in the expression \reef{re:6pt-4b2f}, since each only appears in one term and not the other and thus cannot cancel. But have no fear, these are really local poles in disguise! To see this, we rewrite them as (see Exercise \ref{ex:spec3d})
\eq
\frac{1}{\langle 1 \vert P_{123} \vert 4 \rangle - i \langle 23 \rangle \langle 56 \rangle }=\frac{\langle 1 \vert P_{123} \vert 4 \rangle +i \langle 23 \rangle \langle 56 \rangle }{\langle 1 \vert P_{123} \vert 4 \rangle^2 +\langle 23 \rangle^2 \langle 56 \rangle^2 }=\frac{\langle 1 \vert P_{123} \vert 4 \rangle +i \langle 23 \rangle \langle 56 \rangle}{P^2_{123}P^2_{234} }\,.
\label{Magic}
\eqe 
Thus each spurious-looking pole in \reef{re:6pt-4b2f} is really a product of local poles. Note that this tells us that the two terms in the BCFW result \reef{re:6pt-4b2f} are individually local and free of spurious poles! The reason behind this will be discussed further at in Sections \ref{s:DCabjm} and \ref{s:orthograss}.
\exercise{ex:spec3d}{The final manipulation in \reef{Magic} made use of the identity 
$$\langle i |p_j+p_k| l\rangle^2-(p_i+p_j+p_k+p_l)^2\langle jk\rangle^2=(p_i+p_j+p_k)^2(p_j+p_k+p_l)^2\,,$$
which holds for any four massless vectors $p_i,p_j,p_k,p_l$ in 3d. Prove it.}
\exercise{}{Although the two terms in \reef{re:6pt-4b2f} are individually local, they actually need to come in the combination in \reef{re:6pt-4b2f}: show that the relative minus sign is necessary for the amplitude to have the correct little-group properties. }
%

%%%%%%%%%%%%%%%%%%%%%%%%%%%%%%%%%%%%%%%%%%%%%%%%%%%
\subsubsection{ABJM and dual conformal symmetry}
\label{s:DCabjm}
%%%%%%%%%%%%%%%%%%%%%%%%%%%%%%%%%%%%%%%%%%%%%%%%%%%

Let us dive straight into the deep end and define 3d dual variables $y_i^{ab}$ and $\theta_{iA}^{a}$ such that $y^{ab}_{i}-y^{ab}_{i+1}=p_{i}^{ab}$ and 
$\theta_{iA}^{a}-\theta_{i+1,A}^{a}=\tilde{q}_{iA}^{a}$. Momentum and 
supermomentum delta functions for a 4-point superamplitude are then
\eq
\delta^{3}\big(P\big)\delta^{(\mathcal{N})}\big(\tilde{Q}\big)
\rightarrow 
\delta^3\big(y_{1}-y_{5}\big)\delta^{(\mathcal{N})}\big(\theta^{a}_{1}-\theta^{a}_{5}\big)\,.
\label{3d4ptdeltas}
\eqe
We define {\em dual conformal inversion} on the variables $y_i$ and $\theta_i$ the same way in any spacetime dimension, namely as in \reef{InvertRule}. It then follows from \reef{3d4ptdeltas} that the inversion weights of the momentum and supermomentum delta function exactly cancel for $\mathcal{N}=6$ supersymmetry. Using
\eq
y^2_{i,i+2}=s_{i,i+1}=\langle i,i+1\rangle^2\,
\eqe
we can deduce (as in  Exercise \ref{ex:dci}) the dual inversion rule for a  3d angle bracket
\eq
I[\langle i,i+1\rangle]=\frac{\langle i,i+1\rangle}{\sqrt{y_i^2y_{i+2}^2}}\,.
\eqe
For the 4-point superamplitude \reef{ABJM4pt} of $\cn=6$ ABJM theory, this then implies 
\eq
I\left[\mathcal{A}_4(\bar{\Psi}_1\Phi_2\bar{\Psi}_3\Phi_4)\right]=\sqrt{y_1^2y_2^2y_3^2y_4^2}\;\mathcal{A}_4(\bar{\Psi}_1\Phi_2\bar{\Psi}_3\Phi_4)\,.
\eqe
It can be shown \cite{Gang} using the $\cn=6$ super-BCFW recursion relations that dual inversion on the $n$-point tree-level superamplitude gives
\eq
I\left[\mathcal{A}_n\right]=\bigg( \prod_{i=1}^n\sqrt{y_i^2} \bigg)\;\mathcal{A}_n\,.
\label{ABJMInv}
\eqe
Thus the 3d ABJM tree-level superamplitudes are dual conformal \emph{covariant} with 
uniform inversion weight $\frac{1}{2}$ on each leg. 

Under dual conformal inversion, the superamplitudes of 4d $\cn=4$ SYM transform covariantly with uniform inversion weight $1$ on each leg. In Section \ref{s:emDCS} we argued that as a result, the dual conformal boosts $\mathcal{K}^\mu$  annihilate the superamplitudes only after the non-trivial weights have been compensated by a shift of $\mathcal{K}^\mu$, as below  \reef{DCAnom}. This shift is crucial for defining the dual superconformal symmetry and extending it together with the ordinary superconformal symmetry to the $SU(2,2|4)$ Yangian  of the 4d planar $\cn=4$ SYM  superamplitudes. 

In 3d, the dual conformal symmetry can be enlarged into the dual superconformal symmetry group $OSp(6|4)$~\cite{HuangLipstein}. The symmetry group acts on the dual space that consists of coordinates $(y_i^{ab}, \theta_{iA}^{a}, r_{iAB})$, where the extra R-symmetry coordinate $r_{iAB}$ is defined by:
\eq
r_{i{AB}}-r_{i+1,{AB}}=\eta_{iA}\eta_{iB}\,.
\eqe
The group $OSp(6|4)$ is also the supergroup for the ordinary superconformal symmetry of the ABJM Lagrangian in \reef{ABJML}. The  combination of the dual and ordinary superconformal symmetries forms an infinite dimensional $OSp(6|4)$ Yangian algebra~\cite{Bargheer}, very similar in nature to the $SU(2,2|4)$ Yangian symmetry of 4d planar $\mathcal{N}=4$ SYM.

As an example, the super-BCFW construction \reef{SuperBCFW6} gives the tree-level 6-point superamplitude in terms of two Yangian invariants $Y_1$ and $Y_2$,
\be
  \mathcal{A}^{\rm tree}_6(\bar{\Psi}_1\Phi_2\bar{\Psi}_3\Phi_4\bar{\Psi}_5\Phi_6) = Y_1 + Y_2\,.
  \label{3dsuperA6}
\ee
The two Yangian invariants  $Y_1$ and $Y_2$ arise precisely from the two BCFW-terms in \reef{SuperBCFW6}. We will not need their explicit form; they can be found in \cite{Gang}.  

It will be relevant for us to also consider the tree amplitude with shifted sites, 
\be
  \label{A6sh}
  \mathcal{A}^\textrm{tree}_\textrm{6,shifted} = 
\mathcal{A}^{\rm tree}_6(\Phi_1\bar{\Psi}_2\Phi_3\bar{\Psi}_4\Phi_5\bar{\Psi}_6)= 
\mathcal{A}^{\rm tree}_6(\bar{\Psi}_2\Phi_3\bar{\Psi}_4\Phi_5\bar{\Psi}_6 \Phi_1)\,. 
\ee
It has a  super-BCFW representation that can be written as
\be
  \mathcal{A}^\textrm{tree}_\textrm{6,shifted} = 
\mathcal{A}^{\rm tree}_6(\Phi_1\bar{\Psi}_2\Phi_3\bar{\Psi}_4\Phi_5\bar{\Psi}_6)
   = Y_1 - Y_2\,.
   \label{3dsuperA6sh}
\ee
Now the important point  is that two physical objects,  $\mathcal{A}^\textrm{tree}_6$ and  $\mathcal{A}^\textrm{tree}_\textrm{6,shifted}$, are written as distinct linear combinations of the same two Yangian invariants: this is only possible if each of the two Yangian invariants are local, i.e.~free of spurious poles. We already noted the locality for the particular component amplitude \reef{re:6pt-4b2f}. Now you see why it was needed. Note that this contrasts the 3d ABJM theory from 4d $\cn=4$ SYM where the dual superconformal invariant 5-brackets had spurious poles. 

%%%%%%%%%%%%%%%%%%%%%%%%%%%%%%%%%%%%%%%%%%%%%%%%%%%
\subsubsection{Loops and  on-shell diagrams  in ABJM}
%%%%%%%%%%%%%%%%%%%%%%%%%%%%%%%%%%%%%%%%%%%%%%%%%%%

The loop-level superamplitudes can be explored using unitarity methods (Section \ref{s:loops}). Using the dual inversion property of the tree-level superamplitudes, it can be shown that the planar loop superamplitudes of ABJM, prior to integration, are dual conformal covariant, i.e.~they satisfy \reef{ABJMInv}. Thus perturbatively, planar ABJM has a  structure very similar to planar $\mathcal{N}=4$ SYM, they are almost baby brothers/sisters. This is rather surprising given that the two theories have very distinct Lagrangians and live in different spacetime dimensions. Moreover, in quantum field theory textbooks, one learns that $D<4$ theories generically have more severe IR-divergences compared to $D=4$. Thus one might expect that although planar ABJM is very similar to $\mathcal{N}=4$ SYM at the pre-integrated level, the similarity would be completely scrambled by the potentially severe IR-divergence in $D=3$. 

To see if this is the case,  let us take a look at the planar loop amplitudes in detail. The 1-loop amplitudes in ABJM are purely rational 
functions~\cite{Bargheer:2012cp,Bianchi:2012cq,ABJMOneL}. This can be understood as a consequence of dual conformal symmetry, since the only dual conformal covariant scalar integral is the massive triangle integral, and it integrates to
\eq
 \vcenter{\hbox{\includegraphics[scale=0.45]{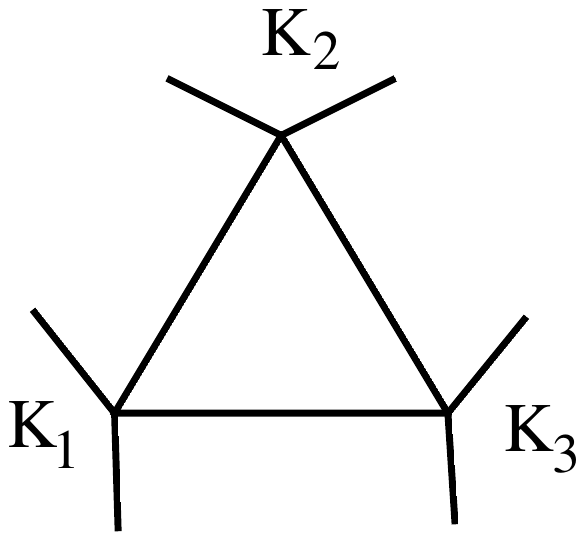}}}
 ~~~
 I_3(K_1,K_2,K_3)= 
 -\frac{i\pi}{2}\frac{1}{\sqrt{-K_1^2}\sqrt{-K_2^2}\sqrt{-K_3^2}}\,,
\label{tri3d}
\eqe
where $K_1$, $K_2$, $K_3$ are the sums of the external momenta going out of each of the three corners and $K_i^2 \ne 0$. There are no triangle diagrams with massless corners $K_i^2=0$; this follows from generalized unitarity methods using that 3-point loop amplitudes vanish. The integrated result \reef{tri3d}  has transcendentality 1 thanks to the factor of $\pi$.

Since 6-point is the lowest multiplicity at which the triangle integral \reef{tri3d} contributes after integration, we conclude that the {\bf \em 4-point 1-loop amplitude vanishes} up to ${O}(\epsilon)$ in dimensional regularization $D=3-2\epsilon$.\footnote{The 4-point 1-loop integrand is non-trivial. It is given by a loop-momentum dependent integrand that integrates to zero up to ${O}(\epsilon)$~\cite{ABJMTwoL41}. } 

The {\bf \em6-point 1-loop superamplitude} is~\cite{Bianchi:2012cq, Bargheer:2012cp, Brandhuber:2012un}\footnote{This result is only valid up to ${O}(\epsilon)$. There are additional integrands, whose coefficient is proportional to the tree-amplitude, that integrate to zero up to ${O}(\epsilon)$. }
\eq
\mathcal{A}_6^\text{1-loop}=-i
\left(\frac{N}{k}\right)
\mathcal{A}^\textrm{tree}_\textrm{6,shifted}
\Big(\langle 12\rangle\langle 34\rangle\langle 56\rangle \,I_3(P_{12},P_{34},P_{56})+ \langle 23\rangle\langle 45\rangle\langle 61\rangle \,I_3(P_{23},P_{45},P_{61})\Big)\,,
\label{OneLoopABJM6pt0}
\eqe
where $N$ comes from the gauge group $U(N) \times U(N)$, and $k$ is the Chern-Simons level of \reef{ABJML}. 
The tree superamplitude 
$\mathcal{A}^\textrm{tree}_\textrm{6,shifted}$ was defined in \reef{A6sh}. 
Using the integrated result \reef{tri3d} for the scalar triangle integrals $I_3$, we find that the 1-loop 6-point superamplitude is 
\be
  \mathcal{A}_6^\text{1-loop}=
  -\frac{\pi}{2}\left(\frac{N}{k}\right)
  \mathcal{A}^\textrm{tree}_\textrm{6,shifted}
  \Big(
  \sgn(\langle 12\rangle)\, \sgn(\langle 34\rangle)\, \sgn(\langle 56\rangle)
  + \sgn(\langle 23\rangle)\, \sgn(\langle 45\rangle)\,\sgn(\langle 61\rangle) 
   \Big)\,.
\label{OneLoopABJM6pt}
\ee
Here we have introduced
\be
  \sgn(\langle ij\rangle)\equiv 
  {\frac{\langle ij\rangle}{\sqrt{-K^2_{ij}}}}
  =\frac{\langle ij\rangle}{|\langle ij\rangle|}\,,
\ee
which equals $\pm1$ depending on the kinematics. Thus, remarkably, the 1-loop 6-point superamplitude can be either zero or non-vanishing depending on the kinematics! This peculiar behavior has to do with an interesting topological feature of lightlike momenta in 3 dimensions. In 3d Minkowski space, a lightlike vector can be written as $p_i^\mu=E_i(1, \cos\theta_i, \sin\theta_i)$. This means that lightlike vectors can be projected to points on a circle $S^1$.
From
\eq
\langle ij\rangle=\sqrt{-2 p_i\cdot p_j}=i\sqrt{E_i\,E_j} \,\sin\left(\frac{\theta_i-\theta_j}{2}\right)
\eqe
we see that the sign of $\langle ij\rangle$  changes whenever the two points that represent $p_i$ and $p_j$ cross each other on the $S^1$. Thus the 1-loop amplitude encounters a sudden jump,  from zero to non-vanishing or vice versa, whenever two points on the $S^1$ cross each other:
\be
\raisebox{-12mm}{\includegraphics[scale=0.45]{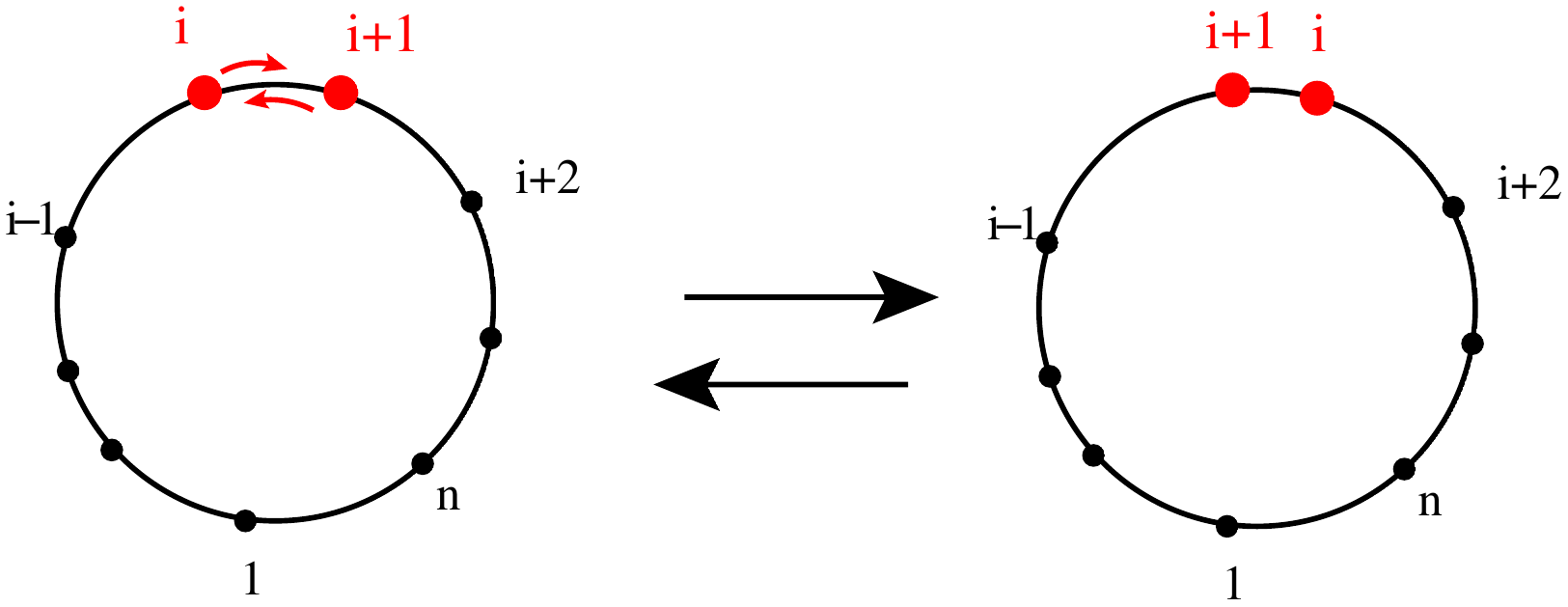}}
\label{3dcirc}
\ee
The two configurations \reef{3dcirc} are topologically inequivalent.\footnote{This can be made more precise. By judiciously adding $2\pi$ to the angles $\theta_i$, one can arrange the angles such that a given kinematics configuration has all angles strictly increasing according to their color ordering, i.e. $0<\theta_{i+1}-\theta_i<2\pi$. This gives a well defined ``winding number" $w=(\theta_n-\theta_1)/(2\pi)$. Now as two points cross each other, the winding number  changes by one, indicating a distinct topological sector.} Thus this sudden jolt is the amplitude way of telling us that we are changing the topology of our momentum space! 
\exercise{}{From 4d, we are familiar with loop-amplitudes being proportional to the tree amplitudes, so it may seem odd that here in 3d the 1-loop 6-point amplitude \reef{OneLoopABJM6pt} is proportional to the shifted tree-amplitude. Verify that the LHS and RHS of \reef{OneLoopABJM6pt} have the same little group scaling thanks to the sign-functions. 
}
The first non-trivial loop contribution to the {\bf \em 4-point amplitude} enters at {\bf \em 2-loop order} and it is given by~\cite{ABJMTwoL41,ABJMTwoL42}
\eq
\mathcal{A}_4^\text{2-loop}=
\left(\frac{N}{k}\right)^2\!
\mathcal{A}_4^\text{tree}\left[
-\frac{
  \big(\!-\!\m^{-2} y_{13}^2 \big)^{-\eps} 
  + \big(\!-\!\m^{-2} y_{24}^2 \big)^{-\eps}}{(2\eps)^2} 
  + 
  \frac{1}{2} \ln^2\Big( \frac{y_{13}^2}{y_{24}^2} \Big) 
+4\zeta_2-3\ln^22+\mathcal{O}(\epsilon)\right].
\eqe
Notice that the IR divergent part is equivalent to that of the 1-loop 4-point superamplitude \reef{4pt1loopA} of $\mathcal{N}=4$ SYM, with $\epsilon\rightarrow 2\epsilon$ because this is 2-loops. Not only is the IR-structure of this theory identical to that of $\mathcal{N}=4$ SYM, but so is the 
$\ln^2\big( {y_{13}^2}/{y_{24}^2} \big)$ piece!  

Moving on to the {\bf \em 6-point 2-loop amplitude}, 
one finds~\cite{ABJMTwoL6}
\begin{equation}
\begin{split}
  \mathcal{A}_6^{\textrm{2-loop}}&=\left(\frac{N}{ k}\right)^2 \bigg\{\frac{\mathcal{A}_6^{\textrm{tree}}}{2}\bigg[ \text{BDS}_6+R_6\bigg]
  +\frac{\mathcal{A}^\textrm{tree}_\textrm{6,shifted}} {4i}
    \bigg[\ln\frac{u_2}{u_3}\ln\chi_1+{\rm cyclic\times2}\bigg]  \bigg\}\,.
    \end{split}
    \label{2LoopAnsw}
\end{equation}   
Here BDS$_6$ is the 1-loop MHV amplitude \reef{BDSansatz} for $\mathcal{N}=4$ SYM, again with proper rescaling of the regulator $\eps \to 2 \eps$ to account being at 2-loops. As the remaining pieces are finite, the BDS ansatz captures the IR-divergent as well as the resulting non-dual-conformal part of the amplitude. So once again, we observe  that the IR structure of planar ABJM theory is identical to that of $\mathcal{N}=4$ SYM!  The ``remainder" function $R_6$ in \reef{2LoopAnsw} is 
\be
R_6=-2\pi^2+\sum_{i=1}^3\left[\textrm{Li}_2(1-u_i)+\frac12\ln u_i\ln u_{i{+}1} +(\arccos\sqrt{u_i})^2\right],
\label{3dremainder}
\ee
where the $u_i$'s are the dual conformal cross-ratios defined in \reef{u123}. 
The shifted tree $\mathcal{A}^\textrm{tree}_\textrm{6,shifted}$ was encountered in \reef{OneLoopABJM6pt0}. Finally, the function $\chi_1$ in \reef{2LoopAnsw} is 
\eq
 \chi_1=\frac{\langle12\rangle\langle45\rangle + i\langle3|P_{123}|6\rangle}{\langle12\rangle\langle45\rangle-i \langle3|P_{123}|6\rangle}\,,
\eqe
while ``cyclic$\times2$" means we sum over all cyclic rotations by two sites, $i\rightarrow i+2$.
\exercise{}{Seeing both $\mathcal{A}^{\rm tree}_6$ and $\mathcal{A}^\textrm{tree}_\textrm{6,shifted}$ in the same amplitude means that you should check that the other factors in \reef{2LoopAnsw} indeed compensate for the  little group weight difference.}

Now that we have seen explicit examples of planar loop-amplitudes in ABJM theory, let us turn to the subject of {\bf \em Leading Singularities} and {\bf \em on-shell diagrams}. We studied these for planar 4d $\cn=4$ SYM in Section \ref{s:loops3}. Because of the dual superconformal symmetry of the loop-integrands, multi-loop amplitudes of ABJM theory can be calculated with Leading Singularity methods. In 3d, a maximal cut takes three propagators on-shell for each loop-momentum.  At 1-loop order,  the only dual conformal scalar integral is the massive triangle, so this plays the equivalent role of the box-diagram in 4d.  In 4d, we built the Leading Singularity on-shell diagrams from vertices that are the fundamental 3-point MHV and anti-MHV superamplitudes. These vanish in 3d, so here we use the 4-point superamplitudes instead. In ABJM theory, 
the first non-trivial 1-loop Leading Singularity is the 6-point diagram
\be
  \raisebox{-6mm}{\includegraphics[scale=0.45]{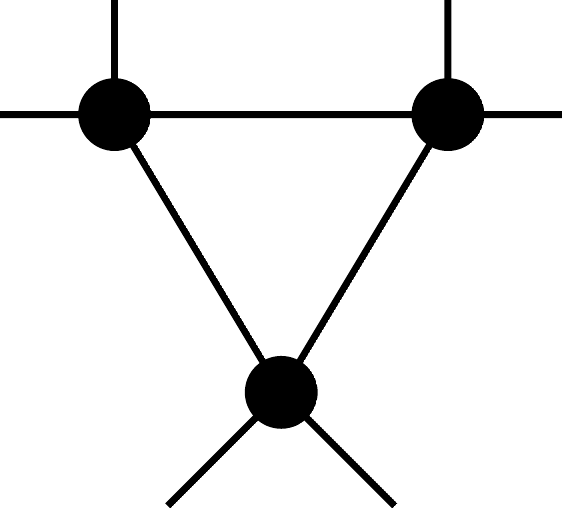}}\,.
  \label{6pttri}
\ee
As noted above,  each vertex represents an on-shell 4-point superamplitude of ABJM theory. There is no distinction of ``black" and ``white" vertices because there is only one type of 4-point superamplitude in ABJM.

In 4d $\mathcal{N}=4$ SYM, we found that the 4-point Leading Singularity box diagram represents the 4-point tree amplitudes (see Section \ref{s:onshelldiag}).
Similarly, in ABJM, it turns out that the 6-point Leading Singularity triangle diagram \reef{6pttri} reproduces the tree-level 6-point superamplitude. 
To see this, we isolate the 3rd vertex in \reef{6pttri} and parameterize the on-shell legs as 
\eq
\vcenter{\hbox{\includegraphics[scale=0.45]{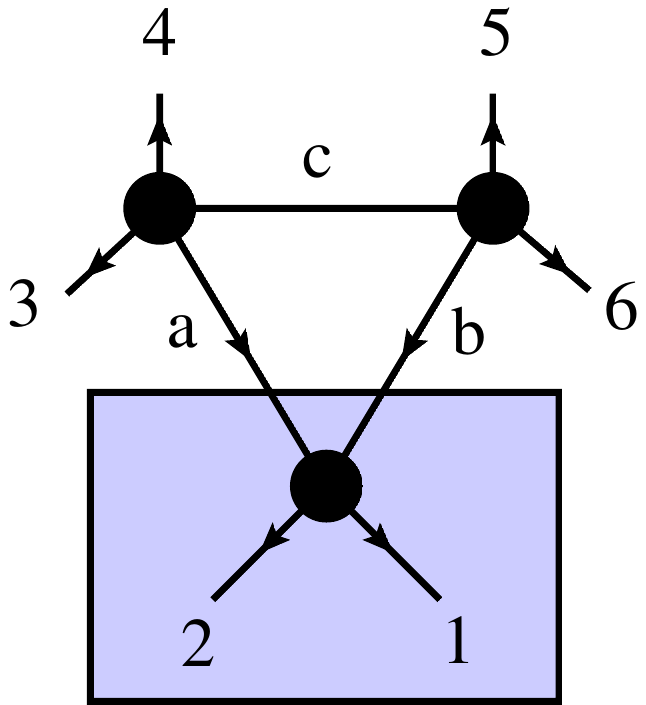}}}\,.
\label{6pttri2}
\eqe
The internal momenta $a$ and $b$ are on-shell, so they each have 2 degrees of freedom. Of the combined $4=2+2$ degrees of freedom in $a$ and $b$ 3 are fixed in terms of momenta 1 and 2 by the momentum conservation delta function  at the bottom vertex in \reef{6pttri2}. Thus the spinor variables of $a$ and $b$ can be parametrized in terms of $|1\>$ and $|2\>$ using a single free variable. 
With a little thought --- or, even better, a little calculation --- one finds that the following parameterization solves the momentum conservation constraints
\eq
|a\>=\cos\theta\,|1\>-\sin\theta\,|2\>\,,
\hspace{1cm} 
|b\>=\sin\theta\,|1\>+\cos\theta\,|2\>\,.
\eqe
This is exactly the BCFW deformation \reef{3DBCFW} of legs $1$ and $2$.  Indeed, the final on-shell condition $p_c^2=0$, becomes the factorization condition that the parameter $\theta$ (i.e.~$z$) must satisfy. 
\exercise{}{Verify that the supermomentum delta function on the bottom vertex enforces the following identification 
$\eta_a=\cos\theta\,\eta_1-\sin\theta\,\eta_2$ and 
$\eta_b=\sin\theta\,\eta_1+\cos\theta\,\eta_2$.
} 
Now it is very tempting to conclude that the  on-shell diagram \reef{6pttri2} can be also understood as a BCFW diagram for the 6-point tree superamplitude in ABJM. This is true, but we have to make sure that we produce the BCFW recursion formula \reef{fact-limit}, including the weight-factor $H(a,b)$ defined in \reef{Hfunction} and the propagator of the factorization channel. Taking into account the Jacobian factors associated with the triple cut and the bottom vertex,  it has been shown \cite{Brandhuber:2012un} that  $H(a,b)$ and the factorization propagator are indeed produced. Thus we have
\be
\raisebox{-15mm}{{\includegraphics[scale=0.45]{ABJMLS1}}} 
~~= 
~ \int d^3 \eta  \bigg(  \mathcal{A}_{4}\big(\Phi \bar{\Psi}\Phi\bar{\Psi}\big)  
 \frac{H( z^*_1 , z^*_{2})}{(P_{234})^2} 
 \mathcal{A}_{4}\big(\bar{\Psi}\Phi\bar{\Psi}\Phi\big)
 +( z^*_1 \leftrightarrow z^*_2 ) \bigg)
 ~=~
 \ca_6^\text{tree}
 \,.
 \label{3d6ptLS}
\ee
Recall that in 4d, the Leading Singularity is closely related to the integral coefficients in expressions like \reef{IntBasis}. Previously we have seen that the 1-loop 6-point amplitude is proportional to $\mathcal{A}^\textrm{tree}_\textrm{6,shifted}$, so it is puzzling that the 6-point Leading Singularity \reef{3d6ptLS} is just $\mathcal{A}^\textrm{tree}$. This has to do with a subtlety of the Jacobian factors. Recall that the integral coefficients can be determined by unitarity cuts. When we apply unitarity cuts, we are substituting the propagators with delta functions, as discussed in Section \ref{s:genunit}. As we solve the delta function constraints, we generate a Jacobian factor with an absolute value. On the other hand, when we are computing the Leading Singularity, we treat the delta functions as contour integrals, thus while localizing on a pole, the Jacobian factor does not come with an absolute value. In the 
1-loop cases that we encountered in 4d, the Jacobian factor for the two 
loop-momentum solutions are identical, so the presence of an absolute value did not make a difference. However in 3d, the Jacobian factors for the two 
loop-momentum solutions differ by a sign, so whether or not there is an absolute value on the Jacobian makes a big difference~\cite{ABJMTwoL6}. The result of this is that the 1-loop 6-point amplitude is proportional to $\mathcal{A}^\textrm{tree}_\textrm{6,shifted}$ while the 6-point Leading Singularity is just $\mathcal{A}^\textrm{tree}$.
\exercise{}{The above discussion indicates that if we had a relative plus sign for  the two BCFW terms on the RHS of \reef{re:6pt-4b2f}, the result would be 
$\mathcal{A}^\textrm{tree}_\textrm{6,shifted}$ instead of $\mathcal{A}^\textrm{tree}_\textrm{6}$. Verify that with a relative plus sign, \reef{re:6pt-4b2f} has the correct little group property of $\mathcal{A}^\textrm{tree}_\textrm{6,shifted}$. [Hint: You need to take into account that the coefficient for the  $\eta$-polynomial corresponds to a different component amplitude in the shifted amplitude.]}
Instead of \reef{6pttri2}, we could have computed the on-shell diagram 
\eq
\vcenter{\hbox{\includegraphics[scale=0.45]{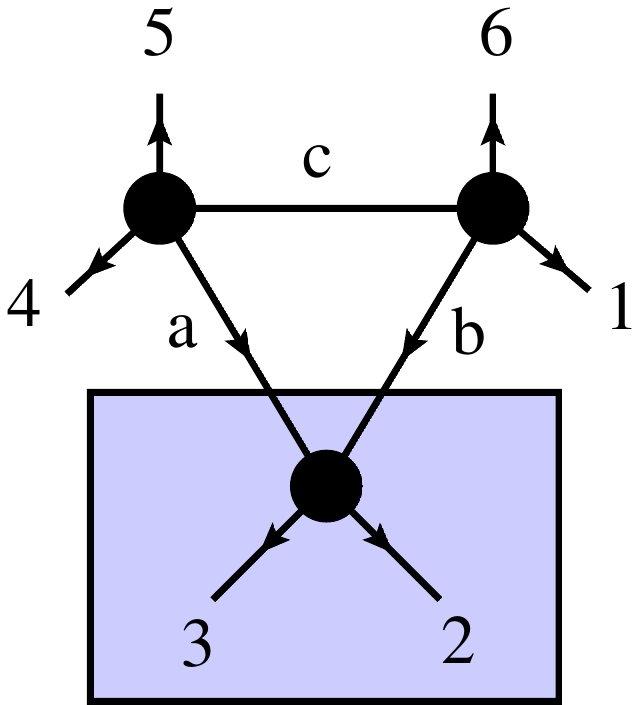}}}
\eqe 
and the result would have been exactly the same, namely 
$\mathcal{A}^\text{tree}_\text{6}$. 
This gives us the ABJM equivalent of the ``square move" \reef{sqmove2} in 4d $\mathcal{N}=4$ SYM. The ABJM ``triangle-move" is
\eq
\vcenter{\hbox{\includegraphics[scale=0.45]{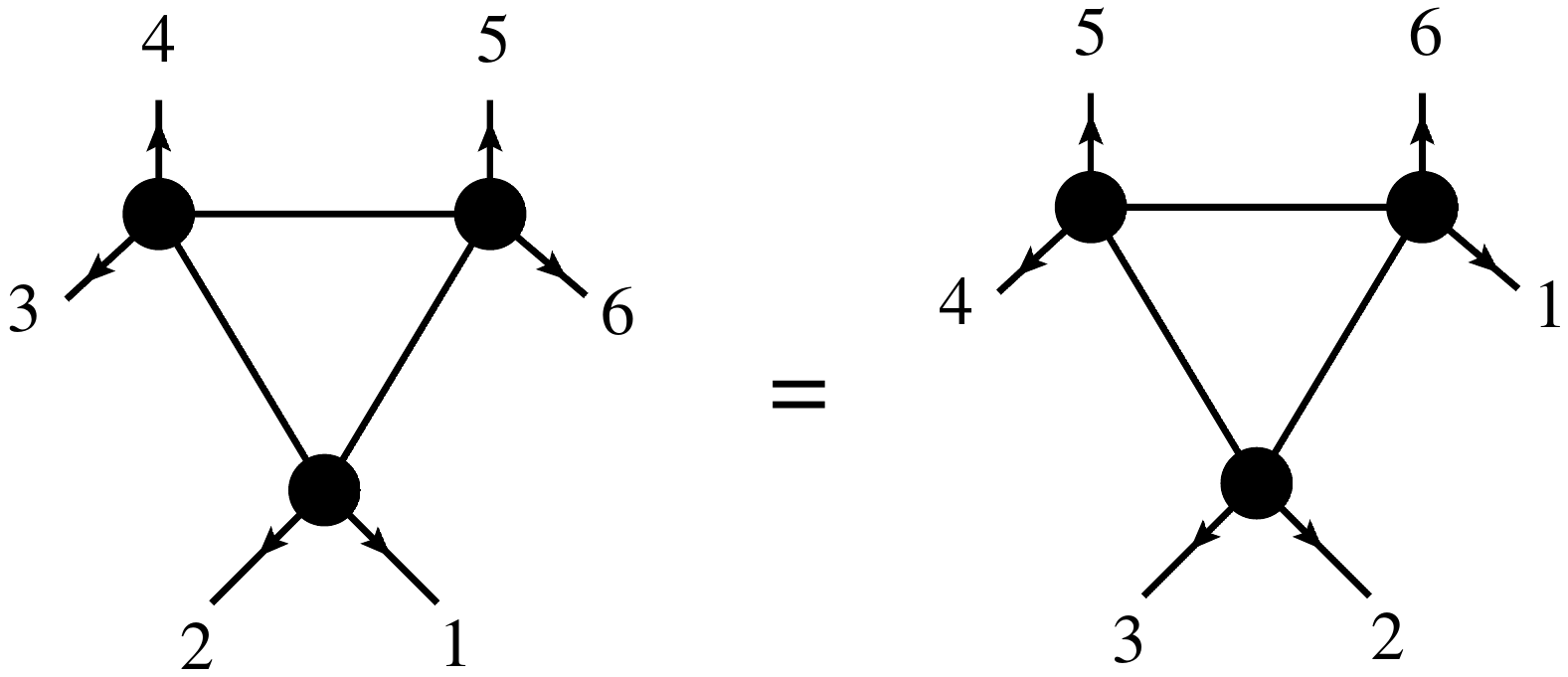}}}\,.
\eqe 
This is dubbed the {\bf \em Yang-Baxter move}, because it is precisely the graphical representation of the Yang-Baxter equation that plays an important role in integrable theory. It is usually represented as  
\eq
\vcenter{\hbox{\includegraphics[scale=0.45]{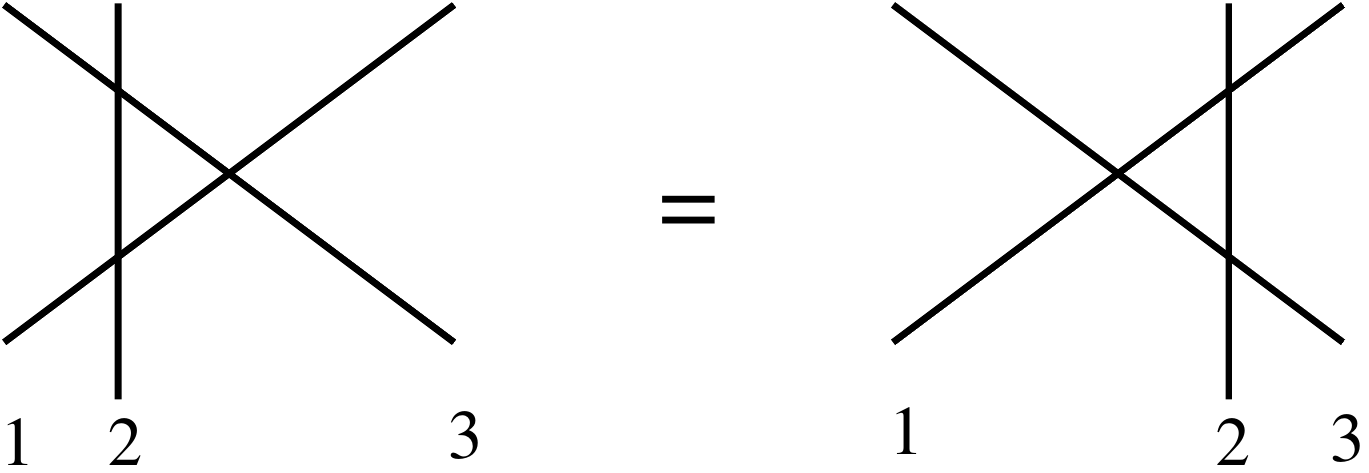}}}\,.
\eqe

Just as in  $\mathcal{N}=4$ SYM, the on-shell diagrams at higher multiplicity have an interesting structure. This is a current area of research and you can learn more from the recent paper \cite{ArkaniHamed:2012nw}.

%%%%%%%%%%%%%%%%%%%%%%%%%%%%%%%%%%%%%%%%%%%%%%%%%%%
\subsubsection{The orthogonal Grassmannian}
\label{s:orthograss}
%%%%%%%%%%%%%%%%%%%%%%%%%%%%%%%%%%%%%%%%%%%%%%%%%%%
Let us see if we can reason our way to a proper Grassmannian formula for the 3d superamplitudes of ABJM theory. We begin with 
\eq
\prod_{\mathsf{a}=1}^k
\delta^{2|3}\Big(\sum_{i}C_{\mathsf{a}i}\Lambda_i\Big)\equiv
\prod_{\mathsf{a}=1}^k
\delta^{2}\Big(\sum_{i}C_{\mathsf{a}i} |i\>\Big)
\delta^{(3)}\Big(\sum_{i}C_{\mathsf{a}i}\eta_i\Big)\,,
\label{Delta}
\eqe
where $\Lambda_i = (|i\>,\eta_i)$. As in the 4d case, treated in Section \ref{s:grassmannia}, the function  \reef{Delta} is invariant under generators in \reef{3DGen} which are linear in derivatives. If we consider the generators that are quadratic in derivatives, for example the conformal boost generator, we find
\be
\bigg(\sum_{i=1}^n\frac{\partial}{\partial |i\>^a}\frac{\partial}{\partial |i\>^b}\bigg)
\prod_{\mathsf{a}=1}^k
\delta^{2}\Big(\sum_{j}C_{\mathsf{a}j}|j\>\Big)
=
\Big(\sum_{i=1}^nC_{\mathsf{a}'i}C_{\mathsf{b}'i}\Big)f_{ab}
\prod_{\mathsf{a}\neq \mathsf{a}',\mathsf{b}'}
\delta^{2}\Big(\sum_{j}C_{\mathsf{a}j}|i\>\Big)\,.
\label{Kguess}
\ee
Here $f_{ab}$ is a function that includes either single derivative or double derivatives of the delta functions, depending on whether $a=b$ or not. The important piece in \reef{Kguess} is the prefactor: it tells us that to ensure invariance under conformal boosts, we need to dress \reef{Delta} with an extra term that enforces $C C^T=0$, i.e.
\be
\delta^{\frac{k(k+1)}{2}}\Big(C C^T\Big)
\prod_{\mathsf{a}=1}^k\delta^{2|3}\Big(\sum_{i}C_{\mathsf{a}i}\Lambda_i\Big)\,.
\label{Delta2}
\ee
The product $C C^T$ is a symmetric 
$k\times k$ matrix, so setting it to zero takes 
$k(k+1)/2$ constraints, as indicated in the delta function. 
\exercise{}{Show that the combination in \reef{Delta2} is also invariant under the multiplicative generators such as $P^{ab}$.} 
Now we can readily write down an Ansatz for an $\mathcal{N}=6$ superconformal invariant integral over a Grassmannian $Gr(k,n)$  subject to the orthogonal constraint $CC^T=0$:
\eq
\int d^{n \times k}C
\;f(M)\;
\delta^{\frac{k(k+1)}{2}}\Big(C C^T\Big)
\prod_{\mathsf{a}=1}^k
\delta^{2|3}\Big(\sum_{i}C_{\mathsf{a}i}\Lambda_i\Big)\,,
\label{Delta3}
\eqe
where $f(M)$ is a function that only depends on the minors of the Grassmannian, so that it preserves $SL(k)$ invariance. In order to interpret \reef{Delta3} as an integral over a Grassmannian manifold, it has to be $GL(k)$ invariant. All terms in \reef{Delta3} are $SL(k)$ invariant, and the $GL(1)$ weight count of the delta functions plus the measure gives $n k-k(k+1)-2k+3k=k(n -k)$. This tells us that the function $f(M)$ needs to have $GL(1)$ weight $-k(n -k)$ .  

We need more input to fix $f(M)$ and the extra information  comes from little group analysis. Under the $\mathbb{Z}_2$ little group, we have $|i\> \to - |i\>$ and $\eta_i \to -\eta_i$, so invariance of the delta functions in \reef{Delta3}  requires  $C_{\mathsf{a}i}\rightarrow -C_{\mathsf{a}i}$. For an amplitude with a $\bar{\Psi}$-supermultiplet  on the odd-sites, the superamplitude should pick up a minus sign whenever we perform a $\mathbb{Z}_2$ transformation on the odd-numbered legs, while it should be inert for  the even legs with their $\Phi$-supermultiplet. Take  $n=2k$, and $k=$\,even: then the product of $k$ consecutive minors, 
\be
f(M)= \prod_{i=1}^k\frac{1}{M_i}\,
\label{Delta4}
\ee
indeed satisfies the little group criteria. (Exercise \ref{ex:orthograss} helps you see this.) Furthermore, since $n=2k$, the function \reef{Delta4}  has $GL(1)$ weight $-k^2$, precisely as needed for overall $GL(1)$ invariance.
\exercise{ex:orthograss}{For $k\!=\!3$ (and hence $n\!=\!6$), verify that \reef{Delta4} indeed picks up a minus sign under little group scaling for odd legs, and invariant for even legs. Show that for $k=$\,odd, the function  $f(M)= \prod_{i=2}^{k+1}\frac{1}{M_i}$ does the right job.}

We conclude that the 3d Grassmannian formula for ABJM theory is given by an orthogonal Grassmannian integral \cite{SLee} which for $k=$ even is
\eq
\mathcal{L}^\text{O}_{2k,k}
=
\int \frac{d^{2k^2}C
}{GL(k)}\; 
\bigg(\prod_{i=1}^k\frac{1}{M_i}\bigg)\;
\delta^{\frac{k(k+1)}{2}}\Big(C C^T\Big)
\prod_{\mathsf{a}=1}^k\delta^{2|3}
\Big(\sum_{i}C_{\mathsf{a}i}\Lambda_i\Big)\,
\label{ABJMGrass}
\eqe
The superscript ``O" indicates it is an {\bf \em orthogonal Grassmannian}. 
When $k\!=$\,odd, the product of minors is replaced by $\prod_{i=2}^{k+1}\frac{1}{M_i}$, as shown in Exercise \ref{ex:orthograss}. Some comments are in order.
\begin{itemize}
\item {\em Momentum conservation} is enforced in \reef{ABJMGrass} in a slightly differently manner than in the 4d version \reef{TheGrassMom} of the Grassmannian integral because  we only have the $|i\>$-spinors in 3d. Here is how it goes. The orthogonality condition forces the Grassmannian to be a collection of null vectors in an $n$-dimensional space. The bosonic delta function 
$\delta^2\big(C\cdot |i\>\big)$ requires the two $n$-dimensional vectors 
$\big\{ |i\> \big\}$ to be orthogonal to $C$. 
This means that $\big\{ |i\> \big\}$ lies in the complement of $C$, which is nothing but $C^T$, and thus  $\big\{ |i\> \big\}$ must also be null: $\sum_i |i\>^a |i\>^b=0$. 
 \item {\em Two-site cyclicity?}
The integral $\mathcal{L}^{\rm O}_{2k,k}$ does not appear to have the correct cyclic invariance by two sites  discussed \reef{ABJMCycle}. However, thanks to the orthogonality condition it can be shown that 
\eq
M_{i}M_{i+1}=(-1)^kM_{i+k}M_{i+1+k}\,.
\label{MShift}
\eqe
Therefore the formula \reef{ABJMGrass} is indeed invariant under cyclic rotation by two sites up to a factor of $(-1)^{k-1}$, as required.  
\exercise{}{Show that \reef{MShift} is indeed true at 4-points: using $GL(2)$ invariance and the orthogonality condition, we can choose to fix the $2\times4$ matrices $C$ to take the form 
\eq
C=\bigg(\begin{array}{cccc}1 & 0 & i\sin\theta &  -i\cos\theta \\0 & 1 & i\cos\theta & i\sin\theta\end{array}\bigg)\,.
\eqe
Verify that $CC^T=0$ and that \reef{MShift} holds.
}
\item
{\em The dimension of the integral}  \reef{ABJMGrass} is found by counting how many free variables are left after localization by the delta functions. To start with, there are a total of $2k^2$ integration variables. The bosonic delta functions fix $k(k+1)/2+2k-3$ constraints, with the $-3$ coming from the removal of the constraints that enforce momentum conservation. Subtracting the $k^2$ redundancy of $GL(k)$, the dimension of the integral is then $\frac{(k-2)(k-3)}{2}$. Thus for 4- and 6-point amplitudes ($k=2,3$), the delta functions completely localize the Grassmannian integral. 

\end{itemize}

Let us now  take a closer look at \reef{ABJMGrass} for $n=4$. Gauge fix the  $GL(2)$ by taking 
\be
C = \begin{pmatrix}
c_{21} & 1 & c_{23} & 0 \\
c_{41} & 0 & c_{43} & 1
\end{pmatrix} \,.
\ee
This leaves 4 parameters that can be fixed by the  4 delta functions in 
$\d\big(C\cdot |i\>\big)$. Denote the solutions to $C\cdot |i\> = 0$ by 
$c^*_{\bar{r}s}$, with barred labels indicating even legs and un-barred odd legs.
Then the delta functions can be rewritten as
\be
\d^4\big(C\cdot |i\>\big) = \frac{1}{\langle 13 \rangle^2}\prod_{\bar{r}, s} \d^4( c_{\bar{r}s} - c^*_{\bar{r}s})   \,, \qquad
\begin{pmatrix}
c^*_{21} & c^*_{23} \\
c^*_{41} & c^*_{43}
\end{pmatrix}
=
-\frac{1}{\langle 13 \rangle}
\begin{pmatrix}
\langle 2 3 \rangle & \langle 1 2 \rangle \\
\langle 4 3 \rangle & \langle 1 4 \rangle
\end{pmatrix} .
\label{4pt-1}
\ee
\exercise{}{Which property of the external momenta does it take for the above solution $c^*_{\bar{r}s}$ to solve the orthogonality constraint? (Show it!)
}
Localizing the Grassamannian integral on to $c^*_{\bar{r}s}$, we then find
\be
\d^{3}\big(C C^T\big) 
=  \frac{\langle 13 \rangle^6}{\langle 24 \rangle^3}\, \d^3\big(P\big) \,,
\hspace{6mm}
\d^6\big(C\cdot \eta\big) = 
\frac{\langle 24 \rangle^3}{\langle 13 \rangle^6}\,
\d^{(6)}\big(\widetilde{Q}\big) \,,
\hspace{6mm}
\frac{1}{M_1 M_2} 
= \frac{\langle 13 \rangle^2}{\langle 1 4 \rangle \langle 34 \rangle} \,.
\label{4pt-2}
\ee
Combining \reef{4pt-1} and \reef{4pt-2}, we recover the superamplitude \reef{ABJM4pt} of ABJM theory.

For $n=6$, the integral \reef{ABJMGrass} is again completely localized by the bosonic delta functions onto two solutions, each corresponding to one of the BCFW terms in \reef{SuperBCFW6}. Recall that these two terms are individually local (see \reef{re:6pt-4b2f}). We can now understand why. At 6-point these are the only possible invariants produced by the Grassmannian integral, so this means that {\em if the orthogonal Grassmannian integral produces all possible dual conformal invariants}\footnote{Using on-shell diagrams, one can show that the Leading Singularities obtained from the result of loop-level recursion can always be identified with residues of the Orthogonal Grassmannian integral~\cite{LSOG}.} the Leading Singularity of the 6-point amplitude must be some linear combination of them. However, we already know that there are two distinct local rational  functions for $n=6$, namely $\ca_n^\text{tree}$ and 
$\ca_n^\text{1-loop} \propto \ca_{n,\text{shifted}}^\text{tree}$. As we noted at the end of Section \ref{s:DCabjm}, since they are distinct, this can only mean one thing, namely that the two terms in \reef{SuperBCFW6} are individually local and free of spurious poles. 

We conclude this Section with a comparison of the 3d and 4d Grassmannians. In the 4d  Grassmannian, the choice of contour that gives the tree amplitude forces the Grassmannian $Gr(k,n)$  to be localized to a $Gr(2,n)$ submanifold. For $n=6$ we saw how  this is intimately related to locality, since the tree contour selected three residues  whose sum was free of spurious poles. In 3d, on the other hand, the orthogonal Grassmannian integral localizes completely for $n=6$ and gives us two local objects without any need for us to pick any contour. Does this mean that the localization to the $Gr(2,n)$ submanifold is not really related to locality? The answer turns out to be `no' in an interesting way. It was found in \cite{ABJMString} that for $n=6$, the orthogonality constraint indeed enforces the Grassmanian to localize to a $Gr(2,n)$ submanifold. Thus, the Grassmannian for $\mathcal{N}=4$ SYM achieves locality for the 6-point NMHV amplitude by choosing a particular ``tree-contour", while for 6-point ABJM amplitudes, the Grassmannian achieves locality by subjecting itself to the orthogonal constraint. The invariant between the two cases is the $Gr(2,n)$ 
submanifold, which was previously \cite{Spradlin:2009qr,ArkaniHamed:2009dg} linked to Witten's twistor string formulation \cite{Witten:2003nn}. So is there perhaps a 3d twistor string theory? A twistor-like string theory with target space $SU(2,3|5)$ was constructed in \cite{RaduTwistor} and it reproduces the Gr$(2,n)$ formula of the ABJM amplitudes. It is quite fascinating how the study of scattering amplitudes reveals the existence of a new twistor string theory!

%%%%%%%%%%%%%%%%%%%%%%%%%%%%%%% 
%%%%%%%%%%%%%%%%%%%%%%%%%%%%%%% 
%%%%%%%%%%%%%%%%%%%%%%%%%%%%%%% 
\newpage
\setcounter{equation}{0}
\section{Supergravity amplitudes}
\label{s:sugra}
%%%%%%%%%%%%%%%%%%%%%%%%%%%%%%% 
%%%%%%%%%%%%%%%%%%%%%%%%%%%%%%% 
%%%%%%%%%%%%%%%%%%%%%%%%%%%%%%% 
%%%%%%%%%%%%% 
We have seen that on-shell methods are particularly powerful for theories with gauge redundancy. Gravity has in a sense even more redundancy because of the diffeomorphism invariance. So perhaps there are hidden structures waiting to be discovered in gravity amplitudes?  In this Section, we discuss what is currently known about the scattering amplitudes in perturbative  supergravity theories, including their UV behavior, and we  review the interesting connections between gauge theory amplitudes and gravity amplitudes, relations that are often phrased loosely as ``gravity = (gauge theory)$^2$''. 

%%%%%%%%%%%%%%%%%%%%%%%%%%%%%%% 
%%%%%%%%%%%%%%%%%%%%%%%%%%%%%%% 
\subsection{Perturbative gravity}
\label{s:gravity}
%%%%%%%%%%%%%%%%%%%%%%%%%%%%%%% 
%%%%%%%%%%%%%%%%%%%%%%%%%%%%%%% 
In a typical course on General Relativity you learn about Einstein's equation and its solutions, for example the Schwarzchild black hole and Friedmann-Robertson-Walker cosmology. (If you have a hot course, you'll also learn about black rings.) These are solutions to the classical equations of motion of gravity, just as the point-particle Coulomb potential, electromagnetic waves, or Dirac monopoles are solutions to the Maxwell equations in electromagnetism. Here we are interested in the scattering of perturbative states at weak coupling. 
From the point of view of perturbation theory,  monopoles and black holes are considered non-perturbative states that are typically suppressed by powers $e^{-1/g^2}$ in the weak-coupling $g \ll 1$ scattering processes.

As you know well from your QFT courses, scattering amplitudes are obtained after quantization of the field theory: start with the Lagrangian, extract the Feynman rules, and off we go to calculate scattering  perturbatively. Of course, in the previous 190-something pages, we have tried to convince you that recursion relations and other on-shell methods offer much more insight and efficiency than the good old Feynman rules, but to understand what we mean by {\bf \em perturbative gravity}, let us start with the Lagrangian approach and Feynman rules. This will also give us  a greater appreciation for powers of the modern on-shell methods. 

The Einstein equation, $G_{\mu\n} = 8\pi T_{\m\n}$,  is the classical equation of motion that follows from the variational principle applied to the {\bf \em Einstein-Hilbert action}
\be
   S_\text{EH}
    = \frac{1}{2\kappa^2} \int d^D x \,\sqrt{-g}\, R 
   ~+~S_\text{matter}\,,
   \label{EHaction}
\ee
where $R$ is the Ricci scalar and $2\kappa^2 = 16 \pi G_N$. We have written the action in $D$ spacetime dimensions with a $D$-dimensional Newton's constant $G_N$. The metric $g_{\m\n}(x)$ is a field in the field theory \reef{EHaction}. The  variation $\delta g_{\m\n}$ of $\sqrt{-g}\,R$ gives (after partial integration and a little work \cite{WeinbergGR,Wald:1984rg,Carroll:2004st}) the Einstein tensor part, $G_{\mu\n} = R_{\mu\n} - \tfrac{1}{2} g_{\mu\n} R$,  of Einstein's equation, while the metric variation of the ``matter" action in \reef{EHaction} it gives the stress-tensor part, $T_{\m\n} = \frac{2}{\sqrt{-g}} \frac{\delta S_\text{matter}}{\delta g^{\mu\nu}}$. In the following, we use the term {\bf \em pure gravity} to describe the field theory \reef{EHaction} without matter fields, $S_\text{matter} = 0$.

Quantum field theory in curved spacetime is a highly non-trivial and interesting subject which has important consequences such as Hawking radiation of black holes. But this is not what we are going to discuss here. Our focus is the application of standard quantum field theory in flat spacetime to scattering of the particles associated with the quantization of the gravitational field $g_{\m\n}$. More precisely, we expand the gravitational field around flat space $g_{\m\n} = \eta_{\m\n} + \kappa h_{\m\n}$ and regard the fluctuating field $h_{\m\n}$ as the {\bf \em graviton field}. 
To start with, let us just consider pure gravity without matter and expand the Einstein-Hilbert action in powers of $\kappa h_{\m\n}$. Since the Ricci-scalar $R$ involves two derivatives, every term in the expansion has two derivatives. Suppressing the increasingly intricate index-structure, we write these terms schematically  as $h^{n-1}\partial^2 h$ for $n=2,3,4,\dots$, so that the action becomes
\be
    S_\text{EH}
    = \frac{1}{2\kappa^2} \int d^D x \,\sqrt{-g}\, R 
    =  \int d^D x \,
    \Big[ 
      h \pa^2 h + \kappa  \,h^2 \pa^2 h
      + \kappa^2  \,h^3 \pa^2 h
      + \kappa^3  \,h^4 \pa^2 h + \dots
    \Big]\,.
    \label{EHaction2}
\ee
There are infinitely many terms. There are two reasons for this:  
(a) in $R$, the series expansion of the inverse metric  generates an infinite series, and 
(b) the expansion of the determinant  $g =\det g_{\m\n} $ is finite, but the square root in $\sqrt{-g}$ generates an infinite series.\footnote{By a field redefinition, we can use $g_{\m\n} = e^{-h_{\m\n}}$ instead; this brings the metric and its inverse on an equal footing and therefore offers a simpler expansion \cite{unpubl-ef}.} There are no mass terms in \reef{EHaction2}, so the particles associated with quantization of the gravitational field $h_{\m\n}$  are massless: they have spin-2 and are called {\bf\em gravitons}. 

In order to extract Feynman rules from \reef{EHaction2} we first have to gauge fix the action. A typical choice is {\bf \em de Donder gauge}, $\pa^\m h_{\m\n} = \frac{1}{2} \pa_\n h_\m{}^\m$, which brings the quadratic terms in the action to the form
\be
   h \pa^2 h 
   ~\to~
    -\frac{1}{2} h_{\m\n} \Box h^{\m\n}
    + \frac{1}{4} h_{\m}{}^\m \Box  h_\n{}^\n \,.
\ee
The propagator resulting from these quadratic terms is
\be
  P_{\m_1 \n_1, \m_2 \n_2}
   = -\frac{i}{2} 
   \Big(
     \eta_{\m_1\m_2}  \,\eta_{\n_1\n_2}
   +\eta_{\m_1\n_2}  \,\eta_{\n_1\m_2}
   - \frac{2}{D-2} \, \eta_{\m_1 \n_1}\, \eta_{\m_2 \n_2}
   \Big)
   \frac{1}{\,\,k^2}\,.
   \label{dedonder}
\ee
Each graviton leg is labelled by two Lorentz-indices. The external line rule is to dot in graviton polarization vectors. In 4d, the polarizations encode the two helicity $h=\pm 2$ physical graviton states. They can be constructed as products of the spin-1 polarization vectors \reef{shpolar}:
\be
  e_-^{\m\n}(p_i) = \eps^\m_-(p_i) \eps^\n_-(p_i) \,,
  \hspace{1cm}
  e_+^{\m\n}(p_i) = \eps^\m_+(p_i) \eps^\n_+(p_i) \,.
  \label{gravpol}
\ee
Note that this ensures the correct little group scaling $t^{-2h_i}$ of the on-shell graviton scattering amplitude. 

The infinite set of 2-derivative interaction terms $h^{n-1} \pa^2 h$ yield Feynman rules for $n$-graviton vertices for {\em any} $n=3,4,5,\dots$. For example, the de Donder gauge 3-vertex takes the form
\be
  V_{3}(p_1,p_2,p_3) = p_1^{\m_3} p_2^{\n_3} \eta^{\m_1 \n_2} \eta^{\m_2 \n_1}
   + \text{(many other terms with various index-structures)}\,.
\ee
You can look up the full expression for the 3-vertex in \cite{gravityQFT}.

The 3-term de Donder propagator \reef{dedonder} and the infinite set of complicated interaction terms should make it clear that calculation of even tree-level graviton scattering amplitudes from Feynman diagrams is not a business for babies. The 4-point graviton tree amplitude was calculated brute force with Feynman diagrams in \cite{sannan} where each of the four contributing Feynman diagrams is about a page or so of elaborate  index-delight. Nonetheless, the final result can be brought to a very simple form: in 4d, it can be written in spinor helicity formalism as
\be  
   M^\text{tree}_4(1^- 2^- 3^+ 4^+)  ~=~ \frac{\<12\>^7 [12]}{\<13\>\<14\>\<23\>\<24\>\<34\>^2}
   ~=~
   \frac{\<12\>^4[34]^4}{stu}
   \,.
  \label{gravM4}
\ee
We already encountered this expression in Exercise \ref{ex:game}. We will be using $M_n$ to denote (super)gravity amplitudes to distinguish them from (super) Yang-Mills  amplitudes $A_n$.

Of course, you already know where we are headed: on-shell methods and recursion relations make the calculation of tree-level graviton scattering amplitudes much more fun and efficient --- and it has the power to reveal structures in the amplitudes that were not visible at the level of the Lagrangian. The short version of the story is that little group scaling fixes the possible 3-graviton amplitudes and recursion then allows you to compute all other tree-level  graviton processes. Loop-level amplitudes can be addressed with unitarity techniques (Section \ref{s:loops}). Thus the infinite set of interaction terms in the Lagrangian are not needed from the point of view of the on-shell scattering amplitudes: their role in life is to ensure off-shell diffeomorphism invariance of the gravitational action. It is an interesting aspect of on-shell recursion relations that they eliminate the need for infinitely many interaction terms.

Let us specialize to $D=4$ and  be more explicit about the graviton scattering amplitudes. 
Dimensional analysis and little group scaling fix the 3-point graviton amplitudes  to be
\be
 \begin{split}
 M_3(1^- 2^- 3^+)  &=~ \frac{\<12\>^6}{\<23\>^2 \<31\>^2} = A_3[1^-2^-3^+]^2 \,,\\
  M_3(1^+ 2^+ 3^-)  &=~ \frac{[12]^6}{[23]^2 [31]^2} ~\,= A_3[1^+2^+3^-]^2 \,.
 \end{split}
 \label{M3}
\ee
The graviton amplitudes with all-plus or all-minus helicity arrangements vanish in pure gravity at tree-level as do those with just one $\pm$-helicity:
\be
   M^\text{tree}_n(1^+ 2^+ \dots n^+) = M^\text{tree}_n(1^- 2^+ \dots n^+)
   = M^\text{tree}_n(1^+ 2^- \dots n^-) = M^\text{tree}_n(1^- 2^- \dots n^-) 
   = 0\,.
   \label{gravallplus}
\ee
This is most easily proven using the supersymmetry Ward identities, just as we did in \reef{susypm}-\reef{YMallplustree} for gluon amplitudes. The tree gravity amplitudes have to obey these same Ward identities as in a supergravity theory because the supersymmetric partners couple quadratically; hence it is only at loop-level the pure graviton amplitudes can distinguish themselves from the supergravity amplitudes. In particular, \reef{gravallplus} has to hold at tree-level. 
\exercise{}{For simplicity, we dropped the explicit powers of the gravitational coupling $\kappa$ in \reef{gravM4} and \reef{M3}, and we continue to do so henceforth. What is the mass dimension of $\kappa$ in 4d? Show that the 4-graviton amplitude \reef{gravM4} has the correct mass dimension (cf.~\reef{massdim4d}).
}
We categorize graviton amplitudes the same way as gluon amplitudes with designation N$^K$MHV. An important difference is that the graviton scattering amplitudes are not color-ordered. 
Using BCFW recursion relations, relatively compact graviton amplitudes can be found for the MHV sector. One of the earliest formulas is BGK (Berends, Giele and Kuijf) \cite{Berends:1988zp} written here in the form presented in \cite{Bern:2007xj} valid for $n>4$:
\be
   M_n^\text{tree}(1^- 2^- 3^+ \dots n^+)
   =\!\!
   \sum_{P(3,4,\dots,n-1)} 
  \frac{
  \<12\>^8 \prod_{l=3}^{n-1} \<n|2+3+\dots+(l-1)|l]}
  {\big(\prod_{i=1}^{n-2} \<i,i+1\>\big) \<1,n-1\> \<1n\>^2 \<2 n\>^2
    \big(\prod_{l=3}^{n-1} \<l n\>\big)}\,.
\ee
The sum is over all permutations of the labels $(3,4,\dots,n-1)$.

Another form of the same MHV graviton amplitude makes the relationship with gauge theory ``squared" more manifest:
\be
  M_n^\text{tree}(1^- 2^- 3^+ \dots n^+)
  =\!\!
   \sum_{P(i_3,i_4,\dots,i_n)} 
   s_{1i_n}
   \bigg(
      \prod_{k=4}^{n-1} \beta_k 
   \bigg)
   A_n^\text{tree}\big[1^- 2^- i_3^+ i_4^+\dots i_n^+\big]^2
   \,,
   \label{mmrec}
\ee
where $n \ge 4$ and 
\be
  \beta_k = 
  - \frac{\<i_k \, i_{k+1}\>}{\<2  i_{k+1}\>} 
  \<2| i_3 + i_4 + \dots + i_{k-1}|i_k]\,.
\ee
The result \reef{mmrec} can be derived \cite{Bedford:2005yy,Elvang:2007sg} using a $[-,-\>$ BCFW-shift. 

There are also other graviton MHV formulas available in the literature, for example the ``soft-factor" formula \cite{Nguyen:2009jk}. You may find that these MHV expressions are terribly complicated compared with Parke-Taylor; however, they are remarkably simple when compared with the mess a Feynman diagram calculation would produce.

Beyond the MHV level, one can readily use BCFW to calculate explicit results. You might be curious if there is also a CSW-like expansion for gravity amplitudes. The MHV vertex expansion \cite{BjerrumBohr:2005jr} based on the Risager-shift (see discussion below Exercise \ref{ex:cswshift}) works for NMHV graviton amplitudes with $n<12$ particles. It fails  \cite{Bianchi:2008pu} for $n\ge 12$ because the large-$z$ falloff of the $n$-point amplitude under Risager-shift  is $1/z^{12-n}$ and the Cauchy contour deformation argument needed to derive the recursion relations therefore picks up a term at infinity for $n>11$. The all-line shift discussed in Section \ref{s:csw} also fails (for interesting reasons \cite{CEK}). For further discussion of CSW for gravity, see \cite{Bianchi:2008pu,Benincasa:2007qj,Conde:2012ik,CEK}.

The relation between gravity and gauge theory amplitudes is clearly visible in the 4d MHV expressions \reef{M3} and \reef{mmrec}, but are there are more general relations available. The first such example are the 
{\bf \em KLT relations}, derived in string theory by Kawai, Lewellen and Tye \cite{Kawai:1985xq}: the KLT relations state that the $n$-point tree-level closed string scattering amplitude is related to a sum over products of $n$-point open string string partial amplitudes, with coefficients that depend on the  kinematic variables as well as the string tension $1/(2\pi \alpha')$. This is natural, albeit non-trivial, since the closed string vertex operators are products of open string vertex operators. The non-triviality of the KLT relations is that the factorization into open string amplitudes survives the integrals over the insertion points of the vertex operators.  In the limit of infinite tension, $\alpha' \to 0$, the closed string amplitudes with massless spin-2 string external states become the regular graviton scattering amplitudes $M_n$ we have discussed above. And in this limit, the open-string partial amplitudes with external  massless spin-1 states become the color-ordered gluon amplitudes $A_n$. Thus, in the limit $\alpha' \to 0$,  KLT offers a relationship between tree-level $M_n$ and $A_n$ for each $n$. For $n=4,5$, the field theory KLT relations are
\be
  \begin{split}
  M^\text{tree}_4(1234) 
  &=~ - s_{12}  \,A^\text{tree}_4[1234] \,A^\text{tree}_4[1243]\,,\\
  M^\text{tree}_5(12345) &=~  
  s_{23} s_{45}  \,A^\text{tree}_5[12345] \,A^\text{tree}_5[13254] + (3 \lra 4)\,,\\
  M^\text{tree}_6(123456) &=~    
  - s_{12} s_{45}  A^\text{tree}_6[123456] 
  \Big(
    s_{35} A^\text{tree}_6[153462] 
    +(s_{34}+s_{35})A^\text{tree}_6[154362] 
  \Big) \\
  & \hspace{1cm}
    + \mathcal{P}(2,3,4)\,.
  \end{split}
  \label{KLT45}
\ee  
In the 6-point case, $\mathcal{P}(2,3,4)$ stands for the sum of all permutations of legs $2,3,4$. 
At 7-point and higher, the KLT relations are more complicated; they can be found in Appendix A of \cite{Bern:1998sv}. The relation between gravity and gluon scattering is not at all visible in the Lagrangian \reef{EHaction2}, although field redefinitions and clever gauge choices can bring the first few terms in the gravitational action into a more KLT-like form; see \cite{Bern:1999ji,Bern:2000mf,Siegel:1993xq} and the review \cite{Bern:2002kj}.

Note that there is no specification of helicities of the external states in \reef{KLT45}: this is because the above relation is valid in $D$-dimensions. In 4d, the KLT relations work for any helicity assignments of the gravitons on the LHS; if the $i$'th graviton has helicity $h_i = +2$, then the  gluons labeled $i$ in the amplitudes on the RHS have helicity $h_i = + 1$; similarly for negative helicity. This ensures that the little group scaling works out on the both sides of the KLT relations. We may then say that KLT in 4d uses
\be
  \text{graviton}^{\pm 2}(p_i) 
  =  \text{gluon}^{\pm 1}(p_i) \otimes   \text{gluon}^{\pm 1}(p_i) \,.
  \label{gravglue}
\ee 
This is also encoded in the graviton polarizations \reef{gravpol}. 

Inspecting the relationship \reef{gravglue} between gravitons and gluons, we could also ask what happens when if we combine gluons of opposite helicity in the KLT relations.  
The result is something that has the little group scaling of a scalar on the gravity side. In fact, what you get is the {\bf \em dilaton and axion}:
\be
  \begin{array}{l}
  \text{dilaton} \\ \text{axion}
  \end{array}
  \bigg\}  
  =  \text{gluon}^{\pm 1}(p_i) \otimes   \text{gluon}^{\mp 1}(p_i) \,.
  \label{dilaxglue}
\ee 
This is completely natural from the string theory point of view where the graviton state comes together with an antisymmetric tensor $B_{\m\n}$ and a scalar ``trace'' mode. The latter is the dilaton and the former has a 3-form field strength $H=dB$ which means that in 4d it is dual to a scalar, the axion. Therefore we can write the relation between the spectra
\be
  \text{4d axion-dilaton gravity}
  ~=~
  \text{(YM theory)} \otimes \text{(YM theory)}\,. 
  \label{dilaxgrav}
\ee 

Given the attention we have poured into the study of amplitudes in (planar) $\cn=4$ SYM in 4d, you may also be curious about what we would get if we tensor'ed the $2^4$ states of $\cn=4$ SYM {\em a la}  \reef{dilaxgrav}. The answer is a very good one: we get the $2^8$ states of $\cn=8$ supergravity, which is the 4d supergravity theory with maximal supersymmetry. Supergravity amplitudes, especially those in  $\cn=8$ supergravity, are the main focus in the following. We return to the study of ``gravity = (gauge theory)$^2$'' in Section \ref{s:BCJ}, though you will see more of it in the following sections too.

%%%%%%%%%%%%%
\subsection{Supergravity}
\label{s:sugraspec}
Supergravity is the beautiful union of gravity and supersymmetry. It is the result of making the supersymmetry transformations local in the sense that the SUSY parameter $\eps$ is spacetime dependent. If you have not previously studied supergravity, you should immediately read \cite{Freedman:1994pr} and then textbooks such as \cite{Freedman:2012zz,Wess:1992cp,Gates:1983nr}.

The supersymmetry partner of the graviton is called the {\bf \em gravitino}. It has spin-$\tfrac{3}{2}$ and (when  supersymmetry is unbroken) it is massless. In 4d we characterize a gravitino by its two helicity states $h=\pm \tfrac{3}{2}$; its Feynman rule for the external line simply combines a $\pm$-helicity spin-1 polarization vector with the $\pm$-helicity spin-$\tfrac{1}{2}$ fermion wavefunction. 

In a 4d supergravity theory with $\cn$ supercharges, $Q^A$ and $\widetilde{Q}_A$, the graviton has $\cn$ gravitino-partners. We can construct the spectrum by starting with the negative helicity graviton $h^-$ as the highest-weight state and apply the supercharges $\widetilde{Q}_A$. Each $\widetilde{Q}_A$ raises the helicity by $\tfrac{1}{2}$ at each step, so when $\widetilde{Q}_A$ is applied to $h^-$ it produces a negative helicity gravitino  $\psi^-_A$.  If $\cn=1$, then the  process terminates because of the Grassmann nature of the supercharge. So  the 
{\bf \em $\cn=1$ pure supergravity} multiplet consists of the two CPT conjugate pairs of graviton+gravitino:
\be
  \mathcal{N}=1~\text{supergravity:}~~~~~
  (h^{-}, \psi^-) ~~~~\text{and}~~~~(\psi^+, h^{+})\,.
\ee 
When we say {\bf \em pure supergravity} we mean that there are no other matter-supermultiplets included; we only have the states that are related to the graviton via supersymmetry. 

Pure {\bf \em $\cn=2$ supergravity} has $2 \times 2^2$ states
\be
  \mathcal{N}=2~\text{supergravity:}~~~~~
  (h^{-}, \psi_A^-,v^-) ~~~~\text{and}~~~~(v^+,\psi^{A+}, h^{+})\,,
\ee 
where the two gravitinos $\psi_A^-$ and $\psi^{A+}$ are labeled by $A=1,2$ and $v^\pm$ denotes the two helicity states of the spin-1 {\bf \em gravi-photon}. 

Fast-forward to pure {\bf \em  $\cn=4$ supergravity}. Its $2 \times 2^4$ states can be characterized as 
\be
      \mathcal{N}=4~\text{supergravity}
      = \big(\cn=4~\text{SYM}\big) \otimes \big(\cn =0~\text{(S)YM}\big)
      \,.
      \label{N4SGspectrum}
\ee
By $\cn =0$~(S)YM we just mean pure Yang-Mills theory.
The spectrum \reef{N4SGspectrum} should be read as follows: the 2 gravitons are given in terms of the gluon states as in \reef{gravglue}. Using the spectrum \reef{n4ops} of $\cn=4$ SYM, we find:
\be
  \begin{array}{lcl}
   \text{gravitons:}&&h^{^\pm} = g^\pm \otimes g^\pm\\
   \text{gravitinos:}&& \psi^{A+} =  \lambda^{A+} \otimes g^+
     ~~\text{and}~~\psi_{A}^- =  \bar{\lambda}_{A}^- \otimes g^{-}\\
   \text{gravi-photons:}&& v_{AB}^{\pm}=S_{AB} \otimes g^\pm\\
    \text{gravi-photinos:}&& \psi^{A-} =  \lambda^{A+} \otimes g^- 
      ~~~\text{and}~~\psi_{A}^+ =  \bar{\lambda}_{A}^- \otimes g^+\\
    \text{scalars (dilaton-axion):}&&   g^\pm \otimes g^\mp\,,
  \end{array}
\ee
where $g^\pm$ are gluons, $\bar{\lambda}_{A}^-$ and  $\lambda^{A+}$ are gluinos, and $S_{AB}$ are the 6 scalars of $\cn=4$ SYM.  Totaling up the states, we get $2 \times 1 + 2 \times 4  + 2 \times 6 + 2 \times 4 +2 = 32$.
\exercise{}{Identify the supermultiplets in the theory whose spectrum is 
$\big(\cn=2~\text{SYM}\big) \otimes \big(\cn =0~\text{YM}\big)$. 

What is the difference between the two $\cn=4$ supergravity theories whose spectra are $\big(\cn=4~\text{SYM}\big) \otimes \big(\cn =0~\text{(S)YM}\big)$ and $\big(\cn=2~\text{SYM}\big) \otimes \big(\cn =2~\text{SYM}\big)$?}
Applying the $\cn$ supersymmetry generators $\widetilde{Q}_A$ to the graviton top state $h^-$ we see that if $\cn>8$ we cannot avoid states with spin greater than 2. There are no consistent interactions in flat space for particles with spin greater than 2, so that tells us that maximal supersymmetry in 4d is $\cn=8$. The  {\bf \em $\cn=8$ supergravity} theory is unique: the ungauged theory, which is our focus here,  was first written down in \cite{deWit:1977fk,CremmerJulia}.\footnote{The gauged $\cn=8$ supergravity theory was presented in \cite{deWit:1982ig}.} Its spectrum of $2^{8}$ states  form a CPT-self-conjugate supermultiplet (just like in $\cn=4$ SYM). As noted at the end of Section \ref{s:gravity}, the spectrum can be characterized as
\be
      \mathcal{N}=8~\text{supergravity}
      \,=\, \big(\cn=4~\text{SYM}\big) \otimes \big(\cn=4~\text{SYM}\big)
      \,.
      \label{N8SGspectrum}
\ee

In any supergravity theory, there are supersymmetry Ward identities that restrict the amplitudes, just as in discussed for gauge theories in Sections \ref{s:SWI} and \ref{s:N4sym}. In particular, the {\em  graviton amplitudes in supergravity} satisfy
\be
   M_n(1^+ 2^+ \dots n^+) = M_n(1^- 2^+ \dots n^+)
   = M_n(1^+ 2^- \dots n^-) = M_n(1^- 2^- \dots n^-) 
   = 0\,
   \label{gravallplus2}
\ee
at all orders in perturbation theory. There are also simple Ward identities  among graviton and gravitino MHV amplitudes that give
\be
  M_n(1^- \psi^- \psi^+  4^+ \dots n^+) 
  = 
  \frac{\<13\>}{\<12\>}
  M_n(1^- 2^- 3^+ 4^+ \dots n^+) \,,
\ee
just as for gluons and gluinos. In extended ($\cn>1$) supergravity there are  further relations, as you will see shortly from the superamplitudes in $\cn=8$ supergravity.

%%%%%%%%%%%%%%%%%%%%%%%%%%%%%%%%%%%%%%%
\subsection{Superamplitudes in $\mathcal{N}=8$ supergravity}
\label{s:N8SG}
%%%%%%%%%%%%%%%%%%%%%%%%%%%%%%%%%%%%%%%

The spectrum \reef{N8SGspectrum} of $\cn=8$ supergravity consists of 128 bosons and 128 fermions. Organized by helicity $h= 2, \tfrac{3}{2}, 1,  \tfrac{1}{2},0,- \tfrac{1}{2},-1, -\tfrac{3}{2},-2$, we can write it out as
\be
  \begin{array}{c}
     \text{1 graviton}~h^+, ~~~~
     \text{8 gravitinos}~\psi^{A}, ~~~~
     \text{28 gravi-photons}~v^{AB},    \\[2mm]
     \text{56 gravi-photinos}~\chi^{ABC}, ~~~~    
     \text{70 scalars}~S^{ABCD}, ~~~~   
     \text{56 gravi-photinos}~\chi^{ABCCDE}, \\[2mm]
     \text{28 gravi-photons}~v^{ABCDEF}, ~~~~    
     \text{8 gravitinos}~\psi^{ABCDEFG}, ~~~~
     \text{1 graviton}~h^- = h^{12345678}.
  \end{array}
      \label{N8SGspectrum2}
\ee
Here $A,B,\ldots =1,2,\ldots,8$ are $SU(8)$ R-symmetry indices and each state above is fully antisymmetric in these labels; this simply reflects that the helicity-$h$ state transforms in the rank $r=4-2h$ fully antisymmetric irrep of $SU(8)$
 and the multiplicity given in \reef{N8SGspectrum2} is the dimension  of the irrep. The 70 scalars are self-dual and satisfy $\overline{S}_{ABCD} = \tfrac{1}{4!} \eps_{ABCDEFGH}S^{EFGH}$. Supersymmetry generators $Q^A$ and $\widetilde{Q}_A$ act on the states in an obvious generalization of  \reef{n4tQ}. 

Just as in $\cn=4$ SYM it is highly convenient to combine the states into a superfield, or super-wavefunction, with the help of an on-shell superspace with Grassmann variables $\eta_{iA}$ whose $i=1,\dots,n$ is a particle label and $A=1,2,\dots,8$ is a fundamental $SU(8)$ R-symmetry index. The $\cn=8$ superfield is then
\be
 \Phi_i = h^+ +  \eta_{iA} \,\psi^{A} 
 -\frac{1}{2} \eta_{iA}\eta_{iB}\,v^{AB}
 +\ldots 
 + \eta_{i1}\eta_{i2} \eta_{i3} \eta_{i4} \eta_{i5} \eta_{i6}
 \eta_{i7} \eta_{i8}\, h^-\,.
 \label{N8sfield}
\ee
The $SU(8)$ R-symmetry requires that the superamplitudes are degree $8k$ polynomials in the Grassmann variables. This directly gives us the $\cn=8$ supergravity version of the N$^K$MHV classification: the $K$'th sector contains the superamplitudes of degree $8(K+2)$ polynomials in the $\eta_{iA}$'s. It should be clear from \reef{N8sfield} that the MHV sector ($K=0$) includes the graviton component  amplitude  $M_n(1^- 2^- 3^+ \dots n^+)$.

The super-Poincare generators --- momentum $P^{\dot{a}b}$,  rotations/boosts, and the supercharges $Q^A$ and $\widetilde{Q}_A$ are given in 
\reef{PMM} and \reef{bigQTQ}, with the only difference that now  $A=1,2,\dots,8$. Momentum- and supermomentum conservation requires that the general superamplitudes in $\cn=8$ supergravity are  
\be
  \mathcal{M}^\text{N$^K$MHV}_n = \delta^{4}\big(P\big) \, 
  \delta^{(16)}  \big( \widetilde{Q}\big)\,
  P_{8K}\,,
  \label{N8samp}
\ee
where $P_{8K}$ is annihilated by $Q^A$ which acts by differentiation: $Q^A P_{8K} = 0$.

At the MHV level, we are already home safe. The Grassmann delta function \label{N8samp} eats up all 16 fermionic variables, so $P_{0}$ is $\eta$-independent. It can be fixed by requiring that $\mathcal{M}_n$ projects out the correct pure graviton MHV amplitude $M_n(1^- 2^- 3^+ \dots n^+)$. This is easily accomplished:
\be
    \mathcal{M}_n^\text{MHV} = \delta^{4}\big(P\big) \, 
  \delta^{(16)}  \big( \widetilde{Q}\big)\,
  \frac{M_n(1^- 2^- 3^+ \dots n^+)}{\<12\>^8}\,.
  \label{N8sampMHV}
\ee
Beyond the MHV level, one can solve the supersymmetric Ward identities 
$Q^A \mathcal{M}_n = \widetilde{Q}_A  \mathcal{M}_n = 0$ (just as in the $\cn=4$ SYM case) to find a basis of input-amplitudes that completely determine the full superamplitude. The basis can be labeled by the $K \times 8$ rectangular Young tableaux of $SU(n-4)$ irreps \cite{Elvang:2009wd}. 
Another approach is to use the super-BCFW recursion relations; they are valid for super-shift of any two lines \cite{ArkaniHamed:2008gz,Cheung:2008dn}.\footnote{A super-shift version of CSW was discussed in \cite{Kiermaier:2009yu} and while it works for all tree superamplitudes in $\cn=4$ SYM, it has more limited validity in $\cn=8$ supergravity.}

The set-up for the MHV superamplitude \reef{N8sampMHV} is perhaps a bit ``cheap" because the component amplitude $M_n(1^- 2^- 3^+ \dots n^+)$, as we have seen in Section \ref{s:gravity}, does not take a particularly compact form and it does not clearly reflect symmetries such as full permutation symmetry of identical external states. So there has been quite a lot of effort towards building an MHV superamplitude that more clearly encodes the symmetries. One representation \cite{Hodges:2012ym} of the superamplitude builds on a super-BCFW shift in the $\cn=7$ formulation of $\cn=8$ supergravity.\footnote{Just as $\cn=3$ SYM is identical to $\cn=4$ SYM, so is  $\cn=7$ supergravity identical to $\cn=8$ supergravity. The validity of the super-BCFW shifts in $\cn=7$ supergravity was proven in \cite{Elvang:2011fx}.} Other MHV formulas use the Grassmannian representations \cite{Cachazo:2012pz,He:2012er} or the twistor string  \cite{Cachazo:2012da,Skinner:2013xp}. Finally, very recently new compact formulas for both Yang-Mills and gravity amplitudes were proposed to be valid in any spacetime dimensions \cite{Cachazo:2013gna,Cachazo:2013hca}.
This is currently a subject of active research.

$\cn=8$ supergravity has, as we have noted above, a {\bf \em global $SU(8)$ R-symmetry}. This symmetry is realized linearly, as you can see on the spectrum and on the amplitudes which vanish unless the external states form an $SU(8)$ singlet. However, the theory also has a `hidden' symmetry: the equations of motion of $\cn = 8$ supergravity have a {\bf \em continuous global $E_{7(7)}(\mathbb{R})$ symmetry}. The group $E_{7(7)}$ is a non-compact version of the exceptional group $E_7$; its maximal compact subgroup is $SU(8)$. It has rank 7 and is 133 dimensional. It is \emph{not} a symmetry of the action of $\cn=8$ supergravity. The best way to think of this is that the $E_{7(7)}$ is spontaneously broken to $SU(8)$. There are $133-63 = 70$ broken generators, giving 70 Goldstone bosons. Those are exactly the 70 scalars $S^{ABCD}$ in the spectrum \reef{N8SGspectrum2}. 

As a spontaneously broken symmetry, $E_{7(7)}$ is not linearly realized on the on-shell scattering amplitudes, but instead it manifests itself via {\bf \em low-energy theorems}.\footnote{Low-energy theorems were originally developed in pion-physics \cite{Adler:1964um}. For a review, see \cite{Coleman:1974hr}.} If the momentum of an external scalar $S^{ABCD}$ is taken soft, then the amplitude must vanish because the Goldstone scalars are derivatively-coupled. Basically this says that the moduli space $E_{7(7)}/SU(8)$ is homogeneous: it does not matter what the vevs of the scalars are, all points on moduli space are equivalent. The soft scalar limit probes the neighborhood of a point in moduli space and since the moduli space is homogeneous, the soft scalar limit vanishes. There are also double-soft limits that involve the commutator of two coset generators and these therefore directly reveal, from the on-shell point of view, the coset structure $E_{7(7)}/SU(8)$. 
\exercise{ex:coulomb}{Project the amplitude $M_4\big(S^{1234} S^{5678} h^- h^+  \big)$ out from the MHV superamplitude \reef{N8sampMHV}. Show that 
\be
  \lim_{p_1 \to 0} M_4\big(S^{1234} S^{5678} h^- h^+\big)  = 0 \,.
\ee
In contrasts, note that the scalars in $\cn=4$ SYM are not Goldstone bosons, so the soft-scalar limits do not have to vanish. For example, show 
\be
   \lim_{p_1 \to 0} A_4\big[S^{12} g^- S^{34}  g^+\big] \ne 0\,.
\label{Ansoft}
\ee
The soft-limit explores the points of moduli space in the neighborhood of the origin: away from the origin, the $\cn=4$ SYM theory is on the Coulomb branch, part of the gauge group is broken, and some of the $\cn=4$ supermultiplets become massive.
The non-vanishing limit \reef{Ansoft} has a nice interpretation. Set $p_1 = \eps \, q$ for some lightlike $q = -|q\> [q|$ and take the soft limit as $\eps \to 0$. The limit \reef{Ansoft} then depends on $|q\>$. The soft limit $p_1 \to 0$ leaves an object with momentum conservation on 3 particles: the result can be interpreted as the  small-mass limit of the Coulomb branch amplitude 
$A_3\big[W^- S^{34}\,  W^+ \big]$, where $W^\pm$ are the longitudinal modes of the massive spin-1  $W$-bosons of a massive $\cn=4$ supermultiplet and $S^{34}$ is a massless scalar. From this point of view $q$ is a reference vector that allows us to project the massive momenta of $W^\pm$ such that the corresponding angle spinors are well-defined. This is actually also needed to define the helicity basis because helicity is not a Lorentz-invariant concept for massive particles; but $q$ breaks Lorentz-invariance and allows us to define a suitable $q$-helicity basis \cite{CEK,Craig:2011ws}. 

The 3-point amplitude $A_3\big[W^- S^{34}\,  W^+ \big]$ violates the $SU(4)$ R-symmetry of $\cn=4$ SYM at the origin of moduli space. This is fine, because the Coulomb branch breaks the R-symmetry. Minimally, one has 
$SU(4) \sim SO(6) \to SO(5) \sim Sp(4)$.

The moral of the story is that single-scalar soft limits for $\cn=8$ supergravity amplitudes  {\em vanish} because the 70 scalars are Goldstone bosons of $E_{7(7)} \to SU(8)$.
And that single-scalar soft limits for $\cn=4$ SYM are {\em non-vanishing} and reproduce the small-mass limit of the Coulomb branch amplitudes \cite{Craig:2011ws}. One can in fact re-sum  the entire small-mass expansion from multiple-soft-scalar limits and recover the general-mass Coulomb branch amplitudes \cite{Craig:2011ws,Kiermaier:2011cr}.
}

The single-soft scalar limits of tree-amplitudes in $\cn=8$ supergravity were first studied in \cite{Bianchi:2008pu}. Single- and double soft limits were discussed extensively and clarified in \cite{ArkaniHamed:2008gz}. The soft-scalar limits play a key role for us in Section \ref{s:CTs}.

%%%%%%%%%%%%%%%%%%%%%%%%%%%%%%%%%%%%%%%
\subsection{Loop amplitudes in supergravity}
\label{s:UVsg1}
%%%%%%%%%%%%%%%%%%%%%%%%%%%%%%%%%%%%%%%
It is taught in all good kindergartens that a point-particle theory of gravity  is badly UV divergent and non-renormalizable. This means that it is not a good quantum theory. 
So what is perturbative gravity all about?

Naive power-counting gives a clear indication that gravity with its 2-derivative interactions generically has worse UV behavior than for example Yang-Mills theory with its 1- and 0-derivative interactions. 
Consider for example a generic 1-loop $m$-gon diagram. In gravity, the numerator of the loop-integrand can have up to $2m$ powers of momenta, while in Yang-Mills theory it is at most $m$. Both have $m$ propagators, so in gravity this gives
\bea
 \text{gravity 1-loop $m$-gon diagram}  
  \sim
  \int^\Lambda d^4 \ell \,\frac{(\ell^2)^m}{(\ell^2)^m} 
  \sim 
  \Lambda^4 \,.
\eea
This is power-divergent as the UV cutoff $\Lambda$ is taken to $\infty$ for all $m$. On the other hand, for Yang-Mills theory the $m$-gon integral has at most $\ell^m$ in the numerator, so it is manifestly UV finite for $m>4$.

Now, the power-counting is too naive. There can be cancellations  within each diagram. Moreover, we have learned that we should not take individual Feynman diagrams seriously if they are not gauge invariant. So cancellations of UV divergences can take place in the sum of diagrams, rendering the on-shell amplitude better behaved than naive power-counting indicates.

In fact, pure gravity in 4d is actually finite at 1-loop order \cite{'tHooft:1974bx}: all the 1-loop UV divergences cancel! This is difficult to see by direct Feynman diagram calculations, but it follows trivially by absence of any valid counterterms. We will review this approach in detail in Section \ref{s:CTs}. 

At 2-loop order, it has been demonstrated by Feynman diagram calculations that pure gravity indeed has a divergence \cite{Goroff:1985sz,vandeVen:1991gw}. In Yang-Mills theory we are not too scared of divergences because we know how to treat them with the procedure of renormalization. However, in gravity, it would take an infinite set of local counterterms to absorb the divergences and hence the result is unpredictable: pure gravity is a non-renormalizable theory.

So what is the theory described by the Einstein-Hilbert action? Because it is non-renormalizable, it is not a well-defined theory of quantum gravity. Instead, we should regard the field theory defined by the Einstein-Hilbert action as an {\bf\em effective field theory}, valid at scales much smaller than the Planck scale $M_\text{Planck} \sim 10^{19}$\,GeV. To see this, recall that  the 4d gravitational coupling $\kappa$ has mass dimension $-1$. So when we  do perturbation theory, we should really use the dimensionless coupling $E \kappa$ where $E$ is the characteristic energy of the process. At high enough energies, this dimensionless coupling is no longer small and we cannot trust perturbation theory. So we should not extrapolate to such high energies. In energy units, $\kappa^{-1} \sim G_N^{-1/2} = M_\text{Planck}$, so this tells us to use gravity, as described by the Einstein-Hilbert action, for energies $E \ll M_\text{Planck}$. 
As a classical effective field theory, though, General Relativity is enormously successful and captures classical gravitational phenomena stunningly as shown by experimental tests.

Regarding gravity as an effective theory, we can study the low-energy perturbative amplitudes: the tree-amplitudes capture the classical physics and there are no divergences to worry about. At 1-loop level, we have mentioned that pure gravity is finite. Could we imagine adding matter to gravity in such a way that its higher-loop amplitudes were also finite?   Gravity with generic matter is 1-loop divergent \cite{'tHooft:1974bx,Deser:1974cz}, but  we know from gauge theories that supersymmetry improves the UV behavior of loop-amplitudes, even to such an extreme extent that the maximally supersymmetric Yang-Mills theory, $\cn = 4$ SYM, is UV finite: the UV divergences  cancel completely at each order in the loop expansion. Could something like that also happen in supergravity? If it did, it would eliminate the need for renormalization and the problems of non-renormalizability would be obsolete. There would still be important questions unresolved about non-perturbative aspects of supergravity; finiteness does not mean that the theory is UV complete.  The question of perturbative UV finiteness of (maximal) supergravity in 4d has received increased attention in the past few years and the on-shell amplitude techniques have facilitated multiple explicit calculations of supergravity loop amplitudes. It should be emphasized that 
 whether or not the perturbative calculations eventually encounter a divergence, one should appreciate that the study of loop amplitudes in supergravity has resulted in a number of new insights, of independent value, about gravity scattering amplitudes. An example is the connection between gravity and Yang-Mills amplitudes via the so-called BCJ dualities (see Section \ref{s:BCJ}).

Pure supergravity in 4d is better behaved in the UV than pure gravity: {\bf \em all pure supergravity theories in 4d are finite at 1-loop} \cite{Grisaru:1976ua} {\bf \em  and 2-loop order} \cite{Grisaru:1976nn,Tomboulis:1977wd,Deser:1977nt}, i.e.~the first possible UV divergence can appear only at 3-loop order, improving on the  2-loop UV divergence of pure gravity \cite{Goroff:1985sz,vandeVen:1991gw}. 
In the spirit of ``the more supersymmetry, the better", it is natural to focus on maximal supersymmetry, i.e.~$\cn=8$ supergravity in 4d. An explicit calculation, using the generalized unitarity method, demonstrated that the 3-loop 4-graviton amplitude is UV finite in $\cn=8$ supergravity in 4d \cite{Bern:2006kd,Bern:2007hh,Bern:2008pv}. This and related observations of unexpected cancellations motivated Bern, Dixon, and Roiban \cite{Bern:2006kd} to ask if $\cn=8$ supergravity in 4d is UV finite? They further proposed that the critical dimension $D_c$ for the first UV divergence of maximal supergravity  in $D$-dimensions follows the same pattern as for maximal super Yang-Mills theory \cite{Bern:1998sv,Howe:2002ui}, namely 
\be
   D_c(L) = \frac{6}{L} + 4 
   ~~~~\text{for}~~~~L>1\,.
   \label{YM-Dc}
\ee
It was then shown \cite{Bern:2009kd,Bern:2010tq} that the 4-loop 4-graviton amplitude is UV finite in $\cn=8$ supergravity in 4d and that it follows the pattern \reef{YM-Dc}. How about 5-loops? A pure-spinor based argument \cite{Bjornsson:2010wm} leads to \reef{YM-Dc} for $L=2,3,4$, but implies that the critical dimension for $L=5$ is $D=24/5$ and not $26/5$ as \reef{YM-Dc} predicts.
This question can be settled by direct computation: at the time of writing, the 5-loop calculation is still in progress, so you'll have to watch the ArXiv for the resolution.

For $D=4$, the symmetries of $\cn=8$ supergravity can be used to establish that 
 \emph{all} amplitudes of the theory are UV finite for $L\le 6$: this explains the finiteness of the 3- and 4-loop 4-graviton amplitudes and predicts that no UV divergence appears in any other amplitude for  $L\le 6$. For $L \ge 7$, the known symmetries do not suffice to rule out UV divergences. 
The following Section reviews how these results are obtained using an on-shell amplitude-based approach \cite{Elvang:2010jv,Elvang:2010kc,Beisert:2010jx,Elvang:2010xn} to counterterms in $\cn=8$ supergravity. We then provide in Section \ref{s:UVsg2} an overview of the current status of the UV behavior of supergravity as a function of dimensions  $D$, supersymmetries $\cn$, and loop order $L$.

%%%%%%%%%%%%%%%%%%%%%%%%%%%%%%%%%%%%%%%
\subsection{$\cn=8$ supergravity: loops and counterterms}
\label{s:CTs}
%%%%%%%%%%%%%%%%%%%%%%%%%%%%%%%%%%%%%%%

Suppose that a supergravity has its first UV divergence in an $n$-point amplitude at $L$-loop order. Then the effective action for the theory must have a {\bf \em local diffeomorphism invariant counterterm} constructed from $n$ fields (corresponding to the $n$ external states) and $(2L+2)$ derivatives. The latter statement follows from dimensional analysis because the gravitational coupling $\kappa$ has mass dimension $-1$: for given $n$, the ratio of the $L$-loop supergravity amplitude to the tree-amplitude has an overall factor of $\kappa^{2L}$, so the corresponding local counterterm has to make up the mass-dimension by having $2L$ more derivatives than the 2-derivative tree-level theory.\footnote{To be a little more precise, the above statement is true for amplitudes with purely bosonic fields. Since external fermions dress the amplitude with dimensionful wavefunctions, each pair of fermions count one derivative for the purpose of dimensional analysis.} 
\exercise{}{Show that the $n$-graviton $1$-loop amplitude has an overall factor of $\kappa^2$ compared with the $n$-graviton tree amplitude. 
}
If we consider just pure gravity, the possible local diff-invariant counterterms must be Lorentz scalars formed from contractions of Riemann-tensors and possibly covariant derivatives. Each Riemann tensor contributes 2-derivatives, so at 1-loop 
(4-derivatives), the possible candidates are $\sqrt{-g}R^2$, 
$\sqrt{-g}R_{\m\n} R^{\m\n}$, and 
$\sqrt{-g} R_{\m\n\rho\sigma} R^{\m\n\rho\sigma}$. 
If we suppress the index contractions, we can write schematically  
\be
  \begin{array}{rcl}
  S_\text{eff} &=&
  \displaystyle
   \frac{1}{2\kappa^2}
  \displaystyle
  \int d^4 x \, \sqrt{-g} \,\Big(
  \underbrace{\,R\,}_{L=0}
  \,+\, \underbrace{\,\,\kappa^2 R^2\,}_{L=1}  
 \,+\, \underbrace{\,\,\kappa^4 R^3\,}_{L=2}  
  \,+\, \underbrace{\,\,\kappa^6 R^4\,}_{L=3}  
 \,+\,\underbrace{\,\,\kappa^6 \big( D^2 R^4 + R^5\big) \,}_{L=4}  \\
 &&\hspace{3cm}
 \,+\,\underbrace{\,\,\kappa^8 \big( D^4 R^4 + D^2 R^5 + R^6 \big) \,}_{L=5}  
   \,+\,\dots
  \Big) ,
  \end{array}
  \label{Seff}
\ee
where $R$ denote Riemann tensors and $D$ covariant derivatives. This should be viewed as a list of possible candidate counterterms; the operators in \reef{Seff} are not necessarily generated in perturbation theory.
\exercise{}{Show that operators of the form $D^{2k} R^3$ have vanishing 3-point matrix elements for $k\ge 1$.}
Now since we consider on-shell amplitudes, we can enforce the equations of motion on the candidate counterterms. In pure gravity, the Einstein equation gives $R_{\m\n} = 0$, so this leaves $\sqrt{-g} R_{\m\n\rho\sigma} R^{\m\n\rho\sigma}$ as the only possibility at 1-loop. However, we are free to add zero to convert $\sqrt{-g} R_{\m\n\rho\sigma} R^{\m\n\rho\sigma}$ to the Gauss-Bonnet term $\sqrt{-g} \big( R_{\m\n\rho\sigma} R^{\m\n\rho\sigma} - 4 R_{\m\n} R^{\m\n} + R^2\big)$ which equals a total derivative. Therefore there is 
{\bf \em no local counterterm for pure gravity at 1-loop order!} And since there is no counterterm, {\bf \em  pure gravity is not UV divergent at 1-loop}. A cleaner way to say this is that one can  do a field redefinition that changes $\sqrt{-g} R_{\m\n\rho\sigma} R^{\m\n\rho\sigma}$ to the Gauss-Bonnet term, and since a field redefinition does not change the amplitude there cannot be a 1-loop divergence. 

At 2-loop order, the candidate counterterm has to be composed of some index contractions of 3 Riemann tensors --- let us denote it $R^3$, here and henceforth leave the $\sqrt{-g}$ implicit. The $R^3$ counterterm is present for pure gravity which (as noted in Section \ref{s:UVsg1}) is 2-loop divergent. 

In supergravity, the counterterms also have to respect the non-anomalous symmetries of the theory. Supersymmetry is preserved at loop-level so any  counterterm candidate must be supersymmetrizable. We showed in Exercise \ref{ex:grav3ptR3} that a matrix element produced by $R^3$ is fixed by little group scaling to be
\be
  M_3(  1^- 2^- 3^-)_{R^3}= \text{constant} \times \<12\>^2 \<23\>^2 \<13\>^2\,.
  \label{M3allminus}
\ee
But we also know from  \reef{gravallplus2}
that this violates the  supersymmetry Ward identities. So this means that any operator that produces a non-vanishing matrix element 
$M_3( 1^- 2^- 3^-)_{R^3}$ violates supersymmetry. Since $R^3$ produces a supersymmetry-violating amplitude, we conclude that $R^3$ cannot be supersymmetrized \cite{vanNieuwenhuizen:1976vb,Grisaru:1976nn}. Therefore $R^3$ is not a viable counterterm and hence {\bf \em any pure supergravity must be 2-loop finite!}

In the above argument, you may object that we may not have to care about the 3-point amplitude \reef{M3allminus} since it vanishes in real kinematics. But it is easy to show (using for example an all-line shift \cite{CEK}) that a non-vanishing all-minus 3-point amplitude implies that there is a non-vanishing 4-graviton amplitude 
$M_4(1^- 2^- 3^- 4^-)$ and clearly this violates supersymmetry. Another objection could be: ``what if the constant in \reef{M3allminus} is zero, then there is no contradiction with supersymmetry?" That is true, but if the matrix element vanishes that means that the 3-field part of $R^3$ is a total derivative, and then we don't care about it anyway because there are no available  gravity diff-invariant 4-field operators. So, either way, pure supergravity is finite at 2-loops.

Let us now specialize to $\cn=8$ supergravity in 4d. The candidate counterterms have to respect  $\cn=8$ supersymmetry and also be $SU(8)$ invariant, since the global R-symmetry  is non-anomalous \cite{Marcus:1985yy,diVecchia:1984jh}. Moreover, they should be compatible with the `hidden' $E_{7(7)}$ symmetry \cite{Bossard:2010dq}; we will come back to $E_{7(7)}$ later --- for now, we explore what constraints supersymmetry and R-symmetry impose on the candidate counterterms in $\cn=8$ supergravity. 

It is in general difficult to analyze the candidate counterterm operators directly: a full field theory $\cn=8$ supersymmetrization of the independent contractions of $R$'s and $D$'s is complicated in component form; for $R^4$ it has been done explicitly at the linearized level only  \cite{Freedman:2011uc}. A better approach is to use superfield formalism. There is no off-shell superfield formalism for $\cn=8$ supergravity, but harmonic superspace techniques have been used to constrain the possible counterterms in supergravity theories in various dimensions. We are going to highlight some of the results of the superspace approach in Section \ref{s:UVsg2}, but otherwise we do not discuss these methods here:  this is a review of amplitudes and that will be the path we take.  

The supersymmetry and R-symmetry constraints on the candidate counterterm operators translate into Ward-identity constraints on the matrix elements produced by counterterms. Let us list the translation of constraints between an operator, whose lowest interaction-term is an $n$-vertex, and the corresponding $n$-point matrix element:
\be
 \begin{array}{lcl}
 \text{$n$-field operator} &  & \text{$n$-point matrix element} \\
 \hline
 \text{local with $2L+2$ derivatives} &\lra & \text{polynomial in $\<ij\>$ and $[kl]$ of degree $2L+2$} \\
 \text{$\cn = 8$ SUSY} &\lra &  \text{$\cn = 8$ SUSY Ward identities} \\
 \text{$SU(8)$ R-symmetry} &\lra &  \text{$SU(8)$ Ward identities} \\
 \end{array}
 \label{N8constraints}
\ee
($E_{7(7)}$ constraints will be treated separately, starting on page \pageref{pE77}.)
In addition, the matrix elements have to respect Bose/Fermi symmetry under exchange of identical external states.

The condition that the matrix element is polynomial follows from the locality of the operator and the insistence that it corresponds to the {\em leading} (i.e.~first) UV divergence in the theory; with other operators present, there could be pole terms. The matrix element we consider here is strictly the amplitude calculated from the $n$-point vertex of the given $n$-field operator, and therefore it cannot have any poles, i.e.~it must be a polynomial in the kinematic variables $\<ij\>$ and $[kl]$. The degree of the polynomial  follows from dimensional analysis. 

Let us be clear about what our approach is: for a given $L$, we ask if the first UV divergence could appear in an $n$-point amplitude. If this is so, then there must be a corresponding 
$n$-field $(2L+2)$-derivative 
counterterm. For example, for $L=3$ the lowest-$n$ candidate counterterm would be an $SU(8)$-invariant $\cn=8$ supersymmetrization of $R^4$.
To analyze if such an operator exists, we write down all possible matrix elements satisfying the constraints \reef{N8constraints}. 
If there are no such matrix elements, we conclude there is no corresponding $\cn=8$ SUSY and $SU(8)$-invariant operator and therefore the first divergence in the theory cannot be in the $n$-point $L$-loop amplitude. On the other hand, if one or more such matrix elements exist, then the corresponding operator respects linearized $\cn=8$ supersymmetry and $SU(8)$ and we may consider it as a candidate counterterm. That does not mean that  perturbation theory actually produces the corresponding UV divergence; that would have to be settled by other means, such as an explicit $L$-loop computation. Thus, the approach here is to use the matrix elements to exclude counterterm operators as well as characterize candidate counterterms as operators that respect  $\cn=8$ SUSY and $SU(8)$-symmetry at the linearized level.  

To illustrate the idea, consider $R^4$.\footnote{The relevant contraction of 4 Riemann tensors is the square of the Bel-Robinson tensor \cite{Deser:1978br}.} Its 4-point matrix element  $M_4( 1^- 2^- 3^+ 4^+ )_{R^4}$ has to be a degree 8 polynomial in angle and square brackets. Taking into account the little group scaling, there is only one option: the matrix element has to be 
$M_4(  1^- 2^- 3^+ 4^+ )_{R^4} = c_{R^4}  \<12\>^4 [34]^4$, where $c_{R^4}$ is some undetermined constant. By the same arguments as above, dimensional analysis and little group scaling, we know that 
$M_4(  1^- 2^+ 3^- 4^+ )_{R^4} = c_{R^4}  \<13\>^4 [24]^4$. We can then check the MHV-level  $\cn=8$ SUSY Ward identity which at $n$-point reads
\be
  \text{$\cn=8$ supergravity:}~~~~~
  M_n( 1^+ \dots  i^- \dots  j^- \dots n^+ )_{\mathcal{O}} =
  \frac{\<ij\>^8}{\<12\>^8}
  M_n(  1^- 2^- 3^+ \dots n^+ )_{\mathcal{O}} \,.
  \label{MnSUSYWI}
\ee
This identity follows directly from the MHV superamplitude \reef{N8sampMHV}.
For our 4-point $R^4$ matrix elements, we have
\be
  M_4(  1^- 2^+ 3^- 4^+ )_{R^4} 
  = \frac{\<13\>^8}{\<12\>^8}  M_4(  1^- 2^- 3^+ 4^+ )_{R^4} 
  = \frac{\<13\>^8}{\<12\>^8} c \<12\>^4 [34]^4
  = c_{R^4} \<13\>^4 [24]^4\,,
\ee
 thanks to momentum conservation 
$\<13\>[34] = -\<12\>[24]$. It is not hard to see that the 4-point super-matrix-element
\be
   \mathcal{M}_4(1234)_{R^4} 
   = \delta^{4}\big(P\big) \, 
  \delta^{(16)}  \big( \widetilde{Q}\big)\,
  \frac{[34]^4}{\<12\>^4}\,
  \label{N8sampMHVme}
\ee
fulfills all criteria \reef{N8constraints}.\footnote{For details of how the R-symmetry acts on the superamplitudes, see Section 2.2 of \cite{Elvang:2009wd}.}
 This means that linear $\cn=8$ supersymmetry and $SU(8)$ do not rule out  $R^4$. But it does not mean that it will occur in perturbation theory: in fact, we  know from the explicit 3-loop calculation \cite{Bern:2006kd} that the 4-graviton amplitude is finite, so $R^4$ does not occur. Why not? Well, read on, we'll get to that later in this section.

Let us now see an example of how the on-shell matrix elements can be used to rule out a counterterm. At 4-loop order, we can write down two pure gravity 10-derivative operators, $D^2 R^4$ and $R^5$. The first one, $D^2 R^4$, stands for the possible scalar contractions of 2 covariant derivatives acting (in some way) on four Riemann tensors. Its matrix element turns out to be proportional to $(s+t+u) \mathcal{M}_4(1234)_{R^4}$, so it vanishes. This means that the 4-point interaction in the operator $D^2 R^4$ is a total derivative when evaluated on the equations of motion. 
So we can rule out the 4-loop 4-graviton amplitude as the first instance of a UV divergence in $\cn=8$ supergravity. The second 10-derivative operator $R^5$ is a little more interesting.

The 5-point MHV matrix element of $R^5$ is  fixed uniquely by dimensional analysis and little group scaling up to an overall constant:
\be
  M_5(1^-2^- 3^+ 4^+ 5^+)_{R^5}
  =
  a_{R^5} \<12\>^4 [34]^2 [45]^2 [53]^2\,.
\ee
Now we check the supersymmetry Ward identity \reef{MnSUSYWI} in much the same way as for $R^4$. But now we find
\bea
  \nonumber
  M_5(1^-2^+ 3^- 4^+ 5^+)_{R^5}
  &= \hspace{-3.5mm}\raisebox{3mm}{?}&
   \frac{\<13\>^8}{\<12\>^8}
  M_5(  1^- 2^- 3^+ 4^+ 5^+ )_{R^5} \\ 
  \implies~~~~~
  a_{R^5} \<13\>^4 [24]^2 [45]^2 [52]^2
  &= \hspace{-3mm}\raisebox{3mm}{!}&
   a_{R^5}  \frac{\<13\>^8  [34]^2 [45]^2 [53]^2}{\<12\>^4} \,.
   \label{5ptR5}
\eea
This time momentum conservation doesn't save us. The LHS and RHS of \reef{5ptR5}  are not equal, in particular the LHS is local (i.e.~does not have any poles) while the RHS has a pole $1/\<12\>^4$. This is a contradiction that can only be resolved when $a_{R^5}=0$. So that means that the operator $R^5$ does not have an $\cn=8$ supersymmetrization. And that in turn rules out that the 5-point 4-loop amplitude would be the first UV divergence in $\cn=8$ supergravity. One can further argue \cite{Drummond:2010fp,Elvang:2010jv} that there are no other possible $10$-derivative operators compatible with $\cn=8$ supersymmetry and $SU(8)$, so this means that $\cn=8$ supergravity cannot have its first UV divergence at 4-loop order.

Equation \reef{5ptR5} illustrates a conflict between supersymmetry and locality, a conflict that can be exploited to rule out potential counterterms. We will describe the method for operators of the form $D^{2k} R^5$, then outline the general results. The strategy is to construct the most general matrix element $M_5(1^- 2^- 3^+ 4^+ 5^+)_{D^{2k} R^5}$ that respects the little group scaling, has mass dimension $2k+10$, and is Bose symmetric in exchange of same-helicity gravitons. Then ask if there exists a  linear combination that respects the SUSY Ward identities. Practically, this was done in \cite{Elvang:2010jv} using Mathematica. (For more advanced cases, Gr\"obner basis techniques are very useful \cite{Beisert:2010jx}, and the results found can be reproduced and extended by an analysis based on the superconformal group $SU(2,2|8)$  \cite{Beisert:2010jx}.) 

\example{As an example of the procedure in \cite{Elvang:2010jv}, consider $D^{2} R^5$. First we first find that there are 40 angle-square bracket monomials of degree 12 that have the correct little group scaling of $M_5(1^- 2^- 3^+ 4^+ 5^+)_{D^{2} R^5}$. This does not take into account redundancy of Schouten or momentum conservation. 
Now take linear combinations of the 40 monomials to enforce Bose symmetry: this leaves 6 polynomials as candidates for $M_5(1^- 2^- 3^+ 4^+ 5^+)_{D^{2} R^5}$. Then impose Schouten and momentum conservation and one finds that only one polynomial survives: this means that the MHV matrix element of 
$D^{2} R^5$ is unique; it takes the form
\be
   M_5(1^- 2^- 3^+ 4^+ 5^+)_{D^{2} R^5}
   =
   a_{D^{2} R^5}\, s_{12}\, \<12 \>^4 [34]^2 [45]^2 [53]^2\,.
\ee
This is actually just $s_{12}$ times  $M_5(1^- 2^- 3^+ 4^+ 5^+)_{R^5}$.
Now repeat the SUSY test  \reef{5ptR5} for $M_{5,D^{2} R^5}$ to find that the RHS has a pole $1/\<12\>^3$, contradicting the locality of the LHS. So the operator $D^{2} R^5$ is excluded as a counterterm for $\cn=8$ supergravity, which means that the first divergence cannot be in the 6-loop 5-graviton amplitude.

Let us summarize the result  of the process outlined above for a few more operators in the same class --- after each step given in the first column, we list how many polynomials  remain:
\be
\begin{array}{rccccccc}
   & R^5 & D^{2} R^5 & D^{4} R^5 & D^{6} R^5 & D^{8} R^5 \\
   \hline 
   \text{little grp} & 1 & 40 & 595 & 4983 & 29397  \\
   \text{Bose symmetry} & 1 & 6 & 63 & 454 & 2562 \\
   \text{Schouten, mom-cons} & 1 & 1 & 6 & 9 & 24 \\
   \text{weakest pole in SUSY Ward id} & \<12\>^{-4} & \<12\>^{-3} & \<12\>^{-2} & \<12\>^{-1} & \text{no pole} 
\end{array}
\ee
It follows from the last line that the $D^{2k} R^5$ operators are excluded as counterterm candidates for $k=0,1,2,3$, but not for $k=4$ where there is one unique matrix element that solves the supersymmetry Ward identities. Thus there is a unique operator $D^{8} R^5$ that passes the tests of linearized SUSY; if present, this would correspond to a first UV divergence at 8-loop order in the 5-graviton amplitude.
}
\exercise{}{Show that for the operator $R^n$ with $n \ge 3$ there are no $n$-point MHV matrix elements that are compatible with the $\cn=8$ supersymmetry Ward identities.}
For operators with $n>5$ fields, one has to distinguish between the different N$^K$MHV sectors. Beyond MHV level, this can be done using the solutions to the SUSY Ward identities \cite{Elvang:2009wd} in $\cn=8$ supergravity. As an example of a non-MHV result, the lowest order NMHV-level operator is $D^4 R^6$ at 7-loops. Actually, there are two independent NMHV matrix elements, so this means that there are two independent linearly-supersymmetrizable operators $D^4 R^6$. 

A detailed analysis of possible counterterm operators was carried out in \cite{Elvang:2010jv} (see also \cite{Drummond:2010fp}) and it was found that {\em below 7-loop order, the only operators compatible with linearized $\cn=8$ supersymmetry and $SU(8)$ R-symmetry are}
\be
  \underbrace{\,R^4\,}_\text{3-loop}\,,~~~~~
  \underbrace{\,D^4 R^4\,}_\text{5-loop}\,,~~~~~  
  \underbrace{\,D^6 R^4\,}_\text{6-loop}\,.~~~~~
  \label{D2kR4}
\ee
At {\em 7-loop order, an infinite tower of linearized $\cn=8$ supersymmetry and $SU(8)$ R-symmetry permissible operators were found}
\be
  \text{7-loops:}~~~~~
  D^8 R^4\,,~~~
  D^{4} R^6\,,~~~
  R^8\,,~~~
  \phi^2 R^8 \,,~~~
  \phi^4 R^8 \,,~~~
  \dots
\ee
The operators $\phi^{2k} R^8$ should be viewed as representatives for the linearized $\cn=8$ supersymmetrization of some contraction of 8 Riemann tensors multiplied by an $SU(8)$-singlet combination of $2k$ scalars $S^{ABCD}$. These do not have purely gravitational $(8\!+\!2k)$-point matrix elements, so they cannot be in the MHV or anti-MHV sector. For example, it was shown \cite{Beisert:2010jx} that $\phi^{2} R^8$ only gives N$^3$MHV matrix elements (4 distinct ones). 

{\em At 8-loop order and beyond, there are infinite towers over operators that respect linearized $\cn=8$ supersymmetry and $SU(8)$.}
You can find a detailed characterization of the counterterms in Table 1 of  \cite{Beisert:2010jx}.

There is one symmetry we did not use to restrict the candidate counterterms operators in the above discussion, and that is the \label{pE77}%  
{\bf \em $E_{7(7)}$ `hidden' symmetry}.\footnote{See also \cite{Bossard:2011ij,alsoE77} for related aspects of $E_{7(7)}$ symmetry.}
As discussed at the end of Section \ref{s:N8SG}, it manifests itself in the amplitudes  of $\cn=8$ supergravity through {\bf \em low-energy theorems}. These also have to apply to the matrix elements of any acceptable candidate counterterm operator $\mathcal{O}$. In particular, the single-soft scalar limit must vanish, for example
\be
  \lim_{p_1 \to 0} 
  M_6\big(S^{1234} S^{5678}  h^- h^- h^+ h^+ \big)_{\mathcal{O}}  = 0 \,.
  \label{softM6}
\ee
If the matrix element of an operator does not pass the single-soft scalar  test, then it is not compatible with $E_{7(7)}$ symmetry. If it does pass the test, then we conclude nothing: it could be  $E_{7(7)}$ at play or just a coincidence. 

It turns out that the single-soft scalar test is non-trivial starting at $n=6$. So precisely \reef{softM6} can be used to test the operators that survived the $\cn=8$ supersymmetry and $SU(8)$ constraints, for example the $L=3,5,6$ operators in \reef{D2kR4}. But it is not easy to use Feynman rules to calculate the 6-point matrix elements of, say, $R^4$. However,
$M_6\big(S^{1234} S^{5678}  h^- h^- h^+ h^+ \big)_{R^4}$ can be extracted from the $\alpha'$-expansion of the closed superstring theory tree amplitude. This may bother you, because there are no continuous global symmetries in string theory, and here we are interested in testing global continuous $E_{7(7)}$. However, the 4d tree-level superstring amplitudes have an accidental global $SU(4) \times SU(4)$ symmetry. This is a consequence of T-duality when 10-dimensional superstring theory is reduced to 4d by compactifying it on a 6-torus. The easiest way to see the $SU(4) \times SU(4)$ symmetry is  through the KLT relations: the two open string tree amplitudes on one side of KLT have  $SU(4)$ symmetry, so the  closed tree string amplitude on the other side of KLT inherits  $SU(4) \times SU(4)$;  the $SU(4) \times SU(4)$ is enhanced to $SU(8)$ only in the $\alpha' \to 0$ limit \cite{Bianchi:2008pu}.

\begin{figure}[t]
\centerline{
\includegraphics[width=15.5cm]{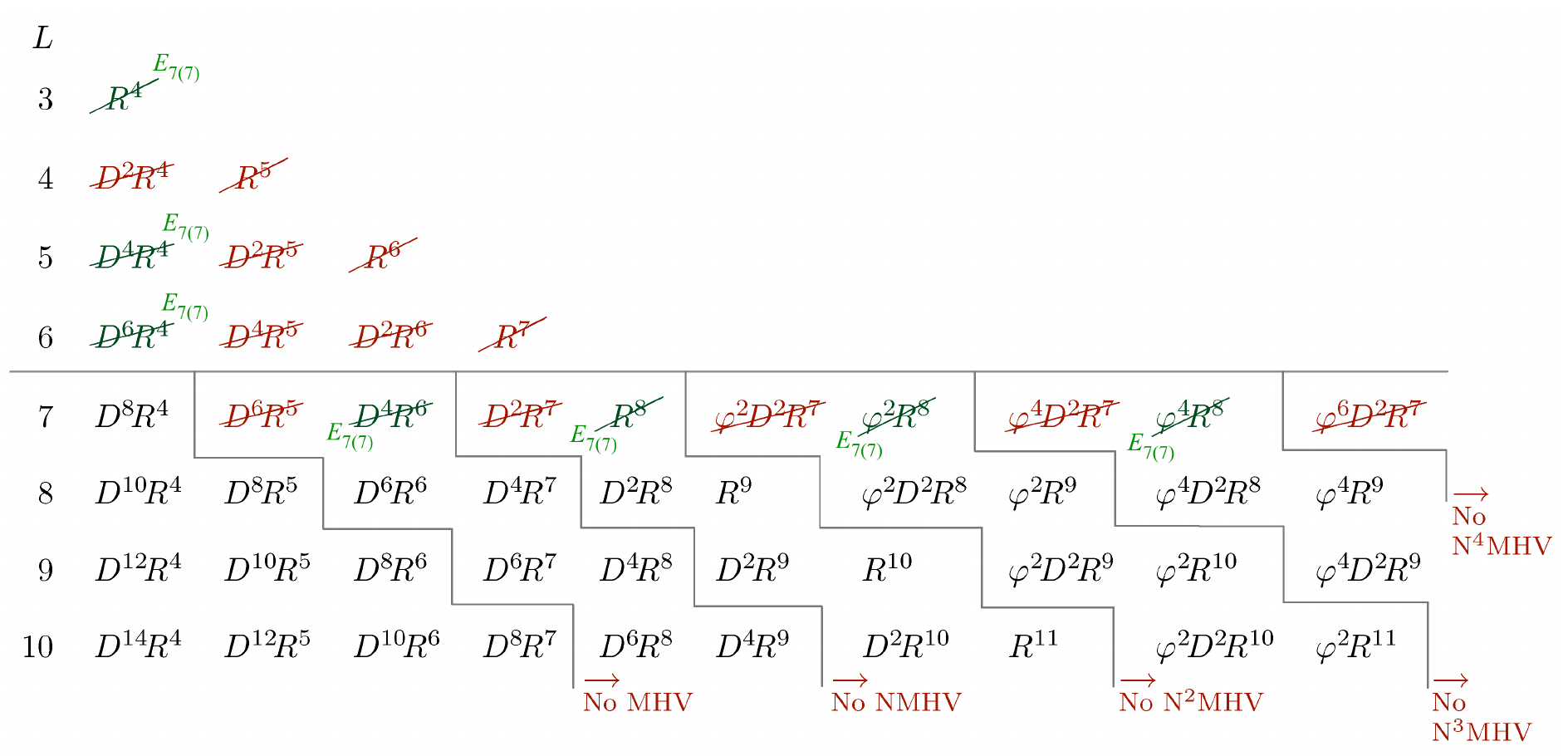}}
\caption{\small Overview of exclusion of counterterm operators in $\cn=8$ supergravity in 4d.}
\label{fig:CTs}
\end{figure}

 Now, we are going to use the $\alpha'$ contributions from the superstring tree amplitudes, but their $SU(4) \times SU(4)$ is not good enough, we need $SU(8)$. So we average the string amplitude over all 35 independent embeddings of $SU(4) \times SU(4)$ into $SU(8)$ to get an $SU(8)$-invariant answer.  For example, the leading $\alpha'$-correction to the closed superstring amplitude is order $\alpha'^3$. Dimensional analysis implies that this comes from an 8-derivative effective operator with $\cn=8$ supersymmetry and (after averaging) $SU(8)$-invariance. But we know from the previous analysis that there is only one such operator, namely $R^4$. So after making it $SU(8)$-invariant, the $\alpha'^3$ contribution from the  $SU(8)$-averaged superstring amplitude must be identical (up to an overall constant) to the matrix element of the operator $R^4$ in  $\cn=8$ supergravity \cite{Elvang:2010kc}!\,\footnote{In the Einstein frame, the effective operator in superstring theory at order $\alpha'^3$ is actually $e^{-6\phi}R^4$, where $\phi$ is the dilaton. The presence of the dilaton (which can be identified as a certain linear combination of the 70 scalar scalars in the $\cn=8$ supergravity spectrum) operator breaks $SU(8)$ to $SU(4) \times SU(4)$.} 
The open superstring tree amplitudes are known in the literature \cite{Stieberger:2007am}, so 
pulling them through KLT, the $M_6\big(S^{1234} S^{5678}  h^- h^- h^+ h^+ \big)_{R^4}$ matrix element can be extracted, $SU(8)$-averaged, and then subjected to the single-soft scalar test \reef{softM6}.\footnote{An earlier test \cite{Brodel:2009hu} did not involve the $SU(8)$-average.} And $R^4$ fails this test: so {\bf \em $R^4$ is not compatible with $E_{7(7)}$}. Hence $E_{7(7)}$ excludes $R^4$ as a candidate counterterm and {\em this explains why the 3-loop 4-graviton amplitude is finite.} The on-shell matrix element technique made it possible to show  3-loop divergence of $\cn=8$ supergravity could be excluded without doing any loop-amplitudes calculations \cite{Elvang:2010kc}.\footnote{Note that this is {\em not} a string theory argument; string theory amplitudes were used but they correspond exactly to certain effective operators in field theory in $\alpha'$-expansion, so the argument in \cite{Elvang:2010kc} is field theoretical.}

The analysis outlined above can be repeated \cite{Beisert:2010jx} for $D^4 R^4$ and $D^6 R^8$ and both are shown to be excluded by $E_{7(7)}$. (See also \cite{StringyCounterterms} for a string-based argument for the absence of $R^4$, $D^4 R^4$ and $D^6 R^8$.) This means that {\em the symmetries of $\cn=8$ supergravity excludes the divergences in any amplitudes below 7-loop order} \cite{Beisert:2010jx}. Explicit calculations of the $L=5,6$ amplitudes are not yet available, but they are expected to yield finite results in 4d.

An overview of possible counterterms in $\cn=8$ supergravity is presented in Figure \ref{fig:CTs}. Note that at 7-loop order, all but the $D^8 R^4$ operator are excluded. This means that calculation of the 4-graviton 7-loop amplitude can completely settle the question of finiteness at 7-loop order. But this is not known to be the case at 8-loops or higher.

In following section, we give a brief overview of the current status of the UV behavior of loop-amplitudes in supergravity theories. 

%%%%%%%%%%%%%%%%%%%%%%%%%%%%%%%%%%%%%%%
%%%%%%%%%%%%%%%%%%%%%%%%%%%%%%%%%%%%%%%
%%%%%%%%%%%%%%%%%%%%%%%%%%%%%%%%%%%%%%%
\subsection{Supergravity divergences for various $\cn$, $L$, and $D$}
\label{s:UVsg2}
%%%%%%%%%%%%%%%%%%%%%%%%%%%%%%%%%%%%%%%
%%%%%%%%%%%%%%%%%%%%%%%%%%%%%%%%%%%%%%%
%%%%%%%%%%%%%%%%%%%%%%%%%%%%%%%%%%%%%%%

Current approaches to examining the possible UV-divergences of perturbative (super)gravity can be categorized as follows:
\begin{itemize}
  \item {\bf \em  Direct computation.} The explicit computations of loop-amplitudes are made possible by increasingly sophisticated applications of the generalized unitarity method; this in itself advances the technical tools for attacking higher-loop computations in general field theories. For the purpose of exploring UV divergences in supergravity theories, most efforts focus on the 4-graviton amplitude; this is because (as we have seen in the previous section) the lowest counterterm for pure supergravity is of the form $D^{2k}R^4$. See~\cite{ck4l}  for a discussion on the various details of obtaining multi-loop integrands and extracting the UV-divergences.   
\item {\bf \em Symmetry Analysis.} 
The leading UV-divergence of theory has a corresponding local gauge-invariant operator that must respect all non-anomalous global symmetries of the theory. Analyzing the symmetry properties of operators, one can rule out UV-divergences and identify candidate counterterms. In Section \ref{s:CTs}, we took an approached based on the on-shell matrix elements of the candidate operators, and used it to rule out divergences in 4d $\cn=8$ supergravity for $L\le 6$. Alternatively, one can also use extended-superspace to construct possible invariant operators; for early construction, see \cite{Howe:1980th,Kallosh:1980fi}. If the invariant operator can be expressed as a superspace integral over a subset of superspace coordinates, then it is considered a ``BPS" operator and it is subject to non-renormalization theorems. If it is given as a full superspace integral, then it is non-BPS and expected to receive quantum corrections, thus serving as a candidate counterterm. The distinction of between BPS and non-BPS invariants relies on subtle assumptions about the number of supersymmetries that can be linearly realized off-shell. This lies outside the scope of this review, so we refer you to \cite{VanishingVolume} and references therein.   
    \item {\bf \em Pure spinor formalism.} 
    The ``pure-spinor" formalism~\cite{PureSpinor} is a first-quantized approach (in contrast, QFT is a second-quantized approach) to scattering processes in  10d maximal supersymmetric theories. Using the 10d loop-integrand obtained in the pure-spinor formalism, one can infer properties of the loop-amplitudes in $D\le 10$ and this can be helpful for assessing potential UV divergences. For more details of this approach to multi-loop amplitudes, see \cite{Bjornsson:2010wm}. 
    \item {\bf \em Role of non-perturbative states?} The $2^8$ massless states of ungauged $\cn=8$ supergravity in 4d matches exactly the spectrum of massless states of closed Type IIB superstring theory compactified on a six-torus $T^6$. At the classical level, $\cn=8$ supergravity in 4d can be viewed as the low-energy ($\alpha' \to 0$) limit of Type IIB superstring theory on $T^6$. However, it was pointed out in \cite{Green:2007zzb} that one cannot obtain \emph{perturbative} $\cn=8$ supergravity in 4d as a consistent truncation of the string spectrum: in the limit $\alpha' \to 0$, keeping the 4d coupling small forces 
    infinite towers of additional states to become light, e.g.~Kaluza-Klein states, winding modes, KK monopoles, and/or wrapped branes. Thus one obtains from string theory not just the spectrum of $\cn=8$ supergravity, but a slew of additional massless states. This argument is independent of whether $\cn=8$ supergravity is finite in 4d or not. However, it does mean that if pure $\cn=8$ supergravity in 4d were to be a well-defined theory of quantum gravity, its UV completion would not be Type IIB string theory.
    
A related objection \cite{Banks:2012dp}\footnote{See also \cite{Bianchi:2009wj}.} to the program of studying finiteness of perturbative supergravity is that even the 4d $\cn=8$ supergravity  theory itself contains non-perturbative states, namely BPS black holes, that in certain regions of moduli-space become light enough   that they may enter the perturbative expansion \cite{Banks:2012dp}. Such contributions would never enter the unitarity method approach to explicit calculation of amplitudes. 
\end{itemize}

As we have noted, there are certainly examples of divergences in various perturbative (non-super) gravity theories in 4d: 1-loop in gravity with matter \cite{'tHooft:1974bx,Deser:1974cz}, 2-loop in pure gravity \cite{Goroff:1985sz,vandeVen:1991gw}, and at 1-loop \cite{Bern:2013yya} in dilaton-axion gravity \reef{dilaxgrav}. The first example of a UV divergence in pure supergravity was found at 4-loop order in the 4-graviton amplitude of $\cn=4$ supergravity \cite{Bern:2013uka}. 

We end this Section by summarizing  what explicit computations of supergravity loop amplitudes have revealed so far about  the critical dimension $D_c$ of supergravity with various numbers of supersymmetry: 

{\bf Maximal supergravity} (32 supercharges)
\eq
  \centering \begin{tabular}{|c|c|c|c|c|c|c|}
\hline
  Loop-order & 1 & 2 & 3 & 4 \\ \hline
  $D_c$ & 8~\cite{Green:1982sw} & 7~\cite{N=42Loop2} & 6~\cite{Bern:2007hh} & $\frac{11}{2}$~\cite{Bern:2009kd}  \\
\hline
\end{tabular}\,\,.
 \eqe
{\bf Half-maximal supergravity} (16 supercharges)
\eq
  \centering \begin{tabular}{|c|c|c|c|c|c|c|}
\hline
  Loop-order & 1 & 2 & 3 & 4  \\ \hline
  $D_c$ & 8~\cite{N4D81L} & 6~\cite{N=4SG1} & $>4$~\cite{N=4SG2}  & $\le 4$~\cite{Bern:2013uka} \\
\hline
\end{tabular}\,\,.
 \eqe
{\bf  Half-maximal supergravity with matter} (both with 16 supercharges)
\eq
  \centering \begin{tabular}{|c|c|c|c|c|c|c|}
\hline
  Loop-order & 1 & 2 & 3  \\ \hline
  $D_c$ & 4~\cite{Fischler} & $\leq 4$~\cite{N4Matter23L} & $\leq 4$~\cite{N4Matter23L}  \\
\hline
\end{tabular}\,\,.
 \eqe
In the above, ``$\leq4$'' indicates an upper bound for the critical dimension. 

The absence of UV divergence for half-maximal supergravity at 3-loops in 
4d~\cite{N=4SG2} as well as 2-loops in 5d~\cite{N=4SG1}, was \emph{not} anticipated by superspace-based analyses. This unexpected result prompted a conjecture of the existence of an off-shell formalism that preserves the full 16 supersymmetries \cite{BossardHoweStelle5D}. This would imply finiteness for half-maximal supergravity with matter at 2-loop in 5d, but explicit calculations \cite{N4Matter23L} have shown that a UV-divergence is actually present, thus contradicting  the conjecture. 

The study of UV structure of perturbative supergravity theories in diverse dimensions has resulted in some interesting insights about the relation between gravity and gauge theory scattering amplitudes: this includes the color-kinematics duality that is the subject of the next Section. It is relevant to note that no matter what one thinks of the program to study the UV-behavior of supergravity, these new insights would have been difficult to come by without the effort put into explicit calculations and the lessons learned in the process.

%%%%%%%%%%%%%%%%%%%%%%%%%%%%%%% 
%%%%%%%%%%%%%%%%%%%%%%%%%%%%%%% 
%%%%%%%%%%%%%%%%%%%%%%%%%%%%%%% 
\newpage
\setcounter{equation}{0}
\section{A colorful duality }
\label{s:BCJ}
%%%%%%%%%%%%%%%%%%%%%%%%%%%%%%% 
%%%%%%%%%%%%%%%%%%%%%%%%%%%%%%% 
%%%%%%%%%%%%%%%%%%%%%%%%%%%%%%% 
A recurring theme in our  discussion of perturbative supergravity in Section   \ref{s:sugra} is captured by the abstract formula ``$\text{gravity}=(\text{gauge theory})^2$". It enters in the context of the spectrum of states, for example as
\be
      \mathcal{N}=8~\text{supergravity}
      \,=\, \big(\cn=4~\text{SYM}\big) \otimes \big(\cn=4~\text{SYM}\big)
      \,,
      \label{N8SGspectrum13}
\ee
and also carries over to the scattering amplitudes. For instance, the gravity 3-point amplitude equals the square of Yang-Mills 3-gluon amplitude \reef{M3}. That is a special case of the KLT formula \reef{KLT45} which expresses the tree-level $n$-graviton amplitude as sums of products of color-ordered $n$-gluon Yang-Mills amplitudes.  The KLT formula extends to all tree-level amplitudes in $\cn=8$ supergravity following the prescription \reef{N8SGspectrum13} for `squaring' the spectrum. 

While the KLT formula follows the  ``$\text{gravity}=(\text{gauge theory})^2$" storyline, it is unsatisfactory in some respects. First, the formula becomes  tangled at higher points, as it involves nested permutation sums and rather complicated kinematic invariants. Second, since it involves products of different color-ordered amplitudes, it is not really a squaring relation (except at 3-points). Finally,  it is only valid at tree-level. You may think that it is asking too much to have gravity amplitudes, arising from the complicated Einstein-Hilbert Lagrangian, closely related to amplitudes of the much simpler  Yang-Mills theory. But one lesson we have learned so far is not to let the Lagrangians get into our way! 
In this section, we explore a form of  ``$\text{gravity}=(\text{gauge theory})^2$" that makes the amplitude squaring relation more direct and has been proposed to be valid at both tree- and loop-level.

We begin by answering a simple question: why is the KLT formula so complicated? In our study of color-ordered amplitudes, we often exploit that the allowed physical poles are those that involve adjacent momenta, e.g.~$1/P^2_{i,i+1,\ldots,j-1,j}$. This is a special feature linked to the color-ordered Feynman rules. But for gravity, there is no color-structure and hence no canonical sense of ordering of the external states. Thus the poles that appear in a gravity amplitude can involve any combination of external momenta. This tells us that in order to faithfully reproduce the pole structure of a gravity amplitude from ``$\text{gravity}=(\text{gauge theory})^2$", we need color-ordered Yang-Mills  amplitudes {\em with different ordering}. For higher points, the proliferation of physical poles in the gravity amplitude forces us to include more and more Yang-Mills amplitudes with distinct ordering. This is why the KLT formula involves a sum over a growing number of different color-ordered  Yang-Mills amplitudes. The complicated kinematic factors in KLT are needed to cancel double poles.
\exercise{}{Justify the  kinematic factors and distinct orderings of the  Yang-Mills amplitude in the KLT relations \reef{KLT45}.}
The above discussion suggests that for the comparison with gravity, it may be more useful to consider the fully color-dressed Yang-Mills amplitude instead of the color-ordered partial amplitudes. The former has the same  physical poles as the gravity amplitude. Indeed, this is a productive path, so to get started, we  review some useful properties of the color-structure of Yang-Mills amplitudes.

%%%%%%%%%%%%%%%%%%%%%%%%%%%%%%%%%%%%%%%%%%%%%%%%%%
\subsection{The color-structure of Yang-Mills theory}
\label{s:colorYM}
%%%%%%%%%%%%%%%%%%%%%%%%%%%%%%%%%%%%%%%%%%%%%%%%%%
The full color-dressed $n$-point tree amplitude of Yang-Mills theory can be conveniently organized in terms of diagrams with only cubic vertices, such as
\be
  \raisebox{-15mm}{\includegraphics[scale=0.45]{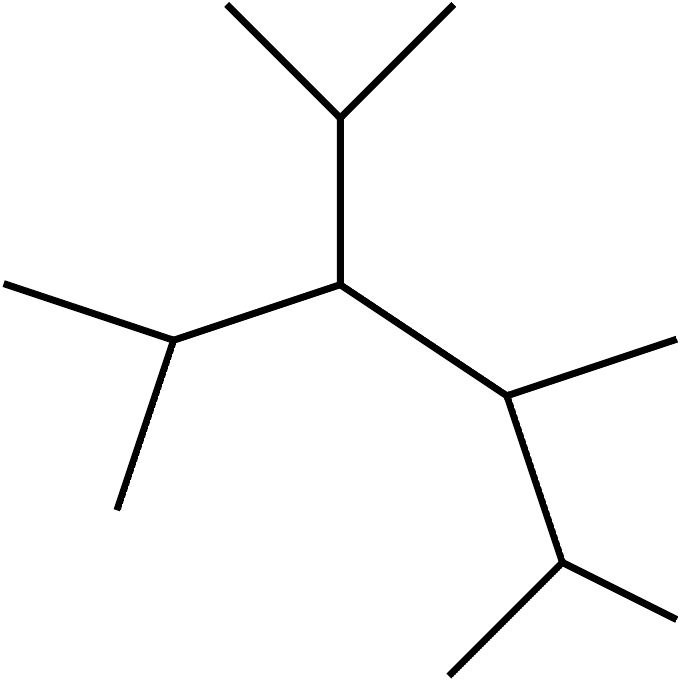}}~~.
\ee
The amplitude is then written as a sum over all distinct trivalent diagrams, labeled by $i$,
\eq
A^\text{tree}_n= \sum_{i\in {\rm trivalent}}\frac{c_in_i}{\prod_{\alpha_i}p^2_{\alpha_i}}\,.
\label{fullAmp}
\eqe
The denominator is given by the product of all propagators (labeled by $\alpha_i$) of a given diagram. The numerators factorize into a group-theoretic color-part $c_i$, which is a polynomial of structure constants $f^{abc}$, and a purely kinematic part $n_i$, which is a polynomial of Lorentz-invariant contractions of polarization vectors $\epsilon_i$ and momenta $p_i$. As an example, the 4-point amplitude is
\eq
A^\text{tree}_4~~=~
\raisebox{-11mm}{\includegraphics[scale=0.45]{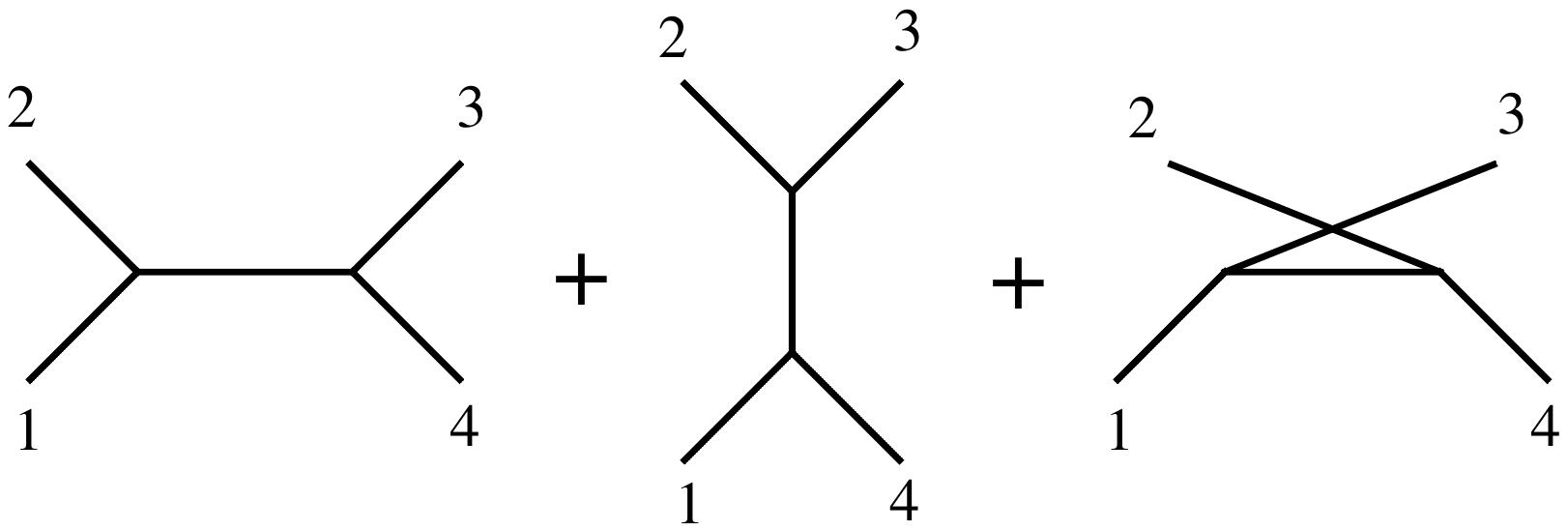}}
~=~~
\frac{c_sn_s}{s}+\frac{c_un_u}{u}+\frac{c_tn_t}{t}\,,
\label{4ptnc}
\eqe
where 
\be
  c_s \equiv \tilde{f}^{a_1 a_2 b}\tilde{f}^{b\, a_3 a_4}\,,
  ~~~~~
  c_t \equiv \tilde{f}^{a_1 a_3 b}\tilde{f}^{b\, a_4 a_2}\,,
  ~~~~~
  c_u \equiv \tilde{f}^{a_1 a_4 b}\tilde{f}^{b\, a_2 a_3}\,,
  \label{4pointcolor}
\ee
as already introduced in \reef{csctcu}. The normalization of the structure constants $\tilde{f}^{abc}$ was discussed in footnote \ref{footiefabc} on page \pageref{footiefabc}.

The numerators $n_i$ can be constructed straightforwardly using Feynman rules. Feynman diagrams with only cubic vertices directly contribute terms of the form $\frac{c_i n_i}{\prod_{\alpha_i}p^2_{\alpha_i}}$. The Yang-Mills 4-point contact terms can be `blown up' into $s$-, $t$- or $u$-channel 3-vertex pole diagrams by trivial multiplication by $1=t/t=s/s=u/u$. Note that since $c_s+c_t+c_u=0$, this does not give a unique prescription for how to assign a given contact term into the cubic diagrams, so  the numerators in \reef{fullAmp} are not uniquely defined. 

We can actually deform the numerators $n_i$ in several ways without changing the  result of the amplitude. For example, one can trivially shift the polarization vectors as $\epsilon_i(p_i) \to  \epsilon_i(p_i)+ \alpha_i p_i$; this changes the kinematic numerator factors $n_i$, but not the overall amplitude because it is gauge invariant. A more non-trivial deformation uses the color factor Jacobi identity $c_s+c_t+c_u=0$: taking $n_s\rightarrow n_s+s\Delta$, $n_t\rightarrow n_t+t\Delta$, and $n_u\rightarrow n_u+u\Delta$, where $\Delta$ is an arbitrary function, leaves the amplitude is invariant since the net deformation is proportional to $c_s+c_t+c_u$. 

\begin{figure}
\begin{center}
\includegraphics[scale=0.5]{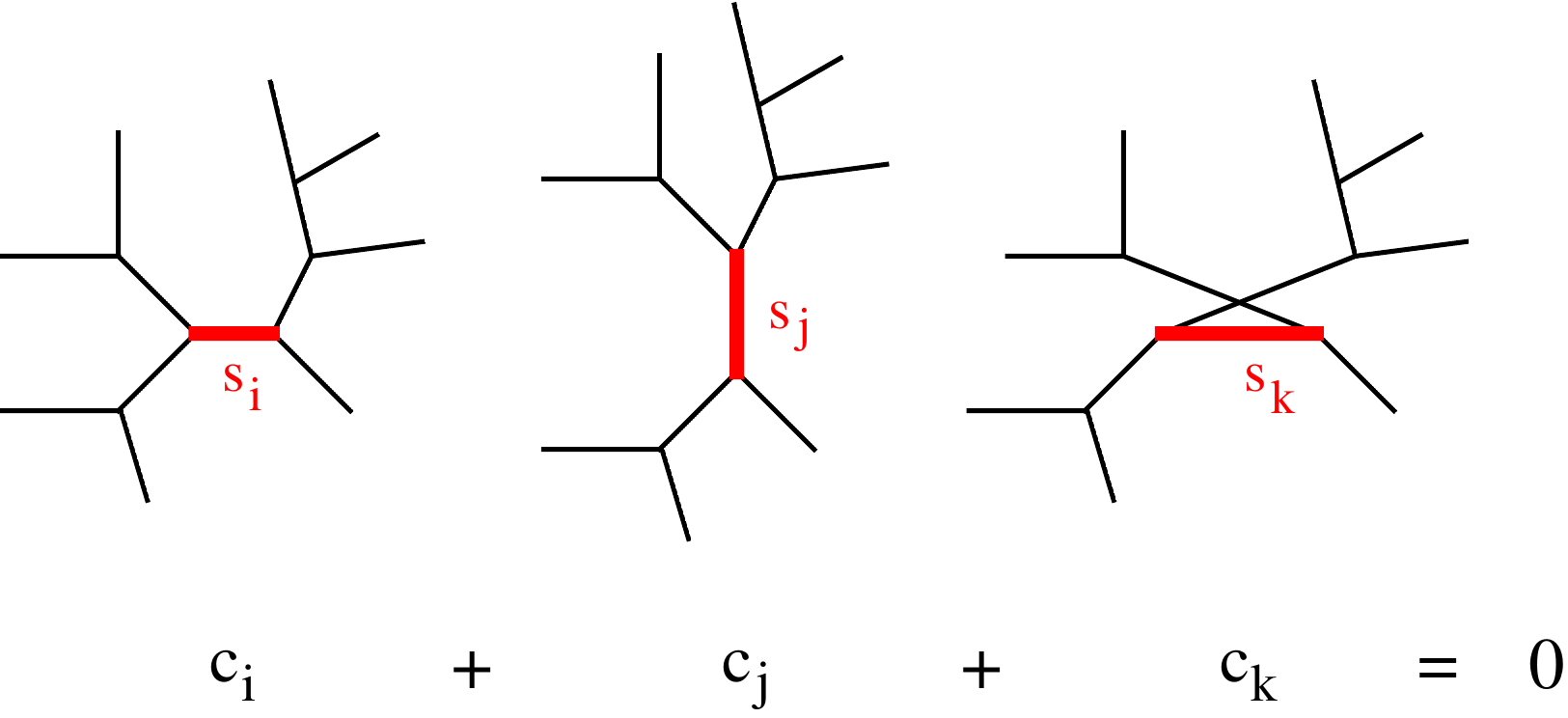}
\caption{\small Three trivalent diagrams whose color factors $c_i$, $c_j$, and $c_k$ are related by the Jacobi identity. Note that the diagrams share the same propagators except one, indicated by a solid red line. We denote the unshared inverse propagators as $s_i$, $s_j$, and $s_k$.}
\label{colorj}
\end{center}
\end{figure}

In general, for any set of three trivalent diagrams whose color factors are related through a Jacobi identity,
\eq
c_i+c_j+c_k=0\,,
\label{GenJacobi}
\eqe 
 the following numerator-deformation  leaves the amplitude invariant:
\eq
n_i\rightarrow n_i+s_i\Delta,\quad n_j\rightarrow n_j+s_j\Delta,\quad n_k\rightarrow n_k+s_k\Delta\,.
\label{ninjnk}
\eqe
Here $1/s_i$, $1/s_j$ and $1/s_k$ are the unique propagators that are \emph{not} shared among the 3 diagrams, as shown in Figure \ref{colorj}. 
Since $\Delta$ can be an arbitrary function, it is similar to a gauge parameter, except that now it is not a transformation of the gauge field, but rather a transformation of the numerator factors $n_i$. Because of this similarity,  the freedom \reef{GenJacobi}-\reef{ninjnk} is often called {\bf \em generalized gauge transformation}~\cite{BCJ}. It plays an important role in linking Yang-Mills and gravity. 

The fact that the numerators $n_i$ are not unique nor gauge invariant should not raise any alarm. After all, the individual Feynman diagrams are not physical observables. For practical purposes, it is useful to focus on gauge invariant quantities. Note that if the color factors are organized in a basis that is independent under Jacobi identities, the coefficient in front of each basis element is necessarily gauge invariant. These coefficients then serve as ``partial-amplitudes" that constitute part of the full amplitude, but are fully gauge invariant. 

A straightforward way to obtain such partial-amplitudes is to start with the full color-dressed amplitude in \reef{fullAmp} and use the color Jacobi identity to systematically disentangle the color factors. This can be achieved in a graphical fashion introduced in \cite{DDM}: start with the color factor of an arbitrary Feynman diagram (with all contact vertices blown up into two cubic vertices as discussed earlier) and convert it by repeated use of the Jacobi identity into a sum of color factors in {\em multi-peripheral form}
\eq
\vcenter{\hbox{\includegraphics[scale=0.5]{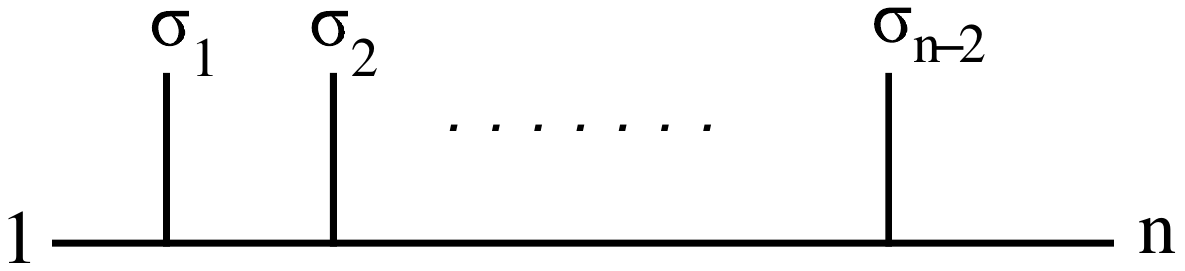}}}\quad \rightarrow \quad 
\tilde{f}^{a_1a_{\sigma_1}b_1}
\tilde{f}^{b_1a_{\sigma_2}b_2}\cdots 
\tilde{f}^{b_{n-3}a_{\sigma_{n-2}}a_n}\,,
\label{DDM}
\eqe
where the positions of legs 1 and $n$ are fixed and $\sigma$ represents  a permutation of the remaining $n\!-\!2$ legs.
As an example, consider a color diagram that has a Y-fork extending from the baseline. Applying the Jacobi identity on the propagator in the Y-fork, the diagram is converted to a linear combination of two diagrams in multi-peripheral form:     
\be
   \raisebox{-5mm}{\includegraphics[scale=0.55]{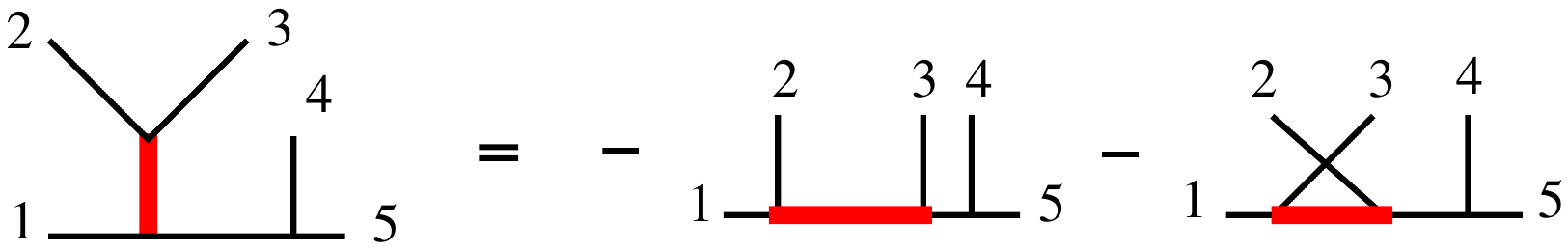}}\,.
\ee
Any trivalent diagram can be cast into a linear combination of diagrams of multi-peripheral form \reef{DDM}. The important point is then that the color factors in multi-peripheral form are not related by any Jacobi identities, so there are a total of $(n-2)!$ independent color factors at a given $n$. 
The full color-dressed tree amplitude can be expressed in terms of this color basis and then the coefficient of each color factor is a gauge invariant quantity, denoted for now by $\tilde{A}_n$. We write full color-dressed amplitude in the multi-peripheral basis as
\be
A^\text{tree}_n=\sum_{\sigma \in S_{n-2}}
\tilde{f}^{a_1a_{\sigma_1}b_1}
\tilde{f}^{b_1a_{\sigma_2}b_2}\cdots 
\tilde{f}^{b_{n-3}a_{\sigma_{n-2}}a_n}
\,\tilde{A}_{n}(1,\sigma_1,\sigma_2,\ldots,\sigma_{n-2},n)\,,
\label{ColorBasis}
\ee
where the sum is over all permutations of lines $2,3,\ldots, n-1$. 

Recall from Section~\ref{s:YM} that we introduced an alternative, manifestly crossing symmetric, representation that uses trace factors of  generators as the basis for the color factor. In this trace-basis the color-dressed amplitude is
\eq
A^\text{tree}_n=\sum_{\sigma \in S_{n-1}}{\rm Tr}(T^{a_{\sigma_1}}T^{a_{\sigma_2}}\cdots T^{a_{\sigma_{n-1}}}T^{a_n})A_{n}\big[\sigma_1,\sigma_2,\ldots,\sigma_{n-1},n\big]\,,
\label{traceDecomp}
\eqe
where one sums over all permutations of lines $1,2,\ldots, n-1$ and $A_{n}[\ldots]$ is our familiar {\bf \em color-ordered amplitude}. Note that there are $(n-1)!$ distinct traces in \reef{traceDecomp}, but since there are only $(n-2)!$ independent color factors, the trace `basis' is over-complete and the color-ordered partial amplitudes satisfy special linear relations. These linear relations are the Kleiss-Kuijf relations~\cite{KK}, the simplest of which is the $U(1)$ decoupling identity shown in \reef{u1dec}. These relations were already discussed in Section \ref{s:YM}; they reduce the number of independent color-ordered amplitudes from $(n\!-\!1)!$ to $(n\!-\!2)!$.

The $(n\!-\!2)!$ partial amplitudes $\tilde{A}_n$ are exactly the color-ordered partial amplitudes that are independent under the Kleiss-Kuijf relations. $\tilde{A}_n$ is not unique, since we could have chosen any other two legs to replace $1$ and $n$ as reference legs in the multi-peripheral color-basis \reef{ColorBasis}. This reflects the fact that there are no unique choice of independent color-ordered partial amplitudes under Kleiss-Kuijf relations. 

In summary, we have reviewed that for a given Yang-Mills $n$-point tree amplitude, there are a total of $(n\!-\!2)!$ color factors that are independent under the Jacobi identities. A convenient choice of independent color factors are those that appear in a multi-peripheral representation, and they can be chosen to be a suitable subset of the $(n-1)!$ color-ordered amplitudes $A_{n}[\ldots]$.

 %%%%%%%%%%%%%%%%%%%%%%%%%%%%%%%%%%%%%%%%%%%%%%%%%%
\subsection{Color-kinematics duality: BCJ, the tree-level story}
%%%%%%%%%%%%%%%%%%%%%%%%%%%%%%%%%%%%%%%%%%%%%%%%%%
The discussion in Section \ref{s:colorYM} may appear to be a deviation from our path to gravity, but it is a useful detour, as we will see shortly. We begin with the only amplitude we know where gravity is given as a direct square of Yang-Mills, namely the 3-point amplitude. We would like to understand what is so special about the 3-point amplitude that is not shared by its higher-point counter parts. The 3-point superamplitude of $\cn=4$ SYM is
\eq
\mathcal{A}_3=\frac{\delta^{(8)}\big(\tilde{Q}\big)}{\langle12\rangle\langle23\rangle\langle31\rangle }\,.
\eqe  
This amplitude is cyclic invariant, as required for a color-ordered superamplitude. But note that it is also totally antisymmetric, exactly as the 3-point color factor $f^{abc}$. Hence, the 3-point superamplitude has kinematics that reflect the structure of the color factor of a 3-vertex. Taking a leap of faith, we might wish to generalize this to higher-points, such that the kinematics of each individual trivalent diagram satisfies the same properties as its color factor, including Jacobi identities as in Figure \ref{colorj}. But what do we mean by the kinematics of each diagram? As we have seen, this is not a gauge invariant statement. It certainly cannot include propagators, as you can see from Figure~\ref{colorj}. Thus we jump to the conclusion that perhaps the numerator factors $n_i$ in \reef{fullAmp} can be arranged to have the same properties as the corresponding color factors $c_i$?

The  {\bf \em Color-Kinematics-Duality} was first proposed for Yang-Mills theories by Bern, Carrasco, and Johansson (BCJ)~\cite{BCJ}. The duality states that scattering amplitudes of Yang-Mills theory, and its supersymmetric extensions, can be given in a representation where the numerators $n_i$ have the {\em same} algebraic properties of the corresponding color factors $c_i$. More precisely, using the representation \reef{fullAmp}, the BCJ proposal is that one can always find a representation such that the following parallel relations hold for the color and kinematic factors:
\be
\begin{array}{rcl}
   c_i= -c_j~~&\Leftrightarrow&~~n_i= -n_j\,\\
   c_i+c_j+c_k=0~~&\Leftrightarrow&~~n_i+n_j+n_k=0\,.
\end{array}
\label{dualityEqns}
\ee
The duality does \emph{not} state that the numerator factors \reef{dualityEqns} have to be local; they are allowed to have poles. 

To illustrate the identity in the first line of \reef{dualityEqns}, consider the two diagrams
\be
  \raisebox{-10mm}{\includegraphics[scale=0.5]{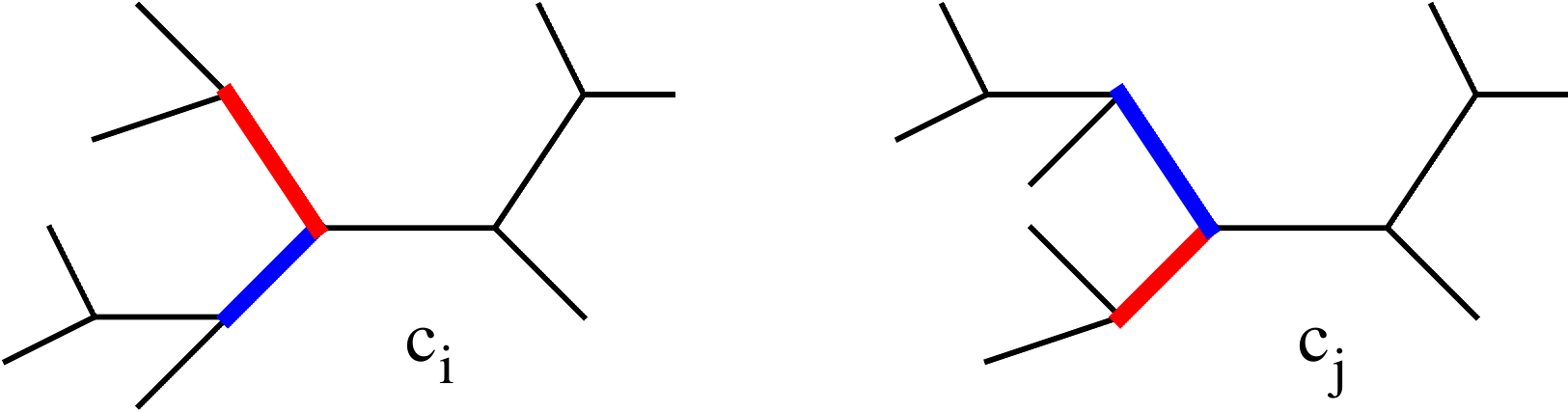}}\,.
  \label{ciminuscj}
\ee  
They are related  by simply switching two lines on a 3-point vertex, highlighted in blue and red in \reef{ciminuscj}. The color factors of the diagrams are related by a minus sign: $c_i=-c_j$, so the color-kinematics duality  \reef{dualityEqns}  states that there exists a representation where the numerator factors of the two diagrams respect the same antisymmetry property: $n_i=-n_j$. 

The second line in \reef{dualityEqns} signifies that the numerator factors must satisfy exactly the same linear relations as their associated color factors. For example, since the color factors of the three diagrams in Figure \ref{colorj} satisfy the Jacobi relation $c_i+c_j+c_k=0$, the color-kinematics duality states that there is a representation of the numerators of the three diagrams such that $n_i+n_j+n_k=0$. 

At first sight this duality may seem implausible. While the underlying reason for the Jacobi identity to hold for the $c_i$'s is the non-abelian gauge algebra defined by two-brackets, $[T^a,T^b]=f^{ab}\,_cT^c$, there appears to be no reason for  the kinematic numerators to satisfy the same relations. As we now show, the color-kinematics duality is not as impossible as it seems. 

Recall that there are $(n-2)!$ independent color factors under the Jacobi identities. If we require that the numerator factors $n_i$ satisfy the same Jacobi identities, then there will only be $(n-2)!$ independent numerators as well. Since there are also only $(n-2)!$ independent color-ordered partial amplitudes, we can express the set of linearly independent partial amplitudes in terms of the 
$(n-2)!$ numerators: 
\be
A_{(i)}=\sum_{j=1}^{(n-2)!} \Theta_{ij}\,\hat n_{j}\,.
\label{Thetaij}
\ee
Here $i,j=1,\ldots , (n-2)!$, and $A_{(i)}$ and $\hat n_{j}$ are the independent color-ordered amplitudes and numerators, respectively. The  $(n-2)!\times (n-2)!$ matrix $\Theta_{ij}$ is comprised solely of massless scalar propagators.  (This matrix was first introduced in \cite{Vaman:2010ez} as the ``propagator matrix".) As an example, for $n=4$ we choose $A_4[1,2,3,4]$ and $A_4[1,3,2,4]$ as the two independent color-ordered amplitudes. Expanding the color factors in \reef{4ptnc} in terms of traces (as in Exercise \ref{ex:earlyBCJ}), we find that 
\eq 
  A_4[1,2,3,4]= -\frac{n_s}{s}+\frac{n_u}{u},
  \hspace{1cm}
 A_4[1,3,2,4]=-\frac{n_u}{u}+\frac{n_t}{t}\,.
\label{4ptncExp}
\eqe 
Enforcing the color-kinematics duality \reef{dualityEqns} on the numerators gives $n_t=-n_s-n_u$. Choosing $n_s,n_u$ as $(\hat{n}_1,\hat{n}_2)$ in \reef{Thetaij}, we can now identify the $2\times2$ matrix $\Theta_{ij}$ from:
\eq
\bigg(\begin{array}{c}A_4[1,2,3,4] \\ A_4[1,3,2,4]\end{array}\bigg)
=
\bigg(\begin{array}{cc} - \frac{1}{s} & \frac{1}{u} \\ -\frac{1}{t} & -\frac{1}{u}-\frac{1}{t}\end{array}\bigg)
\bigg(\begin{array}{c}\hat{n}_1 \\ \hat{n}_2\end{array}\bigg)\,
~~~\longrightarrow~~~
\Theta_{ij}
=\bigg(\begin{array}{cc} - \frac{1}{s} & \frac{1}{u} \\ -\frac{1}{t} & -\frac{1}{u}-\frac{1}{t}\end{array}\bigg)\,.
\label{Theta4pt}
\eqe
As advertised, the matrix $\Theta_{ij}$ is comprised of propagators. The construction generalizes to higher-points. The explicit form of the $6\times6$ matrix $\Theta_{ij}$ for $n=5$ was given in \cite{Vaman:2010ez}. For related work, see \cite{Boels:2012sy}.

Inverting the matrix  $\Theta_{ij}$ would give us numerators  $\hat{n}_i$ expressed in terms of the color-independent amplitudes and from the Jacobi's  one can generate the rest of the numerator factors, thus trivially obtaining a representation that satisfies the color-kinematics duality. If this were true, the color-kinematic duality would be trivial and this doesn't quite smell right. And it isn't: in 4d Yang-Mills theory, the matrix $\Theta_{ij}$ has lower rank, so it cannot be inverted, and we do not have unique numerators $\hat{n}_i$. Indeed looking back at our 4-point example in \reef{Theta4pt}, one can easily verify that the $2\times2$ matrix $\Theta_{ij}$ only has rank 1.

There is also another way to see that the numerator factors cannot be uniquely determined. Suppose we have obtained a set of numerators that satisfy color-kinematic identity. Let us assume that we have achieved this at 5-points and then add to the following term~\cite{Bern:2010yg} to the Yang-Mills action:
\begin{align}
    \mathcal{D}_5
     &=  f^{a_1a_2b}f^{ba_3c}f^{ca_4a_5}\nonumber \\
 &\quad \times \left( \partial^{\phantom{a_1}}_{(\mu} A^{a_1}_{\nu)} A^{a_2}_\rho \frac{\square}{\square} A^{a_3\mu} +
\partial^{\phantom{a_2}}_{(\mu} A^{a_2}_{\nu)}  A^{a_1\mu} \frac{\square}{\square}A^{a_3}_\rho +A^{a_1}_\rho A^{a_2\mu}\frac{\square}{\square}
\partial^{\phantom{a_3}}_{(\mu} A^{a_3}_{\nu)}  \right)
\frac{1}{\square}(A^{a_4\nu} A^{a_5\rho}) \,.
\label{DFive}
\end{align}
The presence of the trivial $1=\frac{\square}{\square}$ facilitates the identification of $n_i$'s in the cubic diagram expansion. The term \reef{DFive} is identically zero thanks for the Jacobi identity, so it does not change the theory. It does, however, modify the Feynman rules so it changes the numerator factors, but in such a way that they still satisfy the color-kinematic Jacobi identity. We conclude that the duality-satisfying numerators $\hat{n}_i$ cannot be unique; and this is why the matrix $\Theta_{ij}$ is not invertible in 4d Yang-Mills theory.

Given that the matrix $\Theta_{ij}$ has lower rank, there must be linear relations among the color-independent partial amplitudes. To expose such relations for $n=4$, use the first row of \reef{Theta4pt} to express $\hat{n}_1$ in terms of the partial amplitude $A_4[1,2,3,4]$ and $\hat{n}_2$, 
\eq
\hat{n}_1= -s A_4[1,2,3,4]+\frac{s}{u}\hat{n}_2\,.
\label{4ptBCJ1}
\eqe  
Substituting this solution into the second row of \reef{Theta4pt}, we find that 
\eq
A_4[1,3,2,4]
~=~ 
-\frac{\hat{n}_1}{t}-\frac{\hat{n}_2}{t}-\frac{\hat{n}_2}{u}
~=~ 
\frac{s}{t}A_4[1,2,3,4]-\left(\frac{s}{ut}+\frac{1}{t}+\frac{1}{u}\right)\hat{n}_2\,.
\label{4ptbcj0}
\eqe
The coefficient of $\hat{n}_2$ is proportional to $s+t+u=0$. Thus, imposing color-kinematics duality, gives the following relation among color-ordered amplitudes:
\eq
t\, A_4[1,3,2,4]\,=\,s \, A_4[1,2,3,4]\,.
\label{4ptBCJ2}
\eqe
This is an example of the {\bf \em BCJ relations} that we discussed earlier in Section \ref{s:YM}. In fact, what we did above is equivalent to your calculation in Exercise \ref{ex:earlyBCJ}.
Since the matrix $\Theta_{ij}$ is defined with respect to color-ordered amplitudes that are independent under the Kleiss-Kuijf relations, the BCJ amplitude relations are new relations beyond the consequences of the color-structure. These novel relations reflect that color-kinematics duality exists in Yang-Mills theory. It is known \cite{BCJ} that for general $n$, the matrix $\Theta_{ij}$ has rank $(n-3)!$, thus implying $(n-2)!-(n-3)!=(n-3)(n-3)!$ BCJ relations among $n$-point color-ordered amplitudes.  The simplest type of such relations (sometimes called {\bf \em fundamental BCJ relations}) can be nicely condensed to the form~\cite{BCJ}
\eq	
\sum_{i=3}^n\bigg(\sum_{j=3}^i s_{2j}\bigg)A_n\big[1,3,\ldots,i,2,i+1,\ldots,n\big] =0\,.
\label{FundBCJ}
\eqe
In our 4-point example, $\hat{n}_2$ dropped out of the final equation \reef{4ptbcj0}. This means that no consistency conditions can be put on $\hat{n}_2$, so we can take $\hat{n}_2$ to be anything we want without affecting the physical amplitude. Such free numerators are sometimes referred to as ``pure-gauge". In practice, it is often convenient to set them to zero. 

Now you may say that this is all very interesting, but have we  lost sight of our original motivation, to get gravity amplitudes from Yang-Mills theory!? No worries, we are already there. A remarkable proposed consequence of the color-kinematics duality is that once duality-satisfying numerators $n_i$ are obtained, the formula 
\eq
M_n= \sum_{i\in {\rm cubic}}\frac{n_i^2}{\prod_{\alpha_i}p^2_{\alpha_i}}\,.
\label{BCJSugra}
\eqe
calculates the $n$-point tree amplitude in the (super)gravity whose spectrum is given by squaring the (super) Yang-Mills spectrum. 
That is, we simply take the Yang-Mills amplitude formula in \reef{fullAmp} and replace each color factor $c_i$ with the corresponding duality-satisfying numerator $n_i$. And, boom, that is gravity! This relation is called the {\bf \em BCJ double-copy relation}. The formula \reef{BCJSugra} manifestly reproduces all possible poles that should appear in the gravity amplitude. Furthermore, the mass-dimension matches on both sides of the equation. 
\example{Let us check \reef{BCJSugra} at 4-points. From \reef{BCJSugra}, we find that 
\eq
M_4
\,=\,
\bigg(\frac{n^2_s}{s}+\frac{n^2_u}{u}+\frac{n_t^2}{t}\bigg)
\,=\, 
\bigg(\frac{n^2_s}{s}+\frac{n^2_u}{u}+\frac{(n_s+n_u)^2}{t}\bigg)\,,
\eqe
where in the second equality, we have used the duality to set $n_t=-n_s-n_u$. Now use \reef{4ptBCJ1} and \reef{4ptBCJ2}, remembering that $(\hat{n}_1,\hat{n}_2)$ is identified with $(n_s,n_u)$ and that we can freely set $\hat{n}_2=0$: we then find
\eq
M_4\,=\, -\frac{su}{t}\,A_4[1,2,3,4]^2\,=\,
- u\, A_4[1,2,3,4]\, A_4[1,3,2,4]\,.
\eqe
This is just a different form of the KLT formula we encountered  previously in \reef{KLT45}! A more involved 5-point example was worked out in details in~\cite{BCJ}. Thus, by reproducing the correct KLT relations, the validity of \reef{BCJSugra} is verified at 4- and 5-points. 
}
\exercise{}{What if we choose a gauge where $\hat{n}_2$ is not zero? Show that when substituting  \reef{4ptBCJ1} and \reef{4ptBCJ2} into \reef{BCJSugra}, $\hat{n}_2$ drops out in the final result. This shows that the gravity formula \reef{BCJSugra} is gauge invariant.} 
The BCJ squaring relations \reef{BCJSugra} can be exploited to determine an explicit representation of color-kinematics duality-satisfying numerators of the tree amplitude \cite{Tree,Treetoo}. Recall that by using the color Jacobi relations, we can convert the fully dressed amplitude in \reef{fullAmp}  into the multi-peripheral form~\reef{DDM}. Assuming that we have duality-satisfying numerators, the double copy representation of the gravity amplitude in \reef{BCJSugra} can now go through exactly the same steps as those that converted \reef{fullAmp}  into~\reef{DDM}, and obtain a multi-peripheral form of gravity amplitude
\eq
M_n= \sum_{\sigma \in S_{n-2}}n_{1| \sigma_1,\sigma_2,\ldots,\sigma_{n-2}| n}\,A_{n}(1,\sigma_1,\sigma_2,\ldots,\sigma_{n-2},n)\,,
\label{MPGravi}
\eqe 
where $n_{1|\sigma_1,\sigma_2,\ldots,\sigma_{n-2}| n}$ are the duality-satisfying numerators for the cubic diagrams in multi-peripheral form with legs 1 and $n$ held fixed. Thus we have an expression for the gravity amplitude in terms of a sum of Yang-Mills color-ordered amplitudes times kinematic factors. Note that it looks a lot like the KLT formula. Indeed, as realized first in \cite{Tree}, by lining up a copy of the KLT formula with the color-ordered amplitudes in \reef{MPGravi}, one can readily read-off duality-satisfying $n_i$'s. Take, for example, the $n=4,5$ KLT formulas \reef{KLT45}
\be
  \begin{split}
  M_4(1234) &=~ - s_{12}  \,A_4[1234] \,A_4[1243]\,,\\
  M_5(12345) &=~  s_{23} s_{45}  \,A_5[12345] \,A_5[13254] + (3 \lra 4)\,.
  \end{split}
\ee  
Choosing legs $1$ and $n$ to be fixed in our multi-peripheral form, we can readily read off:
\eqa
\begin{array}{llll}
\text{$n=4$:}~~~~ &n_{1|2,3|4}= - s_{12} \,A_4[1243], 
& \hspace{-2.2cm}
n_{1| 3,2 |4}=0\,.\\[1.5mm]
\text{$n=5$:}~~~~ &n_{1|2,3,4|5}=s_{23} s_{45} \,A_5[13254]\, , 
& \hspace{-2.2cm} n_{1|2,4,3|5}=s_{24} s_{35} \,A_5[14253]\,,\\
& n_{1|3,4,2|5}=n_{1|4,2,3|5}=n_{1|4,3,2|5}=n_{1|3,2,4|5}=0\,.
\end{array}
\eqae
From these $(n-2)!$ independent numerators, all remaining numerators can be obtained by applying the Jacobi identities.

If you think that the BCJ double-copy relation \reef{BCJSugra} is too good to be true, have no fear: things are about to get even better! It turns out, that the squaring relation \reef{BCJSugra} can be generalized to 
\eq
M_n= \sum_{i\in {\rm cubic}}\frac{n_i\tilde{n}_i}{\prod_{\alpha_i}p^2_{\alpha_i}}\,,
\label{BCJSugra2}
\eqe
where the gravity numerators are given as the product of two possibly \emph{distinct} Yang-Mills numerators. Only one set of numerators, say, $n_i$ has to satisfies the color-kinematics duality \reef{dualityEqns}, while the other copy, $\tilde{n}_i$, can be an arbitrary representation of the Yang-Mills amplitude. To understand why this is so, let us assume that $n_i$ respects the duality  \reef{dualityEqns} while $\tilde{n}_i$ does not. Define the difference of the two distinct numerators to be
\eq
\Delta_i\equiv n_i-\tilde{n}_i\,.
\eqe
Since $n_i$ and $\tilde{n}_i$ are both valid representations of the same Yang-Mills amplitude, it follows from \reef{fullAmp} that 
\eq
 \sum_{i\in {\rm cubic}}\frac{c_i\Delta_i}{\prod_{\alpha_i}p^2_{\alpha_i}}=0\,.
\label{GenNumShift}
\eqe 
In the discussion so far, we have not specified the gauge group, just that it is non-abelian with structure constants that satisfy the Jacobi identities. Thus the only property of the color factors $c_i$ that can make \reef{GenNumShift} hold is the Jacobi relation. Since the $n_i$'s satisfy color-kinematics duality, they satisfy the exact same algebraic properties as the $c_i$'s, so we conclude that
\eq
 \sum_{i\in {\rm cubic}}\frac{n_i\Delta_i}{\prod_{\alpha_i}p^2_{\alpha_i}}=0\,.
 \label{FundBCJx}
\eqe 
This  establishes the equivalence of \reef{BCJSugra} and \reef{BCJSugra2}.

Why bother with the existence of \reef{BCJSugra2} vs.~\reef{BCJSugra}? --- We need one set of duality-satisfying numerators for \reef{BCJSugra2} anyway, so why not simply square them and just use \reef{BCJSugra}? Recall from Section \ref{s:sugraspec} that the spectrum of many supergravity theories can be obtained from tensor'ing two different Yang-Mills theories. For example, the spectrum of pure $\mathcal{N}=4$ supergravity is the product of  $\mathcal{N}=4$ SYM and pure Yang-Mills theory \reef{N4SGspectrum}. The point is then that the BCJ double-copy relation \reef{BCJSugra2} can be used to construct the supergravity scattering amplitude by using the numerators of two distinct (S)YM theories. And importantly, only one copy of the numerators needs to satisfy the duality, not both. Thus, if we have a set of duality-satisfying numerators for $\mathcal{N}=4$ SYM, by simply combining them in \reef{BCJSugra2} with the numerators of ordinary Yang-Mills, say obtained from explicit Feynman diagram computation, we directly get the scattering amplitudes of  $\mathcal{N}=4$ supergravity. This convenient result has powerful consequences as we move on to loop amplitudes in Section  \ref{s:BCJLoops}.

We end this section with some concluding remarks regarding the tree-level BCJ color-kinematics duality \reef{dualityEqns} and the double-copy relations  \reef{BCJSugra} and  \reef{BCJSugra2}. {\em First}, the existence of numerators that satisfy \reef{dualityEqns} was exemplified for any $n$ in \cite{Tree,Treetoo} (see also \cite{Tree2}). 

{\em Second}, the BCJ relations in \reef{FundBCJ} have been successfully derived from string theory using monodromy relations~\cite{stringtheoryBCJ, StringBCJ} and in field theory using the improved large-$z$ fall-off of non-adjacent BCFW shifts \cite{Feng:2010my}. An elegant derivation was given in~\cite{CachazoBCJ}. In our discussion, the BCJ relations were a consequence of imposing color-kinematic duality on the numerator. Given that the BCJ relations can be proven via string- and field-theory arguments, one can reverse the argument and show that the existence of BCJ relations and Kleiss-Kuijf identities give rise to algebraic relations on the kinematics~\cite{BCJAlgebra}.

{\em Third}, assuming that there exists a duality-satisfying set of {\em local} numerators for the Yang-Mills amplitude, one can  rigorously prove that the doubling-relation  \reef{BCJSugra} produces the correct gravity amplitude for any $n$ \cite{Bern:2010yg}. The proof is established inductively by showing that the difference between \reef{BCJSugra} and the gravity amplitude obtained from BCFW recursion, vanishes if one assumes  \reef{BCJSugra} holds for all lower-point amplitudes.

{\em Finally}, you may wonder if duality-satisfying numerators can be obtained directly from the Feynman rules of some Lagrangian. In 4d, this can be done for MHV amplitudes~\cite{SelfDualBCJ}, but difficulties arise beyond MHV. In general dimensions, a straightforward construction of the cubic diagrams using Feynman rules does not give duality-satisfying numerators, even if the freedom of how to assign contact terms is taken into account. However, modification the action by non-local terms can give duality-satisfying numerators straight from the Feynman rules of the deformed action \cite{Bern:2010yg} 
\begin{equation}
  \mathcal{L}_{YM} = \mathcal{L}+\mathcal{L}'_5+ \mathcal{L}'_6+\ldots .
\end{equation}
Here $\mathcal{L}$ is the conventional Yang-Mills Lagrangian and
$\mathcal{L}'_n,\,n>4$ are terms that involve $n$ fields and vanish by the Jacobi identity so that the theory is actually not changed. As an example, the quintic terms are
\begin{equation}
\mathcal{L}'_5\sim \Tr\,[A^{\nu} ,A^{\rho}]
\frac{1}{\square}\Big(\big[[\partial_{\mu} A_{\nu}, A_{\rho}], A^{\mu}\big]
+\big[[A_{\rho}, A^{\mu}],\partial_{\mu} A_{\nu}\big]
+\big[[A^{\mu},\partial_{\mu} A_{\nu} ], A_{\rho}\big]\Big)\,.
\end{equation}
Even though the deformation is non-local, it is completely harmless: 
$\mathcal{L}'_5$ is simply zero because the terms in the parenthesis vanish by the Jacobi identity. Thus by adding a particular zero to the action, one can expose the intricate relation between gravity and Yang-Mills theory. 
(One may say that this points to a curious deficiency of the action, namely that it treats all zeroes in the same way.)
A systematic approach to generating explicit higher-order deformations $\mathcal{L}'_n$ is given in \cite{BCJEL}.

We have seen how the tree-level squaring relation between gravity and Yang-Mills is more straightforward when phrased in terms of the non-gauge-invariant numerators $n_i$ than in terms of the gauge-invariant partial color-ordered Yang-Mills amplitudes, as in KLT. The true power of this is revealed when it is applied to loop-integrands: we will see that the BCJ squaring relations survive at loop-level while this is not the case for the KLT formula.

 %%%%%%%%%%%%%%%%%%%%%%%%%%%%%%%%%%%%%%%%%%%%%%%%%%
\subsection{Color-kinematics duality: BCJ, the loop-level story}\label{s:BCJLoops}
%%%%%%%%%%%%%%%%%%%%%%%%%%%%%%%%%%%%%%%%%%%%%%%%%%
We begin with color-kinematic duality for loop amplitudes of Yang-Mills theory. 
Any diagram involving the 4-point contact term can be blown up into cubic vertices, as discussed in Section \ref{s:colorYM}, so we consider only trivalent loop-diagrams.
The full $L$-loop color-dressed Yang-Mills amplitude can then be written as
\be
  {\cal A}^{L\text{-loop}}_n  =  
\sum_{j\in {\rm cubic}}{\int{
 \Big( \prod_{l = 1}^L \frac{d^D \ell_l}{ (2 \pi)^D} \Big)
  \frac{1}{S_j}  
 \frac {n_j \,c_j}{\prod_{\alpha_j}{p^2_{\alpha_j}}}}}\,, 
 \label{loopCKdual}
\ee
where the notation follows that defined for \reef{fullAmp} and $S_j$ is the symmetry factor of the diagram. 
It was proposed in \cite{BCJLoop} that there exists representations \reef{loopCKdual} where the kinematic numerators $n_i$ satisfy the same algebraic relations as that of color factors, i.e.~\reef{dualityEqns}. And once such numerators are found, the gravity amplitude is given by the double-copy formula \cite{BCJLoop}   
\be
  {\cal M}^{L\text{-loop}}_n  = 
\sum_{j\in {\rm cubic}}{\int{
 \Big( \prod_{l = 1}^L \frac{d^D \ell_l}{ (2 \pi)^D} \Big)
  \frac{1}{S_j}  
 \frac {n_j \,\tilde{n}_j}{\prod_{\alpha_j}{p^2_{\alpha_j}}}}}\,,
 \label{GraviLoop}
\ee
in which only one of the two copies $n_i,\tilde{n}_i$ is required to satisfy color-kinematics duality. 

The validity of \reef{GraviLoop} can  be justified through unitarity cuts~\cite{UnitarityMethod}:  assuming that
gauge-theory numerators $n_i$ satisfy the duality, the gravity integrand built
by taking double copies of numerators has the correct cuts in all channels. To see this, consider a set of generalized unitarity cuts that break the loop amplitude down to products of tree-amplitudes. On the cut, the gauge-theory integrand factorizes into products of tree amplitudes whose numerator factors satisfy all color-kinematics dualities relevant for each tree amplitude, because they are merely a subset of the relations required by the loop-level duality. As an example consider the following unitarity cut of the 3-loop cubic diagram:
\eq
\vcenter{\hbox{\includegraphics[scale=0.42]{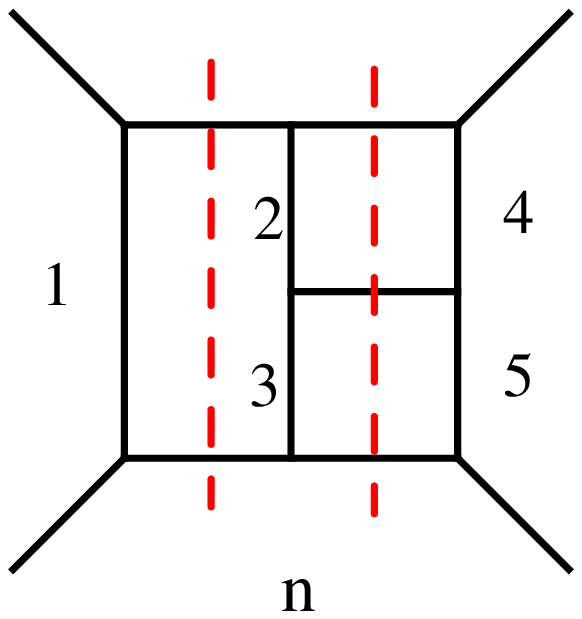}}}=~~ \frac{n}{p^2_1p^2_2p^2_3p^2_4p^2_5}\bigg|_{\rm cut}~ = ~~ \vcenter{\hbox{\includegraphics[scale=0.42]{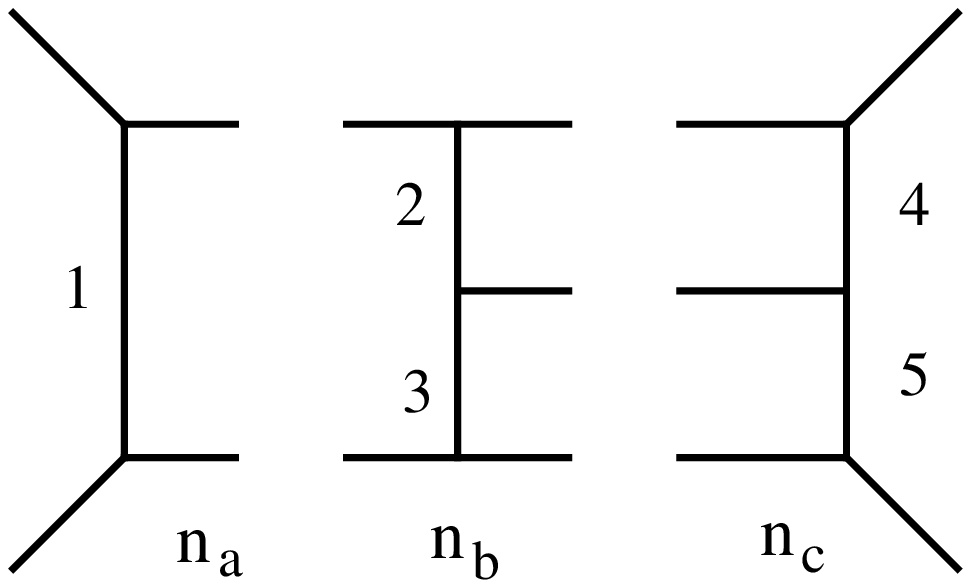}}}~=~~ \sum_{\rm states} \frac{n_a}{p^2_1}\;\frac{n_b}{p^2_2p^2_3}\;\frac{n_c}{p^2_4p^2_5}\,.
\eqe
We have labeled the {\em uncut} propagators $1/p^2_i$ by $i=1,\ldots,5$ and the numerator of the diagram is denoted by $n$. On the cut, the LHS must be equivalent to the RHS, which is the product of kinematic factors of the factorized tree-diagrams. Since the numerator $n$ satisfy all Jacobi identities associated with the parent diagram, the numerators $n_a$, $n_b$, and $n_c$ must satisfy the Jacobi identities of the individual tree-diagrams. Now squaring the duality-satisfying numerators in the Yang-Mills tree amplitude, one obtains the gravity tree amplitude. Thus, squaring the Yang-Mills loop-numerators, one is guaranteed to obtain the correct cut for the gravity 
loop-amplitude. In other words, \reef{GraviLoop} is guaranteed to satisfy all unitarity cuts and therefore give the correct answer.

If any readers have come all the way with us here to page \label{pagealive}\pageref{pagealive}, then they may question if the above argument only justifies \reef{GraviLoop} as the correct answer for the cut-constructible part of the (super)gravity loop-amplitude: what about rational terms that are not 
cut constructible? 
Recall that rational terms can be obtained by considering the unitarity cuts in higher-dimensions, where the extra-dimensional momenta can be interpreted as the regulator (see \reef{rationals}). But in the discussion so far, there was no specification of the spacetime dimension: the color-kinematics duality and the double-copy relation is valid in arbitrary dimensions! In other words, \reef{GraviLoop} produces the correct cut in any spacetime dimension, and thus it also faithfully reproduces the rational terms.

Now the validity of \reef{GraviLoop} relies on the existence of duality-satisfying numerators. Do we know that there always exists such a representation --- and, if so, how to systematically construct it? Indeed,  this is the million dollar question. Unlike at tree-level, there is currently not a formal proof of the existence of duality-satisfying numerators. However, we have explicit examples of such numerators in multiple cases, as will be summarized in Section \ref{s:summaryBCJ}. For now, let us see some non-trivial examples.

%%%%%%%%%%%%%%%%%%%%%%%%%%%%%%%%%%%%%%%%%%%%%%%%%%%%%%%%%%%%
\subsubsection{1-loop 4-point $\mathcal{N}=4$ SYM} 
\label{s:BCJ1L}
%%%%%%%%%%%%%%%%%%%%%%%%%%%%%%%%%%%%%%%%%%%%%%%%%%%%%%%%%%%%
In Section \ref{s:1loopUni}, we used the unitarity method to compute the color-ordered 1-loop 4-point superamplitude of $\mathcal{N}=4$ SYM and found the result to be
\eq
\mathcal{A}_4^\text{1-loop}[1234] 
~=~ 
su \,\mathcal{A}_4^{\rm tree}[1234]\,I_4(p_1,p_2,p_3,p_4)\,,
\eqe 
where $I_4(p_1,p_2,p_3,p_4)$ is the scalar box integral. The fully color-dressed 1-loop amplitude can be written in terms of color-ordered amplitudes as \cite{Brink:1976bc,N=42Loop1}
\eq
\mathcal{A}_4^\text{1-loop,\,full}
~=~
c^{(1)}_{1234}\,\mathcal{A}_4^\text{1-loop}[1234] 
+c^{(1)}_{1342}\,\mathcal{A}_4^\text{1-loop}[1342]
+c^{(1)}_{1423}\,\mathcal{A}_4^\text{1-loop}[1423]\,,
\label{OneLoopFull}
\eqe
with $c^{(1)}_{ijkl}$ the 1-loop color factor of a box diagram with consecutive external legs $(i, j, k, l)$, e.g.~ 
\eq
c^{(1)}_{1234}=\tilde{f}^{e a_1b}\tilde{f}^{ba_2c}\tilde{f}^{ca_3d}\tilde{f}^{da_4e}\,.
\eqe
Now we show that \reef{OneLoopFull} actually satisfies color-kinematic duality. Take one of the four propagators in the box diagram and apply the Jacobi identity to, say, the propagator between legs $1$ and $2$:
\be
  \raisebox{-10mm}{\includegraphics[scale=0.5]{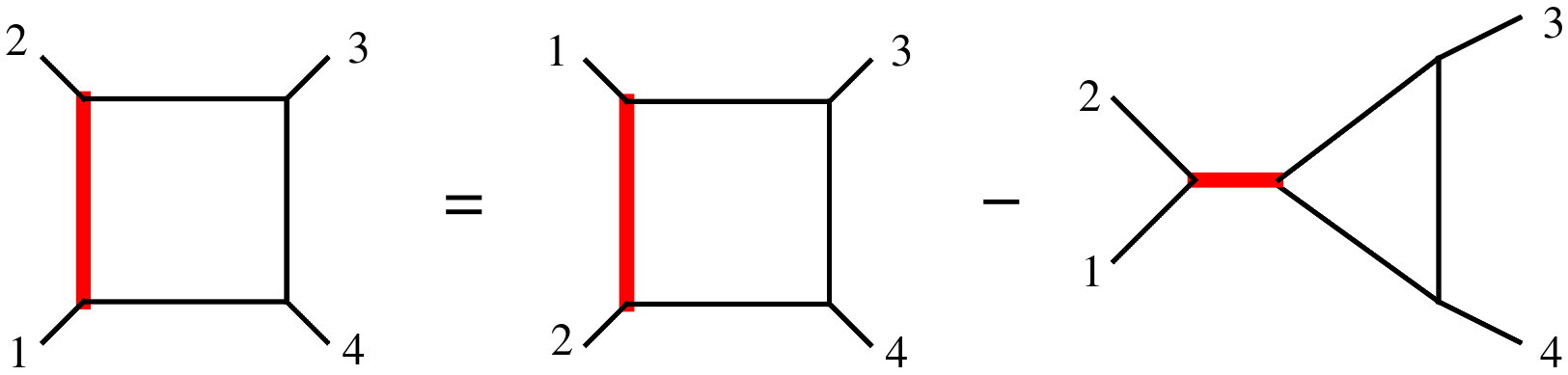}}
\label{OneLoopJacobi0}
\ee
The relevant Jacobi identity is
\eq
c^{(1)}_{1234}-c^{(1)}_{2134}+c^{(1)}_{\text{Tri}:34}=0\,,
\eqe
where $c^{(1)}_{\text{Tri}:34}\equiv \tilde{f}^{a_1ba_2}\tilde{f}^{ebc}\tilde{f}^{ca_3d}\tilde{f}^{da_4e}$ is the color factor for a triangle diagram. The duality then states that the numerators of the integrals are related as 
\be
 n^{(1)}_{1234}-n^{(1)}_{2134}+n^{(1)}_{\text{Tri}:34}=0\,.
\label{OneLoopJacobi}
\ee
Looking back at the 1-loop color-dressed result \reef{OneLoopFull}, we  immediately identify
\eq
n^{(1)}_{1234}=su \,\mathcal{A}_4^{\rm tree}[1234],\qquad 
n^{(1)}_{2134}= st \,\mathcal{A}_4^{\rm tree}[2134],\qquad 
n^{(1)}_{\text{Tri}:34}=0\,.
\label{OneLoopNumerator}
\eqe
Since $su \,\mathcal{A}_4^{\rm tree}[1234]$ is fully permutation invariant, it equals $st \,\mathcal{A}_4^{\rm tree}[2134]$, and therefore the numerator Jacobi identity \reef{OneLoopJacobi} is trivially satisfied! Applying the Jacobi identity to  any other propagator of the box diagram, one arrives at the same result.

In principle we should also check the Jacobi identity
\be
  \raisebox{-10mm}{\includegraphics[scale=0.33]{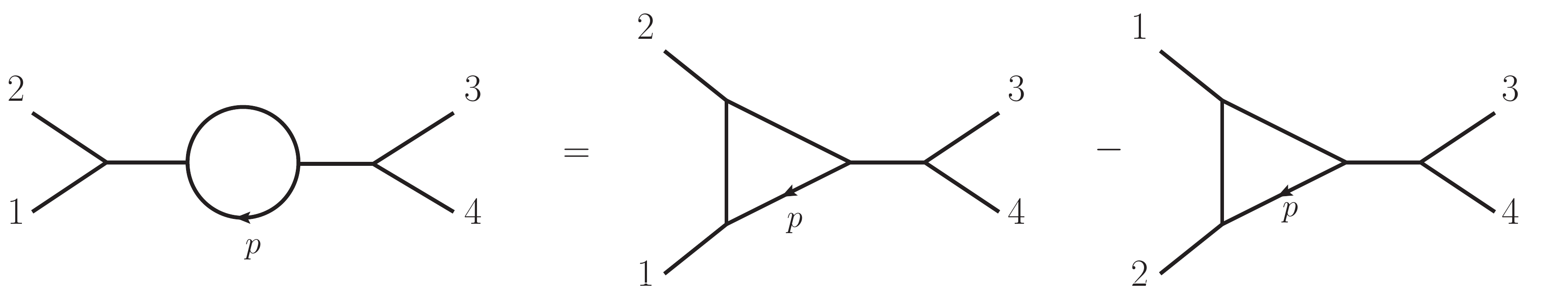}} \,.
\ee
However, for $\mathcal{N}=4$ SYM, this is trivially satisfied since the numerator associated with each of the above diagrams is zero.

%%%%%%%%%%%%%%%%%%%%%%%%%%%%%%%%%%%%%%%%%%%%%%%%%%%%%%
\subsubsection{2-loop 4-point $\mathcal{N}=4$ SYM}
\label{s:BCJ2L} 
%%%%%%%%%%%%%%%%%%%%%%%%%%%%%%%%%%%%%%%%%%%%%%%%%%%%%%
At 2-loop order, the color-dressed Yang-Mills amplitude is 
\cite{N=42Loop1,N=42Loop2} 
\bea
\nonumber\mathcal{A}_4^\text{2-loop,\,full}
&=&
\Big(c^{\rm P}_{1234}\,\mathcal{A}_4^\text{P}[1234] 
+c^{\rm P}_{3421}\,\mathcal{A}_4^\text{P}[3421]
+c^{\rm NP}_{1234}\,\mathcal{A}_4^\text{NP}[1234] 
+c^{\rm NP}_{3421}\,\mathcal{A}_4^\text{NP}[3421]\Big)\\
&&~+~\text{cyclic}(2,3,4)\,,
\label{TwoLoopFull}
\eea
where ``cyclic(2, 3, 4)" indicates a sum over the remaining two cyclic permutations of legs
2, 3 and 4. 
The color factors $c^{\rm P}_{1234}$ and $c^{\rm NP}_{1234}$ are obtained by dressing the planar and nonplanar double-box diagrams with structure constants $\tilde{f}^{abc}$:
\eq
\raisebox{-7mm}{\includegraphics[scale=0.5]{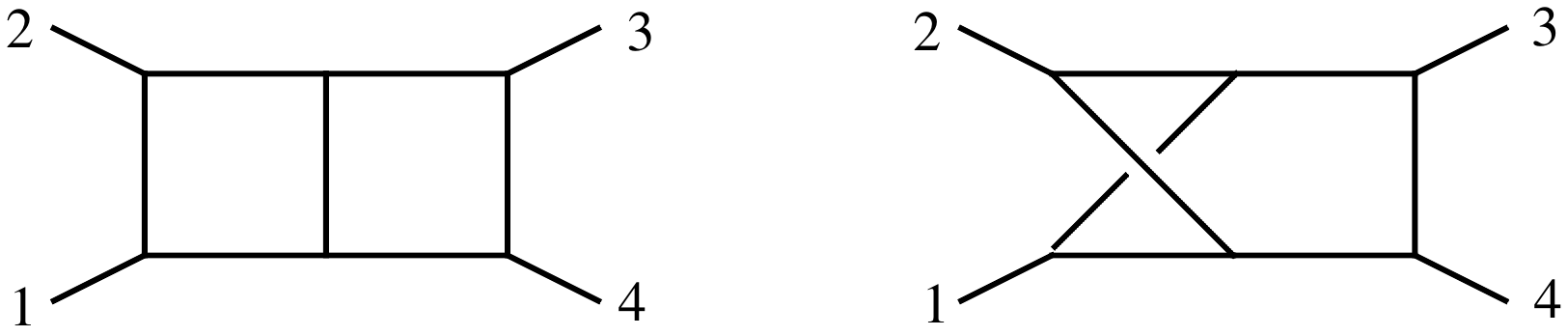}}~~\,.
\label{doubleboxes}
\eqe
The explicit amplitudes $\mathcal{A}_4^\text{P}[1234]$ 
and $\mathcal{A}_4^\text{NP}[1234]$ is given by the 
2-loop scalar-box integrals corresponding to the diagrams \reef{doubleboxes}, multiplied the numerator factors
\eq
n^{\rm P}_{1234}=s^2u\,\mathcal{A}_4^{\rm tree}[1234]\,,
\qquad n^{\rm NP}_{1234}=s^2t\,\mathcal{A}_4^{\rm tree}[1234]\,.
\label{TwoLoopNum}
\eqe
Color-kinematic duality now imposes the following linear relation among the numerators of the scalar integrals:
\eq
\begin{split}
&
\,\raisebox{-5mm}{\includegraphics[scale=0.65]{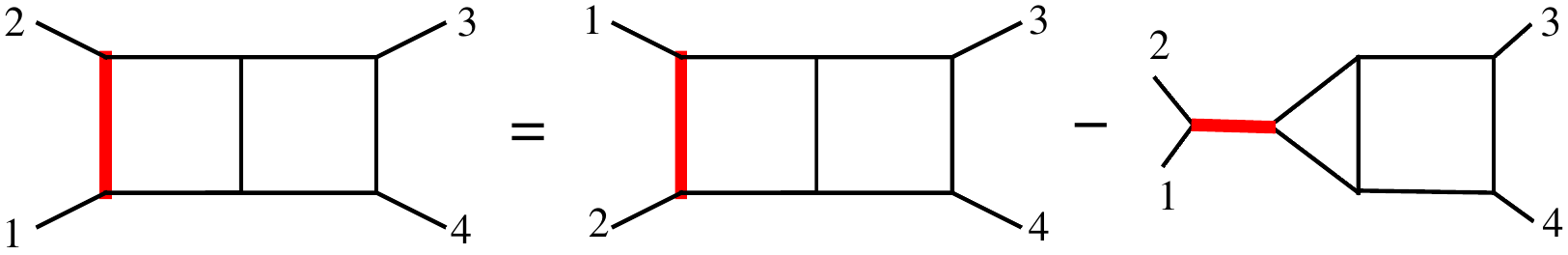}}~,\\[1mm]
&
\raisebox{-5mm}{\includegraphics[scale=0.65]{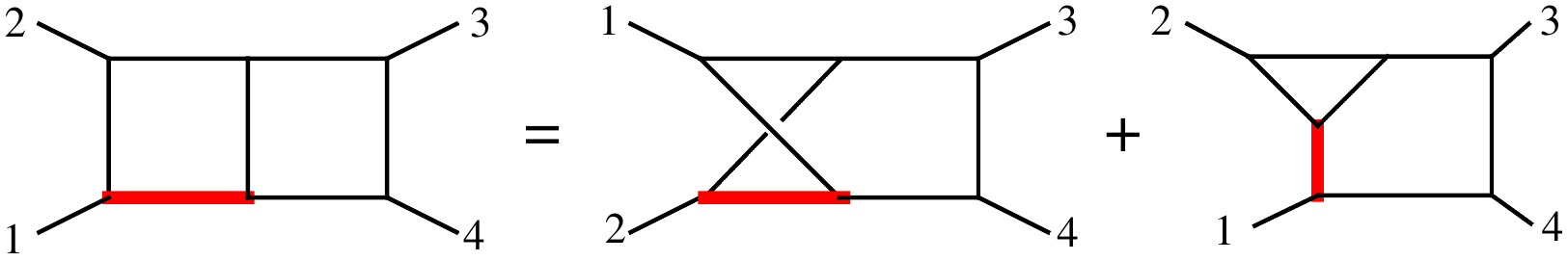}}~\,.
\end{split}
\eqe
The above two identities are satisfied by the numerators in \reef{TwoLoopNum} by virtue of the absence of integrals with triangle sub-loops as well as the permutation invariance of $su\,\mathcal{A}_4^{\rm tree}[1234]$.
\exercise{}{What is the identity associated with the Jacobi relation applied to the red propagator in the following diagram:
\be
   \raisebox{-8mm}{\includegraphics[scale=0.45]{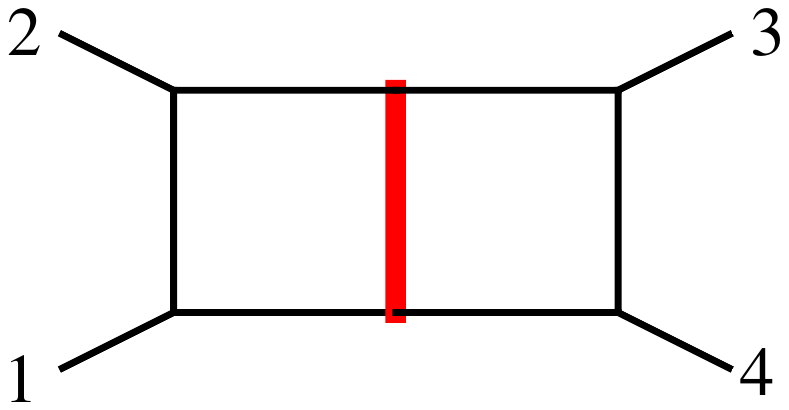}}~~~?
\ee
Do the numerators in \reef{TwoLoopNum} satisfy this identity?}

%%%%%%%%%%%%%%%%%%%%%%%%%%%%%%%%%%%%%%%%%%%%%%%%%%%%%%
\subsubsection{3-loop 4-point $\mathcal{N}=4$ SYM}   
\label{s:BCJ3L}
%%%%%%%%%%%%%%%%%%%%%%%%%%%%%%%%%%%%%%%%%%%%%%%%%%%%%%
Thus far, the duality-satisfying numerators $n_i$ been independent of the loop-momenta. At 3-loop order, the representation given in \cite{BCJLoop} 
 for the 4-point amplitude uses duality-satisfying numerators that do depend on the loop-momenta. The scalar integrals that participate in the 3-loop answer are 
\eq
\raisebox{-35mm}{\includegraphics[scale=0.43]{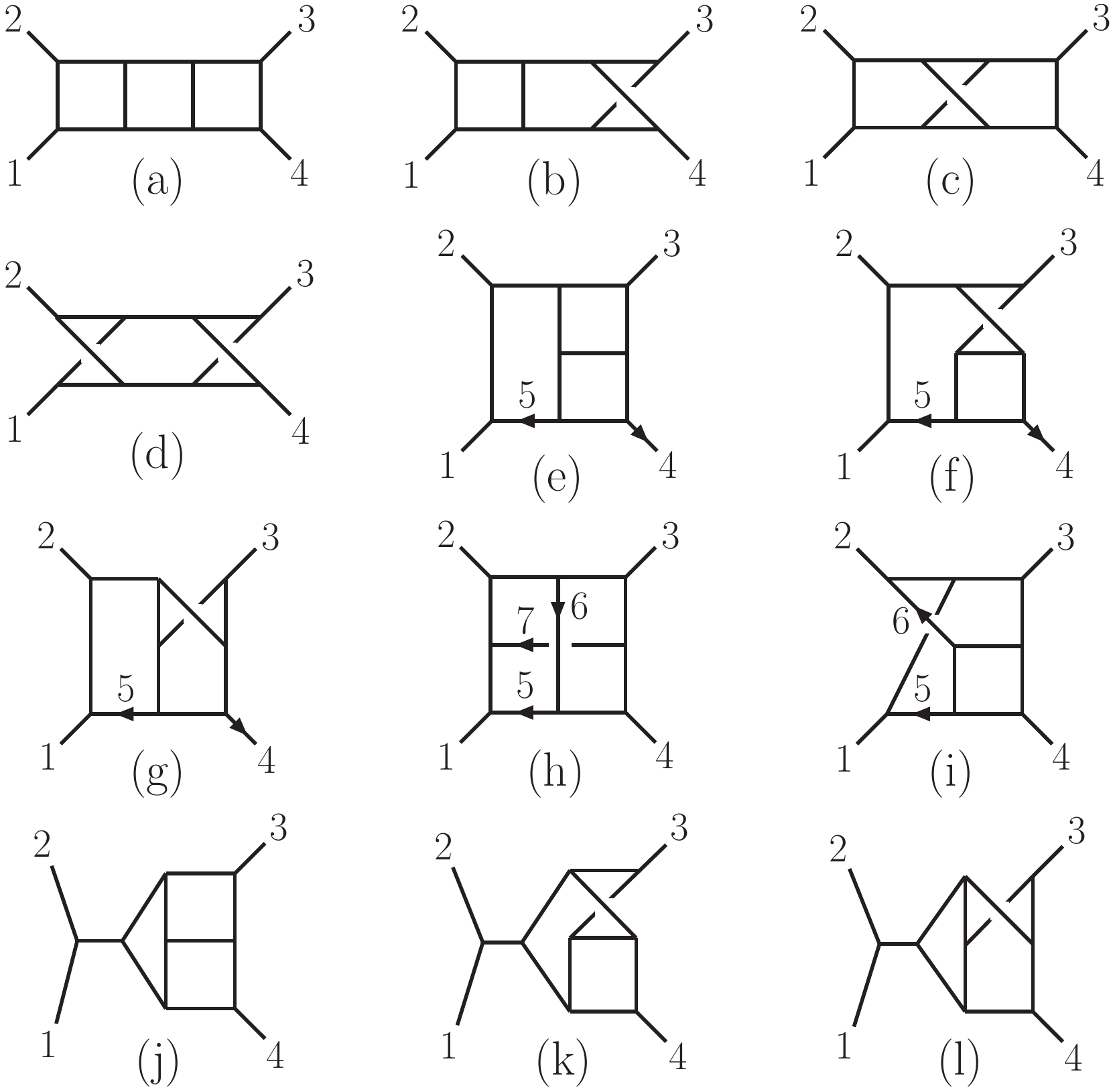}}
\label{3LoopIntegrals}
\eqe
The full amplitude is given as a sum of these integrals (along with permutations of the external legs), their associated color factors and kinematic numerators, suitably normalized by the symmetry factor of the diagram.  The kinematic numerators are 
\be
\begin{tabular}{||c|c||}
\hline
Integral $I^{(x)}$ &  $\cn=4$ SYM  numerator  \\
\hline
\hline
(a)--(d) &  $s^2$   \\
\hline 
(e)--(g) & $\big(\,s \left(-\tau _ {3 5}+\tau _ {4 5}+u \right)- u \left(\tau _ {2 5}+\tau _ {4 5}\right)+
 t \left(\tau _ {2 5}+\tau _ {3 5}\right)-s^2 \, \big)/3$   \\
\hline
(h)& $ \big(\, s \left(2 \tau _ {1 5}-\tau _ {1 6}+2 \tau _ {2 6}-\tau _ {2 7}+2 \tau _ {3 5}+\tau _ {3 6}+\tau _ {3 7}-t \right)$\\
&$+ u \left(\tau _ {1 6}+\tau _ {2 6}-\tau _ {3 7}+2\tau _ {3 6}-2 \tau _ {1 5}-2\tau _ {2 7}-2\tau _ {3 5}-3 \tau _ {1 7}\right)+s^2 
\,\big)/3$\\
\hline
(i)& $\big(\, s \left(-\tau _ {2 5}-\tau _ {2 6}-\tau _ {3 5}+\tau _ {3 6}+\tau _ {4 5}+2 u \right)$\\
&$+ u \left(\tau _ {2 6}+\tau _ {3 5}+ 2\tau _ {3 6}+2\tau _ {4 5}+3 \tau _ {4 6}\right)+ t\,\tau _ {2 5}+s^2 \,\big)/3$\\
\hline
(j)-(l) & $s (u-t)/3 $
 \\
\hline
\end{tabular}
\label{3Lnums}
\ee
An overall factor of $s u \mathcal{A}_4^{\rm tree}$ has
been removed, and $\tau_{i j} = 2 k_i\cdot
l_j$, where $k_i$ and $l_j$ are momenta as labeled in each diagram above. The numerators in the table satisfy all Jacobi identities of the corresponding color factors. For example, 
\be
  \raisebox{-10mm}{\includegraphics[scale=0.5]{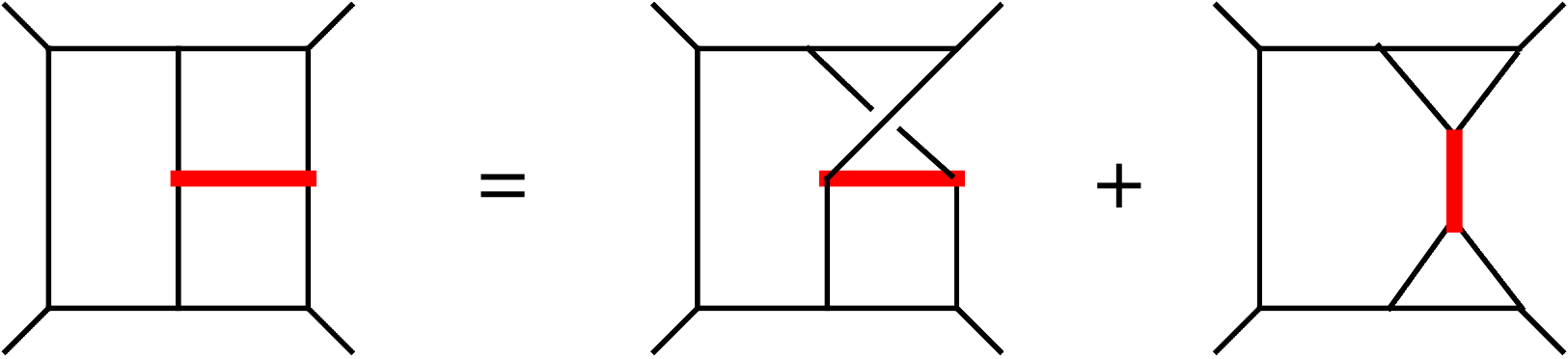}}
  \label{3loopbcjid}
\ee
is trivially satisfied because diagram (e) and (f) have the same numerator factor and the third diagram in \reef{3loopbcjid} vanishes.
\exercise{}{Verify that the numerators \reef{3Lnums} satisfy the following identity:
\be
  \raisebox{-10mm}{\includegraphics[scale=0.5]{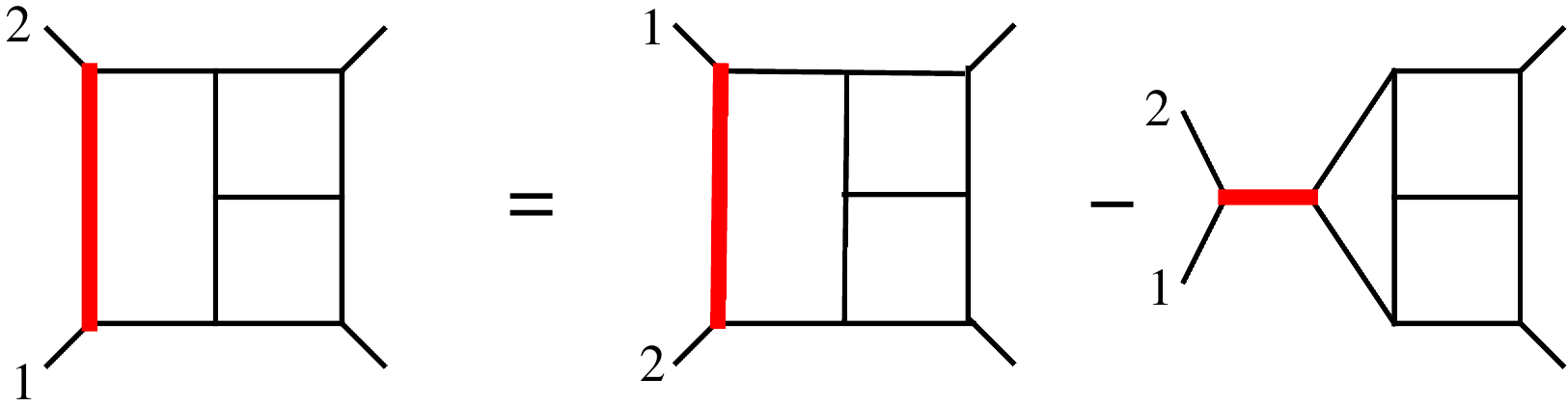}}\,.
\ee
Note that now all three numerators are non-zero.}
%

%%%%%%%%%%%%%%%%%%%%%%%%%%%%%%%%%%%%%%%%%%%%%%%%%%%%%%
\subsubsection{Summary}   
\label{s:summaryBCJ}
%%%%%%%%%%%%%%%%%%%%%%%%%%%%%%%%%%%%%%%%%%%%%%%%%%%%%%
While there is no formal proof that duality-satisfying numerators can always be found for loop amplitudes in Yang-Mills, there is considerable evidence in the favor of this property. We present here a list of non-trivial examples for which the BCJ duality-satisfying numerators have been constructed:
\begin{itemize}
   \item Up to 4-loops for 4-point in $\mathcal{N}=4$ SYM \cite{BCJLoop,ck4l}. 
   \item Up to 2-loops for 5-point in $\mathcal{N}=4$ SYM  \cite{N4Five}. 
   \item At 1-loop up to 7-points in $\mathcal{N}=4$ SYM \cite{Tristan}.
   \item Up to 2-loops for 4-point for the all-plus pure Yang-Mills amplitude \cite{BCJLoop}.
	 \item  1-loop 4-point for pure Yang-Mills theory in arbitrary dimensions \cite{Bern:2013yya}.
	\item 1-loop $n$-point all-plus or single-minus helicity amplitudes in pure Yang-Mills theory \cite{OConnellRational}.       
        \item 1-loop 4-point amplitudes in theories with less than maximally supersymmetry~\cite{OneLoopN1Susy}.
        \item 1-loop 4-point for an abelian orbifold of $\cn=4$ SYM \cite{Chiodaroli:2013upa}.
        \item 1-loop 4-point Yang-Mills theory with matter \cite{Nohle:2013bfa}.
\end{itemize}
Although most progress has been made for $\mathcal{N}=4$ SYM, the examples are not restricted to the maximal supersymmetric theory or to 4d.
%%%%%%%%%%%%%%%%%%%%%%%%%%%%%%%%%%%%%%%%%%%%%%%%%%%%%%%%%
\subsection{Implications for UV behavior of supergravity}   
%%%%%%%%%%%%%%%%%%%%%%%%%%%%%%%%%%%%%%%%%%%%%%%%%%%%%%%%%%
With duality-satisfying numerators for ($\cn=4$) super Yang-Mills amplitudes, we do not need to do much work to compute supergravity amplitudes! We now give several examples of this application of BCJ.

\compactsubsection{$\cn=8$ supergravity}
By squaring the 
duality-satisfying numerators of the 1-, 2- and 3-loop 4-point amplitudes in Sections \ref{s:BCJ1L}-\ref{s:BCJ3L}, we immediately obtain the integrands of 
the 1-, 2- and 3-loop 4-point amplitudes in $\mathcal{N}=8$ supergravity. At 1-  and 2-loop order, the $\cn=4$ SYM numerators are independent of the loop-momentum, so since the scalar box and the scalar double box integrals are UV finite in 4d, we  immediately see that $\mathcal{N}=8$ supergravity is finite in 4d at 1- and 2-loops. At 3-loops, the numerators in \reef{3Lnums} depend on the loop-momenta, but by simple power-counting, one finds that $\mathcal{N}=8$ supergravity is manifestly finite in 4d. Thus, as promised, without further calculations, we have just reproduced the result that $\mathcal{N}=8$ supergravity is finite in 4d up to and including 3-loop order. Of course, this agrees with previous explicit calculations \cite{Grisaru:1976ua,Grisaru:1976nn,Tomboulis:1977wd,Deser:1977nt,Bern:2006kd,Bern:2007hh,Bern:2008pv} and the counterterm analysis discussed in Section \ref{s:CTs}.

\compactsubsection{Critical dimension for maximal pure supergravity}
As mentioned in Section \ref{s:UVsg1}, the critical dimension for UV divergences of maximal supergravity is proposed \cite{Bern:2006kd} to match that of maximal  SYM, 
\be
   D_c(L) = \frac{6}{L} + 4 
   ~~~~\text{for}~~~~L>1\,.
   \label{YM-Dc2}
\ee
At 1- and 2-loop orders, the duality-satisfying numerators of $\cn=4$ SYM are loop-independent, so the critical dimension is simply determined by the scalar integrals; it is therefore universal between $\mathcal{N}=4$ SYM and 
$\mathcal{N}=8$ supergravity. 
At 3-loops, one can use power-counting to see that the integrals with the  worst UV behavior are the three diagrams in the last line of \reef{3LoopIntegrals}, 
i.e.~diagrams $(j), (k),(l)$. These diagrams dictate the critical dimension at 
3-loops for $\mathcal{N}=4$ SYM. But it follows from \reef{3Lnums} that the numerator factors of these three diagrams are loop-independent, so squaring them to get the $\cn=8$ supergravity amplitude does not change the critical dimension. One can check  that the none of the other diagrams have worse behavior than $(j), (k),(l)$ after squaring. 
Hence the relation \reef{YM-Dc2} for the critical dimension also holds for $\cn=8$ supergravity at 3-loop order. 
A similar BCJ argument extends this result to 4-loop order~\cite{ck4l}.
\exercise{}{Use the explicit integrands given in Sections \ref{s:BCJ2L} and \ref{s:BCJ3L} to verify that the critical dimension at 2- and 3-loops is  \reef{YM-Dc2} for both $\mathcal{N}=4$ SYM and $\mathcal{N}=8$ supergravity. 
What is the critical dimension at 1-loop order?}
\compactsubsection{$\mathcal{N}\geq4$ supergravity} 
We can obtain $\mathcal{N}\leq8$ supergravity amplitudes by tensor'ing two sets of numerators from loop amplitudes in $\cn\le 4$ supersymmetric Yang-Mills theories. Only one of the two copies of numerators need to satisfy the color-kinematic duality, as discussed around \reef{GenNumShift}. Since we already have duality-satisfying numerators for the 1-, 2- and 3-loops 4-point $\mathcal{N}=4$ SYM amplitudes, we can just tensor them with any Yang-Mills or SYM numerators we like, and obtain $\mathcal{N}\geq4$ supergravity amplitudes.  Since only a small number of cubic diagrams have non-vanishing numerators in the $\mathcal{N}=4$ SYM copy, we only need a few of the numerators of the other copy.

We begin at 1-loop. Suppose we have an explicit representation of the 1-loop integrand of $\mathcal{N}\leq 4$ SYM computed from Feynman rules. Such a representation usually involves triangles and bubbles as well as diagrams that are not 1-particle-irreducible, but it can be converted into a representation that only involves the box integrals. The price one pays is that the numerators will in general be non-local, but that is not a problem for our application. As an example, the following triangle- and bubble diagrams
can be converted to boxes by introducing inverse propagators:
\be
  \nonumber
  \raisebox{-30mm}{\includegraphics[scale=0.5]{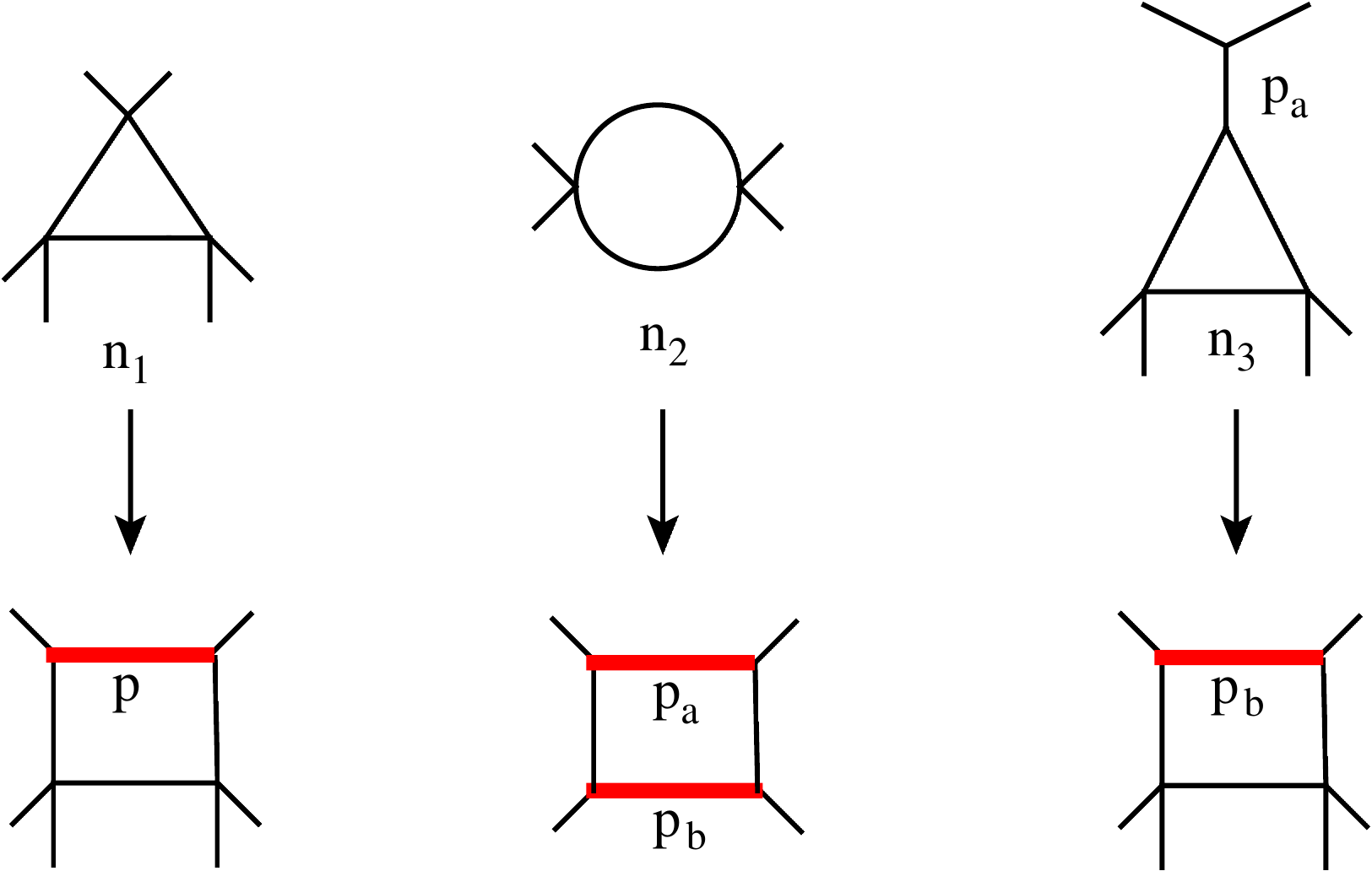}}
\ee
\be
    {}~~~n_1' = n_1 \,p^2\,,~~~~~~~~~~
    n_2' = n_1 \,p_a^2\,p_b^2\,,~~~~~~~~
    n_3' = n_1\, \frac{p_b^2}{p_a^2}\,.    
\ee
 To obtain the $\mathcal{N}\leq 4$  supergravity amplitude, we `tensor' the new numerators $n_i'$ with the duality-satisfying $\cn=4$ SYM numerators $n_i$ in \reef{OneLoopNumerator}. Specializing to 4-point, recall that the numerators in \reef{OneLoopNumerator} are not only loop-momentum independent, but also permutation invariant.  Independence on the loop-momentum means that the $n_i$'s move outside the box-integral. Permutation invariance for the $n_i$'s then tells us that the $n_i$-factor a uniform for each box-integral. In other words, it is just an overall factor, $s u\, A^{\rm tree}_{4,Q=16}[1,2,3,4]$, multiplying the entire 1-loop $\mathcal{N}\leq 4$ SYM amplitude! Thus we have found the following very simple formula for the 4-point 1-loop amplitude in  $\mathcal{N}\geq 4$ supergravity~\cite{N>=4SG},
\begin{equation}
 M^{(1)}_{4,Q+16}=
s u\, A^{\rm tree}_{4,Q=16}[1,2,3,4]
\bigg[A^{(1)}_{4,Q}[1,2,3,4] + A^{(1)}_{4,Q}[1,3,4,2]
 + A^{(1)}_{4,Q}[1,4,2,3] \bigg]\,.
\label{FourPointSupergravity}
\end{equation} 
The subscript $Q$ indicates the number of supercharges, with $Q=16$ corresponding to $\mathcal{N}=4$ supersymmetry. It is remarkable that the 1-loop amplitude of a non-renormalizable gravity theory can be given by a sum of 1-loop amplitudes of a renormalizable one. Note that this relation was exposed only after imposing color-kinematics duality. 

Consider now the UV structure in 4d. The 1-loop amplitudes of $\mathcal{N}=0,1,2$ SYM have UV-divergences. In 4d SYM, the UV divergence must be proportional to the tree amplitude, otherwise it would imply that a new operator is needed to renormalize the 1-loop divergence. 
Thus we conclude that  
\begin{equation}
 M^{(1)}_{4,Q+16}\bigg|_{D=4 \rm\, div.} \sim~
s u \,A^{\rm tree}_{4,Q=16}[1,2,3,4]
\bigg[A^{\rm tree}_{4,Q}[1,2,3,4] + A^{\rm tree}_{4,Q}[1,3,4,2]
 + A^{\rm tree}_{4,Q}[1,4,2,3] \bigg]
 ~=~0\,.
\label{FourPointSupergravityUV} 
\end{equation} 
This result vanishes due the $U(1)$ decoupling identity \reef{u1dec} for the Yang-Mills color-ordered tree amplitudes. Therefore, with the help of BCJ color-kinematics duality, we have shown that pure $\mathcal{N}\geq 4$ supergravity is finite at 1-loop in 4d.

\compactsubsection{1-loop UV divergence in $\mathcal{N}=4$ supergravity-matter theory} 
The simple relation \reef{FourPointSupergravity} can also be used to demonstrate UV divergences. Consider $\mathcal{N}=4$ supergravity coupled to $\mathcal{N}=4$ Maxwell theory (i.e.~$\mathcal{N}=4$ SYM with $U(1)$ gauge group). The spectrum of this $\mathcal{N}=4$ supergravity-matter theory is given by tensor'ing $\mathcal{N}=4$ SYM with Yang-Mills minimally coupled to an adjoint scalar. The 1-loop amplitude is exactly the same as \reef{FourPointSupergravity}, except that $A^{(1)}_{4,Q}$ is now the 1-loop amplitude of the Yang-Mills-scalar theory. The 4-point 1-loop amplitude in Yang-Mills-scalar theory is UV divergent and is renormalized by a 4-scalar 
counterterm
\be
\Delta \mathcal{L}=c^{(1)}_{abcd}\,\phi^{a}\phi^b\phi^c\phi^d\,,
\label{scalarOneLoop}
\ee
where $c^{(1)}_{abcd}$ is the color factor for the box-diagram. Putting this divergence into \reef{FourPointSupergravity} the sum of the three terms is now non-vanishing because the color-structure of \reef{scalarOneLoop} is not that of a tree-amplitude. This then shows that there is a 1-loop UV divergence in the $\mathcal{N}=4$ supergravity-matter model.
\exercise{}{Use \reef{FourPointSupergravity} to show that the 1-loop UV divergence of the  $\mathcal{N}=4$  gravity-matter system  corresponds to an $F^4 = (F_{\m\n}F^{\m\n})^2$ operator. }

\compactsubsection{Color-kinematics constraints on candidate counterterms} 
In Section \ref{s:CTs}, we approached UV divergences of supergravity from the viewpoint of local counterterms: we ruled out  candidate counterterms based on the known symmetries in $\cn=8$ supergravity. It is reasonable to say that ``everything not forbidden is compulsory"\footnote{Also known as Gell-Mann's Totalitarian Principle (from T.~H.~White's ``The Once and Future King").} and therefore expect that if the known symmetries do not rule out a certain counterterm, then it will likely appear in the perturbation theory.  In this section, we have studied a new structure, BCJ color-kinematics duality, that is very different in nature from the other symmetries imposed on the local counterterms. In examples, we have seen how the BCJ doubling relation  reveals the ``true" power-counting (at least ``truer'' than Feynman diagrams) for UV divergences in supergravity amplitudes. So one may now wonder if it possible that there exist counterterms that respect all known `ordinary' symmetries of the theory and yet is ruled out by color-kinematics duality? We now present such a case.

The 2-loop duality-satisfying numerators \reef{TwoLoopNum} of $\cn=4$ SYM are momentum independent. Following the same arguments that gave us the 1-loop result \reef{FourPointSupergravity}, we find that the 2-loop  4-point amplitude of $\mathcal{N}\geq 4$ supergravity is given as a sum of  2-loop (S)YM amplitudes:
\be
\begin{split}
M^{(2)}_{4,Q+16}[1,2,3,4] &=~
su\,A^{\rm tree}_{4,Q=16}[1,2,3,4]
\bigg[s \Bigl( A^{\rm P}_{4,Q}[1,2,3,4] + A^{\rm NP}_{4,Q}[1,2,3,4]  \\
& \hskip 2 cm \null 
 +  A^{\rm P}_{4,Q}[3,4,2,1] + A^{\rm NP}_{4,Q}[3,4,2,1] \Bigr) 
 + {\rm cyclic}(2,3,4)\bigg]\,,  
\end{split}
\label{TwoLoopSugraBCJ}
\ee
Now let us see what  \reef{TwoLoopSugraBCJ} says about the  UV divergence of supergravity. Consider $\mathcal{N}=4$ supergravity, hence $Q=0$ for $A^{\rm P}_{4,Q}$ and $A^{\rm NP}_{4,Q}$. In 4d Yang-Mills theory, no counterterm operators are needed to regularize the UV divergence of the 4-point 2-loop amplitude, so the coefficient of the divergence must be generated by the $F^2$ operator of the classical action. In particular, the 2-loop UV divergence must have tree-level color factors. 
In 5d, dimension-counting shows that  $F^3$ is the only allowed counterterm; it again only has tree-level color factors. This is because the divergence in 5d is renormalized by a tree diagram with one $F^3$ counterterm insertion,
\eq
 \raisebox{-5mm}{\includegraphics[scale=0.5]{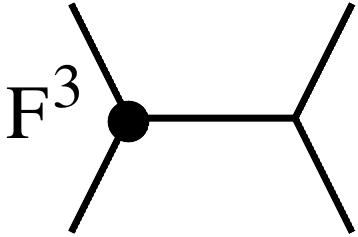}}\,.
\eqe  
With the intention of using this information, we expand the color-structure $c^{\rm P}$ and $c^{\rm NP}$ in \reef{TwoLoopSugraBCJ} into a basis of color factors that are independent under the Jacobi relation. Such a basis is given in  Appendix B of \cite{Bern:2010tq}, and it  consists of two tree, one 1-loop, and two 2-loop color factors. So casting  \reef{TwoLoopSugraBCJ} into said basis, we know that the  coefficients of the 1- and 2-loop color factors have to vanish for the UV-divergent part, since in 4d and 5d only the tree color structure is generated. The requirement of vanishing 2-loop color factors is \cite{N=4SG1}
\begin{eqnarray}
0 &=& 
\bigg[ t \Big( A^{\rm P}_Q[1,3,4,2] + A^{\rm P}_Q[1,4,2,3] + A^{\rm P}_Q[3,1,4,2]
  + A^{\rm P}_Q[3,2,1,4] \nonumber \\
&& \hspace{8mm}
 +  A^{\rm NP}_Q[1,3,4,2] + A^{\rm NP}_Q[1,4,2,3] + A^{\rm NP}_Q[3,1,4,2] 
 + A^{\rm NP}_Q[3,2,1,4] \Big) \nonumber  \\
&& \hspace{3mm}
+ s\Big(A^{\rm P}_Q[1,3,4,2] + A^{\rm P}_Q[3,1,4,2] + A^{\rm NP}_Q[1,3,4,2]
 + A^{\rm NP}_Q[3,1,4,2]\Big) \bigg] \bigg|_{D=4,5 \, \rm div.} \,, 
    \nonumber\\[2mm] 
0 & = & 
\bigg[ s \Big(A^{\rm P}_Q[1,2,3,4] + A^{\rm P}_Q[1,3,4,2] + A^{\rm P}_Q[3,1,4,2]
  + A^{\rm P}_Q[3,4,2,1]   \\
&& \hspace{8mm}
  + A^{\rm NP}_Q[1,2,3,4] + A^{\rm NP}_Q[1,3,4,2]
  + A^{\rm NP}_Q[3,1,4,2] +  A^{\rm NP}_Q[3,4,2,1] \Big) \nonumber  \\
&&  \hspace{3mm}
+ t \Big(A^{\rm P}_Q[1,3,4,2] + A^{\rm P}_Q[3,1,4,2] + A^{\rm NP}_Q[1,3,4,2] +
 A^{\rm NP}_Q[3,1,4,2] \Big) \bigg]
\bigg|_{D=4,5\, \rm div.} \,. \nonumber 
\label{ColorConstraint}
\end{eqnarray}
Substituting the above into \reef{TwoLoopSugraBCJ}, we immediately find that 
\eq
M^{(2)}_{4,16}[1,2,3,4] \Big|_{D=4,5\, \rm div.} = 0 \,.
\label{M2div}
\eqe
So the 2-loop 4-point amplitude in pure $\cn \ge 4$ supergravity is finite in both 4d and 5d.

In 4d, the 2-loop $R^3$ operator can be ruled out by supersymmetry, as we have seen in Section \ref{s:CTs} from the spinor helicity violating amplitude it generates. This is of course perfectly compatible with the UV finiteness of the 2-loop amplitude \reef{M2div}.

However, in 5d the relevant operator at 2-loop order is $R^4$, and it is compatible with supersymmetry. It has also been argued to be duality invariant \cite{GrisaruSiegel,BossardHoweStelle5D}. So this is an explicit 5d example  where a counterterm appears to respect all known symmetries of the theory, yet is not generated because the corresponding 2-loop 4-point amplitude is UV finite. 

In 4d, it has been argued \cite{VanishingVolume}  that the 3-loop candidate counterterm operator $R^4$ is allowed by all symmetries of $\mathcal{N}=4$ supergravity. (It was ruled out in $\cn=8$ supergravity by $E_{7(7)}$.) Yet, by explicit computation, utilizing  color-kinematic duality, it has been shown that $\mathcal{N}=4$ supergravity is actually finite at 3-loop order \cite{N=4SG2}. So what is going on? The absence of a divergence could be coincidence. Or they could indicate that there is a hidden symmetry that is violated by the would-be-counterterm \cite{Ferrara}. The issue is not settled and you can now go ahead and do calculations yourself to help understand this better. 

%%%%%%%%%%%%%%%%%%%%%%%%%%%%%%%%%%%%%%%%%%%%%%%%%%%%%%%%%%
\subsection{Extensions}
%%%%%%%%%%%%%%%%%%%%%%%%%%%%%%%%%%%%%%%%%%%%%%%%%%%%%%%%
We end this section with brief mention of other applications of the BCJ color-kinematics duality. Color-kinematic duality has been extended to scattering amplitudes involving higher-dimension operators \cite{BroedelDixon} and also to form factors \cite{FormFactors}. 

In Section \ref{s:BLG}, we encountered the Lie 3-algebra 
for 3d Chern-Simons matter theory (BLG): it involved 4-index structure constants that, in place of the usual Jacobi identity, satisfy a 4-term fundamental identity \reef{FundamentalId}. Surprisingly, color-kinematic duality can also be established for such 3-algebra theories \cite{Till} with the basic diagrams  built from 4-point vertices only. Which supergravity amplitude is calculated by the BCJ double-copy of duality-satisfying numerators from the 3-algebra Chern-Simons matter theory? 
At first sight it seems that the answer has to be different from the  supergravity amplitude obtained from `squaring' 3d Yang-Mills amplitudes, 
because diagrams built from quartic vertices must have an even number of external legs $n$, while trivalent diagrams can have even or odd $n$. But one can use the double-copy based on 3d Yang-Mills theory (or the KLT formula) to show that the odd-$n$ supergravity amplitudes vanish in 3d, eventhough the odd-$n$ Yang-Mills amplitudes are non-vanishing. 
It has in fact been shown that applying the double-copy trick to 3d Yang-Mills and 3d BLG theory remarkably results in the same supergravity amplitudes~\cite{Till, HenrikYt}.

By dimensional analysis, 3d gravity is non-renormalizable.  It is curious that 3d supergravity amplitudes can be constructed from two distinct color-kinematic dualities. Perhaps this puts constraints on the UV behavior of supergravity in 3d.

%%%%%%%%%%%%%
%%%%%%%%%%%%%
%%%%%%%%%%%%%
%%%%%%%%%%%%%
%%%%%%%%%%%%%
%%%%%%%%%%%%%
%%%%%%%%%%%%%
%%%%%%%%%%%%%
%%%%%%%%%%%%%
%%%%%%%%%%%%%

%%%%%%%%%%%%%
%%%%%%%%%%%%%
%%%%%%%%%%%%%
%%%%%%%%%%%%%
%%%%%%%%%%%%%

%%%%%%%%%%%%%%%%%%%%%%%%%%%%%%% 
%%%%%%%%%%%%%%%%%%%%%%%%%%%%%%% 
%%%%%%%%%%%%%%%%%%%%%%%%%%%%%%% 
\newpage 
%%%%%%%%%%%%%%%%%%%%%%%%%%%%%%% 
%%%%%%%%%%%%%%%%%%%%%%%%%%%%%%% 
%%%%%%%%%%%%%%%%%%%%%%%%%%%%%%% 

%%%%%%%%%%%%%%%%%%%%%%%%%%%%%%% 
%%%%%%%%%%%%%%%%%%%%%%%%%%%%%%% 
%%%%%%%%%%%%%%%%%%%%%%%%%%%%%%%  
\setcounter{equation}{0}
\section{Further reading}
%heAug6
\label{s:reading}
In this section, we list references to other reviews and we highlight a few subjects that were not covered in the main text.

\noindent {\bf Reviews on on-shell methods for scattering amplitudes }
\begin{itemize}
\item {\em Introduction to on-shell methods}

The QFT textbooks by Srednicki \cite{MSqft}, Zee \cite{Zee:2003mt}, and Schwartz \cite{Schwartz:2013pla} provide brief introductions to the spinor helicity formalism and on-shell recursion relations. In addition, notes \cite{HennPlefka} offer a comprehensive introduction to on-shell recursion and to loop-integrals and also covers some aspects of scattering amplitudes in $\cn=4$ SYM. 

The following review cover various aspects of on-shell methods (the most recent reviews are listed first):

``A brief introduction to modern amplitude methods" \cite{Dixon:2013uaa}.

``Scattering amplitudes: the most perfect microscopic structures in the universe'' \cite{Dixon:2011xs}.

``Hidden Simplicity of Gauge Theory Amplitudes'' \cite{Drummond:2010ep}.   

``A First Course on Twistors, Integrability and Gluon Scattering Amplitudes'' \cite{Wolf:2010av}.

``On-Shell Methods in Perturbative QCD'' \cite{Bern:2007dw}. 

``Calculating scattering amplitudes efficiently'' \cite{dixon}.

``Multiparton amplitudes in gauge theories'' \cite{Mangano:1990by}. 

%%%%%  
\item {\em Numeric methods \& applications in phenomenology}

``Susy Theories and QCD: Numerical Approaches'' \cite{Review3}.

``One-loop calculations in quantum field theory: from Feynman diagrams to unitarity cuts''  \cite{Ellis:2011cr}. 

``Simplifying Multi-Jet QCD Computation,'' \cite{Peskin:2011in}.

``Loop Amplitudes in Gauge Theories: Modern Analytic Approaches''  \cite{Review4}.

%%%%%  
\item {\em Gravity}

``Introduction to the effective field theory description of gravity'' 
\cite{Donoghue:1995cz}.

``Perturbative quantum gravity and its relation to gauge theory'' \cite{Bern:2002kj}.

``Ultraviolet Behavior of $\cn=8$ Supergravity'' \cite{Dixon:2010gz}.

\end{itemize}

\compactsubsection{Less supersymmetry: $1\leq\mathcal{N}<4$}
An obvious arena for on-shell methods is theories with non-maximal supersymmetry. 
Scattering amplitudes in $\mathcal{N}=1$ supersymmetric theories have been of particular interest in particle phenomenology and on-shell methods were used in such studies \cite{UnitarityMethod,Bern:2009xq,SusyPheno,Bidder:2005ri,Britto:2005ha}. The superamplitude and on-shell superspace formalism generalizes to $1\leq\mathcal{N}<4$ SYM~\cite{Elvang:2011fx} and has also been used for SYM coupled with matter~\cite{NoTria3,Lal:2010qq}. It is expected that the on-shell diagrams approach to planar amplitudes of $\mathcal{N}=4$ SYM has a natural extension to $\mathcal{N}<4$ theories~\cite{ArkaniHamed:2012nw}.

\compactsubsection{Amplitudes with massive particles}
We have focused on amplitudes with massless particles, but there are also efficient on-shell methods available for amplitudes involving massive particles. The spinor helicity formalism for massless particles, introduced in Section \ref{s:sh}, can be generalized to massive particles. There are two approaches to this. In the first, one studies the eigenvectors of the momentum matrix $p_{\alpha\dot{\beta}} = p_\mu \sigma^\mu_{\alpha\dot{\beta}}$ to directly get solutions to the massive Dirac equation. In the second approach,  the time-like momentum $p_i$ is decomposed along two lightlike directions by introducing a null reference vector $q_i$ for each state: 
$p_i^\mu = p_{i\perp}^\mu - \frac{m_i^2}{2 q_i\cdot p_i} \,q^\mu$. 
The familiar spinor helicity formalism can then be used for $q_i = -|q_i\>[q_i|$ and $p_{i\perp}= -|i_\perp \>[i_\perp|$. Helicity is only a Lorentz-invariant quantity for massless particles, but $q_i$ breaks Lorentz-invariance and can therefore be used to define a helicity basis in which we can calculate helicity amplitudes. For an introduction to both approaches, see \cite{Dittmaier:1998nn}.  The papers \cite{Boels:2008du,Boels:2009bv,Boels:2010mj} studied applications of BCFW recursion relations with massive particles.
A recent discussion of the latter approach, as well as applications to CSW-like recursion relations, was given in \cite{CEK}. 
Finally, let us mention that there are simple amplitudes with a pair of massive particles. Examples of such `towers' of amplitudes ---  in a sense massive versions of Parke-Taylor --- were presented for scattering processes on the Coulomb branch of $\cn=4$ SYM in \cite{Craig:2011ws,Kiermaier:2011cr} (see also \cite{Ferrario:2006np,Forde:2005ue,Rodrigo:2005eu}).

\compactsubsection{Extensions of recursion relations}
Attempts have been made to generalize various forms of recursion relations  
 beyond gauge theory and gravity, for example to string theory 
 amplitudes~\cite{Cheung:2010vn,Boels:2010bv} and to non-linear sigma models~\cite{Kampf:2012fn,Kampf:2013vha}. Recursion relations can also be utilized to obtain rational functions that appear at 1-loop~\cite{RationalRecurssion}. A review of various recursion relations are given in \cite{FengReview}.

\compactsubsection{Triality: Wilson loop, correlation function, amplitude}
In Section \ref{s:emDCS} we discussed the emergence of dual superconformal symmetry  in planar $\mathcal{N}=4$ SYM.  It states that 
the scattering amplitude in the dual coordinates is superconformal covariant. Could we define  $\mathcal{N}=4$ SYM directly in these new coordinates? In dual coordinates, the kinematic setup for the on-shell momenta is a polygon with null edges. A similar physical quantity is a Wilson-loop specialized to  a null-polygon contour. So could the amplitude in momentum space be dual to a null-polygon Wilson-loop in the dual space? Indeed it is. This duality has been established at strong coupling by Alday and Maldacena~\cite{Alday:2007hr} as well as at weak-coupling by Drummond, Korchemsky and Sokatchev~\cite{Drummond:2007aua}. 

At strong coupling, the duality can be understood as a consequence of T-duality in string theory~\cite{Berkovits:2008ic,Beisert:2008iq}. 
At weak coupling, evidence for such duality was first reported by \cite{Drummond:2007aua,Brandhuber:2007yx}. It was later proven by defining the action of $\cn=4$ SYM directly in supertwistor space  \cite{Mason:2010yk}. Remarkably using the duality, the first computation of the six-point two-loop MHV amplitude was done by computing the six-edged Wilson loop~\cite{DelDuca:2009au}.  The amplitude/Wilson-loop duality was first established between the bosonic Wilson-loop and the ratio of the MHV scattering amplitude, divided by the MHV tree-amplitude. It can be generalized to N$^K$MHV amplitudes by supersymmetrizing the bosonic Wilson-loop~\cite{Mason:2010yk,Caron-Huot:2010ek,Eden:2011}. 

Another extension is the realization that the super Wilson-loop in $\mathcal{N}=4$ SYM is also dual to correlation function of operators with lightlike separation~\cite{Alday:2010zy}. This can also be proven in super-twistor space~\cite{Adamo:2011dq}.  For a review on the (MHV)amplitude/(bosonic)Wilson-loop duality see~\cite{WilsonReview1,WilsonReview3,WilsonReview4}, for the general amplitude/super-Wilson loop duality in the framework of supertwistor space, see~\cite{WilsonReview2}.  

Based on an operator-product-expansion approach first developed in the perturbative computation of null Wilson-loops \cite{Alday:2010ku}, a non-perturbative formulation of the $S$-matrix/Wilson-loop for planar $\cn=4$ SYM has been proposed in \cite{Basso:2013vsa, Basso:2013aha, Basso:2014koa}. Its
perturbative expansion gives important predictions for the explicit loop-amplitude,
and as a result the final functional form of the 6-point 2- and 3-loop MHV amplitude for planar $\cn=4$ SYM \cite{Dixon:2013eka,Dixon:2014voa} was found.

\compactsubsection{Twistors}
Standard reviews of twistor space include~\cite{Penrose:1986ca,WardWells,HuggettTod}. For amplitude-friendly reviews, we suggest~\cite{Cachazo:2005ga} as well as \cite{Witten:2003nn,Mason}.

\compactsubsection{What's next?}
%Well, now you know as much as we do. 
\mbox{We hope you have found this review useful. 
Now you go make the future.}

%%%%%%%%%%%%%%%%%%%%%%%%%%%%%%% 
%%%%%%%%%%%%%%%%%%%%%%%%%%%%%%% 
%%%%%%%%%%%%%%%%%%%%%%%%%%%%%%% 

\newpage

%%%%%%%%%%%%%

%%%%%%%%%%%%%

%%%%%%%%%%%%%%%%%%%%%%%%%%%%%%% 
%%%%%%%%%%%%%%%%%%%%%%%%%%%%%%% 
%%%%%%%%%%%%%%%%%%%%%%%%%%%%%%% 
\appendix
%%%%%%%%%%%%%%%%%%%%%%%%%%%%%%% 
%%%%%%%%%%%%%%%%%%%%%%%%%%%%%%% 
%%%%%%%%%%%%%%%%%%%%%%%%%%%%%%% 
\setcounter{equation}{0}
\section{Conventions for 4d spinor helicity formalism}
\label{app:conv}
%%%%%%%%%%%%%%%%%%%%%%%%%%%%%%% 
%%%%%%%%%%%%%%%%%%%%%%%%%%%%%%% 
%%%%%%%%%%%%%%%%%%%%%%%%%%%%%%% 
%%%%%%%%%%%%%%%%%%%%%%%%%%%%%%% 
The conventions of these notes follow those in Srednicki's QFT textbook \cite{MSqft}.

\label{s:gammas}
We use a ``mostly-plus'' metric, 
$\eta_{\m\n} = \diag(-1,+1,+1,+1)$  and define
\be
  (\sigma^\m)_{a\dot{b}} = (1,\sigma^i)_{a\dot{b}} \, ,~~~~~~~~
  (\bar{\sigma}^\m)^{\dot{a}b} = (1,-\sigma^i)^{\dot{a}b} 
\ee
with Pauli matrices
\be
  \sigma^1 = 
   \left( 
     \begin{array}{cc}
        0 & 1 \\
        1 & 0 
      \end{array}
   \right) \, ,~~~~~
  \sigma^2 = 
   \left( 
     \begin{array}{cc}
        0 & -i \\
        i & 0 
      \end{array}
   \right) \, ,~~~~~
  \sigma^3 = 
   \left( 
     \begin{array}{cc}
        1 & 0 \\
        0 & -1 
      \end{array}
   \right) 
   \, .
   \label{pauliM}
\ee
Two-index spinor indices are raised/lowered using
\be
   \label{epsies}
   \varepsilon^{ab} 
   ~=~ \varepsilon^{\dot{a}\dot{b}} 
   ~=~ \left( 
     \begin{array}{cc}
        0 & 1 \\
        -1 & 0 
      \end{array}
   \right) 
   ~=~
   -\varepsilon_{ab} 
   ~=~- \varepsilon_{\dot{a}\dot{b}} 
   \, ,
\ee
which obey $\varepsilon_{ab} \varepsilon^{bc} = \delta_a{}^c$.

We list the following properties
\bea
  (\bar{\sigma}^\m)^{\dot{a}a} 
  &=& \varepsilon^{ab} \varepsilon^{\dot{a}\dot{b}} 
    (\sigma^\m)_{b\dot{b}} \,, \\[1mm]
  (\sigma^\m)_{a\dot{a}} (\sigma_\m)_{b\dot{b}}  
  &=& - 2 \varepsilon_{ab} \varepsilon_{\dot{a}\dot{b}} \,, \\[1mm]
  \big( \sigma^\m \bar{\sigma}^\n 
  + \sigma^\n \bar{\sigma}^\m \big)_a{}^b
   &=& - 2 \eta^{\m\n} \d_a{}^b \,, \\[1mm]
  \label{Trss}
  \text{Tr}\, (\sigma^\m \bar{\sigma}^\n) 
  &=&  \text{Tr}\, (\bar{\sigma}^\m \sigma^\n)
  ~=~
  - 2 \eta^{\m\n}\,.
\eea
Define $\g$-matrices:
\be
 \label{gammamatrices}
  \g^\mu = 
  \left( 
     \begin{array}{cc}
        0 & (\sigma^\mu)_{a\dot{b}} \\
        (\bar{\sigma}^\mu)^{\dot{a}b} & 0 
      \end{array}
   \right) \, ,
  ~~~~~~~
  \{\g^\m , \g^\n \} = -2 \eta^{\m\n} \,,
\ee
and 
\be
  \label{g5proj}
  \g_5 \equiv i \g^0 \g^1 \g^2 \g^3 
  =  
  \left( 
     \begin{array}{cc}
        -1 & 0 \\
        0 & 1 
      \end{array}
   \right) \, ,
   \hspace{1cm}
  P_L =\frac{1}{2} (1 - \g_5 )\, ,~~~~~~
  P_R =\frac{1}{2} (1 + \g_5 )\, .
\ee
For a momentum 4-vector $p^\mu = (p^0,p^i) = (E,p^i)$ with $ p^\m p_\m = - m^2$, 
we define momentum bi-spinors
\bea
  p_{a\dot{b}} ~\equiv~ p_\m\, (\sigma^\mu)_{a \dot{b}} \, , ~~~~~
  p^{\dot{a}b} ~\equiv~ p_\m\, (\bar{\sigma}^\mu)^{\dot{a} b} \, .
\eea
For example, 
\be
 p_{a\dot{b}}  = 
  \left(
    \begin{array}{cc}
      -p^0 + p^3 & p^1 - i p^2 \\
      p^1 + i p^2 & - p^0 - p^3 \\
    \end{array}
  \right)\,.
\ee
Taking the determinant of this 2$\times$2 matrix gives
\be
   \det p = - p^\m p_\m = m^2\,.
\ee
The Dirac conjugate $\overline{\Psi}$ is defined as 
\be
 \label{Dconj}
 \overline{\Psi} = \Psi^\dagger \beta\,,
 ~~~~~
 \beta =
 \left(
    \begin{array}{cc}
      0 & \delta^{\dot{a}}{}_{\dot{b}} \\
      \delta_a{}^b & 0
    \end{array}
 \right)
\ee
The $4 \times 4$ matrix $\beta$ is the same as $\gamma^0$ but has a different index structure.

For convenience, we collect here some useful spinor helicity identities 
\bea
  \label{rlp2a}
  \begin{array}{rclcrcl}
  [p|^a &=& \eps^{ab} |p]_b \, ,
  &&
  |p]_a&=& \eps_{ab} [p|^b\,,
  \\[2mm]
  |p\>^{\da} &=&  \eps^{\da\db} \<p|_{\db}
  &&
  \<p|_{\da} &=& \eps_{\da\db}  |p\>^{\db}\, ,
  \\[4mm]
   p_{a \db}&=& -|p]_a\, \<p|_{\db} \, ,
   &&
  p^{\da b}&=&- |p\>^{\da}\, [p|^{b}\,,
  \\[4mm]
  [p|^a &=& (|p\>^{\da})^*\,, 
  && 
  \<p|_{\da} &=& (|p]_a)^*\,, ~~~  \leftarrow \text{for real momenta}
  \\[4mm]
  \< p \, q \> &=&{\<p|}_{\da}\, |q\>^{\da} \, ,
  &&
  [ p \, q ]  &=& [p|^{a} \,|q]_a \, ,\\[4mm]
  \< p \, q \>\, [ p \, q ] &=&  2 \, p \cdot q \, ~=~ (p+q)^2\,,\hspace{-8mm}
  \\[4mm]
   [k | \g^\m | p \> 
  &=& \< p | \g^\m | k ] \, , 
  &&
  [k | \g^\m | p \>^* &=& [p | \g^\m | k \>  ~~\text{for real momenta}\,,
  \\[4mm]
  \<p| P |k]   &=& \<p|_{\da}\, P^{\da b} \,|k]_b\,,
  &&
  \<p| y_1 y_2  |k\> &=& \<p|_{\da} (y_1)^{\da b} (y_2)_{b \dot{c}}  |k\>^{\dot{c}}\,,
  \\[4mm]
    \<p| q |k] 
  &=& - \<pq\> [q k] \,,
  &&
  \< 1 | \g^\m | 2 ] \< 3 | \g_\m| 4 ]
   &=& 2 \< 1 3 \> [2 4]
  \end{array}
\eea
We also use the analytic continuation
\be
  |-p\> \,=\, - |p\> \,,~~~~~
   |-p] \,=\, +|p]\,.
\ee
These identities are used multiple places in the text and exercises.

%%%%%%%%%%%%%%%%%%%%%%%%%%
%%%%%%%%%%%%%%%%%%%%%%%%%%
%%%%%%%%%%%%%%%%%%%%%%%%%%
%%%%%%%%%%%%%%%%%%%%%%%%%%

\newpage
%%%%%%%%%%%%%%%%%%%%%%%%%%%%%%%%%%%%
\setcounter{equation}{0}
\section{Very brief introduction to twistors}
\label{app:twistor}
%%%%%%%%%%%%%%%%%%%%%%%%%%%%%%%%%%%%%
The conformal group in $D=4$ dimensions is $SO(2,4)$. To have a linear representation, it is convenient to interpret $SO(2,4)$ as the Lorentz group of a 6d space with signature $(-,-,+,+,+,+)$. This way conformal symmetry is realized as Lorentz symmetry if we  embed the 4d spacetime into 6 dimensions. Consider a null-subspace in 6d defined by $X\cdot X=0$ where $X^\mu$ is a 6d vector. As the null constraint is invariant if we rescale $X\rightarrow r X$, it is natural to identify $X\sim r X$ on the null-space. Since the constraint and the projective nature leaves $6-2=4$ degrees of freedom, the 4d space can indeed be identified as this null-space in 6d. This is the so called ``embedding formalism" that was fist introduced by Dirac in 1937~\cite{Dirac} (see \cite{Warren} for a recent discussion).

We now spinorize the above discussion. Since the $SO(2,4)\sim SU(2,2)$, we can rewrite the 6d vector $X^\mu$ as a bi-spinor $X^{IJ}$. This $4\times4$ antisymmetric matrix transforms in the {\bf 6} irrep of $SU(2,2)$, so $X^{IJ} = -X^{JI}$.

The null condition now translates to:
\eq
X^2~=~\frac{1}{2}\epsilon_{IJKL}X^{IJ}X^{KL}~=~0\,.
\eqe  
This implies that $X^{IJ}$ has rank $2$, and therefore we can write it as 
\eq
X^{IJ}=Z_i^{[I}Z_{j}^{J]}\,.
\label{XZZ}
\eqe
where the 4-component spinors $Z^{I}$ are called {\bf \em twistors}. From \reef{XZZ} we see that a point $X$ is defined by the line formed by two twistor variables $(Z_i,Z_j)$. Since $X$ is defined projectively, we identify $Z_i\sim tZ_i$ and therefore the twistor-space is really 
$\mathbb{CP}^{3}$. The $SU(2,2)$ covariant form of the incidence relation in \reef{DualDef} is simply: 
\eq
X^{[IJ}Z_i^{K]}=0\,.
\label{Xincidence}
\eqe
To see this, note that any point in four-dimensions that satisfy \reef{Xincidence} must have $Z_i$ as one of its twistors (in the representation of \reef{XZZ}). Since \reef{ID3} tells us that 
\eq
y^2_{ij}=-\frac{\langle i,i-1,j,j-1\rangle}{\langle i,i-1\rangle\langle j,j-1\rangle}\,,
\eqe
any two point $X_i,X_j$ that share a common twistor must be null separated as $y^2_{ij}=0$. Thus \reef{Xincidence} indeed defines a null line in the four-dimensional space, precisely the same as \reef{DualDef}.

%%%%%%%%%%%%%%%%%%%%%%%%%%
%%%%%%%%%%%%%%%%%%%%%%%%%%
%%%%%%%%%%%%%%%%%%%%%%%%%%
%%%%%%%%%%%%%%%%%%%%%%%%%%

%%%%%
%%%%%


\newpage
\begin{thebibliography}{99}

%\cite{Mangano:1990by}
\bibitem{Mangano:1990by} 
  M.~L.~Mangano and S.~J.~Parke,
  ``Multiparton amplitudes in gauge theories,''
  Phys.\ Rept.\  {\bf 200}, 301 (1991)
  [hep-th/0509223].
  %%CITATION = HEP-TH/0509223;%%


\bibitem{MSqft}
%\cite{Srednicki:2007qs}
  M.~Srednicki,
  ``Quantum field theory,''
  Cambridge, UK: Univ. Pr. (2007) 641 p

%\cite{Dixon:1996wi}
%\bibitem{Dixon:1996wi} 
\bibitem{dixon}
  L.~J.~Dixon,
  ``Calculating scattering amplitudes efficiently,''
  In *Boulder 1995, QCD and beyond* 539-582
  [hep-ph/9601359].
  %%CITATION = HEP-PH/9601359;%%

%\cite{Dixon:2011xs}
\bibitem{Dixon:2011xs} 
  L.~J.~Dixon,
  ``Scattering amplitudes: the most perfect microscopic structures in the universe,''
  J.\ Phys.\ A {\bf 44}, 454001 (2011)
  [arXiv:1105.0771 [hep-th]].
  %%CITATION = ARXIV:1105.0771;%%

%\cite{Kampf:2013vha}
\bibitem{Kampf:2013vha} 
  K.~Kampf, J.~Novotny and J.~Trnka,
  ``Tree-level Amplitudes in the Nonlinear Sigma Model,''
  JHEP {\bf 1305}, 032 (2013)
  [arXiv:1304.3048 [hep-th]].
  %%CITATION = ARXIV:1304.3048;%%

\bibitem{KK} 
  R.~Kleiss and H.~Kuijf,
  ``Multi-Gluon Cross-sections And Five Jet Production At Hadron Colliders,''
  Nucl.\ Phys.\ B {\bf 312}, 616 (1989).
  %%CITATION = NUPHA,B312,616;%%  
  
\bibitem{DDM}
V.~Del Duca, L.~J.~Dixon and F.~Maltoni,
  ``New color decompositions for gauge amplitudes at tree and loop level,''
  Nucl.\ Phys.\ B {\bf 571}, 51 (2000)
  [hep-ph/9910563].
  %%CITATION = HEP-PH/9910563;%%  
  
%\cite{Bern:2008qj}
\bibitem{BCJ} 
  Z.~Bern, J.~J.~M.~Carrasco and H.~Johansson,
  ``New Relations for Gauge-Theory Amplitudes,''
  Phys.\ Rev.\ D {\bf 78}, 085011 (2008)
  [arXiv:0805.3993 [hep-ph]].
  %%CITATION = ARXIV:0805.3993;%%

\bibitem{alsogluonfusion}
%\cite{Berger:2006sh}
%\bibitem{Berger:2006sh} 
  C.~F.~Berger, V.~Del Duca and L.~J.~Dixon,
  ``Recursive Construction of Higgs-Plus-Multiparton Loop Amplitudes: The Last of the Phi-nite Loop Amplitudes,''
  Phys.\ Rev.\ D {\bf 74}, 094021 (2006)
  [Erratum-ibid.\ D {\bf 76}, 099901 (2007)]
  [hep-ph/0608180].
  %%CITATION = HEP-PH/0608180;%%  
 
%\cite{Badger:2007si}
%\bibitem{Badger:2007si} 
  S.~D.~Badger, E.~W.~N.~Glover and K.~Risager,
  ``One-loop phi-MHV amplitudes using the unitarity bootstrap,''
  JHEP {\bf 0707}, 066 (2007)
  [arXiv:0704.3914 [hep-ph]].
  %%CITATION = ARXIV:0704.3914;%%

%\cite{Dixon:2009uk}
%\bibitem{Dixon:2009uk} 
  L.~J.~Dixon and Y.~Sofianatos,
  ``Analytic one-loop amplitudes for a Higgs boson plus four partons,''
  JHEP {\bf 0908}, 058 (2009)
  [arXiv:0906.0008 [hep-ph]].
  %%CITATION = ARXIV:0906.0008;%%

%\cite{Badger:2009hw}
%\bibitem{Badger:2009hw} 
  S.~Badger, E.~W.~Nigel Glover, P.~Mastrolia and C.~Williams,
  ``One-loop Higgs plus four gluon amplitudes: Full analytic results,''
  JHEP {\bf 1001}, 036 (2010)
  [arXiv:0909.4475 [hep-ph]].
  %%CITATION = ARXIV:0909.4475;%%
%gluon fusion refs
  
%\cite{Dixon:2013uaa}
\bibitem{Dixon:2013uaa} 
  L.~J.~Dixon,
  ``A brief introduction to modern amplitude methods,''
  arXiv:1310.5353 [hep-ph].
  %%CITATION = ARXIV:1310.5353;%%


\bibitem{BGrecrels}
%\cite{Berends:1987me}
%\bibitem{Berends:1987me} 
  F.~A.~Berends and W.~T.~Giele,
  ``Recursive Calculations for Processes with n Gluons,''
  Nucl.\ Phys.\ B {\bf 306}, 759 (1988).
  %%CITATION = NUPHA,B306,759;%%

%\cite{Berends:1989hf}
%\bibitem{Berends:1989hf} 
  F.~A.~Berends, W.~T.~Giele and H.~Kuijf,
  ``Exact And Approximate Expressions For Multi-Gluon Scattering,''
  Nucl.\ Phys.\ B {\bf 333}, 120 (1990).
  %%CITATION = NUPHA,B333,120;%%

\bibitem{bcf}
  R.~Britto, F.~Cachazo and B.~Feng,
  ``New recursion relations for tree amplitudes of gluons,''
  Nucl.\ Phys.\ B {\bf 715}, 499 (2005)
  [hep-th/0412308].
  %%CITATION = HEP-TH/0412308;%%
    
\bibitem{bcfw} 
%\cite{Britto:2005fq}
%\bibitem{Britto:2005fq} 
  R.~Britto, F.~Cachazo, B.~Feng and E.~Witten,
  ``Direct proof of tree-level recursion relation in Yang-Mills theory,''
  Phys.\ Rev.\ Lett.\  {\bf 94}, 181602 (2005)
  [hep-th/0501052].
  %%CITATION = HEP-TH/0501052;%%

\bibitem{cswref}
%\cite{Cachazo:2004kj}
%\bibitem{Cachazo:2004kj} 
  F.~Cachazo, P.~Svrcek and E.~Witten,
  ``MHV vertices and tree amplitudes in gauge theory,''
  JHEP {\bf 0409}, 006 (2004)
  [hep-th/0403047].
  %%CITATION = HEP-TH/0403047;%%

%\cite{Feng:2009ei}
\bibitem{Feng:2009ei} 
  B.~Feng, J.~Wang, Y.~Wang and Z.~Zhang,
  ``BCFW Recursion Relation with Nonzero Boundary Contribution,''
  JHEP {\bf 1001}, 019 (2010)
  [arXiv:0911.0301 [hep-th]].
  %%CITATION = ARXIV:0911.0301;%%
  
%\cite{Conde:2012ik}
\bibitem{Conde:2012ik} 
  E.~Conde and S.~Rajabi,
  ``The Twelve-Graviton Next-to-MHV Amplitude from Risager's Construction,''
  JHEP {\bf 1209}, 120 (2012)
  [arXiv:1205.3500 [hep-th]].
  %%CITATION = ARXIV:1205.3500;%%   
  

%\cite{ArkaniHamed:2008yf}
\bibitem{ArkaniHamed:2008yf} 
  N.~Arkani-Hamed and J.~Kaplan,
  ``On Tree Amplitudes in Gauge Theory and Gravity,''
  JHEP {\bf 0804}, 076 (2008)
  [arXiv:0801.2385 [hep-th]].
  %%CITATION = ARXIV:0801.2385;%%  

%\cite{Benincasa:2007qj}
\bibitem{Benincasa:2007qj} 
  P.~Benincasa, C.~Boucher-Veronneau and F.~Cachazo,
  ``Taming Tree Amplitudes In General Relativity,''
  JHEP {\bf 0711}, 057 (2007)
  [hep-th/0702032 [HEP-TH]].
  %%CITATION = HEP-TH/0702032;%%
  
%\cite{Kawai:1985xq}
\bibitem{Kawai:1985xq} 
  H.~Kawai, D.~C.~Lewellen and S.~H.~H.~Tye,
  ``A Relation Between Tree Amplitudes of Closed and Open Strings,''
  Nucl.\ Phys.\ B {\bf 269}, 1 (1986).
  %%CITATION = NUPHA,B269,1;%% 
    
 \bibitem{sannan}  
 %\cite{Sannan:1986tz}
%\bibitem{Sannan:1986tz} 
  S.~Sannan,
  ``Gravity As The Limit Of The Type II Superstring Theory,''
  Phys.\ Rev.\ D {\bf 34}, 1749 (1986).
  %%CITATION = PHRVA,D34,1749;%%
 
%\cite{ArkaniHamed:2009dn}
\bibitem{ArkaniHamed:2009dn} 
  N.~Arkani-Hamed, F.~Cachazo, C.~Cheung and J.~Kaplan,
  ``A Duality For The S Matrix,''
  JHEP {\bf 1003}, 020 (2010)
  [arXiv:0907.5418 [hep-th]].
  %%CITATION = ARXIV:0907.5418;%%

%\cite{Hodges:2009hk}
\bibitem{Hodges} 
  A.~Hodges,
  ``Eliminating spurious poles from gauge-theoretic amplitudes,''
  JHEP {\bf 1305}, 135 (2013)
  [arXiv:0905.1473 [hep-th]].
  %%CITATION = ARXIV:0905.1473;%% 


%\cite{Bern:2007dw}
\bibitem{Bern:2007dw} 
  Z.~Bern, L.~J.~Dixon and D.~A.~Kosower,
  ``On-Shell Methods in Perturbative QCD,''
  Annals Phys.\  {\bf 322}, 1587 (2007)
  [arXiv:0704.2798 [hep-ph]].
  %%CITATION = ARXIV:0704.2798;%%

\bibitem{bonusrel}  
  %\cite{Spradlin:2008bu}
%\bibitem{Spradlin:2008bu} 
  M.~Spradlin, A.~Volovich and C.~Wen,
  ``Three Applications of a Bonus Relation for Gravity Amplitudes,''
  Phys.\ Lett.\ B {\bf 674}, 69 (2009)
  [arXiv:0812.4767 [hep-th]].
  %%CITATION = ARXIV:0812.4767;%%

\bibitem{CEK}  
%\cite{Cohen:2010mi}
%\bibitem{Cohen:2010mi} 
  T.~Cohen, H.~Elvang and M.~Kiermaier,
  ``On-shell constructibility of tree amplitudes in general field theories,''
  JHEP {\bf 1104}, 053 (2011)
  [arXiv:1010.0257 [hep-th]].
  %%CITATION = ARXIV:1010.0257;%%

%\cite{Craig:2011ws}
\bibitem{Craig:2011ws} 
  N.~Craig, H.~Elvang, M.~Kiermaier and T.~Slatyer,
  ``Massive amplitudes on the Coulomb branch of N=4 SYM,''
  JHEP {\bf 1112}, 097 (2011)
  [arXiv:1104.2050 [hep-th]].
  %%CITATION = ARXIV:1104.2050;%%  

%\cite{Kiermaier:2011cr}
\bibitem{Kiermaier:2011cr} 
  M.~Kiermaier,
  ``The Coulomb-branch S-matrix from massless amplitudes,''
  arXiv:1105.5385 [hep-th].
  %%CITATION = ARXIV:1105.5385;%%

%\cite{Elvang:2008na}
\bibitem{Elvang:2008na} 
  H.~Elvang, D.~Z.~Freedman and M.~Kiermaier,
  ``Recursion Relations, Generating Functions, and Unitarity Sums in N=4 SYM Theory,''
  JHEP {\bf 0904}, 009 (2009)
  [arXiv:0808.1720 [hep-th]].
  %%CITATION = ARXIV:0808.1720;%%

%\cite{Risager:2005vk}
\bibitem{Risager:2005vk} 
  K.~Risager,
  ``A Direct proof of the CSW rules,''
  JHEP {\bf 0512}, 003 (2005)
  [hep-th/0508206].
  %%CITATION = HEP-TH/0508206;%%  
  
%\cite{Elvang:2008vz}
\bibitem{Elvang:2008vz} 
  H.~Elvang, D.~Z.~Freedman and M.~Kiermaier,
  ``Proof of the MHV vertex expansion for all tree amplitudes in N=4 SYM theory,''
  JHEP {\bf 0906}, 068 (2009)
  [arXiv:0811.3624 [hep-th]].
  %%CITATION = ARXIV:0811.3624;%%  

%\cite{Dixon:2004za}
\bibitem{Dixon:2004za} 
  L.~J.~Dixon, E.~W.~N.~Glover and V.~V.~Khoze,
  ``MHV rules for Higgs plus multi-gluon amplitudes,''
  JHEP {\bf 0412}, 015 (2004)
  [hep-th/0411092].
  %%CITATION = HEP-TH/0411092;%%  

%\cite{Badger:2004ty}
\bibitem{Badger:2004ty} 
  S.~D.~Badger, E.~W.~N.~Glover and V.~V.~Khoze,
  ``MHV rules for Higgs plus multi-parton amplitudes,''
  JHEP {\bf 0503}, 023 (2005)
  [hep-th/0412275].
  %%CITATION = HEP-TH/0412275;%%

%  
%\cite{Brandhuber:2011ke}
\bibitem{Brandhuber:2011ke} 
  A.~Brandhuber, B.~Spence and G.~Travaglini,
  ``Tree-Level Formalism,''
  J.\ Phys.\ A {\bf 44}, 454002 (2011)
  [arXiv:1103.3477 [hep-th]].
  %%CITATION = ARXIV:1103.3477;%%  

\bibitem{MHVlagr}  
%  
%\cite{Gorsky:2005sf}
%\bibitem{Gorsky:2005sf} 
  A.~Gorsky and A.~Rosly,
  ``From Yang-Mills Lagrangian to MHV diagrams,''
  JHEP {\bf 0601}, 101 (2006)
  [hep-th/0510111].
  %%CITATION = HEP-TH/0510111;%%

%\cite{Mansfield:2005yd}
%\bibitem{Mansfield:2005yd} 
  P.~Mansfield,
  ``The Lagrangian origin of MHV rules,''
  JHEP {\bf 0603}, 037 (2006)
  [hep-th/0511264].
  %%CITATION = HEP-TH/0511264;%%
 
 %\cite{Ettle:2006bw}
%\bibitem{Ettle:2006bw} 
  J.~H.~Ettle and T.~R.~Morris,
  ``Structure of the MHV-rules Lagrangian,''
  JHEP {\bf 0608}, 003 (2006)
  [hep-th/0605121].
  %%CITATION = HEP-TH/0605121;%%
  
  %\cite{Ettle:2007qc}
%\bibitem{Ettle:2007qc} 
  J.~H.~Ettle, C.~-H.~Fu, J.~P.~Fudger, P.~R.~W.~Mansfield and T.~R.~Morris,
  ``S-matrix equivalence theorem evasion and dimensional regularisation with the canonical MHV Lagrangian,''
  JHEP {\bf 0705}, 011 (2007)
  [hep-th/0703286].
  %%CITATION = HEP-TH/0703286;%%
  
%\cite{Feng:2006yy}
%\bibitem{Feng:2006yy} 
  H.~Feng and Y.-t.~Huang,
  ``MHV Lagrangian for N=4 super Yang-Mills,''
  JHEP {\bf 0904}, 047 (2009)
  [hep-th/0611164].
  %%CITATION = HEP-TH/0611164;%%  

%\cite{Boels:2007qn}
\bibitem{Boels:2007qn} 
  R.~Boels, L.~J.~Mason and D.~Skinner,
  ``From twistor actions to MHV diagrams,''
  Phys.\ Lett.\ B {\bf 648}, 90 (2007)
  [hep-th/0702035].
  %%CITATION = HEP-TH/0702035;%%

%\cite{BjerrumBohr:2005jr}
\bibitem{BjerrumBohr:2005jr} 
  N.~E.~J.~Bjerrum-Bohr, D.~C.~Dunbar, H.~Ita, W.~B.~Perkins and K.~Risager,
  ``MHV-vertices for gravity amplitudes,''
  JHEP {\bf 0601}, 009 (2006)
  [hep-th/0509016].
  %%CITATION = HEP-TH/0509016;%%
  
%\cite{Bianchi:2008pu}
\bibitem{Bianchi:2008pu} 
  M.~Bianchi, H.~Elvang and D.~Z.~Freedman,
  ``Generating Tree Amplitudes in N=4 SYM and N = 8 SG,''
  JHEP {\bf 0809}, 063 (2008)
  [arXiv:0805.0757 [hep-th]].
  %%CITATION = ARXIV:0805.0757;%%  

%\cite{Wess:1992cp}
\bibitem{Wess:1992cp} 
  J.~Wess and J.~Bagger,
  ``Supersymmetry and supergravity,''
  Princeton, USA: Univ. Pr. (1992) 259 p    

\bibitem{SWI}
M. T. Grisaru and H. N. Pendleton, ÒSome Properties Of Scattering Amplitudes In Supersymmetric Theories,Ó Nucl. Phys. B 124, 81 (1977).

M. T. Grisaru, H. N. Pendleton and P. van Nieuwenhuizen, ÒSupergravity And The S Matrix,Ó Phys. Rev. D 15, 996 (1977).

%\cite{Brink:1976bc}
\bibitem{Brink:1976bc} 
  L.~Brink, J.~H.~Schwarz and J.~Scherk,
  ``Supersymmetric Yang-Mills Theories,''
  Nucl.\ Phys.\ B {\bf 121}, 77 (1977).
  %%CITATION = NUPHA,B121,77;%%    

\bibitem{Ferber1977qx}
  A.~Ferber,
  ``Supertwistors And Conformal Supersymmetry,''
  Nucl.\ Phys.\  B {\bf 132}, 55 (1978).
  %%CITATION = NUPHA,B132,55;%%

%\cite{Elvang:2009wd}
\bibitem{Elvang:2009wd} 
  H.~Elvang, D.~Z.~Freedman and M.~Kiermaier,
  ``Solution to the Ward Identities for Superamplitudes,''
  JHEP {\bf 1010}, 103 (2010)
  [arXiv:0911.3169 [hep-th]].
  %%CITATION = ARXIV:0911.3169;%%    
    
%\cite{Elvang:2010xn}
\bibitem{Elvang:2010xn} 
  H.~Elvang, D.~Z.~Freedman and M.~Kiermaier,
  ``SUSY Ward identities, Superamplitudes, and Counterterms,''
  J.\ Phys.\ A {\bf 44}, 454009 (2011)
  [arXiv:1012.3401 [hep-th]].
  %%CITATION = ARXIV:1012.3401;%%    


%\cite{Kiermaier:2009yu}
\bibitem{Kiermaier:2009yu} 
  M.~Kiermaier and S.~G.~Naculich,
  ``A Super MHV vertex expansion for N=4 SYM theory,''
  JHEP {\bf 0905}, 072 (2009)
  [arXiv:0903.0377 [hep-th]].
  %%CITATION = ARXIV:0903.0377;%%    

\bibitem{nimatalk} 
N.~Arkani-Hamed, "What is the Simplest QFT?," talk given at the Paris \emph{Workshop Wonders of Gauge Theory and Supergravity}, June 24, 2008.    

%\cite{Brandhuber:2008pf}
\bibitem{Brandhuber:2008pf} 
  A.~Brandhuber, P.~Heslop and G.~Travaglini,
  ``A Note on dual superconformal symmetry of the N=4 super Yang-Mills S-matrix,''
  Phys.\ Rev.\ D {\bf 78}, 125005 (2008)
  [arXiv:0807.4097 [hep-th]].
  %%CITATION = ARXIV:0807.4097;%%
    
%\cite{ArkaniHamed:2008gz}
\bibitem{ArkaniHamed:2008gz} 
  N.~Arkani-Hamed, F.~Cachazo and J.~Kaplan,
  ``What is the Simplest Quantum Field Theory?,''
  JHEP {\bf 1009}, 016 (2010)
  [arXiv:0808.1446 [hep-th]].
  %%CITATION = ARXIV:0808.1446;%%    

%\cite{Cheung:2008dn}
\bibitem{Cheung:2008dn} 
  C.~Cheung,
  ``On-Shell Recursion Relations for Generic Theories,''
  JHEP {\bf 1003}, 098 (2010)
  [arXiv:0808.0504 [hep-th]].
  %%CITATION = ARXIV:0808.0504;%%    

%\cite{Bern:2009xq}
\bibitem{Bern:2009xq} 
  Z.~Bern, J.~J.~M.~Carrasco, H.~Ita, H.~Johansson and R.~Roiban,
  ``On the Structure of Supersymmetric Sums in Multi-Loop Unitarity Cuts,''
  Phys.\ Rev.\ D {\bf 80}, 065029 (2009)
  [arXiv:0903.5348 [hep-th]].
  %%CITATION = ARXIV:0903.5348;%%
  
%\cite{Drummond:2008cr}
\bibitem{Drummond:2008cr} 
  J.~M.~Drummond and J.~M.~Henn,
  ``All tree-level amplitudes in N=4 SYM,''
  JHEP {\bf 0904}, 018 (2009)
  [arXiv:0808.2475 [hep-th]].
  %%CITATION = ARXIV:0808.2475;%%    

%\cite{Drummond:2008vq}
\bibitem{Drummond:2008vq} 
  J.~M.~Drummond, J.~Henn, G.~P.~Korchemsky and E.~Sokatchev,
  ``Dual superconformal symmetry of scattering amplitudes in N=4 super-Yang-Mills theory,''
  Nucl.\ Phys.\ B {\bf 828}, 317 (2010)
  [arXiv:0807.1095 [hep-th]].
  %%CITATION = ARXIV:0807.1095;%% 
    
%\cite{Drummond:2009ge}
\bibitem{Drummond:2009ge} 
  J.~M.~Drummond, M.~Spradlin, A.~Volovich and C.~Wen,
  ``Tree-Level Amplitudes in N=8 Supergravity,''
  Phys.\ Rev.\ D {\bf 79}, 105018 (2009)
  [arXiv:0901.2363 [hep-th]].
  %%CITATION = ARXIV:0901.2363;%%
    
%\cite{Drummond:2010ep}
\bibitem{Drummond:2010ep} 
  J.~M.~Drummond,
  ``Hidden Simplicity of Gauge Theory Amplitudes,''
  Class.\ Quant.\ Grav.\  {\bf 27}, 214001 (2010)
  [arXiv:1010.2418 [hep-th]].
  %%CITATION = ARXIV:1010.2418;%%    

\bibitem{Penrose} 
  R.~Penrose,
  ``Twistor algebra,''
  J.\ Math.\ Phys.\  {\bf 8}, 345 (1967).
  %%CITATION = JMAPA,8,345;%%

%\cite{Witten:2003nn}
\bibitem{Witten:2003nn} 
  E.~Witten,
  ``Perturbative gauge theory as a string theory in twistor space,''
  Commun.\ Math.\ Phys.\  {\bf 252}, 189 (2004)
  [hep-th/0312171].
  %%CITATION = HEP-TH/0312171;%%

\bibitem{Drummond} 
  J.~M.~Drummond, J.~Henn, V.~A.~Smirnov and E.~Sokatchev,
  ``Magic identities for conformal four-point integrals,''
  JHEP {\bf 0701}, 064 (2007)
  [hep-th/0607160].
  %%CITATION = HEP-TH/0607160;%%

  
  \bibitem{Drummond3} 
  J.~M.~Drummond, J.~M.~Henn and J.~Plefka,
  ``Yangian symmetry of scattering amplitudes in N=4 super Yang-Mills theory,''
  JHEP {\bf 0905}, 046 (2009)
  [arXiv:0902.2987 [hep-th]].
  %%CITATION = ARXIV:0902.2987;%%

 \bibitem{Mason} 
  L.~J.~Mason and D.~Skinner,
  ``Dual Superconformal Invariance, Momentum Twistors and Grassmannians,''
  JHEP {\bf 0911}, 045 (2009)
  [arXiv:0909.0250 [hep-th]].
  %%CITATION = ARXIV:0909.0250;%%




\bibitem{UnitarityMethod}
Z.~Bern, L.~J.~Dixon, D.~C.~Dunbar and D.~A.~Kosower,
``One loop n point gauge theory amplitudes, unitarity and collinear limits,''
Nucl.\ Phys.\ B {\bf 425}, 217 (1994)
[hep-ph/9403226];
%%CITATION = NUPHA,B425,217;%%

Z.~Bern, L.~J.~Dixon, D.~C.~Dunbar and D.~A.~Kosower,
``Fusing gauge theory tree amplitudes into loop amplitudes,''
Nucl.\ Phys.\ B {\bf 435}, 59 (1995)
[hep-ph/9409265].
%%CITATION = NUPHA,B435,59;%%    


\bibitem{Review1} 
  J.~J.~M.~Carrasco and H.~Johansson,
  ``Generic multiloop methods and application to N=4 super-Yang-Mills,''
  J.\ Phys.\ A {\bf 44}, 454004 (2011)
  [arXiv:1103.3298 [hep-th]].
  %%CITATION = ARXIV:1103.3298;%%  

\bibitem{Review2} 
  Z.~Bern and Y.-t.~Huang,
  ``Basics of Generalized Unitarity,''
  J.\ Phys.\ A {\bf 44}, 454003 (2011)
  [arXiv:1103.1869 [hep-th]].
  %%CITATION = ARXIV:1103.1869;%%

\bibitem{Review3} 
  H.~Ita,
  ``Susy Theories and QCD: Numerical Approaches,''
  J.\ Phys.\ A {\bf 44}, 454005 (2011)
  [arXiv:1109.6527 [hep-th]].
  %%CITATION = ARXIV:1109.6527;%%  

 \bibitem{Review4} 
  R.~Britto,
  ``Loop Amplitudes in Gauge Theories: Modern Analytic Approaches,''
  J.\ Phys.\ A {\bf 44}, 454006 (2011)
  [arXiv:1012.4493 [hep-th]].
  %%CITATION = ARXIV:1012.4493;%% 

 
\bibitem{Scalar1} 
  W.~L.~van Neerven and J.~A.~M.~Vermaseren,
  ``Large Loop Integrals,''
  Phys.\ Lett.\ B {\bf 137}, 241 (1984).
  %%CITATION = PHLTA,B137,241;%%

\bibitem{Scalar2} 
  Z.~Bern, L.~J.~Dixon and D.~A.~Kosower,
  ``Dimensionally regulated one loop integrals,''
  Phys.\ Lett.\ B {\bf 302}, 299 (1993)
  [Erratum-ibid.\ B {\bf 318}, 649 (1993)]
  [hep-ph/9212308].
  %%CITATION = HEP-PH/9212308;%%

\bibitem{Scalar3} 
  Z.~Bern, L.~J.~Dixon and D.~A.~Kosower,
  ``Dimensionally regulated pentagon integrals,''
  Nucl.\ Phys.\ B {\bf 412}, 751 (1994)
  [hep-ph/9306240].
  %%CITATION = HEP-PH/9306240;%%
  %292 citations counted in INSPIRE as of 16 Feb 2013
  
\bibitem{Scalar4}
 L.~M.~Brown and R.~P.~Feynman,
  ``Radiative corrections to Compton scattering,''
  Phys.\ Rev.\  {\bf 85}, 231 (1952);
  %%CITATION = PHRVA,85,231;%%

G.~Passarino and M.~J.~G.~Veltman,
  ``One Loop Corrections for e+ e- Annihilation Into mu+ mu- in the Weinberg Model,''
  Nucl.\ Phys.\ B {\bf 160}, 151 (1979);
  %%CITATION = NUPHA,B160,151;%%

%%%%%%%%
 G.~'t Hooft and M.~J.~G.~Veltman,
  ``Scalar One Loop Integrals,''
  Nucl.\ Phys.\ B {\bf 153}, 365 (1979).
  %%CITATION = NUPHA,B153,365;%%

\bibitem{HenrikLoopsLegs} 
 H.~Johansson, D.~A.~Kosower and K.~J.~Larsen,
  ``An Overview of Maximal Unitarity at Two Loops,''
  PoS LL {\bf 2012}, 066 (2012)
  [PoS LL {\bf 2012}, 066 (2012)]
  [arXiv:1212.2132 [hep-th]].
  %%CITATION = ARXIV:1212.2132;%%

\bibitem{OneLoopMethods}
C.~Anastasiou, R.~Britto, B.~Feng, Z.~Kunszt and P.~Mastrolia,
``D-dimensional unitarity cut method,''
Phys.\ Lett.\  B {\bf 645}, 213 (2007)
[hep-ph/0609191];

R.~Britto and B.~Feng,
``Integral Coefficients for One-Loop Amplitudes,''
JHEP {\bf 0802}, 095 (2008)
[0711.4284 [hep-ph]];
%%CITATION = JHEPA,0802,095;%%

R.~Britto and B.~Feng,
``Unitarity cuts with massive propagators and algebraic expressions for
coefficients,''
Phys.\ Rev.\  D {\bf 75}, 105006 (2007)
[hep-ph/0612089];
%%CITATION = PHRVA,D75,105006;%%
%

G.~Ossola, C.~G.~Papadopoulos and R.~Pittau,
``Reducing full one-loop amplitudes to scalar integrals at the integrand
level,''
Nucl.\ Phys.\  B {\bf 763}, 147 (2007)
[hep-ph/0609007];
%%CITATION = NUPHA,B763,147;%%
%

R.~Britto, B.~Feng and P.~Mastrolia,
``Closed-Form Decomposition of One-Loop Massive Amplitudes,''
Phys.\ Rev.\  D {\bf 78}, 025031 (2008)
[0803.1989 [hep-ph]];
%%CITATION = PHRVA,D78,025031;%%

D.~Forde,
``Direct extraction of one-loop integral coefficients,''
Phys.\ Rev.\  D {\bf 75}, 125019 (2007)
[0704.1835 [hep-ph]];
%%CITATION = PHRVA,D75,125019;%%
%

\bibitem{DDimUnitarity}
Z.~Bern and A.~G.~Morgan,
``Massive Loop Amplitudes from Unitarity,''  
Nucl.\ Phys.\ B {\bf 467}, 479 (1996)
[hep-ph/9511336];
%%CITATION = HEP-PH 9511336;%%

Z.~Bern, L.~J.~Dixon and D.~A.~Kosower,
``Progress in one-loop QCD computations,''
Ann.\ Rev.\ Nucl.\ Part.\ Sci.\  {\bf 46}, 109 (1996)
[hep-ph/9602280].
%%CITATION = HEP-PH 9602280;%%
%

\bibitem{Rational}
S.~D.~Badger,
``Direct Extraction Of One Loop Rational Terms,''
JHEP {\bf 0901}, 049 (2009)
[0806.4600 [hep-ph]].
%%CITATION = JHEPA,0901,049;%%

%\cite{Green:1982sw}
\bibitem{Green:1982sw} 
  M.~B.~Green, J.~H.~Schwarz and L.~Brink,
  ``N=4 Yang-Mills and N=8 Supergravity as Limits of String Theories,''
  Nucl.\ Phys.\ B {\bf 198}, 474 (1982).
  %%CITATION = NUPHA,B198,474;%%

\bibitem{NoTria2}
  Z.~Bern, N.~E.~J.~Bjerrum-Bohr and D.~C.~Dunbar,
  ``Inherited twistor-space structure of gravity loop amplitudes,''
  JHEP {\bf 0505}, 056 (2005)
  [hep-th/0501137];
  %%CITATION = HEP-TH/0501137;%%
  
N.~E.~J.~Bjerrum-Bohr, D.~C.~Dunbar, H.~Ita, W.~B.~Perkins and K.~Risager,
  ``The No-Triangle Hypothesis for N=8 Supergravity,''
  JHEP {\bf 0612}, 072 (2006)
  [hep-th/0610043];
  %%CITATION = HEP-TH/0610043;%%  
 
N.~E.~J.~Bjerrum-Bohr and P.~Vanhove,
  ``Absence of Triangles in Maximal Supergravity Amplitudes,''
  JHEP {\bf 0810}, 006 (2008)
  [arXiv:0805.3682 [hep-th]].
  %%CITATION = ARXIV:0805.3682;%%  

%\cite{Bern:2007xj}
\bibitem{Bern:2007xj} 
  Z.~Bern, J.~J.~Carrasco, D.~Forde, H.~Ita and H.~Johansson,
  ``Unexpected Cancellations in Gravity Theories,''
  Phys.\ Rev.\ D {\bf 77}, 025010 (2008)
  [arXiv:0707.1035 [hep-th]].
  %%CITATION = ARXIV:0707.1035;%%
 %%We have this ref above too.   

 \bibitem{NoTria3} 
  S.~Lal and S.~Raju,
  ``The Next-to-Simplest Quantum Field Theories,''
  Phys.\ Rev.\ D {\bf 81}, 105002 (2010)
  [arXiv:0910.0930 [hep-th]].
  %%CITATION = ARXIV:0910.0930;%% 
 
  \bibitem{NoTria4} 
  D.~C.~Dunbar, J.~H.~Ettle and W.~B.~Perkins,
  ``Perturbative expansion of $N<8$ Supergravity,''
  Phys.\ Rev.\ D {\bf 83}, 065015 (2011)
  [arXiv:1011.5378 [hep-th]].
  %%CITATION = ARXIV:1011.5378;%% 
  
%\cite{Elvang:2011fx}
\bibitem{Elvang:2011fx} 
  H.~Elvang, Y.-t.~Huang and C.~Peng,
  ``On-shell superamplitudes in $N<4$ SYM,''
  JHEP {\bf 1109}, 031 (2011)
  [arXiv:1102.4843 [hep-th]].
  %%CITATION = ARXIV:1102.4843;%%  

%\cite{Huang:2012aq}
\bibitem{Huang:2012aq} 
  Y.-t.~Huang, D.~A.~McGady and C.~Peng,
  ``One-loop renormalization and the S-matrix,''
  arXiv:1205.5606 [hep-th].
  %%CITATION = ARXIV:1205.5606;%%
  

%\cite{Roiban:2010kk}
\bibitem{Roiban:2010kk} 
  R.~Roiban,
  ``Review of AdS/CFT Integrability, Chapter V.1: Scattering Amplitudes - a Brief Introduction,''
  Lett.\ Math.\ Phys.\  {\bf 99}, 455 (2012)
  [arXiv:1012.4001 [hep-th]].
  %%CITATION = ARXIV:1012.4001;%%
    
%\cite{Marcus:1985yy}
\bibitem{Marcus:1985yy} 
  N.~Marcus,
  ``Composite Anomalies In Supergravity,''
  Phys.\ Lett.\ B {\bf 157}, 383 (1985).
  %%CITATION = PHLTA,B157,383;%%    
    
%\cite{diVecchia:1984jh}
\bibitem{diVecchia:1984jh} 
  P.~di Vecchia, S.~Ferrara and L.~Girardello,
  ``Anomalies Of Hidden Local Chiral Symmetries In Sigma Models And Extended Supergravities,"
  Phys.\ Lett.\ B {\bf 151}, 199 (1985).
  %%CITATION = PHLTA,B151,199;%%   


%\cite{Drummond:2008bq}
\bibitem{Drummond:2008bq} 
  J.~M.~Drummond, J.~Henn, G.~P.~Korchemsky and E.~Sokatchev,
  ``Generalized unitarity for N=4 super-amplitudes,''
  Nucl.\ Phys.\ B {\bf 869}, 452 (2013)
  [arXiv:0808.0491 [hep-th]].
  %%CITATION = ARXIV:0808.0491;%%
  %99 citations counted in INSPIRE as of 18 Apr 2013

    
%\cite{Brandhuber:2009xz}
\bibitem{Brandhuber:2009xz} 
  A.~Brandhuber, P.~Heslop and G.~Travaglini,
  ``One-Loop Amplitudes in N=4 Super Yang-Mills and Anomalous Dual Conformal Symmetry,''
  JHEP {\bf 0908}, 095 (2009)
  [arXiv:0905.4377 [hep-th]].
  %%CITATION = ARXIV:0905.4377;%%

%\cite{Elvang:2009ya}
\bibitem{Elvang:2009ya} 
  H.~Elvang, D.~Z.~Freedman and M.~Kiermaier,
  ``Dual conformal symmetry of 1-loop NMHV amplitudes in N=4 SYM theory,''
  JHEP {\bf 1003}, 075 (2010)
  [arXiv:0905.4379 [hep-th]].
  %%CITATION = ARXIV:0905.4379;%%    
    
%\cite{Korchemsky:2009hm}
\bibitem{Korchemsky:2009hm} 
  G.~P.~Korchemsky and E.~Sokatchev,
  ``Symmetries and analytic properties of scattering amplitudes in N=4 SYM theory,''
  Nucl.\ Phys.\ B {\bf 832}, 1 (2010)
  [arXiv:0906.1737 [hep-th]].
  %%CITATION = ARXIV:0906.1737;%%    


\bibitem{2LoopBasis} 
  J.~Gluza, K.~Kajda and D.~A.~Kosower,
  ``Towards a Basis for Planar Two-Loop Integrals,''
  Phys.\ Rev.\ D {\bf 83}, 045012 (2011)
  [arXiv:1009.0472 [hep-th]].
  %%CITATION = ARXIV:1009.0472;%%  

  
\bibitem{2LoopBasisCut} 
%\cite{Kosower:2011ty}
%\bibitem{Kosower:2011ty} 
  D.~A.~Kosower and K.~J.~Larsen,
  ``Maximal Unitarity at Two Loops,''
  Phys.\ Rev.\ D {\bf 85}, 045017 (2012)
  [arXiv:1108.1180 [hep-th]].
  %%CITATION = ARXIV:1108.1180;%%

%\bibitem{Johansson:2012zv} 
  H.~Johansson, D.~A.~Kosower and K.~J.~Larsen,
  ``Two-Loop Maximal Unitarity with External Masses,''
  Phys.\ Rev.\ D {\bf 87}, 025030 (2013)
  [arXiv:1208.1754 [hep-th]].
  %%CITATION = ARXIV:1208.1754;%%


%\cite{Badger:2012dp}
\bibitem{Badger:2012dp} 
  S.~Badger, H.~Frellesvig and Y.~Zhang,
  ``Hepta-Cuts of Two-Loop Scattering Amplitudes,''
  JHEP {\bf 1204}, 055 (2012)
  [arXiv:1202.2019 [hep-ph]].
  %%CITATION = ARXIV:1202.2019;%%

%\cite{Zhang:2012ce}
\bibitem{Zhang:2012ce} 
  Y.~Zhang,
  ``Integrand-Level Reduction of Loop Amplitudes by Computational Algebraic Geometry Methods,''
  JHEP {\bf 1209}, 042 (2012)
  [arXiv:1205.5707 [hep-ph]].
  %%CITATION = ARXIV:1205.5707;%%

%\cite{Sogaard:2013yga}
\bibitem{Sogaard:2013yga} 
  M.~S¿gaard,
  ``Global Residues and Two-Loop Hepta-Cuts,''
  JHEP {\bf 1309}, 116 (2013)
  [arXiv:1306.1496 [hep-th]].
  %%CITATION = ARXIV:1306.1496;%%

%\cite{Smirnov:2010hn}
\bibitem{Smirnov:2010hn} 
  A.~V.~Smirnov and A.~V.~Petukhov,
  ``The Number of Master Integrals is Finite,''
  Lett.\ Math.\ Phys.\  {\bf 97}, 37 (2011)
  [arXiv:1004.4199 [hep-th]].
  %%CITATION = ARXIV:1004.4199;%%


\bibitem{N42Loop} 
  C.~Anastasiou, Z.~Bern, L.~J.~Dixon and D.~A.~Kosower,
  ``Planar amplitudes in maximally supersymmetric Yang-Mills theory,''
  Phys.\ Rev.\ Lett.\  {\bf 91}, 251602 (2003)
  [hep-th/0309040].
  %%CITATION = HEP-TH/0309040;%%  

  \bibitem{N43Loop} 
  Z.~Bern, L.~J.~Dixon and V.~A.~Smirnov,
  ``Iteration of planar amplitudes in maximally supersymmetric Yang-Mills theory at three loops and beyond,''
  Phys.\ Rev.\ D {\bf 72}, 085001 (2005)
  [hep-th/0505205].
  %%CITATION = HEP-TH/0505205;%%

%\cite{Bern:1997nh}
\bibitem{N=42Loop1} 
  Z.~Bern, J.~S.~Rozowsky and B.~Yan,
  ``Two loop four gluon amplitudes in N=4 superYang-Mills,''
  Phys.\ Lett.\ B {\bf 401}, 273 (1997)
  [hep-ph/9702424].
  %%CITATION = HEP-PH/9702424;%%

%\cite{Bern:2006vw}
\bibitem{Bern:2006vw} 
  Z.~Bern, M.~Czakon, D.~A.~Kosower, R.~Roiban and V.~A.~Smirnov,
  ``Two-loop iteration of five-point N=4 super-Yang-Mills amplitudes,''
  Phys.\ Rev.\ Lett.\  {\bf 97}, 181601 (2006)
  [hep-th/0604074].
  %%CITATION = HEP-TH/0604074;%%


%\cite{Cachazo:2006tj}
\bibitem{Cachazo:2006tj} 
  F.~Cachazo, M.~Spradlin and A.~Volovich,
  ``Iterative structure within the five-particle two-loop amplitude,''
  Phys.\ Rev.\ D {\bf 74}, 045020 (2006)
  [hep-th/0602228].
  %%CITATION = HEP-TH/0602228;%%

%\cite{Alday:2007he}
\bibitem{Alday:2007he} 
  L.~F.~Alday and J.~Maldacena,
  ``Comments on gluon scattering amplitudes via AdS/CFT,''
  JHEP {\bf 0711}, 068 (2007)
  [arXiv:0710.1060 [hep-th]].
  %%CITATION = ARXIV:0710.1060;%%

%\cite{Alday:2007hr}
\bibitem{Alday:2007hr} 
  L.~F.~Alday and J.~M.~Maldacena,
  ``Gluon scattering amplitudes at strong coupling,''
  JHEP {\bf 0706}, 064 (2007)
  [arXiv:0705.0303 [hep-th]].
  %%CITATION = ARXIV:0705.0303;%%

%\cite{Bern:2008ap}
\bibitem{Bern:2008ap}
  Z.~Bern, L.~J.~Dixon, D.~A.~Kosower, R.~Roiban, M.~Spradlin, C.~Vergu and A.~Volovich,
  ``The Two-Loop Six-Gluon MHV Amplitude in Maximally Supersymmetric Yang-Mills Theory,''
  Phys.\ Rev.\ D {\bf 78} (2008) 045007
  [arXiv:0803.1465 [hep-th]].
  %%CITATION = ARXIV:0803.1465;%%     
   
%\cite{Cachazo:2008hp}
\bibitem{Cachazo:2008hp} 
  F.~Cachazo, M.~Spradlin and A.~Volovich,
  ``Leading Singularities of the Two-Loop Six-Particle MHV Amplitude,''
  Phys.\ Rev.\ D {\bf 78}, 105022 (2008)
  [arXiv:0805.4832 [hep-th]].
  %%CITATION = ARXIV:0805.4832;%%      
      
%\cite{DelDuca:2009au}
\bibitem{DelDuca:2009au} 
  V.~Del Duca, C.~Duhr and V.~A.~Smirnov,
  ``An Analytic Result for the Two-Loop Hexagon Wilson Loop in N=4 SYM,''
  JHEP {\bf 1003}, 099 (2010)
  [arXiv:0911.5332 [hep-ph]].
  %%CITATION = ARXIV:0911.5332;%%      
      
%\cite{DelDuca:2010zg}
\bibitem{DelDuca:2010zg} 
  V.~Del Duca, C.~Duhr and V.~A.~Smirnov,
  ``The Two-Loop Hexagon Wilson Loop in N=4 SYM,''
  JHEP {\bf 1005}, 084 (2010)
  [arXiv:1003.1702 [hep-th]].
  %%CITATION = ARXIV:1003.1702;%%      

%\cite{Goncharov:2010jf}
\bibitem{Goncharov:2010jf} 
  A.~B.~Goncharov, M.~Spradlin, C.~Vergu and A.~Volovich,
  ``Classical Polylogarithms for Amplitudes and Wilson Loops,''
  Phys.\ Rev.\ Lett.\  {\bf 105}, 151605 (2010)
  [arXiv:1006.5703 [hep-th]].
  %%CITATION = ARXIV:1006.5703;%%


\bibitem{SingleCut1}
 E.~W.~Nigel Glover and C.~Williams,
  ``One-Loop Gluonic Amplitudes from Single Unitarity Cuts,''
  JHEP {\bf 0812}, 067 (2008)
  [0810.2964 [hep-th]].
  %%CITATION = JHEPA,0812,067;%%

 I.~Bierenbaum, S.~Catani, P.~Draggiotis and G.~Rodrigo,
  ``A Tree-Loop Duality Relation at Two Loops and Beyond,''
  JHEP {\bf 1010}, 073 (2010)
  [arXiv:1007.0194 [hep-ph]].
  %%CITATION = ARXIV:1007.0194;%%


%\cite{Elvang:2011ub}
%\bibitem{Elvang:2011ub} 
  H.~Elvang, D.~Z.~Freedman and M.~Kiermaier,
  ``Integrands for QCD rational terms and N=4 SYM from massive CSW rules,''
  JHEP {\bf 1206}, 015 (2012)
  [arXiv:1111.0635 [hep-th]].
  %%CITATION = ARXIV:1111.0635;%%


\bibitem{SingleCut2}
S.~Caron-Huot,
  ``Loops and trees,''
  JHEP {\bf 1105}, 080 (2011)
  [arXiv:1007.3224 [hep-ph]].
  %%CITATION = ARXIV:1007.3224;%%

\bibitem{SingleCut3} 
  N.~Arkani-Hamed, J.~L.~Bourjaily, F.~Cachazo, S.~Caron-Huot and J.~Trnka,
  ``The All-Loop Integrand For Scattering Amplitudes in Planar N=4 SYM,''
  JHEP {\bf 1101}, 041 (2011)
  [arXiv:1008.2958 [hep-th]].
  %%CITATION = ARXIV:1008.2958;%%  


\bibitem{RutgerBCFW} 
  R.~H.~Boels,
  ``On BCFW shifts of integrands and integrals,''
  JHEP {\bf 1011}, 113 (2010)
  [arXiv:1008.3101 [hep-th]].
  %%CITATION = ARXIV:1008.3101;%%  


\bibitem{MaximalCut}
Z.~Bern, J.~J.~M.~Carrasco, H.~Johansson and D.~A.~Kosower,
``Maximally supersymmetric planar Yang-Mills amplitudes at five loops,''
Phys.\ Rev.\  D {\bf 76}, 125020 (2007)
[0705.1864 [hep-th]].
%%CITATION = PHRVA,D76,125020;%%

\bibitem{LS}
R.~Britto, F.~Cachazo and B.~Feng,
``Generalized unitarity and one-loop amplitudes in N=4 super-Yang-Mills,''
Nucl.\ Phys.\  B {\bf 725}, 275 (2005)
[hep-th/0412103];
%%CITATION = NUPHA,B725,275;%%

E.~I.~Buchbinder and F.~Cachazo,
``Two-loop amplitudes of gluons and octa-cuts in N=4 super Yang-Mills,''
JHEP {\bf 0511}, 036 (2005)
[hep-th/0506126].
%%CITATION = JHEPA,0511,036;%%


\bibitem{UnitLS}
  N.~Arkani-Hamed, J.~L.~Bourjaily, F.~Cachazo and J.~Trnka,
  ``Local Integrals for Planar Scattering Amplitudes,''
  JHEP {\bf 1206}, 125 (2012)
  [arXiv:1012.6032 [hep-th]].
  %%CITATION = ARXIV:1012.6032;%%
  
For extensions, see \\
%\cite{Bourjaily:2013mma}
%\bibitem{Bourjaily:2013mma} 
  J.~L.~Bourjaily, S.~Caron-Huot and J.~Trnka,
  ``Dual-Conformal Regularization of Infrared Loop Divergences and the Chiral Box Expansion,''
  arXiv:1303.4734 [hep-th].
  %%CITATION = ARXIV:1303.4734;%%
  

%\cite{ArkaniHamed:2012nw}
\bibitem{ArkaniHamed:2012nw} 
  N.~Arkani-Hamed, J.~L.~Bourjaily, F.~Cachazo, A.~B.~Goncharov, A.~Postnikov and J.~Trnka,
  ``Scattering Amplitudes and the Positive Grassmannian,''
  arXiv:1212.5605 [hep-th].
  %%CITATION = ARXIV:1212.5605;%%
    
\bibitem{Sharpening} 
  F.~Cachazo,
  ``Sharpening The Leading Singularity,''
  arXiv:0803.1988 [hep-th].
  %%CITATION = ARXIV:0803.1988;%%

\bibitem{DrummondTDual} 
  J.~M.~Drummond and L.~Ferro,
  ``Yangians, Grassmannians and T-duality,''
  JHEP {\bf 1007}, 027 (2010)
  [arXiv:1001.3348 [hep-th]].
  %%CITATION = ARXIV:1001.3348;%%
  
\bibitem{YangianFix}
 J.~M.~Drummond and L.~Ferro,
  ``The Yangian origin of the Grassmannian integral,''
  JHEP {\bf 1012}, 010 (2010)
  [arXiv:1002.4622 [hep-th]];
  %%CITATION = ARXIV:1002.4622;%%
  
  G.~P.~Korchemsky and E.~Sokatchev,
  ``Superconformal invariants for scattering amplitudes in N=4 SYM theory,''
  Nucl.\ Phys.\ B {\bf 839}, 377 (2010)
  [arXiv:1002.4625 [hep-th]].
  %%CITATION = ARXIV:1002.4625;%%

\bibitem{Roiban:2004yf} 
  R.~Roiban, M.~Spradlin and A.~Volovich,
  ``On the tree level S matrix of Yang-Mills theory,''
  Phys.\ Rev.\ D {\bf 70}, 026009 (2004)
  [hep-th/0403190].
  %%CITATION = HEP-TH/0403190;%%
 %YtTwistor  
   
  %\cite{Spradlin:2009qr}
\bibitem{Spradlin:2009qr} 
  M.~Spradlin and A.~Volovich,
  ``From Twistor String Theory To Recursion Relations,''
  Phys.\ Rev.\ D {\bf 80}, 085022 (2009)
  [arXiv:0909.0229 [hep-th]].
  %%CITATION = ARXIV:0909.0229;%%
  
 
 \bibitem{ArkaniHamed:2009dg} 
  N.~Arkani-Hamed, J.~Bourjaily, F.~Cachazo and J.~Trnka,
  ``Unification of Residues and Grassmannian Dualities,''
  JHEP {\bf 1101}, 049 (2011)
  [arXiv:0912.4912 [hep-th]].
  %%CITATION = ARXIV:0912.4912;%%
    
\bibitem{RaduTwistor} 
  O.~T.~Engelund and R.~Roiban,
  ``A twistor string for the ABJ(M) theory,''
  arXiv:1401.6242 [hep-th].
  %%CITATION = ARXIV:1401.6242;%%  
  
  
\bibitem{Bourjaily:2010kw} 
  J.~L.~Bourjaily, J.~Trnka, A.~Volovich and C.~Wen,
  ``The Grassmannian and the Twistor String: Connecting All Trees in N=4 SYM,''
  JHEP {\bf 1101}, 038 (2011)
  [arXiv:1006.1899 [hep-th]].
  %%CITATION = ARXIV:1006.1899;%%

%\cite{Bullimore:2009cb}
\bibitem{Bullimore:2009cb} 
  M.~Bullimore, L.~J.~Mason and D.~Skinner,
  ``Twistor-Strings, Grassmannians and Leading Singularities,''
  JHEP {\bf 1003}, 070 (2010)
  [arXiv:0912.0539 [hep-th]].
  %%CITATION = ARXIV:0912.0539;%%
  %43 citations counted in INSPIRE as of 11 Apr 2013  
  

%\cite{Dolan:2011za}
\bibitem{Dolan:2011za} 
  L.~Dolan and P.~Goddard,
  ``Complete Equivalence Between Gluon Tree Amplitudes in Twistor String Theory and in Gauge Theory,''
  JHEP {\bf 1206}, 030 (2012)
  [arXiv:1111.0950 [hep-th]].
  %%CITATION = ARXIV:1111.0950;%%  
   

\bibitem{NimaPoly} 
  N.~Arkani-Hamed, J.~L.~Bourjaily, F.~Cachazo, A.~Hodges and J.~Trnka,
  ``A Note on Polytopes for Scattering Amplitudes,''
  JHEP {\bf 1204}, 081 (2012)
  [arXiv:1012.6030 [hep-th]].
  %%CITATION = ARXIV:1012.6030;%%
    
\bibitem{Bullimore:2010pj} 
  M.~Bullimore, L.~J.~Mason and D.~Skinner,
  ``MHV Diagrams in Momentum Twistor Space,''
  JHEP {\bf 1012}, 032 (2010)
  [arXiv:1009.1854 [hep-th]].
  %%CITATION = ARXIV:1009.1854;%%

\bibitem{NT} 
%\cite{Arkani-Hamed:2013jha}
%\bibitem{Arkani-Hamed:2013jha} 
  N.~Arkani-Hamed and J.~Trnka,
  ``The Amplituhedron,''
  arXiv:1312.2007 [hep-th].
  %%CITATION = ARXIV:1312.2007;%%

%\cite{Arkani-Hamed:2013kca}
%\bibitem{Arkani-Hamed:2013kca} 
  N.~Arkani-Hamed and J.~Trnka,
  ``Into the Amplituhedron,''
  arXiv:1312.7878 [hep-th].

%Talks given by Nima Arkani-Hamed at Strings 2013 (Seoul, South Korea, June 24-29, 2013) and by Jaroslav Trnka at Amplitudes 2013 (Tegernsee, Germany, April 28 - May 3, 2013).


\bibitem{MasonSkinnerPoly} 
  L.~Mason and D.~Skinner,
  ``Amplitudes at Weak Coupling as Polytopes in AdS$_5$,''
  J.\ Phys.\ A {\bf 44}, 135401 (2011)
  [arXiv:1004.3498 [hep-th]].
  %%CITATION = ARXIV:1004.3498;%%

\bibitem{NonPlanarPoly} 
  H.~Nastase and H.~J.~Schnitzer,
  ``Twistor and Polytope Interpretations for Subleading Color One-Loop Amplitudes,''
  Nucl.\ Phys.\ B {\bf 855}, 901 (2012)
  [arXiv:1104.2752 [hep-th]].
  %%CITATION = ARXIV:1104.2752;%%


\bibitem{Fields} 
  W.~Siegel,
  ``Fields,''
  hep-th/9912205.
  %%CITATION = HEP-TH/9912205;%%

\bibitem{Boels:2009bv}
  R.~Boels,
  ``Covariant representation theory of the Poincare algebra and some of its extensions,''
  JHEP {\bf 1001}, 010 (2010)
  [arXiv:0908.0738 [hep-th]].
  %%CITATION = ARXIV:0908.0738;%%

\bibitem{CaronHuot:2010rj} 
  S.~Caron-Huot and D.~O'Connell,
  ``Spinor Helicity and Dual Conformal Symmetry in Ten Dimensions,''
  JHEP {\bf 1108}, 014 (2011)
  [arXiv:1010.5487 [hep-th]].
  %%CITATION = ARXIV:1010.5487;%%

\bibitem{Boels:2012ie} 
  R.~H.~Boels and D.~O'Connell,
  ``Simple superamplitudes in higher dimensions,''
  JHEP {\bf 1206}, 163 (2012)
  [arXiv:1201.2653 [hep-th]].
  %%CITATION = ARXIV:1201.2653;%%

\bibitem{6DSYM1} 
Z.~Bern, J.~J.~Carrasco, T.~Dennen, Y.-t.~Huang and H.~Ita,
  ``Generalized Unitarity and Six-Dimensional Helicity,''
  Phys.\ Rev.\ D {\bf 83}, 085022 (2011)
  [arXiv:1010.0494 [hep-th]].
  %%CITATION = ARXIV:1010.0494;%%
 
 \bibitem{Scott} 
  S.~Davies,
  ``One-Loop QCD and Higgs to Partons Processes Using Six-Dimensional Helicity and Generalized Unitarity,''
  Phys.\ Rev.\ D {\bf 84}, 094016 (2011)
  [arXiv:1108.0398 [hep-ph]].
  %%CITATION = ARXIV:1108.0398;%%
 
 
 
\bibitem{CheungO'Connell} 
  C.~Cheung and D.~O'Connell,
  ``Amplitudes and Spinor-Helicity in Six Dimensions,''
  JHEP {\bf 0907}, 075 (2009)
  [arXiv:0902.0981 [hep-th]].
  %%CITATION = ARXIV:0902.0981;%%

\bibitem{Dennen:2009vk} 
  T.~Dennen, Y.-t.~Huang and W.~Siegel,
  ``Supertwistor space for 6D maximal super Yang-Mills,''
  JHEP {\bf 1004}, 127 (2010)
  [arXiv:0910.2688 [hep-th]].
  %%CITATION = ARXIV:0910.2688;%%

\bibitem{6DSYM2} 
  A.~Brandhuber, D.~Korres, D.~Koschade and G.~Travaglini,
  ``One-loop Amplitudes in Six-Dimensional (1,1) Theories from Generalised Unitarity,''
  JHEP {\bf 1102}, 077 (2011)
  [arXiv:1010.1515 [hep-th]].
  %%CITATION = ARXIV:1010.1515;%%

 C.~Saemann, R.~Wimmer and M.~Wolf,
  ``A Twistor Description of Six-Dimensional N=(1,1) Super Yang-Mills Theory,''
  JHEP {\bf 1205}, 020 (2012)
  [arXiv:1201.6285 [hep-th]].
  %%CITATION = ARXIV:1201.6285;%%

\bibitem{6DSYMDC} 
  T.~Dennen and Y.-t.~Huang,
  ``Dual Conformal Properties of Six-Dimensional Maximal Super Yang-Mills Amplitudes,''
  JHEP {\bf 1101}, 140 (2011)
  [arXiv:1010.5874 [hep-th]].
  %%CITATION = ARXIV:1010.5874;%%

\bibitem{6DSC}
 T.~Chern,
  ``Superconformal Field Theory In Six Dimensions And Supertwistor,''
  arXiv:0906.0657 [hep-th].
  %%CITATION = ARXIV:0906.0657;%%

 M.~Chiodaroli, M.~Gunaydin and R.~Roiban,
  ``Superconformal symmetry and maximal supergravity in various dimensions,''
  JHEP {\bf 1203}, 093 (2012)
  [arXiv:1108.3085 [hep-th]].
  %%CITATION = ARXIV:1108.3085;%%

 L.~J.~Mason, R.~A.~Reid-Edwards and A.~Taghavi-Chabert,
  ``Conformal Field Theories in Six-Dimensional Twistor Space,''
  J.\ Geom.\ Phys.\  {\bf 62}, 2353 (2012)
  [arXiv:1111.2585 [hep-th]].
  %%CITATION = ARXIV:1111.2585;%%

%\cite{Saemann:2011nb}
  C.~Saemann and M.~Wolf,
  ``On Twistors and Conformal Field Theories from Six Dimensions,''
  J.\ Math.\ Phys.\  {\bf 54}, 013507 (2013)
  [arXiv:1111.2539 [hep-th]].
  %%CITATION = ARXIV:1111.2539;%%
  

\bibitem{M5}
B.~Czech, Y.-t.~Huang and M.~Rozali,
  ``Amplitudes for Multiple M5 Branes,''
  JHEP {\bf 1210}, 143 (2012)
  [arXiv:1110.2791 [hep-th]].
  
\bibitem{M52}
 C.~Saemann and M.~Wolf,
  ``Non-Abelian Tensor Multiplet Equations from Twistor Space,''
  arXiv:1205.3108 [hep-th].
  %%CITATION = ARXIV:1205.3108;%%
  
  
\bibitem{M5a}
 Y.-t.~Huang and A.~E.~Lipstein,
  ``Amplitudes of 3D and 6D Maximal Superconformal Theories in Supertwistor Space,''
  JHEP {\bf 1010}, 007 (2010)
  [arXiv:1004.4735 [hep-th]].
  
\bibitem{Alday:2009zm} 
  L.~F.~Alday, J.~M.~Henn, J.~Plefka and T.~Schuster,
  ``Scattering into the fifth dimension of N=4 super Yang-Mills,''
  JHEP {\bf 1001}, 077 (2010)
  [arXiv:0908.0684 [hep-th]].
  %%CITATION = ARXIV:0908.0684;%%
  
\bibitem{TristanBeisert} 
  A.~Agarwal, N.~Beisert and T.~McLoughlin,
  ``Scattering in Mass-Deformed $N\ge 4$ Chern-Simons Models,''
  JHEP {\bf 0906}, 045 (2009)
  [arXiv:0812.3367 [hep-th]].
  %%CITATION = ARXIV:0812.3367;%%   
    
 \bibitem{BLG1} 
  A.~Gustavsson,
  ``Algebraic structures on parallel M2-branes,''
  Nucl.\ Phys.\ B {\bf 811}, 66 (2009)
  [arXiv:0709.1260 [hep-th]].
  %%CITATION = ARXIV:0709.1260;%%

\bibitem{BLG2} 
  J.~Bagger and N.~Lambert,
  ``Gauge symmetry and supersymmetry of multiple M2-branes,''
  Phys.\ Rev.\ D {\bf 77}, 065008 (2008)
  [arXiv:0711.0955 [hep-th]].
  %%CITATION = ARXIV:0711.0955;%%

\bibitem{Schwarz} 
  M.~A.~Bandres, A.~E.~Lipstein and J.~H.~Schwarz,
  ``N = 8 Superconformal Chern-Simons Theories,''
  JHEP {\bf 0805}, 025 (2008)
  [arXiv:0803.3242 [hep-th]].
  %%CITATION = ARXIV:0803.3242;%%

\bibitem{Bargheer} 
  T.~Bargheer, F.~Loebbert and C.~Meneghelli,
  ``Symmetries of Tree-level Scattering Amplitudes in N=6 Superconformal Chern-Simons Theory,''
  Phys.\ Rev.\ D {\bf 82}, 045016 (2010)
  [arXiv:1003.6120 [hep-th]].
  %%CITATION = ARXIV:1003.6120;%%

\bibitem{ABJM} 
  O.~Aharony, O.~Bergman, D.~L.~Jafferis and J.~Maldacena,
  ``N=6 superconformal Chern-Simons-matter theories, M2-branes and their gravity duals,''
  JHEP {\bf 0810}, 091 (2008)
  [arXiv:0806.1218 [hep-th]].
  %%CITATION = ARXIV:0806.1218;%%
  

%\cite{Bargheer:2012cp}
\bibitem{Bargheer:2012cp} 
  T.~Bargheer, N.~Beisert, F.~Loebbert and T.~McLoughlin,
  ``Conformal Anomaly for Amplitudes in $\mathcal{N}=6$ Superconformal Chern-Simons Theory,''
  J.\ Phys.\ A {\bf 45}, 475402 (2012)
  [arXiv:1204.4406 [hep-th]].
  %%CITATION = ARXIV:1204.4406;%%
  
%\cite{Bianchi:2012cq}
\bibitem{Bianchi:2012cq} 
  M.~S.~Bianchi, M.~Leoni, A.~Mauri, S.~Penati and A.~Santambrogio,
  ``One Loop Amplitudes In ABJM,''
  JHEP {\bf 1207}, 029 (2012)
  [arXiv:1204.4407 [hep-th]].
  %%CITATION = ARXIV:1204.4407;%%
    
\bibitem{ABJMOneL}  
A.~Brandhuber, G.~Travaglini and C.~Wen,
  ``All one-loop amplitudes in N=6 superconformal Chern-Simons theory,''
  JHEP {\bf 1210}, 145 (2012)
  [arXiv:1207.6908 [hep-th]].
  %%CITATION = ARXIV:1207.6908;%%
  
\bibitem{ABJML} 
  M.~Benna, I.~Klebanov, T.~Klose and M.~Smedback,
  ``Superconformal Chern-Simons Theories and AdS(4)/CFT(3) Correspondence,''
  JHEP {\bf 0809}, 072 (2008)
  [arXiv:0806.1519 [hep-th]].
  %%CITATION = ARXIV:0806.1519;%%
  

\bibitem{ABJML2} 
  M.~A.~Bandres, A.~E.~Lipstein and J.~H.~Schwarz,
  ``Studies of the ABJM Theory in a Formulation with Manifest SU(4) R-Symmetry,''
  JHEP {\bf 0809}, 027 (2008)
  [arXiv:0807.0880 [hep-th]].
  %%CITATION = ARXIV:0807.0880;%%
   
 \bibitem{ABJM3} 
   A.~Gustavsson,
  ``Selfdual strings and loop space Nahm equations,''
  JHEP {\bf 0804}, 083 (2008)
  [arXiv:0802.3456 [hep-th]];
  %%CITATION = ARXIV:0802.3456;%% 

  J.~Bagger and N.~Lambert,
  ``Three-Algebras and N=6 Chern-Simons Gauge Theories,''
  Phys.\ Rev.\ D {\bf 79}, 025002 (2009)
  [arXiv:0807.0163 [hep-th]].
  %%CITATION = ARXIV:0807.0163;%%  
  
\bibitem{Gang} 
  D.~Gang, Y.-t.~Huang, E.~Koh, S.~Lee and A.~E.~Lipstein,
  ``Tree-level Recursion Relation and Dual Superconformal Symmetry of the ABJM Theory,''
  JHEP {\bf 1103}, 116 (2011)
  [arXiv:1012.5032 [hep-th]].
  %%CITATION = ARXIV:1012.5032;%%

\bibitem{HuangLipstein} 
  Y.-t.~Huang and A.~E.~Lipstein,
  ``Dual Superconformal Symmetry of N=6 Chern-Simons Theory,''
  JHEP {\bf 1011}, 076 (2010)
  [arXiv:1008.0041 [hep-th]].
  %%CITATION = ARXIV:1008.0041;%%


\bibitem{ABJMTwoL41}  
W.~-M.~Chen and Y.-t.~Huang,
  ``Dualities for Loop Amplitudes of N=6 Chern-Simons Matter Theory,''
  JHEP {\bf 1111}, 057 (2011)
  [arXiv:1107.2710 [hep-th]];
  %%CITATION = ARXIV:1107.2710;%%

\bibitem{Brandhuber:2012un} 
  A.~Brandhuber, G.~Travaglini and C.~Wen,
  ``A note on amplitudes in N=6 superconformal Chern-Simons theory,''
  JHEP {\bf 1207}, 160 (2012)
  [arXiv:1205.6705 [hep-th]].
  %%CITATION = ARXIV:1205.6705;%%

\bibitem{ABJMTwoL42}   
M.~S.~Bianchi, M.~Leoni, A.~Mauri, S.~Penati and A.~Santambrogio,
  ``Scattering Amplitudes/Wilson Loop Duality In ABJM Theory,''
  JHEP {\bf 1201}, 056 (2012)
  [arXiv:1107.3139 [hep-th]].
  %%CITATION = ARXIV:1107.3139;%%

\bibitem{ABJMTwoL6}
 S.~Caron-Huot and Y.-t.~Huang,
  ``The two-loop six-point amplitude in ABJM theory,''
  JHEP {\bf 1303}, 075 (2013)
  [arXiv:1210.4226 [hep-th]].
  %%CITATION = ARXIV:1210.4226;%%



\bibitem{SLee} 
  S.~Lee,
  ``Yangian Invariant Scattering Amplitudes in Supersymmetric Chern-Simons Theory,''
  Phys.\ Rev.\ Lett.\  {\bf 105}, 151603 (2010)
  [arXiv:1007.4772 [hep-th]].
  %%CITATION = ARXIV:1007.4772;%%

\bibitem{LSOG} 
  Y.~-t.~Huang, C.~Wen and D.~Xie,
  ``The Positive orthogonal Grassmannian and loop amplitudes of ABJM,''
  arXiv:1402.1479 [hep-th].
  %%CITATION = ARXIV:1402.1479;%% 


\bibitem{ABJMString} 
  Y.-t.~Huang and S.~Lee,
  ``A new integral formula for supersymmetric scattering amplitudes in three dimensions,''
  Phys.\ Rev.\ Lett.\  {\bf 109}, 191601 (2012)
  [arXiv:1207.4851 [hep-th]].
  %%CITATION = ARXIV:1207.4851;%%




%\cite{Wald:1984rg}
\bibitem{WeinbergGR} 
  S.~Weinberg,
  ``Gravitation and Cosmology: Principles and Applications of the General Theory of Relativity,''
   USA, John Wiley \& Sons (1972).
    
%\cite{Wald:1984rg}
\bibitem{Wald:1984rg} 
  R.~M.~Wald,
  ``General Relativity,''
  Chicago, Usa: Univ. Pr. ( 1984) 491p.    

%\cite{Carroll:2004st}
\bibitem{Carroll:2004st} 
  S.~M.~Carroll,
  ``Spacetime and geometry: An introduction to general relativity,''
  San Francisco, USA: Addison-Wesley (2004) 513p.

\bibitem{unpubl-ef}
H.~Elvang and D.~Z.~Freedman, unpublished notes (2007).    
    
\bibitem{gravityQFT}    
%\cite{DeWitt:1967ub}
%\bibitem{DeWitt:1967ub} 
  B.~S.~DeWitt,
  ``Quantum Theory of Gravity. 2. The Manifestly Covariant Theory,''
  Phys.\ Rev.\  {\bf 162}, 1195 (1967).
  %%CITATION = PHRVA,162,1195;%%

%\cite{DeWitt:1967uc}
%\bibitem{DeWitt:1967uc} 
  B.~S.~DeWitt,
  ``Quantum Theory of Gravity. 3. Applications of the Covariant Theory,''
  Phys.\ Rev.\  {\bf 162}, 1239 (1967).
  %%CITATION = PHRVA,162,1239;%%

%\cite{Veltman:1975vx}
%\bibitem{Veltman:1975vx} 
  M.~J.~G.~Veltman,
  ``Quantum Theory of Gravitation,''
  Conf.\ Proc.\ C {\bf 7507281}, 265 (1975).
  %%CITATION = CONFP,C7507281,265;%%
%\cite{BjerrumBohr:2005jr}

%\cite{Berends:1988zp}
\bibitem{Berends:1988zp} 
  F.~A.~Berends, W.~T.~Giele and H.~Kuijf,
  ``On relations between multi-gluon and multigraviton scattering,''
  Phys.\ Lett.\ B {\bf 211}, 91 (1988).
  %%CITATION = PHLTA,B211,91;%%

%\cite{Bedford:2005yy}
\bibitem{Bedford:2005yy} 
  J.~Bedford, A.~Brandhuber, B.~J.~Spence and G.~Travaglini,
  ``A Recursion relation for gravity amplitudes,''
  Nucl.\ Phys.\ B {\bf 721}, 98 (2005)
  [hep-th/0502146].
  %%CITATION = HEP-TH/0502146;%%

%\cite{Elvang:2007sg}
\bibitem{Elvang:2007sg} 
  H.~Elvang and D.~Z.~Freedman,
  ``Note on graviton MHV amplitudes,''
  JHEP {\bf 0805}, 096 (2008)
  [arXiv:0710.1270 [hep-th]].
  %%CITATION = ARXIV:0710.1270;%%
 
 %\cite{Nguyen:2009jk}
\bibitem{Nguyen:2009jk} 
  D.~Nguyen, M.~Spradlin, A.~Volovich and C.~Wen,
  ``The Tree Formula for MHV Graviton Amplitudes,''
  JHEP {\bf 1007}, 045 (2010)
  [arXiv:0907.2276 [hep-th]].
  %%CITATION = ARXIV:0907.2276;%%

%\cite{Bern:1998sv}
\bibitem{Bern:1998sv} 
  Z.~Bern, L.~J.~Dixon, M.~Perelstein and J.~S.~Rozowsky,
  ``Multileg one loop gravity amplitudes from gauge theory,''
  Nucl.\ Phys.\ B {\bf 546}, 423 (1999)
  [hep-th/9811140].
  %%CITATION = HEP-TH/9811140;%%

%\cite{Bern:1999ji}
\bibitem{Bern:1999ji} 
  Z.~Bern and A.~K.~Grant,
  ``Perturbative gravity from QCD amplitudes,''
  Phys.\ Lett.\ B {\bf 457}, 23 (1999)
  [hep-th/9904026].
  %%CITATION = HEP-TH/9904026;%%    

%\cite{Bern:2000mf}
\bibitem{Bern:2000mf} 
  Z.~Bern, L.~J.~Dixon, D.~C.~Dunbar, A.~K.~Grant, M.~Perelstein and J.~S.~Rozowsky,
  ``On perturbative gravity and gauge theory,''
  Nucl.\ Phys.\ Proc.\ Suppl.\  {\bf 88}, 194 (2000)
  [hep-th/0002078].
  %%CITATION = HEP-TH/0002078;%%

%\cite{Siegel:1993xq}
\bibitem{Siegel:1993xq} 
  W.~Siegel,
  ``Two vierbein formalism for string inspired axionic gravity,''
  Phys.\ Rev.\ D {\bf 47}, 5453 (1993)
  [hep-th/9302036].
  %%CITATION = HEP-TH/9302036;%%

%\cite{Bern:2002kj}
\bibitem{Bern:2002kj} 
  Z.~Bern,
  ``Perturbative quantum gravity and its relation to gauge theory,''
  Living Rev.\ Rel.\  {\bf 5}, 5 (2002)
  [gr-qc/0206071].
  %%CITATION = GR-QC/0206071;%%  
    
%\cite{Freedman:1994pr}
\bibitem{Freedman:1994pr} 
  D.~Z.~Freedman,
  ``Some beautiful equations of mathematical physics,''
  In *ICTP (ed.): The Dirac medals of the ICTP 1993* 25-53, and CERN Geneva - TH.-7367 (94/07,rec.Sep.) 19 p
  [hep-th/9408175].
  %%CITATION = HEP-TH/9408175;%%    

%\cite{Freedman:2012zz}
\bibitem{Freedman:2012zz} 
  D.~Z.~Freedman and A.~Van Proeyen,
  ``Supergravity,''
  Cambridge, UK: Cambridge Univ. Pr. (2012) 607 p

%\cite{Gates:1983nr}
\bibitem{Gates:1983nr} 
  S.~J.~Gates, M.~T.~Grisaru, M.~Rocek and W.~Siegel,
  ``Superspace Or One Thousand and One Lessons in Supersymmetry,''
  Front.\ Phys.\  {\bf 58}, 1 (1983)
  [hep-th/0108200].
  %%CITATION = HEP-TH/0108200;%%

%\cite{deWit:1977fk}
\bibitem{deWit:1977fk} 
  B.~de Wit and D.~Z.~Freedman,
  ``On SO(8) Extended Supergravity,''
  Nucl.\ Phys.\ B {\bf 130}, 105 (1977).
  %%CITATION = NUPHA,B130,105;%%
  
  
%\cite{Cremmer:1978ds}
\bibitem{CremmerJulia} 
  E.~Cremmer and B.~Julia,
  ``The N=8 Supergravity Theory. 1. The Lagrangian,''
  Phys.\ Lett.\ B {\bf 80}, 48 (1978).
  %%CITATION = PHLTA,B80,48;%%
  
%\cite{Cremmer:1979up}
%\bibitem{Cremmer:1979up} 
  E.~Cremmer and B.~Julia,
  ``The SO(8) Supergravity,''
  Nucl.\ Phys.\ B {\bf 159}, 141 (1979).
  %%CITATION = NUPHA,B159,141;%%
  
%\cite{de Wit:1982ig}
\bibitem{deWit:1982ig} 
  B.~de Wit and H.~Nicolai,
  ``N=8 Supergravity,''
  Nucl.\ Phys.\ B {\bf 208}, 323 (1982).
  %%CITATION = NUPHA,B208,323;%%  
  
 
%\cite{Hodges:2012ym}
\bibitem{Hodges:2012ym} 
  A.~Hodges,
  ``A simple formula for gravitational MHV amplitudes,''
  arXiv:1204.1930 [hep-th].
  %%CITATION = ARXIV:1204.1930;%%    
    
%\cite{Cachazo:2012pz}
\bibitem{Cachazo:2012pz} 
  F.~Cachazo, L.~Mason and D.~Skinner,
  ``Gravity in Twistor Space and its Grassmannian Formulation,''
  arXiv:1207.4712 [hep-th].
  %%CITATION = ARXIV:1207.4712;%%  
  
%\cite{He:2012er}
\bibitem{He:2012er} 
  S.~He,
  ``A Link Representation for Gravity Amplitudes,''
  arXiv:1207.4064 [hep-th].
 %%CITATION = ARXIV:1207.4064;%%  
  
%\cite{Cachazo:2012da}
\bibitem{Cachazo:2012da} 
  F.~Cachazo and Y.~Geyer,
  ``A 'Twistor String' Inspired Formula For Tree-Level Scattering Amplitudes in N=8 SUGRA,''
  arXiv:1206.6511 [hep-th].
 %%CITATION = ARXIV:1206.6511;%%  
  
%\cite{Skinner:2013xp}
\bibitem{Skinner:2013xp} 
  D.~Skinner,
  ``Twistor Strings for N=8 Supergravity,''
  arXiv:1301.0868 [hep-th].
  %%CITATION = ARXIV:1301.0868;%% 

%\cite{Cachazo:2013gna}
\bibitem{Cachazo:2013gna} 
  F.~Cachazo, S.~He and E.~Y.~Yuan,
  ``Scattering Equations and KLT Orthogonality,''
  arXiv:1306.6575 [hep-th].
  %%CITATION = ARXIV:1306.6575;%%

%\cite{Cachazo:2013hca}
\bibitem{Cachazo:2013hca} 
  F.~Cachazo, S.~He and E.~Y.~Yuan,
  ``Scattering of Massless Particles in Arbitrary Dimension,''
  arXiv:1307.2199 [hep-th].
  %%CITATION = ARXIV:1307.2199;%%

%\cite{Adler:1964um}
\bibitem{Adler:1964um} 
  S.~L.~Adler,
  ``Consistency conditions on the strong interactions implied by a partially conserved axial vector current,''
  Phys.\ Rev.\  {\bf 137}, B1022 (1965).
  %%CITATION = PHRVA,137,B1022;%%

%\cite{Coleman:1974hr}
\bibitem{Coleman:1974hr} 
  S.~R.~Coleman,
  ``Secret Symmetry: An Introduction to Spontaneous Symmetry Breakdown and Gauge Fields,''
  Subnucl.\ Ser.\  {\bf 11}, 139 (1975).
  %%CITATION = SUSEE,11,139;%%

%\cite{'tHooft:1974bx}
\bibitem{'tHooft:1974bx} 
  G.~'t Hooft and M.~J.~G.~Veltman,
  ``One loop divergencies in the theory of gravitation,''
  Annales Poincare Phys.\ Theor.\ A {\bf 20}, 69 (1974).
  %%CITATION = AHPAA,A20,69;%%


%\cite{Goroff:1985sz}
\bibitem{Goroff:1985sz}
  M.~H.~Goroff and A.~Sagnotti,
  ``Quantum Gravity At Two Loops,''
  Phys.\ Lett.\ B {\bf 160}, 81 (1985).
  %%CITATION = PHLTA,B160,81;%%

%\cite{vandeVen:1991gw}
\bibitem{vandeVen:1991gw} 
  A.~E.~M.~van de Ven,
  ``Two loop quantum gravity,''
  Nucl.\ Phys.\ B {\bf 378}, 309 (1992).
  %%CITATION = NUPHA,B378,309;%%

\bibitem{Deser:1974cz} 
  S.~Deser and P.~van Nieuwenhuizen,
  ``One Loop Divergences of Quantized Einstein-Maxwell Fields,''
  Phys.\ Rev.\ D {\bf 10}, 401 (1974).
  %%CITATION = PHRVA,D10,401;%%


%\cite{Grisaru:1976ua}
\bibitem{Grisaru:1976ua} 
  M.~T.~Grisaru, P.~van Nieuwenhuizen and J.~A.~M.~Vermaseren,
  ``One Loop Renormalizability of Pure Supergravity and of Maxwell-Einstein Theory in Extended Supergravity,''
  Phys.\ Rev.\ Lett.\  {\bf 37}, 1662 (1976).
  %%CITATION = PRLTA,37,1662;%%

%\cite{Grisaru:1976nn}
\bibitem{Grisaru:1976nn} 
  M.~T.~Grisaru,
  ``Two Loop Renormalizability of Supergravity,''
  Phys.\ Lett.\ B {\bf 66}, 75 (1977).
  %%CITATION = PHLTA,B66,75;%%


  
%\cite{Tomboulis:1977wd}
\bibitem{Tomboulis:1977wd} 
  E.~Tomboulis,
  ``On the Two Loop Divergences of Supersymmetric Gravitation,''
  Phys.\ Lett.\ B {\bf 67}, 417 (1977).
  %%CITATION = PHLTA,B67,417;%%

%\cite{Deser:1977nt}
\bibitem{Deser:1977nt} 
  S.~Deser, J.~H.~Kay and K.~S.~Stelle,
  ``Renormalizability Properties of Supergravity,''
  Phys.\ Rev.\ Lett.\  {\bf 38}, 527 (1977).
  %%CITATION = PRLTA,38,527;%%
  
%\cite{Bern:2006kd}
\bibitem{Bern:2006kd} 
  Z.~Bern, L.~J.~Dixon and R.~Roiban,
  ``Is N = 8 supergravity ultraviolet finite?,''
  Phys.\ Lett.\ B {\bf 644}, 265 (2007)
  [hep-th/0611086].
  %%CITATION = HEP-TH/0611086;%%
  
%\cite{Bern:2007hh}
\bibitem{Bern:2007hh} 
  Z.~Bern, J.~J.~Carrasco, L.~J.~Dixon, H.~Johansson, D.~A.~Kosower and R.~Roiban,
  ``Three-Loop Superfiniteness of N=8 Supergravity,''
  Phys.\ Rev.\ Lett.\  {\bf 98}, 161303 (2007)
  [hep-th/0702112].
  %%CITATION = HEP-TH/0702112;%%  
  
%\cite{Bern:2008pv}
\bibitem{Bern:2008pv} 
  Z.~Bern, J.~J.~M.~Carrasco, L.~J.~Dixon, H.~Johansson and R.~Roiban,
  ``Manifest Ultraviolet Behavior for the Three-Loop Four-Point Amplitude of N=8 Supergravity,''
  Phys.\ Rev.\ D {\bf 78}, 105019 (2008)
  [arXiv:0808.4112 [hep-th]].
  %%CITATION = ARXIV:0808.4112;%%  
  
%\cite{Howe:2002ui}
\bibitem{Howe:2002ui} 
  P.~S.~Howe and K.~S.~Stelle,
  ``Supersymmetry counterterms revisited,''
  Phys.\ Lett.\ B {\bf 554}, 190 (2003)
  [hep-th/0211279].
  %%CITATION = HEP-TH/0211279;%%  

%\cite{Bern:2010tq}
\bibitem{Bern:2010tq} 
  Z.~Bern, J.~J.~M.~Carrasco, L.~J.~Dixon, H.~Johansson and R.~Roiban,
  ``The Complete Four-Loop Four-Point Amplitude in N=4 Super-Yang-Mills Theory,''
  Phys.\ Rev.\ D {\bf 82}, 125040 (2010)
  [arXiv:1008.3327 [hep-th]].
  %%CITATION = ARXIV:1008.3327;%%  
    
%\cite{Bjornsson:2010wm}
\bibitem{Bjornsson:2010wm} 
  J.~Bjornsson and M.~B.~Green,
  ``5 loops in 24/5 dimensions,''
  JHEP {\bf 1008}, 132 (2010)
  [arXiv:1004.2692 [hep-th]].
  %%CITATION = ARXIV:1004.2692;%%       
    
%\cite{Elvang:2010jv}
\bibitem{Elvang:2010jv} 
  H.~Elvang, D.~Z.~Freedman and M.~Kiermaier,
  ``A simple approach to counterterms in N=8 supergravity,''
  JHEP {\bf 1011}, 016 (2010)
  [arXiv:1003.5018 [hep-th]].
  %%CITATION = ARXIV:1003.5018;%%    
    
%\cite{Elvang:2010kc}
\bibitem{Elvang:2010kc} 
  H.~Elvang and M.~Kiermaier,
  ``Stringy KLT relations, global symmetries, and $E_{7(7)}$ violation,''
  JHEP {\bf 1010}, 108 (2010)
  [arXiv:1007.4813 [hep-th]].
  %%CITATION = ARXIV:1007.4813;%%    
    
%\cite{Beisert:2010jx}
\bibitem{Beisert:2010jx} 
  N.~Beisert, H.~Elvang, D.~Z.~Freedman, M.~Kiermaier, A.~Morales and S.~Stieberger,
  ``E7(7) constraints on counterterms in N=8 supergravity,''
  Phys.\ Lett.\ B {\bf 694}, 265 (2010)
  [arXiv:1009.1643 [hep-th]].
  %%CITATION = ARXIV:1009.1643;%%
  
%\cite{vanNieuwenhuizen:1976vb}
\bibitem{vanNieuwenhuizen:1976vb} 
  P.~van Nieuwenhuizen and C.~C.~Wu,
  ``On Integral Relations for Invariants Constructed from Three Riemann Tensors and their Applications in Quantum Gravity,''
  J.\ Math.\ Phys.\  {\bf 18}, 182 (1977).
  %%CITATION = JMAPA,18,182;%%    


%\cite{Bossard:2010dq}
\bibitem{Bossard:2010dq} 
  G.~Bossard, C.~Hillmann and H.~Nicolai,
  ``E7(7) symmetry in perturbatively quantised N=8 supergravity,''
  JHEP {\bf 1012}, 052 (2010)
  [arXiv:1007.5472 [hep-th]].
  %%CITATION = ARXIV:1007.5472;%%

%\cite{Freedman:2011uc}
\bibitem{Freedman:2011uc} 
  D.~Z.~Freedman and E.~Tonni,
  ``The $D^{2k} R^4$ Invariants of $N=8$ Supergravity,''
  JHEP {\bf 1104}, 006 (2011)
  [arXiv:1101.1672 [hep-th]].
  %%CITATION = ARXIV:1101.1672;%%

%\cite{Deser:1978br}
\bibitem{Deser:1978br} 
  S.~Deser and J.~H.~Kay,
  ``Three Loop Counterterms For Extended Supergravity,''
  Phys.\ Lett.\ B {\bf 76}, 400 (1978).
  %%CITATION = PHLTA,B76,400;%%    

%\cite{Drummond:2010fp}
\bibitem{Drummond:2010fp} 
  J.~M.~Drummond, P.~J.~Heslop and P.~S.~Howe,
  ``A Note on N=8 counterterms,''
  arXiv:1008.4939 [hep-th].
  %%CITATION = ARXIV:1008.4939;%%

%\cite{Bossard:2011ij}
\bibitem{Bossard:2011ij} 
  G.~Bossard and H.~Nicolai,
  ``Counterterms vs. Dualities,''
  JHEP {\bf 1108}, 074 (2011)
  [arXiv:1105.1273 [hep-th]].
  %%CITATION = ARXIV:1105.1273;%%

\bibitem{alsoE77}
%\cite{Kallosh:2008rr}
%\bibitem{Kallosh:2008rr} 
  R.~Kallosh and T.~Kugo,
  ``The Footprint of E(7(7)) amplitudes of N=8 supergravity,''
  JHEP {\bf 0901}, 072 (2009)
  [arXiv:0811.3414 [hep-th]].
  %%CITATION = ARXIV:0811.3414;%%

%\cite{Kallosh:2011dp}
%\bibitem{Kallosh:2011dp} 
  R.~Kallosh,
  ``$E_{7(7)}$ Symmetry and Finiteness of N=8 Supergravity,''
  JHEP {\bf 1203}, 083 (2012)
  [arXiv:1103.4115 [hep-th]].
  %%CITATION = ARXIV:1103.4115;%%
  
%\cite{Kallosh:2011qt}
%\bibitem{Kallosh:2011qt} 
  R.~Kallosh,
  ``N=8 Counterterms and $E_{7(7)}$ Current Conservation,''
  JHEP {\bf 1106}, 073 (2011)
  [arXiv:1104.5480 [hep-th]].
  %%CITATION = ARXIV:1104.5480;%%

%\cite{Kallosh:2012yy}
%\bibitem{Kallosh:2012yy} 
  R.~Kallosh and T.~Ortin,
  ``New E77 invariants and amplitudes,''
  JHEP {\bf 1209}, 137 (2012)
  [arXiv:1205.4437 [hep-th]].
  %%CITATION = ARXIV:1205.4437;%%

%\cite{Gunaydin:2013pma}
%\bibitem{Gunaydin:2013pma} 
  M.~Gunaydin and R.~Kallosh,
  ``Obstruction to $E_{7(7)}$ Deformation in N=8 Supergravity,''
  arXiv:1303.3540 [hep-th].
  %%CITATION = ARXIV:1303.3540;%%

%\cite{Carrasco:2013qia}
%\bibitem{Carrasco:2013qia} 
  J.~J.~M.~Carrasco and R.~Kallosh,
  ``Hidden Supersymmetry May Imply Duality Invariance,''
  arXiv:1303.5663 [hep-th].
  %%CITATION = ARXIV:1303.5663;%%

%\cite{Stieberger:2007am}
\bibitem{Stieberger:2007am} 
  S.~Stieberger and T.~R.~Taylor,
  ``Complete Six-Gluon Disk Amplitude in Superstring Theory,''
  Nucl.\ Phys.\ B {\bf 801}, 128 (2008)
  [arXiv:0711.4354 [hep-th]].
  %%CITATION = ARXIV:0711.4354;%%

%\cite{Brodel:2009hu}
\bibitem{Brodel:2009hu} 
  J.~Broedel and L.~J.~Dixon,
  ``R**4 counterterm and E(7)(7) symmetry in maximal supergravity,''
  JHEP {\bf 1005}, 003 (2010)
  [arXiv:0911.5704 [hep-th]].
  %%CITATION = ARXIV:0911.5704;%%

\bibitem{StringyCounterterms}

%\cite{Berkovits:2006vc}
%\bibitem{Berkovits:2006vc}
  N.~Berkovits,
  ``New higher-derivative R**4 theorems,''
  Phys.\ Rev.\ Lett.\  {\bf 98}, 211601 (2007)
  [arXiv:hep-th/0609006].
  %%CITATION = PRLTA,98,211601;%%

%\cite{Green:2006gt}
%\bibitem{Green:2006gt}
  M.~B.~Green, J.~G.~Russo and P.~Vanhove,
  ``Non-renormalisation conditions in type II string theory and maximal
  supergravity,''
  JHEP {\bf 0702}, 099 (2007)
  [arXiv:hep-th/0610299].
  %%CITATION = JHEPA,0702,099;%%

  %\cite{Green:2006yu}
%\bibitem{Green:2006yu}
  M.~B.~Green, J.~G.~Russo and P.~Vanhove,
  ``Ultraviolet properties of maximal supergravity,''
  Phys.\ Rev.\ Lett.\  {\bf 98}, 131602 (2007)
  [arXiv:hep-th/0611273].
  %%CITATION = PRLTA,98,131602;%%


%\cite{Green:2008bf}
%\bibitem{Green:2008bf}
  M.~B.~Green, J.~G.~Russo and P.~Vanhove,
  ``Modular properties of two-loop maximal supergravity and connections with
  string theory,''
  JHEP {\bf 0807}, 126 (2008)
  [arXiv:0807.0389 [hep-th]].
  %%CITATION = JHEPA,0807,126;%%


%\cite{Berkovits:2009aw}
%\bibitem{Berkovits:2009aw}
  N.~Berkovits, M.~B.~Green, J.~G.~Russo and P.~Vanhove,
  ``Non-renormalization conditions for four-gluon scattering in supersymmetric
  string and field theory,''
  JHEP {\bf 0911}, 063 (2009)
  [arXiv:0908.1923 [hep-th]].
  %%CITATION = JHEPA,0911,063;%%



%\cite{Vanhove:2010nf}
%\bibitem{Vanhove:2010nf}
  P.~Vanhove,
  ``The critical ultraviolet behaviour of N=8 supergravity amplitudes,''
  arXiv:1004.1392 [hep-th].
  %%CITATION = ARXIV:1004.1392;%%

\bibitem{ck4l}
  Z.~Bern, J.~J.~M.~Carrasco, L.~J.~Dixon, H.~Johansson and R.~Roiban,
  ``Simplifying Multiloop Integrands and Ultraviolet Divergences of Gauge Theory and Gravity Amplitudes,''
  Phys.\ Rev.\ D {\bf 85}, 105014 (2012)
  [arXiv:1201.5366 [hep-th]].
  %%CITATION = ARXIV:1201.5366;%%    

%\cite{Howe:1980th}
\bibitem{Howe:1980th} 
  P.~S.~Howe and U.~Lindstrom,
  ``Higher Order Invariants In Extended Supergravity,''
  Nucl.\ Phys.\ B {\bf 181}, 487 (1981).
  %%CITATION = NUPHA,B181,487;%%

%\cite{Kallosh:1980fi}
\bibitem{Kallosh:1980fi} 
  R.~E.~Kallosh,
  ``Counterterms in extended supergravities,''
  Phys.\ Lett.\ B {\bf 99}, 122 (1981).
  %%CITATION = PHLTA,B99,122;%%
  
\bibitem{VanishingVolume}
G.~Bossard, P.~S.~Howe, K.~S.~Stelle and P.~Vanhove,
``The vanishing volume of D=4 superspace,''
Class.\ Quant.\ Grav.\  {\bf 28}, 215005 (2011)
[arXiv:1105.6087 [hep-th]].
%%CITATION = ARXIV:1105.6087;%%
  
\bibitem{PureSpinor} 
  N.~Berkovits,
  ``Super Poincare covariant quantization of the superstring,''
  JHEP {\bf 0004}, 018 (2000)
  [hep-th/0001035].
  %%CITATION = HEP-TH/0001035;%%

%\cite{Green:2007zzb}
\bibitem{Green:2007zzb} 
  M.~B.~Green, H.~Ooguri and J.~H.~Schwarz,
  ``Nondecoupling of Maximal Supergravity from the Superstring,''
  Phys.\ Rev.\ Lett.\  {\bf 99}, 041601 (2007)
  [arXiv:0704.0777 [hep-th]].
  %%CITATION = ARXIV:0704.0777;%%

%\cite{Banks:2012dp}
\bibitem{Banks:2012dp} 
  T.~Banks,
  ``Arguments Against a Finite N=8 Supergravity,''
  arXiv:1205.5768 [hep-th].
  %%CITATION = ARXIV:1205.5768;%%

%\cite{Bianchi:2009wj}
\bibitem{Bianchi:2009wj} 
  M.~Bianchi, S.~Ferrara and R.~Kallosh,
  ``Perturbative and Non-perturbative N =8 Supergravity,''
  Phys.\ Lett.\ B {\bf 690}, 328 (2010)
  [arXiv:0910.3674 [hep-th]].
  %%CITATION = ARXIV:0910.3674;%%

\bibitem{Bern:2013yya} 
  Z.~Bern, S.~Davies, T.~Dennen, Y.-t.~Huang and J.~Nohle,
  ``Color-Kinematics Duality for Pure Yang-Mills and Gravity at One and Two Loops,''
  arXiv:1303.6605 [hep-th].
  %%CITATION = ARXIV:1303.6605;%%

%\cite{Bern:2013uka}
\bibitem{Bern:2013uka} 
  Z.~Bern, S.~Davies, T.~Dennen, A.~V.~Smirnov and V.~A.~Smirnov,
  ``The Ultraviolet Properties of N=4 Supergravity at Four Loops,''
  Phys.\ Rev.\ Lett.\  {\bf 111}, 231302 (2013)
  [arXiv:1309.2498 [hep-th]].
  %%CITATION = ARXIV:1309.2498;%%

\bibitem{N=42Loop2}    
  Z.~Bern, L.~J.~Dixon, D.~C.~Dunbar, M.~Perelstein and J.~S.~Rozowsky,
  ``On the relationship between Yang-Mills theory and gravity and its implication for ultraviolet divergences,''
  Nucl.\ Phys.\ B {\bf 530}, 401 (1998)
  [hep-th/9802162].
  %%CITATION = HEP-TH/9802162;%%

%\cite{Bern:2009kd}
\bibitem{Bern:2009kd} 
  Z.~Bern, J.~J.~Carrasco, L.~J.~Dixon, H.~Johansson and R.~Roiban,
  ``The Ultraviolet Behavior of N=8 Supergravity at Four Loops,''
  Phys.\ Rev.\ Lett.\  {\bf 103}, 081301 (2009)
  [arXiv:0905.2326 [hep-th]].
  %%CITATION = ARXIV:0905.2326;%%  
  
\bibitem{N4D81L} 
  D.~C.~Dunbar, B.~Julia, D.~Seminara and M.~Trigiante,
  ``Counterterms in type I supergravities,''
  JHEP {\bf 0001}, 046 (2000)
  [hep-th/9911158].
  %%CITATION = HEP-TH/9911158;%%

\bibitem{N=4SG1}
 Z.~Bern, S.~Davies, T.~Dennen and Y.-t.~Huang,
  ``Ultraviolet Cancellations in Half-Maximal Supergravity as a Consequence of the Double-Copy Structure,''
  Phys.\ Rev.\ D {\bf 86}, 105014 (2012)
  [arXiv:1209.2472 [hep-th]].
  %%CITATION = ARXIV:1209.2472;%%  

\bibitem{N=4SG2} 
Z.~Bern, S.~Davies, T.~Dennen and Y.-t.~Huang,
  ``Absence of Three-Loop Four-Point Divergences in N=4 Supergravity,''
  Phys.\ Rev.\ Lett.\  {\bf 108}, 201301 (2012)
  [arXiv:1202.3423 [hep-th]];
  %%CITATION = ARXIV:1202.3423;%%

 \bibitem{Fischler} 
  M.~Fischler,
  ``Finiteness Calculations For O(4) Through O(8) Extended Supergravity And O(4) Supergravity Coupled To Selfdual O(4) Matter,''
  Phys.\ Rev.\ D {\bf 20}, 396 (1979).
  %%CITATION = PHRVA,D20,396;

\bibitem{N4Matter23L} 
  Z.~Bern, S.~Davies and T.~Dennen,
  ``The Ultraviolet Structure of Half-Maximal Supergravity with Matter Multiplets at Two and Three Loops,''
  arXiv:1305.4876 [hep-th].
  %%CITATION = ARXIV:1305.4876;%%

\bibitem{BossardHoweStelle5D}
G.~Bossard, P.~S.~Howe and K.~S.~Stelle,
  ``Invariants and divergences in half-maximal supergravity theories,''
  arXiv:1304.7753 [hep-th].
  %%CITATION = ARXIV:1304.7753;%%


\bibitem{Vaman:2010ez} 
  D.~Vaman and Y.~-P.~Yao,
  ``Constraints and Generalized Gauge Transformations on Tree-Level Gluon and Graviton Amplitudes,''
  JHEP {\bf 1011}, 028 (2010)
  [arXiv:1007.3475 [hep-th]].
  %%CITATION = ARXIV:1007.3475;%%

%\cite{Boels:2012sy}
\bibitem{Boels:2012sy} 
  R.~H.~Boels and R.~S.~Isermann,
  ``On powercounting in perturbative quantum gravity theories through color-kinematic duality,''
  JHEP {\bf 1306}, 017 (2013)
  [arXiv:1212.3473].
  %%CITATION = ARXIV:1212.3473;%%

\bibitem{Bern:2010yg} 
  Z.~Bern, T.~Dennen, Y.-t.~Huang and M.~Kiermaier,
  ``Gravity as the Square of Gauge Theory,''
  Phys.\ Rev.\ D {\bf 82}, 065003 (2010)
  [arXiv:1004.0693 [hep-th]].
  %%CITATION = ARXIV:1004.0693;%%  
  
\bibitem{Tree}
 M. Kiermaier, talk at Amplitudes 2010, May 2010 at QMUL, London, UK. 
 http://www.strings.ph.qmul.ac.uk/$\sim$theory/Amplitudes2010/
 
\bibitem{Treetoo} 
 N.~E.~J.~Bjerrum-Bohr, P.~H.~Damgaard, T.~Sondergaard and P.~Vanhove,
  ``The Momentum Kernel of Gauge and Gravity Theories,''
  JHEP {\bf 1101}, 001 (2011)
  [arXiv:1010.3933 [hep-th]].
  %%CITATION = ARXIV:1010.3933;%%
   
  
 \bibitem{Tree2}
  C.~R.~Mafra, O.~Schlotterer and S.~Stieberger,
  ``Explicit BCJ Numerators from Pure Spinors,''
  JHEP {\bf 1107}, 092 (2011)
  [arXiv:1104.5224 [hep-th]];
  %%CITATION = ARXIV:1104.5224;%%

 C.-H.~Fu, Y.-J.~Du and B.~Feng,
  ``An algebraic approach to BCJ numerators,''
  JHEP {\bf 1303}, 050 (2013)
  [arXiv:1212.6168 [hep-th]];
  %%CITATION = ARXIV:1212.6168;%%  
  
\bibitem{stringtheoryBCJ}
N.~E.~J.~Bjerrum-Bohr, P.~H.~Damgaard and P.~Vanhove,
``Minimal Basis for Gauge Theory Amplitudes,''      
Phys.\ Rev.\ Lett.\  {\bf 103}, 161602 (2009)
[0907.1425 [hep-th]];
%%CITATION = PRLTA,103,161602;%%

S.~Stieberger,
  ``Open \& Closed vs. Pure Open String Disk Amplitudes,''
  arXiv:0907.2211 [hep-th];
  %%CITATION = ARXIV:0907.2211;%%

  C.~R.~Mafra and O.~Schlotterer,
  ``The Structure of n-Point One-Loop Open Superstring Amplitudes,''
  arXiv:1203.6215 [hep-th];
  %%CITATION = ARXIV:1203.6215;%%

O.~Schlotterer and S.~Stieberger,
  ``Motivic Multiple Zeta Values and Superstring Amplitudes,''
  arXiv:1205.1516 [hep-th];
  %%CITATION = ARXIV:1205.1516;%%

J.~Broedel, O.~Schlotterer and S.~Stieberger,
  ``Polylogarithms, Multiple Zeta Values and Superstring Amplitudes,''
  arXiv:1304.7267 [hep-th].
  %%CITATION = ARXIV:1304.7267;%%

\bibitem{StringBCJ}
   S.~H.~Henry Tye and Y.~Zhang,
  ``Dual Identities inside the Gluon and the Graviton Scattering Amplitudes,''
  JHEP {\bf 1006}, 071 (2010)
  [Erratum-ibid.\  {\bf 1104}, 114 (2011)]
  [arXiv:1003.1732 [hep-th]].
  %%CITATION = ARXIV:1003.1732;%%

\bibitem{Feng:2010my} 
  B.~Feng, R.~Huang and Y.~Jia,
 ``Gauge Amplitude Identities by On-shell Recursion Relation in S-matrix Program,''
  Phys.\ Lett.\ B {\bf 695}, 350 (2011)
  [arXiv:1004.3417 [hep-th]].
  %%CITATION = ARXIV:1004.3417;%%      
  
\bibitem{CachazoBCJ} 
  F.~Cachazo,
  ``Fundamental BCJ Relation in N=4 SYM From The Connected Formulation,''
  arXiv:1206.5970 [hep-th].
  %%CITATION = ARXIV:1206.5970;%%

\bibitem{BCJAlgebra} 
  N.~E.~J.~Bjerrum-Bohr, P.~H.~Damgaard, R.~Monteiro and D.~O'Connell,
  ``Algebras for Amplitudes,''
  JHEP {\bf 1206}, 061 (2012)
  [arXiv:1203.0944 [hep-th]].
  %%CITATION = ARXIV:1203.0944;%%



\bibitem{SelfDualBCJ} 
  R.~Monteiro and D.~O'Connell,
  ``The Kinematic Algebra From the Self-Dual Sector,''
  JHEP {\bf 1107}, 007 (2011)
  [arXiv:1105.2565 [hep-th]].
  %%CITATION = ARXIV:1105.2565;%%

 \bibitem{BCJEL}  
   M.~Tolotti and S.~Weinzierl,
  ``Construction of an effective Yang-Mills Lagrangian with manifest BCJ duality,''
  arXiv:1306.2975 [hep-th].
  %%CITATION = ARXIV:1306.2975;%%

\bibitem{BCJLoop}
Z.~Bern, J.~J.~M.~Carrasco and H.~Johansson,
``Perturbative Quantum Gravity as a Double Copy of Gauge Theory,''
Phys.\ Rev.\ Lett.\  {\bf 105}, 061602 (2010)
[arXiv:1004.0476 [hep-th]].
%%CITATION = PRLTA,105,061602;%%

\bibitem{N4Five}
J.~J.~.Carrasco and H.~Johansson,
``Five-Point Amplitudes in N=4 Super-Yang-Mills Theory and N=8 Supergravity,''
Phys.\ Rev.\ D {\bf 85}, 025006 (2012)
[arXiv:1106.4711 [hep-th]].
%%CITATION = ARXIV:1106.4711;%%  

\bibitem{Tristan}
N.~E.~J.~Bjerrum-Bohr, T.~Dennen, R.~Monteiro and D.~O'Connell,
``Integrand Oxidation and One-Loop Colour-Dual Numerators in N=4 Gauge Theory,''
arXiv:1303.2913 [hep-th].
  %%CITATION = ARXIV:1303.2913;%%

%\cite{Boels:2013bi}
\bibitem{OConnellRational} 
  R.~H.~Boels, R.~S.~Isermann, R.~Monteiro and D.~O'Connell,
  ``Colour-Kinematics Duality for One-Loop Rational Amplitudes,''
  JHEP {\bf 1304}, 107 (2013)
  [arXiv:1301.4165 [hep-th]].
  %%CITATION = ARXIV:1301.4165;%%
  
%\cite{Carrasco:2012ca}
\bibitem{OneLoopN1Susy} 
  J.~J.~M.~Carrasco, M.~Chiodaroli, M.~GŸnaydin and R.~Roiban,
  ``One-loop four-point amplitudes in pure and matter-coupled $N \le 4$ supergravity,''
  JHEP {\bf 1303}, 056 (2013)
  [arXiv:1212.1146 [hep-th]].
  %%CITATION = ARXIV:1212.1146;%%  

\bibitem{Chiodaroli:2013upa} 
  M.~Chiodaroli, Q.~Jin and R.~Roiban,
  ``Color/kinematics duality for general abelian orbifolds of N=4 super Yang-Mills theory,''
  JHEP {\bf 1401}, 152 (2014)
  [arXiv:1311.3600 [hep-th], arXiv:1311.3600].

\bibitem{Nohle:2013bfa} 
  J.~Nohle,
  ``Color-Kinematics Duality in One-Loop Four-Gluon Amplitudes with Matter,''
  arXiv:1309.7416 [hep-th].
  %%CITATION = ARXIV:1309.7416;%%

\bibitem{N>=4SG} 
  Z.~Bern, C.~Boucher-Veronneau and H.~Johansson,
  ``$N\ge  4$ Supergravity Amplitudes from Gauge Theory at One Loop,''
  Phys.\ Rev.\ D {\bf 84}, 105035 (2011)
  [arXiv:1107.1935 [hep-th]];
  %%CITATION = ARXIV:1107.1935;%%
  
  C.~Boucher-Veronneau and L.~J.~Dixon,
  ``$N \ge4$ Supergravity Amplitudes from Gauge Theory at Two Loops,''
  JHEP {\bf 1112}, 046 (2011)
  [arXiv:1110.1132 [hep-th]].
  %%CITATION = ARXIV:1110.1132;%%
  
\bibitem{GrisaruSiegel}
M.~T.~Grisaru and W.~Siegel,
``Supergraphity. 2. Manifestly Covariant Rules and Higher Loop Finiteness,''
Nucl.\ Phys.\ B {\bf 201}, 292 (1982)
[Erratum-ibid.\ B {\bf 206}, 496 (1982)].
%%CITATION = NUPHA,B201,292;%%  
  

\bibitem{Ferrara} 
  S.~Ferrara, R.~Kallosh and A.~Van Proeyen,
  ``Conjecture on Hidden Superconformal Symmetry of N=4 Supergravity,''
  Phys.\ Rev.\ D {\bf 87}, 025004 (2013)
  [arXiv:1209.0418 [hep-th]].
  %%CITATION = ARXIV:1209.0418;%%

\bibitem{BroedelDixon} 
  J.~Broedel and L.~J.~Dixon,
  ``Color-kinematics duality and double-copy construction for amplitudes from higher-dimension operators,''
  JHEP {\bf 1210}, 091 (2012)
  [arXiv:1208.0876 [hep-th]].
  %%CITATION = ARXIV:1208.0876;%%

 \bibitem{FormFactors} 
    R.~H.~Boels, B.~A.~Kniehl, O.~V.~Tarasov and G.~Yang,
  ``Color-kinematic Duality for Form Factors,''
  JHEP {\bf 1302}, 063 (2013)
  [arXiv:1211.7028 [hep-th]];
  %%CITATION = ARXIV:1211.7028;%%

\bibitem{Till} 
  T.~Bargheer, S.~He and T.~McLoughlin,
  ``New Relations for Three-Dimensional Supersymmetric Scattering Amplitudes,''
  Phys.\ Rev.\ Lett.\  {\bf 108}, 231601 (2012)
  [arXiv:1203.0562 [hep-th]].
  %%CITATION = ARXIV:1203.0562;%%
  
\bibitem{HenrikYt} 
Y.-t.~Huang and H.~Johansson,
``Equivalent D=3 Supergravity Amplitudes from Double Copies of Three-Algebra and Two-Algebra Gauge Theories,''
arXiv:1210.2255 [hep-th].
  %%CITATION = ARXIV:1210.2255;%%  
  
\bibitem{Dirac} 
P.~A.~M.~Dirac,
``Wave equations in conformal space,''
Annals Math.\  {\bf 37}, 429 (1936).
%%CITATION = ANMAA,37,429;%%

\bibitem{Warren}
W.~Siegel,
``Embedding versus 6D twistors,''
arXiv:1204.5679 [hep-th].
%%CITATION = ARXIV:1204.5679;%%

%\cite{Zee:2003mt}
\bibitem{Zee:2003mt} 
  A.~Zee,
  ``Quantum field theory in a nutshell,''
  Princeton, UK: Princeton Univ. Pr. (2010) 576 p

%\cite{Schwartz:2013pla}
\bibitem{Schwartz:2013pla} 
  M.~D.~Schwartz,
  ``Quantum Field Theory and the Standard Model,''
  %%CITATION = INSPIRE-1276589;%%

\bibitem{HennPlefka}
J.~M.~Henn and J.~C.~Plefka,
``Scattering Amplitudes in Gauge Theories," 
Lecture Notes in Physics 883 (2014), Springer.


%\cite{Wolf:2010av}
\bibitem{Wolf:2010av} 
  M.~Wolf,
  ``A First Course on Twistors, Integrability and Gluon Scattering Amplitudes,''
  J.\ Phys.\ A {\bf 43}, 393001 (2010)
  [arXiv:1001.3871 [hep-th]].
  %%CITATION = ARXIV:1001.3871;%%  


\bibitem{Ellis:2011cr} 
 R.~K.~Ellis, Z.~Kunszt, K.~Melnikov and G.~Zanderighi,
  ``One-loop calculations in quantum field theory: from Feynman diagrams to unitarity cuts,''
  Phys.\ Rept.\  {\bf 518}, 141 (2012)
  [arXiv:1105.4319 [hep-ph]].
  %%CITATION = ARXIV:1105.4319;%%

%\cite{Peskin:2011in}
\bibitem{Peskin:2011in} 
  M.~E.~Peskin,
  ``Simplifying Multi-Jet QCD Computation,''
  arXiv:1101.2414 [hep-ph].
  %%CITATION = ARXIV:1101.2414;%%

%\cite{Donoghue:1995cz}
\bibitem{Donoghue:1995cz} 
  J.~F.~Donoghue,
  ``Introduction to the effective field theory description of gravity,''
  gr-qc/9512024.
  %%CITATION = GR-QC/9512024;%%

%\cite{Dixon:2010gz}
\bibitem{Dixon:2010gz} 
  L.~J.~Dixon,
  ``Ultraviolet Behavior of N = 8 Supergravity,''
  arXiv:1005.2703 [hep-th].
  %%CITATION = ARXIV:1005.2703;%%  

%%%%%%%%


\bibitem{SusyPheno} 
  Z.~Bern, P.~Gondolo and M.~Perelstein,
  ``Neutralino annihilation into two photons,''
  Phys.\ Lett.\ B {\bf 411}, 86 (1997)
  [hep-ph/9706538]
  
  Z.~Bern, A.~De Freitas and L.~J.~Dixon,
  ``Two loop helicity amplitudes for gluon-gluon scattering in QCD and supersymmetric Yang-Mills theory,''
  JHEP {\bf 0203}, 018 (2002)
  [hep-ph/0201161]
  
   Z.~Bern, A.~De Freitas and L.~J.~Dixon,
  ``Two loop helicity amplitudes for quark gluon scattering in QCD and gluino gluon scattering in supersymmetric Yang-Mills theory,''
  JHEP {\bf 0306}, 028 (2003)
  [hep-ph/0304168]
  
\bibitem{Bidder:2005ri} 
  S.~J.~Bidder, N.~E.~J.~Bjerrum-Bohr, D.~C.~Dunbar and W.~B.~Perkins,
  ``One-loop gluon scattering amplitudes in theories with N < 4 supersymmetries,''
  Phys.\ Lett.\ B {\bf 612}, 75 (2005)
  [hep-th/0502028].
  %%CITATION = HEP-TH/0502028;%%
 
\bibitem{Britto:2005ha} 
  R.~Britto, E.~Buchbinder, F.~Cachazo and B.~Feng,
  ``One-loop amplitudes of gluons in SQCD,''
  Phys.\ Rev.\ D {\bf 72}, 065012 (2005)
  [hep-ph/0503132].
  %%CITATION = HEP-PH/0503132;%%
  
  
  \bibitem{Lal:2010qq} 
  S.~Lal and S.~Raju,
 ``Rational Terms in Theories with Matter,''
  JHEP {\bf 1008}, 022 (2010)
  [arXiv:1003.5264 [hep-th]].
  %%CITATION = ARXIV:1003.5264;%%

%\cite{Dittmaier:1998nn}
\bibitem{Dittmaier:1998nn} 
  S.~Dittmaier,
  ``Weyl-van der Waerden formalism for helicity amplitudes of massive particles,''
  Phys.\ Rev.\ D {\bf 59}, 016007 (1998)
  [hep-ph/9805445].
  %%CITATION = HEP-PH/9805445;%%

%\cite{Boels:2008du}
\bibitem{Boels:2008du} 
  R.~Boels and C.~Schwinn,
  ``CSW rules for massive matter legs and glue loops,''
  Nucl.\ Phys.\ Proc.\ Suppl.\  {\bf 183}, 137 (2008)
  [arXiv:0805.4577 [hep-th]].
  %%CITATION = ARXIV:0805.4577;%%

 %\cite{Boels:2010mj}
\bibitem{Boels:2010mj} 
  R.~H.~Boels,
  ``No triangles on the moduli space of maximally supersymmetric gauge theory,''
  JHEP {\bf 1005}, 046 (2010)
  [arXiv:1003.2989 [hep-th]].
  %%CITATION = ARXIV:1003.2989;%% 
  
%\cite{Ferrario:2006np}
\bibitem{Ferrario:2006np} 
  P.~Ferrario, G.~Rodrigo and P.~Talavera,
  ``Compact multigluonic scattering amplitudes with heavy scalars and fermions,''
  Phys.\ Rev.\ Lett.\  {\bf 96}, 182001 (2006)
  [hep-th/0602043].
  %%CITATION = HEP-TH/0602043;%%  
  
%\cite{Forde:2005ue}
\bibitem{Forde:2005ue} 
  D.~Forde and D.~A.~Kosower,
  ``All-multiplicity amplitudes with massive scalars,''
  Phys.\ Rev.\ D {\bf 73}, 065007 (2006)
  [hep-th/0507292].
  %%CITATION = HEP-TH/0507292;%%  
  
%\cite{Rodrigo:2005eu}
\bibitem{Rodrigo:2005eu} 
  G.~Rodrigo,
  ``Multigluonic scattering amplitudes of heavy quarks,''
  JHEP {\bf 0509}, 079 (2005)
  [hep-ph/0508138].
  %%CITATION = HEP-PH/0508138;%%  

\bibitem{Cheung:2010vn} 
  C.~Cheung, D.~O'Connell and B.~Wecht,
  ``BCFW Recursion Relations and String Theory,''
  JHEP {\bf 1009}, 052 (2010)
  [arXiv:1002.4674 [hep-th]].
  %%CITATION = ARXIV:1002.4674;%%  
  
\bibitem{Boels:2010bv} 
  R.~H.~Boels, D.~Marmiroli and N.~A.~Obers,
  ``On-shell Recursion in String Theory,''
  JHEP {\bf 1010}, 034 (2010)
  [arXiv:1002.5029 [hep-th]].
  %%CITATION = ARXIV:1002.5029;%%
 

%\cite{Kampf:2012fn}
\bibitem{Kampf:2012fn} 
  K.~Kampf, J.~Novotny and J.~Trnka,
  ``Recursion Relations for Tree-level Amplitudes in the SU(N) Non-linear Sigma Model,''
  Phys.\ Rev.\ D {\bf 87}, 081701 (2013)
  [arXiv:1212.5224 [hep-th]].
  %%CITATION = ARXIV:1212.5224;%% 


 
% \bibitem{Kampf:2012fn}
%  K.~Kampf, J.~Novotny and J.~Trnka,
%  ``Recursion Relations for Tree-level Amplitudes in the SU(N) Non-linear Sigma Model,''
%  arXiv:1212.5224 [hep-th];\\
%  ``Tree-level Amplitudes in the Nonlinear Sigma Model,''
%  JHEP {\bf 1305}, 032 (2013)
%  [arXiv:1304.3048 [hep-th]].
%  %%CITATION = ARXIV:1212.5224;%% 
  
\bibitem{RationalRecurssion}
Z.~Bern, L.~J.~Dixon and D.~A.~Kosower,
  ``The last of the finite loop amplitudes in QCD,''
  Phys.\ Rev.\ D {\bf 72}, 125003 (2005)
  [hep-ph/0505055].
  %%CITATION = HEP-PH/0505055;%%

\bibitem{FengReview} 
  B.~Feng and M.~Luo,
  ``An Introduction to On-shell Recursion Relations,''
  arXiv:1111.5759 [hep-th].
  %%CITATION = ARXIV:1111.5759;%%



%%%


\bibitem{Drummond:2007aua}
G.~P.~Korchemsky, J.~M.~Drummond and E.~Sokatchev,
  ``Conformal properties of four-gluon planar amplitudes and Wilson loops,''
  Nucl.\ Phys.\ B {\bf 795}, 385 (2008)
  [arXiv:0707.0243 [hep-th]].
  %%CITATION = ARXIV:0707.0243;%%

 \bibitem{Berkovits:2008ic} 
  N.~Berkovits and J.~Maldacena,
  ``Fermionic T-Duality, Dual Superconformal Symmetry, and the Amplitude/Wilson Loop Connection,''
  JHEP {\bf 0809}, 062 (2008)
  [arXiv:0807.3196 [hep-th]].
  %%CITATION = ARXIV:0807.3196;%%

  \bibitem{Beisert:2008iq} 
  N.~Beisert, R.~Ricci, A.~A.~Tseytlin and M.~Wolf,
  ``Dual Superconformal Symmetry from AdS(5) x S**5 Superstring Integrability,''
  Phys.\ Rev.\ D {\bf 78}, 126004 (2008)
  [arXiv:0807.3228 [hep-th]].
  %%CITATION = ARXIV:0807.3228;%%

\bibitem{Drummond:2007aua} 
  J.~M.~Drummond, G.~P.~Korchemsky and E.~Sokatchev,
  ``Conformal properties of four-gluon planar amplitudes and Wilson loops,''
  Nucl.\ Phys.\ B {\bf 795}, 385 (2008)
  [arXiv:0707.0243 [hep-th]].
  %%CITATION = ARXIV:0707.0243;%%

\bibitem{Brandhuber:2007yx} 
  A.~Brandhuber, P.~Heslop and G.~Travaglini,
  ``MHV amplitudes in N=4 super Yang-Mills and Wilson loops,''
  Nucl.\ Phys.\ B {\bf 794}, 231 (2008)
  [arXiv:0707.1153 [hep-th]].
  %%CITATION = ARXIV:0707.1153;%%

  \bibitem{Mason:2010yk}
 L.~J.~Mason and D.~Skinner,
  ``The Complete Planar S-matrix of N=4 SYM as a Wilson Loop in Twistor Space,''
  JHEP {\bf 1012}, 018 (2010)
  [arXiv:1009.2225 [hep-th]].
  %%CITATION = ARXIV:1009.2225;%%

\bibitem{DelDuca:2009au} 
  V.~Del Duca, C.~Duhr and V.~A.~Smirnov,
  ``An Analytic Result for the Two-Loop Hexagon Wilson Loop in N = 4 SYM,''
  JHEP {\bf 1003}, 099 (2010)
  [arXiv:0911.5332 [hep-ph]].
  %%CITATION = ARXIV:0911.5332;%%

\bibitem{Caron-Huot:2010ek}
 S.~Caron-Huot,
  ``Notes on the scattering amplitude / Wilson loop duality,''
  JHEP {\bf 1107}, 058 (2011)
  [arXiv:1010.1167 [hep-th]].
  %%CITATION = ARXIV:1010.1167;%%

\bibitem{Eden:2011}
B.~Eden, P.~Heslop, G.~P.~Korchemsky and E.~Sokatchev,
  ``The super-correlator/super-amplitude duality: Part I,''
  Nucl.\ Phys.\ B {\bf 869}, 329 (2013)
  [arXiv:1103.3714 [hep-th]];\\
  Nucl.\ Phys.\ B {\bf 869}, 378 (2013)
  [arXiv:1103.4353 [hep-th]].
  %%CITATION = ARXIV:1103.3714;%%

\bibitem{Alday:2010zy}
L.~F.~Alday, B.~Eden, G.~P.~Korchemsky, J.~Maldacena and E.~Sokatchev,
  ``From correlation functions to Wilson loops,''
  JHEP {\bf 1109}, 123 (2011)
  [arXiv:1007.3243 [hep-th]].
  %%CITATION = ARXIV:1007.3243;%%
 
 \bibitem{Adamo:2011dq} 
  T.~Adamo, M.~Bullimore, L.~Mason and D.~Skinner,
  ``A Proof of the Supersymmetric Correlation Function / Wilson Loop Correspondence,''
  JHEP {\bf 1108}, 076 (2011)
  [arXiv:1103.4119 [hep-th]].
  %%CITATION = ARXIV:1103.4119;%%
  
\bibitem{WilsonReview1} 
  L.~F.~Alday and R.~Roiban,
  ``Scattering Amplitudes, Wilson Loops and the String/Gauge Theory Correspondence,''
  Phys.\ Rept.\  {\bf 468}, 153 (2008)
  [arXiv:0807.1889 [hep-th]].
  %%CITATION = ARXIV:0807.1889;%%
  
  \bibitem{WilsonReview3} 
  R.~M.~Schabinger,
  ``One-loop N=4 super Yang-Mills scattering amplitudes in d dimensions, relation to open strings and polygonal Wilson loops,''
  J.\ Phys.\ A {\bf 44}, 454007 (2011)
  [arXiv:1104.3873 [hep-th]].
  %%CITATION = ARXIV:1104.3873;%%
  
  
\bibitem{WilsonReview4} 
%\cite{Henn:2009bd}
%\bibitem{Henn:2009bd} 
  J.~M.~Henn,
  ``Duality between Wilson loops and gluon amplitudes,''
  Fortsch.\ Phys.\  {\bf 57}, 729 (2009)
  [arXiv:0903.0522 [hep-th]].
  %%CITATION = ARXIV:0903.0522;%%  

\bibitem{WilsonReview2} 
  T.~Adamo, M.~Bullimore, L.~Mason and D.~Skinner,
  ``Scattering Amplitudes and Wilson Loops in Twistor Space,''
  J.\ Phys.\ A {\bf 44}, 454008 (2011)
  [arXiv:1104.2890 [hep-th]].
  %%CITATION = ARXIV:1104.2890;%%
  
\bibitem{Alday:2010ku} 
  L.~F.~Alday, D.~Gaiotto, J.~Maldacena, A.~Sever and P.~Vieira,
  ``An Operator Product Expansion for Polygonal null Wilson Loops,''
  JHEP {\bf 1104}, 088 (2011)
  [arXiv:1006.2788 [hep-th]].
  %%CITATION = ARXIV:1006.2788;%%

%\cite{Basso:2013vsa}
\bibitem{Basso:2013vsa}
  B.~Basso, A.~Sever and P.~Vieira,
  ``Space-time S-matrix and Flux-tube S-matrix at Finite Coupling,''
  Phys.\ Rev.\ Lett.\  {\bf 111}, 091602 (2013)
  [arXiv:1303.1396 [hep-th]].
  %%CITATION = ARXIV:1303.1396;%%


\bibitem{Basso:2013aha} 
  B.~Basso, A.~Sever and P.~Vieira,
  ``Space-time S-matrix and Flux tube S-matrix II. Extracting and Matching Data,''
  JHEP {\bf 1401}, 008 (2014)
  [arXiv:1306.2058 [hep-th]].
  %%CITATION = ARXIV:1306.2058;%%
  
%\cite{Basso:2014koa}
\bibitem{Basso:2014koa}
  B.~Basso, A.~Sever and P.~Vieira,
  ``Space-time S-matrix and Flux-tube S-matrix III. The two-particle
contributions,''
  arXiv:1402.3307 [hep-th].
  %%CITATION = ARXIV:1402.3307;%% 

%\cite{Dixon:2013eka}
\bibitem{Dixon:2013eka}
  L.~J.~Dixon, J.~M.~Drummond, M.~von Hippel and J.~Pennington,
  ``Hexagon functions and the three-loop remainder function,''
  JHEP {\bf 1312}, 049 (2013)
  [arXiv:1308.2276 [hep-th]].
  %%CITATION = ARXIV:1308.2276;%%

%\cite{Dixon:2014voa}
\bibitem{Dixon:2014voa}
  L.~J.~Dixon, J.~M.~Drummond, C.~Duhr and J.~Pennington,
  ``The four-loop remainder function and multi-Regge behavior at
NNLLA in planar N=4 super-Yang-Mills theory,''
  arXiv:1402.3300 [hep-th].


  
  
 \bibitem{Penrose:1986ca}
R.~Penrose and W.~Rindler, {\em Spinors and Space-Time}, vol.~2.
\newblock Cambridge University Press, 1986.

\bibitem{WardWells}
R.~Ward and R.~Wells, {\em {Twistor Geometry and Field Theory}}.
\newblock CUP, 1990.

\bibitem{HuggettTod}
S.~Huggett and P.~Tod, {\em {An Introduction to Twistor Theory}}.
\newblock Student Texts 4. London Mathematical Society, 1985.

\bibitem{Cachazo:2005ga} 
  F.~Cachazo and P.~Svrcek,
  ``Lectures on twistor strings and perturbative Yang-Mills theory,''
  PoS RTN {\bf 2005}, 004 (2005)
  [hep-th/0504194].
  %%CITATION = HEP-TH/0504194;%%



  

  
  
  
\end{thebibliography}
\end{document}